\documentclass[12pt,a4paper,titlepage,pdftex]{book}
\title{\Huge \scshape TESIS }
\author{\scshape Beatriz Seoane}
\date{ }

\usepackage{tesis}
\usepackage{epstopdf}

\def\dt{\delta_\mathrm{t}}
\def\gmc{\varGamma_\mathrm{g}}
\def\eNliq{e_N^\mathrm{L}}
\def\eNsol{e_N^\mathrm{S}}
\def\bcN{\beta_\mathrm{c}^N}

\newcommand{\bx}{{\boldsymbol x}}
\newcommand{\by}{{\boldsymbol y}}
\newcommand{\bk}{{\boldsymbol k}}
\newcommand{\tw}{t_{\mathrm w}}

\newcommand{\I}[0]{\mathrm{i}}
\newcommand{\E}{\mathrm{e}}
\newcommand{\D}{\mathrm{d}}
\newcommand{\V}[1]{\boldsymbol{#1}}
\newcommand{\RM}[1]{{\rm {#1}}}
\newcommand{\Int}[2]{\int_{#1}^{#2}}
\newcommand{\Tr}{\mathrm{Tr}}

\newcommand{\be}[0]{\begin{equation}}
\newcommand{\ee}[0]{\end{equation}}
\newcommand{\bea}[0]{\begin{eqnarray}}
\newcommand{\eea}[0]{\end{eqnarray}}
\newcommand{\mean}[1]{\left\langle {#1} \right\rangle}
\newcommand{\paren}[1]{\left( {#1} \right)}
\newcommand{\caja}[1]{\left[  {#1} \right]}
\newcommand{\lazo}[1]{\left\{  {#1} \right\}}
\newcommand{\q}[0]{Q_6}
\newcommand{\hq}[0]{\hat Q_6}
\newcommand{\hc}[0]{\hat C}
\newcommand{\p}[0]{p_\mathrm{co}}
\newcommand{\ho}[0]{\hat{o}}

\newcommand{\aO}{\textsf{\itshape O\/}}
\newcommand{\aob}{\textsf{\itshape o\/}}

\newcommand{\grad}[0]{\boldsymbol{\nabla}}

\newcommand{\hsiz}[0]{\hat{\sigma}_i^z}
\newcommand{\hsix}[0]{\hat{\sigma}_i^x}
\newcommand{\siz}[0]{\sigma_i^z}
\newcommand{\six}[0]{\sigma_i^x}
\newcommand{\sump}[0]{\sum_{i=1}^N}

\newindex[myThePage]{default}{idx}{ind}{Alphabetic index}
\makenomenclature

    \DeclareCaptionFont{blue}{\color{blue}} 

  \usepackage{caption}
\DeclareCaptionFont{white}{\color{white}}
\DeclareCaptionFormat{listing}{\colorbox[gray]{0.5}{\parbox{\textwidth}{\hspace{15pt}#1#2#3}}}
\renewcommand{\sectionmark}[1]{\markright{\thesection\ --- #1}}
\fancypagestyle{plain}{%
\fancyhf{} 
\fancyfoot[C]{\oldstylenums\thepage} 

}

\makeatletter
\preto\@tabular{\fontfamily{pplx}\selectfont}
\makeatother

\titleformat{\chapter}[display]
{\bfseries\LARGE} {\filleft\MakeUppercase{\chaptertitlename}
\Huge\Roman{chapter}} {2ex} {\titlerule
\vspace{1.5ex}%
\filright}
[\vspace{1.5ex}%
]

\titleformat{\section}[block]
{\Large\normalfont}
{\bfseries\thesection}{.5em}{\titlerule\\[.8ex]\bfseries}

\begin{document}


\acrodef{SG}[SG]{spin glass}
\acrodef{MF}[MF]{mean field}
\acrodef{MC}[MC]{Monte Carlo}
\acrodef{TL}[TL]{thermodynamic limit}
\acrodef{SK}[SK]{Sherrington-Kirkpatrick}
\acrodef{EA}[EA]{Edwards-Anderson}
\acrodef{RSB}[RSB]{Replica Symmetry Breaking}
\acrodef{TC}[TC]{temperature chaos}
\acrodef{pdf}[pdf]{probability distribution function}
\acrodef{PT}[PT]{Parallel Tempering}
\acrodef{FDT}[FDT]{Fluctuation-Dissipation Theorem}
\acrodef{TNT}[TNT]{trivial-non-trivial}
\acrodef{FSS}{finite size scaling}
\acrodef{FCC}{face centered cubic}
\acrodef{BCC}{body centered cubic}
\acrodef{ECSD}{exponential critical slowing down}
\acrodef{EDSD}{exponential dynamic slowing down}
\acrodef{HS}{hard spheres}
\acrodef{EMCS}{elementary Monte Carlo step}
\acrodef{dof}{degrees of freedom}
\acrodef{PSS}{polydisperse soft spheres}


\thispagestyle{empty}
\vspace*{2cm}

\begin{center}
{\Huge \bfseries
Spin glasses, the quantum annealing, colloidal glasses and
crystals: exploring complex free-energy landscapes\\}
 \vspace*{0.3cm}
{\large \em
Vidrios de esp\'{i}n, computaci\'on cu\'antica adiab\'atica, vidrios y cristales coloidales:
explorando paisajes complejos de energ\'{i}a libre
}

\vspace*{0.8cm}
{\large 
Memoria de tesis doctoral presentada por\\
\textsc{Beatriz Seoane Bartolom\'{e}}
\vspace*{1cm}

Directores\\
\textsc{Luis Antonio Fern\'{a}ndez P\'{e}rez}\\
\textsc{V\'{i}ctor Mart\'{i}n Mayor}
\vspace*{2.5cm}}

\includegraphics[scale=0.5]{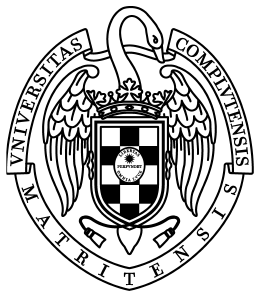}
\vspace*{0.6cm}

Universidad Complutense de Madrid\\
Facultad de Ciencias F\'{i}sicas\\
Departamento de F\'{i}sica Te\'{o}rica I\\
MMXII

\end{center}


\newpage
\thispagestyle{empty}
$ $
\newpage

\thispagestyle{empty}
\vspace*{7cm}
\begin{flushright}
\itshape A mis padres,\vspace*{0.8cm}

and to the memory of Paolo Verrocchio. \\ I will always be thankful for your help,\\ hard work and inspiration.
\\ Addio amico mio.

\end{flushright}
\clearpage

\thispagestyle{empty}
\vspace*{7cm}
{\em
Paolo Verrocchio passed away two days after the defense of this
thesis, when he was only 42 years old. We wish to honor here his memory.\vspace*{0.3cm}

Paolo was one of the most important collaborators for the work
reported in this thesis. He provided both "inspiration" and
"perspiration" to our research. He also hosted in Trento the author of
this thesis (Beatriz), and organized the conference where she gave her
first talk. Even when seriously ill, Paolo managed to contribute to
our joint work. We would have wished to maintain and reinvigorate our
collaboration, pursuing some of the projects started in this
thesis. Alas, this is no longer possible.\vspace*{0.3cm}

With this note, we aim to show our support to his widow, Stefania, and
his children Giovanni and Irene.  In a few years, maybe Irene and
Giovanni will want to know more about his father's work. This thesis
is part of Paolo's legacy.

}

\vspace*{1cm}

\begin{flushright}
Luis Antonio Fern\'andez, V\'ictor Mart\'in Mayor and Beatriz Seoane.
\end{flushright}
\chapter{Acknowledgements}

Con estas líneas pongo punto y final a esta tesis doctoral. Pero
no puedo darla por terminada sin antes agradecer el apoyo recibido durante
estos cuatros años a toda la gente sin la que jamás tendría este texto en mis manos.

En primer lugar, mi más sentido agradecimiento para mis dos directores de
tesis, Luis Antonio Fernández y Víctor Martín. Sin su constante dedicación y
esfuerzo no sé si habría llegado hasta el final. Desde que comenzamos a
trabajar juntos allá por el 2007, siempre he sentido que el éxito de mi 
trabajo ha sido una de sus prioridades. Nunca olvidaré las horas que han
pasado conmigo revisando código, haciendo papeleo o esmerándose en que todo
estuviera listo para la fecha señalada, aún a costa de perder muchas horas de su
vida.  Tengo total convicción de que siempre han hecho todo lo que estaba en
su mano para que pudiera tener la mejor carrera investigadora posible. Por
todo ello, no puedo decir otra cosa que gracias.  Dentro de mi grupo de
investigación, también me gustaría agradecer a David Yllanes, mi antiguo
compañero de doctorado, toda la ayuda prestada. Me gustaría hacer una
especial mención al apoyo de estos últimos meses, ya que sin su ayuda técnica la
escritura de la tesis habría sido bastante más complicada. 

Durante estos años he sido miembro del Departamento de Física Teórica I de la
Universidad Complutense de Madrid, donde siempre me he sentido particularmente
a gusto e integrada. Dentro del departamento, siento un especial
agrade\-cimiento a su director Antonio Muñoz por todas las horas que hemos
compartido. También me gustaría mencionar a su inolvidable secretaria, Chon, a
quien debo multitud de favores, y a David, el informático, por su
imprescindible ayuda técnica. También quiero agradecer a Víctor Martín, Ramón
Álvarez, Diego Porras y Antonio Muñoz la oportunidad de
introducirme en la docencia como profesora ayudante de las asignaturas de
Fenómenos Colectivos, Física para biólogos, Mecánica Cuántica y Física Cuántica.

También quiero aprovechar para agradecer a la JANUS collaboration el esfuerzo
y trabajo conjunto. Además, me gustaría agradecer su predisposición a dejarme
reutilizar las configuraciones de~\cite{janus:10,janus:10b} para el estudio de
caos en temperatura. Dentro de la colaboración, me gustaría dar las gracias a
Juan Jesús Ruíz, por el curso intensivo de vidrios de espín que se molestó en
preparar hace unos años para los jóvenes miembros de la colaboración. Tampoco
quiero olvidar a Sergio Pérez o a Enzo Marinari por su enorme ayuda durante mi
período en Roma. Y por supuesto, no puedo dejar de mencionar a Giorgio Parisi,
con quien comparto varios trabajos y del que he aprendido muchísimo.

Por otro lado, quiero agradecer al Instituto de Biocomputación y Física de los
Sistemas Complejos (BIFI) de la Universidad de Zaragoza, institución de la que
soy miembro, la gran cantidad de recursos computacionales puestos a mi
disposición durante estos años. El mismo agradecimiento se lo tengo que mandar
a la red Española de Supercomputación por haberme permitido utilizar el
ordenador {\em Mare Nostrum}.

En especial, quiero agradecer a Paolo Verrocchio todo su
cariño y esfuerzo. Comenzamos a trabajar juntos casi al final de mi
licenciatura y desde entonces hemos compartido varios trabajos y estancias.
Me da mucha pena pensar en las dificultades que han surgido en los últimos
tiempos, y que no vaya a poder estar presente el día que presente mi
tesis. Quiero aprovechar estas páginas para mandarle un fuerte abrazo y mucho
apoyo.  También se merece una especial mención Hidetoshi Nishimori, quién me
aceptó sin conocerme unos meses en su facultad en Tokio y me hizo sentir como
en casa, de quién he aprendido mucho y a quién debo gran parte de la ilusión
por continuar ahora con este gran proyecto de la investigación.

No puedo olvidar tampoco mis inicios en investigación con Miguel Ángel
Rodríguez durante la licenciatura, ya que a él le debo mi determinación a
comenzar el doctorado. De la misma forma tengo mucho que agradecer a Jacobo
Ruiz de Elvira, porque fue con él con quién comencé mi sueño de dedicarme a la
física y de continuar después con la investigación, y sin él, no sé muy bien
dónde estaría ahora, pero estoy segura de que no sería aquí.

Quiero dedicar los párrafos finales de estos agradecimientos a la parte más
personal, a toda la gente que ha estado a mi lado durante estos años a las
buenas y a las malas. No considero que el doctorado haya sido una etapa fácil
de mi vida, ha estado tan llena de ilusión como de desencanto. Lo que sí que
puedo decir es que pese a todo, han sido probablemente los años más felices de
mi vida, algo que lo debo a mucha gente. Los nueve años y medio que he vivido
en esta facultad no habrían sido lo mismo sin la existencia de Hypatia, del
Club Deportivo o del equipo de fútbol femenino de físicas, sin ellos la
facultad habría sido un lugar mucho más triste. Tampoco habría sido lo mismo
sin todos mis compañeros de doctorado: Jenifer, Ricardo, Jacobo, Giovanni,
Guillermo, Édgar, Joserra, Jose, Alejandro, Lourdes, Álvaro, Jose Alberto,
Davide, Nikos, Markus, Alexandre, Óscar, Diego, etc. Entre ellos, les debo una
especial mención a mis compañeros del DomiLab: Santos, Juantxo, David... y
sobretodo a Domingo, Vivy y Marco, porque a ellos les debo los mejores
momentos de esta tesis, y porque nuestra relación va mucho más allá de meros
compañeros de despacho. Acabo esta etapa con la seguridad de que no encontraré
nunca mejores compañeros de viaje.

Siguiendo con los agradecimientos, no puedo olvidar a mis dos incondicionales:
José Ramón Vázquez y Javier Andrade, porque no sólo habéis compartido conmigo
incontables horas de trabajo y ocio, sino que también habéis aguantado todas
mis penas, quejas y dudas, y porque sin vosotros dos, nada habría sido lo
mismo. También me gustaría agradecer a Jesús Díaz su compañía durante las
largas horas de escritura de esta tesis, así como su interés a la hora de leer
y comentar críticamente este texto.  Y bueno, antes hablaba de los años más
felices de mi vida, y ésto se lo debo sin duda a mis amigos, siempre
dispuestos a sonreír ante cualquier mal, o a celebrar todo lo incelebrable. Sé
que me dejo a muchos importantes, pero quiero mandar un fuerte abrazo a
Izarra, Javi Campos, Amalia, Rubio, Jorge, Alvarito, Marta, Carla, Diego,
Punky, Berto, Pitufa, Itxi, Jerbo, Iria, Lon, Champi, Agus, Hugo, Illo, Elena,
Eze y Óscar. Tampoco quiero olvidar mandar un cariñoso agradecimiento al grupo
Scout Kimball 110, ya que, durante estos últimos años, ha aportado a mi vida
una parte solidaria, alegre y completamente diferente a todo lo demás.

Y dejo para el final a los más importantes. El agradecimiento más grande es
para mi familia. Entre ellos, a mi hermano, uno de los mejores amigos que
tengo y tendré. Y por supuesto, a mis padres.
 Sin su exigencia desde pequeña, sin su cariño, sin su ayuda, y sin su
incondicional apoyo a cualquier empresa, no estaría donde estoy ni sería la
persona la que soy. A ellos va dedicada esta tesis.

Durante esta tesis he estado financiada primero por una beca del BIFI, luego
por una beca de la Caixa, que me dio la oportunidad de cursar el máster en
Física Fundamental, y finalmente, por una beca FPU del Ministerio de
Educación, Cultura y Deporte. Por otro lado he recibido apoyo de los proyectos 
 FIS2009-12648-C03 del MICINN y de los Grupos UCM - Banco Santander.

\begin{flushright}
\textsc{Beatriz Seoane Bartolomé}\\
\itshape
Universidad Complutense, Madrid, noviembre de 2012
\end{flushright}


\clearpage
\tableofcontents
\phantomsection





\chapter{General introduction}\label{chap:intro} \index{complexity}

Traditionally, the step-forwards in physics are obtained with the ``divide and
conquer'' strategy. In other words, one normally splits up the system
in small parts and tries to infer the behavior of the whole by understanding
the parts. But what if the system is that interacting or that complex that
there is no way  to understand the overall problem by the knowledge of
the individuals?  What if the whole is a lot more than the summation of the
parts or something completely different? Many systems in nature can only be
studied from a collective point of view, this is the case of a variety of
systems such as, for instance, earthquakes, neural networks, protein folding,
turbulence, glasses...

This research field has suffered a major boost in the last decades with the
upcoming and development of computers. Indeed, computers have allowed
scientists to simulate large systems under complicate interactions or with
induced disorder, and to study their emerging properties. Furthermore, thanks
to the computing improvements, now it is possible to collect and analyze
unprecedentedly large amounts of data coming from both from experiments or
simulations. Because of that, complex systems have become a whole field by
itself, but an interdisciplinary field shared by physicists, biologists,
mathematicians, etc.

The most successful theory to approach the equilibrium state of a system
composed of a large amount of particles is the statistical mechanics. In this
theory, it is assumed that, even though each of one components of the system
describes a chaotic behavior, the resulting macroscopic equilibrium state is
extremely simple if the system is big enough. Somehow the individual chaotic
behavior cancels out when the equilibrium is achieved. But what happens if the
system evolves as slowly as the relevant state in nature is out of
equilibrium?  Then, the traditional notions of thermodynamics do not hold and
new surprising phenomena emerge. The traditional control parameters, such as
the temperature or the pressure do not longer describe the system by
themselves. In fact, one needs to track not only the time elapsed in
experiments but also the age of the system in these complex phases.  Then, the
evolution of the system depends on their whole history, which results in a
very striking behavior: event an inert material chunk (such as a spin glass)
ages, rejuvenizes or has memory.  Besides, these systems react drastically to
slight changes in the external conditions, which is known as chaos (see
Chapter~\ref{chap:chaos} in this thesis).

Up to this point, this kind of materials are very discouraging.  However not
everything is bad news. In fact, when one studies the collective properties of
various of these complex systems, one realizes that they exhibit somehow a
kind of new universal behavior. Indeed, the same prescriptions seem to work
for completely different systems, no matter the properties of the individuals
that compose them.

One example of these complex materials are everyday glasses. Macroscopically
they behave as solids, but microscopically, they look very much like a
fluid. Actually, they present no long range order, but the particles are so
packed that the flow is impeded. In fact, glasses are often obtained by
cooling fluids very quickly. Normally these fluids  would become
crystalline if they were frozen slowly enough. Even nowadays, after thousand
of years manipulating glasses, the nature of the glassy phase is not
understood. In fact, to determine whether the glass transition is a real phase
transition or not, is one of the most important open questions in solid-state
physics.  The glassy phase is characterized by a extremely high
viscosity. When the temperature is lowered down the viscosity grows
dramatically with the temperature, and then, the particles have no room to
move, which results in diverging characteristic flow times. In fact, the
system evolves so slowly near the transition point (defined purely
dynamically, as the temperature at which the viscosity reaches $10^{13}$
poises) that one must consider it to be always out of equilibrium.

However, this extremely slow evolution of the dynamic variables associated
with disorder is not peculiar to the particle positions in structural
glasses. In fact, there are some magnetic allows (known as spin glasses, see
Part \ref{part:sg} in this dissertation) that present a similar frozen phase
in their magnetic moments. Indeed, the spin glass phase has a
vanishing total magnetization (in absence of magnetic field) but at variance
with the paramagnetic phase, each spin in the lattice is frozen in time but in
seemingly random spatial pattern.  This spin glass phase and the ordinary
glass phase share many not understood phenomena, even though their nature is
completely different.  Indeed, in spin glasses the interaction between
particles is magnetic, and particle diffusion does not play any role.

Spin glasses, at least up to now, are useless materials. However, they still
carry the fundamental origin of the glass phase. It is hoped that the
theoretical treatment will be simpler in spin glasses. Indeed, among other
simplifications, particles can be placed in lattice nodes (since no diffusion
is involved), which encourages notably the analytical and numerical
computations.  For this reason, even though structural glasses would be more
interesting for practical applications, spin glasses are nowadays the usual
benchmark to investigate complex behavior, and most of our intuition about
glasses comes precisely from spin glass studies.

This thesis is centered on the numerical study of complex systems. As
discussed above, although their fauna is broad, the inner mechanism causing
their striking effects, as well as the tools we use to study all of them, are
very well interchangeable from one system to the other. For this reason, in
this thesis we worked both with spin glasses (Part~\ref{part:sg} of this
dissertation) and colloidal systems (Part~\ref{part:colloids}) as two faces of
the same coin.  When concerning equilibrium in a computer simulation of these
kind of systems, the problem is definitely time (or computer resources). As
mentioned above, nearby the glass transition, the inner system' dynamics gets
stuck in these kind of materials. Indeed, from an experimental point of view,
they are permanently out of equilibrium. This freezing in the evolution is
also observed in the simulations, which is translated in a divergence on the
exponential autocorrelation times that makes equilibrium unreachable in human
times for relatively small systems.  From the point of view of experiments,
the relevant state is out of equilibrium. However, from the theoretical point
of view, the limited theories available for these materials correspond to the
equilibrium state. This is were the computer simulations come to play. With
simulations, we are able to investigate non-perturbatively the equilibrium phase
on a system, but also can explore its nonequilibrium behavior (which is
relevant to analyze experiments).

In addition to the slow behavior associated to glasses, the numerical study of
any phase transition is always hard. Indeed, any phase transition introduces a
divergence in the simulation times with the number of particles $N$ (the
physical mechanism is related to the growth of one phase into the other). This
limits strongly the system sizes that can be equilibrated in a simulation. The
problems we are considering here combine both kind of problems, an extremely
slow dynamics induced by disorder, and the presence of phase transitions,
which makes these problems intrinsically hard in the computer science language
(see Chapter~\ref{chap:QA}). For this reason, the research on optimized
algorithms or the construction of dedicated supercomputers for these problems
is also mandatory in the field. Indeed, although the final research goal is
physics, computer and algorithms are important. In fact, no progress is
possible we are not able to approach the equilibrium state or to simulate systems
big enough to display the desired phenomena.

Roughly a half of this thesis is devoted to fast computation
strategies and new algorithms. This emphasis is less strong in
Part~\ref{part:sg} devoted to spin glasses, where the Monte Carlo algorithms
used are rather standard and the progress relies on either the implementation
of multi-spin coding (see Appendix~\ref{app:multispin}) or thanks to the JANUS
supercomputer. In Part~\ref{part:colloids} we deal with colloidal polydisperse
systems, systems that combine both a structural glass transition and a first
order solidification transition in a narrow region in the space of
parameters. First order transitions come together with an divergence
(exponential in the number of particles $N$) of computational times within the
normal approaches. For this reason, our main goal has been to beat this
divergence. Continuing in the strategy of searching new optimized algorithms
to approach glasses, I moved to quantum mechanics (see Part~\ref{part:QA})
during the last months of my Ph.D. In particular, I started to work with
the new and promising quantum annealing (also known as adiabatic computation)
algorithms.

Most of the results collected in this thesis have appeared in international
journals and were presented in international conferences. We take the chance
to summarize here all of them. We start with Part~\ref{part:sg}, the part
devoted to spin glasses.  The general introduction in
Chapter~\ref{chap:intro-sg} relays heavily on Refs.~\cite{janus:10,janus:10b}
(by the Janus collaboration to which I belong). However, no original results
are presented in Chapter~\ref{chap:intro-sg}.  Chapter~\ref{chap:hypercube} is
mainly based on \cite{fernandez:09f}. I had the chance to expound this work in
an oral presentation in the most important conference in the field, the {\em
  STATPHYS 24, the XXIV International Conference on Statistical Physics} that
took place in Cairns (Australia) in July 2010. In addition, I also presented a
talk about it in the {\em CompPhys09, 10th International NTZ-Workshop on New
  Developments in Computational Physics} in Leipzig (Germany) in November
2009. Chapter~\ref{chap:chaos} is based on~\cite{fernandez:12b} (currently
under review).
Part~\ref{part:colloids} concerns to colloidal
systems. Chapter~\ref{chap:micro} is based on~\cite{fernandez:09e}. I
presented talks about this work both in {\em the XII International Workshop on
  Complex Systems} in Andalo (Italy) in March 2010 and {\em the International
  Workshop on Complex Energy Landscapes} in Zaragoza (Spain) in June 2010.  On
the other hand, Chapter~\ref{chap:HS} is based
on~\cite{fernandez:12,martin-mayor:11}. I gave a talk about this work in a
satellite meeting to STATPHYS 24, {\em Monte Carlo Algorithms in Statistical
  Physics} in Melbourne (Australia) in July 2010, and in a poster session in
{\em Viscous Liquids III} in Rome (Italy) in March 2011.
Finally, Chapter~\ref{chap:QA} is based on~\cite{seoane:12b}. I presented a
poster on this subject in the conference {\em Quantum Information meets
  Statistical Mechanics} in Innsbruck (Austria) in September 2012.

It is also important to acknowledge that this work has been supported by MECCD
(Spain) through the FPU program, and by MICINN (Spain) through research
contracts No. FIS2009-12648-C03.

\part{Spin Glasses\label{part:sg}}

\chapter{General description of spin glasses}\label{chap:intro-sg}\index{spin
  glass|(} 

It was late in the 1960s when the first unusual effects on spin glasses were
detected in  experiments. These effects appeared in the now known as  canonical
spin glasses; the traditional and simple magnetic alloys composed by the
mixture of noble-metals and transition metals (such as Au-Fe or Cu-Mn). Indeed,
by that time, researchers were wondering what would happen after introducing
magnetic impurities into a non-magnetic matrix. In such a mixture, the
magnetic moments coming from the impurities would be dissolved on a sea of
conducting electrons, and the direct question was: does the magnetism remain?

The experiments were shocking. The remanent magnetization roughly disappeared
at low temperatures, but at the same time the susceptibility presented a broad
maximum. Besides, the magnetization and its hysteresis were completely
different to what expected for a ferromagnetic phase. Rather they were more
similar to the result for a mixture of mutually interacting ferromagnetic and
antiferromagnetic domains.  On the other hand, experiments pointed out some
kind of magnetic random order at low temperature, different from everything
known up to that moment.

The name {\em spin glass} did not appear until 1970s, and it was coined when
linking the problem of localization in disordered systems with the magnetic
alloy problem.

It was not until around 1975 when theorists became interested in the problem,
and when the spin glass boom really started. Since then, a lot of simplified
models and theories have been presented, leading to a great progress in the
understanding of these ``weird'' materials. But not only, this knowledge
supposed also a break-through in the field of disordered systems and
statistical mechanics. Nowadays, results or techniques obtained in spin
glasses are applied to many different fields, from biology to computer
science.  However, even after $50$ years of intensive spin glass study, many
of their most striking properties remain to be explained, and the debate about
its equilibrium low temperature phase still remains open.

In this chapter, I will discuss what a spin glass is and will try to outline
some outstanding results, including experimental and theoretical
work. With this aim, I will summarize the most important experiments in spin
glasses and discuss the most popular simplified theoretical models as well as
the main competing theories for the equilibrium spin glass phase.
 I will end
the chapter by an introduction to the numerical techniques in spin glasses,
defining the observables that will be used in the following chapters. Finally,
I will discuss one of the main progresses recently achieved by means of
numerical simulations, the time-length dictionary that finds a quantitative
relation between the worlds of equilibrium (where theory is developed) and the
nonequilibrium (the one relevant to experiments).

I would like to stress that the results summarized in this introductory
chapter are not original. They are based on the general spin glass literature,
mainly on~\cite{mydosh:93,young:97,mezard:87,vincent:96}. The
review of numerical simulations, I also include some recent results taken
from~\cite{janus:10,janus:10b}. 

\section{What is a spin glass?}\label{sec:whatsg}

The \ac{SG} is a new state of magnetism, completely different from the
traditional ordered ferromagnetic and anti-ferromagnetic phases, but still
with a co-operative and collective nature in the low temperature phase. This
phase is characterized by the following properties. Below some critical
temperature, $T_\mathrm{c}$, the spins are frozen in time. That means that, below
$T_\mathrm{c}$, the local magnetization at each site of the lattice $\V{x}$ is
$\mean{\V{s}_{\V{x}}}_t\neq 0$, being $\mean{\cdots}_t$ the average over the
experimental time. However, though frozen, the orientation of the spins seems to be completely random,
leading to a vanishing total magnetization when summing up over all the
system,
\begin{equation}\label{eq:SG-total-M}
\boldsymbol M= \frac{1}{N} \sum_\bx  \braket{\boldsymbol S_\bx}_t = 0.
\index{long-range order}
\end{equation}
The condition is even stronger. Indeed, there is no long range order of any
kind, i.e.
\begin{equation}\label{eq:SG-long-range}
\boldsymbol M_\bk = \frac{1}{N} \sum_\bx \E^{-\I \bk\cdot \boldsymbol x} \braket{\boldsymbol S_\bx}_t = 0,\qquad \forall \bk.
\index{long-range order}
\end{equation}
This last expression includes both the ferromagnetic, $\V{k}=(0,0,0)$ and
antiferromagnetic, $\V{k}=(\pi,\pi,\pi)$, order parameters.\footnote{Indeed, when the
interaction is ferromagnetic, all the spins tend to align in the same
direction, whereas in the antiferromagnetic case, nearby spins point to
alternate opposite directions to minimize the energy.}

It is interesting to point out the difference between this {\em frozen}
disordered phase, and a normal disordered phase, like a paramagnetic phase. In
this latter, there is also a total absence of long range order but spins
fluctuate randomly due to thermal excitations, leading to a vanishing local
magnetization when one averages over a long time.

Note that this new magnetic phase resembles a normal glass very much. Indeed,
in these materials, the particle positions are apparently random but do not
evolve with time (structural glasses are characterized by a extreme slow
flow). In fact, as mentioned above, the term {\em spin glass} comes precisely
from this similarity between the frozen random orientation of spins and the
frozen location of particles in ordinary glasses.

Nowadays we know that the existence of a glassy phase in spin glasses occurs
as a consequence of a combination of three basic ingredients: randomness,
mixed interactions and frustration. Let us explain briefly each term. The
disorder or randomness in the interactions is introduced in spin glasses by
randomizing either the distance between the magnetic moments, namely {\em site
  randomness}, or the nearest neighbors interaction in a regular lattice,
known as {\em bond randomness}. In addition, these interactions must be, not
only random in strength, but also of mixed ferromagnetic and antiferromagnetic
nature (for a pair of spins the interaction can be either ferromagnetic, which
favors a parallel orientation of both, or antiferromagnetic, which results in
an antiparallel layout). The combination of randomness and competing
interactions causes {\em frustration}.

\begin{figure}
\centering
\includegraphics[width=0.7\linewidth]{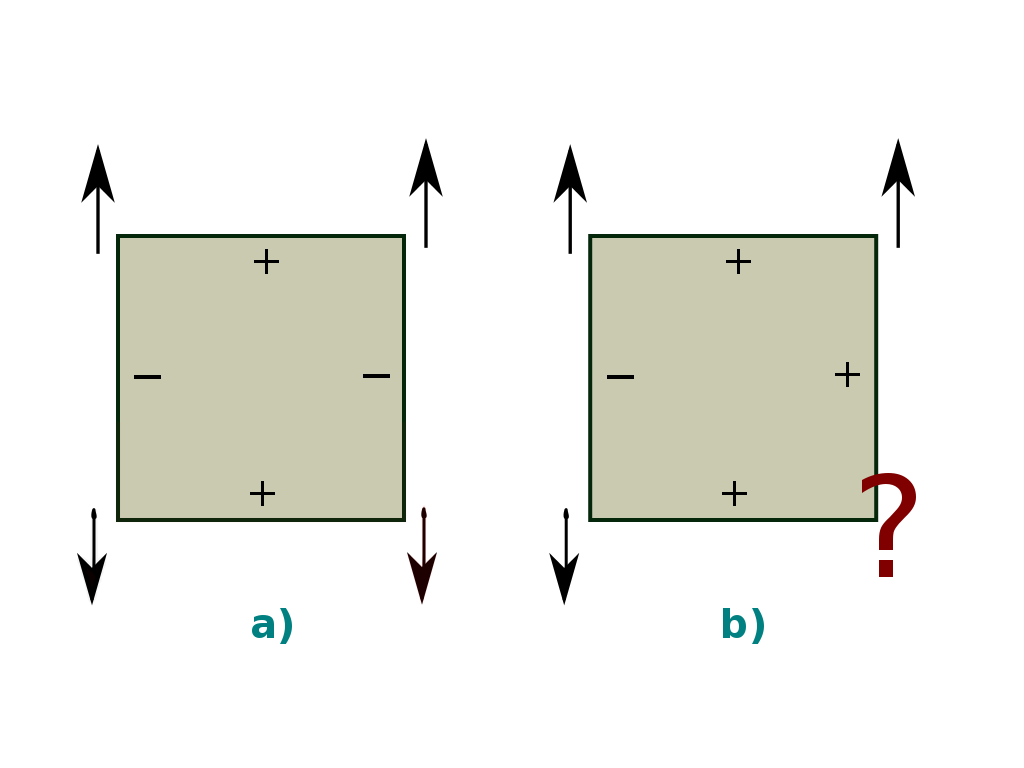}
\caption[Frustrated plaquette]{Examples of {\bf (a)} unfrustrated and
  {\bf (b)} frustrated  plaquettes.
\index{frustration}
\label{fig:SG-plaquette}}
\end{figure}

The idea of frustration is exemplified in Figure \ref{fig:SG-plaquette}. Let
us consider four spins each lying on the four vertices of a plaquette. Each
spin is connected with only two neighbors and the nature of the interaction is
represented through the signs in the edges. When the coupling between two
spins is positive (ferromagnetic), the spins minimize their energy by aligning
parallel. On the contrary, if the interaction is negative
(antiferromagnetic), they ``want'' to align anti-parallel. In Figure
\ref{fig:SG-plaquette}-left, there is no frustration, all the spins can
minimize their energy at the same time. On the other hand, in a frustrated
plaquette, such as the one shown in Figure \ref{fig:SG-plaquette}-right, this
is not possible. Indeed, let us follow the following procedure. One chooses
randomly an orientation for the spin-$1$ placed in the upper right corner. The
election is, for instance, spin up ($\uparrow$). Now, since the interaction is
ferromagnetic, its nearest left neighbor (spin-$2$) will align parallel with
spin-$1$, that is, also up ($\uparrow$). Afterwards, we consider the spin in
the left bottom corner (spin-$3$). The interaction with spin-$2$ is
antiferromagnetic, so it will orientate down ($\downarrow$) in order to
satisfy the coupling. Finally, let us consider the spin in the bottom right
corner, where the question mark is. The decision problem appears when spin-$4$
has to decide its orientation: if considers the coupling with spin-$3$, it
should be down ($\downarrow$) (parallel to spin-$3$), but if considers the
interaction with spin-$1$ should point up ($\uparrow$) (parallel to
spin-$1$). Then, spin-$4$ cannot satisfy simultaneously both couplings. This
absence of a ``everybody happy'' configuration, is precisely what the term
frustration refers to.

The presence of frustration draws a rugged free-energy valley, with many
minima and large barriers. Each of these minima corresponds to a frozen state
where the system hardly evolves due to the constant competition between the
interactions. This ``confused'' ground state is the origin of the interesting
and unique properties of \ac{SG}. However, frustration is not the only
necessary condition for a \ac{SG}; it must be combined with the randomness and
the competition between interactions discussed above. In fact, the
antiferromagnetic, regular triangular lattice is a fully frustrated system,
but has no co-operative freezing. In fact, frustration is a direct consequence
of the disorder and mixed interactions, but while a necessary condition to
induce a spin glass phase, it is not a sufficient one.

\section{Real spin glasses}\label{sec:realsg}

Now the question is which kind of materials develop a \ac{SG} phase.  As
discussed in the introduction, the first \ac{SG} were found accidentally in
binary allows. In these materials, the magnetic impurities (bearing
magnetic moments or localized spins) occupy random sites in a non-magnetic
host metal. The concentration of these magnetic impurities, $x$, can be
controlled during the manufacture.  The archetypal specimens of the metallic
(site random) spin glass are Cu$_{1-x}$Mn$_x$ or Au$_{1-x}$Fe$_x$. These
noble-metal alloys are known as canonical spin glasses. Indeed, the
dissolution of the magnetic solute in the non-magnetic solvent occurs
completely randomly, with no particular atomic or chemical short-range
order. Then, the system can be treated statistically and modeled using
Gaussian probabilities.  However, more complicated alloys can be manufactured
as well.  For instance, it is possible to have \ac{SG} which are both
insulating and conducting.  In these materials, one of the non-magnetic
sub-lattices is substituted by a magnetic one. As an example we can cite a
semiconductor, such as Eu$_{x}$Sr$_{1-x}$S or a metal
La$_{1-x}$Gd$_{x}$Al$_2$.

Another way of creating site disorder, is to start with an intermetallic
compound, e.g. GdAl$_2$ and to destroy its crystalline form by making it
amorphous. This can be done with many different techniques such as
splat-cooling, quench-condensation or sputtering.

However, as discussed, the randomness in the interactions is not only created
through a random distribution of sites, it can be also synthesized in a regular
lattice by randomizing the sign of the couplings. In fact,
Rb$_{2}$Cu$_{1-x}$Co$_x$F$_4$ and Fe$_{1-x}$Mn$_{x}$TiO$_3$ can be modeled up
to very good approximation on a perfect lattice with only $\pm J$ couplings
(see, for instance Eq.~\eqref{eq:maginteraction}).

\subsection{Magnetic interactions}

As usual, the magnetic interactions are written in terms of a exchange
potential. Let us consider two spins placed at $\bx$ and $\by$, then, the
interaction between them two is given by a spin Hamiltonian
\begin{equation}\label{eq:maginteraction}
\mathcal H_{\bx\by} = -J_{\bx\by}\ \boldsymbol S_\bx\cdot \boldsymbol S_\by,
\end{equation}
where the $J_{\bx\by}$ are the {\em couplings}. As discussed above, a
necessary condition for spin glass behavior is that the couplings $J_{\bx\by}$
can take both positive and negative values. This condition is fulfilled by
different kinds of interactions as reviewed in ~\cite{mydosh:93}.

 We will only discuss here the classical solution found in the magnetic alloys
 where the conduction electrons create an indirect exchange interaction known
 as the Ruderman-Kittel-Kasuya-Yosida (RKKY)
 interaction~\cite{ruderman:54,kasuya:56,yosida:57}, whose Hamiltonian
 is\index{RKKY interaction} $\mathcal H_{\bx,\by} = J(|\bx-\by|)\ \boldsymbol
 S_\bx\cdot \boldsymbol S_{\by}$. In these materials, the sea of conducting
 electrons with oscillating spins induce an oscillating interaction between
 the impurities magnetic moments located at $\V{x}$ and $\V{y}$ that depends
 on their separation $r=|\bx-\by|$. For large separations within the
 impurities, the coupling strength is given by
\begin{equation}\label{eq:SG-RKKY}
J(r) \simeq J_0 \frac{\cos (2k_\text{F} r + \phi)}{(k_\text{F} r)^3},
\end{equation}
where  $k_\text{F}$ is the Fermi momentum
of the metal and the phase $\phi$ accounts for the charge difference between
the impurity and the host.

The coupling $J(r)$ is thus an oscillating function of the distance between
spins.  Now, these distances are determined by the position of the impurities,
which are random. Then, the interaction between spins oscillates randomly from
positive to negative interactions, as needed to produce a spin glass.

\section{Experimental spin glasses}\label{sec:expSG}
As discussed in Section~\ref{sec:whatsg}, the spin glass phase is
characterized by a frozen random configuration of spins that hardly evolves
with time. In fact, as in other glassy systems, one of its main features is
that the relaxation times become exceedingly long at low temperatures. For this
reason, at least to discuss the experimental work, \ac{SG} must be considered
to be always out of equilibrium.

\subsection{Aging}\label{sec:aging}
\begin{figure}
\centering
\includegraphics[width=.8\linewidth]{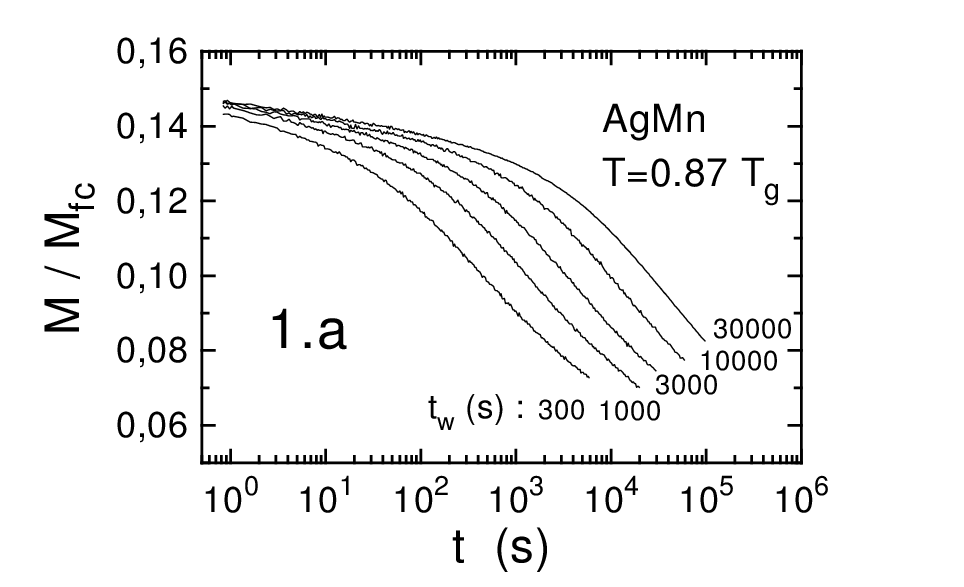}
\includegraphics[width=.8\linewidth]{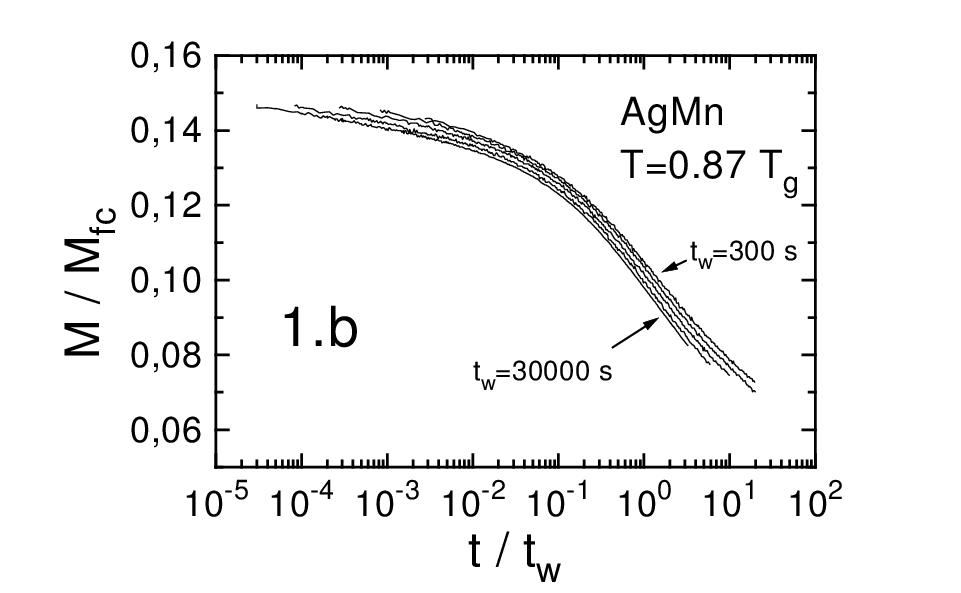}
\caption[Full Aging in the thermo-remanent magnetization]{
{\bf (Top)} Thermo-remanent magnetization $M$ normalized by the field-cooled
value $M_\mathrm{fc}$, vs.  time $t$. {\bf (Bottom)} Same curve but presented
as function of $t/\tw$, the full aging scaling. Figures taken from~\cite{vincent:96}.\label{fig:fullaging-intro}}
\end{figure}

One of the most studied consequences the nonequilibrium nature of spin glasses
is the {\em aging}~\cite{vincent:96,berthier:02b}.  Let us discuss how aging
shows up in the simplest experimental protocol, the {\em direct quench}. The
system is cooled down very fast below the critical temperature $T_c$ in
presence of a magnetic field, and it is let to equilibrate from $t=0$ (the
time of the quench) for a {\em waiting time}, $\tw$. At $t=\tw$ the field is
suddenly switched off. The relaxation of the ``Thermo-remanent magnetization''
(TRM), $M$, is measured at a later time $t+\tw$, see Figure
\ref{fig:fullaging-intro}--top. It can be decomposed as
\begin{eqnarray}\label{eq:TRM}
M(t+\tw)=M_\mathrm{ST}(t)+M_\mathrm{AG}(t+\tw,\tw),&M_\mathrm{ST}(t)\equiv\lim_{\tw\to\infty}M(t+\tw,\tw),\end{eqnarray}
then, there is a fast stationary contribution $M_\mathrm{ST}(t)$ independent
from $\tw$, and an aging part, which, to good approximation, is a function of
the quotient $t/\tw$, see Figure \ref{fig:fullaging-intro}--bottom, at least
for $10^{-3}<t/\tw<10$ and $\tw$ in the range
$50$s---$10^4$s~\cite{rodriguez:03}. This suggests that the effective
relaxation time of the system is of the order of its age. This effect is known
as {\em Full Aging}. Nowadays, there is some controversy about the validity of
this natural time value. In fact, it has been proposed to use $t/\tw^\mu$ with
$\mu\lesssim 1$~\cite{dupuis:05}. At any rate, the moral is that the only
relevant time scale in spin glasses seems to be $\tw$, that is, the age of the
system in the SG phase.

Similar consequences are observed when looking to the response to the system
to an oscillating field. Indeed, let us consider we cool the system from $T\gg
T_\mathrm{c}$ to the working temperature $T<T_\mathrm{c}$ at $\tw=0$. Then, we
apply a very small oscillating field, and measure the a.c. susceptibility
$\chi$ at certain frequency $\omega$. What is observed is that the amplitude
of $\chi$ decreases with $\tw$ (the age of the system as a \ac{SG}). In other words, the
response of the system to the perturbation depends on its thermal history. In
fact, $\chi$ is both a function of $\omega$ and $\tw$.  To a good approximation
it can be written as
\begin{eqnarray}
\chi(\omega,\tw)=\chi_\mathrm{ST}(\omega)+\chi_\mathrm{AG}(\omega\tw).
\end{eqnarray}
Again, there is a stationary part $\chi_\mathrm{ST}$, independent of $\tw$,
and an aging one $\chi_\mathrm{AG}$ that scales roughly on the scaling
variable $\omega\tw$. Note that $M$ and $\chi$ are essentially the Fourier
transform one from the other in the linear response theory, so the full aging
$t/\tw$ found in $M$ translates to $\omega\tw$ in the frequency space.

\begin{figure}
\centering
\includegraphics[width=0.7\linewidth]{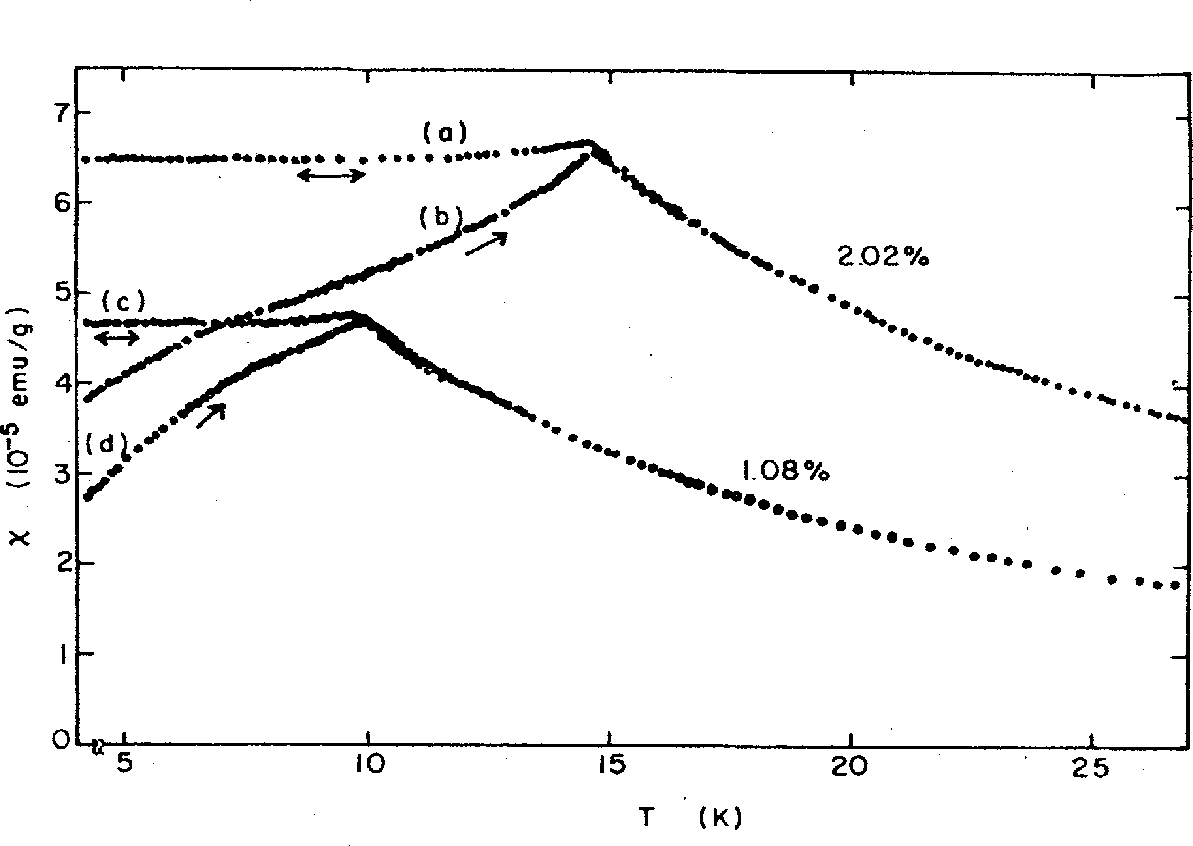}
\caption[Field-cooled and zero-field-cooled  susceptibilities]{%
Comparison of the 
field-cooled [(a), (c)] and zero-field-cooled [(b), (d)] susceptibilities
for two samples of \emph{Cu}Mn with different
concentration of impurities. Figure from~\cite{nagata:79}, as
quoted in~\cite{mydosh:93}.
\index{susceptibility!spin glass|indemph}
\index{susceptibility!FC and ZFC|indemph}
\label{fig:SG-FC-ZFC}}
\end{figure}

Let us consider another aging experiment, but now concerning more complicate
protocols.  This is the case of the response of spin glasses to temperature
cycles with or without the influence of a small constant magnetic field
$H$. We investigate the behavior of the dc susceptibility,
$\chi_\text{dc}$,\footnote{As a matter of fact, experimentalists refers to
  $\chi_\text{dc}$ as $\chi_\text{DC}= M/H$.} under two different cooling
procedures. In both protocols, we start on the paramagnetic phase $T_0\gg
T_\mathrm{c}$ and end in the spin glass phase at a working temperature
$T_1<T_\mathrm{c}$. In the first protocol, named field cooling (FC), the field
$H$ is applied constantly during all the cooling process. On the second case,
on the zero field cooling (ZFC), the field is only switched on once reached
$T_1$.  Figure~\ref{fig:SG-FC-ZFC} shows the temperature dependency of
$\chi_\text{dc}$ for CuMn (1 and 2 at. \%) with a field of 6 gauss. Let us
discuss the two different behaviors.  First, when one performs the
field-cooling [curves (a) and (c)], $\chi_\text{dc}^\text{FC}$ increases as
the temperature decreases in the paramagnetic phase up to a point from which
it remains constant with temperature, that is, in the spin glass phase
region. Now, if one considers the inverse heating cycle, the curve in
$\chi_\text{dc}^\text{FC}$ is roughly reversible.  On the other hand, in the
ZFC procedure, one cools the sample up to $T_1<T_\mathrm{c}$ with no
field. Once at $T_1$, the field is switched on, and the susceptibility
$\chi_\text{dc}^\text{ZFC}(t)$ evolves with time. It starts from the initial
value zero and grows with time. In the infinite time limit (not achieved in
experiments), this susceptibility would reach the FC curve, i.e.
$\chi_\text{dc}^\text{ZFC}(t\to\infty)\approx \chi_\text{dc}^\text{FC}$. Now
we let the sample relax some time at fixed temperature until it reaches the
curves (b) and (d) in Figure~\ref{fig:SG-FC-ZFC}. If we then increase the
temperature keeping also fixed the field, the susceptibility starts to grow
until it reaches the FC curve at $T_\mathrm{c}$. From that point, the FC and
ZFC curves overlap. Finally, if we restart to cool the system again, the
curves that are reproduced are again (a) and (c), that is, the FC curves. That
means that the process is not reversible and curves (b) and (d) can only be
obtained during the heating of a sample cooled by ZFC, in the direction marked
by the flags in Figure~\ref{fig:SG-FC-ZFC}.

\subsection{Fluctuation-Dissipation Relations}\label{sec:FDT}
Another consequence from the nonequilibrium nature of \ac{SG} is the violation
of the \ac{FDT}.

\begin{figure}
\centering
\includegraphics[width=0.5\linewidth]{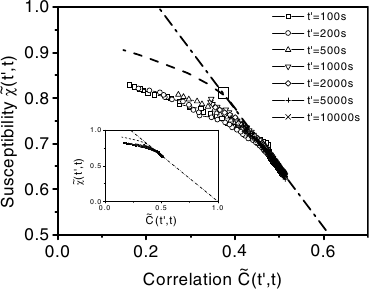}
\caption[Violation of fluctuation-dissipation
in an experimental spin glass]{Experimental realization of the violation 
of the fluctuation-dissipation theorem (linear relation \eqref{eq:fdt2} in
dotted-dashed line) in a spin glass (from~\cite{herisson:02}, 
see text for discussion).
In the figure, $t'$ corresponds to our $\tw$ and $t$ 
stands for our $t+\tw$.
\index{fluctuation-dissipation|indemph}
\label{fig:SG-FDT}}
\end{figure}

In equilibrium, the response $R_O(t+\tw,\tw)$ to an external field $h$
conjugate to any observable $O$ is related to the two-time
autocorrelation function $C_O(t+\tw,\tw)\equiv\mean{O(t+\tw)O(\tw)}$ by means of
the \ac{FDT}~\cite{bouchaud:97},
\begin{eqnarray}\label{eq:fdt1}
  R_O(t+\tw,\tw)\equiv\left.\frac{\delta \mean{O(t+\tw)O(\tw)}}{\delta h(\tw)}\right|_{h=0}=R_{O,\mathrm{EQ}}(t)=
-\frac{1}{T}\frac{\partial
  C_{O,\mathrm{EQ}}(t)}{\partial t
}.
\end{eqnarray}
If we introduce the integrated response $\chi(t+\tw,\tw)=\int_{\tw}^{\tw+t}
R(t+\tw,t')\D t'$, which in equilibrium is nothing
but the magnetic susceptibility, the \ac{FDT} reads
\begin{eqnarray}\label{eq:fdt2}
\chi(t+\tw,\tw)=\chi_{O,\mathrm{EQ}}(t)=\frac{C_{O,\mathrm{EQ}}(0)-C_{O,\mathrm{EQ}}(t)}{T}.
\end{eqnarray}
One can check the validity of this relation by making a parametric plot of
$\chi(t+\tw,\tw)$ vs. $C_O(t+\tw,\tw)$ as shown in Figure~\ref{fig:SG-FDT}.
The linear relation \eqref{eq:fdt2} is only fulfilled for a system in
equilibrium, which means that one should only recover the straight line of
slope $-1/T$ (dashed straight line in Figure~\ref{fig:SG-FDT}) if $\tw\gg
t_\mathrm{EQ}$, where $t_\mathrm{EQ}$ is the equilibration time.

The \ac{FDT} is normally violated in nonequilibrium
systems. In general, the  \ac{FDT} violation can be parameterized by
introducing a violation factor $X(t,t')$ in \eqref{eq:fdt2}, defined as
\begin{eqnarray}\label{eq:neqfdt}
R_O(t,t')\equiv-\frac{X_O(t,t')}{T}\frac{\partial C_O(t,t')}{\partial t'
}.\end{eqnarray} In analytic studies in spin glasses, it is shown that for
large times, this $X_O$ depends on $t$ and $t'$ always through the value of
the correlation function, i.e. $X_O(t,t')=X[C_O(t,t')]$.  Then, since the
different theoretical models for spin glasses predict different behaviors of
$C_O(t,t')$, the different theories (see Section
~\ref{sec:SG-scenarios}) predict different violation factors that can be
compared with experiments.

.

\subsection{Memory and rejuvenation effects}\label{sec:memory-rejuvenation}
Among the surprising experiments concerning \ac{SG}, the experiments of memory
and rejuvenation are probably the most striking ones. Besides, these two
phenomena are purely a glassy feature, not just a nonequilibrium one.  Notice
that the concept of aging also applies to the coarsening dynamics
\index{coarsening} in a ferromagnet~\cite{bray:94}, while no memory or rejuvenation effect has been found in these systems.

\begin{figure}
\centering
\includegraphics[width=.6\linewidth]{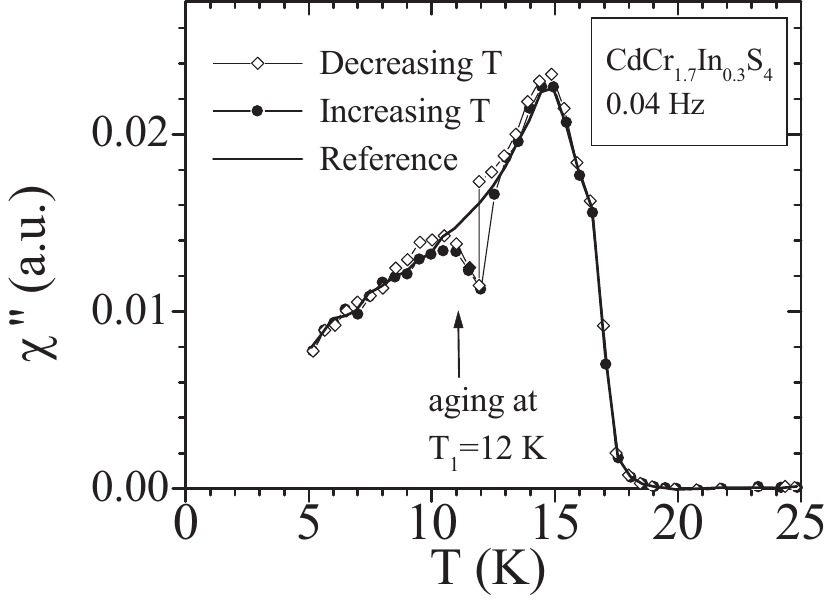}
\caption[Memory and rejuvenation in spin glasses]{%
Memory and rejuvenation in an experimental
spin-glass (Figure taken from~\cite{jonason:98}). Reference line
corresponds to experiment 1, while diamonds and circles correspond to
the cooling and heating part of experiment 2 respectively.
\index{memory|indemph} \index{rejuvenation|indemph}
\index{aging|indemph} \index{susceptibility!ac|indemph}
\label{fig:SG-memory}
}
\end{figure}

We consider the experiment studied in~\cite{jonason:98} shown in Figure
\ref{fig:SG-memory}. In it, the imaginary part of the a.c. susceptibility
$\chi''$ is measured as a function of the temperature, under the influence of a
low frequency $\omega/2\uppi = 0.04$~Hz magnetic field.  We consider the two
following experiments:
\begin{enumerate}
\item One starts at a temperature in the paramagnetic phase, that is well
  above $T_\text{c}$, and cools the system at a constant slow rate of $0.1$
  K/min (small as compared to the $\omega/2\uppi = 0.04$~Hz frequency to
  ensure one stays in the $t\ll\tw$ regime). The $\chi''(T)$ initially
  increases while $T>T_\text{c}$, then describes a cusp at the transition
  temperature $T_\text{c}\approx 15$~K and finally decreases monotonically in
  the SG phase. If afterwards the reverse cycle is repeated but now heating
  the system, the resulting $\chi''(T)$ describes roughly the cooling
  curve. In other words, the process is essentially reversible. This
  experiment is represented in Figure \ref{fig:SG-memory} by a thick black
  line.
\item This time (curve with empty diamonds), we consider the same cooling
  procedure, but this time we make a stop of few hours when the sample reaches
  an intermediate temperature $T_1=12\text{K}(<T_\text{c})$ (within the SG
  phase). The system relaxes (ages at $T_1$), which produces a dip in the $\chi''$ curve. However, if one restarts the cooling again at the same
  original cooling rate, the susceptibility quickly returns to the reference
  curve obtained with experiment 1, as if the cooling had never stopped. This
  astonishing effect is known as {\em rejuvenation}. \index{rejuvenation} Now, as
  before, one heats the system again at constant rate until the highest
  temperature, but this time without making any stop on the path (curve with
  black diamonds). Even though no stop is made at $T_1$, the susceptibility
  remembers the dip and reproduces the curve in empty diamonds. This
  phenomenon is called {\em memory}.
\end{enumerate}

The theoretical description for spin glasses will be discussed later
on. Although, let us anticipate that there are two possible scenarios to
rationalize these experiments: 
\begin{itemize}
\item The first one is to relate them to the so-called {\em temperature chaos}
  predicted for spin glasses~\cite{bray:87}. The Chapter \ref{chap:chaos} is
  fully devoted to this effect. In this approach, the frozen spin pattern
  strongly varies with the temperature. In this picture, the aging at $T_1$
  would not be affected by the aging at $T_2<T_1$, since the spin pattern at
  $T_1$ would look completely random for the eyes of the system at $T_2$.
  Rejuvenation is very natural on this scheme, but memory is unexplained.  We
  will come back to this discussion in Chapter \ref{chap:chaos}.

\item On the other hand, there is a much simpler approach that assumes length
  scale separation at different temperatures and fast
  modes~\cite{berthier:02,berthier:03}. Indeed, in all theories, as we shall
  see, the aging in the spin glass phase is explained as a process where
  coherence domains grow with time. In Chapter \ref{chap:hypercube}, we will
  study numerically this coherence length. This growth is expected to be
  slower the lower is the temperature.  Then, if one assumes that this speed
  varies sharply with temperature, hand-waving, one can explain
  both the rejuvenation and the memory effects at least in very simple
  protocols. In order to illustrate this idea, let us consider the cycling
  experiment $E\to A\to B\to A$, where $T_E\gg T_\mathrm{c}$ ($E$ in the
  paramagnetic phase), and $T_B<T_A< T_\mathrm{c}$ ($A$ and $B$ in the SG
  phase). Now, starting from $E$, the system is quickly quenched to $T_A$ and
  is let to age for a time $0<t<t_A$. Then, the coherence length grows with
  $t$, and at every time it will be $\ell(t,T_A)$. That means that, at $t$,
  the system will be equilibrated up to length scales $\ell<\ell(t_A,T_A)$,
  but will continue out of equilibrium for larger scales, evolving still from
  the state at $T_E$. If now the system is again cooled to $T_B<T_A$ at $t_A$
  (but $T_A-T_B$ large), all length scales are out of equilibrium again
  (assuming sensibility of the equilibrium phase to external conditions, but
  not necessarily such a strong sensibility as in the temperature
  chaos). Then, the rejuvenation is due to the re-equilibration of the small
  length scales below the new coherence length $\ell(t,T_B)$ that starts to
  grow for $t_A<t<t_B$. Now, if the growing speed is a lot slower than at
  $T_A$, and $t_B-t_A\approx t_A$ as normally happens in experiments, at
  $\ell(t_B-t_A,T_B)\ll\ell(t_A,T_A)$. Then, if now the system in heated again
  to $T_A$, the intermediate lengths $\ell(t_B-t_A,T_B)<\ell<\ell(t_A,T_A)$
  will be already equilibrated for $T_A$ from the previous aging, they {\em
    remember} the previous ordering. That would be the explanation to memory.
 
We should stress, however, that the numerical methods of~\cite{berthier:02}
can be applied as well to  disordered ferromagnets. In fact, one finds as well
``memory''  and ``rejuvenation''~\cite{jimenez:05}, although they do not
appear on experimental ferromagnets. One must thus question whether the
``memory'' and ``rejuvenation'' on~\cite{berthier:02} is related to the
experimental effects.

\end{itemize}
\section{Spin glass models}\label{sec:SG-EA}\index{Edwards-Anderson model|(}
All the description up to know was purely experimental. For a theoretical
approach a simple model to work with is needed. In nature there are many
different kinds of magnetic interactions that lead to the qualitatively
similar \ac{SG} behavior. The only obvious common features have been
mentioned: randomness, mixed interactions and frustration.  With this idea in
mind, the goal for theoretical physicists is to find a model simple enough
that it allows analytical treatment but yet complex enough to display the
surprising effects observed in experiments.
\subsection{Edwards-Anderson model}\label{sec:EA}
Concerning theoretical physics of \ac{SG}, on 1975 ``all hell broke loose''
with the proposal of Edwards and Anderson of a very simple
model~\cite{edwards:75}. In it, the spins $\V{s}_i$ ($i=1,\ldots,N$) lie on a
regular, translationally invariant lattice, and the couplings $J_{ij}$ are
random. The Hamiltonian is given by \be {\cal
  H}=-\sum_{i,j}J_{ij}\V{s}_i\V{s}_j, \ee where the $\V{s}_i$ are unitary
vectors of three components (in Heisenberg spin glasses), two components ($XY$
spin glasses) or with only one component (Ising spin glasses). In principle,
the interactions $J_{ij}$ are random variables with a distribution that
depends on the distance between the spins $|\V{R}_i-\V{R}_j|$.  However, among
all the possible options, the most popular election is the one where the
interactions occur only between nearest neighbors. Actually, this model is the
one often known as the \ac{EA} model. The Hamiltonian is now
\be\label{eq-intro:EAham} {\cal H}=-\sum_{\mean{i,j}}J_{ij}\V{s}_i\V{s}_j, \ee
where $\mean{i,j}$ indicates a nearest neighbors summation. The $J_{ij}$ are
generally extracted from a probability distribution such that
$\overline{J_{ij}}=0$. The most popular elections are Gaussian and bimodal
($\pm J$) couplings. Actually, the shape of the distribution seems not to be
very important.

Edwards and Anderson also came up with a proposal of order parameter for the
spin glass phase. Concerning all what discussed in the previous section, this
parameter cannot be long-ranged since the spin glass phase has no long-range
order, and must depend on the temperature if one assumes temperature chaos.
Their proposal was
\begin{equation}\label{eq:overlap-def0}
q_\mathrm{EA} = \lim_{t\to\infty} \frac{1}{N} \sum_i \braket{s_i(0) s_i(t)}_t,
\nomenclature[q]{$q$}{Spin overlap}
\end{equation}
namely the overlap between the spin configurations at two different distant
times (in equilibrium). We discussed before that time average of the local
magnetization is non-zero in the spin glass frozen phase.  In particular,
$q=1$ at $T=0$ (no evolution at all) and since the transition is second order,
we should expect $q\to0$ when $T\to T_\text{c}$ as in a paramagnetic phase. As
usual, Eq. \eqref{eq:overlap-def0} can be simplified
\begin{equation}\label{eq:overlap1}
q_\mathrm{EA} =\frac{1}{N} \sum_i \mean{s_i}^2.
\nomenclature[q]{$q$}{Spin overlap}
\end{equation}
We will come back
to this parameter in Section \ref{sec:observables}.

\subsection{Quenched averages and replicas}\label{sec:replicatrick}
Before introducing analytical derivations, it is interesting to
discuss how to deal with disorder averages.

In the disordered magnetic systems we are considering here, as in the EA model
just defined, the Hamiltonian $H_J(\lazo{s_i})$ depends on two kinds of
variables: the spins, $\lazo{s_i}$, and the couplings
$J\equiv\lazo{J_{ij}}$. Now, one notes that the diffusion time for impurities
(think of Mn atoms on Cu$_{1-x}$Mn$_x$) is huge as compared with the timescale
for \textit{spin-flip} (picoseconds). This suggests to work in the  {\em quenched approximation}: spins cannot have any kind
of influence over the material impurities.  Then, the set of coupling
constants in a particular realization of $J$, namely {\em sample}, will
be considered random variables distributed according to certain
probability distribution $P(\lazo{J})$ known in theoretical
models.  The free energy density within each sample is then also
a random variable, and is given by \be\label{eq:fj}
f_J=-\frac{1}{\beta N}\log Z_J, \ee where \be
Z_J=\sum_{\lazo{s}}\E^{-\beta H_J(\lazo{s})}, \ee is the partition
function for this sample.

However, ordinary statistical mechanics tells us how to compute the free
energy for a given set of $J$'s. But what if we do not know which is the
actual set of $J$'s because they are random? how do we compute $f_J$? Indeed,
the only thing we know about these $J$'s is their probability distribution
function. Fortunately, if one considers the $N\to\infty$ limit, thermodynamic
magnitudes such as the energy density must take the same value in all the
samples (this property is known as self-averaging). That means that the
randomness in the samples leads to fluctuations of order $1/N$\be
\overline{f_J^2}-\paren{\overline{f_J}}^2=\mathcal{O}\paren{\frac{1}{N}}, \ee where
$\overline{(\ldots)}$ refers to average over samples $J$'s, i.e.
\be\label{eq:fj2} \overline{f}\equiv\overline{f_J}=\sum_J P({J})f_J.  \ee

According to this last statement, for finite system sizes, the best way of
inferring the thermodynamic limit is to average over all the samples. Indeed,
fluctuations will be reduced by $1/N_s$, being $N_s$ the number of samples.
That means that from now on we will be interested in the
averaged magnitudes. As usual in statistical mechanics, the central
magnitude is the free energy $\overline{f}$ defined in \eqref{eq:fj2}.

This magnitude can be computed easily using the so-called {\em replica
  method}. Technically, it is computed as an analytical continuation of the
disorder average of the partition function of $n$ uncoupled replicas of the
system. Before using this trick, it is useful to introduce some
definitions,
\begin{align}\label{eq:Znfn}
Z_n\equiv\sum_J P(\lazo{J})\paren{Z_J}^n=\overline{\paren{Z_J}^n},& &f_n\equiv-\frac{1}{n\beta N}\log Z_n.
\end{align}

Now, using the relation $A^n\approx 1+ n\ \log A$ valid for $n\approx
0$ and the usual $\sum_J P(\lazo{J})=1$, we get \be
\log\overline{\paren{Z_J}^n}\approx \log\paren{1+ n\ \overline{\log
    Z_J}}\approx n\ \overline{\log Z_J}, \ee for $n\approx 0$. Then,
it is clear that the desired averaged free energy is \be
\overline{f}=\lim_{n\to 0} f_n.  \ee Here comes the so-called replica
trick. If one considers $n$ to be an integer, $\paren{Z_J}^n$ can be easily
computed by means of $n$ uncoupled replicas of the same system
(evolving under the same set of $J$'s), \be
\paren{Z_J}^n=\sum_{\lazo{s_i^{(1)}}}\sum_{\lazo{s_i^{(2)}}}\cdots\sum_{\lazo{s_i^{(n)}}}\E^{-\beta\sum_{a=1}^n
  H_J\paren{\lazo{s_i^{(a)}}}}, \ee where the spins $s_i^{(a)}$ carry two
indices: the upper is the replica index, running from $1$ to $n$, and
the lower labels the site of the spin, running from $1$ to $N$.

We now use this approach to obtain the famous solution to the Sherrington
Kirkpatrick model, the mean field version of the \ac{EA} model already
discussed.

\subsection{The mean-field spin glass: the Sherrington Kirkpatrick
  model}\label{sec:SK} \index{mean field}

At variance with ferromagnets, the \ac{MF} approximation
in spin glasses is highly non trivial. We will discuss in this section
the mean field solution to the EA spin-glass. As we shall see,
although \ac{MF} allows an exact analytical description, the emergent
picture is by no means, simpler. In fact, it is not even clear if it is
simpler than the unperturbed problem.

In this section we present a sketch of the derivation of the mean-field
solution for the EA spin glass. For a full derivation see,
e.g. \cite{dotsenko:01,mezard:87}. We will just concentrate on the necessary
information to understand its predictions for the spin-glass phase.

The most important \ac{MF} model in spin glasses is the \ac{SK}
model~\cite{sherrington:75}, which is both the first and the most
studied model. However, more realistic mean field models have been
proposed in the last decades, we will discuss some of them in Section
\ref{sec:hypercube}, as well as define a new \ac{MF} model, called
the Hypercube model.

The \ac{SK} model is the fully connected version of the \ac{EA}
model~\cite{sherrington:75}.  In it, all the spins interact with all the other
spins in the system, and the strength of these interactions is random, with no
relation with the distance between them. In this sense, this is quite an
unnatural model since no distance or, at least, notion of neighborhood
exists. In addition, spins are considered to be Ising variables, that is, only
two orientations are possible.  The Hamiltonian is thus defined as
\index{Sherrington-Kirkpatrick model|(}
\begin{equation}\label{eq:SG-SK}
\mathcal H = -\sum_{i<k} J_{ik} s_i s_k,
\end{equation}
where the couplings are Gaussian distributed with mean $\overline{J_{ik}} = 0$
and variance $\overline{J^2_{ik}} = \frac{1}{N}$, that is,\be\label{eq:pJ}
P(J_{ik})=\prod_{i<j}\sqrt{\frac{N}{2\pi}}\E^{-\frac{N}{2}J_{ik}^2}.  \ee With
this election, the total energy \eqref{eq:SG-SK} is proportional to $N$.

Now we apply the replica approach discussed in Section
\ref{sec:replicatrick}. Our first step is to compute the $Z_n$ introduced in
\eqref{eq:Znfn} using the Hamiltonian \eqref{eq:SG-SK}
\begin{align}
Z_n=\overline{\paren{Z_J}^n}=\sum_J P(\lazo{J})\sum_{\lazo{s}}\E^{\beta\sum_{a=1}^n\sum_{i<k}J_{ik}\ s_i^{(a)}s_k^{(a)}},
\end{align}
where $\sum_{\lazo{s}}$ denotes the sum over all the possible spin configuration in all
the $n$ replicas.  We introduce the \ac{pdf} for the couplings, defined in
\eqref{eq:pJ} and use it to remove the $J$'s dependency. The result is
 \begin{align}
Z_n=\sum_{\lazo{s}}\E^{\frac{\beta^2}{2N}\sum_{i<k}\paren{\sum_{a=1}^n
    s_i^{(a)}s_k^{(a)}}^2}=\sum_{\lazo{s}}\E^{\frac{\beta^2 N n}{4}
  +\frac{\beta^2N}{2} \sum_{1\le a,b\le n}\paren{\sum_{i=1} s_i^{(a)}s_i^{(b)}}^2}.
\end{align}
Finally, one can linearize the sum over the sites using the so-called replica
matrix $Q_{ab}$,
 \begin{align}\label{eq:linearrel}
Z_n=\paren{\prod_{a<b}^n\int \D Q_{ab}}\sum_{\lazo{s}}\exp\caja{\frac{\beta^2 N n}{4}
  -\frac{\beta^2N}{2} \sum_{1\le a,b\le n}Q_{ab}^2+\beta^2\sum_{1\le a,b\le
    n}\sum_i Q_{ab}s_i^{(a)}s_i^{(b)} },
\end{align}
where $Q$ is a $n\times n$ symmetric matrix, with zeros on the diagonal. This
last expression can be simplified so that
\begin{align}
Z_n&=\paren{\prod_{a<b}^n\int \D Q_{ab}}\E^{-N A\paren{\lazo{Q}}},\\
A\paren{\lazo{Q}}&=-\frac{\beta^2 n}{4}
  +\frac{\beta^2}{2} \sum_{1\le a,b\le
    n}Q_{ab}^2-\frac{1}{N}\log{\sum_{\lazo{s}}\exp\caja{-\beta H\paren{\lazo{Q,s}}}},\\
H\paren{\lazo{Q,s}}&=-\beta\sum_{1\le a,b\le
    n}\sum_i Q_{ab}s_i^{(a)}s_i^{(b)}. 
\end{align}

Then, we can use the saddle-point approximation to compute $Z_n$ for the large
$N$ limit, i.e. $Z_n=\min\caja{ A\paren{\lazo{Q}}}$. Therefore, the task is to
find the solution to the $n(n-1)/2$ equations $\partial A/\partial Q_{ab}=0$
. It turns out that the solution is given by
\begin{equation}\label{eq:SG-Qab}
Q_{ab} = \frac1 N \sum_i \braket{s_i^a s_i^b}_Q,\qquad a\neq b,
\end{equation}
where the average $\braket{\cdot}_Q$ is taking using the Hamiltonian
$H\paren{\lazo{Q,s}}$ defined in \eqref{eq:linearrel}.

The function $A\paren{\lazo{Q}}$ is symmetric with respect to the exchange of
rows or columns: all the replicas are equivalent. The only replica symmetric
solution is then
\begin{equation}\label{eq:SG-RS}
Q_{ab} = (1-\delta_{ab}) q.
\end{equation}
However, although this solution reproduces the right phase diagram, it leads to
a negative value of the entropy at low temperatures and the $q=0$ solution
turns to be a maximum in the free energy for $T>T_\mathrm{c}$, which makes no
sense. Besides, the $q=0$ solution below $T_\mathrm{c}$ seems to be more
stable than the spin glass solution $q\neq 0$~\cite{almeida:78}. Furthermore,
it leads to a negative susceptibility which contradicts the experiments and
basic thermodynamic notions.

Some deeper analysis concluded that the conditions $\partial
A/\partial Q_{ab}=0$ did not imply that $A\paren{\lazo{Q}}$ is a
minimum function of $Q$ for all values of $n$. Indeed, the number of
equations, the $n(n-1)/2$ becomes negative when $0<n<1$ and for the
replica trick one needs precisely to take the $n\to 0$ limit.

The solution to this problem was proposed by Parisi some years
later~\cite{parisi:79b,parisi:80}, and implies breaking the replica symmetry,
i.e the solution is not longer \eqref{eq:SG-RS} (see
Fig. \ref{fig:RSBsteps}). The starting point is this symmetrical matrix
\eqref{eq:Q0}. Now, one step of \ac{RSB} consists on dividing the matrix into
constant blocks $[(n/m_1)\times (n/m_1)]$ of size $m_1\times m_1$ and set each
diagonal block as a sub-matrix whose off-diagonal elements are all $q_1$ and
the remaining terms stay how they were, i.e with the value $q_0$, as done in
\eqref{eq:Q1}.  A second \ac{RSB} is taking in the same way, but now
introducing a new overlap $q_2$, see \eqref{eq:Q2}.  This process is continued
indefinitely.
\begin{figure}\begin{center}
\begin{eqnarray}\label{eq:Q0}
\text{replica symmetric solution}&\left(\begin{array}{cccccccc}
0 & & & & \multicolumn{4}{c}{\multirow{4}{*}{\Huge $q_0$}}\\
 & 0 & &  & \\
 &  & 0 &  & \\
 & &  & 0 & \\
\multicolumn{4}{c}{\multirow{4}{*}{\Huge $q_0$}} & 0 &  &  &  \\
 & & & &   & 0 &  &   \\
& & &  &  &  & 0 &   \\
 & & & &  &  &  & 0
\end{array}\right) \quad \longrightarrow \quad &\\\label{eq:Q1}
\text{1 RSB step}&\left(\begin{array}{cccc|cccc}
0 &  &\multicolumn{2}{r|}{\multirow{2}{*}{\Large $\ \ q_1$}} & \multicolumn{4}{c}{\multirow{4}{*}{\Huge $q_0$}}\\
 & 0 & &  & \\
\multicolumn{2}{c}{\multirow{2}{*}{\Large $q_1$}}  & 0 & & \\
 &  &  & 0 & \\
\hline
\multicolumn{4}{c|}{\multirow{4}{*}{\Huge $q_0$}} & 0 & & \multicolumn{2}{c}{\multirow{2}{*}{\Large $q_1$}}\\
 & & & &  & 0 &   &  \\
& & & & \multicolumn{2}{c}{\multirow{2}{*}{\Large $q_1$}}  & 0 &   \\
 & & & &  & &  & 0
\end{array}\right) \quad \longrightarrow \quad& \\\label{eq:Q2}
\text{2 RSB steps}&\left(\begin{array}{cc|cc|cc|cc}
0 & q_2 &\multicolumn{2}{c|}{\multirow{2}{*}{\Large $\ \ q_1$}} & \multicolumn{4}{c}{\multirow{4}{*}{\Huge $q_0$}}\\
q_2 & 0 & &  & \\
\cline{1-4}
\multicolumn{2}{c|}{\multirow{2}{*}{\Large $q_1$}}  & 0 &q_2 & \\
 &  & q_2 & 0 & \\
\hline
\multicolumn{4}{c|}{\multirow{4}{*}{\Huge $q_0$}} & 0 &q_2 & \multicolumn{2}{c}{\multirow{2}{*}{\Large $q_1$}}\\
\multicolumn{4}{c|}{} &q_2  & 0 &   &  \\
\cline{5-8}
\multicolumn{4}{c|}{} & \multicolumn{2}{c|}{\multirow{2}{*}{\Large $q_1$}}  & 0 & q_2  \\
\multicolumn{4}{c|}{} &  & &q_2  & 0
\end{array}\right)  \quad \longrightarrow \quad &\\\nonumber&&\\\nonumber&&
\\\nonumber&{\Huge \cdots}&
\end{eqnarray}
\end{center}\caption{Sketch of the replica symmetry steps for a problem of
$n=8$ replicas (see main text for discussion).}\label{fig:RSBsteps}
\end{figure}

Note that with this description, the equivalence between replicas is still
conserved. In fact, all the rows or columns have the same components, although
the order of appearance of the $q_i$ is different.

The \ac{pdf} is given by
\begin{align}
p(q) &= \frac{1}{n(n-1)} \sum_{a\neq b} \delta(Q_{ab} - q)\\
   &= \frac{n}{n(n-1)} \bigl[ (n-m_1) \delta(q-q_0)
+ (m_1-m_2) \delta (q-q_1)  \\
& \qquad  \qquad\qquad+(m_2-m_3) \delta(q-q_2) + \ldots
\bigr],
\end{align}
and taking the $n\to0$ limit one gets
\begin{equation}
p(q)  = m_1 \delta(q-q_0) + (m_2-m_1) \delta(q-q_1) 
+ (m_3-m_2) \delta(q-q_2) + \ldots.
\end{equation}
Note that, although by construction $n\ge m_1\ge m_2\cdots\ge 1$, when taking
the $n\to 0$ limit, it turns around and $0\le m_1\le m_2\le\cdots\le 1$.

Then, in the limit of infinite \ac{RSB} steps, 
\begin{align}
&m_k/m_{k+1}\to 1-\D x/x, &\text{and}& &q_k\to q(x)
\end{align}
 with $x\in [0,1]$, i.e. can take whatever value within this interval. Thus,
 the spin glass order parameter is not a number, but a function. In other
 words, we have now an infinite number of order parameters.  This solution
 suggests that the \ac{SK} has infinite number of ``frozen spin patterns'',
 whose overlap $q'$ can take all values in $q(0)\leq q'\leq q(1)$.

All the description here was quite naive, specially concerning an integer
number of replicas that, at certain point, is analytically continued to
zero. However, although  later in time, it has been rigorously shown that the
RSB scheme produces the correct free energy for the Sherrington-Kirkpatrick
model~\cite{talagrand:06}.

\begin{figure}
\begin{center}
\includegraphics[angle=270,width=0.4\columnwidth,trim=50 0 0 0]{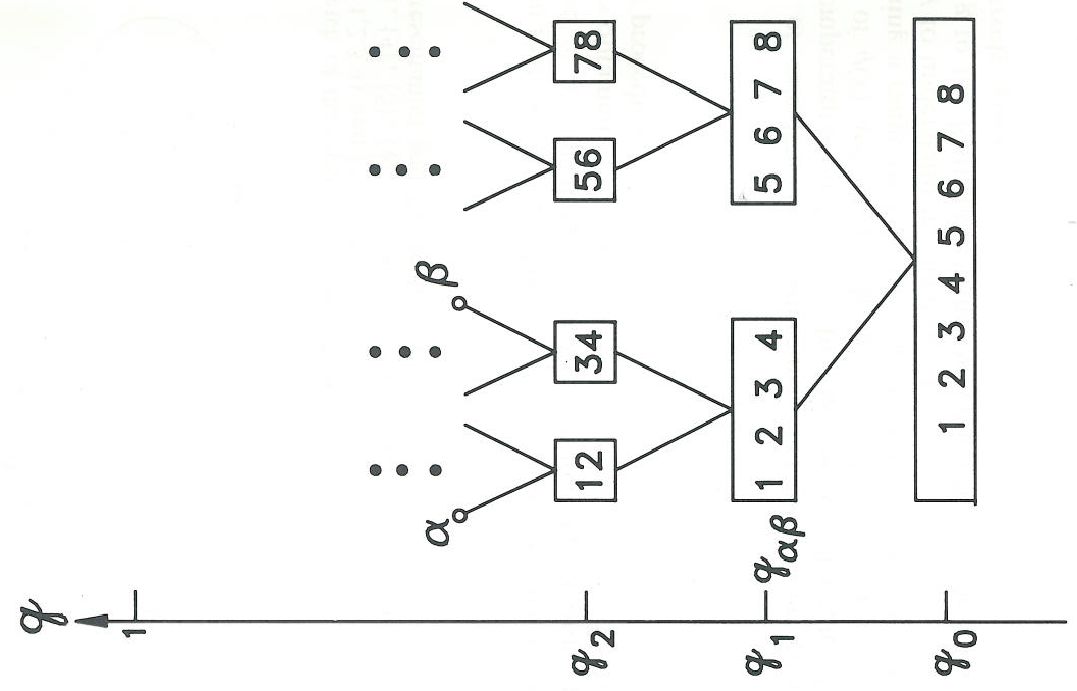}
\includegraphics[angle=270,width=0.5\columnwidth,trim=-200 0 100 0]{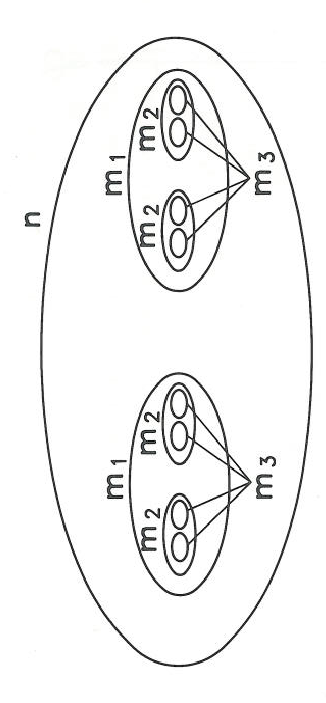}
\end{center}
\caption{ Sketch of the ultrametric space for $n=8$ replicas. The
  trace between $\alpha$ and $\beta$ one must go back to the point
  where their branches join, for instance $q_{\alpha\beta}=q_1$. The
  overlaps of the overlaps within the same row are the same.}
\label{fig:tree}
\end{figure}

In addition, the \ac{RSB} approach leads to an ultrametric distribution of
states \cite{rammal:86}. Indeed, the order parameter matrix $Q_{ab}$ can be
represented by a tree with emanating branches. In order to illustrate this
idea, let us consider the $8\times 8$ matrix presented in Figure
\ref{fig:RSBsteps} of a system of $n=8$ replicas. We can represent this matrix
on a tree as the one shown in Figure \ref{fig:tree}---left. At the root (no
RSB step), all the elements have the same overlap $q_0$. After one RSB step
with $m_1=4$ all elements are divided into two branches $\lazo{1,2,3,4}$ and
$\lazo{5,6,7,8}$, each with overlap $q_1$. The process continues with many
sub-divisions until $q=1$. Then, the overlap between two given replicas
$\alpha$ and $\beta$ in Figure \ref{fig:tree}---left, $q_{\alpha\beta}$ is
given by the level at which the branches coming from each replica join. In the
case of the Figure, $q_{\alpha\beta}=q_1$.  This leads to a hierarchical
distribution of clusters. As in any tree graph, the overlaps fulfill the
ultrametric inequality \be q_{\alpha\beta} \geq
\min\{q_{\alpha\gamma},q_{\beta\gamma}\}.\ee This equality implies that the
space of states can be divided into clusters of a given overlap, and each of
them subdivided in other subclusters and so on, see \ref{fig:tree}---right for
the previous example. Note that there is no overlap between clusters of
similar order, each point lies on one single cluster. An space with an
organization like the one described here is called ultrametric.

One of the most important features of the RSB solution is that the $p(q)$ is
non-trivial, as discussed above.  In addition, another important feature of
the RSB solution of the \ac{SK} model, is that the spin glass phase is not
destroyed in the presence of a magnetic field. Besides, when computing the
susceptibility, it reproduces the constant behavior $\chi(T)$ in the field
cooling (FC) experiments discussed in Section \ref{sec:expSG}.

\section{Theoretical scenarios: the Droplets and the RSB pictures}\label{sec:SG-scenarios}

There are mainly two competing theories for explaining the equilibrium SG
phase: the \textit{droplets} \cite{fisher:86,fisher:87,huse:87,fisher:88b} and
the \ac{RSB} one \cite{marinari:00}. In addition, numerical simulations
suggested an intermediate picture between these
two. This theory is known as the TNT~\cite{krzakala:00,palassini:00} (we will leave its discussion for Section \ref{sec:TNT}).

On one hand, the \textit{droplets} picture is based on the Migdal-Kadanoff
renormalization, exact for $D=1$ EA model. According to the \textit{droplets}
picture, the SG phase would be ferromagnetic-like, with a complicated spin
texture due to the disorder, but essentially with only two equilibrium states
related by spin-flip symmetry.  The dynamics is thus explained in terms of low
lying excitations of compact domains (droplets) of coherently flipped spins
about these states. Since the couplings are disordered, the boundaries of
these domains wander so that they can take advantage of the unsatisfied bonds
and avoid the stronger satisfied ones. This effect results in a non convex
droplet structure. Indeed, droplets are expected to be fractal of dimension
$D-1\le D_\mathrm{s}<D$, thus, not space filling. In addition, this theory
assumes that the lowest energy excitations of spatial extent $\ell$ typically
cost a free energy\be\label{eq:freedroplets} F_\ell\sim \gamma(T)\ell^\theta,
\ee where $\gamma(T)$ is called the {\em stiffness modulus} and $\theta$ the
    {\em stiffness coefficient}, which fulfills \be 0 <\theta<(D-1)/2.  \ee
    Hence, in the thermodynamic limit, an excitation involving a finite fraction
    of the total spins, i.e. $\ell=O(L)$, would cost an infinite free
    energy. Then, this approach only expects excitations of size $\ell\ll L$.  As a
    consequence of this model, the spatial correlations decay with $\theta$,
    \be
    C(r_{ij})=\overline{\mean{s_is_j}^2}-\overline{\mean{s_i}^2\mean{s_j}^2}\sim\frac{1}{r_{ij}^\theta},
    \ee which makes the \ac{pdf} for the overlap trivial, i.e. $P(q)=\delta
    (q^2-q_\mathrm{EA}^2)$, as mentioned before.

On the other hand, the \ac{RSB} is based on the mean-field solution sketched
in Section~\ref{sec:SK}. There is a growing consensus that the RSB is valid
for the EA model for dimensions $D>D_u=6$, with $D_u$ being the upper critical
dimension. In this theory, the $D=3$ EA spin-glass is drawn as a perturbative
extension from the exact solution obtained above for the SK model. The
emerging picture is very similar to the one presented in the previous section:
the equilibrium \ac{SG} phase is composed by an infinite number of degenerate
states organized through an ultrametric structure. Indeed, as in the \ac{SK}
solution, the \ac{pdf} for the order parameter $P(q)$ is not trivial, and all
values for the overlap between $[-q_\mathrm{EA},+q_\mathrm{EA} ]$ are possible
even in the infinite volume limit. The \ac{RSB} theory expects non-compact
domains whose surface is space filling, that is, the fractal dimension
$D_\mathrm{s}$ is equal to the space dimension $D$. In addition, as in the SK
model, there can be excitations that involve flipping an $O(L)$ number of
spins with a finite energy cost (the MF prediction is $\theta_\mathrm{MF}=0$).
In addition, in this scenario, the spin glass survives under the influence of
a magnetic field.

\begin{figure}
\centering
\includegraphics[angle=0,width=.5\linewidth, trim= 30 0 40 0]{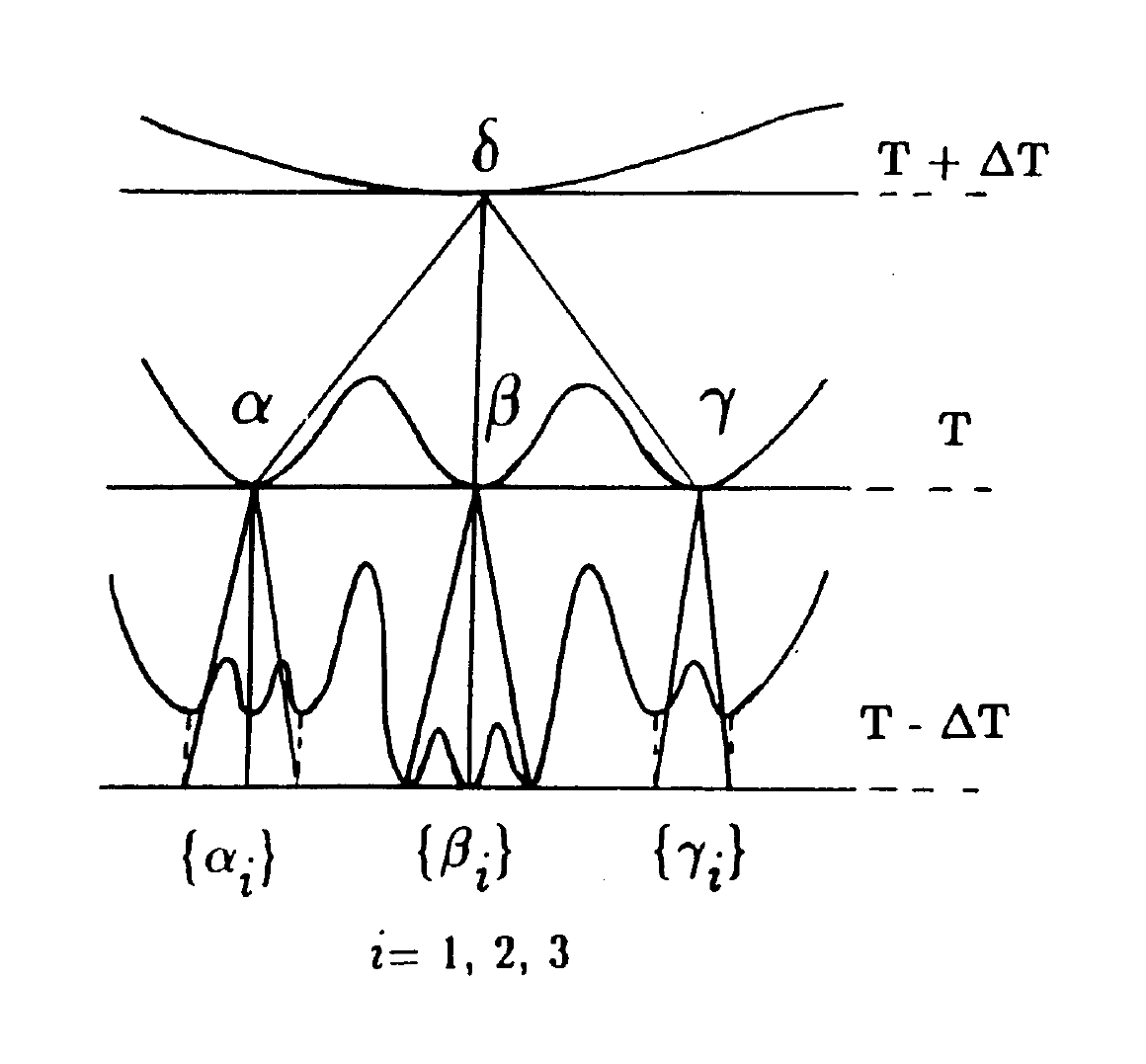}
\caption{Sketch of the hierarchical structure of metastable states as a
  function of temperature. Figure taken from~\cite{vincent:96}.
\label{fig:tree-aging}
}
\end{figure}

 For both theories, \textit{Aging} is a explained as a process where coherent
 domains of low temperature phase grow with time. The characteristic length
 scale for these domains is $\xi(t_\mathrm{w})$, the coherence
 length. The two theories disagree in their predictions for these domains
 properties:
 \begin{itemize}
 \item For the droplets theory, these domains are compact objects: the surface-volume ratio vanishes in the high
 $\xi(t_\mathrm{w})$ limit \cite{fisher:88}. The SG order parameter is non
 zero inside of each domain. 
\item The RSB theory expects space
 filling domains with a surface-volume ratio constant for large
 $\xi(t_\mathrm{w})$. The SG order parameter
 vanishes inside those domains.
\end{itemize}
It is interesting to point out that even though the droplet picture is simpler
compared to the RSB, it still accounts for the complex physics of experimental
spin glasses. A curious example of the diverse explanations of the same
effects appears on the evolution of the spin freezing pattern with temperature
and the apparition of temperature chaos, see Chapter \ref{chap:chaos} (let us
note that temperature chaos has not been directly measured in experiments, but
it is predicted by both theories). In the droplet theory, the compact domains
can suddenly flip due to an infinitesimal change of the temperature, because
of a delicate balance between the free energy \eqref{eq:freedroplets} and the
entropy of the system.  We will discuss this approach in detail in Section
\ref{sec:chaos-intro}. On the contrary, the RSB explains it with a
hierarchical structure of the ground states as function of the temperature, as
shown in Figure \ref{fig:tree-aging}. In it, during the aging at certain
temperature $T$, the system samples the infinitely many metastable states at a
given level of the hierarchical tree. The aging is later restarted upon
lowering the temperature following the subdivisions in possible states in the
free energy of each valley. The system must find the equilibrium state but
always inside the branch already chosen. Within this approach, the
rejuvenation and memory effects discussed in Section
\ref{sec:memory-rejuvenation} are directly explained. Once we lower the
temperature, the aging is reactivated with the subdivision, leading to the
rejuvenation effect. But if the temperature is increased again, the system
returns to the initial valley. In the same sense, temperature chaos is
expected in such a picture, the distribution of valleys in free-energy changes
completely from one temperature to the other.

As mentioned before, even after 40 years of intense study, there is not still
consensus about which is the nature of the equilibrium phase. One of the main
difficulties to test experimentally these different scenarios is precisely the
fact that the real spin glasses are always out of equilibrium. Then, one needs
to find a way of tracking the influence of the equilibrium configuration in
the nonequilibrium evolution of the system. An example of this idea is found
in the violation of the fluctuation-dissipation theorem (FDT) discussed in
Section \ref{sec:FDT}.  Indeed, as discussed above, the violations of the FDT
can be quantified by means of the violation factor $X(t,t')$ introduced in
\eqref{eq:neqfdt}. In equilibrium, $t,\tw\to\infty$ and $C(t,\tw)\to q$, then,
the susceptibility
\begin{eqnarray}\label{eq:fdtneq}
\chi(t+\tw,\tw)=\int_{0}^{t} R(t+\tw,t'+\tw)\D t'\to
-\frac{1}{T}\int_{q_\mathrm{min}}^q \D q`\ P(q').
\end{eqnarray}
That means that the violation of the FDT depends on the equilibrium pdf for
the overlap $P(q)$. Since this pdf is radically different in the two
scenarios, the prediction for $x(q)\equiv\int_{q_\mathrm{min}}^q \D q`\ P(q')$
changes from one theory to the other. This factor can be measured in a
experiment if comparing with the linear behavior with the $1/T$ expected for the FDT, as was
shown in Figure \ref{fig:SG-FDT}. As long as in the droplet theory the overlap
distribution is trivial, one would expect a constant behavior below
$q_\mathrm{EA}$, while in the RSB theory, this function should decrease with
$q$ as shown in the dashed lines in Figure \ref{fig:SG-FDT}.

\section{Spin glasses in numerical simulations}

The two theoretical scenarios for the equilibrium phase discussed above are
both exact in two different oversimplified models for spin glasses. Namely, in
those models where the mean field approximation (in the case of RSB) or the
Migdal-Kadanoff renormalization (in the case of droplets) are exact. However,
their validity for describing real spin glasses or, at least for more
realistic theoretical models, such as the $D=3$ EA model, is not clear.

That is precisely where computer simulations are essential: they can
fill the gap between experiments and theory (obtained in oversimplified
models). Indeed, numerical simulations allow us to explore the SG phase
in more realistic theoretical models which, because of their
complexity, cannot be solved analytically. Furthermore,
simulations allow both equilibrium and non equilibrium studies, which
enable to compare results either with theory  or with experiments.

Basically without exceptions, numerical work in $D\!=\!3$ is best described by
the RSB theory (see~\cite{marinari:00} for a review,
refs.~\cite{contucci:06,contucci:07b,contucci:09,janus:10} for recent work and
refs.~\cite{krzakala:00,palassini:00,jorg:08} for some somewhat dissenting
views). However, the system sizes that can be thermalized in reasonable times
in a simulation are so small that one should question whether the observed
effects are really coming from the nature of the SG phase or are just
finite size effects~\cite{moore:98,bokil:00}. The same situation is observed
in nonequilibrium, where the simulation times are often too far away from
experimental scales to account for the interesting phenomena. In other
words, the computer capacity is currently the real bottleneck in spin glass
studies.

However, the situation has improved notably in the last years, with the
large-scale simulations performed on Janus~\cite{janus:06,janus:08}, a
special-purpose computer designed for the simulation of SG. Indeed, the Janus
computer outperforms the conventional computing architectures by several
orders of magnitude, both in times and in lengths scales. Considering
nonequilibrium simulations~\cite{janus:08b}, it was able to follow the
nonequilibrium dynamics up to times $\sim 0.1 s$, which improves by three
orders of magnitude a conventional computer. This time must be compared with
the experimental window that goes from seconds to hours. Simulations still
cannot reproduce experiments, but Janus is almost there. On the other hand,
with the Janus computer it is possible to thermalize lattices of size
$L\!=\!32$ down to temperatures $T\!\approx \! 0.64 T_\mathrm{c}$. This is not
only a world record, but provides as well the best glimpse on the low
temperature SG phase ever. We will use these unprecedentedly large
configurations, thermalized up to very low temperature for our study of
temperature chaos  in Chapter~\ref{chap:chaos}. The existence of
these configurations will be crucial for the conclusions achieved.

Leaving aside the computer capacity improvements, the finite time/size
problems can be useful. Indeed, the comparison between dynamics and statics
has created a bridge between these two separate worlds. A dictionary between
finite-time nonequilibrium and finite-size equilibrium simulations has been
established, allowing us to directly relate non equilibrium experiments (which
also take place in finite times as compared with the relaxation times in
glasses) with theory, which is almost exclusively concerned with
equilibrium. Before introducing this equivalence, it is convenient to present
the observables usually measured in computer simulations.  These magnitudes
will be used in the following chapters as well.

\subsection{Observables}\label{sec:observables}
The starting point is the Edwards-Anderson model, discussed in Section
\ref{sec:SG-EA}. For the following, we will consider only Ising spins,
which means that our spin variables, $s_i$, can only take two opposite values
$\pm 1$. The Hamiltonian is then,
 \be\label{eq:HEA} {\cal
  H}=-\sum_{\mean{i,j}}J_{ij}s_is_j, \ee 
where $\sum_{\mean{i,j}}$ refers to the summation over the nearest neighbors. 

As in the ferromagnetic case, the Hamiltonian \eqref{eq:HEA} has a
global symmetry $\mathbb{Z}_2$ ($s_{i}\rightarrow-s_{i}$ for all $i$),
which is spontaneously broken in the low temperature phase. Not as
obvious is the gauge symmetry induced by the disorder average over
couplings (see Section \ref{sec:replicatrick}). In fact, we choose
 a random sign for each position, $\varepsilon_{i}\!=\!\pm 1$, the
energy \eqref{eq:HEA} remains invariant under the transformation
\be\label{eq:gi-intro}\begin{array}{cc} s_{i}\rightarrow
  \varepsilon_{i}s_{i}\,,\ & J_{ik}\rightarrow
  \varepsilon_{i}\varepsilon_{k}J_{ik}\,.\\
   \end{array}
\ee Now, since the transformed couplings $
\varepsilon_{i}\varepsilon_{k}J_{ik}$ are just as probable as the
original ones, the quenched mean value of $\overline{O(\{s_{i}\})}$ is
identical to that of its gauge average $\sum_{\{\epsilon_j=\pm 1\}}
\overline{ O(\{\epsilon _j s_i\})}/2^N\,,$ which typically is an
uninteresting constant value.  Then, we need to define observables
that are invariant under the gauge transformation \eqref{eq:gi-intro}. The
Hamiltonian (\ref{eq:HEA}) provides, of course, a first example.  For
the rest of magnitudes, we first form gauge invariant fields. This can
be done by considering two systems at equal time, that evolve
independently with the same set of couplings couplings,
$\{s_{i}^{(1)},s_{i}^{(2)}\}$ (this is nothing but the replicas
introduced as a trick in Section \ref{sec:replicatrick})
or, alternatively,
from a single system considered at two different times:
\be \begin{array}{c}q_{i}(t_\mathrm{w})=s_{i}^{(1)}(t_\mathrm{w})s_{i}^{(2)}(t_\mathrm{w})
  \,\text{, }
  \\c_{i}(t,t_\mathrm{w})=s_{i}^{(1)}(t+t_\mathrm{w})s_{i}^{(1)}(t_\mathrm{w})\,.\end{array}\ee
Indeed, as discussed in Section~\ref{sec:expSG}, relaxation depends on two times.
 One considers pairs of times
$\tw$ and $t+\tw$, with $t,\tw>0$, after a sudden quench from a fully
disordered state to the working temperature $T$.

We discuss first the time-dependent observables to end up with the  equilibrium observables.

\subsubsection{One-time-quantities.}  

The order parameter \be
q(t_\mathrm{w})=\frac{1}{N}\sum_{i}q_{i}(t_\mathrm{w})\,,\ee vanishes in the
nonequilibrium regime (so the system is much bigger than the coherence length,
$\xi(t_\mathrm{w})$).  We define the SG susceptibility as
\be\label{eq:SGsusc}\chi_\mathrm{SG}(t_\mathrm{w})=N\overline{q^2(t_\mathrm{w})}\,.\ee The long
$t_\mathrm{w}$ limit of $\chi_\mathrm{SG}(t_\mathrm{w})$ is proportional to
the non-linear magnetic susceptibility, but only in the paramagnetic phase. In
the SG phase, for an infinite system, $\chi_\mathrm{SG}$ grows with
$t_\mathrm{w}$ without bound (in fact, as a power of $\xi(t_\mathrm{w})$).

The Binder parameter provides us with information about the fluctuations
\be\label{eq:NEQbinder}
B(t_\mathrm{w})=\frac{\overline{q^4(t_\mathrm{w})}}{\overline{q^2(t_\mathrm{w})}^2}\,.\ee
In the Gaussian regime $B\!=\!3$. In a ferromagnetic phase, $B\!=\!1$.  In the
SG phase, the long time limit and the infinite size limit do not commute. If
one takes first the thermodynamic limit, one stays forever in the $q=0$ sector
of the nonequilibrium dynamics. In this regime $B\!=\!3$ since the
fluctuations are Gaussian. On the other hand, if one takes before the limit of
long $t_\mathrm{w}$, thermal equilibrium is reached. $B$ grows with the
temperature from $B\!=\!1$ at $T=0$.  The equilibrium paramagnetic phase is in
Gaussian regime.

\subsubsection{ Two-time-quantities} 
 The correlation spin function tells us about the memory kept by the system at
 time $t+t_\mathrm{w}$, about the configuration at $t_\mathrm{w}$:
 \be\label{eq:Cttw}
 C(t,t_\mathrm{w})=\frac{1}{N}\overline{\sum_{i}c_{i}(t,t_\mathrm{w})}\,.\ee
 As discussed in Section \ref{sec:FDT}, the SG susceptibility and the time
 correlation function are related through the fluctuation-dissipation
 theorem~\eqref{eq:fdt2},
 $\chi(\omega\!=\!2\pi/t,t_\mathrm{w})\!\propto\![1-C(t,t_\mathrm{w})]/T$,
 which is only valid in equilibrium (then, in the SG phase this is true only
 for $t\ll t_\mathrm{w}$ \cite{bouchaud:97}).

On the other hand, when $t_\mathrm{w}$ is fixed, $C(t,t_\mathrm{w})$
is just the thermo-remanent magnetization presented in \eqref{eq:TRM}
and Figure \ref{fig:fullaging-intro}. Indeed, using the gauge transformation
\eqref{eq:gi-intro}, it is possible to rewrite an ordered configuration (by
an external magnetic field, for instance), as the spin configuration
found at time $t_\mathrm{w}$ after a random start.

The link correlation function ($z$ being the connectivity of the
system) \be\label{eq:clink}
C_\mathrm{link}(t,t_\mathrm{w})=\frac{1}{zN}\overline{\sum_{\mean{ik}}\ c_{i}(t,t_\mathrm{w})c_{k}(t,t_\mathrm{w})}\,,\ee
carries the information of the density of the interfaces between
coherent domains at $t_\mathrm{w}$, that at $t+t_\mathrm{w}$ have
flipped. Indeed, the sum $\sum_{\mean{ik}}$ runs only over the
connected spins.

\subsubsection{Spatial correlation functions} \label{sec:spatialcorrdef}
In all the scenarios considered above for the SG phase, the dynamics are
characterized by the growth of coherent domains. For this reason, we introduce separately the spatial
correlation functions.

For the sake of simplicity, for these definitions, we will label the spins in
the lattice by their spatial coordinates $\V{x}$, instead of just the index
$i$ as done before. Then, the spatial correlation function is
\be\label{eq:c4def}
c_4(\V{r},t_\mathrm{w})=\frac{1}{N}\overline{\sum_{\V{x}}q_{\V{x}}(t_\mathrm{w})q_{\V{x}+\V{r}}(t_\mathrm{w})}\,.\ee
The large distance decay defines a coherence length $\xi(\tw)$ through the
scaling of the form \be \label{eq:scalinglongr}c_4(r,\tw)\rightarrow
\frac{1}{r^a}f\paren{\frac{r}{\xi(\tw)}}.  \ee The function $f$ is a
damping function. It must be there, if anything else fails, because of
causality. It is normally assumed an exponential decay.

Note that this $c_4(r,\tw)$ is related to the SG susceptibility [defined in
\eqref{eq:SGsusc}] by means of the relation
\be\label{eq:rel_sus_c4}
\chi_\mathrm{SG}(t_\mathrm{w})=\int \D^D\V{r}\ c_4(r,\tw).
\ee

We define one additional spatial correlation function that takes aging
explicitly into account. For this reason, we introduce the
{\em non-equilibrium spatial correlation function},
\begin{equation}\label{eq:c22}
c_{2+2}(\V{r};t,\tw)= \frac{1}{V}\sum_{\V{x}}\ \overline{\langle s_{\V{x}}(\tw) s_{\V{x}}(t+\tw) s_{\V{x}+\V{r}}(\tw) s_{\V{x}+\V{r}}(t+\tw)\rangle}.
\end{equation}

\subsubsection{Equilibrium Observables}\label{SECT:MODELDEF}
Equilibrium quantities are a straight-forward generalization of the
nonequilibrium ones. In this case the
explicit dependence with time is not longer necessary and magnitudes
are averaged over the time history. We will use now two kinds of
averages, the disorder average $\overline{(\cdots)}$ already introduced,
and the time average, represented by $\mean{\cdots}$.

The Edwards-Anderson order
parameter, the {\em spin overlap}, already defined in \eqref{eq:overlap1}, is
\begin{equation}
q=\frac{1}{N}\sum_i q_i\,,\label{DEF:Q}
\end{equation}
with $q_i$, the overlap field,
\begin{equation}
q_i= s_i^{(1)} s_i^{(2)}\,.\label{Q-FIELD-DEF}
\end{equation}
In particular, it yields the (non-connected) spin-glass susceptibility
\begin{equation}
\chi_\mathrm{NC}(T)= N \overline{\langle q^2\rangle}\,,\label{DEF:CHI}
\end{equation}
that diverges at $T_\mathrm{c}$ with the critical exponent
$\gamma$. For all $T<T_\mathrm{c}$, one expects
$\chi_\mathrm{NC}=\mathcal{O}(N)\,$. We shall also consider the Binder
ratio
\begin{equation}\label{eq:EQbinder}
B(T)=\frac{\overline{\langle q^4\rangle}}{\overline{\langle
    q^2\rangle}^2}\,,
\end{equation}
Which as its nonequilibrium counterpart, takes $\lim_{L\to\infty} B=3$
for all $T>T_c$. Its behavior in the low temperature phase is
controversial. For a {\em disguised ferromagnet} picture one expects
$B$ to approach $1$ in the limit of large lattices. On the other hand,
for an RSB system one expects $1<B<3$ in the SG phase
($T<T_\mathrm{c}$).

The link overlap is  \be\label{eq:qlink}
Q_\mathrm{link}=\frac{1}{zN}\overline{\mean{\sum_{\mean{ik}}
    q_{i}q_{k}}}\,.\ee We will devote the next section to discuss the
implications of this observable.

Finally, we introduce the overlap spatial correlation function in equilibrium
\begin{equation}\label{eq:C4}
c_4(\V{r})= \frac{1}{V}\sum_{\V{x}}\ \overline{\langle q_{\V{x}}\, q_{\V{x}+\V{r}}\rangle}\,.
\end{equation}

\subsection{On the link overlap and the overlap equivalence}\label{sec:overlapequiv}
We devote this section to the link overlap $Q_\mathrm{link}$ \eqref{eq:qlink}
[or its nonequilibrium counterpart $C_\mathrm{link}(t,t_\mathrm{w})$
  \eqref{eq:clink}] and its relation with the spin overlap $q$ \eqref{DEF:Q}
[or $C(t,t_\mathrm{w})$ \eqref{eq:Cttw} in nonequilibrium].  From a
mathematical point of view, the square of the overlap represents the
covariance of the Hamiltonian in the SK model, while the link overlap is the
covariance of the Hamiltonian in the EA model. For this reason, it has been
suggested that the $Q_\mathrm{link}$ should be the fundamental quantity to
describe the SG phase below the upper critical
dimension~\cite{contucci:03,contucci:05,contucci:06}.

When summing over all the spins in the system, a priori, these two overlaps
should lead to different global order parameters.  Indeed, $Q_\mathrm{link}$
refers to the correlation between the links, and $q$ between the
spins. However, in the SK model (defined in Section~\ref{sec:SK}) they are
essentially the same quantity. In fact, it is trivial to check that
$Q_\mathrm{link}=q^2$ [and
  $C_\mathrm{link}(t,t_\mathrm{w})=\caja{C(t,t_\mathrm{w})}^2$]. On the other
hand, when one considers only nearest neighbors interactions, like in the EA
model, these two magnitudes have different behaviors under spin inversion: $q$
undergoes changes of volume sizes after spin flips, while $Q_\mathrm{link}$
suffers only surface changes. Indeed, after a domain flip, $Q_\mathrm{link}$
is only affected by the links that cross its domain's surface. 

According to the previous discussion, in the droplet theory (where the
domains' surface-to-volume ratio vanishes in the large-$L$ limit),
$Q_\mathrm{link}$ should become constant, no relation with $q^2$ should be
observed. On the contrary, in the RSB theory (were the domains are
space-filling) these two magnitudes would be completely correlated, as in the
SK model. In other words, in the RSB theory, the link overlap distribution is
also non trivial.

In fact, this relation between the overlaps is known as {\em overlap
  equivalence}~\cite{parisi:00}. This property states that all
the mutual information about two equilibrium configurations is encoded in the
mutual overlap, and thus, no other definition of overlap (such as the
$Q_\mathrm{link}$) can increase the knowledge of the system. This overlap
equivalence is equivalent to ultrametricity, but a lot simpler to check. Its
validity in MF is straight-forward but there is still a broad discussion about
its validity in the $D=3$ EA model. Indeed, according to the previous
discussion, the overlap equivalence is not fulfilled in the droplets theory.

\subsection{The TNT picture}\label{sec:TNT}

The above discussion about the geometry of the excitations and its relation
with the behavior of the overlap and the link overlap, led to an intense study
of the properties of these two magnitudes in realistic EA models. Simulations
leading to a somehow mixed scenario between the droplets and the RSB
predictions~\cite{krzakala:00,palassini:00}. The emerging picture was named
\ac{TNT}, accounting for a trivial distribution of $Q_\mathrm{link}$ and a non
trivial distribution of $q$. Let us explain this last statement.

The new model would behave like the droplets theory for finite length scales
and like RSB for system sizes excitations.  In that sense, there would be two
transient stiffness coefficients $\theta$, as introduced in
\eqref{eq:freedroplets}. One, $\theta_\mathrm{l}$, valid for local excitations
$\ell\ll L$, whose free energy would scale like droplets' $F(\ell)\propto
\ell^{\theta_\mathrm{l}} $. And another, $\theta_\mathrm{g}\approx 0$ (like
MF), for the global excitations $\ell=O(L)$. This allowed large excitations
lead to a probability distribution of the overlap composed by many valleys,
like in RSB. Then, a non-trivial distribution for $q$. However, concerning
the geometry of these excitations, they should be like the droplets, since
locally this scenario behaves like them. Then, one would expect a vanishing surface-volume
ration and thus, a trivial distribution for the link overlap.
According to this scheme, as in the droplets, no overlap equivalence should be found.

\subsection{Statics-dynamics relation: the time-length dictionary}\label{sec:tldictionary}
As discussed many times in this section, the theory of spin glasses accounts
for an equilibrium phase, which for experimental samples is unreachable in a
laboratory. However, it is assumed that this equilibrium phase still
conditions the nonequilibrium behavior.

However, one should make this above statement quantitative. We recently
established a quantitative relation between the statics and the dynamics
correlation functions~\cite{janus:08b,janus:10}. The proposal
is that the equilibrium correlation functions computed in finite systems
reproduce the nonequilibrium counterparts in the thermodynamic limit but for
finite times. The idea besides this statement is that a system with finite
coherence length $\xi(\tw)$ can be regarded as a collection of finite systems
with $L\sim\xi(\tw)$ in equilibrium. If this relation holds, it is possible to
establish a time-length dictionary $\tw\leftrightarrow L$.

The goal is to relate the equilibrium correlation function $c_4(\V{r})$ in
\eqref{eq:C4} with the two-times spatial correlation function
$c_{2+2}(\V{r},t,\tw)$ in \eqref{eq:c22}. Now, the explicit dependency on
$\tw$ is removed using $L\sim\xi(\tw)$ and the dependency on $t$,
is taken from the two-times correlation function $C(t,\tw)$ defined in
\eqref{eq:Cttw}. Indeed, as aging states, there is one-to-one relation between
$t$ and $C(t,\tw)$ for fixed $\tw$. Furthermore, in order to relate
$c_4(\V{r})$ and $c_{2+2}(\V{r},C(t,\tw),\tw)$ one needs to consider the
equilibrium correlation function conditioned to a fixed value of $q$ (as
an analogy to the dependency on $C(t,\tw)$ in the nonequilibrium
counterpart). Summing up, the sought relation is taken between \be
c_{2+2}(\V{r},C(t,\tw),\tw)\leftrightarrow c_4(\V{r}|q).  \ee 

For the $q$-conditioned $c_4(\boldsymbol r |q)$, the natural
election would be using the \ac{pdf} for the overlap
\begin{equation}
p_1(q)  = \overline{\biggl\langle\delta\biggl(q-\frac1N \sum_\bx q_\bx\biggr)\biggr\rangle},
\end{equation}
however, for finite systems this is a non smooth function composed by a sum of
$N+1$ Dirac deltas. As a solution, we considered a smoother version of
it, the convolution of $p_1(q)$ with a Gaussian of width $1/\sqrt N$
\begin{align}\label{eq:SG-p-q}
p(q=c) &= \int_{-\infty}^\infty q'\ p_1(q') \mathscr G_N(c-q')= \overline{\biggl\langle\mathscr G_N\biggl(c-\frac{1}{N} \sum_\bx q_\bx\biggr)\biggr\rangle}, \\
\mathscr G_N(x) &= \sqrt{\frac{N}{2\pi}} \E^{-N x^2/2}.
\end{align}
Using this $p(q)$, they defined a new conditional 
expectation value for fixed $q$ as
\begin{equation}\label{eq:SG-E}
\mathrm{E}(O|q=c) = \frac{\overline{\biggl\langle O \mathscr G_N\biggl(c-\frac1N \sum_\bx q_\bx\biggr)\biggr\rangle}}{\overline{\biggl\langle \mathscr G_N\biggl(c-\frac1N \sum_\bx q_\bx\biggr)\biggr\rangle}}.
\nomenclature[E(O)]{$\mathrm{E}(O\vert q)$}{Conditional expectation value of $O$ at fixed $q$}
\end{equation}
Using this definition, the  standard expectation values
can be easily computed  from these $\mathrm{E}(O|q)$, by means of
\begin{equation}
\overline{\braket{O}} = \int_{-\infty}^\infty q\ p(q)\mathrm{E}(O|q).
\end{equation}
Finally, the fixed-$q$ correlation function we were looking for is obtained as 
\begin{equation}\label{eq:SG-C4-q}
c_4(\boldsymbol r|q) = \mathrm{E}\left( \frac1N \sum_\bx q_\bx q_{\bx +\boldsymbol r}\middle| q\right).
\end{equation}

\begin{figure}
\centering
\includegraphics[height=.7\linewidth,angle=270]{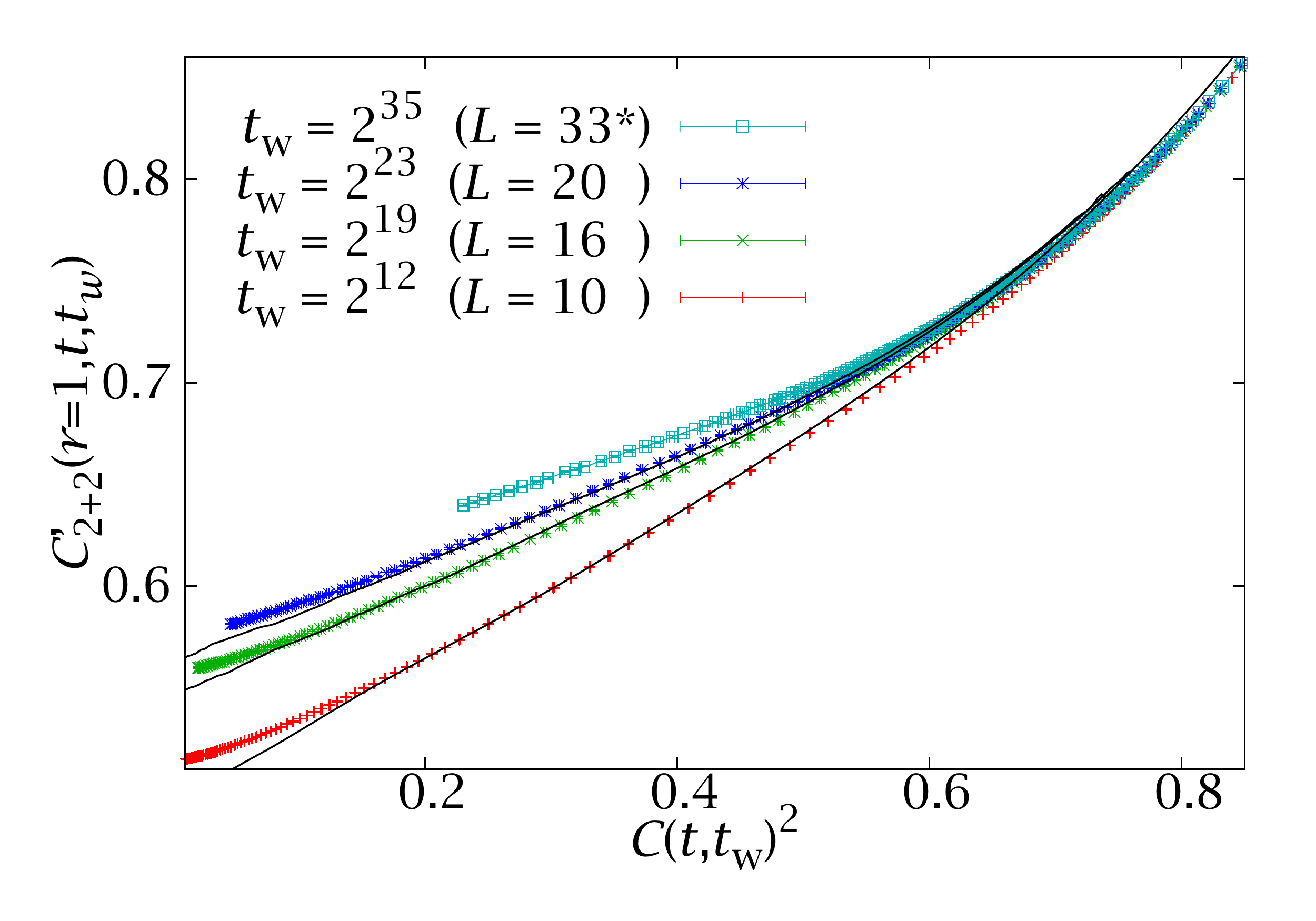}
\caption[Time-length dictionary for $r=1$]{ Comparison of the
  nonequilibrium curve $C_{2+2}'(r=1,C^2,\tw)$ vs. $C^2$ for for
  several waiting times at $T=0.7$ (in points), with the equilibrium $C_4(r=1|q)$ vs. $q^2$ (in continuous lines). 
Figure taken from~\cite{janus:08b}.
\index{correlation function (dynamics)!spatial|indemph}
\index{correlation function (equilibrium)!spatial|indemph}
\index{statics-dynamics equivalence|indemph}
\index{time-length dictionary|indemph}
\label{fig:SG-statics-dynamics-PRL}
}
\end{figure}
\begin{figure}
\centering
\includegraphics[height=.7\linewidth,angle=270]{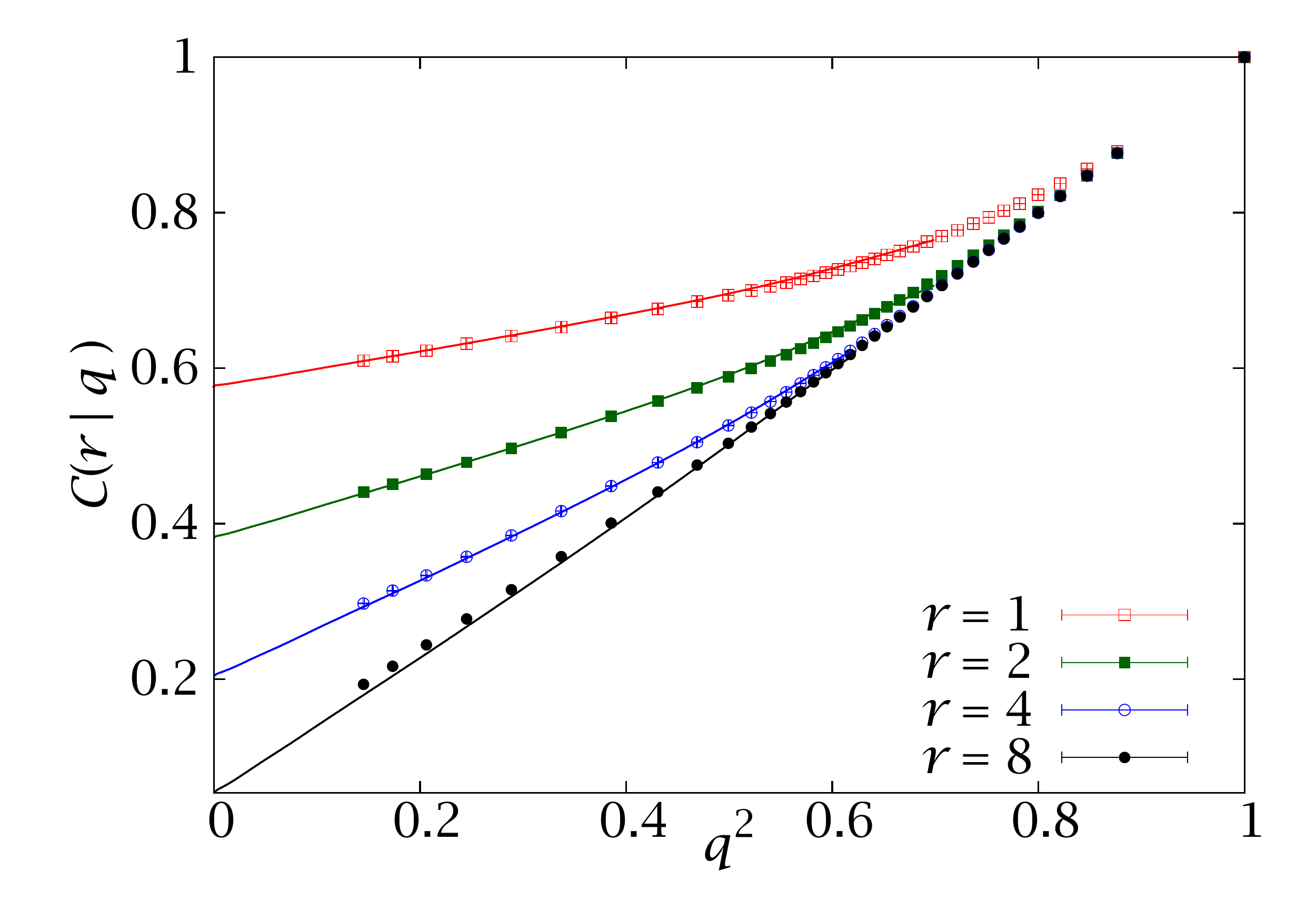}
\vspace*{1.5cm}
\includegraphics[height=.7\linewidth,angle=270]{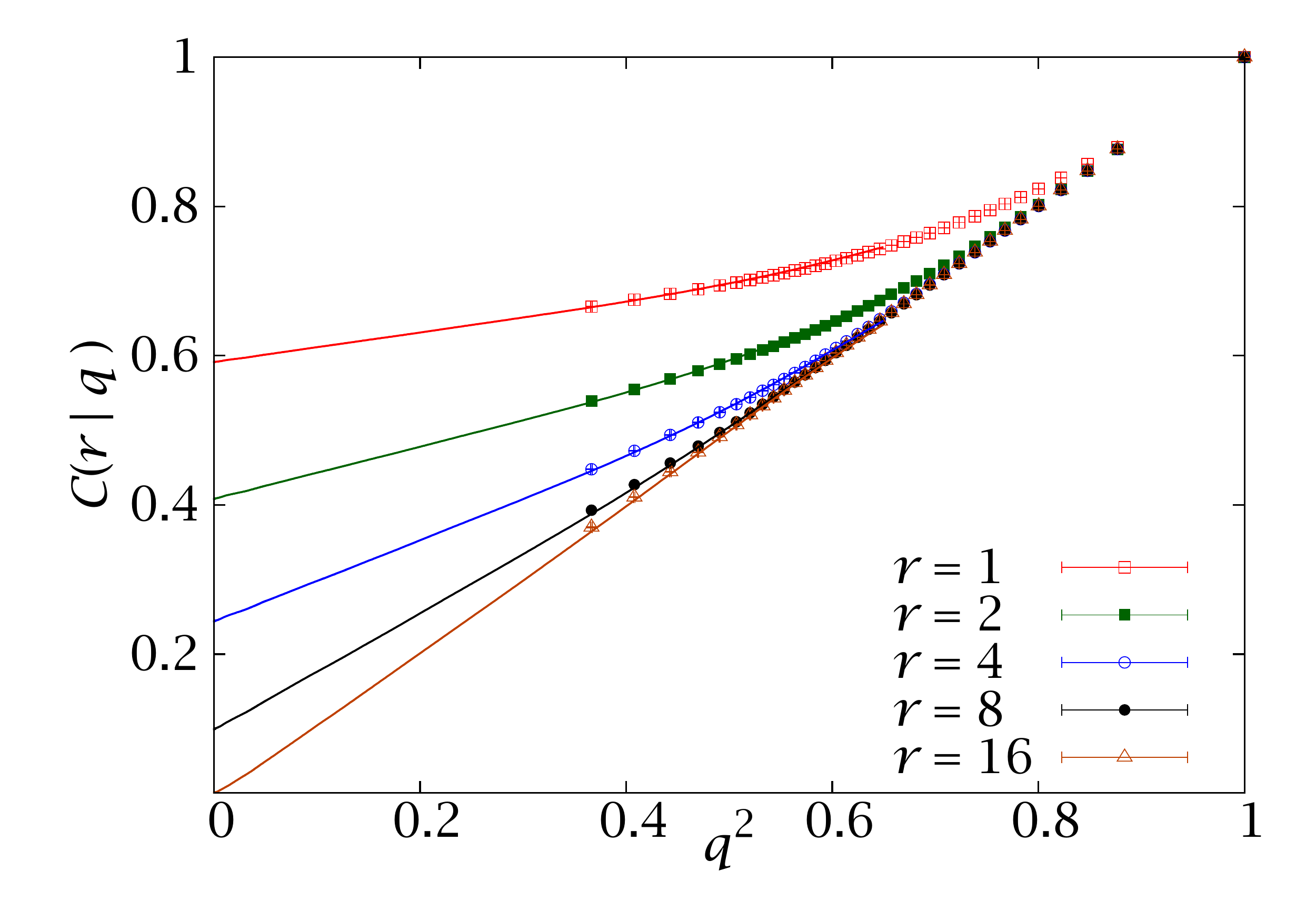}
\caption[The time-length dictionary for $r>1$]{%
Comparison of our non-equilibrium correlation functions
$C'_{2+2}(r,C^2,\tw)$ (points) and  equilibrium $C_4(r|q)$ (lines)
at $T=0.7$. In the top  panel we compare $L=24$ with $\tw=2^{26}$
and on the bottom one we consider $L=32$ and $\tw=2^{31}$. Figure taken from~\cite{janus:10}.\index{statics-dynamics equivalence|indemph}
\label{fig:SG-statics-dynamics-JSTAT}
}
\end{figure}

In the first attempt to establish this time-length dictionary
~\cite{janus:08b}, the non-equilibrium curve $c_{2+2}'(r=1,t,\tw)$ as a
function of $\caja{C(t,\tw)}^2$ was compared to the equilibrium $c_4(r=1|q)$
versus $q^2$. Both curves could be superposed almost perfectly (see Figure
\ref{fig:SG-statics-dynamics-PRL}) if we take as time-length dictionary
\index{time-length dictionary}
\begin{equation}
L(\tw)\approx3.7\xi(\tw).
\end{equation}
In a second attempt in~\cite{janus:10} unprecedentedly large sizes could be
thermalized thanks to \textsc{Janus} computer, reaching $L=32$
(see~\cite{yllanes:11} for details), leading to more precise results. The
equivalent curves are displayed in Figure~\ref{fig:SG-statics-dynamics-JSTAT}
for other values of $r(>1)$.  According to the time-length dictionary, the  $L=32$ equilibrium simulations can be corresponded with non equilibrium simulations with $\tw\approx2^{31}$, while the $L=24$ correspond to $\tw\approx2^{26}$.

The equivalence has some striking consequences. The longest times $\tw$
studied in these numerical works~\cite{janus:08b,janus:10,janus:10b}, are,
thanks to Janus computer, very near to the experimental times. Indeed, these
works reached $2^{32} \text{ MCS}$ which corresponds roughly to $0.1\text{
  seconds}$. This are not that short times in comparison with the experiments,
and still, they correspond to equilibrium simulations for rather small
systems. This leads to the direct question of how important is the
thermodynamic limit in equilibrium for the experiments.  In fact, we
considered a typical experiment, which takes place in $1$ hour, and
extrapolated the correspondent length for equilibrium. The result was that the
relevant length scale for nonequilibrium experiments~\cite{janus:10b} was
$L\sim 100$, which is definitely very far away from the thermodynamic
limit. This relation brings closer the simulations to the experimental
behavior. Indeed, the state-of-the-art in the equilibrium studies is around
$L=32$, then, not that far as one would expect.

\renewcommand{\thesection}{\arabic{chapter}.\arabic{section}}
\renewcommand{\thesubsection}{\texorpdfstring{\oldstylenums{\arabic{chapter}.\arabic{section}.\arabic{subsection}}}{\arabic{chapter}.\arabic{section}.\arabic{subsection}}}
\renewcommand{\thesubsubsection}{\texorpdfstring{\oldstylenums{\arabic{chapter}.\arabic{section}.\arabic{subsection}.\arabic{subsubsection}}}{\arabic{chapter}.\arabic{section}.\arabic{subsection}.\arabic{subsubsection}}}
\renewcommand{\thefigure}{\oldstylenums{\arabic{chapter}.\arabic{figure}}}
\renewcommand{\thetable}{\oldstylenums{\arabic{chapter}.\arabic{table}}}
\renewcommand{\theequation}{\oldstylenums{\arabic{chapter}.\arabic{equation}}}

\chapter{The hypercube model}\label{chap:hypercube} 

\section{Introduction}

As discussed in Chapter \ref{chap:intro-sg}, our understanding of the spin
glass phase comes, to a considerable degree, from analytical results obtained
under mean field (MF) approximations. As discussed then, the mean field
solution to the EA model (the Sherrington-Kirkpatrick (SK) model, see Section
  \ref{sec:SK}) still accounts for  most of the complex physics found spin
glasses, and draws a picture of an intricate structure for the SG
phase characterized by an infinite number of equilibrium states following an
ultrametric organization. Even though the SK model allows analythical
calculations, there are many phenomena which are not yet understood not even
in MF.

In fact, the understanding of the nonequilibrium behavior is not yet well
understood not even in the SK model.  And, as was widely supported all over
the Chapter \ref{chap:intro-sg}, the nonequilibrium is the only relevant
regime for a real spin glass. However, the analythical treatment for the
simplest experiments of aging is difficult, not to say anything about
explaining more complicate temperature protocols such as the one necessary for
the memory and rejuvenation experiments discussed in
Section~\ref{sec:memory-rejuvenation}. Furthermore, not only the
nonequilibrium is not fully understood, some equilibrium effects such as the
temperature chaos (see Chapter \ref{chap:chaos}) are still under investigation~\cite{parisi:10}.

For this reason, even at \ac{MF} level, non perturbative tools, such as
\ac{MC} calculations, are still necessary. Indeed, simulations in \ac{MF}
models, understanding MF model as a model where the MF approximation becomes
exact in the thermodynamic limit, can be most useful considering that our
large theoretical understanding of these models provides us with much
extra information when approaching other unknown  phenomena.

We introduced our first \ac{MF} model in Section \ref{sec:SK} when talking
about the Sherrington-Kirkpatrick model \ac{SK}, which was the simplest
possible MF model for the Edwards-Anderson Hamiltonian \eqref{eq-intro:EAham}.
As explained then, in the SK model all spins are connected, which simplifies a
lot the analytical calculations, but makes simulations unaffordable (energy
calculations for the \ac{MC} tests are $\mathcal{O}(N^2)$). Besides, it lacks either
from a finite coordination number or a notion of neighborhood, which spaces it
out from more realistic models where the spins hardly interact beyond nearest
neighbors.  For these two reasons, a whole family of solvable \ac{MF} models
with finite connectivity has been proposed in the last
years,~\cite{parisi:06,mezard:01}.  Among all the MF models, those formulated
on graphs have become very popular. First, because they allow an analytical
approach based on the statistical mechanics' iterative methods typical for
tree-like structures, and second, because they are deeply connected with the
random optimization problems in computer science (see
Section~\ref{sec:introQA}), which turn out to have finite connectivity too.

We discuss here some popular tree-like lattices with finite connectivity $z$
for the connections between spins before introducing our own new model.  We
start with the so-called {\em Cayley tree}. In this graph, starting from site
$i=0$, one chooses randomly a first shell of $z$ neighbors. Afterwards, each
of these spins in the first shell is connected again with $z-1$ new neighbors
for the second shell and so on, until there are no more new spins to
connect. Thus this graph is a true tree, in the sense that nearest neighbors
are only connected by their common link, there is no overlap between new
neighbors. However, in such construction, there is finite number of spins
lying on the boundary, which makes the system very inhomogeneous. Indeed,
these boundary spins origin properties that are far from the usual finite
dimensional problems. To avoid this problem, the {\em Bethe lattice} is
normally considered. In it, only the first $L'$ shells of the Cayley graph are
considered. This approach works well as long as the graph completely forgets
the information from the boundaries, which is not the case, in general, for
spin glasses, where boundaries still impose some degree of frustration. For
this reason, spin glasses are often defined on other kind of Bethe lattice
structures. Now, let us consider a {\em random graph with fluctuating
  connectivity} known as {\em Erdos-Renyi graph}. In it, each link between the
pair $(i,j)$ is active with probability $z/N$. Then, each spin is connected in
average with $z$ spins. As a subgroup of this last set, one can define the
{\em fixed connectivity random graphs} containing only those graphs where each
spin interacts with exactly $z$ neighbors. We will come back these two models
in further detail later on.

Then, we want to define a spin glass on a graph. As we did when presenting the
SK model, the starting point is the \ac{EA} model discussed in Section
\ref{sec:EA} for Ising spins, i.e. $s_{i}(\pm 1)$. With the sake of
clarity, at variance with the discussion when describing the EA model, we
encode the nearest neighbor summation by introducing a connectivity matrix,
$n_{ik}\!=\!n_{ki}\!=\!1,0$ ($n_{ik}\!=\!1$ as long as spins $i$ and $k$
interact). In addition, we must consider the coupling constants,
$J_{ik}\!=\!J_{ki}$ too carrying the information of the ferromagnetic or
antiferromagnetic character of the interaction (we will consider $J_{ik}\!=1$
for the ferromagnet and $J_{ik}\!=\!\pm 1$ for the SG, which defines our
energy scale). In other words, we consider now two quenched variables (see
Section \ref{sec:replicatrick}), the connectivity matrix $\lazo{n}$ and the
couplings $\lazo{J}$. Using these two kinds of variables, the interaction
energy is now\be\label{eq-hy:H}
\mathcal{H}=-\sum_{i<k}J_{ik}n_{ik}s_{i}s_{k}\,.\ee Now we consider
the Erdos-Renyi graph described above. As defined, this graph is
drawn by connecting each possible couple of spins, $(i,k)$, (among the
$N(N-1)/2$ possible ones) with probability $z/(N-1)$. In terms of the
variables described just above, activating a link means setting
$n_{ik}\!=\!1$.  According to this probability, the number of neighbors of
spin $i$ or coordination number $n_i$, follows a Bernoulli distribution
function \be
p(n_i)=\paren{\begin{array}{c}N-1\\n_i\end{array}}\paren{\frac{z}{N-1}}^n_i\paren{1-\frac{z}{N-1}}^{N-1-n_i},\ee
which tends to a Poisson distribution function with average $z$ (the
connectivity) in the large-$N$ limit, \be
p(n_i)=\frac{z^{n_i}}{n_i!}e^{-z}.\ee We will consider $z\!=\!6$ to mimic a
three dimensional system. This kind of graphs are locally cycle-less: the mean
shortest length among all the closed loops that passes through a given point
is $\mathcal{O}(\log N)$, i.e. the system is still locally tree-like. In order to
support his statement, let us compute the the amount of all possible closed
graphs of length $l$,\footnote{Defining distance between nodes as the minimum
  number of links that must be crossed for going from one node to the other.}
that pass through a given point of the graph, multiplied by the probability of
all the links involved are active,\footnote{Once one spin of the loop is
  chosen, the amount of eligible spins decreases by one, and so on. Indeed, we
talk about a loop of size $l$, but there are only $l-1$ links can be freely
elected. In
  addition, in order to not count the same loop more than once, we must divide
  by $2l$, which takes into account all the possible starting spins within the
  same loop.}
\be\label{eq:loops}\frac{(N-1)(N-2)\cdots(N-l+1)}{2l}\paren{\frac{z}{N-1}}^l=\frac{(N-1)!}{2l(N-l)!}\paren{\frac{z}{N-1}}^l,\ee
taking the $N\rightarrow\infty$ limit, and using the Stirling relation
$n!\approx \sqrt{2\pi n}(n/\E)^n$ for large $n$, we get
\be\frac{z}{2l}\paren{\frac{z}{e}}^{l-1}\frac{1}{N-1}.\ee Then, the only
surviving loops are those whose length is $\mathcal{O}(\log N)$.  In other words, there
are no local loops a the Erdos-Renyi graph. We will see that this condition is
enough for the Bethe approximation to hold.

This spin glass in an Erdos-Renyi graph has a finite connectivity as we needed
for a numerical study, but we still want to step forward and to find a model
that also allows us to define a notion of distance. Indeed, as discussed, the
origin of the \ac{SG} phase is an association of the spins in coherent
domains, and the different theories precisely differ in the properties of
these domains. For this reason, we want a MF model that lets us to explore as
well the growth of a coherence length $\xi(\tw)$.

In this chapter, we will present a new \ac{MF} model for \ac{SG}: the spin
glass on a $D$-dimensional hypercube with {\em fixed}
connectivity~\cite{marinari:95}.  In such a model, as we will discuss later,
the Bethe approximation becomes exact in the thermodynamic limit (which
coincides with the large $D$ limit for this model).  As a consequence, the
statics is of Bethe-lattice type and can be computed.  A nice feature of this
new model, is that it has a natural definition of distance, which allows us to
study the spatial correlations within \ac{MF} approximation.  In other words:
this \ac{MF} model is more similar to a real $D\!=\!3$ system than any of
those considered previously. Indeed, it let us to compute space correlation
functions.

\section{The hypercube model}\label{sec:hypercube}

\begin{figure}
\begin{center}
\includegraphics[scale=0.3,trim=30 0 40 0]{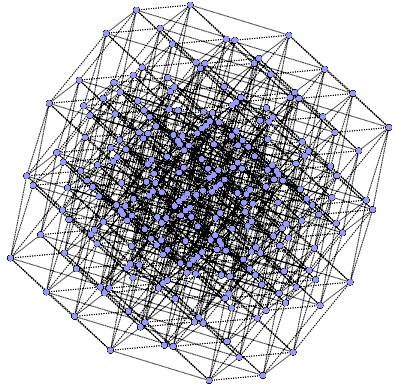}
\caption{Projection of a  $D=8$ hypercube.
Spins are placed in the nodes, represented by blue circles, and links are the edges.}\label{fig:hypercube}
 \end{center}
\end{figure}

A simple alternative consists on formulating the spin model on a
$D$-dimensional unit hypercube, see Figure \ref{fig:hypercube}. Thus, the spins are located in each of the
hypercube vertices $\V{x}$ (then, $N\!=\!2^D$) and the bonds lie on the edges,
${\V{x},\V{\hat\mu}}$, where $\V{\hat\mu}$ labels the $D$ possible unit vectors
in the $D-$dimensional space. We consider periodic boundary conditions, and
then, each spin can be connected with, at most, $D\!=\!\log_2 N$ spins. 
 The
interaction energy \eqref{eq-hy:H} is now written as
\be\label{eq:Hhy}
\mathcal{H}=-\frac{1}{2}\sum_{\V{x},\V{\hat \mu}}J_{\V{x},\V{\hat
    \mu}}n_{\V{x},\V{\hat \mu}}s_{\V{x}}s_{\V{x}+\V{\hat \mu}}\,,\ee
where $n_{\V{x},\V{\hat \mu}}=1$ if spins $s_{\V{x}}$ and
$s_{\V{x}+\V{\hat \mu}}$ interacts, and $n_{\V{x},\V{\hat \mu}}=0$ if not. In the sections below,
we will discuss how to distribute these connectivity variables.

Note that, at variance with other infinite-dimensional graphs, this hypercube
model has at least two natural notions of distance: Euclidean metrics and the
postman metrics. In the postman metrics, the distance between two points,
$\V{x}$ and $\V{y}$, is given by the minimum number of edges, either occupied
or not, that must be covered when joining $\V{x}$ and $\V{y}$. The two
distances are essentially equivalent, since the Euclidean distance between two
sites in the hypercube is merely the square root of the postman distance.

In the following we shall use the postman metrics, which has some amusing
consequences. For instance, our correlation-length will be the {\em square} of
the Euclidean one, thus yielding a critical exponent $\nu=1$, doubling the
expected $\nu_\text{\ac{MF}}=1/2$. Of course, if we use the Euclidean metric we
recover the usual exponent $\nu=1/2$.

\subsection{Random connectivity model}\label{sec:modelos}

\begin{figure}
\begin{center}
\includegraphics[scale=0.3,trim=30 0 40 0]{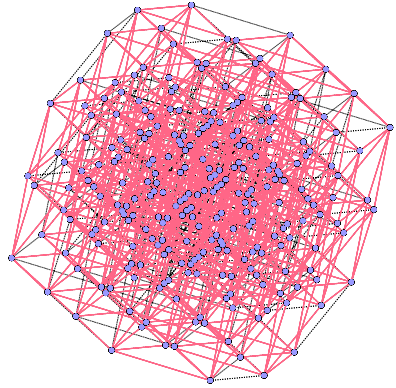}
\caption{Random connectivity hypercube with $z=6$ for a $D=8$
  hypercube. Active links, i.e. $n_{\V{x},\V{\hat{\mu}}}=1$ are represented in
pink.}\label{fig:hypercube_grafo}
 \end{center}
\end{figure}

By analogy with the Poissonian graph, we consider that a link is active
(i.e. $n_{ik}=1$) over each edge with probability $z/D$. We call this model
\textit{random connectivity hypercube}. This model is also a Poisson graph. Indeed,
in the thermodynamic limit ($D\rightarrow\infty$) the probability distribution function for the coordination
number $n_i$ of the $i$th spin is locally
Poissonian, i.e.  \be p(n_i)=\frac{z^{n_i}}{n_i!}e^{-z}.\ee
with average the coordination number $z$. We show in Figure
\ref{fig:hypercube_grafo} an example of this graph for $D=8$.

Again, this graph is also locally cycle-less.  In a hypercube with only
nearest neighbor interactions, for drawing a loop, one must move only along
orthogonal directions. Besides, if one wants to close a loop, each orthogonal
direction can be either covered an even number of times or not covered at
all. Then, when drawing a loop of $l$ links, only half of them can be freely
chosen, the remaining half steps are forced to cover previous
directions. Then, if the length of the loop is $l=2r$, and it contains
movements along $n$ different orthogonal directions ($n\leq r$, directions can
be repeated), the number of possible loops multiplied by the probability that
all the links are active, is \be
f(n)\ D(D-1)\cdots(D-n+1)\paren{\frac{z}{D}}^{2r},\ee where $f(n)$ is a
function only of $n$. Clearly, the less suppressed contribution in $D$
corresponds to $n=r$, i.e. when each direction in the loop is covered only
twice. We may count the number of loops of length $l=2r$ that imply $r$
different orthogonal directions in space:\footnote{Again, once a direction has
  been taken, there are only $D-1$ available directions for the following
  step, and so on. As before, we can cover the same loop beginning in any of
  its nodes, then we have to divide the final expression by a factor
  $l$. However, in contrast to the Poisson graphs discussed in the previous
  section, in an elementary hypercube, a loop can be covered only in one
  orientation due to its periodical boundary conditions (both orientations are
  equivalent).}
\be \label{eq:loopshy}\frac{D!}{(D-r)!}\frac{(r-1)(r-1)!}{2r}\paren{\frac{z}{D}}^{2r}.\ee
Then, for $D\rightarrow\infty$, the number of loops of length $l=2r$ is
\be\frac{(r-1)(r-1)!}{2r}e^{-r}\frac{z^{2r}}{D^r},\ee then, the density of
closed loops of length $l$ decays, at least, with $D^{-2}$ (i.e. with the
squared logarithm of $N$, as it also happens in the Erdos-Renyi graph).
But there can be still closed loops in the graph, those of length
$\mathcal{O}(D)$. Nevertheless, the absence of closed loops of finite length is sufficient
for the Bethe approximation to be exact in the thermodynamic limit, as we
discuss in Appendix \ref{ap:bethe}.

However, it turns out that the random connectivity hypercube suffers a major
disadvantage. The inverse of the critical temperature in a ferromagnet (see
Appendix \ref{ap:bethe} for details of the calculation) or in a SG
\cite{thouless:86} can be computed within the Bethe
approximation:\be\label{eq:thouless:86}
K_\mathrm{c}^\mathrm{FM}=\mathrm{atanh}\frac{1}{\mean{n}_1-1}\ ,\ K_\mathrm{c}^\mathrm{SG}=\mathrm{atanh}\frac{1}{\sqrt{\mean{n}_1-1}}\,.
\ee In this expression $\mean{n}_1$ is a conditional expectation value for
$n$, the coordination number of a given site in the graph.  This conditional
expectation value is computed knowing for sure that our site is connected to
another {\em specific} site (this is different from the average number of
neighbors of a site that has at least one neighbor!). A simple calculation  (see
Appendix \ref{ap:bethe})
shows that $\mean{n}_1\!=\!1+z-\frac{z}{D}$ in the random connectivity
model. Since $D=\log_2 N$, we must expect huge finite size corrections
($\mathcal{O}(1/\log N)$) at the critical point. Note that this problem is far less
dramatic for a Erdos-Renyi graph where $\mean{n}_1\!=\!1+z-\frac{z}{N-1}$.

The cure seems rather obvious: place the occupied links in the hypercube in
such a way that $n=z$ (here, $z=6$). Unfortunately, drawing these graphs poses
a non trivial problem in Computer Science. Our solution to this
problem is discussed in the next subsection.

\subsection{The fixed connectivity hypercube}\label{sec:fixed-con}

We have not found any systematic way of activating links in the hypercube that
respects the fixed connectivity condition. Thus, we have adopted an
operational approach: the distribution of bonds is obtained by means of a
dynamic \ac{MC}.  We must define a \ac{MC} procedure that generates a set of graphs that
remains invariant under all symmetry transformations of the hypercube
group. We include a detailed description of the program used to implement
this \ac{MC} in a computer in Appendix \ref{app:multispin}.

Specifically, we start with an initial condition in which all bonds
along the directions 1 to 6 are activated (of course, this procedure makes sense only for $D\!\ge\!6$). Clearly enough,
the initial condition verifies the constraint $n=6$. We shall modify
the bond distribution by means of movements that do not change $n$. We
perform what we called a ``plaquette'' transformation (a plaquette is
the shortest possible loop in the hypercube, of length 4). 

\begin{figure}
\begin{center}
 \includegraphics[width=0.75\columnwidth,trim=30 0 40 0]{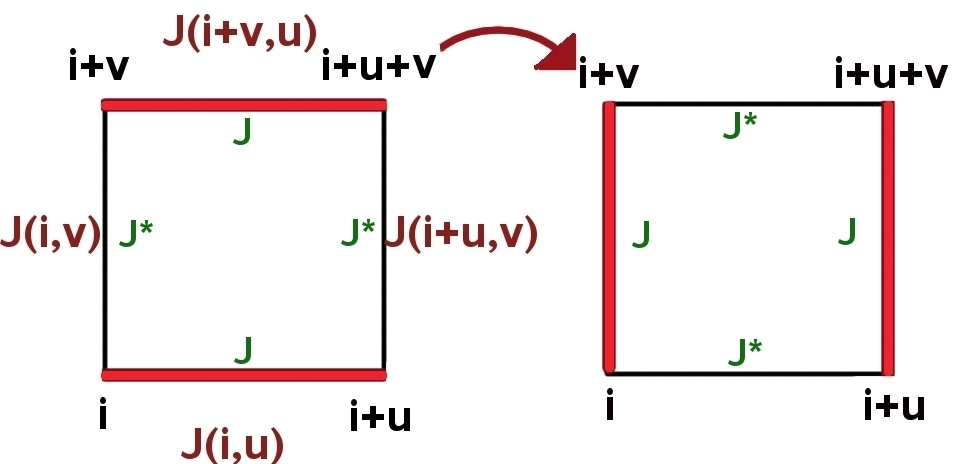}
\caption{Plaquette transformation.}\label{fig:plaquettetrans}
 \end{center}
\end{figure}

 We
randomly pick, with uniform probability, one hypercube plaquette. In
case this plaquette contains only two parallel active links
($n_{ik}=1$), these two links are deactivated at the same time that
the other two are activated. On the opposite case, nothing is
done.\footnote{This movement keeps each vertex connectivity
  unaltered. Besides, a transformation and its opposite are equally
  probable. As a consequence the Detailed Balance Condition is
  satisfied with respect to the uniform measure on the ensemble of
  fixed connectivity graphs. An standard theorem \cite{amit:05} ensures
  that the equilibrium state of this Markov chain is the uniform measure over
  the subset of fixed connectivity hypercubes reachable from the
  initial condition by means of plaquette transformations} This
transformation is illustrated in Figure \ref{fig:plaquettetrans}. This
guarantees that the set of generated graphs is isotropic.

\begin{figure}
\begin{center}
 \includegraphics[angle=270,width=0.8\columnwidth,trim=0 0 0 0]{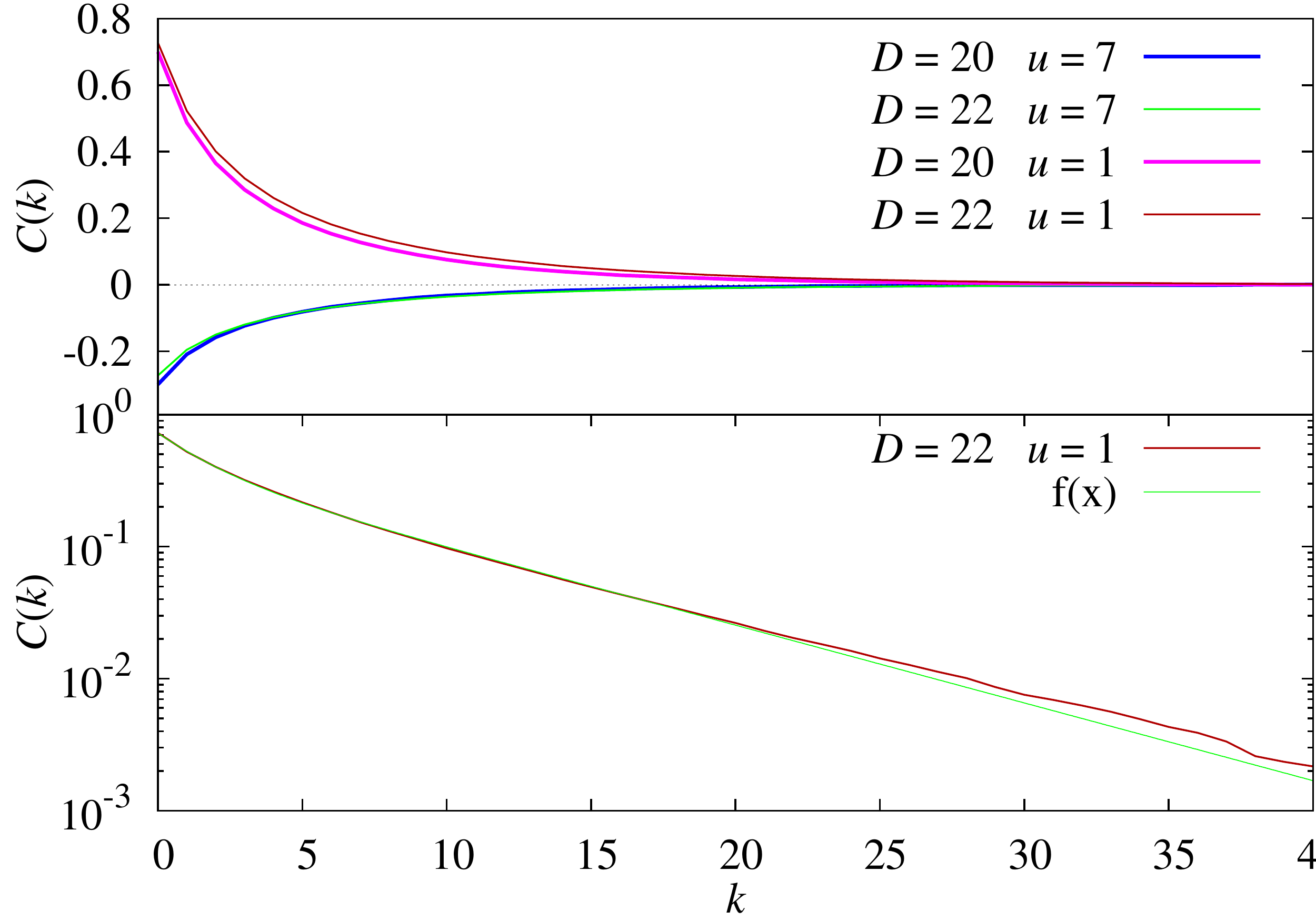}
\caption{ Generation algorithm of fixed
  connectivity graphs: (top) for two system sizes ($D\!=\!20,22$) and two spatial
  directions ($u\!=\!1,7$), we represent the density of occupied edges as
  function of the \ac{MC} time. As \ac{MC} time goes on, the system recover the lost
  isotropy induced by the initial condition. (Bottom) Fit of the upper data
  for $D\!=\!22$ and $u=1$ to a double exponential $A\mathrm{e}^{-x/B}+C\mathrm{e}^{-x/D}$
  obtaining $B\approx 7.37N$ and $D\approx 1.76N$.  }\label{fig:corrtimes}
 \end{center}
\end{figure}

In order to this procedure to be useful, the dynamic \ac{MC} correlations
times must be short. In Figure \ref{fig:corrtimes}, we show the \ac{MC}
evolution of the system isotropy. We make $kN$ plaquette transformations, and
we control the density of occupied bonds in two directions: the first
direction (initially occupied in every vertex) and the seventh direction
(initially unoccupied). As we see, for two different system sizes, we get
short isotropization exponential times (for $D\!=\!22$ we get
$\tau_\mathrm{exp}\!\approx \! 7.4 N$).\footnote{Note that the
  article~\cite{fernandez:09f} we presented $\tau_\mathrm{exp}\!\approx \! 4.2
  N$. However, a refined analysis using two exponentials instead of just one
  for the fit, leaded to this more accurate new result.} For this reason, we
assume that taking $k\!=\!100$ is long enough to ensure that the
configurations obtained are completely independent from the initial condition.

\begin{figure}
\begin{center}
 \includegraphics[angle=90,width=0.7\columnwidth,trim=40 0 40 0]{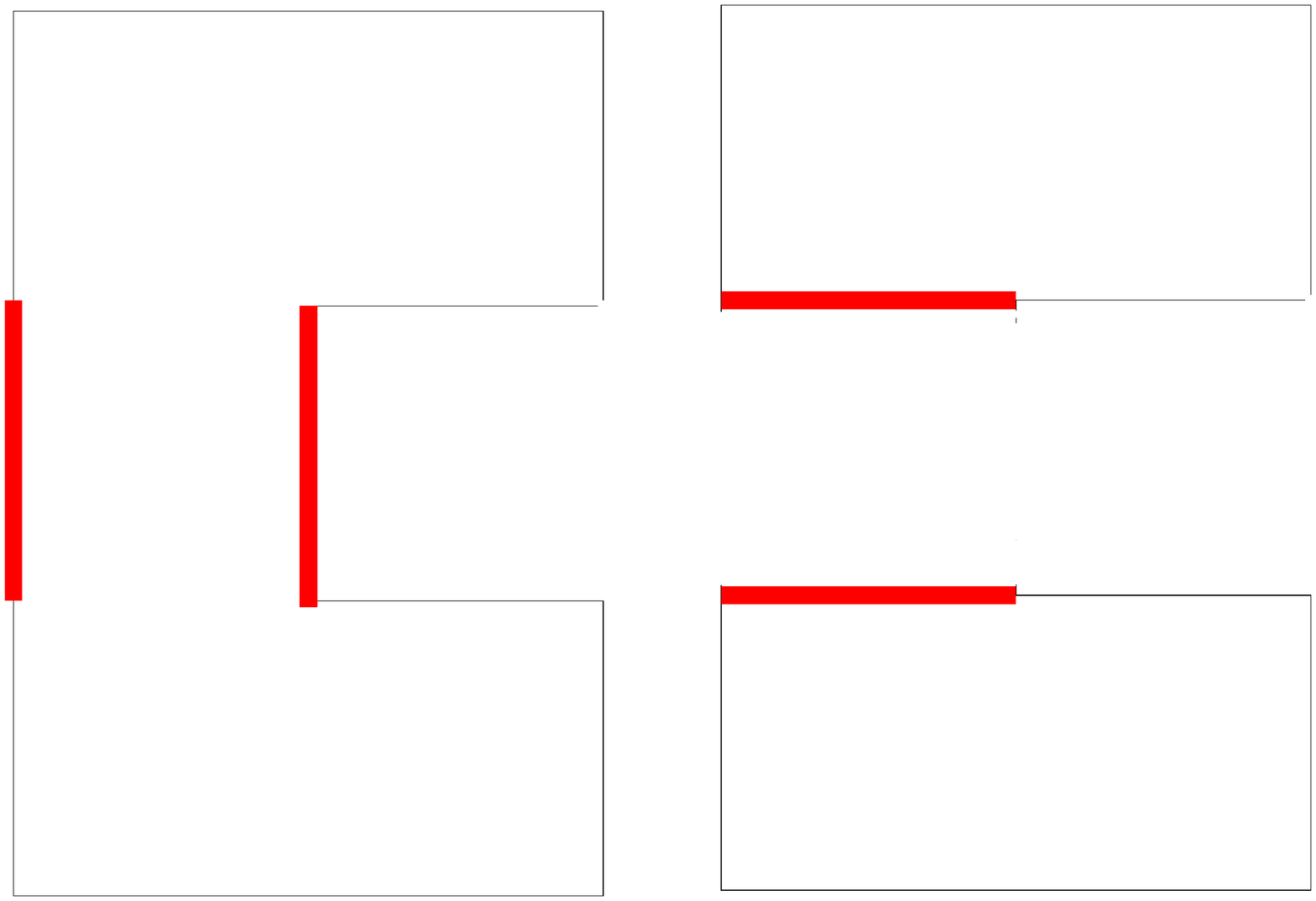}
\caption{Illustration of the effect of a plaquette transformation on a loop of
  active links, as discussed in the main text, it can never open a loop. Here
  we present a projection in two dimensions of a small loop of active links
  (in lines) in the hypercube,
  before and after applying a plaquette transformation represented by thick
  red lines.}\label{fig:loops}
 \end{center}
\end{figure}

At this point, a question arises about the completeness of set of graphs we
can generate by means of this procedure. Can we achieve all the possible
graphs of fixed connectivity in the hypercube? or on the contrary, we only
create graphs within a fixed subgroup. Nevertheless, as we will show below,
most of the sample dispersion is induced by the coupling matrix
$\lazo{J_{ik}}$. One could argue that there the generated set of graphs is
incomplete for a simple reason: the plaquette transformation cannot break
loops. Indeed, when we interchange neighboring links we can only either join
two different loops or split up a loop into two loops as shown in Figure
\ref{fig:loops}. Due to the hypercube boundary conditions, in the initial
configuration all sites belonged to closed loops. This situation cannot be
changed by plaquette transformations. However, this objection does not resist
a close inspection. In fact, a non-closed lattice path formed by occupied
links should have an ending point with an {\em odd} coordination number, which
violates the constraint $n=z$ for any even $z$. Thus, all lattice paths
compatible with our fixed connectivity constraint, do form closed loops.  This
argument, as well as the numerical checks reported below, make us confident
that the set of generated graphs is general enough for our purposes. Actually,
we conjecture that our algorithm generates \textit{all} possible fixed
connectivity graphs with $z$ even.

One may worry as well about the applicability of the Bethe approximation to
the fixed connectivity model, since all loops are closed. Actually, the
crucial point to apply the Bethe approximation is that the probability of
having a closed path of any {\em fixed} length should vanish in the large $D$
limit.  It is easy to prove for the random
connectivity model . In the fixed connectivity case, one may argue as
follows. Let us imagine a walk over the closed path. On the very first step,
the probability that the chosen link is present is $z/D$, whereas in the
following step the probability of finding the link is $(z-1)/(D-1)$ in the
limit of large $D$ (since one of the $z$ links available at the present site
was already used to get there).  This estimate implicitly assumes that the
occupancy of different links is statistically independent. The independence
approximately holds for large $D$ and becomes exact in the $D\to\infty$ limit,
where occupied links form a diluted set. At this point, the estimate of the
number of paths of any given fixed length in the large $D$ limit is analogous
to the one performed for the random-connectivity case but changing the
probability of active link. In other words, one finds that, in the fixed
connectivity case,  the number of closed loops of a given length per site
also decays at least as $\mathcal{O}(1/D^2)$.

In addition to the above considerations, we can compute numerically the
probability of having graphs of a given length in our set of generated graphs
for a finite dimension $D$. The idea is to obtain the length of the second
shortest path that joins two connected nearest neighbors $i_1,\ i_2$ in the
hypercube, and this we can by iterating the connectivity matrix. In fact, we
consider a truncated connectivity matrix, $\tilde{n}$, that coincides with the
true one, $n$, but for the link $i_1-i_2$, which is deactivated: $\tilde
n_{i_1,i_2}=\tilde n_{i_2,i_1}=0$. We take a starting vector $\V{v}^{(0)}$
with all its components set to zero but the component $i_1$ which is set to
one. We iteratively multiply the vector by the truncated connectivity matrix,
i.e. $\V{v}^{(t)}=\tilde n \V{v}^{(t-1)}$, until the $i_2$-th component is
nonzero. The sought length is just the minimum value of $t$ that fulfills the
stopping condition.

\begin{figure}[h]
\begin{center}
 \includegraphics[angle=270,scale=0.45, trim=0 70 0 70]{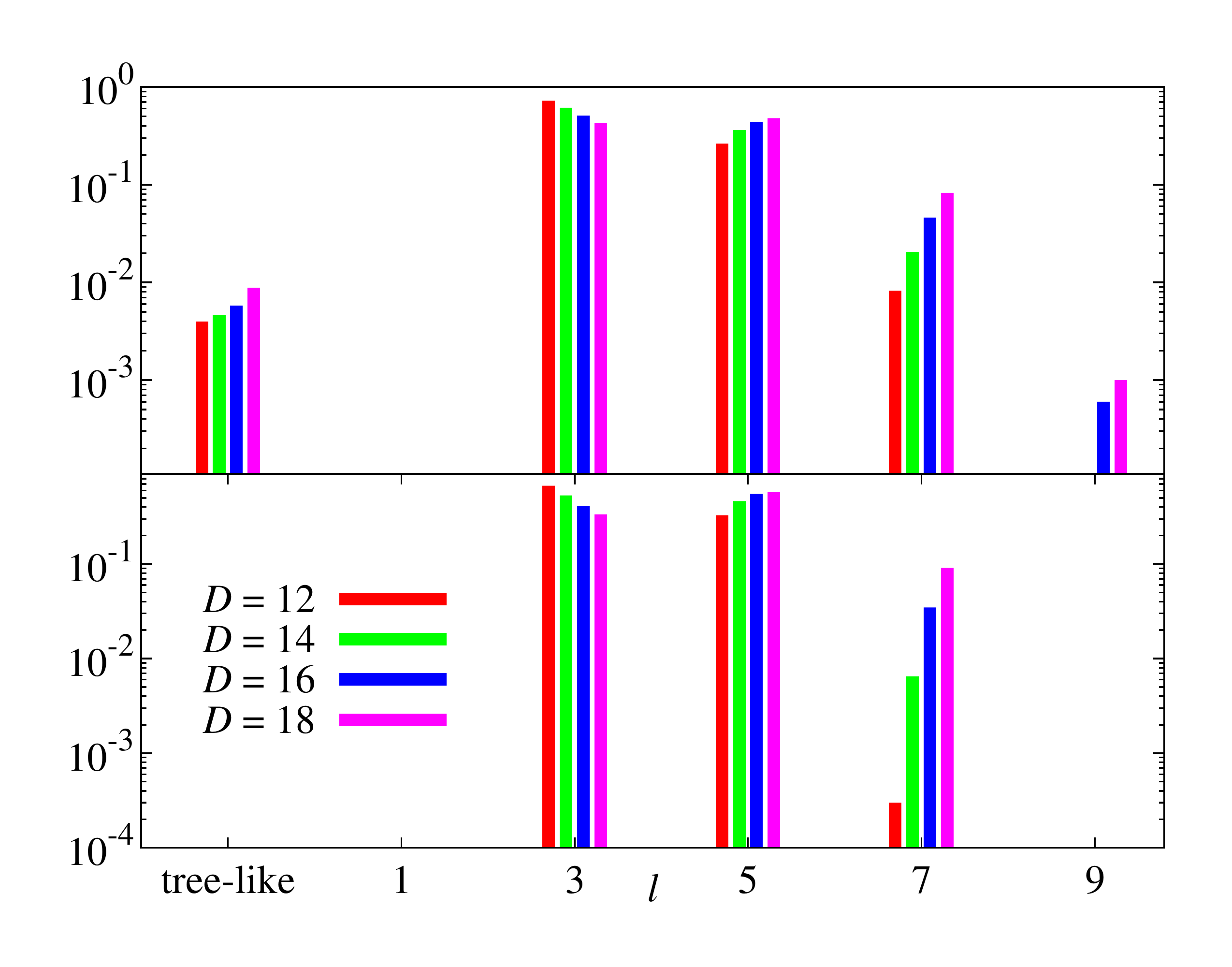}
\caption{ Probability distribution function of the length of the {\em second
    shortest} path joining nearest-neighbors in hypercubes of
  random-connectivity (\textbf{top}) or fixed-connectivity
  (\textbf{bottom}). For both panels, the average connectivity is $z=6$ and
  several system dimensions have been considered. Lines have been slightly
  displaced in order to help the visualization. Mind that the vertical axis is
  shown in logarithmic scale. }\label{fig:min_distance}
\end{center}
\end{figure}

In Figure \ref{fig:min_distance}, we compare the probabilities for the length of
such paths in the random (top) and fixed (bottom) connectivity models, for
different system sizes obtained by averaging over $10^4$ set of graphs. In both cases,
we note that the maximum of the probability shifts to larger length as $D$
grows. We note as well that, for fixed connectivity, no tree-like graph
arises.\footnote{We say that a graph is a tree-graph if, once the link between
  two neighboring spins is removed, there is no way of joining them following
  any other path.}

\subsection{Ising model in the hypercube}
As a first check of the model, we study numerically the Ising model on these
two kinds of hypercube graphs. We will see that the random-connectivity model
suffers from very strong finite size effects that makes it essentially useless
for computer simulations. However, we will show that this problem is
originated precisely in the randomness on the connectivity, and then, the cure
is as simple as to fix the connectivity.

The Ising model is simpler than the spin glass. The interaction now is always
ferromagnetic, i.e. $J_{\V{x},\V{\hat\mu}}=1$, and thus the randomness is only
introduced through the connectivity matrix
$n_{\V{x},\V{\hat\mu}}$. Besides, as explained in Appendix \ref{ap:bethe}, one
can obtain many analytical results for this model, always under the Bethe
approximation.  For this reason, the Ising model is here a perfect benchmark
to study the performance of these newly introduced random graphs, as measured
by the magnitude of finite size effects.

\subsubsection{Simulation details}
For this study, we run simulations using parallel tempering~\cite{hukushima:96}
for the temperature updates, and the spin updates (at constant temperature) is
done with the cluster algorithm~\cite{amit:05}. Thermalizing a ferromagnet is
easy, however we use parallel tempering to correlate the measurements at
different temperatures within the same sample. This correlation helps us to
reduce the error when averaging over disorder.

We display in Tables \ref{tab:details} and \ref{tab:details2} the relevant
simulation parameters for the study in the random-connectivity and the fixed
connectivity graphs respectively. We include the number of temperatures $N_T$
simulated in parallel for each dimension, the number of samples $N_\mathrm{samples}$, and the time
expended for each sample $N_t$. Time lengths are written in terms of an elementary
Monte Carlo step, defined here as the combination of one cluster update and
one single tempering update.
\begin{table}
\begin{center}
\begin{tabular}{|c|ccc|c|cc|}\hline
$D$&$N_T$&$1/T_\mathrm{min}$&$1/T_\mathrm{max}$&$N_\mathrm{samples}$&$N_t^\mathrm{sim}$&$N_t^\mathrm{term}$\\\hline
6&6&0.135&0.275&1000&1000&1000\\
8&7&0.151&0.247&1000&1000&1000\\
10&6&0.171&0.211&1000&1000&1000\\
12&10&0.172&0.208&1000&1000&1000\\
14&10&0.171&0.189&1000&1000&1000\\
16&12&0.175&0.185&1000&1000&1000\\
18&19&0.1735&0.1825&665&1000&1000\\\hline
\end{tabular}
\end{center}
\caption{Technical data for the simulation of the Ising model in random-connectivity graphs.}\label{tab:details}
\end{table}
\begin{table}
\begin{center}
\begin{tabular}{|c|ccc|c|cc|}\hline
$D$&$N_T$&$1/T_\mathrm{min}$&$1/T_\mathrm{max}$&$N_\mathrm{samples}$&$N_t^\mathrm{sim}$&$N_t^\mathrm{term}$\\\hline
6&6&0.135&0.275&1000&1000&1000\\
7&4&0.151&0.247&1000&1000&1000\\
8&7&0.151&0.247&1000&1000&1000\\
9&7&0.1545&0.247&1000&1000&1000\\
10&6&0.19&0.24&1000&1000&1000\\
11&6&0.19&0.22&1000&1000&1000\\
12&11&0.19&0.235&1000&1000&1000\\
13&9&0.184&0.2175&1000&1000&1000\\
14&14&0.185&0.2175&1000&1000&1000\\
15&8&0.195&0.21&1000&1000&1000\\
16&19&0.185&0.215&1000&1000&1000\\
17&8&0.1996&0.2057&1000&1000&1000\\
18&17&0.2&0.21&1000&1000&1000\\\hline
\end{tabular}
\end{center}
\caption{Technical data for the simulation of the Ising model in
  fixed-connectivity graphs.}\label{tab:details2}
\end{table}

\subsubsection{Determination of the critical point}

The final aim of this section, is to reproduce numerically the critical point
obtained in Appendix \ref{ap:bethe} with the Bethe approximation for the two
kinds of graphs: the random connectivity,
\be \label{eq:KcD}K_\mathrm{c}(D)=\tanh^{-1}\paren{\frac{D}{z\paren{D-1}}}.\ee
and the fixed connectivity graph \be\label{eq:kczfixed}
K_\mathrm{c}=\tanh^{-1}\paren{\frac{1}{z-1}}.\ee One must recall that these
two expressions are only exact in the $D\to\infty$ limit, where the Bethe
approximation becomes exact. The presence of short loops, as the ones
presented in Figure \ref{fig:min_distance} should introduce corrections of
$\mathcal{O}(D^{-2})$ for finite dimensions, i.e. logarithmic corrections in $N$. As
mentioned before, we expect the random connectivity model to display as huge
$D$-corrections, to become completely useless for numerical purposes. Indeed,
not only suffers from finite-loops corrections, but the Bethe approximation
itself leads to a $D-$dependent expression for $K_\mathrm{c}$ \eqref{eq:KcD}.

With this purpose, we compute the Binder cumulant. We already introduced this
quotient in \eqref{eq:EQbinder} for a spin glass. Now, since the order
parameter in a ferromagnet is the magnetization \be {\cal M}=\sum_i
s_i,\ee instead of the overlap, we define the Binder parameter as the
dimensionless quotient
\be \label{eq:binderIsing}B=\frac{{\mean{\mathcal{M}^4}}}{\mean{\mathcal{M}^2}^2}.\ee
As discussed then, the expectation value for the Binder parameter is $3$ in
the disordered paramagnetic phase. At variance with the spin glass, the
expectation value for the ferromagnetic phase is strictly $1$. We can use the
cross over between these two limiting behaviors to obtain numerically
$K_\mathrm{c}(D)$.

\begin{figure}
\begin{center}
 \includegraphics[scale=0.45,trim=0 70 0 70]{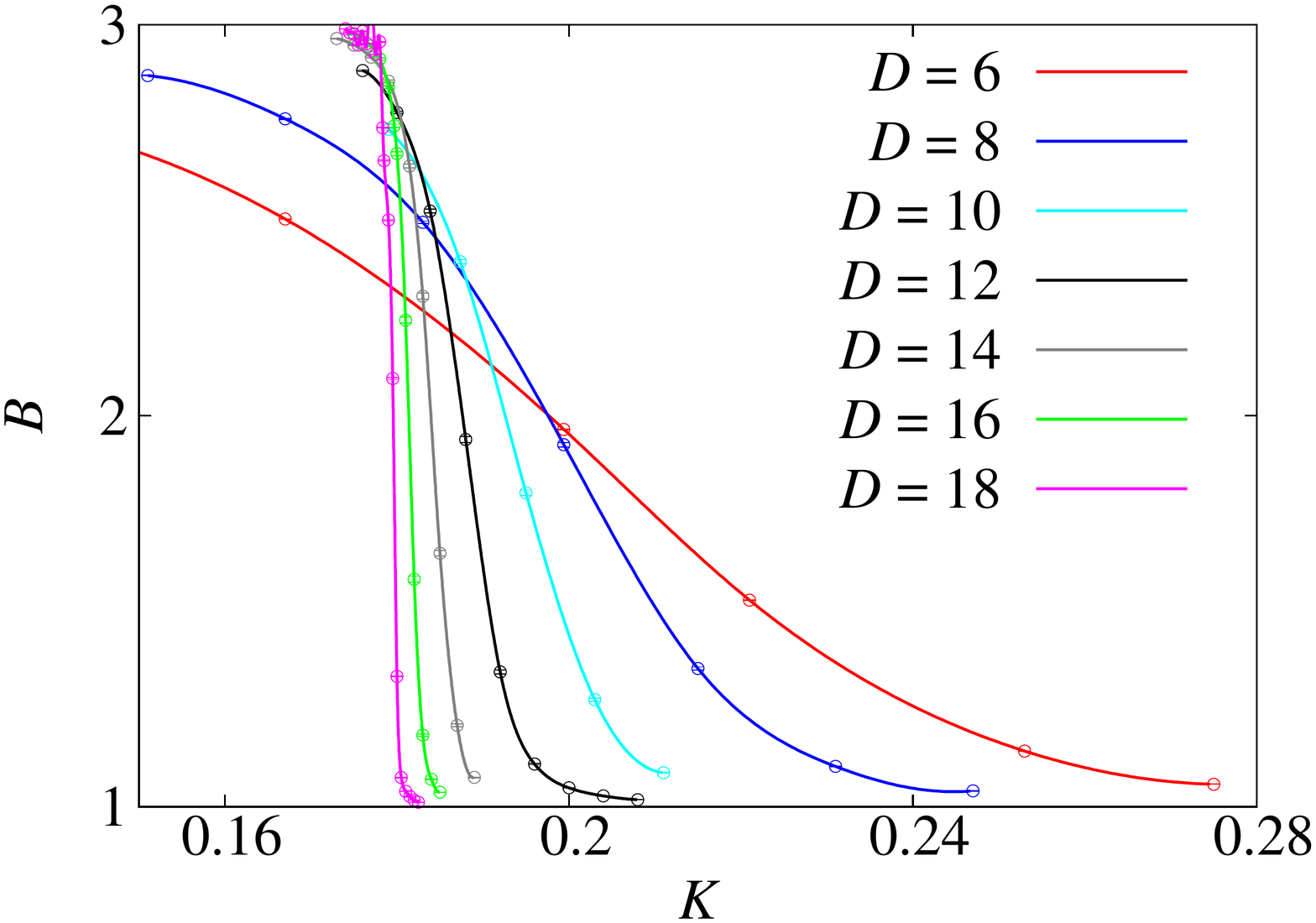}
\caption{\textbf{Random connectivity model in a ferromagnet}. Binder cumulant as function of $K$ for different
  $D$. The lines are interpolated from the simulation points using cubic spline.}\label{fig:observables}
\end{center}
\end{figure}

We show in Figure \ref{fig:observables} the dependence of the Binder cumulant
with the temperature for different system sizes.  As predicted, it drops from
$3$ to $1$ when $K$ increases ($K$ is proportional to the inverse temperature)
but the point at which the fall occurs displaces a lot with the system size,
which makes difficult to obtain $K_\mathrm{c}^\infty$ by means of crossings
between curves, as it is normally done using a finite size
scaling approach~\cite{amit:05}.
 Instead, as an alternative, we compute the average of two values of $K$ where
 $B(K,D)$ takes prescribed values. We refer to this estimate as
 $K_\mathrm{c}^{\mathrm{mp}}$. This $K_\mathrm{c}^{\mathrm{mp}}(D)$ tends to
 the desired $K_\mathrm{c}^\infty$ when $D\to\infty$. We present the results
 obtained for the midpoint defined as \be\label{eq:Kcest}
 K_\mathrm{c}^{\mathrm{mp.}}(D)=\frac{B^{-1}(1.2)+B^{-1}(2.4)}{2}, \ee in
 Table \ref{tab:KcDsim} together with the analytical expectation values
 obtained by means of the Bethe approximation \eqref{eq:KcD}.  As we
 previously discussed, one should not expect to have compatible values for
 finite values of $D$, since both calculations are plagued by finite size
 corrections, they should only be equal in the $D\to\infty$ limit. Indeed, the
 higher $D$, the better mutually agreement is.
\begin{table}
\begin{center}
\begin{tabular}{ccc}
$D$&$K_\mathrm{c}^{\mathrm{Bethe}}$&$K_\mathrm{c}^{\mathrm{mp}}$\\\hline
6&0.20273&0.216(2)\\
8&0.19283&0.207(5)\\
10&0.18735&0.195(2)\\
12&0.18386&0.1889(15)\\
14&0.18145&0.1846(8)\\
16&0.17969&0.1818(4)\\
18&0.17834&0.17967(3)
\end{tabular}
\end{center}
\caption{\textbf{Random connectivity model in a ferromagnet}. Comparison of the analytical estimates of $K_\mathrm{c}(D)$, eq. \eqref{eq:KcD}, with the numerical estimate 
$K_\mathrm{c}^\mathrm{mp}(D)$, eq. \eqref{eq:Kcest}.
}\label{tab:KcDsim}
\end{table}

In order to study more quantitatively this convergence, we fit the
$K_\mathrm{c}^{\mathrm{mp}}(D)$ values to 
\be\label{eq:fitKcD}
K_\mathrm{c}(D)=K_\mathrm{c}^{\infty}+\frac{a_1}{D}+\frac{a_2}{D^2}+\frac{a_3}{D^3},\ee
keeping  $K_\mathrm{c}^{\infty}$ fixed. We summarize the results in Table
\ref{tab:ajuste2}. The conclusion of this fit, though expected, is
devastating.  The large 
coefficients  $a_1$ and $a_2$ show how important the finite size corrections
are. 
\begin{table}
\begin{center}
\begin{tabular}{c|cccc}
$D_\mathrm{min}$&$\chi^2/\mathrm{dof}$&$a_1$&$a_2$&$a_3$\\\hline
6&0.53/4&0.073(13)&2.9(3)&-9.8(15)\\
8&0.0313/3&0.117(7)&1.6(2)&-0.7(13)\\
8&0.0342/4&0.1208(15)&1.53(3)&0\\
\end{tabular}
\end{center}
\caption{\textbf{Random connectivity model in a ferromagnet}. Results of the fits of $K_\mathrm{c}^\mathrm{mp}(D)$ to $K_\mathrm{c}^\mathrm{mp}(D)=K_\mathrm{c}^\infty+\frac{a_1}{D}+\frac{a_2}{D^2}+\frac{a_3}{D^3}$. $K_\mathrm{c}^\infty$ was fixed to the exact value $K_\mathrm{c}^{\infty}=0.16824\ldots$. We include in the fit all data with $D\geq D_\mathrm{min}$. In the last row, $a_3=0$ has been taken.  }\label{tab:ajuste2}
\end{table}

Clearly, the random connectivity hypercube is a disaster even for the Ising
model, so we will not consider it to study the SG, where randomness makes a
lot more difficult to control finite size effects.

\begin{figure}
\begin{center}
 \includegraphics[scale=0.45,trim=0 70 0 70]{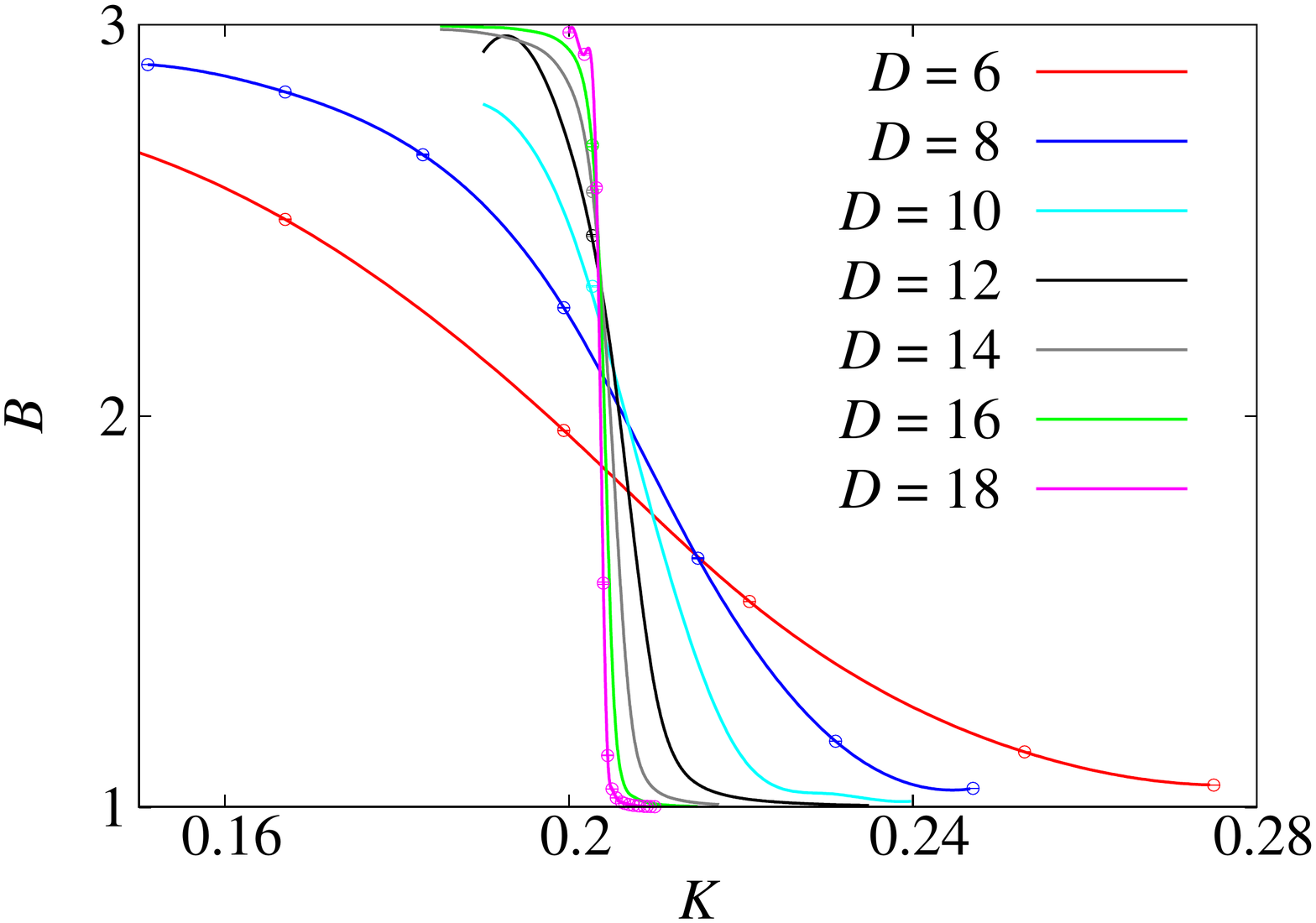}
\caption{\textbf{Fixed connectivity model in a ferromagnet}. Binder cumulant
  as a function of $K$ for different
  dimensions. The lines are interpolated from the simulation points using cubic spline.}\label{fig:observables2}
\end{center}
\end{figure}

On the contrary, if we perform exactly the same study but on the fixed
connectivity graph, we obtain more promising results. Indeed, we plot in
Figure \ref{fig:observables2} the Binder cumulant as a function of $K$. The
finite sizes effects are reduced drastically.

\begin{figure}
\begin{center}
 \includegraphics[angle=270,width=0.8\columnwidth,trim= 40 0 40 0]{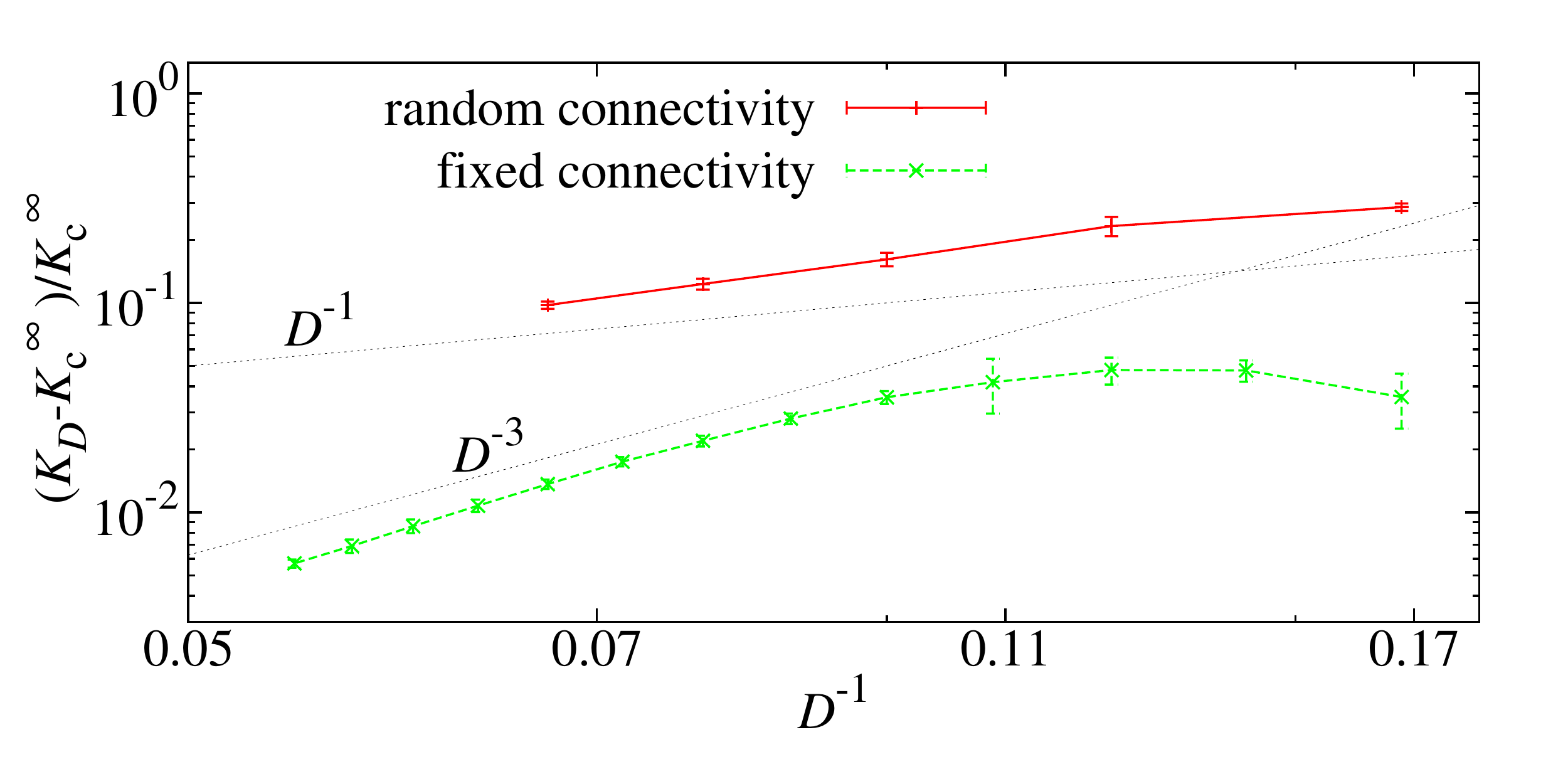}
      \caption{ Comparison of the finite volume effects at the critical point
        in $K_\mathrm{c}^D$ for the ferromagnetic Ising model, both in
        the random \textbf{(red circles)} and fixed \textbf{(green
          circles)} connectivity hypercubes. As a guide to the eye, we have
        included two different scalings with $D$.}\label{fig:KD}
\end{center}
\end{figure}

In this case, we may easily compute the crossings between two consecutive
curves in $B$, and thus obtain the the crossing estimator $K_\mathrm{c}^{D,D+1}$.
We expect a scaling ($K_\mathrm{c}^\infty=0.20273\ldots$) \be
K_\mathrm{c}(D)=K_\mathrm{c}^\infty+\frac{c_2}{D^2}+\frac{c_3}{D^3}+\cdots.\ee
Indeed, as discussed previously, the contribution of closed loops (plaquettes
or larger) are of order $1/D^2$. Hence, the linear term in $1/D$ found for the
random connectivity model arises exclusively from the Bethe equation
\eqref{eq:KcD}. These expectations are confirmed by our numerical data shown
in Figure \ref{fig:KD}. Note that while the random connectivity model
did not reach the asymptotic $1/D$ regime even for $D=18$, for fixed $z$ the
asymptotic $1/D^2$ regime is reached, although strong $1/D^3$ corrections are
visible for $D<14$.  This qualitative picture on scale corrections is
confirmed by a $\chi^2$ test in Table \ref{tab:ajuste3}.
\begin{table}
\begin{center}
\begin{tabular}{c|ccc}
$D_\mathrm{min}$&$\chi^2/\mathrm{dof}$&$c_2$&$c_3$\\\hline
12&6.41/4&0.07(2)&1.3(3)\\
13&0.92/3&0.114(14)&0.66(20)\\
14&0.40/2&0.136(18)&0.3(3)\\
\end{tabular}
\end{center}
\caption{\textbf{Fixed connectivity model  in a ferromagnet}. Results of the fits of $K_\mathrm{c}^{D,D+1}$ to $K_\mathrm{c}^\infty+\frac{c_2}{D^2}+\frac{c_3}{D^3}$. $K_\mathrm{c}^\infty$ is fixed to the exact value $K_\mathrm{c}^\infty=0.20273\ldots$. We include in the fit all data with $D\geq D_\mathrm{min}$. }\label{tab:ajuste3}
\end{table}

A summary of our efforts is shown in Figure \ref{fig:KD}, where we plot the
dependency of the critical point with $D$ for the ferromagnetic Ising model
for the two kinds of graphs. As anticipated several times already, the random
connectivity model suffers very important finite volume corrections which make
it essentially useless for numerical studies. The problem is solved using
fixed connectivity hypercubes instead, where the finite volume effects are
only caused by the residual presence of short closed loops. From now on, we will only consider this second kind of graphs.

\section{Spin glass in the fixed connectivity hypercube}
After this last short preliminary study in the Ising model, we can be
confident about the introduced graphs (and its generation algorithm). Indeed,
numerical results were compatible with the expected analytical determinations
of the critical point at $D\to\infty$. Furthermore, now we know that we should
only consider fixed-connectivity hypercubes if we want to reduce the finite
size effect as much as possible.  Thus, we are ready to move to our final
interest, the numerical study of a spin glass on these graphs.

In particular, we will consider a diluted spin glass where couplings can take
only two possible values $J_{\V{x},\V{\hat \mu}}=\pm 1$, randomly chosen with
probability $1/2$ each. Since we only consider a fixed connectivity graph,
each spin is connected with exactly 6 neighbors (among the $D$ possible 
nearest neighbors).

The structure of this section will be the following. We will begin in
Section~\ref{sec:MN} with a description of the used numerical methods and a
discussion on the particularities of the spatial correlation functions in the
hypercube in Section~\ref{sec:spatial_hy}. Later on in
Section~\ref{sec:res_eq}, we will study the equilibrium behavior (where
we have analythical predictions to compare with), to end up with the final
goal of this chapter, the nonequilibrium study in
Section~\ref{sec:res_feq}

\subsection{Numerical Methods}\label{sec:MN}

All the variables involved in the Hamiltonian \eqref{eq:Hhy}, the spins and
the couplings, are binary. They can thus be coded in the bits of a computer
word, making this model highly parallelizable. In fact, we implement the
so-called \textit{Multi-spin Coding}: we simultaneously codify $64$ systems in
one single $64$ bits computer word.  Besides, being the nodes of the lattice
distributed on an unit hypercube, also sites can be written in term of bits,
see the appendix Section \ref{sec:hypercubecomputer} for a detailed
definition. For the computational point of view, it is a challenge to write a
simulation program that takes fully benefit of the parallelization of bitwise
operations.  For this reason, we have included a section in the Appendix
\ref{app:multispin} where we explain in detail how to do it.

Following this approach, one can simulate $64$ samples in practically the same
time it would take to simulate just one. However, in order to keep the
parallelism, all the samples in the same run share the same connectivity
matrix $n_{\V{x},\V{\hat \mu}}$ (and differ only in the configuration of
couplings $J_{\V{x},\V{\hat \mu}}$). With this common matrix, we find errors
which are $\!\sim 7$ times smaller than those obtained with one single sample
per matrix. This should be compared with the error reduction by a factor $8$,
expected if $64$ truly independent samples were simulated. Our program
needs $0.29\,\text{ns}$/\textit{spin-flip} in an Intel i$7$ at $2.93$GHz (in
Ref. \cite{hasenbusch:08} they report
$\!\sim\!1.2\,\text{ns}$/\textit{spin-flip} on an Opteron at 2.0 GHz, for the
simulation of the $D=3$ \ac{EA} model in the cubic lattice)\footnote{Note that
  we are considering the $z=6$ case. Then, the core of the Metropolis
  algorithm is equivalent to the $D=3$ \ac{EA} model in a cubic lattice.}.

In a nonequilibrium dynamical study such as ours, one computes both one-time and
two-times quantities, see Sect. \ref{sec:observables}. The calculation of
two-times quantities implies the storage on disk of intermediate
configurations. Disk capacity turned out to be the main limiting factor
for the simulation. For this reason, we have worked in parallel with two
program versions: one valid for measuring quantities at one and two
times and another restricted to  the computation of  one-time quantities.

We have computed two-time quantities at temperature $T=0.7T_\text{c}$,
on systems with $D=16,18,20$ and $22$. The number of simulated samples
were $8\!\times\! 64$ samples for each system size (hence, for
self-averaging quantities (see Section \ref{sec:replicatrick}) the statistical quality of our data grow
with $D$).

Besides, since this new model requires intensive testing, we have
computed {\em equilibrium} one-time quantities at
$T/T_\text{c}=0.95,0.97,0.99,1,1.1,1.2,1.3$ and $1.4$. The system
sizes were again $D=16,18,20$ and $22$. The number of simulated
samples was $128\!\times\! 64$ samples per temperature (at
$T_\mathrm{c}$ we computed $256\!\times\! 64$ samples).

\subsection{Spatial correlation functions in the hypercube} \label{sec:spatial_hy}

As mentioned in the objectives of this work, the goal of this project was to
define a MF model that allows to approach the domain growth in spin
glasses. Indeed, we discussed in Section \ref{sec:hypercube} that the
hypercube geometry let us to define a distance based on the minimum number of
edges one needs to cross to join two nodes. Then, as in the 3-$D$ EA model, we
can use the standard spatial correlation functions defined in Section
\ref{sec:spatialcorrdef} but using the postman metrics instead of the
Euclidean one. However, even being basically the same definitions, the
hypercube introduces certain particularities that will be discussed in this
section.

\begin{figure}
\begin{center}
\includegraphics[angle=270,width=0.8\columnwidth,trim=20 0 0 0]{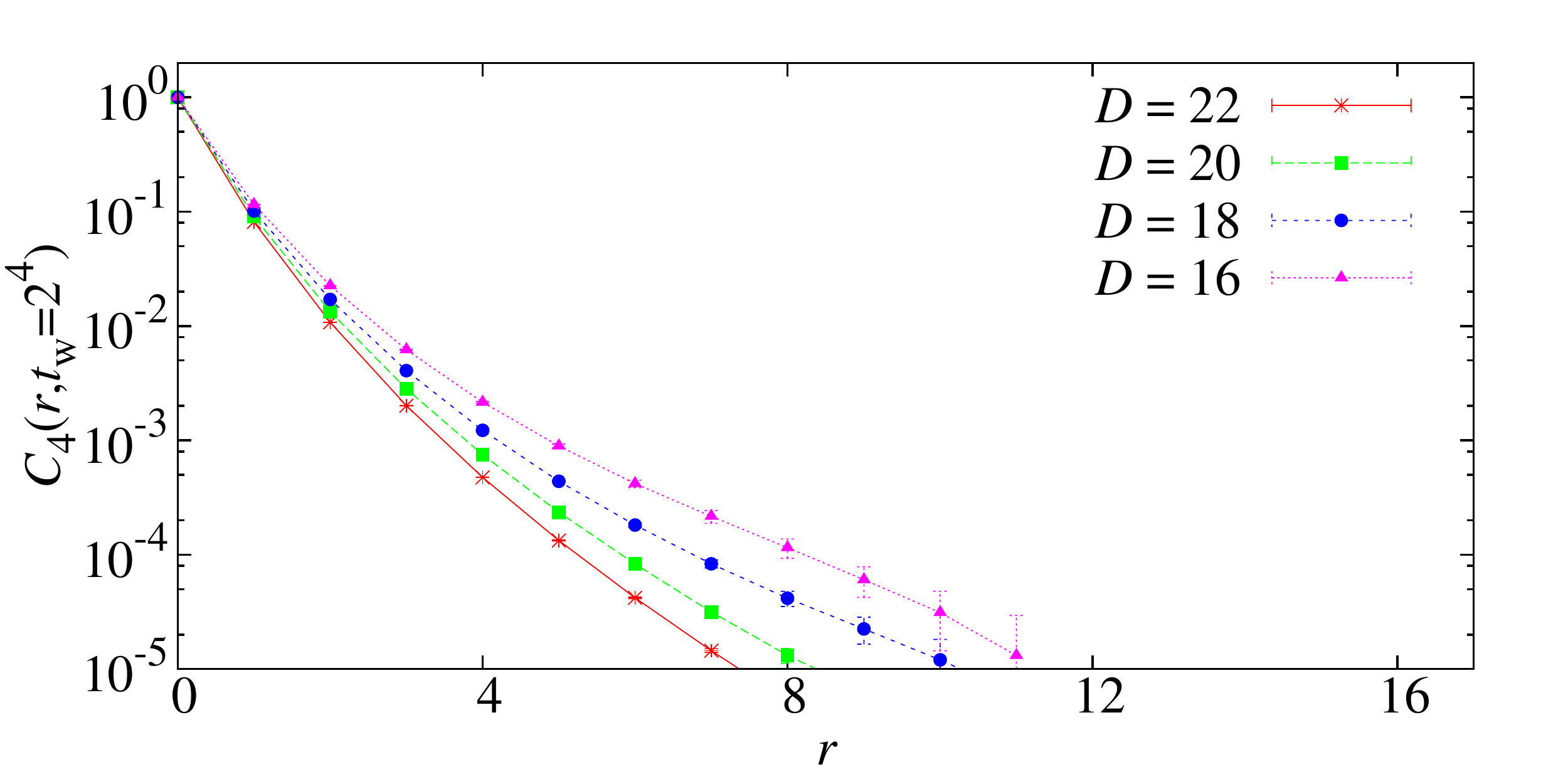}
\includegraphics[angle=270,width=0.8\columnwidth,trim=20 0 0 0]{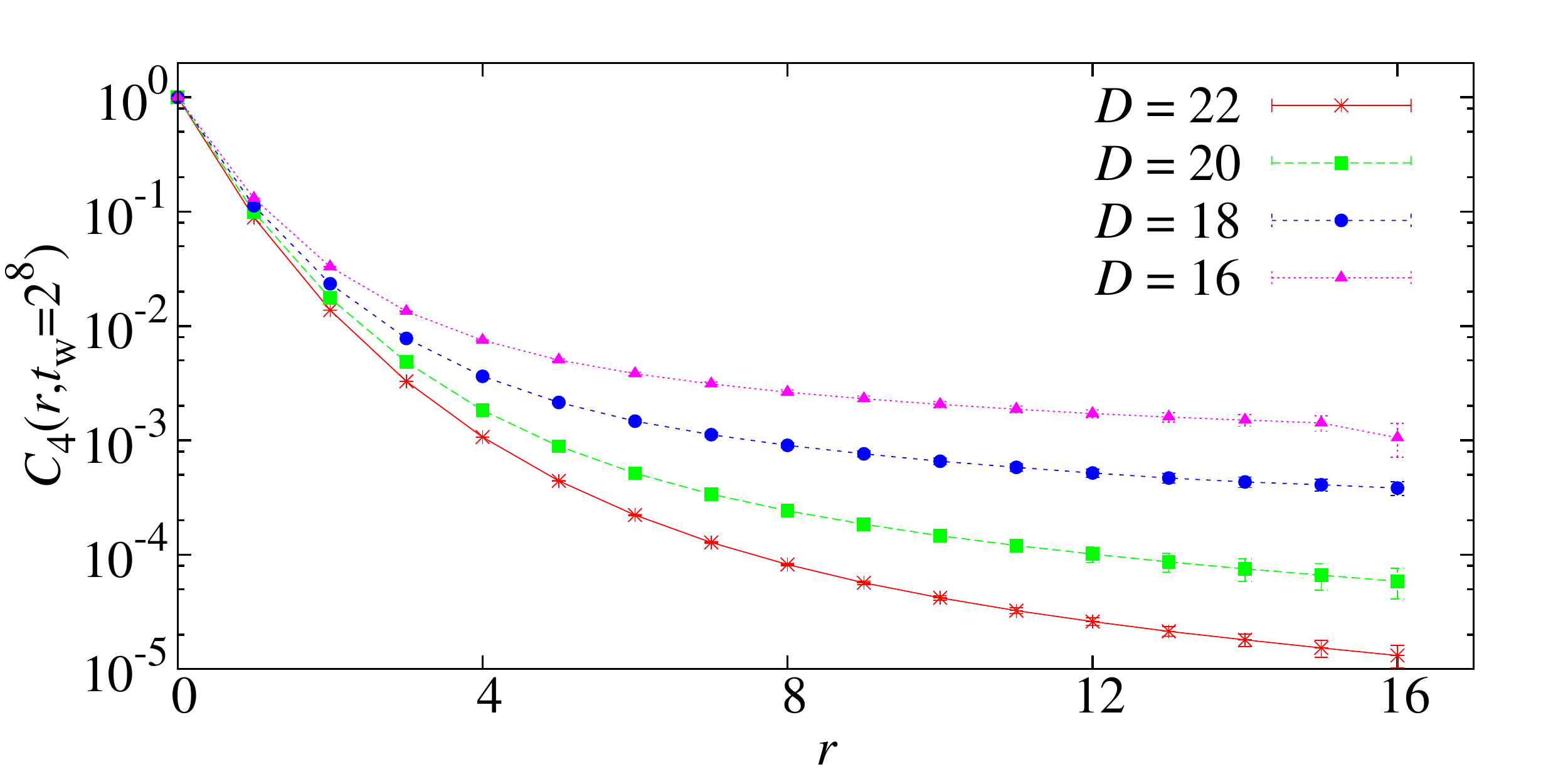}
\includegraphics[angle=270,width=0.8\columnwidth,trim=20 0 0 0]{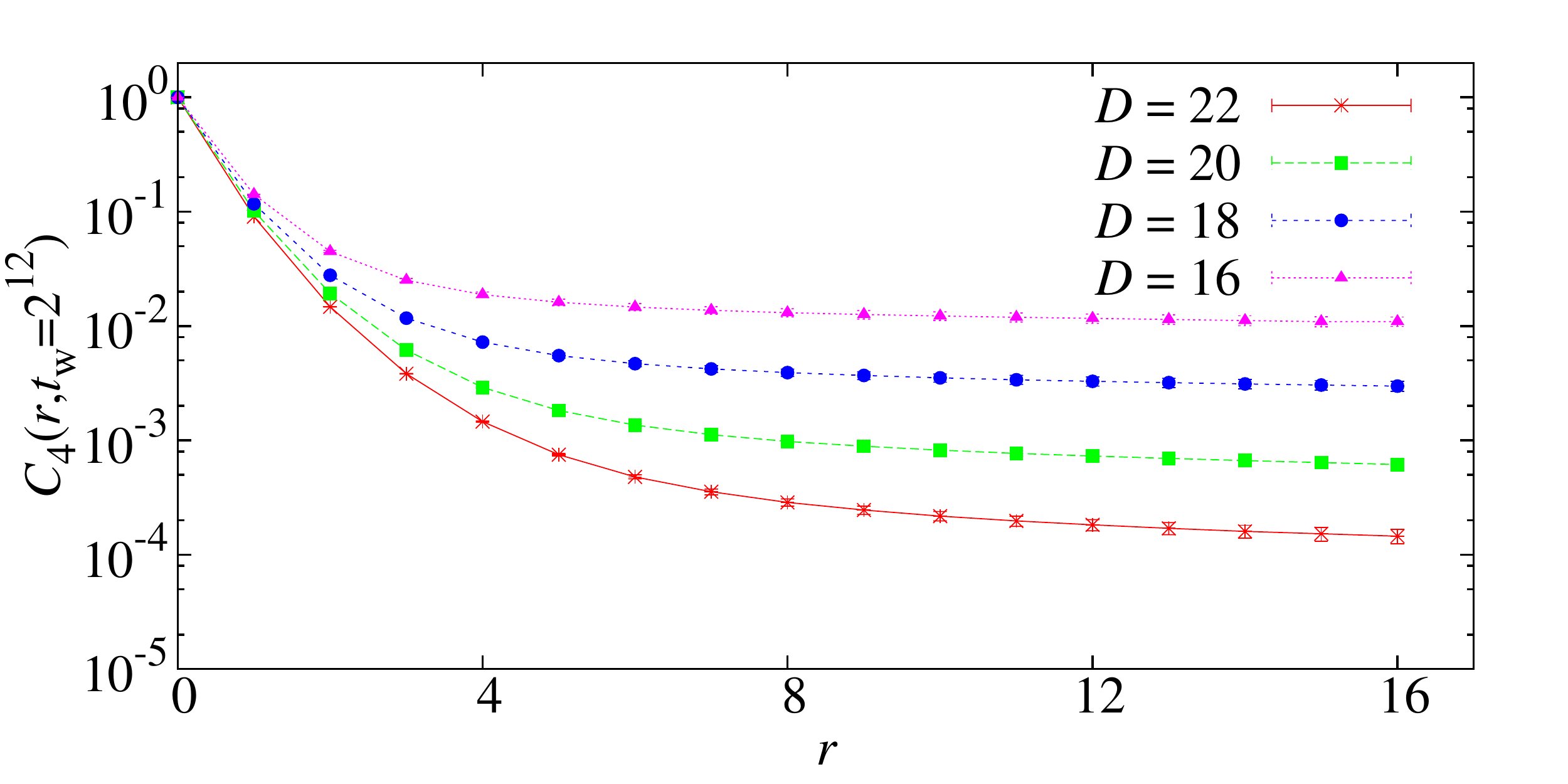}
\caption{ $C_4(r,t_\mathrm{w})$,  \eqref{eq:fcor_sinJ}, for
  (top) $t_\mathrm{w}\!=\!2^4$, (center) $t_\mathrm{w}\!=\!2^8$ and (bottom) $t_\mathrm{w}\!=\!2^{12}$, and different system sizes, $N\!=\!2^D$, at $T\!=\!0.7T_\mathrm{c}$.}\label{fig:CR}
\end{center}
\end{figure}

\begin{figure}
\begin{center}
\includegraphics[angle=270,width=0.8\columnwidth,trim=40 0 40 0]{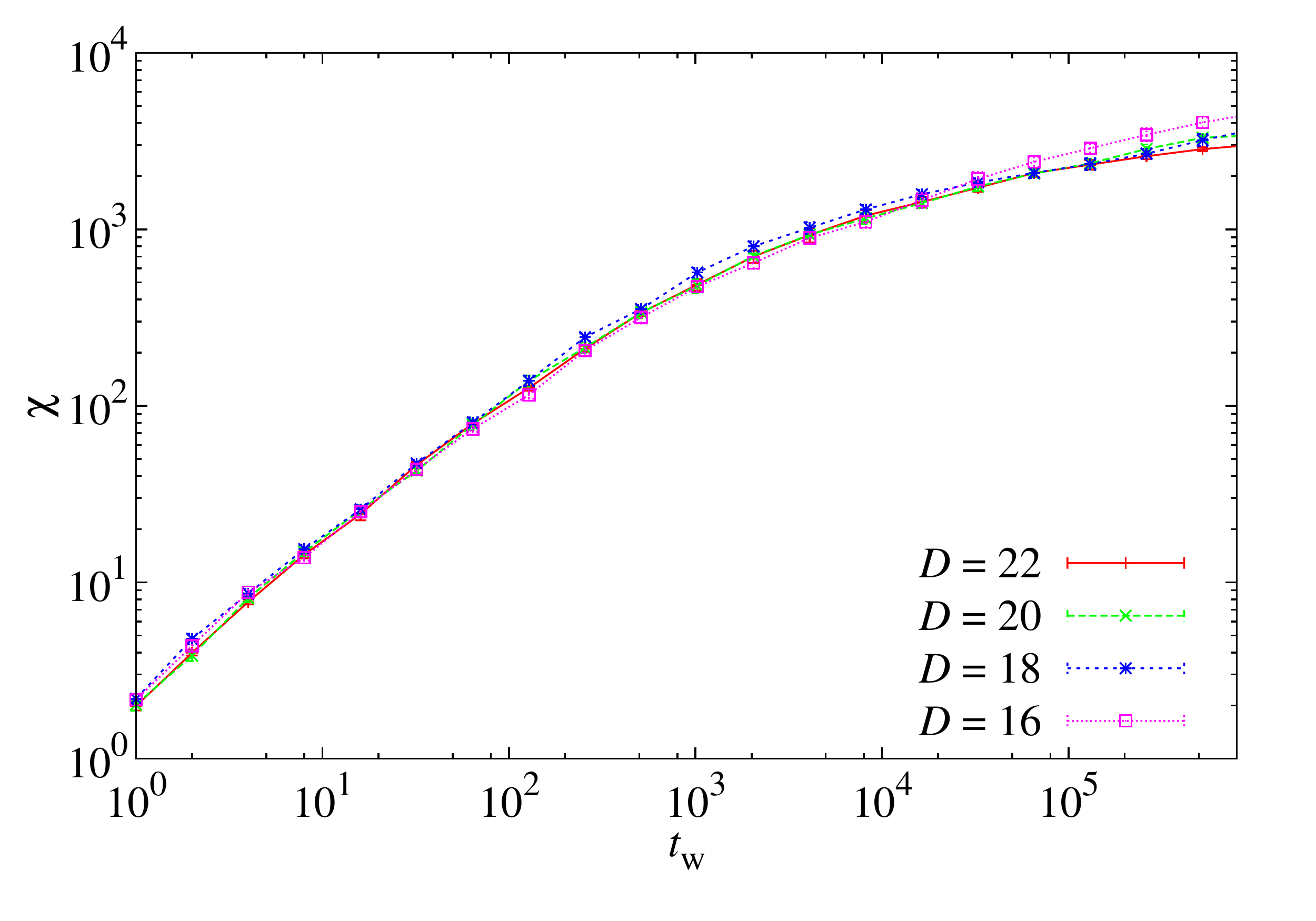}
\caption{ SG susceptibility a $T\!=\!0.7T_\mathrm{c}$ as function of $t_\mathrm{w}$ 
for different system sizes, $N\!=\!2^D$.}\label{fig:chi_07}
\end{center}
\end{figure}

The main problem here appears when averaging $c_4(\V{r},\tw)$ defined in
\eqref{eq:c4def} over all the displacements $\V{r}=r$. In the hypercube, at
variance with the 3-$D$ EA model, the number of spins separated by $r$ depends
strongly on the precise value of $r$. Indeed, a short calculation tells us
that it is given by $N_r\!=\!\binom{D}{r}$. As a consequence, when we consider
the average
\be\label{eq:fcor_sinJ} C_4(r,t_\mathrm{w})=\frac{1}{N_r}\sum_{
\V{r},
|\V{r}|=r
}c_4(\V{r},t_\mathrm{w})\,,\ee
see Figure \ref{fig:CR}, $C_4(r,t_\mathrm{w})$ does not
present a limiting behavior with $D$ for a given $t_\mathrm{w}$.
We can get a clue by looking at $\chi_\mathrm{SG}(t_\mathrm{w})$ (defined in
\eqref{eq:SGsusc}), see Figure \ref{fig:chi_07}, which does reach a
thermodynamic limit. It was discussed in \eqref{eq:rel_sus_c4} that
$\chi_\mathrm{SG}(t_\mathrm{w})$ is nothing but the integral of
$C_4(r,t_\mathrm{w})$ with a Jacobian that here is precisely $N_r$, that is
$\binom{D}{r}$. Then, it seems reasonable to define the following spatial
correlation function instead: \be\label{eq:fcor_conJ} \hat {C_4
}(r,t_\mathrm{w})=\sum_{ \V{r}, |\V{r}|=r }c_4(\V{r},t_\mathrm{w})\,.\ee

\begin{figure}
\begin{center}
\includegraphics[angle=270,width=0.8\columnwidth,trim=20 0 20 0]{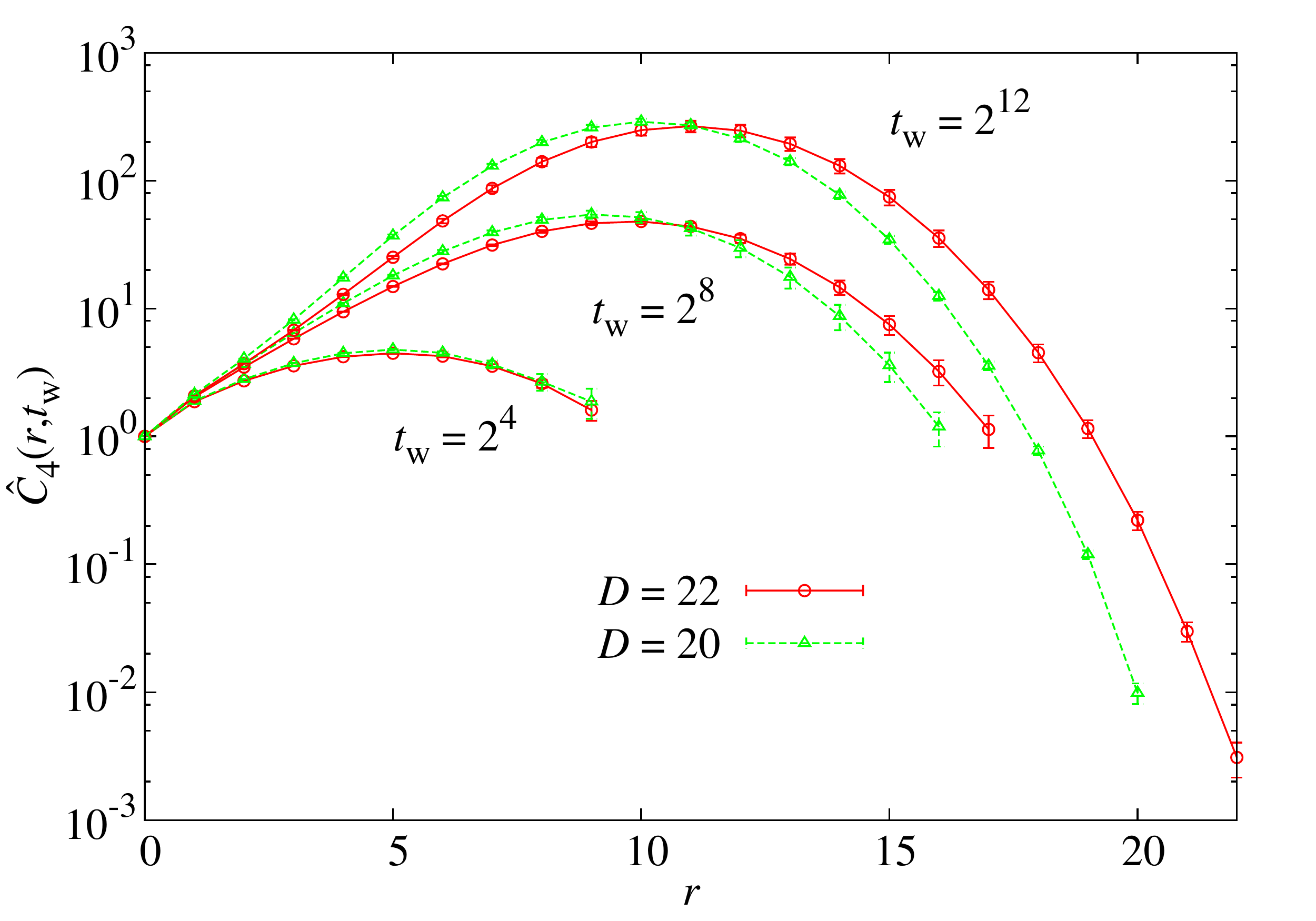}
\caption{ $\hat {C_4 }(r,t_\mathrm{w})$,  \eqref{eq:fcor_conJ},  for
  $D\!=\!20$ and $22$ for $t_\mathrm{w}\!=\!2^4,\ 2^8$ and $2^{12}$ at
  $T\!=\!0.7T_\mathrm{c}$. This has to be compared with the behavior of $C_4(r,t_\mathrm{w})$, Figure \ref{fig:CR}. }\label{fig:TCR}
\end{center}
\end{figure}

We can see that $\hat {C_4 }(r,t_\mathrm{w})$ does reach the high-$D$
limit, Figure \ref{fig:TCR}, at least for short
$t_\mathrm{w}$. Besides, in the paramagnetic phase, it is possible to compute analytically $\hat {C_4 }(r,t_\mathrm{w})$, see Appendix \ref{ap:HTE}, taking first the limit $t_\mathrm{w}\to\infty$ and making afterwards $D\to\infty$. The resulting correlation function, which is only valid in the paramagnetic phase, is a simple exponential. Hence, both the equilibrium and the nonequilibrium computations, suggest that one should focus on $\hat {C_4 }$ rather than on $C_4$.

We note in Figure \ref{fig:TCR}, that in the SG phase, $\hat {C_4 }$ is non monotonically decreasing with $r$, but rather presents a maximum. This maximum moves to bigger $r$ with
$t_\mathrm{w}$, then, the system has a characteristic length that increases
with time.
Thus, we can estimate the coherence length, by means of the integral estimator
$\xi_{0,1}(t_\mathrm{w})$: \be\label{eq:xi}
\xi_{0,1}(t_\mathrm{w})=\frac{\int_0^\infty \mathrm{d}r\ r\ \hat {C_4
  }(r,t_\mathrm{w})}{\int_0^\infty \mathrm{d}r\ \hat {C_4
  }(r,t_\mathrm{w})}\,.\ee When computing numerically this
$\xi_{0,1}(t_\mathrm{w})$, we only summed up to certain cutoff in order to
avoid meaningless noise in the determination of $ \hat {C_4
}(r,t_\mathrm{w})$. Our actual criterion was to sum up contributions while
$\hat {C_4 }(r,t_\mathrm{w})$ was higher than $3$ times its error (obtained
with the fluctuations between samples).
A major advantage of $\xi_{0,1}$ over more
heuristic definitions of the coherence length, is that it is computed from
self-averaging quantities (see details in~\cite{janus:08,janus:09}, we note
that, in this work, we have not tried to estimate the contribution to the
integrals by the noise-induced long distance cutoff).

The existence of such a characteristic length is the main advantage of the hypercube model over other \ac{MF} models.

In addition to this correlation function in the real space, we will also be
interested in its behavior in the Fourier space.
We define the Fourier transform in the standard way. Our wave vectors
are $\V{k}=\pi (n_1,n_2,\ldots,n_D)$ with $n_i=0,1$. The propagator is 
\be
G(\V{k},t_\mathrm{w})=\sum_{\V{r}}e^{\I\V{k}\cdot\V{r}}c_4(\V{r},t_\mathrm{w}).\ee 
In particular, $G(0,t_\mathrm{w})=\sum_{r=0}^D
\hat{C}_4(r,t_\mathrm{w})=\chi_{\mathrm{SG}}(t_\mathrm{w})$. 

Now, because of the disorder average,
$c_4(\V{r},t_\mathrm{w})$ is only a function of $r\!=\!|\V{r}|$ (postman
metrics). It follows from  \eqref{eq:Gk} that $G(\V{k},t_\mathrm{w})$ actually depends
only on $k\!=\!|\V{k}|$. 

The rotational invariance allows us for a major simplification \cite{parisi:06}[with a slight
abuse of notation, we write $c_4(r,t_\mathrm{w})$ rather than $c_4(\V{r},t_\mathrm{w})$],
\be\label{eq:Gkint}
G(k,t_\mathrm{w})=\sum_{r=0}^D K_r(D,k)\ c_4(r,t_\mathrm{w}),\ee
where $K_r(D,k)$ are the \textit{Krawtchouk polynomials}:
\begin{eqnarray}
  K_r(D,k)=\sum_{m=\mathrm{max}(0,r+k-D)}^{\mathrm{min}(k,r)}(-1)^m\binom{k}{m}\binom{D-k}{r-m}.
\end{eqnarray}
It is interesting to point out that neither $K_r(D,k)$ nor
$c_4(r,t_\mathrm{w})$ have a thermodynamic limit, while $G(k,t_\mathrm{w})$
does so. In fact, when $k=0$, $K_r(D,0)=\binom{D}{r}$ is diverging. Thus, we
can rewrite  \eqref{eq:Gkint} in terms of quantities with a well defined
limit, i.e.  \be\label{eq:Gk} G(k,t_\mathrm{w})=\sum_{r=0}^D
\frac{K_r(D,k)}{\binom{D}{r}}\ \hat C_4(r,t_\mathrm{w}).\ee

\subsection{Equilibrium Results}\label{sec:res_eq}
Since the present work is the first study ever made of a \ac{EA} model on a fixed connectivity
hypercube it is necessary to make a few consistency
checks. Equilibrium results are most convenient in this respect, since we have analytical computations (valid only for the large $D$ limit) to compare with.

We will briefly study the spatial correlations in the
paramagnetic phase. In addition, we will check, by approaching to
$T_\mathrm{c}$ from the SG phase, that the SG transition does lie on the
predicted $T_\mathrm{c}$,  \eqref{eq:thouless:86}.

\subsubsection{Paramagnetic Phase}
Our very first check will be the comparison between the Monte Carlo estimate
of the SG susceptibility defined in \eqref{DEF:CHI} (at finite $D$) with the analytical computation for infinite $D$:
\be\label{eq:chi_infty}\chi(K)=1+\frac{z\tanh^2K}{1-(z-1)\tanh^2K},\ee
see Appendix \ref{ap:HTE}. Our results are presented in Table \ref{tab:chi}. We see that finite size effects increase while approaching $T_\mathrm{c}$. For our larger system, $D=22$, the susceptibility significantly deviates from the asymptotic result only in the range $T_\mathrm{c}<T<1.1T_\mathrm{c}$.

\begin{table}
\begin{center}
\begin{tabular*}{\columnwidth}{@{\extracolsep{\fill}}cccc}\hline
$T$&$\chi(T)_{D=\infty}$&$\chi(T)_{D=20}$&$\chi(T)_{D=22}$\\\hline
$1.4T_\mathrm{c}$&2.4497$\ldots$&2.41(3)&2.44(3)\\
$1.3T_\mathrm{c}$&3.0176$\ldots$&2.98(4)&2.98(4)\\
$1.2T_\mathrm{c}$&4.1650$\ldots$&4.08(6)&4.10(7)\\
$1.1T_\mathrm{c}$&7.6344$\ldots$&7.11(13)&7.43(11)\\
$T_\mathrm{c}$&$\infty$&26(2)&98(7)\\\hline
\end{tabular*}
\end{center}
\caption{Comparison between the SG susceptibility in large $D$ limit for the paramagnetic phase,  \eqref{eq:chi_infty}, and numerical results for $D=20,22$.}\label{tab:chi}
\end{table}

\begin{figure}
\begin{center}
\includegraphics[angle=270,width=0.8\columnwidth,trim=0 0 0 0]{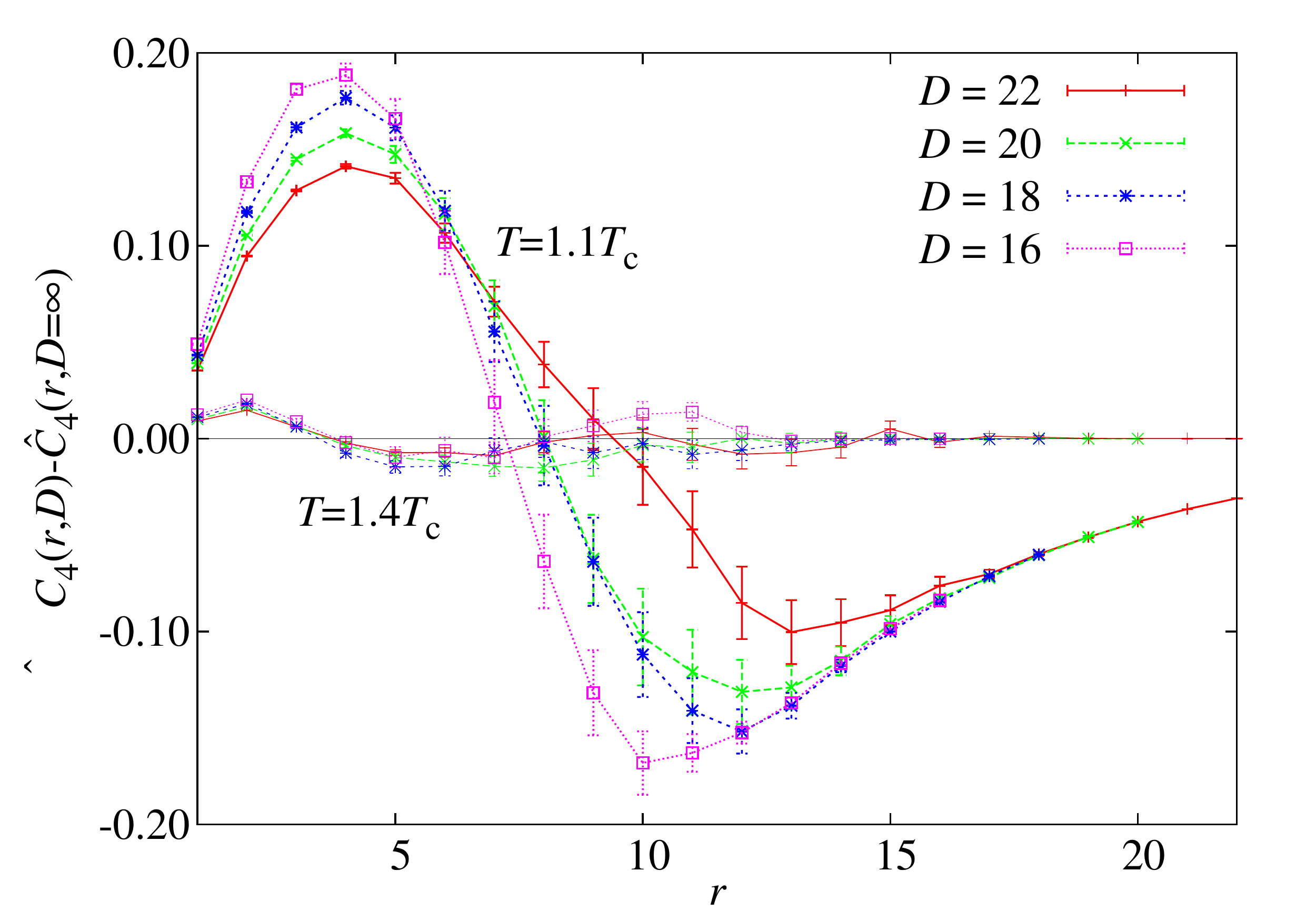}
\caption{ Difference between the numerical and analytical spacial correlation
  function for different system sizes at two temperatures $T=1.1T_\mathrm{c}$
  and $T=1.4T_\mathrm{c}$.}\label{fig:CRpara}
\end{center}
\end{figure}

After the fast convergence to the large $D$ limit observed in the SG
susceptibility, the results for $\hat C_4$ are a little bit disappointing. In
Figure \ref{fig:CRpara} we display $\hat C_4(r,D)-\hat C_4(r,\infty)$ as a
function of $r$ ($\hat C_4(r,\infty)$ is obtained using \eqref{eq:hatCap} in Appendix \ref{ap:HTE}).
We can see that finite size effects become more
important once one approaches $T_\mathrm{c}$.
Besides, finite $D$ corrections as a function of $r$ oscillate between
positive and negative values. This is not surprising: the finite $D$
corrections to the susceptibility are very small, and they are nothing but the
integral under these curves.  More quantitatively, we see in Table
\ref{tab:CRparaD} that the corrections with $D$ for $r=1,2$ are
$\mathcal{O}(D^{-1})$. Indeed, the path counting arguments in Appendix \ref{ap:HTE} are
plagued by corrections of $\mathcal{O}(D^{-1})$.

\begin{table}
\begin{center}
\begin{tabular*}{\columnwidth}{@{\extracolsep{\fill}}ccc|cc}\hline
&\multicolumn{2}{c|}{$r=1$}&\multicolumn{2}{c}{$r=2$}\\
$D$&$T=1.1T_\mathrm c$&$T=1.4T_\mathrm c$&$T=1.1T_\mathrm c$&$T=1.4T_\mathrm c$\\\hline
16&0.783(6)&0.198(5)&2.130(18)&0.320(12)\\
18&0.779(4)&0.201(3)&2.115(11)&0.327(7)\\
20&0.784(2)&0.202(2)&2.109(6)&0.332(4)\\ 
22&0.7776(12)&0.2006(9)&2.083(4)&0.324(2)\\\hline
\end{tabular*}
\end{center}
\caption{$D$ times the difference between $\hat C_4(r)$, for finite $D$ and
  infinite $D$ (using \eqref{eq:hatCap} in Appendix \ref{ap:HTE}), as computed for $r=1,2$. The absence of any $D$ evolution evidences finite-$D$ corrections of order $1/D$.}\label{tab:CRparaD}
\end{table}

\subsubsection{SG phase}
\begin{figure}
\begin{center}
\includegraphics[angle=270,width=0.8\columnwidth,trim=40 0 40 0]{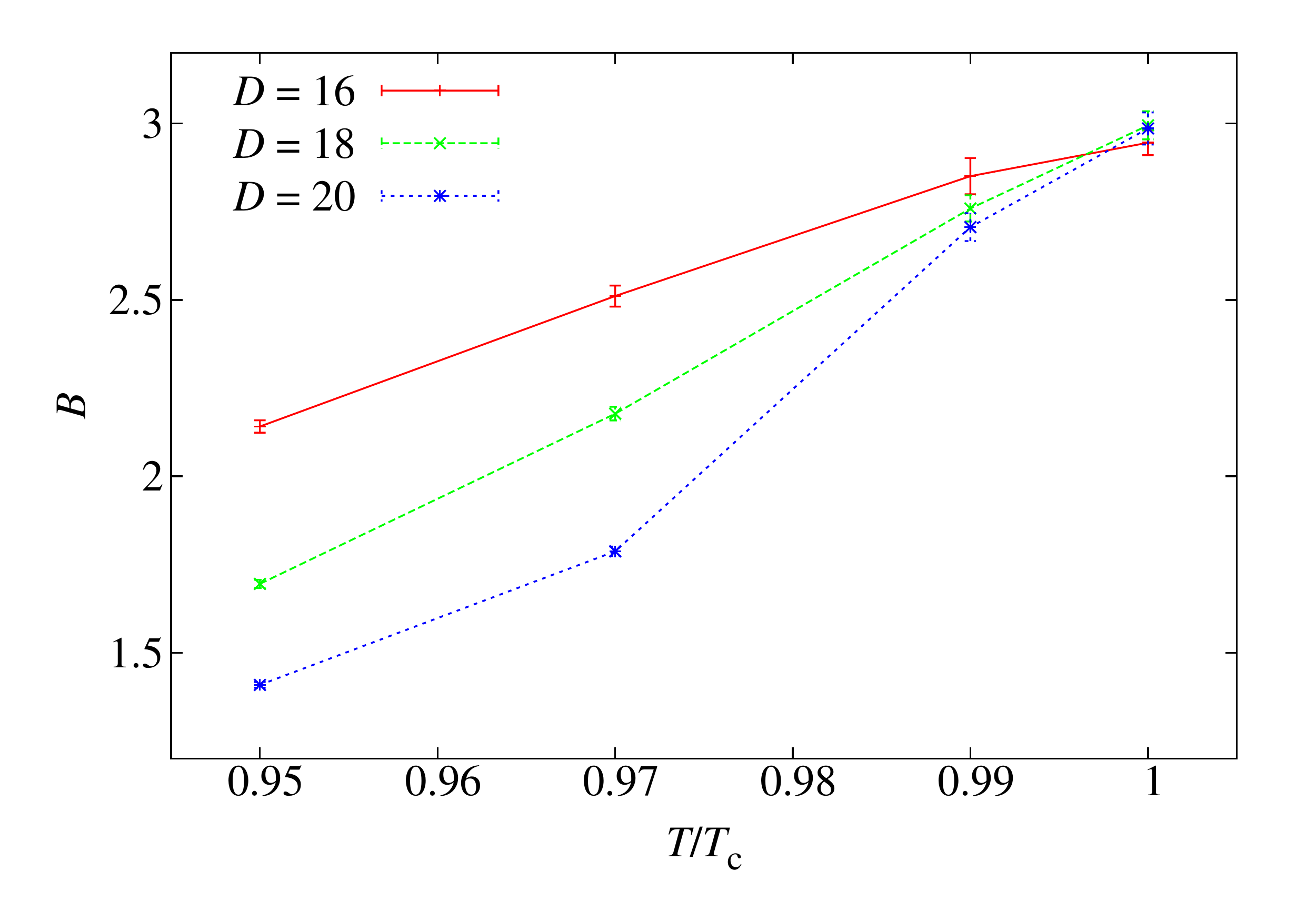}
\caption{ Equilibrium values of the Binder cumulant,  \eqref{eq:EQbinder}, for
  several system sizes, as function of the temperature in units of the exact
  asymptotic value of $T_\mathrm{c}$,  \eqref{eq:thouless:86}, in the SG phase.}\label{fig:binder}
\end{center}
\end{figure}

In the SG phase, our test has been restricted to a check of
\eqref{eq:thouless:86}, that predicts a SG phase transition for the high-$D$
limit. With this aim, we compute the Binder cumulant, $B(T)$, defined in
\eqref{eq:EQbinder}, nearby $T_\mathrm{c}$. For all $T<T_\mathrm{c}$, we
expect $B(T)<3$ for large enough $D$. As we show in Figure \ref{fig:binder},
$B(T)$ decreases with $T$ and shows sizeable finite size effects. In fact, at
$T=0.99T_\mathrm{c}$, we need to simulate lattices as large as $D=20$ to find
values below 3. Right at $T_\mathrm{c}$, the Gaussian value $B(T)=3$ is found
for all the simulated sizes.

\subsection{Nonequilibrium Results}\label{sec:res_feq}
In this section we will address the main features of the nonequilibrium
dynamics obtained in our largest system, $D=22$. The issue of finite $D$
corrections will be postponed to Sect. \ref{sec:TF}.

\subsubsection{The structure of isothermal aging}
We widely presented evidences of isothermal aging in experiments in Section
\ref{sec:aging}. Now we approach it in numerical simulations.  The
picture of isothermal aging dynamics in \ac{MF} models of SG behavior was
largely drawn in~\cite{cugliandolo:94} (see also \cite{young:97}). The
dynamics is ruled by an infinite number of \textit{time-sectors}:
\be\label{eq:timesectors} C(t,t_\mathrm{w})=\sum_{i}f_i\paren{h_i(t_\mathrm
  w)/h_i(t+t_\mathrm w)},  \ee 
where $C(t,t_\mathrm{w})$ is the two-time correlation function introduced in \eqref{eq:Cttw}.
The scaling functions $f_i$ are positive,
monotonically decreasing and normalized, i.e. $1=\sum_i f_i(1)$. The
unspecified functions $h_i$ are such that, in the large $t_\mathrm w$ limit,
$h_i(t_\mathrm w)/h_i(t+t_\mathrm w)$ is $1$ if $t\ll t_\mathrm w^{\mu_i}$,
while it tends to zero if $t\gg t_\mathrm w^{\mu_i}$. In other words, the
decay of $C$ between values $C_i$ and $C_{i+1}$ is ruled by the scaling
function $f_i$ and takes place in the \textit{time-sector} $t\sim t_\mathrm
w^{\mu_i}$.

This picture is radically different to the \textit{Full Aging} often found
both in experiments (recall Figure \ref{fig:fullaging-intro}) and in $3D$ simulations. A full aging dynamics is ruled only by two sectors of time, $\mu_i=0,1$. Nevertheless, recent experimental studies~\cite{kenning:06} show that full aging is no longer fulfilled for $t\!\gg\!t_\mathrm{w}$. Probably more time-sectors must be considered to rationalize these experiments.

However,  \eqref{eq:timesectors} is probably an oversimplification, since the spectrum of exponents $\mu_i$ might be continuous. An explicit realization of this idea was found in the critical dynamics of the trap model~\cite{bertin:02}, where the correlation function behaves for large $t_\mathrm w$ as
\be\label{eq:ultrametric1}C(t,t_\mathrm{w})=f\paren{\alpha(t,t_\mathrm{w})}\,,\quad \alpha(t,t_\mathrm{w})=\log t/\log t_\mathrm{w}\,.\ee
Again, the scaling function $f$ is positive and monotonically decreasing. Clearly enough, in the limit of large $t_\mathrm w$ and for any positive exponent $\mu$, if $t=At_\mathrm w^\mu$, the correlation function takes a value that depends only on $\mu$, no matter the value of the amplitude $A$.

\begin{figure}
\begin{center}
\includegraphics[angle=270,width=0.8\columnwidth,trim=40 0 40 0]{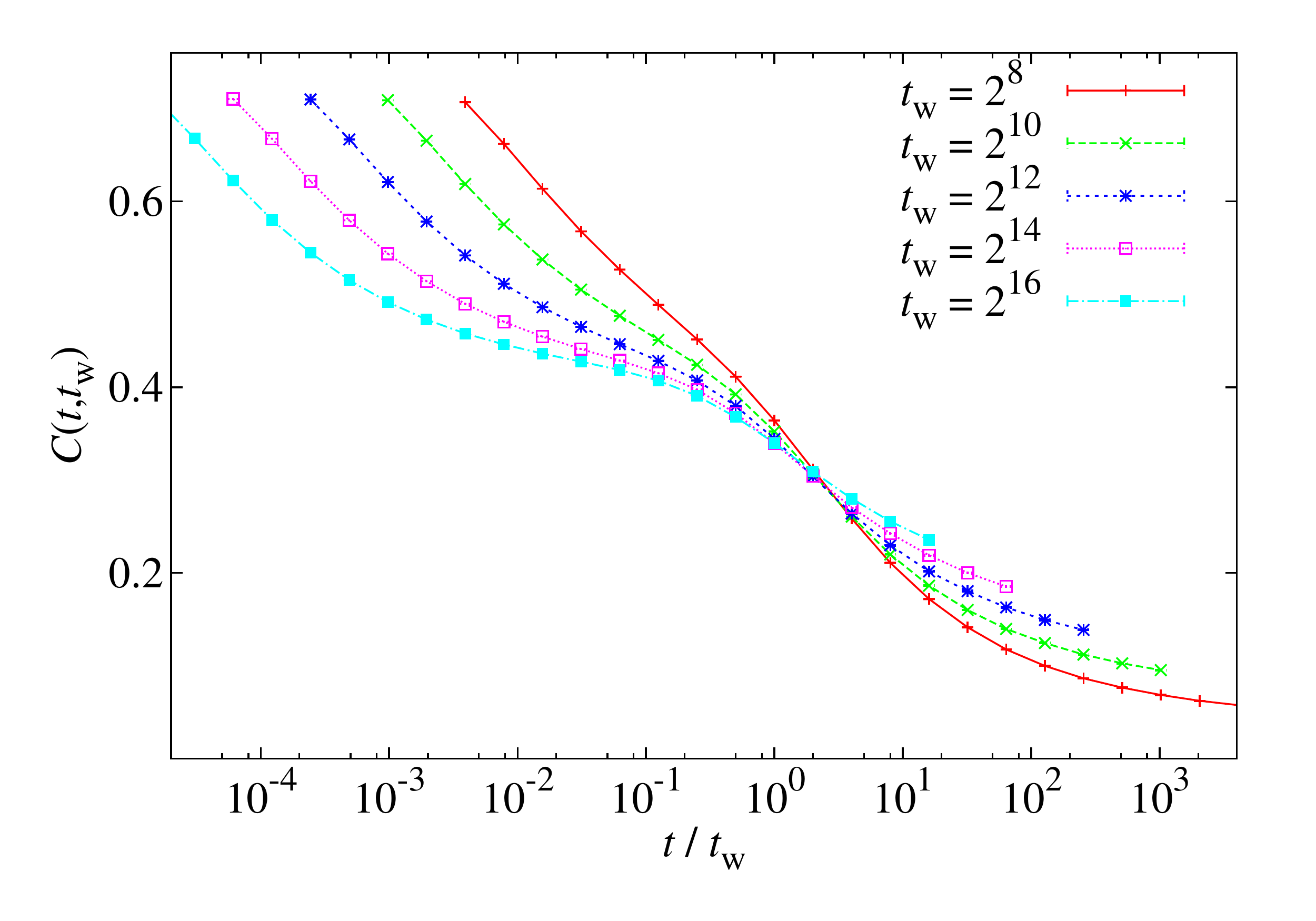}
\caption{ $C(t,t_\mathrm{w})$ over $ t/ t_\mathrm{w}$ for $D\!=\!22$ and $T\!=\!0.7T_\mathrm{c}$.}\label{fig:fullaging}
\end{center}
\end{figure}

\begin{figure}
\begin{center}
\includegraphics[angle=270,width=0.8\columnwidth,trim=40 0 40 0]{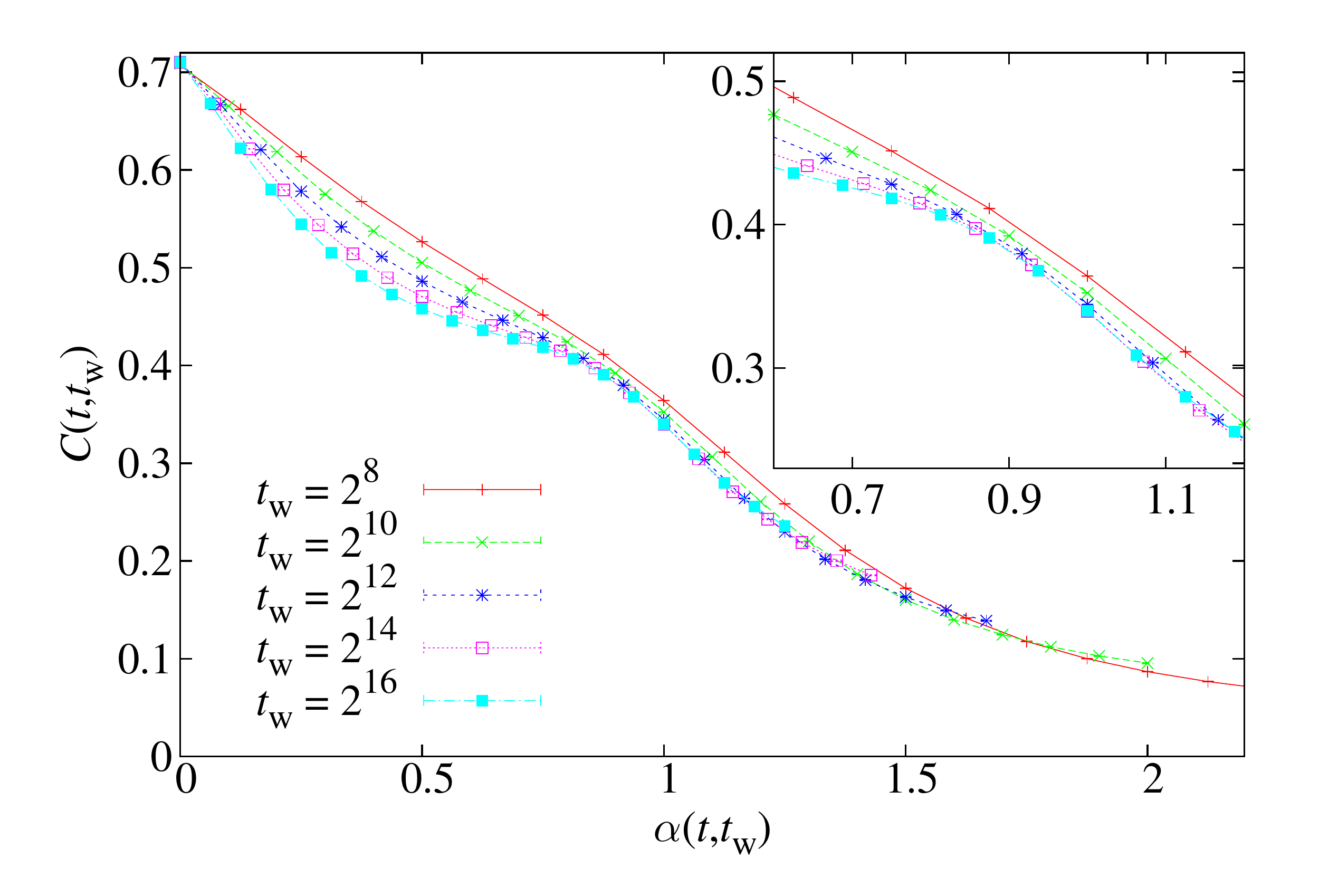}
\caption{ Same data of Figure \ref{fig:fullaging}, as a function of $\alpha(t,t_w)$, defined in  \eqref{eq:ultrametric1}. The window is a zoomed image of the central region.}\label{fig:ultrametric1}
\end{center}
\end{figure}

As expected, $C(t,t_\mathrm{w})$ is clearly not a function of $t/t_\mathrm{w}$ in our model, see Figure \ref{fig:fullaging}. On the contrary, data seem to tend to a constant value when $t_\mathrm w\to \infty$ in any finite range of the variable $t/t_\mathrm{w}$. This is precisely what one would expect in a time-sectors scheme.
On the other hand, if we try (without any supporting argument) the Bertin-Bouchaud scaling,  \eqref{eq:ultrametric1}, see Figure \ref{fig:ultrametric1}, the data collapse is surprisingly good. Therefore, the nonequilibrium dynamics in the SG phase seems ruled by a, not only infinite but continuous, spectrum of time-sectors.

We note \textit{en passant} that the scaling \eqref{eq:ultrametric1} is
ultrametric only if the scaling function reaches a constant value for all
$\alpha(t,t_w)>1$, for details see Appendix \ref{ap:Ultrametricity}. In fact,
dynamic ultrametricity is a geometric property~\cite{cugliandolo:94} that states
that for all triplet of times $t_1\gg t_2\gg t_3$, one has in the limit
$t_3\to\infty$: \be
C(t_1-t_3,t_3)=\min\left\{C(t_1-t_2,t_2),C(t_2-t_3,t_3)\right\}.\ee Finding
dynamical ultrametricity in concrete models has been rather elusive up to
now. An outstanding example is the critical trap model~\cite{bertin:02}, where
$f(\alpha>1)=0$. It is amusing that the trap model is \textit{not} ultrametric
from the point of view of the equilibrium
states, as discussed in Section~\ref{sec:SK}. Thus, the casual connections between
static and dynamic ultrametricity are unclear to us. At any rate, since our
scaling function in Figure \ref{fig:ultrametric1} does not show any tendency to
become constant for $\alpha(t,t_w)>1$, we do not find compelling evidences for
dynamic ultrametricity in this model.

We have also looked directly to the plots of $C(t_1-t_2,t_2)$ versus
$C(t_2-t_3,t_3)$ (see Appendix \ref{ap:Ultrametricity}) and we have not found convincing indications for the onset
of dynamical ultrametricity. In this respect, it is worth to recall similarly
inconclusive numerical investigations of the Sherrington-Kirkpatrick model~\cite{cugliandolo:94,berthier:00}. There are two possible conclusions:
\begin{enumerate}
\item the model does not satisfy dynamical ultrametricity in spite of the
fact that it satisfies (according to the standard wisdom) static
ultrametricity.
\item Dynamical ultrametricity  holds but its onset is terribly slow.
\end{enumerate}
Both conclusions imply that it is rather difficult to use the dynamic
experimental data (or any kind of data) to get conclusions on static
ultrametricity.
\subsection{Aging in $C_\text{link}$}
\begin{figure}
\begin{center}
\includegraphics[angle=270,width=0.8\columnwidth,trim=40 0 40 0]{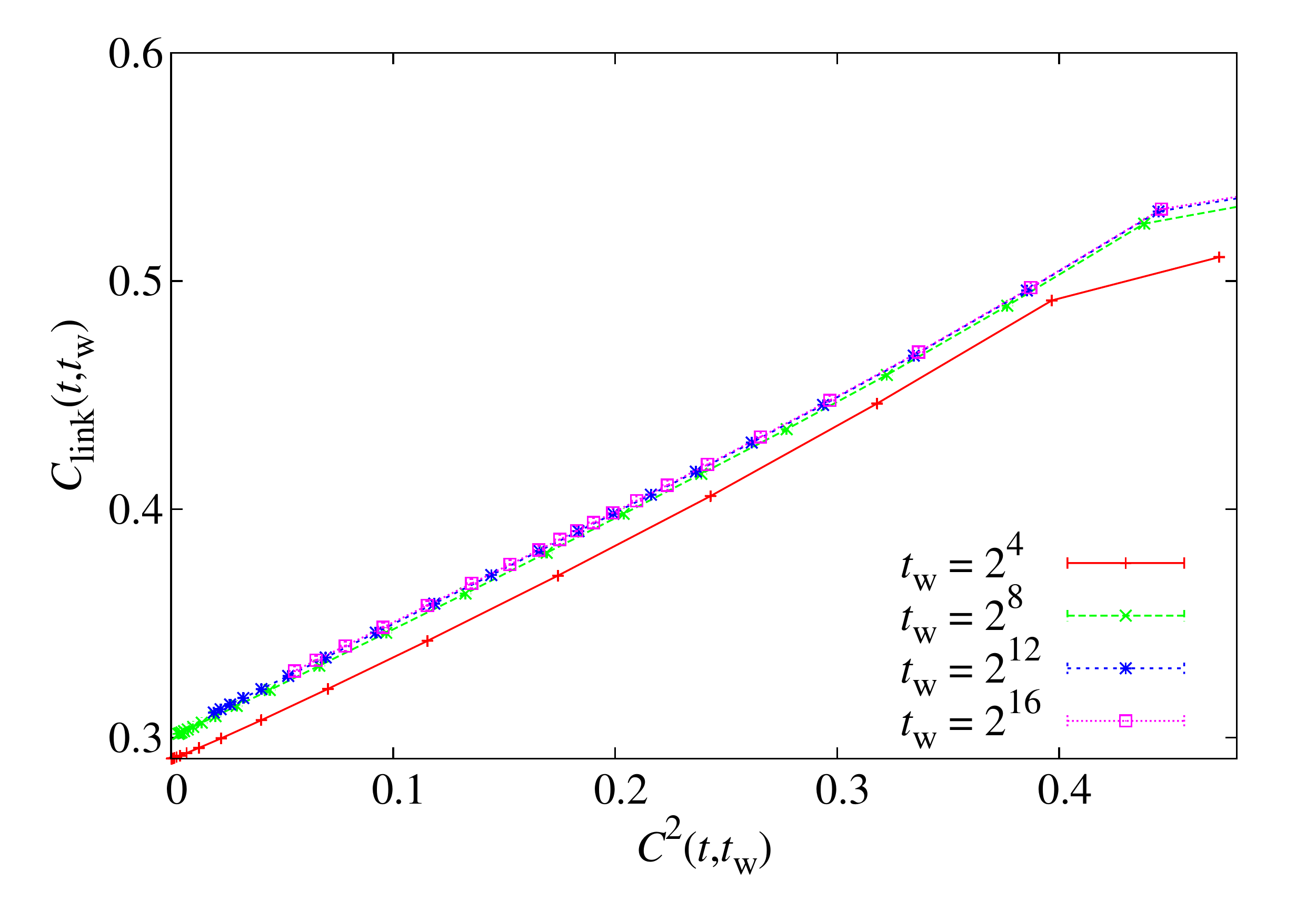}
\caption{ $C_\mathrm{link}$ over $C^2(t,t_\mathrm{w})$ for different
  $t_\mathrm{w}$ at $T\!=\!0.7T_\mathrm{c}$ and for $D\!=\!22$.}\label{fig:Clink}
\end{center}
\end{figure}

Just as in the $3$D case~\cite{janus:08}, the aging dynamics in SG in the
hypercube is a domain-growth process, see Figure \ref{fig:xi_70}. For any such
process, the question of the ratio surface-volume arises. When this ratio
vanishes in the limit of large domain size, as it is the case for any \ac{RSB}
dynamics (recall the discussion on the replica equivalence in Section
\ref{sec:overlapequiv}), one expects a linear relation between
$C_\mathrm{link}(t,t_\mathrm{w})$ (defined in \eqref{eq:clink}) and
$C^2(t,t_\mathrm{w})$. This is precisely what we find in Figure
\ref{fig:Clink}.

It is interesting to point out, that the linear relation found for $D=22$ is
the same one in the rest of system sizes, see Figure \ref{fig:ClinkD}.

\subsection{Thermoremanent magnetization}
\begin{figure}
\begin{center}
\includegraphics[angle=270,width=0.8\columnwidth,trim=0 0 0 0]{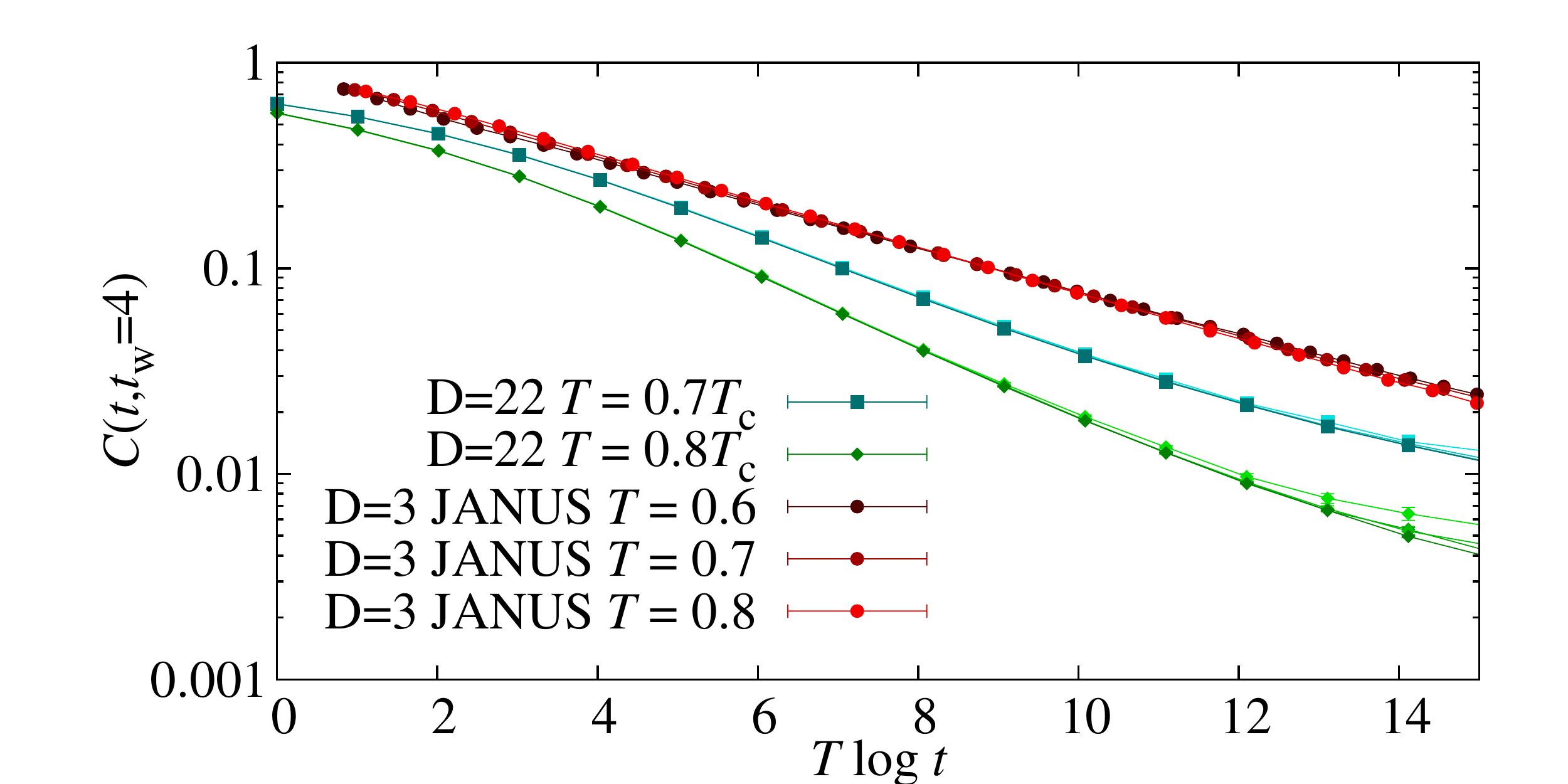}
\caption{ Thermoremanent magnetization over $T\log t$. The JANUS data
  (in red circles), follow a power law with an exponent $\propto\!1/T$. Our
  results for $D\!=\!22$ are shown in dark tonalities (lighter colors: $D\!<\!22$).}\label{fig:termorremanente}
\end{center}
\end{figure}

The experimental work indicates that for
$T\!<\!0.9T_\mathrm{c}$, the thermoremanent magnetization follows a power law with an exponent proportional
to $T_\mathrm{c}/T$  \cite{granberg:87}. The data obtained in JANUS for a
three dimensional SG (see Figure \ref{fig:termorremanente} and \cite{janus:09})
agree with this statement. However, the data obtained in the hypercube model
does not follow such power law, neither can them be rescaled with $T\log t$.

\begin{figure}
\begin{center}
\includegraphics[angle=270,width=0.8\columnwidth,trim= 40 0 40 0]{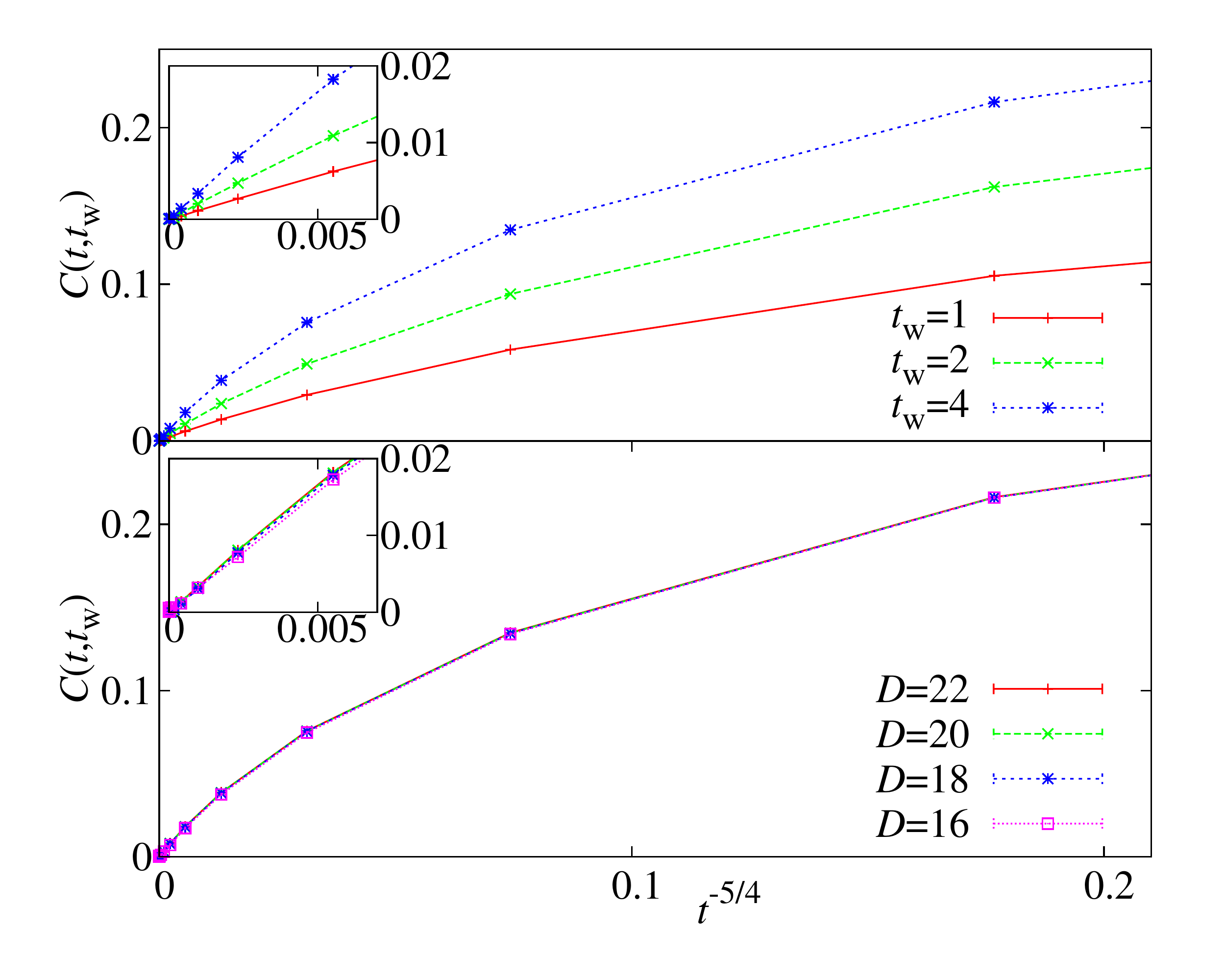}
\caption{ Thermoremanent magnetization at $T_\mathrm{c}$ vs. $t^{-5/4}$, for
  \textbf{(up)} different $t_\mathrm{w}$ and $D=22$, and \textbf{(down)}
  different system sizes for $t_\mathrm{w}=4$. The two insets are close-ups of the origin.}\label{fig:termo}
\end{center}
\end{figure}

This lack of an algebraic decay is surprising on the view of the exact results of Ref. \cite{parisi:97}. Indeed, it was analytically shown there that, at $T_\mathrm{c}$, the thermoremanent magnetization of the \ac{SK} model decays as $t^{-5/4}$. Universality strongly suggests that the same scaling should hold for our model. Although it seems not to be the case, at the first glance, Figure \ref{fig:termo}---top, a closer inspection confirms our expectation. Indeed, when plotted as a function of $t^{-5/4}$
, see inset in Figure \ref{fig:termo}---top, the thermoremanent magnetization curve has a finite non-vanishing slope at the origin. As we show in bottom panel of Figure \ref{fig:termo}, finite size effects do not contradict this claim. In summary, the magnetization decay for the hypercube suffers from quite strong finite time effects, but asymptotically it scales with the proper exponent, at least at $T_\mathrm{c}$.

\subsection{Nonequilibrium Correlation Functions and Finite Size
  Effects}\label{sec:TF}
\begin{figure}
\begin{center}
\includegraphics[angle=270,width=0.8\columnwidth,trim=0 0 0 0]{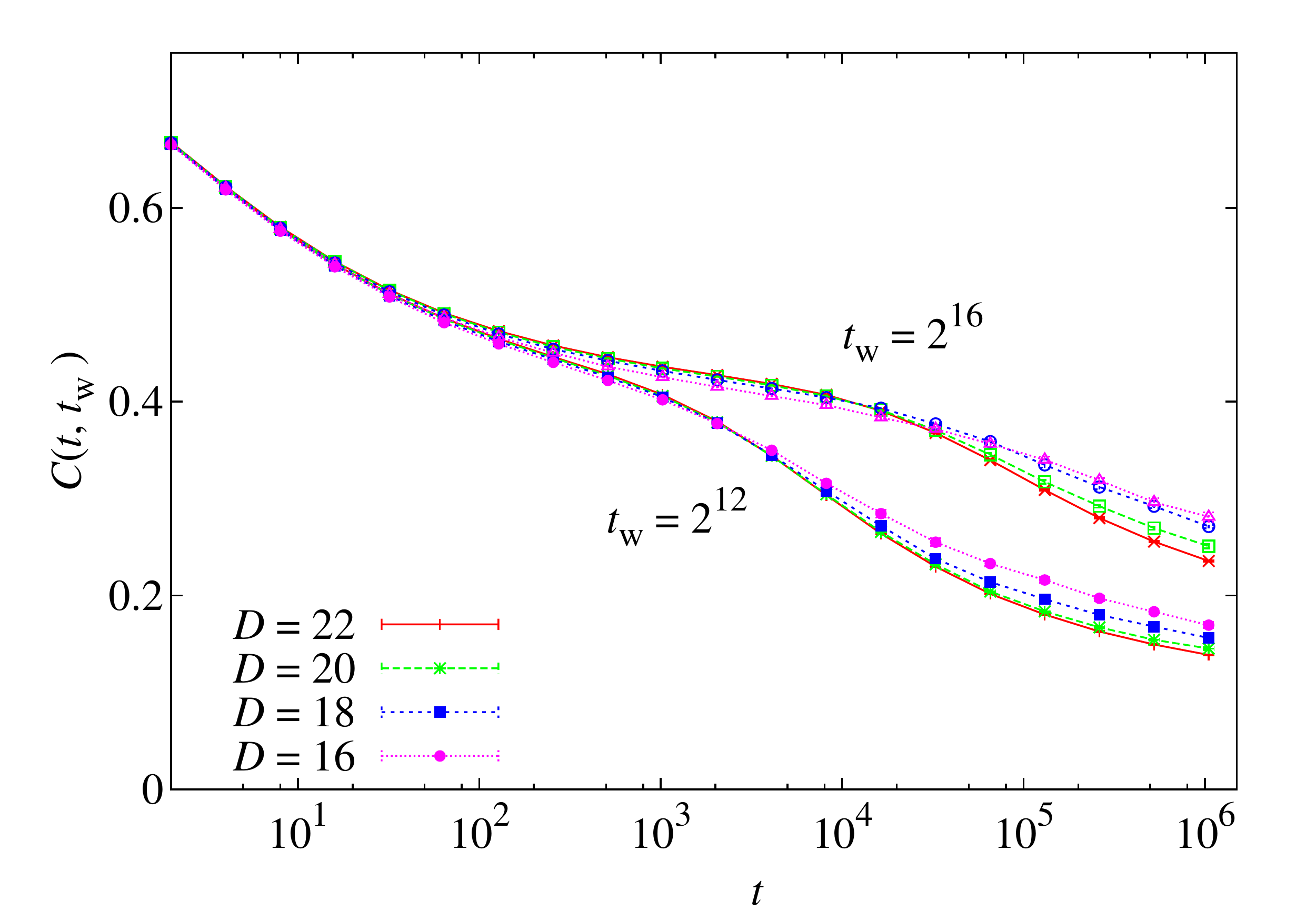}
\caption{ Finite size effects in $C(t,t_\mathrm{w})$ at $T\!=\!0.7T_\mathrm{c}$.}\label{fig:Cttw}
\end{center}
\end{figure}

\begin{figure}
\begin{center}
\includegraphics[angle=270,width=0.9\columnwidth,trim=0 0 0 0]{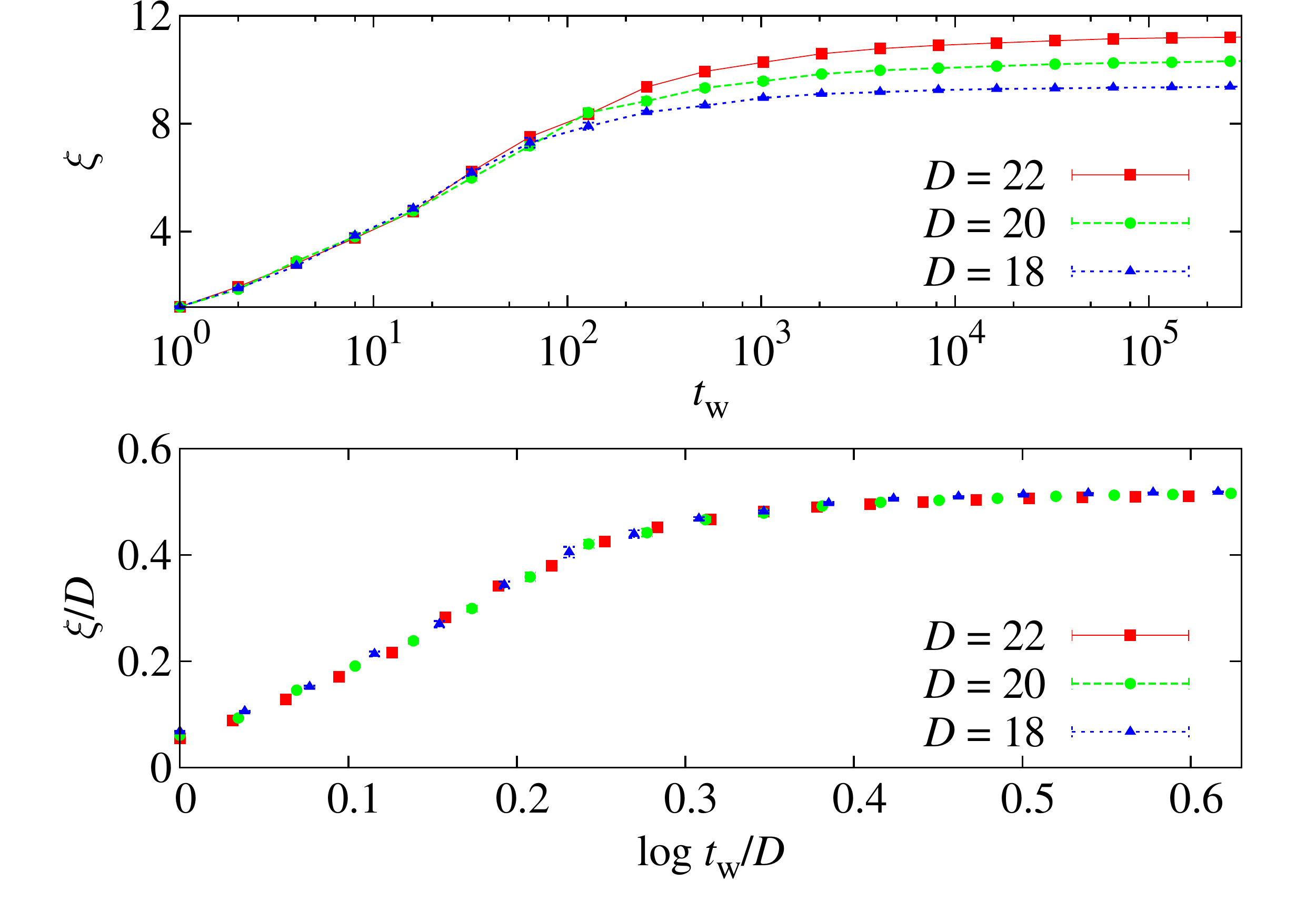}
\caption{ \textbf{(Top)} Coherence length $\xi$ in the SG phase at $T=0.7T_\mathrm{c}$ vs. $t_\mathrm w$ for different system sizes. \textbf{(Bottom)} same data of the top panel rescaled by $D$ as a function of $\log t_\mathrm{w}/D$.}\label{fig:xi_70}
\end{center}
\end{figure}

The importance of finite size effects in nonequilibrium dynamics has been emphasized recently~\cite{janus:08,janus:09}. In our case, we have encountered important size effects, both in $C(t,t_\mathrm w)$, Figure \ref{fig:Cttw}, and in $\xi(t_\mathrm w)$, Figure \ref{fig:xi_70}--top.

\begin{figure}
\begin{center}
\includegraphics[angle=270,width=0.8\columnwidth,trim=0 0 0 0
]{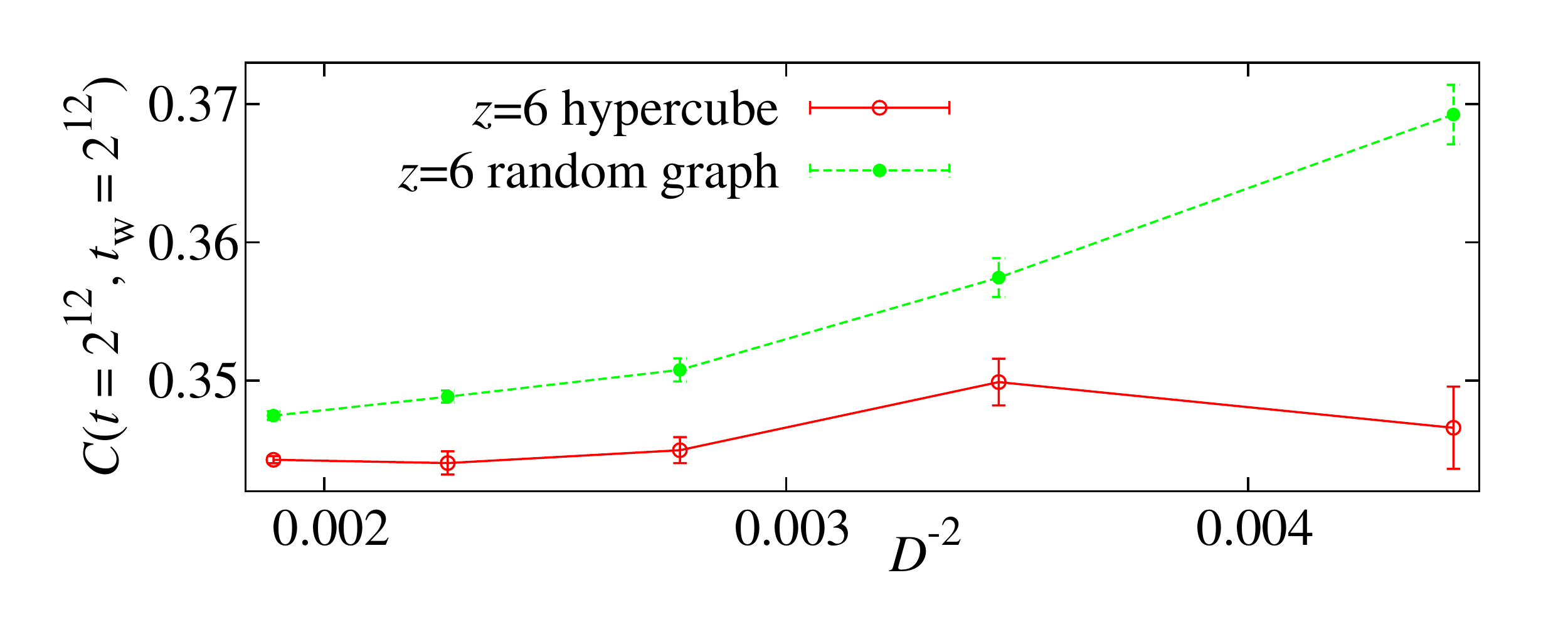}
\caption{ $C(t,t_\mathrm{w})$ at $T\!=\!0.7T_\mathrm{c}$ for $t\!=\!t_\mathrm{w}\!=\!2^{12
}$ vs. $1/D^2$. We compare results obtained with two $z\!=\!6$ models: one
  with hypercubic topology \textbf{(red open circles)} and another in a totally random
  graph \textbf{(green full circles)}.}\label{fig:Cttw_kc_compara}
\end{center}
\end{figure}

We compare in Figure \ref{fig:Cttw_kc_compara} the finite $D$ effects in $C(t,t_\mathrm w)$ for two different \ac{MF} models with fixed connectivity: the hypercube and a previously studied model (the random graph with connectivity $z\!=\!6$, where each spin can interact with any other spin with
uniform probability~\cite{leuzzi:08}). Clearly enough, the effects are much weaker in the hypercube model.

\begin{figure}
\begin{center}
\includegraphics[angle=270,width=0.8\columnwidth,trim=40 0 40 0]{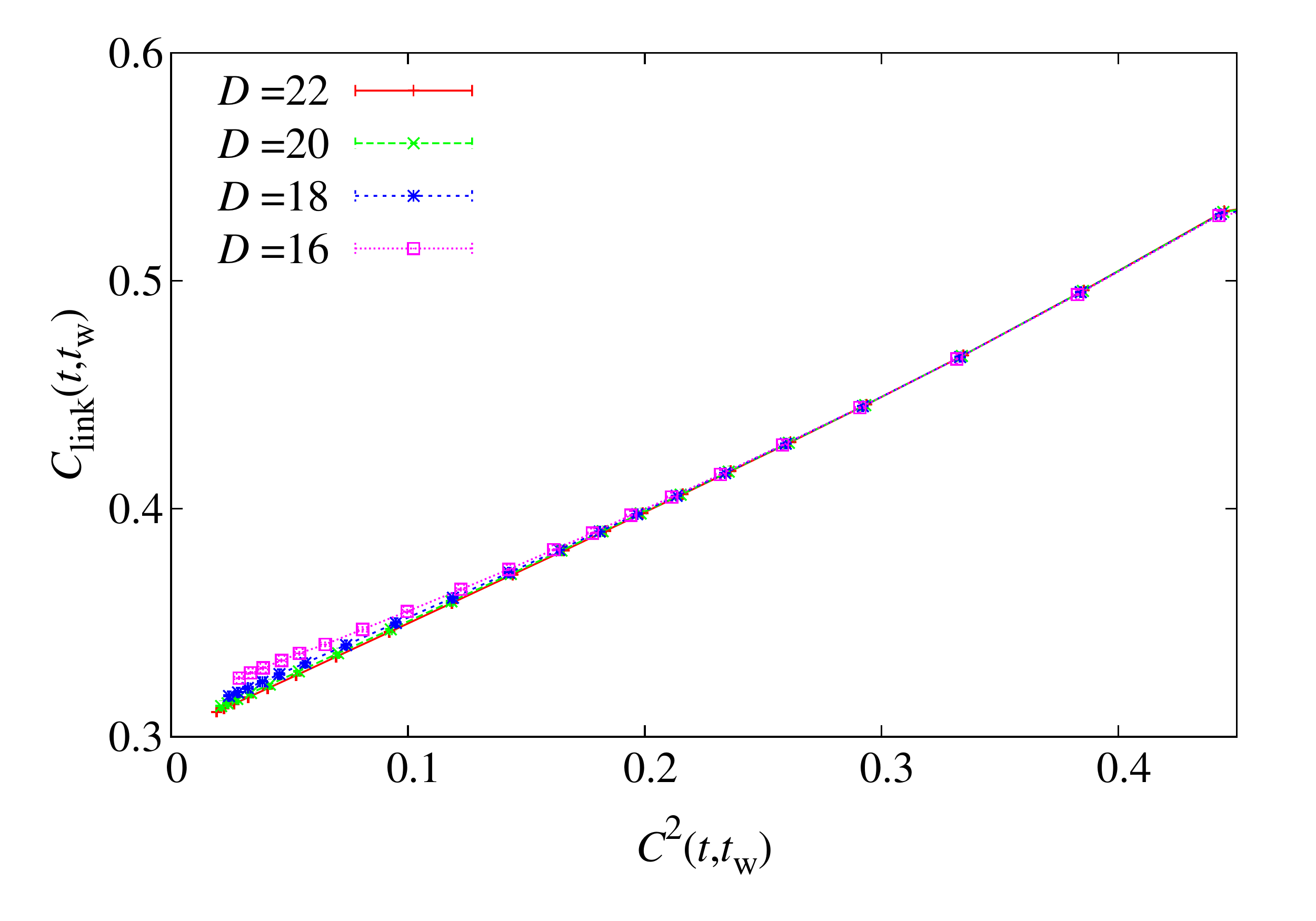}
\caption{ $C_\mathrm{link}$ over $C^2(t,t_\mathrm{w})$ at
  $T\!=\!0.7T_\mathrm{c}$ for $t_\mathrm{w}=2^{12}$ and for different system sizes.}\label{fig:ClinkD}
\end{center}
\end{figure}

It is interesting to point out that, although the finite size effects
seems to be important in $C(t,t_\mathrm{w})$, they are largely absorbed when one eliminates the variable $t$ in favor of $C^2(t,t_\mathrm{w})$, see Figure \ref{fig:ClinkD}. Hence, one of our main findings (the linear behavior of $C_\mathrm{link}$ as function of $C^2$) seems not endangered by finite size effects.

\begin{figure}
\begin{center}
\includegraphics[angle=270,width=0.9\columnwidth,trim=40 0 40 0]{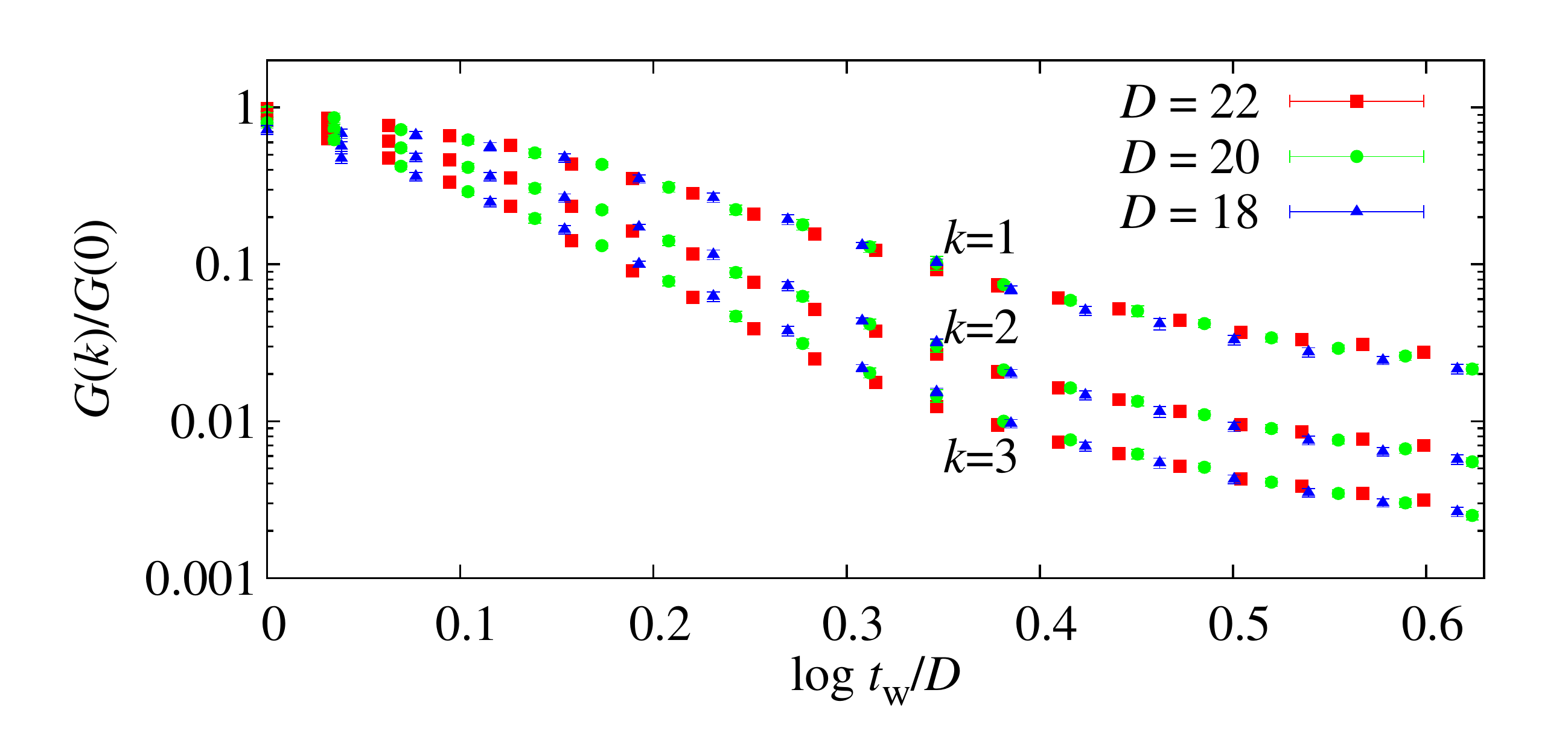}
\caption{ Fourier transform $G(k)$ of $\hat{C}_4(r)$ in units of $G(0)$ as a function of $\log t_\mathrm{w}/D$ for several values of $D$ and $k$ at $T=0.7T_\mathrm{c}$. For each value of $k$, a different scaling function is found.}\label{fig:Gk_scaling}
\end{center}
\end{figure}

A very clear finite size effect is in the coherence length,
$\xi(t_\mathrm{w})$. By definition, it cannot grow beyond $D$. Furthermore,
what we find is that it hardly grows beyond $D/2$,
Figure \ref{fig:xi_70}--top. Nevertheless, at short times, we can identify a
$D$-independent region, where it grows roughly as $\log t_\mathrm{w}$. Hence,
one is tempted to conclude that $\xi_{D=\infty}(t_\mathrm{w})\propto\log
t_\mathrm{w}$. At this point, finite size scaling suggests that both $\xi_D/D$
and $\log t_\mathrm{w}/D$ are dimensionless scaling variables. This is
confirmed in Figure \ref{fig:xi_70}--bottom, where a spectacular data collapse
occurs. This is further confirmed by the Fourier transform $G(k)$, defined in
 \eqref{eq:Gk}. Now, since $k$ can range from $0$ to $D$, it is clearly a dimensionless quantity (a dimensionful momentum would be $p=k/D$). It follows that $G(k)/G(0)$ is a dimensionless quantity that may depend only on a dimensionless variable, such as $\log t_\mathrm{w}/D$. Our data support this expectation, see Figure \ref{fig:Gk_scaling}.

As for the $k$ dependence of $G(k)$, we expect a $1/p^4$
behavior in the range of $1/\xi(t_\mathrm{w})\ll p\ll
1$~\cite{dedominicis:93,dedominicis:98,dedominicis:06}. Indeed, when comparing nonequilibrium with equilibrium
spatial correlation functions, it should be kept in mind that the
nonequilibrium ones correspond to the equilibrium $q=0$ sector
\cite{janus:08,janus:10} (since we take the large $D$ limit at fixed
$t_\mathrm{w}$).

\begin{figure}
\begin{center}
\includegraphics[angle=270,width=0.9\columnwidth,trim=40 0 40 0 ]{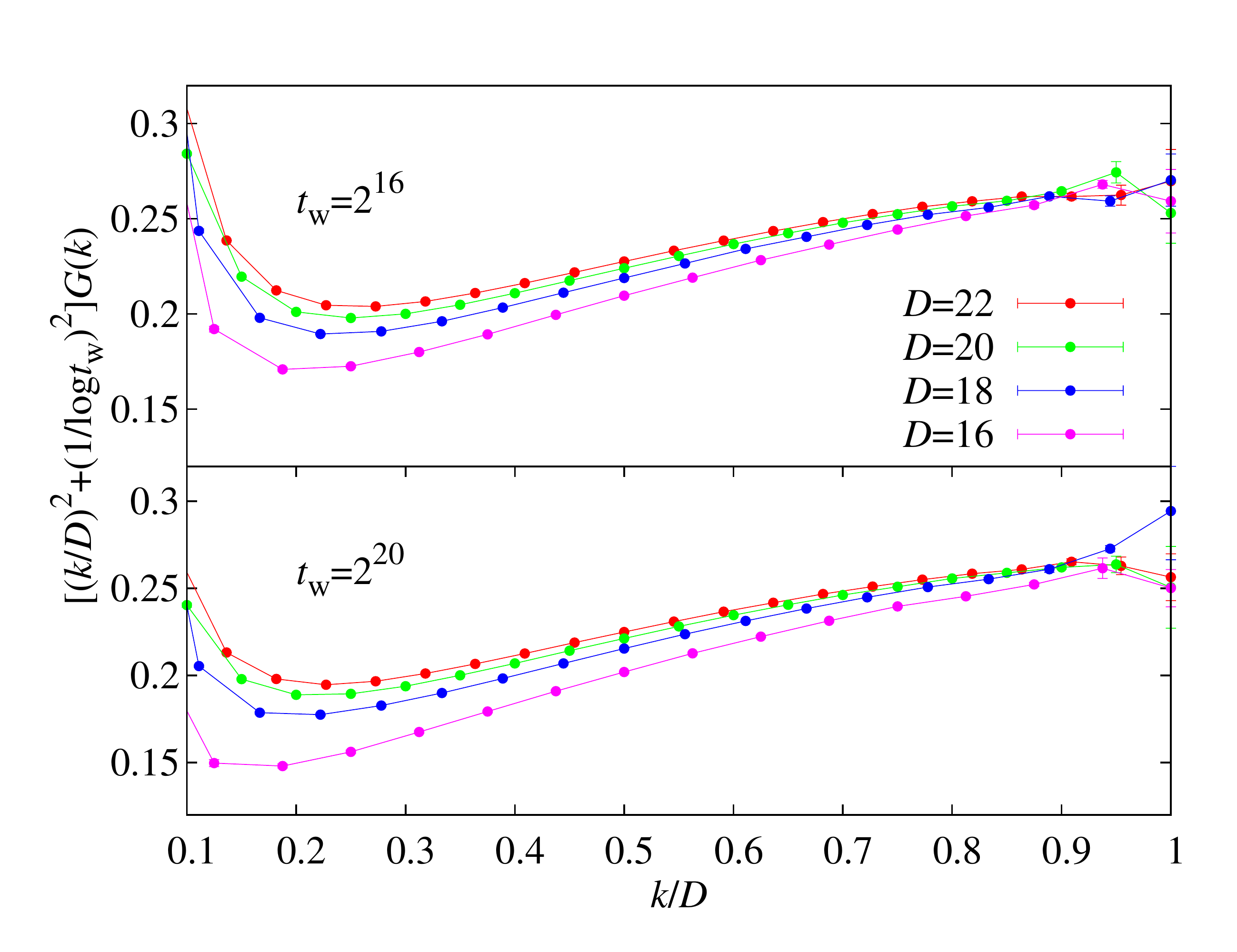}
\caption{ Fourier transform $G(k)$,  \eqref{eq:Gk}, in units of the
propagator [$\paren{p^2+1/\xi^2(t_\mathrm{w})}^{-1}$]~\cite{dedominicis:93,dedominicis:98,dedominicis:06} as a function of $p$,
where the dimensionful momentum is $p=k/D$ and $\xi(t_\mathrm{w})\sim \log
t_\mathrm{w}$. Recall that we are using postman metrics, hence, $p^2$
translates to $p^4$ in the Euclidean metrics. We show results for two waiting times: $t_\mathrm{w}=2^{16}$
\textbf{(top)} and $t_\mathrm{w}=2^{20}$
\textbf{(bottom)}. }\label{fig:Gk_scaling2}
\end{center}
\end{figure}

 Now, it is very important to recall that $p^4$ in
Euclidean metrics translates into $p^2$ in the postman metrics. We have also
seen that the dimensionful $p$ (postman metrics) corresponds to $k/D$. Thus,
since in our range of $t_\mathrm{w}$,
$\xi(t_\mathrm{w})\sim \log t_\mathrm{w}$, the product
$G(k)\paren{p^2+1/\log^2 t_\mathrm{w}}$ should be roughly constant as $D$ grows.
As we show in Figure \ref{fig:Gk_scaling2}, the scaling is better for $p$ of
order 1 ($k\sim D$), although it seems to improve for smaller $p$ as $D$ grows.
As far as we know, this is the first observation of the $p^4$ propagator in a
numerical work.

\chapter{Temperature Chaos \label{chap:chaos}}

\section{Introduction}\label{sec:chaos-intro}

Spin glasses (SG) display an anomalously large response to external
perturbations. This phenomena is known as {\em chaos}. Some of these
instabilities are well established. This is the case of the chaos induced in
the system by a magnetic field~\cite{parisi:84} or by small perturbations in
the bond configurations, known as disorder chaos. However, the temperature
counterpart (the fragility of the equilibrium state of a \ac{SG} when the
the temperature is slightly changed) remains to be
understood. This effect is named {\em temperature chaos} and will be the scope
of this chapter.

In the last years, temperature chaos has attracted a lot of attention
because of its suspected relation with the impressive memory and
rejuvenation experiments (see Sect.~\ref{sec:memory-rejuvenation})
which are still far from being understood. In fact, chaos is one of
the simplest explanations for rejuvenation. Indeed, if the equilibrium
states at two different temperatures, $T_1$ and $T_2$ ($T_2<T_1$),
were completely uncorrelated, the susceptibility would not be affected
by the isothermal aging at the previous temperature $T_1$. Memory is
still compatible with temperature chaos because of the length scale
separation discussed in Sect.~\ref{sec:memory-rejuvenation}, the ghost
domains in droplets scenario~\cite{yoshino:03,jonsson:04} or a
hierarchical organization of states with $T$, as discussed in
Section~\ref{sec:SG-scenarios}.

The property of temperature chaos was predicted a lot time
ago~\cite{bray:87}, but unfortunately, it remains still nowadays an
elusive phenomenon. The analytical work on temperature chaos is based
on two different approaches:
\begin{itemize} 
\item First, by means of scaling arguments and
real renormalization analysis~\cite{bray:87,fisher:88,banavar:87}. 
The scaling approach appears in the droplet theory framework (see
Sect. \ref{sec:SG-scenarios}). As discussed
below, this theory assumes that the lowest-energy excitations of the system
are compact domains of coherently flipped spins, the so-called
droplets. Because of its relevance in the field, we will spend some few lines
in describing how chaos is described phenomenologically using this droplet
picture. However, we will not follow this approach in our work, we will
explain why later.

By definition, the temperature chaos appears if the spin polarizations at two
temperatures $T_1$ and $T_2$ are decorrelated beyond certain characteristic
length, $\xi_\mathrm{C}(T_1,T_2)$, namely the {\em chaotic length}. One can
estimate this length using thermodynamic arguments and the scaling
picture. Indeed, two states will be uncorrelated if from temperature $T_1$ to
$T_2$ a droplet of size higher than this $\xi_\mathrm{C}(T_1,T_2)$ has
flipped. This happens if the free-energy inverts the sign between these two
temperatures. We use thermodynamic arguments to compute this free energy. Let
us consider two states at $T_1$ that differ one from the other by a large
droplet of size $\xi$. Then, using \eqref{eq:freedroplets}, the two free
energies differ by $\Delta F(T_1)\approx \gamma(T_1)\xi^\theta$. Now, we slightly change the temperature to $T_2$, so
that $|T_2-T_1|\ll T_1$. The total change of free-energy will be
\be\label{eq-chaos:dif-free}\Delta F(T_2)\approx
\gamma(T_1)\xi^\theta\pm|T_2-T_1| \sigma(T_1)\xi^{D_\mathrm{s}/2},\ee with
$\sigma (T_1)$ the {\em entropy stiffness} and $D_\mathrm{s}$ the fractal
dimension of the droplet's surface. Here, with the sake of clarity, we
considered only very small variations in temperature in order to neglect the
changes with the temperature in $\gamma$ and $\sigma$ (these effects can be
considered without too much change in the final expression, as done
in~\cite{katzgraber:07}).

According to  \eqref{eq-chaos:dif-free}, if $\theta\le D_\mathrm{s}/2$, as
happens in the droplet theory, the free energy $\Delta F(T_2)$ can have
opposite sign than $\Delta F(T_1)$ because of the entropic term.  This will
occur for scales greater than \be\label{eq:xichaos}
\xi_\mathrm{C}(T_1,T_2)\propto\caja{\frac{\gamma(T_1)}{\sigma(T_1)|T_2-T_1|}}^{1/\zeta},
\ee with $\zeta=D_\mathrm{s}/2-\theta$ being the {\em chaotic exponent}. Summarizing, when the temperatures are
changed, even for very small variations, the equilibrium configurations are expected to differ in scales
higher than this $\xi_\mathrm{C}(T_1,T_2)$.

There is a lot of numerical work available both in
\ac{MF}~\cite{billoire:00,billoire:02} or in more realistic~\ac{EA}
models~\cite{ney-nifle:97,ney-nifle:98,aspelmeier:02,rizzo:03,krzakala:04,sasaki:05,katzgraber:07},
and almost invariably, numerical data were analyzed using a scaling picture.

\item Second, using large deviation functionals and perturbation theory
  in~\ac{MF} models~\cite{billoire:02,rizzo:03,parisi:10}.  In \ac{MF}
  analytical calculations, temperature chaos is described in terms of a
  large-deviation functional (the free-energy of a system constrained to have
  similar spin configurations at two different temperatures in the SG phase,
  $T_1,T_2<T_\mathrm{c}$). Later on, this functional is obtained by means of a
  perturbative approach. The existence of this large-deviation functional
  implies a large fluctuation in the possible overlaps, which anticipates a
  dramatic sample-to-sample variability. 

A full analytical study of temperature chaos has been achieved
recently~\cite{rizzo:03,parisi:10} in mean field models, expected to be
accurate in spatial dimensions above $D\!=\!6$.  Surprisingly, it has been
shown that the most studied \ac{MF} model, the \ac{SK} model (see Section
\ref{sec:SK}) suffers anomalously weak temperature chaos effects, which
explains why it has been that slippery to find it in computer simulations (even
more than in more realistic models)~\cite{billoire:00,billoire:02}. Indeed,
all the lower power terms in the perturbative expansion of the
large-deviation functional pathologically vanish in the particular case of the 
\ac{SK} model. In fact, the temperature chaos has been studied in
diverse Bethe lattices models reaching the conclusion that chaos is stronger
the more heterogeneous the model is.
\end{itemize}

Despite of the intensive numerical work on this topic in the last $15$
years~\cite{billoire:02,ney-nifle:97,ney-nifle:98,krzakala:04,sasaki:05,katzgraber:07}
the numerical confirmation for the scaling picture is rather weak. All the
evidences presented are based on an indirect phenomenological renormalization
approach. Indeed, authors find nice scalings of the data which allow them to
infer the chaos exponent $\zeta$ (for instance in \ac{EA} models,
$\zeta\approx 1.07$ in $D\!=\!3$~\cite{katzgraber:07} and $\zeta\approx 1.12$
in $D\!=\!4$~\cite{sasaki:05}) which seems to be compatible with the accepted
values for $D_\mathrm{s}$ and $\theta$ in each model. However, this method
presents a major caveat: scaling holds also for $T_1,T_2 >
T_\mathrm{c}$, that is, deep in the paramagnetic phase (see
~\cite{ney-nifle:97,ney-nifle:98,katzgraber:07} and Figure
\ref{fig:ajuste}--bottom) where no chaos should be found.

Moreover, apart from this phenomenological renormalization, no numerical work
has succeeded in presenting clear evidences of chaotic behavior, that is, in
the sense of decorrelation between spin configurations at different
temperatures.  This failure has been attributed to a very large
$\xi_\mathrm{C}(T_1,T_2)$, comparable or larger than the simulated system
sizes~\cite{aspelmeier:02}. Summing up, this approach states that chaos should
be there but we are in the border of detecting it. Because of that, the
overall emerging picture is that of a gradual and extremely weak
phenomenon. However, if this $\xi_\mathrm{C}(T_1,T_2)$ were as large as
suggested ($\xi_\mathrm{C}\approx 20$ for $T_1=0.7$ and
$T_2=0.4$~\cite{aspelmeier:02}), the effect of temperature chaos should also
be very weak in experiments which handle with coherent clusters of roughly
$\sim 10^5$ spins~\cite{bert:04} (i.e. $\xi\sim 40$ lattice spacings). On the
other hand, rejuvenation is observed both in experiments~\cite{jonason:98} and
simulations~\cite{jimenez:05}, which means that either there is no connection
between temperature chaos and rejuvenation (as some authors
suggest~\cite{berthier:02,berthier:03} and Section
\ref{sec:memory-rejuvenation}), or there is something wrong in this picture
and the chaos pops up at much shorter length scales. Our analysis suggests
this second scenario.  In fact, in this chapter we shall extend the~\ac{MF}
picture and the large deviation functional approach to the $D\!=\!3$ \ac{EA} model.

However, even if the temperature chaos is an \emph{equilibrium} property, 
experimental SG are out of equilibrium as was widely shown all over Chapter
\ref{chap:intro-sg}. As discussed in Section \ref{sec:tldictionary}, this gap
between theory and experiment has been recently filled for isothermal
aging. The static-dynamics dictionary relates equilibrium properties of
\emph{finite-size} systems, with macroscopic aging samples at
\emph{finite-times}.  Unfortunately, the dictionary presented in Section
\ref{sec:tldictionary} works only for the simplest experimental protocol, in
which you cool the SG as fast as possible to the working temperature, then
keep $T$ constant. Instead, memory and rejuvenation effects are exposed only
by temperature-varying protocols. Static-dynamics dictionaries are yet to be
built for these richer protocols. Experimental attempts to establish them were
very crude~\cite{jonsson:02,jonsson:04,bert:04}. Indeed, a crucial ingredient
was missing: the characterization of equilibrium temperature-chaos and of its
system-size dependence.  Here, we achieve this task, thus paving the way for
extensions of the isothermal time-length dictionary to temperature-varying
experiments.

\section{Simulation set-up}

In this work, we revisit numerically the temperature chaos problem in the
$D\!=\!3$  Edwards-Anderson model studied in~\cite{katzgraber:07} but
using significantly higher systems ($L_\mathrm{max}=32$ here vs. $10$
in~\cite{katzgraber:07}) thermalized up to unprecedentedly low temperatures. For
this purpose, we reanalyze JANUS' equilibrium spin configurations already used
for previous equilibrium studies~\cite{janus:10,janus:10b}.
\begin{table}
\centering
\small
\label{tab:parameters}
\begin{tabular*}{\columnwidth}{@{\extracolsep{\fill}}ccccrlllcc}
\toprule
 $L$ &  $T_{\mathrm{min}}$ &  $T_{\mathrm{max}}$ &
  $\mN_T$ &  $\mN_\mathrm{mes}$ &  $\mN_\mathrm{HB}^\mathrm{min}$ &
\multicolumn{1}{c}{ $\mN_\mathrm{HB}^\mathrm{max}$}& \multicolumn{1}{c}{ $\mN_\mathrm{HB}^\mathrm{med}$} &
 $\mN_\mathrm{s}$ & System\\
\toprule
 8 & 0.245 & 1.575 &  8  & $10^3$ & $1.0\!\times\! 10^6$ & $6.48\!\times\!10^8$    & $2.30\!\times\!10^6$ & 4000 & PC    \\
12 & 0.414 & 1.575 & 12  & $5\!\times\! 10^3$ & $1.0\!\times\! 10^7$ & $1.53\!\times\!10^{10}$ & $3.13\!\times\!10^7$ & 4000 & PC \\
16 & 0.479 & 1.575 & 16  & $10^5$ & $4.0\!\times\! 10^8$ & $2.79\!\times\!10^{11}$ & $9.71\!\times\!10^8$ & 4000 & Janus \\
24 & 0.625 & 1.600 & 28  & $10^5$ & $1.0\!\times\! 10^9$ & $1.81\!\times\!10^{12}$ & $4.02\!\times\!10^9$ & 4000 & Janus \\
32 & 0.703 & 1.549 & 34  & $2\!\times\!10^5$ & $4.0\!\times\! 10^9$ & $7.68\!\times\!10^{11}$ & $1.90\!\times\!10^{10}$ & 1000 & Janus \\
\bottomrule
\end{tabular*}
\caption[Parameters of our parallel tempering
simulations]{Parameters of our spin-glass parallel tempering simulations. \index{parallel tempering|indemph}
 In all cases we have simulated four independent real replicas per
  sample. The $\mathcal N_T$ temperatures are uniformly distributed between
  $T_\mathrm{min}$ and $T_\mathrm{max}$ (except for the runs of the
  first row, which have all the temperatures of the second one plus
  $T=0.150$ and $T=0.340$).  In this table $\mN_\mathrm{mes}$ is the
  number of Monte Carlo Steps between measurements (one MCS consists
  of $10$ heat-bath updates and $1$ parallel-tempering update).
 The table   shows the minimum, maximum and medium simulation times
  ($\mN_\mathrm{HB}$) for each lattice, in heat-bath steps (the length of each
 simulation depends on the sample, for thermalization protocol see~\cite{yllanes:11}).  Lattice
  sizes $L=8,12$ were simulated on conventional PCs, while sizes
  $L=16,24,32$ were simulated on \textsc{Janus}. Whenever we have two runs with
  different $T_\mathrm{min}$ for the same $L$ the sets of simulated samples
  are the same for both. The total spin updates for all lattice sizes sum
  $1.1\times 10^{20}$. 
  \label{tab-chaos:parameters-eq}}
\end{table}

Our Ising spins $s_{\V{x}}=\pm 1$ are placed in the $V=L^D$ nodes $\V{x}$ of a
cubic lattice of linear size $L$, with periodic boundary conditions. The
interaction is restricted to lattice nearest neighbors. The coupling constants
$J_{\V{x},\V{y}}=\pm 1$ are chosen with $50\%$ probability. This model
undergoes a SG transition at
$T_\mathrm{c}=1.109(10)$~\cite{hasenbusch:08b}. We study $4000$ realizations
of disorder, named {\em samples}, for $L=8,\ 12,\ 16$ and $24$ ($1000$ samples
for $L=32$). The minimal temperature in the Parallel Tempering simulation
increased with $L$ (for $L=32$ it was $T_\mathrm{min}= 0.7026$). Simulation
details are summarized in Table \ref{tab-chaos:parameters-eq}.

\section{Selecting the right observables to change the
  paradigm}\label{sec:changeparadigm} As discussed above, the temperature
chaos has been an elusive phenomenon up to now. In this work, we argue that
the reason for its apparently small consequences was in the observables and
the statistical methods used in previous studies. In this section, we will
support that a change of paradigm is necessary: chaos must be treated as a
{\em rare event driven} phenomenon. Since this approach is quite novel, we
will spend some lines discussing which magnitudes are better to detect the
temperature chaos, and to define the concept of {\em chaotic event}.

In analogy with the rest of \ac{SG} studies, the natural parameter 
to approach the temperature chaos is the two temperatures overlap:
\begin{equation}\label{eq-chaos:overlap}
q_{T_1,T_2}=\frac{1}{V}\sum_{\V{x}} s_{\V{x}}^{T_1}s_{\V{x}}^{T_2}\,,
\end{equation}
i.e. the traditional spin overlap (see Eq.~\eqref{DEF:Q}) but mixing
configurations at two different temperatures. As it also happens at one single
temperature, the overlap \eqref{eq-chaos:overlap} is the one preferred
magnitude for mean-field analytical computations~\cite{rizzo:03,parisi:10}. As
an extension, numerical approaches to temperature chaos in \ac{MF} (only available in the
\ac{SK} model) also investigated this overlap~\cite{billoire:00,billoire:02},
obtaining an extremely low chaotic signal. For some time, this signal was so
low that this fact was used to support the non existence of this temperature
chaos phenomenon.  Nowadays we now that, among the mean field models, the
\ac{SK} model is pathological, in the sense that chaos is anomalously weak on
it~\cite{parisi:10} and all terms in perturbation theory below the ninth order
vanish in this precise model~\cite{rizzo:03}.

Here we want to detect chaos in the $D=3$ \ac{EA} model. As a first attempt,
we try to look directly to the spin overlap. According to the chaos
hypothesis, the overlap between equilibrium states at two different
temperatures should be always zero. That means that the \ac{pdf}
$P(q_\mathrm{C})$ should be a delta function centered on $q_\mathrm{C}=0$. Of
course, this the large $L$-limit, for a finite system one would expect a
growing peak with $L$ around $q_\mathrm{C}=0$ in the \ac{pdf}.  However, as we
discuss below, the chaotic signal through this magnitude is still too weak in
our computation, and more sophisticate quantities are needed.

The two-temperatures overlap is a stochastic variable, with two sources of
randomness: the thermal fluctuations, and the choice of the nearest-neighbors
couplings. In practice, for each sample, we have at our disposal four
independent sets of thermalized configurations (each independent set
corresponds to a single parallel-tempering Markov chain)~\cite{janus:10}.
Consider Monte Carlo times $t_A$ and $t_B$, from the Parallel-Tempering chains
$A$ and $B$.\footnote{ For
each chain, we pick a subset of $N_t$ configurations, evenly spaced in Monte
Carlo time (for $L=32$, $N_t=100$).}  In a fully explicit way, the two-temperatures overlap is
computed as
\begin{equation}\label{eq-app:overlap}
q_{T_1,T_2}(t_A,t_B;J)=\frac{1}{V}\sum_{\V{x}} s_{\V{x}}^{T_1}(t_A)
s_{\V{x}}^{T_2}(t_B)\,.
\end{equation}
 Computing such 
a large amount of overlaps in a feasible time was not an easy task,\footnote{For a single sample, we compute $6 N_t^2$ such overlaps (there are six ways of
choosing a pair $A\neq B$ out of four parallel-tempering chains). These
thermal fluctuations will be integrated out for further studies, but here,
instead, we want to explore  what happens when the thermal fluctuations
\emph{are} considered. In addition, there are $N_T^2/2$ possible couples of
temperatures, being $N_T$ the number of temperatures simulated.}
multispin coding techniques were necessary. We include an explanation about
these techniques in Appendix \ref{app:multispin-chaos}.

\begin{figure}[h]
\begin{center}
\includegraphics[angle=270,width=0.9\columnwidth,trim=0 0 0 0]{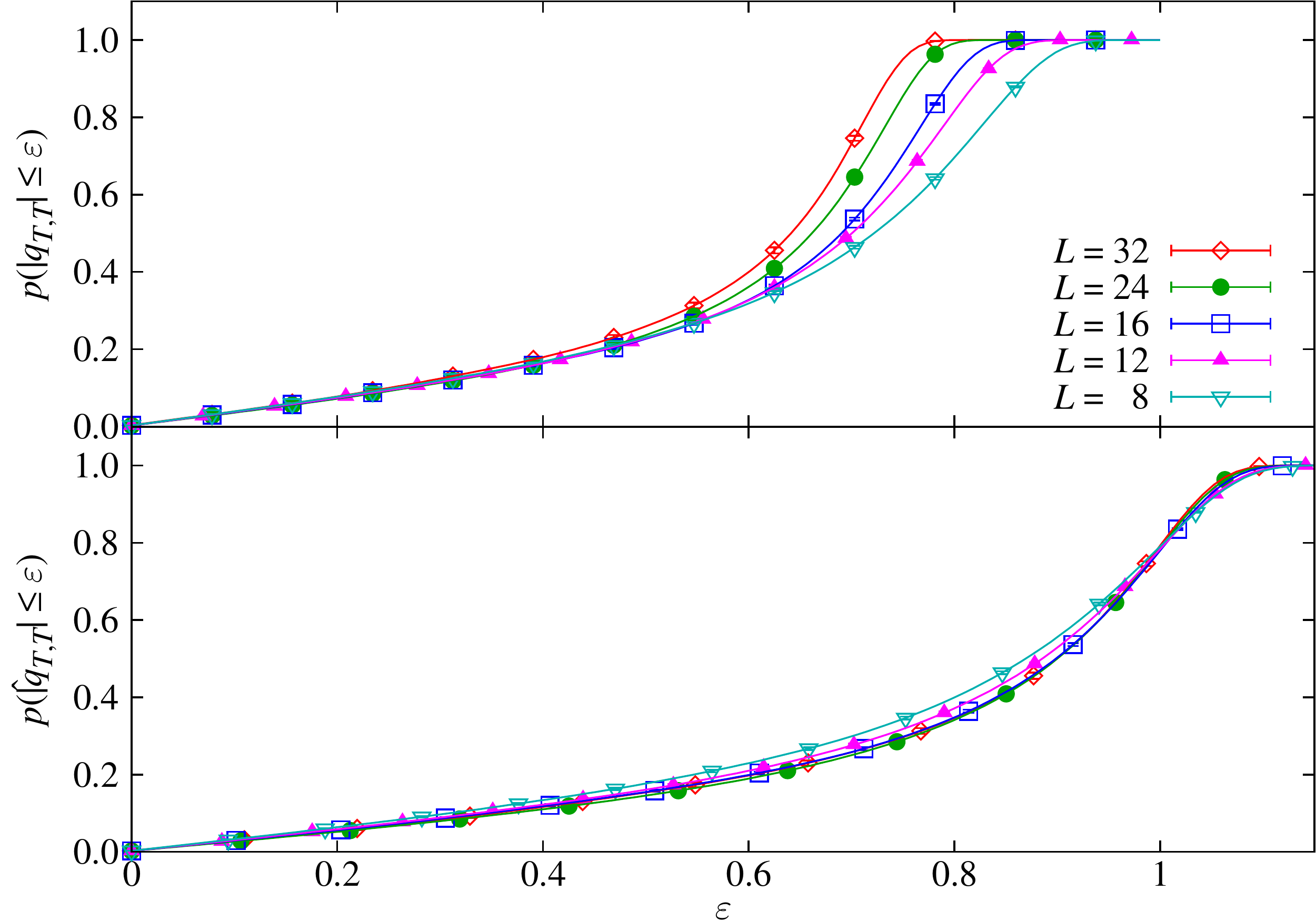}
\end{center}\caption{{\bf (Top)} Probability distribution function for the overlap at the
  same temperature, $T_1=T_2=0.7026$,
  $p(|q_{T_1=0.7026,T_2=0.7026}|\le\varepsilon)$. {\bf (Bottom)} Same as top
  panel, but subtracting system-size effects with the parameter
  $\hat{q}_{T_1,T_2}$, defined in \eqref{eq-app:qhat}. Error bars (smaller
  than the symbol size) are displayed.}
\label{fig:func_dist_q_T_T}
\end{figure}

In Figure \ref{fig:func_dist_q}--top, we show its accumulated \ac{pdf}, namely
the probability of finding a value of $q_{T_1,T_2}(t_A,t_B;J)$ no larger than
$\varepsilon$. In the chaos scenario, one would expect a step function in
$\varepsilon=0$. Clearly, we are far away from this limit, but the evolution
with $L$ seems to approach it.  However, these curves must be compared with
the curves at one single temperature, i.e.  $T_1=T_2$, see Figure
\ref{fig:func_dist_q_T_T}--top, which do not suffer from chaos.  We can see
that the situation is very much the same, these one-temperature curves
displays a strong size-dependency too.

In order to absorb the spurious non-chaotic
finite-size effects, we employ the lattice-size dependent Edwards-Anderson
parameter, $q_{EA}(L,T)$, defined in \eqref{eq:overlap1}, obtained with the same set of data in~\cite{janus:10}. In fact, if we rather compute the \ac{pdf} for
the following modified parameter,
\begin{equation} \label{eq-app:qhat}
\hat{q}_{T_1,T_2}=\frac{q_{T_1,T_2}}{\sqrt{q_\mathrm{EA}(L,T_1)\ q_\mathrm{EA}(L,T_2)}},
\end{equation}
see Figure \ref{fig:func_dist_q}--bottom, we realize that the chaos signal is
basically non-existent for $L\le 16$, and extremely weak for $L=24, 32$.  On
the other hand, if we consider the analogous curve for $T_1=T_2$, see Figure
\ref{fig:func_dist_q_T_T}--bottom, the curves collapse as one would expect.
Clearly enough, other effects with no relation with chaos (such as the global
spin reversal $\mathbb{Z}_2$ symmetry and the non-triviality of the
$P(q,T_1,T_2=T_1)$, at least on small lattices), are responsible for most of
the probability at low $q_{T_1,T_2}$.
\begin{figure}
\begin{center}
\includegraphics[angle=270,width=0.9\columnwidth,trim=0 0 0 0]{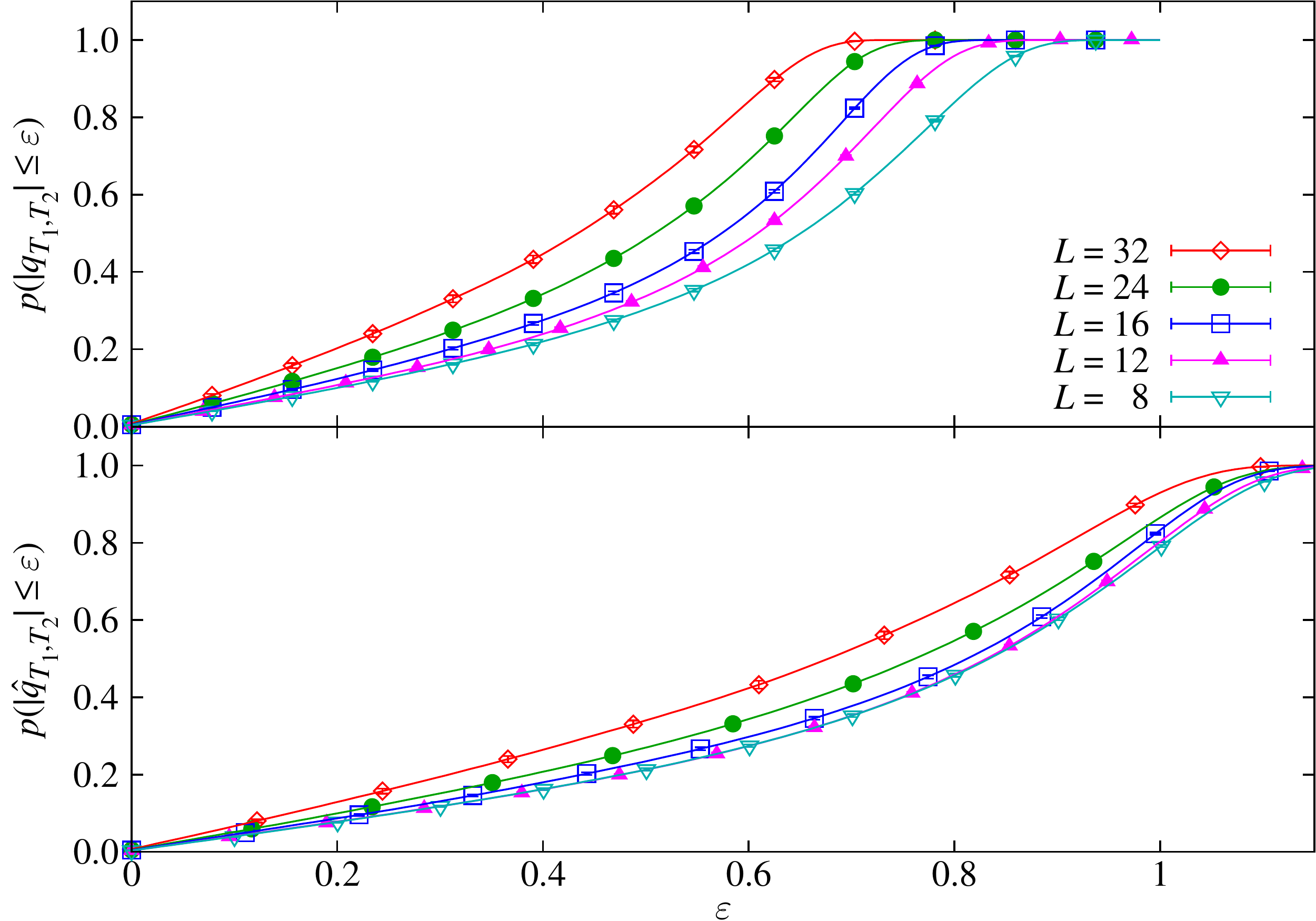}
\end{center}
\caption{{\bf (Top)} Probability distribution function for the overlap defined
  in ~\eqref{eq-app:overlap},
  $p(|q_{T_1=0.7026,T_2=0.90318}|\le\varepsilon)$. {\bf (Bottom)} Same as top
  panel, but subtracting system-size effects with the parameter
  $\hat{q}_{T_1,T_2}$, defined in \eqref{eq-app:qhat}. Error bars (smaller
  than the symbol size) are displayed.}
\label{fig:func_dist_q}
\end{figure}

With this idea idea in mind, now integrating out times and replicas,  we use a popular slight-modification to the
$\hat{q}_{T_1,T_2}$ parameter discussed before, known as the {\em chaotic parameter}~\cite{ney-nifle:97}
\begin{equation} \label{eq-chaos:X12}
X^J_{T_1,T_2}=\frac{\mean{q^2_{T_1,T_2}}_J}{\sqrt{\mean{q^2_{T_1,T_1}}_J\mean{q^2_{T_2,T_2}}_J}}\,.
\end{equation}
Here, $\mean{\cdot}_J$ refers to thermal-averages within the same sample. By
definition, $0\!\leq\! X^J_{T_1,T_2}\!\leq\! 1$. In fact, $X^J_{T_1,T_2}$ is
similar to a correlation parameter (if $X^J_{T_1,T_2}\!=\!1$ 
 two typical spin-configurations at $T_1$ and $T_2$ are
indistinguishable in a particular sample, while $X^J_{T_1,T_2}\!=\!0$ indicates completely different
configurations, then, extreme chaos). This
parameter absorbs many of the spurious effects found in the two-temperatures
overlap, but still, this $ X^J_{T_1,T_2}$ was used in numerical works
before~\cite{ney-nifle:97,ney-nifle:98,katzgraber:07}, and  the standard
analysis (wrongly)  concluded that chaos was very weak. That means that
this parameter~\eqref{eq-chaos:X12} is not enough by itself, and we need something more.

\begin{figure}\begin{center}
\includegraphics[angle=270,width=0.9\columnwidth,trim=30 0 40 0]{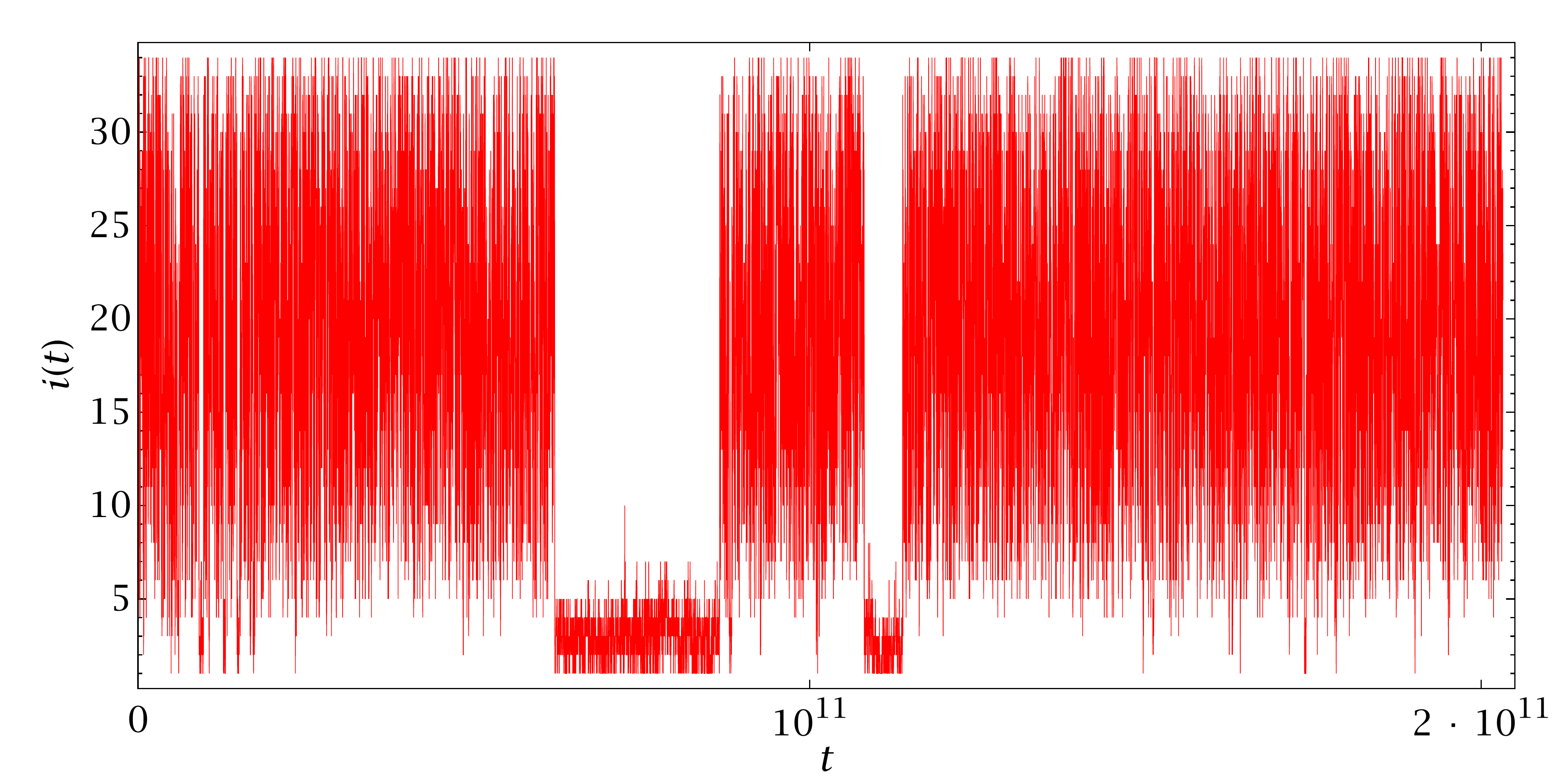}
\end{center}\caption{Temperatures random walk of a single configuration of one
  of the spin glass samples. The temperature index $i(t)$ is plotted as a
  function of the time in units of heat bath updates. The critical temperature
  corresponds to $i_\mathrm{c}=17$ (in the middle range). Clearly there is a
  well defined blockage in the dynamics' ergodicity around $i\sim 5$, our bet
  is that its origin is precisely the temperature chaos.}
\label{fig:RW}
\end{figure}

We look for some intuition.  We seek it in the Monte Carlo dynamics,
specifically in the temperature flow of the
\ac{PT}~\cite{hukushima:96,marinari:98b}. Indeed, if the equilibrium
configuration for two neighboring temperatures are too different (temperature
chaos), a bottleneck in the temperature random-walk should appear. This is
precisely what we find in the simulations, as it is illustrated in
Figure~\ref{fig:RW} for one of our configurations.
The simulation temperature flow can be characterized using the exponential
autocorrelation times, as we already did in~\cite{janus:10} to establish a 
thermalization protocol. Besides, the performance of \ac{PT} deteriorated
dramatically when the system size grows from $L=8$ to $L=32$. In fact, it was
precisely this strong stagnation of the \ac{PT} dynamics in certain samples,
what made us fell that a strong form of temperature chaos was waiting to be
unveiled. This idea of identifying equilibrium properties using the dynamics
is not new, it was used in the glassy context
before~\cite{schulman:07,fernandez:06}.

The temperature-flow dynamics is characterized by its exponential
autocorrelation time, $\tau_\mathrm{exp}$~\cite{sokal:97,janus:10}. Our
standpoint is that the quantity that better correlates with
$\log\tau_\mathrm{exp}$ will also be the most informative about chaos.  The
reason for seeking correlations with $\log \tau_{\mathrm{exp}}$ instead of
just $\tau_{\mathrm{exp}}$ is precisely the large sample to sample
variability. Indeed, given the disparity of times, one must take $\log
\tau_{\mathrm{exp}}$ in order to ensure that familiar concepts from Gaussian
statistics, such as the correlation parameter, make sense.

\begin{figure}\begin{center}
\includegraphics[angle=270,width=0.9\columnwidth,trim=30 0 40 0]{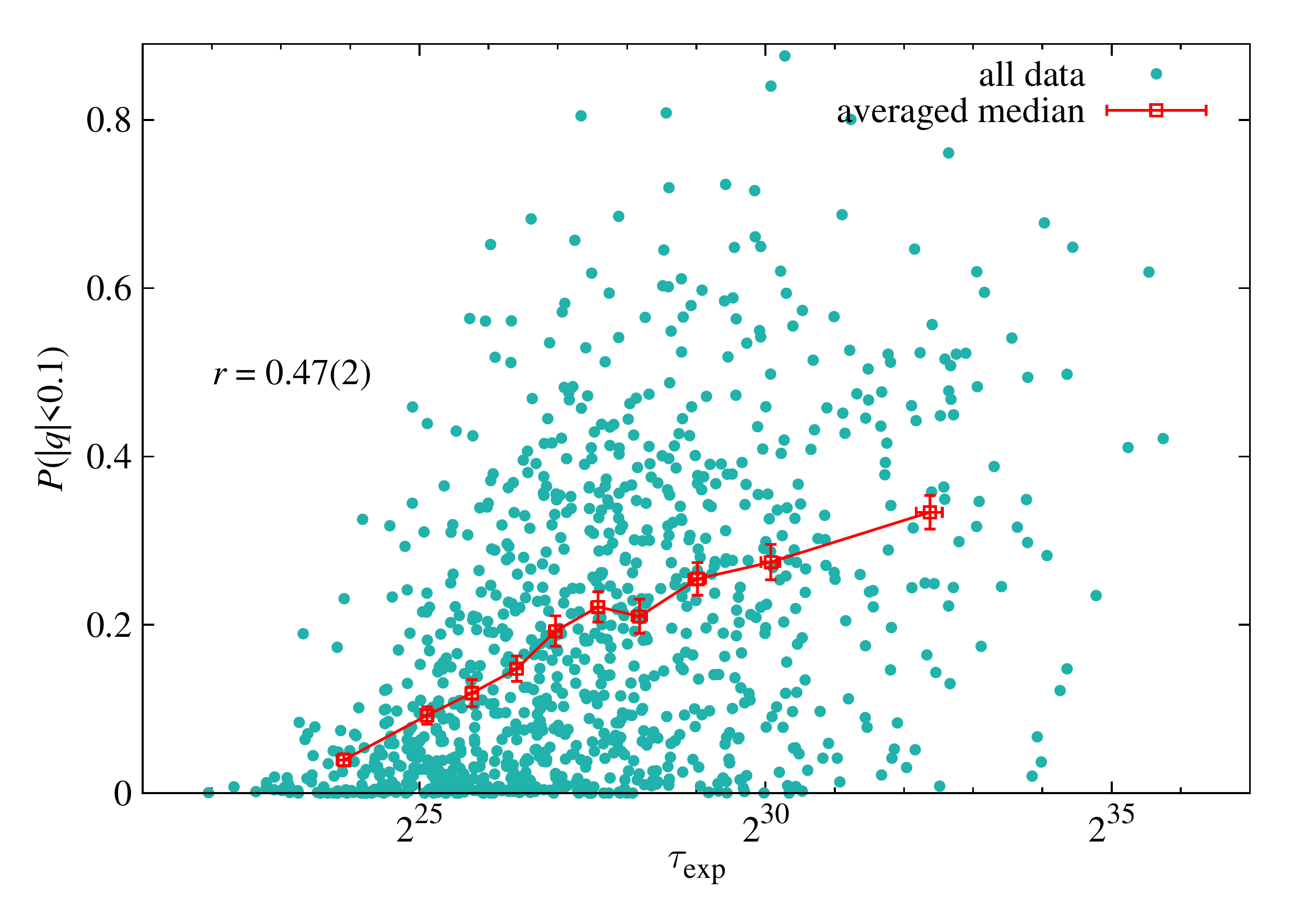}
\end{center}\caption{ Seeking clues about temperature chaos on the parallel
  tempering autocorrelation time $\tau_\mathrm{exp}$~\cite{janus:10}. We show
  the scatter plot $p(|q^J_{T_1=0.7026,T_2=T_\mathrm{c}|}<0.1)$
  [$q^J_{T_1,T_2}$ in ~\eqref{eq-chaos:overlap}] versus
  $\log{\tau_\mathrm{exp}}$. The correlation parameter $r$ is displayed. To
  compute the red lines, we ordered the samples by increasing
  $\log{\tau_\mathrm{exp}}$ and made groups of 100 consecutive samples. Within
  each group, medians were computed (errors from bootstrap).}
\label{fig:ob_vs_logtau_pq0}
\end{figure}

\begin{figure}\begin{center}
\includegraphics[angle=270,width=0.9\columnwidth,trim=30 0 30 0]{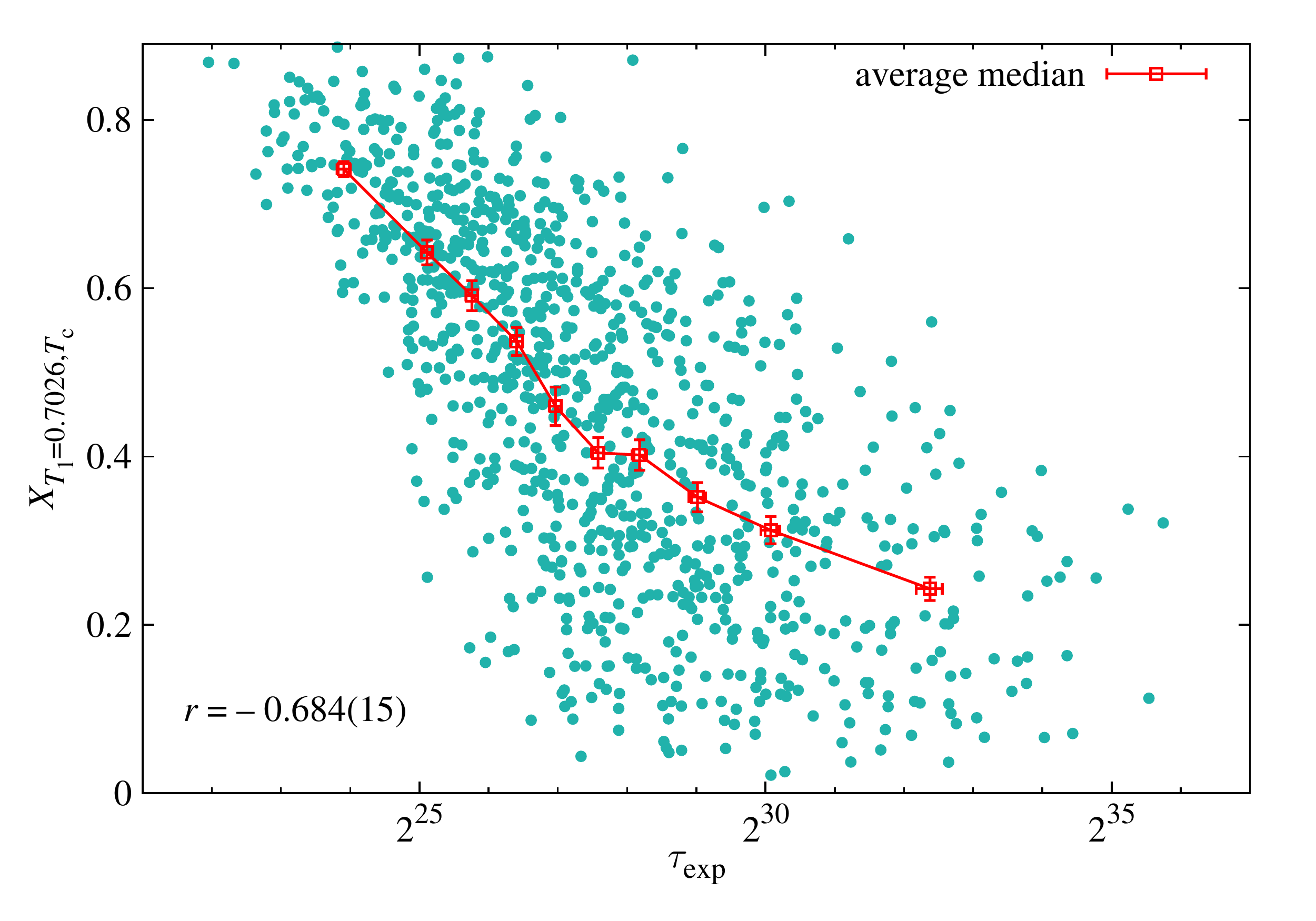}
\end{center}\caption{Scatter plot of
  $X_{T_1=T_\mathrm{min},T_2=T_\mathrm{c}}$ vs. the logarithm of the
  exponential autocorrelation time, $\tau_\mathrm{exp}$. Data for $L=32$. The
  correlation parameter $r$ is displayed.  The line is obtained in the same
  way than in Figure \ref{fig:ob_vs_logtau_pq0}.}
\label{fig:ob_vs_logtau_x1Tc}
\end{figure}

\begin{figure}\begin{center}
\includegraphics[angle=270,width=0.9\columnwidth,trim=50 0 50 0]{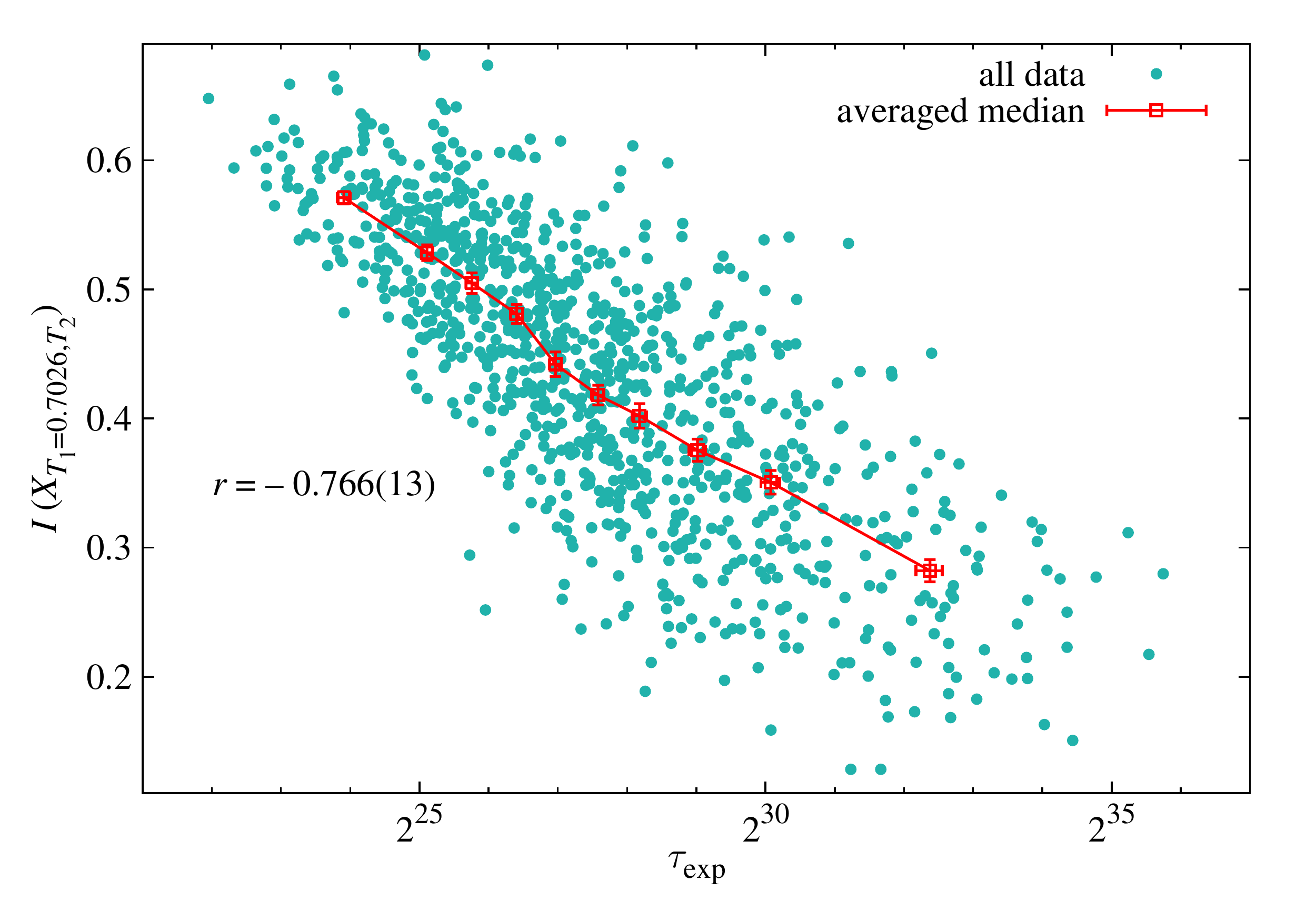}
\end{center}\caption{ 
  Scatter plot of $I\!=\!\int_{T_1}^{T_{\text{max}}}
  X^J_{T_1,T_2}\ \mathrm{d}T_2$ versus ${\tau_\mathrm{exp}}$,
  [$X^J_{T_1,T_2}$ in  \eqref{eq-chaos:X12}, $T_1\!=\!T_{\text{min}}\!=\!0.7026$,
    $T_\mathrm{max}\!=\! 1.549$,
    data for $L\!=\!32$]. The line is obtained in the same
  way than in Figure \ref{fig:ob_vs_logtau_pq0}. }
\label{fig:ob_vs_logtau_integral}
\end{figure}

As a first step, we study the correlation of the probability of finding small
overlaps \eqref{eq-chaos:overlap},
$p(|q^J_{T_1=0.7026,T_2=T_\mathrm{c}}|<0.1)$, with $\log\tau_\mathrm{exp}$ for
each sample in Figure \ref{fig:ob_vs_logtau_pq0}. Indeed, chaotic samples
should have very small overlaps, but as discussed above, the histogram for
$q_{T_1,T_2}$ around $0$, is affected by other non chaotic effects (that do
not hamper thermalization), and thus the correlation with times is poor.  The
situation is very much improved if we consider the correlation of the chaotic
parameter $X^J_{T_1,T_2}$ with $\log\tau_\mathrm{exp}$ instead.  We show in
Figure \ref{fig:ob_vs_logtau_x1Tc} this magnitude computed for
$T_1=T_{\text{min}}$ (our lowest temperature for $L=32$), and
$T_2=T_{\text{c}}$, the critical temperature, versus
$\log\tau_\mathrm{exp}$. The correlation is higher, but still we can find a
better magnitude.  In fact, our optimum is the integral of $X^J_{T_1,T_2}$
with temperature, i.e. \be I=\int_{T_1}^{T_{\text{max}}}
X^J_{T_1,T_2}\ \mathrm{d}T_2\ee, see Figure
\ref{fig:ob_vs_logtau_integral}. This integral will be small in the case that
$X^J_{T_1,T_2}$ suffers a sharp drop at low $T_2$, and as seen, the samples
with small $I$ correspond with those where the temperature flow is likely to
get stuck.  This correlation calls for the notion of {\em chaotic event},
rather than an analysis based on sample-averages.

\begin{figure}\begin{center}
\includegraphics[angle=270,width=0.8\columnwidth,trim=0 0 0 0]{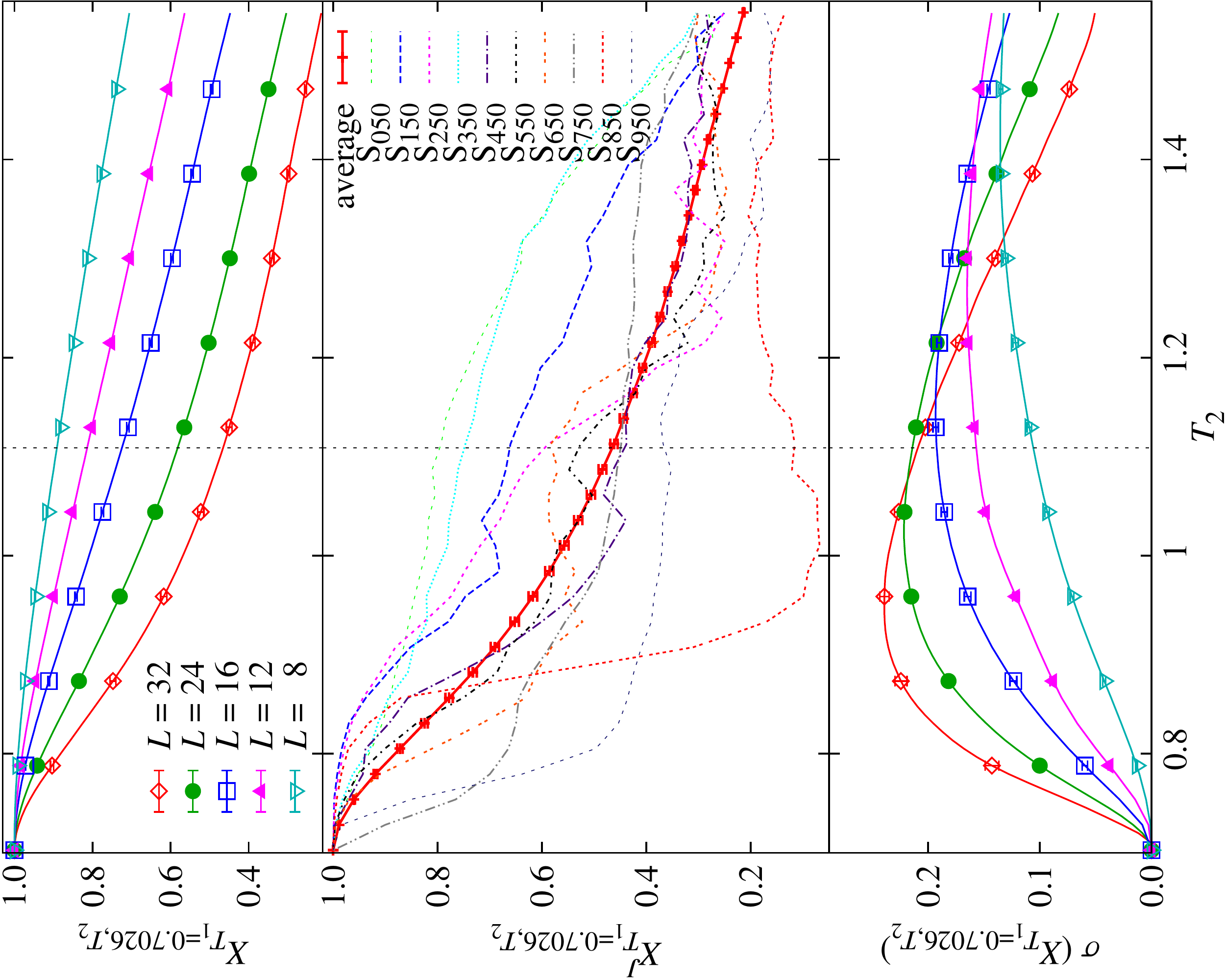}
\end{center}\caption{ Different views on $X^J_{T_1,T_2}$,
  ~\eqref{eq-chaos:X12}, as function of $T_2$ ($T_1\!=\!0.7026$, the vertical
  line is $T_2\!=\!T_\mathrm{c}$).{\bf (Top)} For all our system sizes,
  sample-average of $X^J_{T_1,T_2}$. {\bf (Center)} For $L=32$, we show
  $X^J_{T_1,T_2}$ for ten samples evenly spaced on a list of growing
  $\tau_\mathrm{exp}$, recall Fig~\ref{fig:ob_vs_logtau_pq0}. {\bf (Bottom)} For
  all our system sizes, we show the dispersion (i.e. square root of variance
  over the samples) of $X^J_{T_1,T_2}$.}
\label{fig:x12_y_var}
\end{figure}

Figure \ref{fig:x12_y_var} describes this change of paradigm.  The top panel
shows the standard average over the samples of $X^J_{T_1,T_2}$, as a function
of $T_2$.  In agreement with previous
work~\cite{ney-nifle:97,ney-nifle:98,katzgraber:07}, our simulated sizes are
painfully away from the large-$L$ limit, where the average of $X^J_{T_1,T_2}$
should vanish if $T_2\neq T_1$. Instead, our curves are smooth and cross
$T_\mathrm{c}$ without qualitative changes. This smoothness is a clear
indicator that chaos is not being detected. Indeed, temperature chaos is a
inner property of the spin glass phase, it cannot be found in the paramagnetic
phase.  Yet, the behavior of individual samples is quite different, see
Figure~\ref{fig:x12_y_var}---center. For some samples, $X^J_{T_1,T_2}$ falls
abruptly at well defined temperatures $T_2$. This we name {\em chaotic
  event\/}.  The temperature at which these events occur is random (many
samples do not suffer any). In fact, as $L$ grows, the sample dispersion of
$X^J_{T_1,T_2}$ in the SG phase, see Figure \ref{fig:x12_y_var}--bottom,
approaches $1/\sqrt{12}$ (which is the dispersion of a random variable
uniformly distributed between 0 and 1), whereas it tends to zero if
$T_2>T_\mathrm{c}$. This is quite a remarkable achievement: it is the first
time that a quantitative different behavior is observed between the \ac{SG}
phase (with chaos) and the paramagnetic phase (without chaos).  We conclude
that a statistical analysis based on sample averages (as shown in Figure~\ref{fig:x12_y_var}---top) throws away crucial
information about temperature chaos.

\begin{figure}\begin{center}
\includegraphics[angle=270,width=0.9\columnwidth,trim=50 0 50 0]{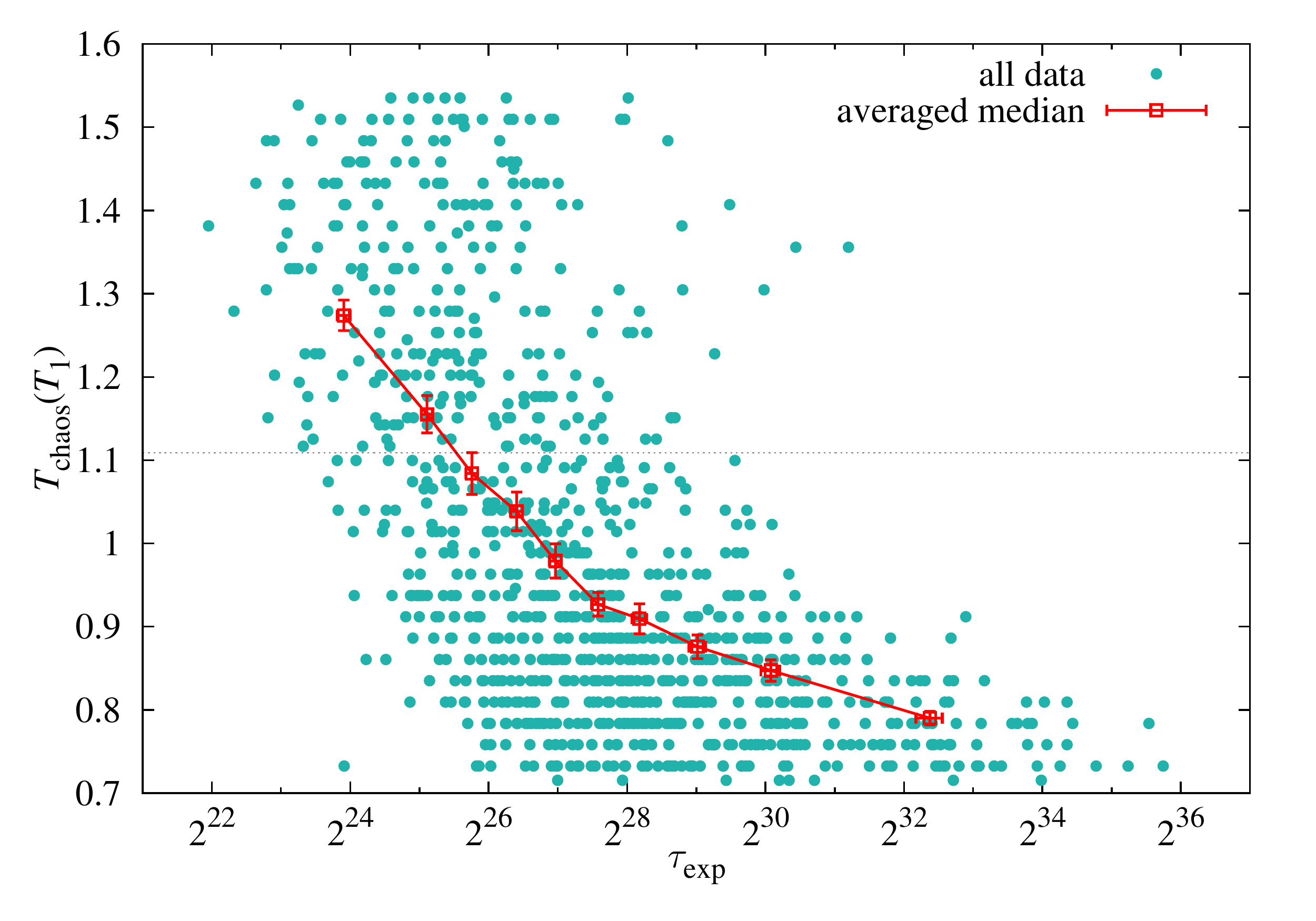}
\end{center}\caption{ Chaos temperature, $T_{\text{chaos}}^J$, defined in the
    text versus the logarithm of the exponential autocorrelation time
    $\tau_\mathrm{exp}$. Data for $L=32$. The red line is obtained in the same
    way than in Figure \ref{fig:ob_vs_logtau_pq0}. Horizontal dashed line
    corresponds to $T_\mathrm{c}=1.109$~\cite{hasenbusch:08b}.}
\label{fig:ob_vs_logtau_t2}
\end{figure}

Now, even though the temperature at which one of these sudden drops takes
place is random distributed within the \ac{SG} phase, it has strong effects in
the Parallel Tempering' performance (as suggests Figure
\ref{fig:ob_vs_logtau_integral}). Indeed, the deeper in the \ac{SG} phase, the
more stagnant the temperature flow is. In order to compute the correlation of
this temperature with the exponential autocorrelation time, we need to define
a method to compute this temperature. Our choice is the following: for each
sample, we consider the dependency of $X^J_{T_1,T_2}$ on $T_2$, keeping fixed
$T_1$ (as shown in Figure \ref{fig:x12_y_var}---center). We compute its
derivative with $T_2$, i.e. $\D X^J_{T_1,T_2}/\D T_2$, and obtain the
temperature at which it reaches its maximum value. This temperature will be
our {\em chaos temperature}, $T_{\text{chaos}}^J(T_1)$. This definition counts
drops for all the samples, even in those whose $X^J_{T_1,T_2}$ displays a soft
behavior without any chaotic effect. However, if this were the case, these
temperatures would, in majority, lay nearby $T_\mathrm{c}$ or on the
paramagnetic phase.  Now we study the correlation of this
$T_{\text{chaos}}^J(T_1)$ with $\log\tau_\mathrm{exp}$, see Figure
\ref{fig:ob_vs_logtau_t2}. The conclusion of this figure is clear, the deeper
in the \ac{SG} phase the drop takes place, the longer the thermalization time.

In summary, our statement is that chaos was not clearly observed in
numerical simulations up to now because the portion of samples that
suffered chaotic events was still too limited in the simulated system
sizes. Then, chaos seemed to be very week because it was very rare. Indeed,
we have seen that, when appearing, it is a strong phenomenon. Of
course, in order to chaos to be relevant in the large-$L$ limit this
portion of chaotic samples must grow with $L$. We devote the next
section to this discussion.

\section{Large-deviation approach}

\begin{figure}\begin{center}
\includegraphics[angle=270,width=0.8\columnwidth,trim=0 0 0 0]{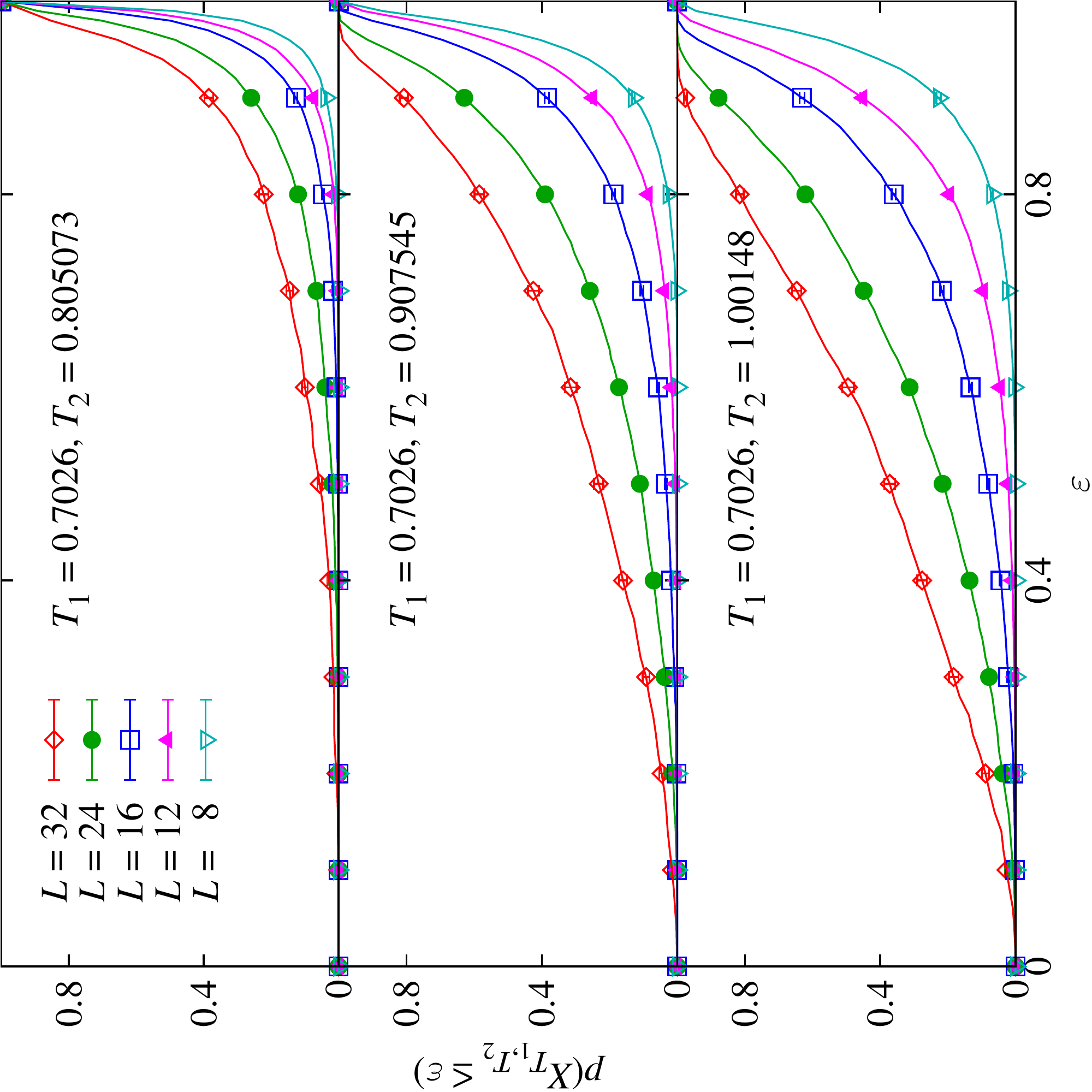}
\end{center}\caption{Probability distribution function for $X^J_{T_1,T_2}$ for $T_1=0.7026$ and {\bf (top)} $T_2=0.805703$, {\bf (center)} $T_2=0.907545$ and {\bf (bottom)} $T_2=1.00148$.}
\label{fig:func_dist_x}
\end{figure}

\begin{figure}\begin{center}
\includegraphics[angle=270,width=0.8\columnwidth,trim=0 0 0 0]{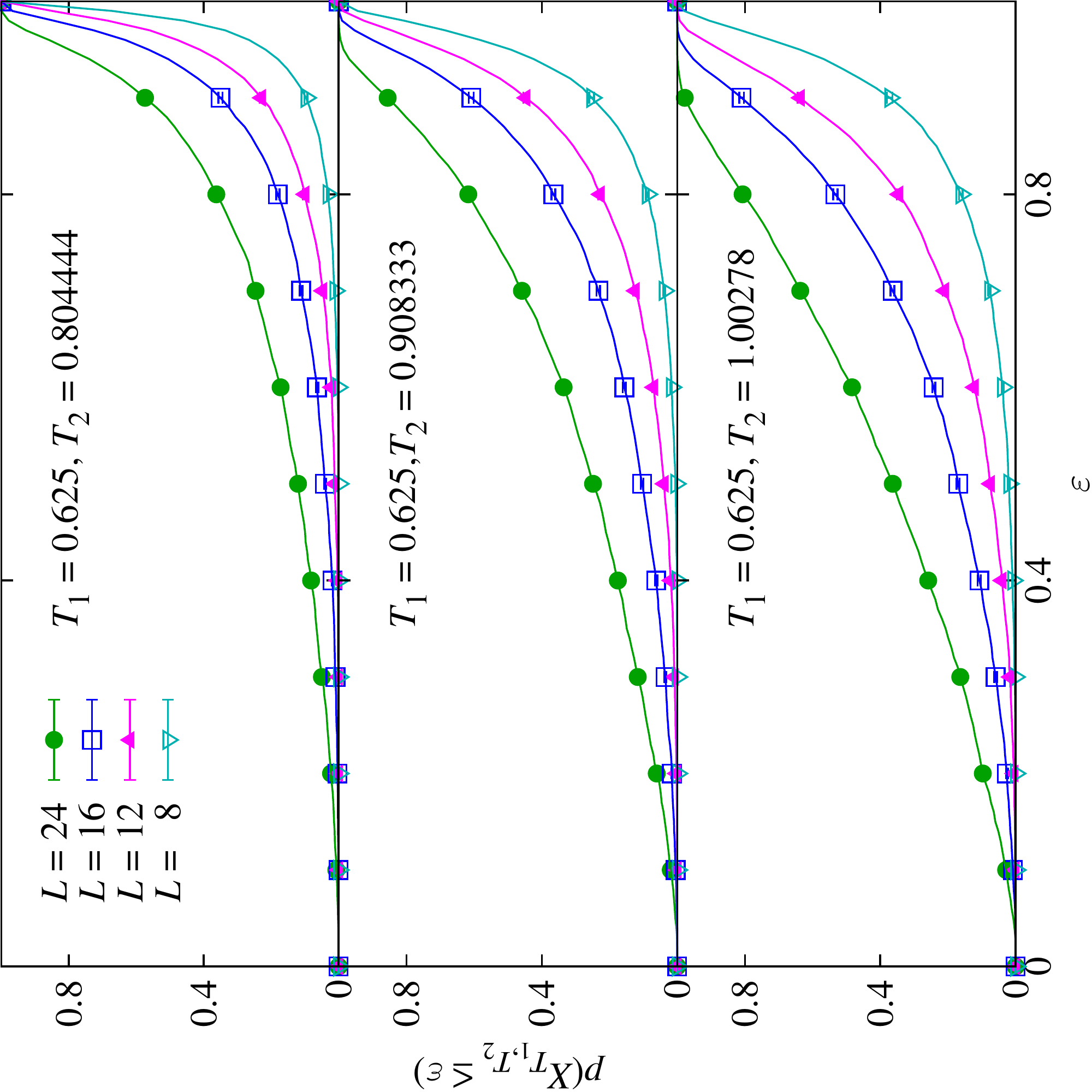}
\end{center}\caption{Probability distribution function for $X^J_{T_1,T_2}$ for $T_1=0.625$ and {\bf (top)} $T_2=0.80444$, {\bf (center)} $T_2=0.908333$ and {\bf (bottom)} $T_2=1.00278$.}
\label{fig:func_dist_x_625}
\end{figure}

In this section, we will check that the fraction of samples that suffer a
chaotic event for any pair of temperatures $T_1,T_2$ ($T_1<T_2<T_\mathrm{c}$)
indeed increases with $L$. With this aim, we compute the cumulative
distribution function for the chaotic parameter $X_{T_1,T_2}^J$ (i.e. the
probability that $X^J_{T_1,T_2}\le \varepsilon$), for several system
sizes. Results are summarized in Figure \ref{fig:func_dist_x} for $T_1=0.7026$
and Figure \ref{fig:func_dist_x_625} for $T_1=0.625$ (this last one is only
available for $L\le 24$, see Table~\ref{tab-chaos:parameters-eq}). According
to these results, the portion of samples whose chaotic parameter is smaller
than a given value seems to increase very fact with the system size no
matter the couple of temperatures considered.

This fact suggests the introduction of a large-deviation potential,
$\varOmega^L_{T_1,T_2}(\varepsilon)$, as the one introduced in \ac{MF}
computations~\cite{rizzo:03,parisi:10}:\footnote{The large-deviation potential
  is normally associated to the probability density, instead of the
  accumulative probability.  Nevertheless both statements are equivalent in
  the large-$L$ limit. From a numerical point of view, computing the
  accumulative probability is easier than the probability.}
\begin{equation}\label{eq-chaos:potential}
\text{Probability}[X^J_{T_1,T_2}> \varepsilon]=\mathrm{e}^{-L^D\varOmega^L_{T_1,T_2}(\varepsilon)}\,,
\end{equation}
note that here we are considering the complementary probability to the one
discussed before.  The notion of a large-deviation potential is useful only if
$\varOmega^L_{T_1,T_2}(\varepsilon)$ becomes $L$-independent for moderate
system sizes. In a chaotic scenario, this probability should vanish in the
thermodynamic limit for $\varepsilon>0$. In terms of the large deviation
potential, $\varOmega_{T_1,T_2}(\varepsilon)$ must remain positive for large
$L$ and all $\varepsilon>0$.

\begin{figure}\begin{center}
\includegraphics[angle=270,width=0.8\columnwidth,trim=0 0 0 0]{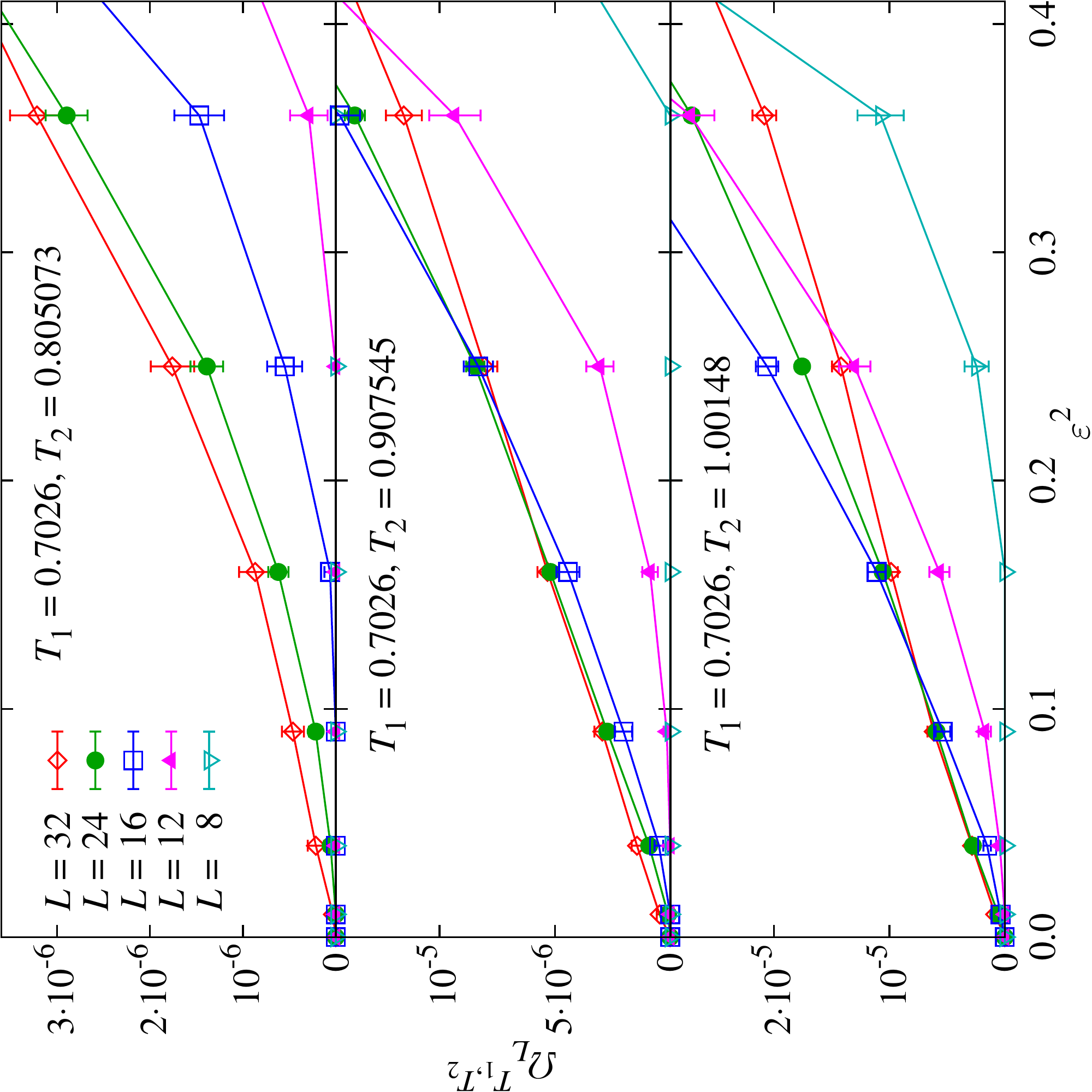}
\end{center}\caption{Large deviation potential for $T_1=0.7026$ and {\bf (top)} $T_2=0.805703$, {\bf (center)} $T_2=0.907545$ and {\bf (bottom)} $T_2=1.00148$.}
\label{fig:potencial_varias_Ts}
\end{figure}

\begin{figure}\begin{center}
\includegraphics[angle=270,width=0.8\columnwidth,trim=0 0 0 0]{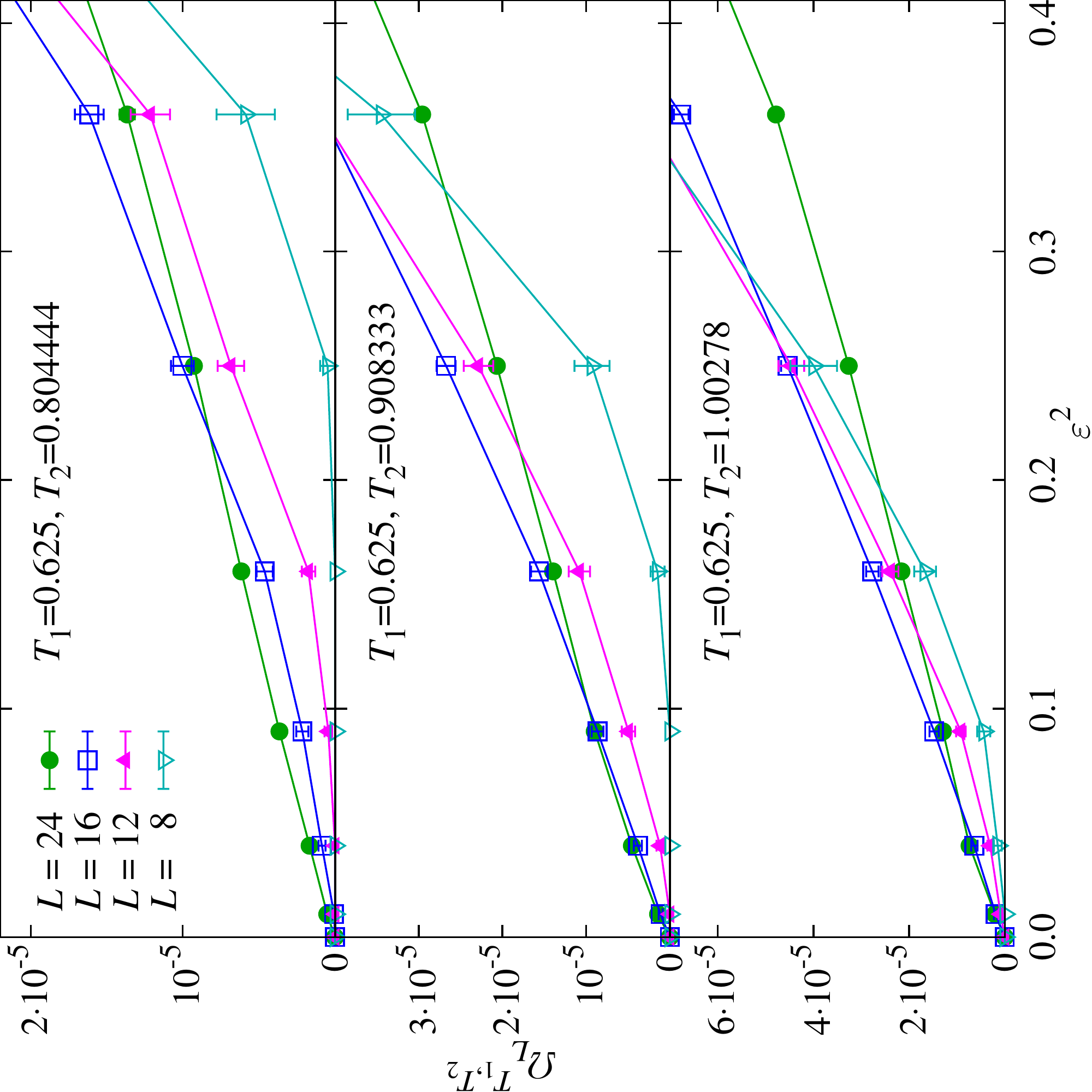}
\end{center}\caption{Large deviation potential  for $T_1=0.625$ and {\bf (top)} $T_2=0.80444$, {\bf (center)} $T_2=0.908333$ and {\bf (bottom)} $T_2=1.00278$.}
\label{fig:potencial_varias_Ts_625}
\end{figure}

We plot $\varOmega_{T_1,T_2}(\varepsilon)$ in
Figs. \ref{fig:potencial_varias_Ts} and \ref{fig:potencial_varias_Ts_625} for
$T_1=0.7026$ and $0.625$ respectively. As expected for chaos,
$\varOmega_{T_1,T_2}(\varepsilon)$ is nonnegative and increases with
$\varepsilon$, but suffers from very strong finite size effects. However,
$\varOmega_{T_1,T_2}(\varepsilon)$ does reach the large-$L$ limit for the
largest systems, at least for small $\varepsilon$. Besides, the lower $T_1$ is
(that is, the deeper in the \ac{SG} phase), the faster the convergence is
achieved. Actually, the large-$L$ limit is reached for $L=24$ for $T_1=0.7026$
and for $L=16$ for $T_1=0.625$.  It is important to point out that this study
only makes sense for low $\varepsilon$. Indeed, for a finite amount of samples
$N_S$, there is always a $X^\mathrm{max}_{T_1,T_2}\le 1$ above which the
potential diverges,
i.e. $\varOmega_{T_1,T_2}(\varepsilon>X^\mathrm{max})=\infty$, and then no fit
makes sense.

Both Figs. \ref{fig:potencial_varias_Ts} and \ref{fig:potencial_varias_Ts_625}
suggest a linear behavior of $\varOmega^L_{T_1,T_2}$ on $\varepsilon^2$, at
least for the large-$L$ limit. Besides, chaos seems to weaken the closer $T_2$
is to $T_1$. These two properties suggest a \ac{MF}-inspired scaling (see
~\cite{parisi:10} and later discussion) \be\label{eq-chaos:potential-scaling}
\varOmega^L_{T_1,T_2}(\varepsilon)\propto |T_2-T_1|^b\varepsilon^\beta.  \ee
For later discussion, we note that the \ac{MF} analytical calculations obtain
a large deviation potential which is sum of several terms like this one, 
with different exponents.

\begin{figure}\begin{center}
\includegraphics[angle=270,width=0.8\columnwidth,trim=0 0 0 0]{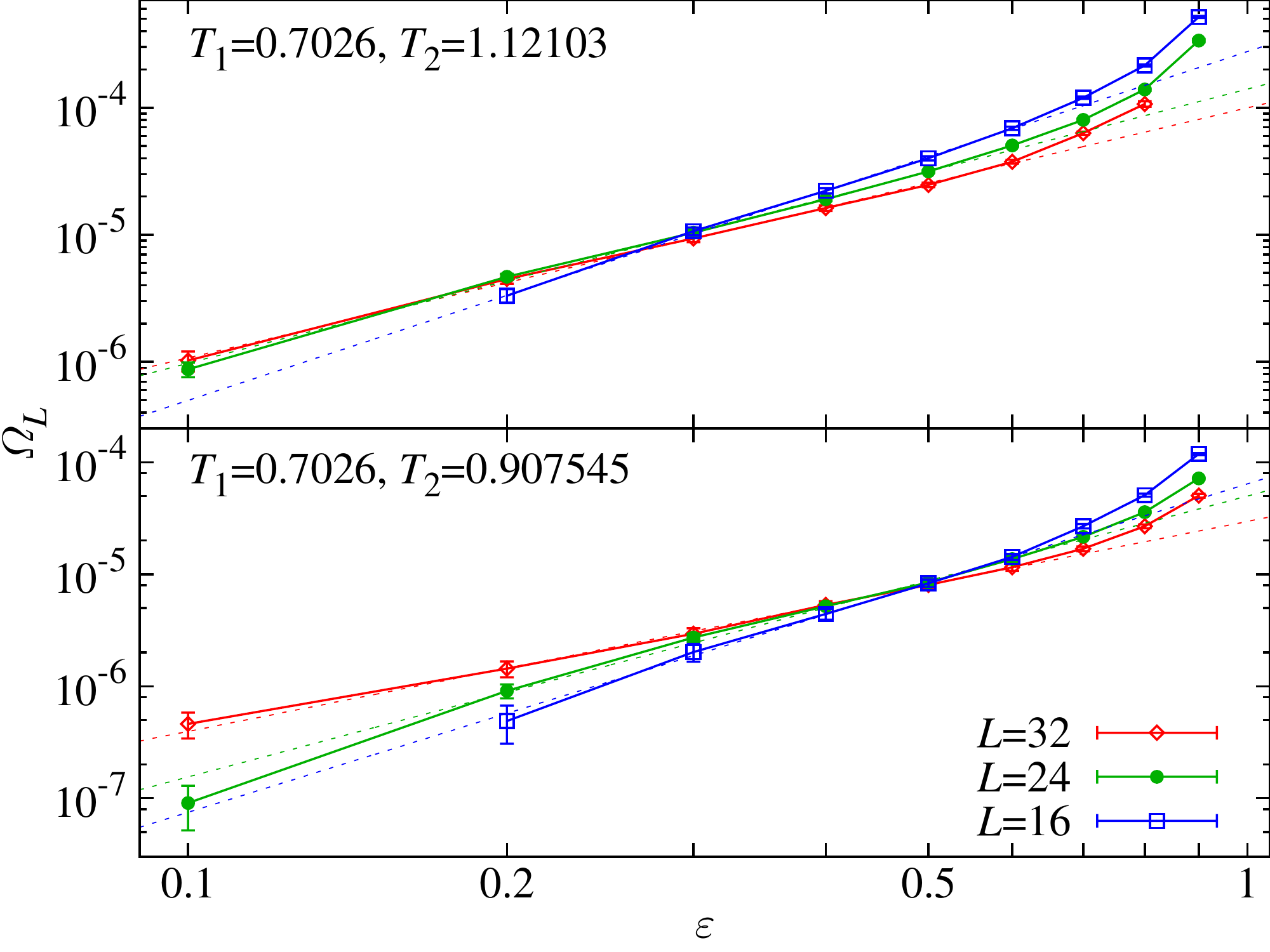}
\end{center}
\caption{Computation of exponent $\beta$, in
  ~\eqref{eq-chaos:potential-scaling}. $\varOmega_L(\varepsilon)$ is
  fitted to $a\, \varepsilon^\beta$ for $L=32,\ 24,$ and $16$ and for two
  different values of $T_2$. The fitted exponents are $\beta^{L=32}=1.94(4)$,
  $\beta^{L=24}=2.16(5)$, and $\beta^{L=16}=2.74(5)$ for $T_2=1.12103$, while
  $\beta^{L=32}=1.88(4)$, $\beta^{L=24}=2.51(15)$ and $\beta^{L=16}=2.93(7)$
  for $T_2=0.907545$. }
\label{fig:exponentes}
\end{figure}

\begin{figure}\begin{center}
\includegraphics[angle=270,width=0.8\columnwidth,trim=0 0 30 0]{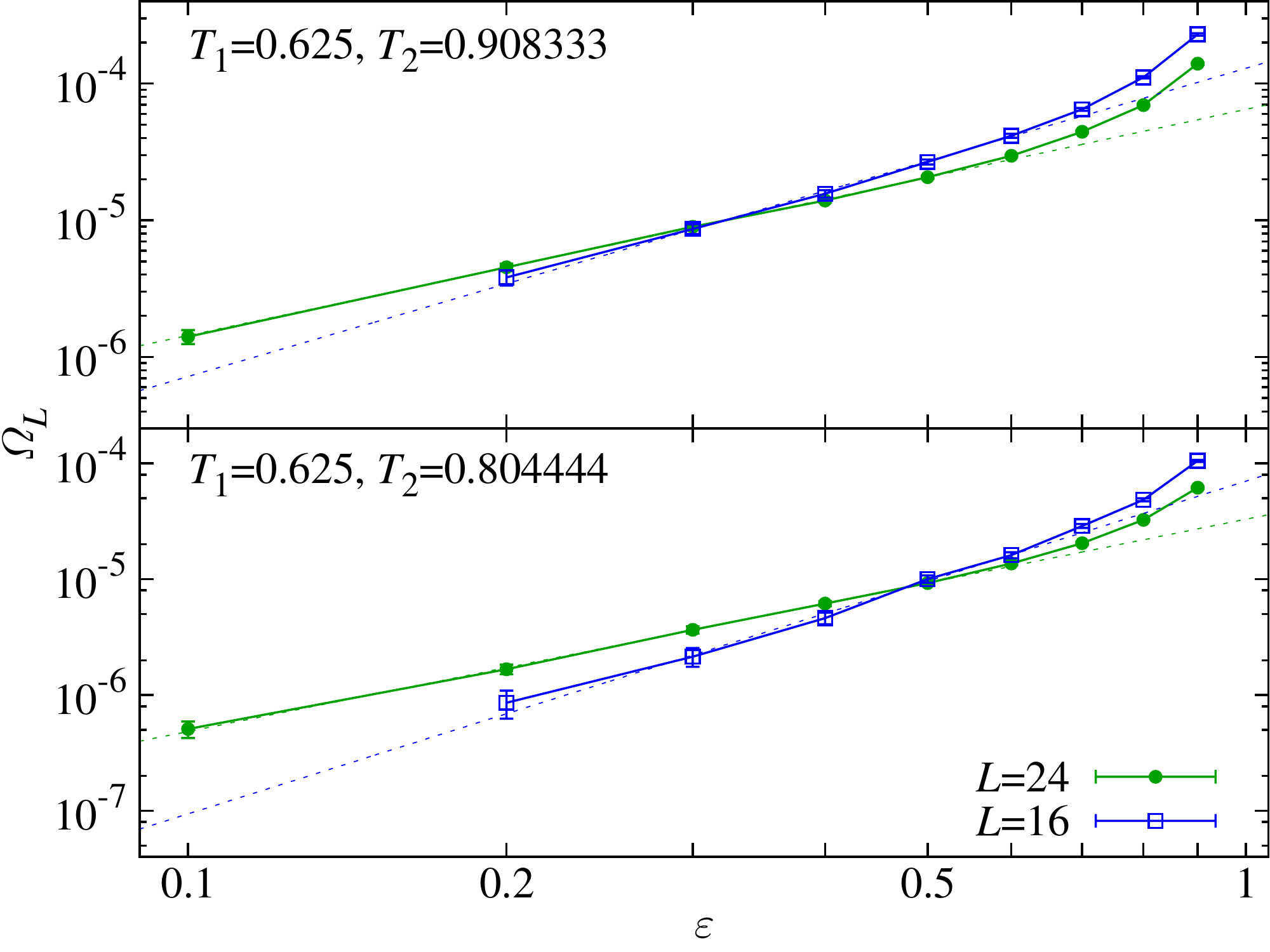}
\end{center}\caption{As in Figure~\ref{fig:exponentes}, for $T_1=0.625$. The
  fitted exponents are $\beta^{L=24}=1.65(2)$ and $\beta^{L=16}=2.27(7)$ for
  $T_2=0.908333$, while $\beta^{L=24}=1.84(2)$ and $\beta^{L=16}=2.87(13)$ for
  $T_2=0.904444$.}
\label{fig:exponentes_625}
\end{figure}

To check this scaling, we fit our data
for different system sizes to \be\label{eq-chaos:potential-scaling1}
\varOmega^L_{T_1,T_2}(\varepsilon)\propto A(|T_1-T_2|)\varepsilon^\beta.\ee
As shown in Figures~\ref{fig:exponentes} and~\ref{fig:exponentes_625} for different pairs of temperatures, data agree very well with one single $\beta$ exponent.
Again, the determination for $\beta$ suffers
from finite size effects, though seems to converge to a finite large-$L$ limit
clearly $\beta>1$ (this will have important consequences later). All the data
fit very well to scaling \eqref{eq-chaos:potential-scaling}. Indeed, the values of
$\chi^2/{\rm dof}$ of the fits are in most cases in the interval [0.25,0.5],
but the different points are very  statistically correlated, then, the errors
in the fitting parameters are probably underestimated.

\begin{figure}\centering
\includegraphics[angle=270,width=0.8\columnwidth,trim=10 0 40 0]{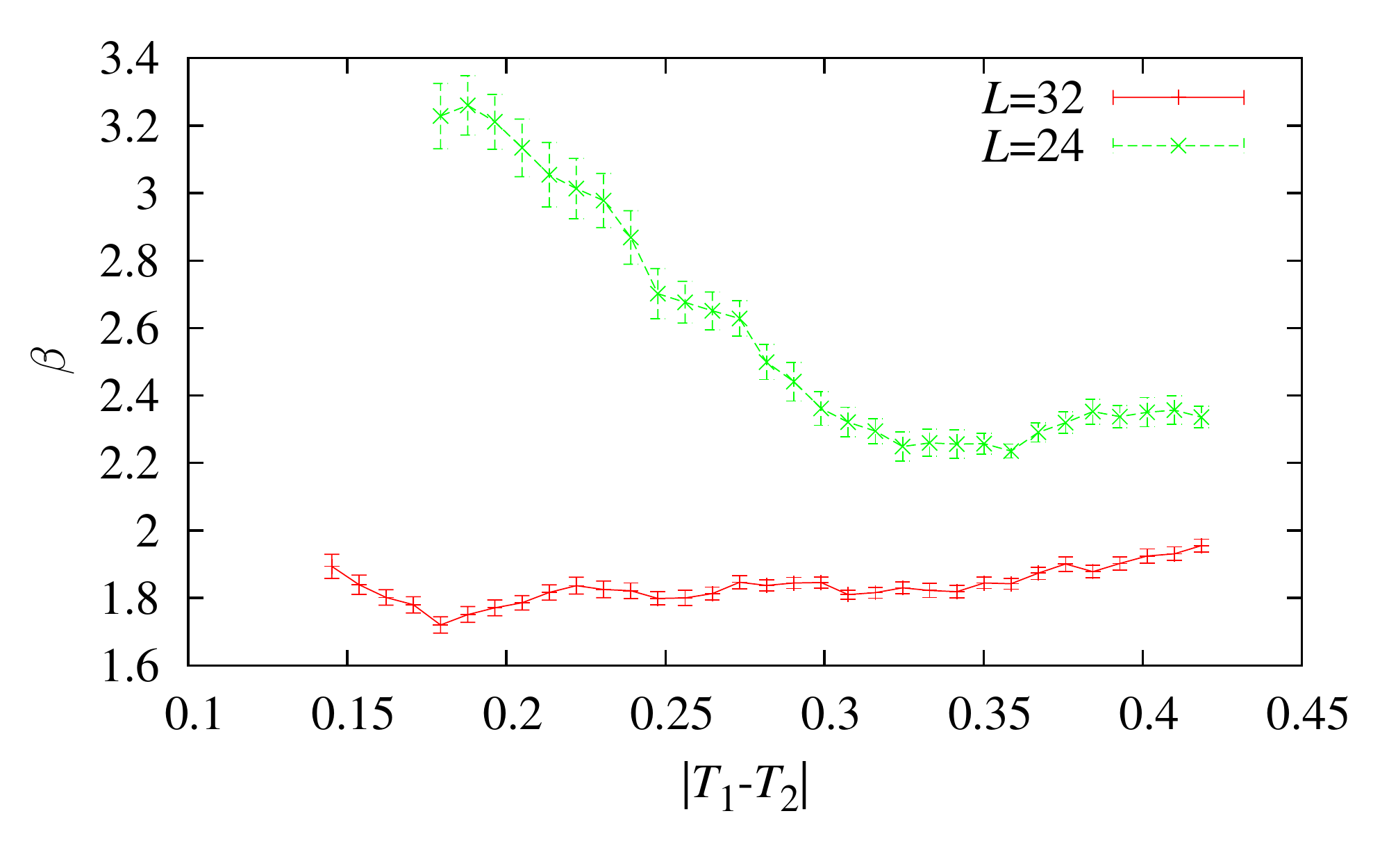}
\caption{Dependency of the exponent $\beta$ with $T_2$ obtained from the fit of
  $\varOmega^L_{T_1,T_2}(\varepsilon)$ for $T_1=0.7026$ and $L=32$ and $24$,
  to the scaling function \eqref{eq-chaos:potential-scaling}. The exponent for $L=32$
  obtained with the data the interval $\varepsilon\in[0.03,0.4]$ obtaining a
$\chi^2/dof\in[2.95/35,11.3/35]$. The data for $L=24$ was fitted in the
interval $\varepsilon\in[0.05,0.25]$. As discussed before, all the data are
very correlationed, which means that errors in this figure are probably underestimated.}
\label{fig:beta_con_T}
\end{figure}

On the other hand, when one compares the two fits for each value of $T_1$,
though similar, the fitted $\beta$ seems to depend on $T_1$ and $T_2$. We
explore this point. With
this aim, we plot the exponent $\beta$ as a function of $T_1-T_2$ (only for
$T_1=0.7026$) in Figure \ref{fig:beta_con_T}. For $L=32$ one obtains quite stable
values. The situation for $L=24$ is very different. It monotonically drops from
$\beta\approx 3$ at low temperature differences, until it reaches certain
temperature, from which, it remains stable. Low temperature differences imply
little chaos, which indicates that the scaling \eqref{eq-chaos:potential-scaling} is
only valid when chaos is present.

\begin{figure}\centering
\includegraphics[angle=270,width=0.8\columnwidth,trim=10 0 40 0]{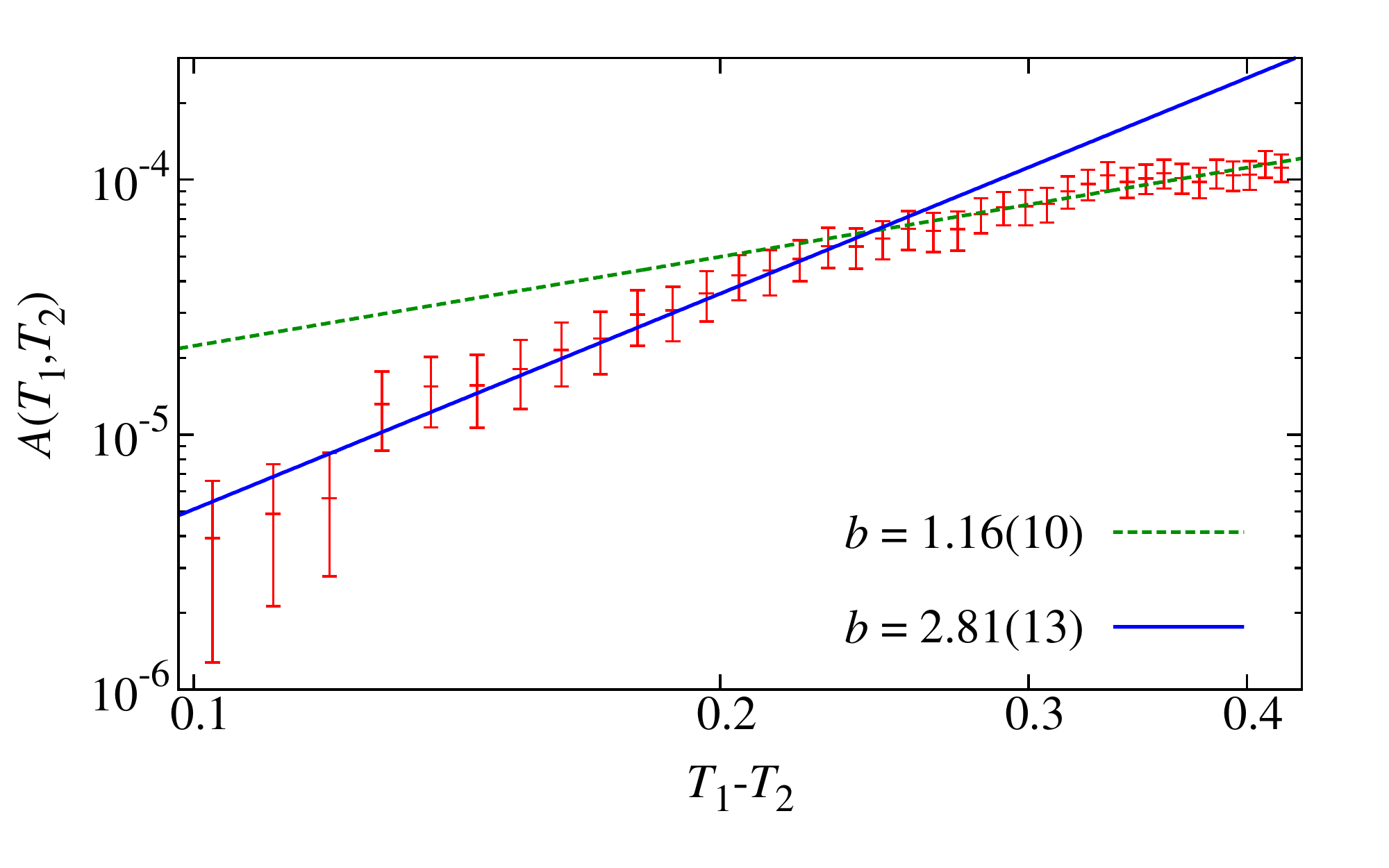}
\caption{ Computation of the $b$ exponent. Data in red represents
  $A(T_1,T_2)$, lines in blue and green corresponds to the low and high values
  of $T_2$. Fitting data is displayed in the label.}
\label{fig:termino_con_T}
\end{figure}

Once obtained $\beta$ we can compute the other exponent $b$ in
\eqref{eq-chaos:potential-scaling}. According to Figure \ref{fig:beta_con_T} only data
for $L=32$ can be considered as a representative from the large-$L$ behavior in
the whole temperature range. For this size, we fix $\beta\sim 2$  and obtain
the temperature dependent part of the potential by fitting the data to
$A(T_1,T_2)\varepsilon^2$. This temperature dependent factor $A(T_1,T_2)$ is shown in
Figure \ref{fig:termino_con_T}. Afterwards, $b$ is obtained by fitting this
factor to 
\be
A(T_1,T_2)\propto|T_2-T_1|^b.
\ee
We distinguish in Figure \ref{fig:termino_con_T} two different regimes. On one
hand, for small temperature differences $b\approx 2.8$, while for bigger ones, 
$b\approx 1.16$. We will come back to this discussion later.
We would like to point out, that the election of $\beta$ to obtain
$A(T_1,T_2)$ is not crucial, one obtains compatible results at least for $b\in[1.7,2]$.

\section{Spatial correlation functions}
As discussed in the introduction, almost all numerical work up to now was
based on the scaling picture. In it, it is possible to identify the $\zeta$
exponent, associated with the correlation length by collapsing the curves
$X_{T_1,T_2}(L)$ (averaged over all the samples in the system). This
approached predicted a correlation length as well as its critical exponent
which seemed to be recovered in simulations. Some authors
argued~\cite{aspelmeier:02}, that the chaotic correlation lengths would be
very large (in the order of the system sizes reached in simulations) to be
observed in numerical work.

According to the above discussion, all measures of the correlation length in
previous studies were obtained in an indirect way. Considering our discussion
in Section~\ref{sec:changeparadigm}, we argued that the averaged curves
studied in previous studies had lost most of the chaos signal, so we did not
believe that this phenomenological scaling study could carry information about
chaos. To prove that, we compute directly the correlation length, by means of
the spatial correlation functions. Indeed, now we defined the concept of
chaotic event, we can compute the actual correlation length in a chaotic
sample.

With this aim, we shall be considering here two types of spatial correlation
functions.  The simplest one is
\begin{equation}\label{eq-chaos:cr}
C^J_{T_1,T_2}(r)=\frac{1}{3V}\sum_{\V{r}=r\V{e}_{x,y,z}}\sum_{\V{x}}\mean{s^{T_1}_{\V{x}}s^{T_2}_{\V{x}}s^{T_1}_{\V{x}+\V{r}}s^{T_2}_{\V{x}+\V{r}}}_J\,,
\end{equation}
which is the two temperatures version of the equilibrium spatial correlation
function $c_4$ introduced in \eqref{eq:C4} but averaged over all the $\V{r}$
of the form $\V{r}=r \V{e}_i$, with $i=x,\ y,\ z$.

Alternatively, in analogy with the chaotic parameter $X^J_{T_1,T_2}$, we may
consider also a renormalized function:
\begin{equation}\label{eq-app:kr}
K^J_{T_1,T_2}(r)=C^J_{T_1,T_2}(r)/\sqrt{C^J_{T_1,T_1}(r) \,C^J_{T_2,T_2}(r)}\,.
\end{equation}
Note that computing these correlation functions is even harder (in number of operations' sense) than computing the overlaps. We discuss in Appendix
\ref{app:multispin-chaos} how to take benefit of multispin coding to obtain these magnitudes.

Considering all the discussion performed before about chaotic events, we
cannot average the two correlation functions over all samples if we want to
keep track of the chaos phenomena.  But we still have disorder and need to
average in order to infer something about the thermodynamic limit. Our
approach is the following: we compute separately the chaotic or the non
chaotic spatial correlation functions by averaging only over the most chaotic
samples or over the less chaotic samples. Of course we need a criterion to
select which samples belong to each sets. We use the chaotic parameter
$X_{T_1,T_2}^J$ for this purpose. Indeed, the lower $X_{T_1,T_2}^J$ the more
chaotic the sample is conversely the higher, the less chaotic. Then, our
choice is to consider as chaotic (non chaotic) samples, the $10\%$ of the
samples for $L=32$ with smallest (higher) $X_{T_1,T_2}^J$ and average within
each group of samples. This selection (in our systems) is equivalent to the
condition $X_{T_1,T_2}^J\le 0.33 $ or $X_{T_1,T_2}^J> 0.93$ for 
$T_1=0.7026$ and $T_2=0.90318$. We argued before
that in thermodynamic limit all samples would be chaotic, then, only the
average over the chaotic-samples is the one really representative of the
thermodynamic limit.  Actually, if one wanted to get a representative over the
convergence to thermodynamic limit, one should average over samples with \be
X_{T_1,T_2}^J \sim 1/L^{D/\beta}|T_2-T_1|^{-b},\ee with $\beta$ and $b$
defined in ~\eqref{eq-chaos:potential-scaling}.

\begin{figure}\begin{center}
\includegraphics[angle=270,width=0.8\columnwidth,trim=0 0 0 0]{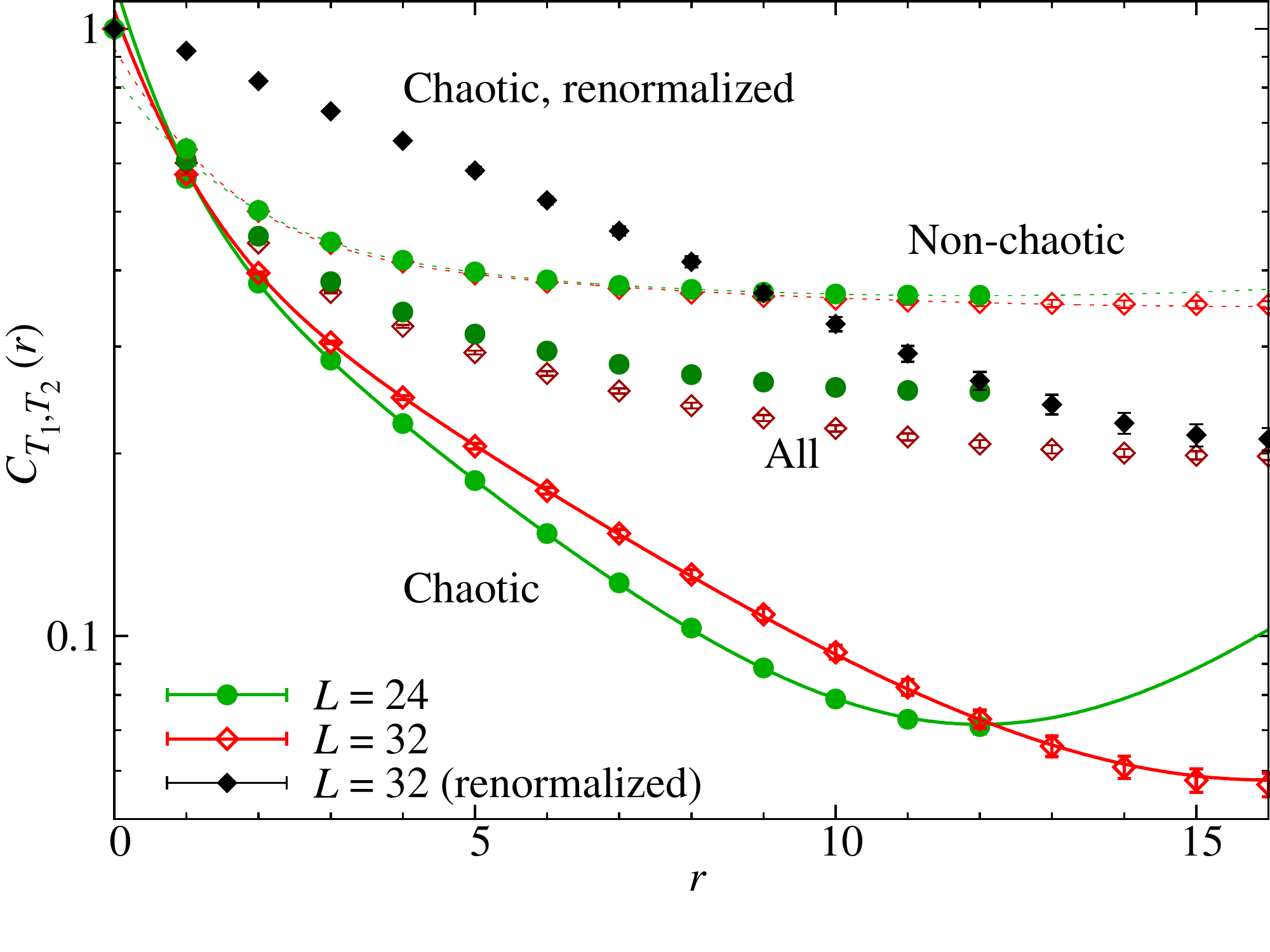}
\end{center}\caption{Spatial correlation function for $T_1=0.70260$ and
  $T_2=0.90318$, ~\eqref{eq-chaos:cr}, as averaged over different sets of
  samples: all samples, non-chaotic samples
  ($X_{T_1,T_2}> 0.93$) and chaotic samples ($X_{T_1,T_2}\le
  0.33$). The filled black diamonds correspond to the average of
  $K^J_{T_1,T_2}(r)$ ~\eqref{eq-app:kr} over the $L=32$ chaotic samples.
  Lines are fits to $\sum_{i=1,2}
  A_i\big(\mathrm{e}^{-x/\xi_i}+\mathrm{e}^{-(L-x)/\xi_i}\big)$. For each fit,
  the largest correlation lengths were $\xi_\mathrm{chaos}^{L=32}=5.69(2)$,
  $\xi_\mathrm{chaos}^{L=24}=4.447(15)$,
  $\xi_\mathrm{non-chaos}^{L=32}=23.7(7)$,
  $\xi_\mathrm{non-chaos}^{L=24}=18.9(3)$.}
\label{fig:spatialcorr}
\end{figure}

We show in Figure~\ref{fig:spatialcorr} $C_{T_1,T_2}(r)$ either averaged over
all the samples, or over the set of  chaotic or  non-chaotic samples. As expected, the behavior is
qualitatively different when chaos is present, and the global behavior is more
similar to the non-chaotic behavior than to the chaotic one.  On the contrary
as what scaling theory predicts, curves for the chaotic samples fall down
at very short lengths, noticeable shorter than when there is no
chaos. According to this, chaos can be detected even at very short
distances (which contradicts all previous knowledge about it).

One can make this discussion quantitative by fitting the curves to decaying
exponentials as expected for long distances \eqref{eq:scalinglongr}, i.e. \be\label{eq:scalingCr}
C^J_{T_1,T_2}(r)\sim \mathrm{exp}[-r/\xi_\mathrm{C}]/r^a,\ee in order to
obtain the correlation length.

We found that all the $C_{T_1,T_2}(r)$ are extremely
well fitted by a sum of two exponentials,
\be \label{eq:scalingCr2} C_{T_1,T_2}(r)\sim\sum_{i=1,2}
A_i\big(\mathrm{e}^{-r/\xi_i}+\mathrm{e}^{-(L-r)/\xi_i}\big),\ee in the range
$2\leq r\leq L/2$, see Figure~\ref{fig:spatialcorr} and Table~\ref{tab-chaos:fitcr}
 for the fitting details. According to our data, there
is no need for a pre-algebraic factor, and thus $c\approx0$. According to the
the $\chi^2$ test, see Table~\ref{tab-chaos:fitcr}, the fits are
extremely good. However, the resulting values of $\chi^2/\mathrm{dof}$ are too
small (smaller than $1/10$ for all four fits). The reason for that, is that we
only use, as usual, the diagonal elements in the covariance matrix to compute
$\chi^2$ and the non-diagonal elements are very important here due to the
large correlation of our data. Hence, the error estimates in the fitted
parameters must be regarded as merely indicative. 

As a summary of the the fits, on non-chaotic samples, the correlation length
is $\xi_{\mathrm{NC}}\approx L/2$ while, for the chaotic ones,
$\xi_{\mathrm{C}}\approx 6$ for $L=32$ (or $\xi_{\mathrm{C}}\approx 4$ for
$L=24$).  Given the disparity of scales, it is not obvious how to estimate the
\emph{single} chaotic length of Refs.~\cite{fisher:86,bray:87}.
\begin{table}
\begin{center}
\begin{tabular}{|c|cc|cc|}\hline
&\multicolumn{2}{c|}{$L=32$}&\multicolumn{2}{c|}{$L=24$}\\
&Chaotic&Non-chaotic&Chaotic&Non-chaotic\\\hline
$A_1$&0.483(3)&0.343(6)&0.531(3)&0.343(4)\\
$A_2$&0.59(6)&0.498(6)&0.67(10)&0.401(7)\\
$\xi_1$&5.687(22)&23.7(7)&4.447(15)&18.9(3)\\
$\xi_2$&0.84(4)&0.343(6)&0.70(5)&1.31(3)\\
$\chi^2/\text{dof}$&1.03/11&3.3/12&0.5/7&0.3/7\\\hline
\end{tabular}
\end{center}\caption{Fitting variables obtained by fitting  $C_{T_1,T_2}(r)$, \eqref{eq-chaos:cr},
  averaged over the chaotic and not chaotic samples as defined in the text, to
  the curve $\sum_{i=1,2}
A_i\big(\mathrm{e}^{-r/\xi_i}+\mathrm{e}^{-(L-r)/\xi_i}\big)$. The longest
length corresponds to the correlation length.}\label{tab-chaos:fitcr}
\end{table}

We include also show in Figure~\ref{fig:spatialcorr} the average over the $L=32$
chaotic samples for $K^J_{T_1,T_2}(r)$. The renormalization with respect to
$C_{T_1,T_2}(r)$ allows to fit to a single-exponential for $6\leq r \leq
L/2$. However, the renormalization causes a change of curvature in the small
$r$ region, that can be fitted for $r\geq 2$ as $(1+r^2)^{0.18/2}$ times a
decaying exponential. Anyhow, we also obtain $\xi_{\mathrm{C}}\approx 6$, in
qualitative agreement with the estimate from $C_{T_1,T_2}(r)$.  According to
that, in the case of the spatial correlation function, there is no particular
improvement in detecting chaos by using $K^J_{T_1,T_2}(r)$ instead of
$C_{T_1,T_2}(r)$. In fact, $K^J_{T_1,T_2}(r)$ makes the fitting more difficult.

It might be quite shocking that the chaotic correlation length, though small
in comparison with $L$, increases with the system size
($\xi_{\mathrm{C}}\approx 6$ for $L=32$ and $\xi_{\mathrm{C}}\approx 4$ for
$L=24$). Our data seem to suggest a non finite correlation length. This
somehow strange result is however what one should expect, as we discuss below.

Indeed, barring normalizations, $\langle q_{T_1,T_2}^2\rangle_J$ is the space
integral of the correlation function~\eqref{eq-chaos:cr}, recall
\eqref{eq:rel_sus_c4}.  Then, if one considers an exponential decay with
$r/\xi_\mathrm{C}$ in \eqref{eq:scalingCr} or \eqref{eq:scalingCr2}, it follows that \be\langle q_{T_1,T_2}^2\rangle_J\propto
(\xi_\mathrm{C}/L)^D.\ee Let us assume that, below $T_\mathrm{c}$, both
$\langle q_{T_1,T_1}^2\rangle_J$ and $\langle q_{T_2,T_2}^2\rangle_J$ are of
order one (the \ac{EA} parameter depends on $L$ but has a finite large-$L$
limit). Then, also \be \label{eq:scalingX1}X_{T_1,T_2}^J\sim
(\xi_\mathrm{C}/L)^D.\ee Using this scaling, we are ready to discuss the size
dependency of the chaotic length $\xi_\mathrm{C}$. Indeed, let us plug
~\eqref{eq-chaos:potential-scaling} with $\beta\approx 1.7$ in
~\eqref{eq-chaos:potential}. If the probability in ~\eqref{eq-chaos:potential}
is to remain of order one for large $L$, then \be\label{eq:scalingX2}
X_{T_1,T_2}^J\sim 1/L^{D/\beta}.\ee Then, combining \eqref{eq:scalingX1} with
\eqref{eq:scalingX2}, one obtains \be X_{T_1,T_2}^J\sim
(\xi_\mathrm{C}/L)^D,\ee which leads to \be\xi_\mathrm{C}\sim L^{d},\ee with
$d=(\beta-1)/\beta\approx 0.4$. In other words, the chaotic correlation length
increases with $L$, but still is very small as compared with the system
size. In other words, $\xi_\mathrm{C}/L\to 0$ for long $L$.

Let us come back to the spatial correlation scaling \eqref{eq:scalingCr}. Our
data suggest that, at least for our system sizes, there is no need of
algebraic pre-factor, i.e. $a\approx 0$. However, in numerical work it is not
possible to distinguish a very small  $a$ from the clean $0$
value. Besides, let us discuss briefly the effect in the previous scaling of
an hypothetical pre-factor.
Then, we consider the case $a\neq 0$. It follows that \be\langle
q_{T_1,T_2}^2\rangle_J\propto \xi_\mathrm{C}^{D-a}/L^D,\ee which would lead to
\be\xi_\mathrm{C}\sim L^{D(\beta-1)/(\beta(D-a))}.\ee

The above conclusion is shocking: the chaotic-length~\cite{fisher:86,bray:87}
is expected to be infinite for large $L$. Actually, only $\beta=1$ in
~\eqref{eq-chaos:potential-scaling} would be compatible with a finite
$\xi_\mathrm{C}$ (recall we obtained $\xi_\mathrm{C}\sim
L^{(\beta-1)/\beta}$), and our data suggests a $\beta$ clearly above this value.  However, mean-field results~\cite{parisi:10} warns
about transient effects.  In fact, in mean-field, for small overlap and
$|T_1-T_2|$, the large-deviations potential scales as
\begin{equation}\label{eq-chaos:mean-field}
\tilde\Omega^{\mathrm{mean-field}}\propto
 A q_{T_1,T_2}^2 |T_1-T_2|^3+B |q_{T_1,T_2}|^3 |T_1-T_2|^2\,,
\end{equation}
 ($A$ and $B$ are constants). Either of the two terms can be dominant for some
region of $q$, $N$ and $T_1-T_2$ ($N$ is the number of spins).  We now let $N$
grow at fixed $T_1-T_2$, and seek $q$ such that $\tilde \Omega\sim 1/N$ [i.e.
  probability of order one, see ~\eqref{eq-chaos:potential}]. We realize
that there is a crossover size $N^*\sim |T_1 -T_2|^{-5}$ such that $q^2\sim
N^{-2/3}$ if $N\ll N^*$. On the other hand, if $N\gg N^*$, $q^2\sim N^{-1}$:
the mean-field prediction for ~\eqref{eq-chaos:potential-scaling} is
$\beta=1$. That means that in \ac{MF} it is possible to define a finite
correlation length. This might be quite strange if one thinks in \ac{SK}
model~\ref{sec:SK}, where no notion of distance or at least neighborhood
exists. However, in this precise model, as discussed in the introduction, all
the coefficients in the perturbation in $q$ below $|q|^9$ ($\beta=4.5$)
vanish~\cite{rizzo:03,parisi:10}. On the other hand, the Eq.
 \eqref{eq-chaos:mean-field} was obtained for Bethe lattices discussed in 
Sect.~\ref{sec:hypercube}, where although distance is not yet well defined,
there exists a notion of neighborhood.

\section{Phenomenological scaling}
 A question arises at this point: if the chaotic-length is not finite (at
 least for our system sizes that yield $\beta\approx 1.7$), what is the chaos
 exponent $\zeta$ computed in previous works~\cite{sasaki:05,katzgraber:07}?
 We argue that this exponent is actually $\zeta=D/b$ [$b$ is the
   temperature-difference exponent in
   ~\eqref{eq-chaos:potential-scaling}]. According to that, the exponent
 $\zeta$ would be unrelated to the chaotic length.

\begin{figure}\begin{center}
\includegraphics[angle=270,width=0.9\columnwidth,trim=30 0 40 0]{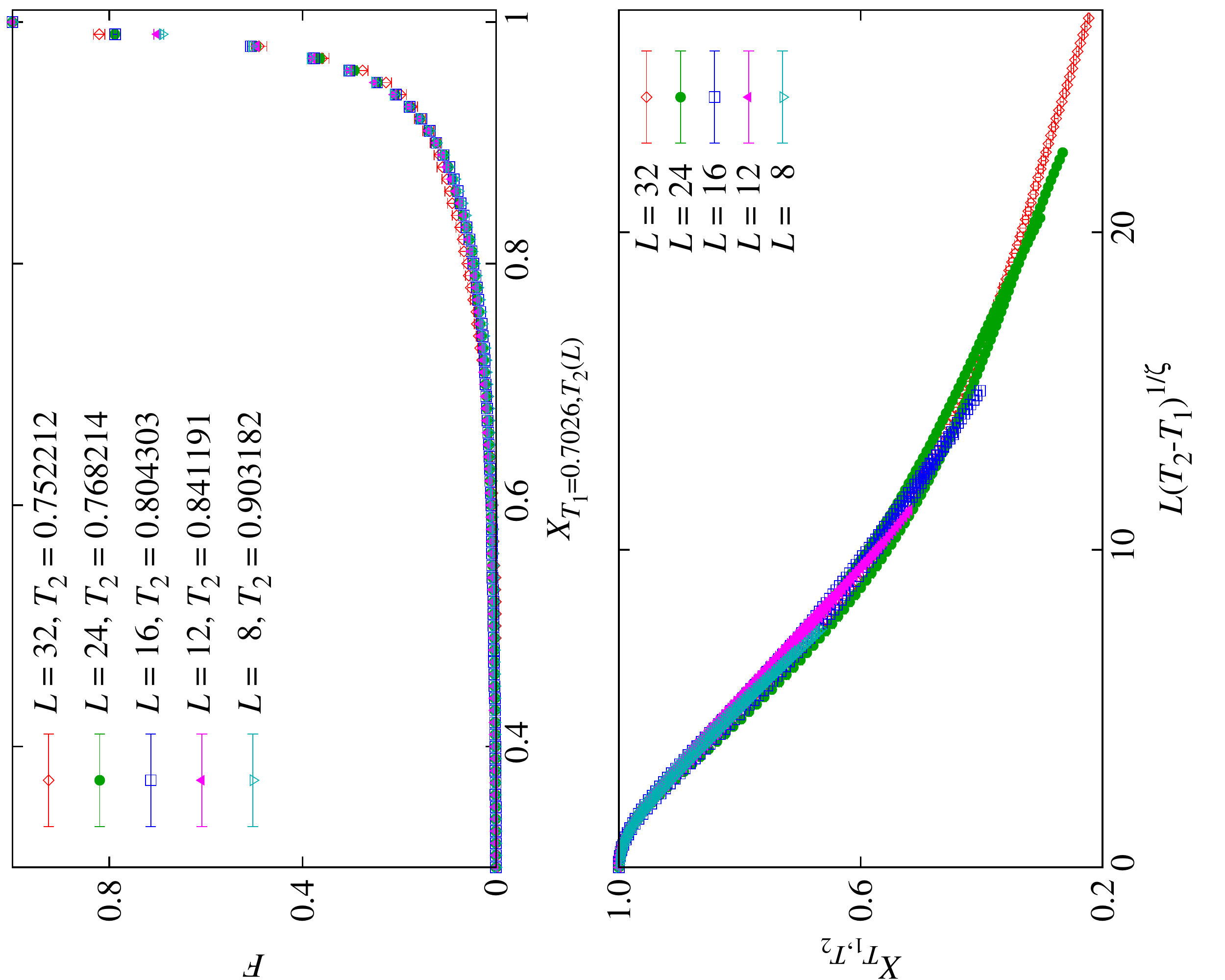}
\end{center}\caption{ 
 {\bf (Top)}
  Phenomenological renormalization: for each $L$, we seek $T_2(L)$ such that
  the distribution function $p\big(X^J_{T_1=0.7026,T_2(L)}\le
  \varepsilon\big)$ best resembles the $L=8$ distribution for $T_1=0.7026$ and
  $T_2=0.90318$.  
{\bf (Bottom)} Sample-averaged $X^J_{T_1,T_2}$
  vs. $L(T_1-T_2)^{1/\zeta}$ using $\zeta=1.06$ (data for $T_1=0.7026$ and,
  when $L\leq 24$, also for $T_2=0.625$). The data collapse is found both for
  $T_2<T_\mathrm{c}$ and for $T_2$ in the paramagnetic phase. }
\label{fig:ajuste}
\end{figure}

Indeed, some reflection reveals that phenomenological
renormalization~\cite{katzgraber:07} can be cast as follows. For the purpose
of discussion, we fix the lowest temperature $T_1$. Then, for each $L$, we
find a $T_2(L)$ such that the probability distribution function for
$X_{T_1,T_2(L)}^J$, becomes $L$-independent, see
Figure~\ref{fig:ajuste}--top. The scaling picture is based on the statement
that $X_{T_1,T_2}=F\paren{\xi_\mathrm{C}(T_1,T_2)/L}$, with $F$ a universal
function and $\xi_\mathrm{C}(T_1,T_2)$ following the scaling 
\be\label{eq:xichaoslast}
\xi_\mathrm{C}(T_1,T_2)\propto\caja{\frac{\gamma(T_1)}{\sigma(T_1)|T_2-T_1|}}^{1/\zeta}.\ee
Then, once collapsed all the curves, the exponent $\zeta$ of
Ref.~\cite{katzgraber:07} follows from $ L\propto |T_1 - T_2(L)|^{1/\zeta}.$
In fact, we can fit our data to \be T_1-T_2(L)\propto 1/L^{\zeta},\ee
obtaining $\zeta=1.02(3)$ (for $L\leq 32$, $T_1=0.7026$ and
$T_2(L=8)=0.90318$, with $\chi^2/\mathrm{dof}=3.57/3$) or $\zeta=1.07(2)$ (for
$L\leq 24$, $T_1=0.625$ and $T_2(L=8)=0.815$, with
$\chi^2/\mathrm{dof}=1.77/2$). These results are compatible to the exponent
$\zeta\approx 1.07$ obtained in~\cite{katzgraber:07}.

On the other hand, if we combine ~\eqref{eq-chaos:potential}
and~\eqref{eq-chaos:potential-scaling}, we obtain that the phenomenological
renormalization amounts to \be L^{D} |T_1 - T_2(L)|^b=\text{constant},\ee
which is equivalent to $\zeta=D/b$.

We have already computed $b$, see the fits in Figure \ref{fig:termino_con_T}. We
discussed then that there seem to be two different regimes. For $|T_1 -
T_2(L)|<0.25$ we got $b=2.81(13)$, which implies $\zeta=1.07(5)$, which in
excellent agreement with~\cite{katzgraber:07} and with the data collapse just
discussed shown in Figure~\ref{fig:ajuste}---bottom.  We note, however, that
this $b$ value only applies for small $|T_1 - T_2(L)|<0.25$, when the chaotic
events are extremely rare for our system sizes. In a nutshell, the
phenomenological renormalization applies where chaos is not present. It is
thus not surprising that their results hold also in the paramagnetic phase.

\section{Overlap equivalence}
\begin{figure}\begin{center}
\includegraphics[angle=270,width=0.8\columnwidth,trim=0 0 0 0]{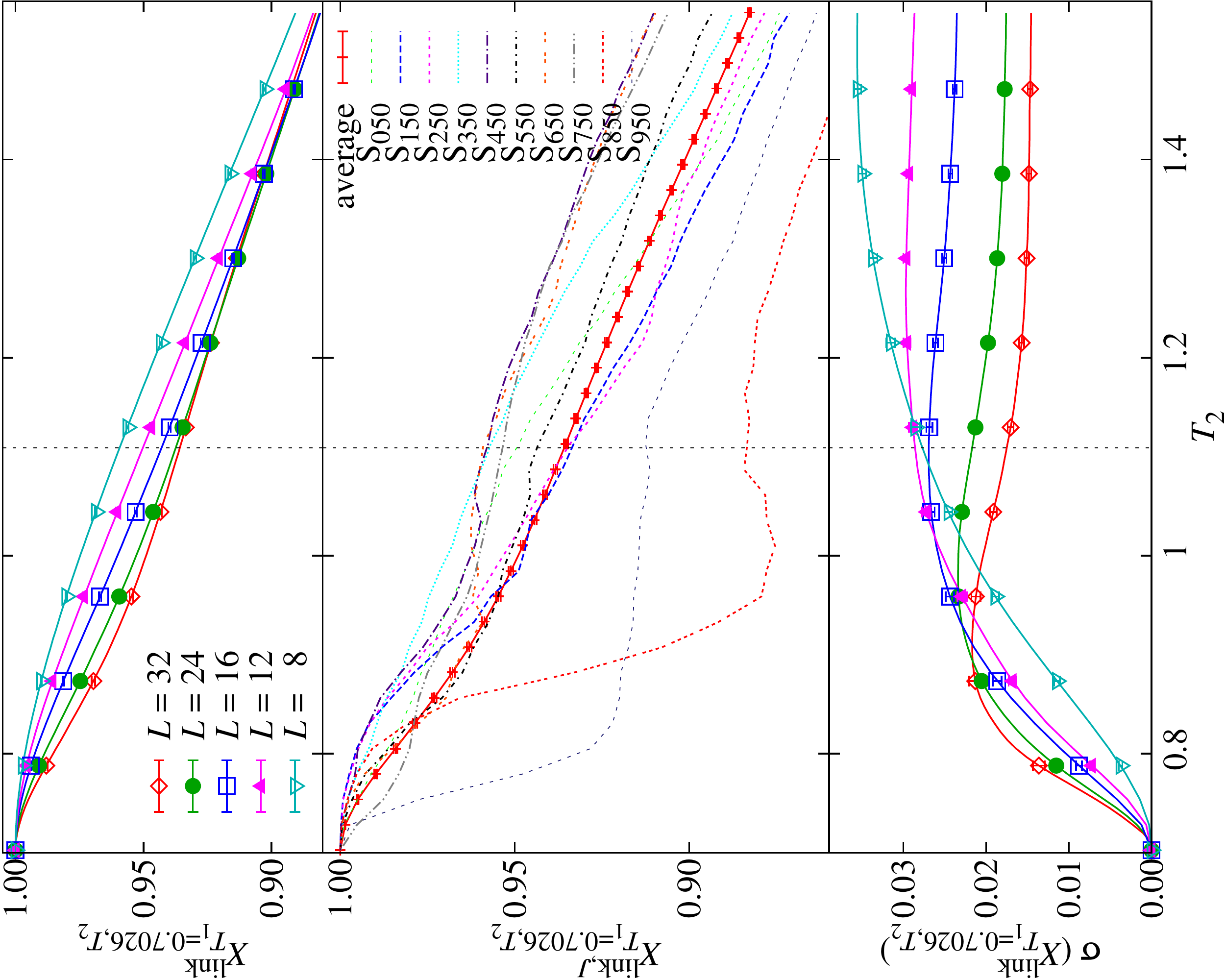}
\end{center}\caption{Analogous of Figure 2 of the main text, but for $X^\mathrm{link}_{T_1,T_2}$.}
\label{fig:xlink_y_var}
\end{figure}

The fact that the chaotic samples have a fairly small correlation length
suggests to ask whether chaos might be detected even at distance $r=1$. This
quantity is nothing but the link-overlap \eqref{eq:qlink} but defined for two
temperatures\be Q_{T_1,T_2}^{\mathrm{link},J}=\frac{1}{zN}\sum_{\mean{ik}}
q_{i}^{T_1 T_2,J}q_{k}^{T_1 T_2,J}\,,\ee
where the sum runs on the nearest neighbors $\mean{ij}$.
With this aim, we perform the same
study we did for $X_{T_1 T_2}$ in Figure \ref{fig:x12_y_var} but this time
for an analogous  $X^\mathrm{link}_{T_1 T_2}$  defined as
\begin{equation} \label{eq-chaos:X12link}
X^{\mathrm{link},J}_{T_1,T_2}=\frac{ \mean{Q_{T_1,T_2}^{\mathrm{link},J}}}{\sqrt{\mean{Q_{T_1,T_1}^{\mathrm{link},J}}\mean{Q_{T_2,T_2}^{\mathrm{link},J}}}},
\end{equation}
as shown in Figure~\ref{fig:xlink_y_var}. The results are very much the same
that the ones discussed in Section \ref{sec:changeparadigm}. In fact, the
notion of \emph{chaotic event} makes just as much sense if we study the link
overlap. This fact suggests the overlap-equivalence property
discussed in Section \ref{sec:overlapequiv}. As discussed then, the overlap
equivalence leads to the notion of ultrametricity, and this time, it is
between overlaps at different temperatures, which somehow contradicts
oversimplified pictures of temperature chaos as a scale-separation
phenomenon~\cite{berthier:02} discussed in Section
\ref{sec:memory-rejuvenation} and suggests an explanation based on a
hierarchical structure of states as the one discussed in Section \ref{sec:SG-scenarios}.

\part{Colloids\label{part:colloids}}
\chapter{Introduction to colloidal systems \label{chap:intro-colloids}}

\section{The problem}\label{sec:problem}

We devoted all the Part \ref{part:sg} of this thesis to the study of the
effect of inherently disordered interactions on magnetic systems.  As
discussed then, the low temperature phase is characterized by a frozen
disordered orientation of the spins. At variance with other problems in
condense matter, the spatial ordering of the impurities is not an important
feature of the spin glass phase (although there is other kind of order, as
already discussed).  This fact let us to model them as lying on a regular
lattice and simplify the theoretical calculations, as well as to speed up the
computer simulations.  For this last reason, the effect of disorder has been
extensively addressed in lattice systems (spin glasses, magnetic materials in
random field, etc.) while the situation in off-lattice systems is much less
understood.  In this Part \ref{part:colloids}, we tackle the same problem but
this time in colloidal or liquid systems. In these systems, disordered
interactions can be induced by for instance, a random distribution of charges,
polymer chain lengths, or particle sizes as we shall consider here.

Let us make some general considerations about disorder in statistical
mechanics. There are two well defined limits~\cite{parisi:92}. The first one,
the {\em quenched} disorder, was introduced in the Part \ref{part:sg} when
studying spin glasses.  As discussed in Section \ref{sec:realsg}, the spin
glasses are alloys, normally a core of metal with a few magnetic impurities that
carry the spins.  These impurities are in random positions after the
synthesize process (inducing the random interactions), but interactions
make them diffuse although at geological times as compared with the spin
evolution characteristic times. For this reason, the {\em quenched}
approximation assumes that the spin configuration has no influence in the
distribution of disorder. Then, we always seek the equilibrium configuration
of the spins within a given fixed realization of disorder (sample). With this
approach we are not considering the much stabler minimum achieved after the
equilibration of the impurities (a ferromagnet dot in a non-magnetic matrix,
for instance) since the time at which it would be relevant is far beyond our
experimental window.  The opposite limit is observed in the fluid phase of the
systems considered here. Now the particles (carrying the disorder on their
random size) can easily diffuse through the total volume but this diffusion is
very influenced by the instantaneous distribution of disorder. Indeed, the
particles tend to crowd with those of similar size since the small ones
diffuse faster than the big ones. This kind of disorder where there is not
clear scale separation between distribution of disorder and the particle
motion, is known as {\em annealed} disorder.

However in colloidal systems, when considering the solid phase, the diffusion
is almost suppressed, and the equilibration of disorder occurs at much longer
times than local formation of crystal clusters. For instance, a big particle
could minimize the free-energy by locating in a conglomerate far away in the
system, but if gets trapped in a crystal structure will hardy ever
move. This situation is better described by the quenched than the annealed
approximation.  This problem is not clearly recognized in literature. In fact,
among the chemical physics' community, it has become fairly common the use of
semi-grand canonical ensembles to approach these kind of systems
~\cite{sollich:10}. These ensembles accelerate the annealing dynamics by
changing the disorder. These algorithms are indeed very powerful to study the
fluid phase, but when applied to the solid phase, the equilibrium obtained
corresponds to the relevant state at much longer time-scales that what can be
found in experimental times (when the disorder equilibrates).  In fact, these
simulations lead to a fractionation scenario (see the discussion below) and
phase diagrams that are not observed in experiments~\cite{liddle:11}.

In spite of this conceptual problems, the solid phases are also
important. Indeed, most fluids become crystalline solids upon cooling or
compression. Then, crystallization is a vast field of research, where a
fruitful exchange is achieved between experiments and theory.  Consider, for
instance, the simplest model of fluid, the ~\acf{HS}. The numerical finding of
a fluid-solid phase transition~\cite{alder:57,wood:57} motivated experiments
on colloidal suspensions~\cite{pusey:86,pusey:89}. Nowadays, an accurate
assessment of the equilibrium phases (and phase-boundaries) for colloids is
crucial to address novel challenges for statistical mechanics, such as
super-solidity (as modeled by quantum \ac{HS}) or self-assembly (the
spontaneous organization of particles into desired arrangements). Moreover,
the custom design of particles with non-spherical interaction potentials
(Janus particles) opens exciting opportunities~\cite{manoharan:03,glotzer:07},
but puts further demands on numerical work~\cite{sciortino:09,romano:11}.

It seems reasonable that if the size dispersion $\delta$ (i.e. the ratio of
the particles' size dispersion with the average, see definition
in ~(\ref{DEF:DELTA})) is very high, it would be difficult to
accommodate the particles in a lattice structure and thus the crystal phase
should somehow destabilize. Experiments confirm this hypothesis. In fact,
crystallization of very viscous colloidal samples with $\delta$ larger than
$12\%$ does not occur, even after several months spent from the sample
preparation~\cite{poly:Pusey86}. This leads to several basic questions about
the equilibrium phase diagram of polydisperse
systems~\cite{poly:Bartlett98,poly:Kofke99,poly:Auer01,poly:Fasolo04,poly:Dullens04,poly:Chaudhuri05,poly:Fernandez07,colloids:Brambilla09,colloids:Zaccarelli09,wilding:10,sollich:10}. Does
enough large polydispersity hinder crystallization?  Is the suppression of
crystallization a dynamical effect arisen from the low diffusivity of large
particles~\cite{evans:01}, the interplay with the glass
transition~\cite{colloids:Brambilla09,colloids:Zaccarelli09,poly:Fernandez07}
or anomalously large barriers~\cite{poly:Auer01}? Is the glass phase stable
rather than only metastable?  And, probably at a more fundamental level, is
thermodynamic equilibrium relevant at all to describe real polydisperse
materials or these are instead inherently off-equilibrium over the
experimental time scales?  Answering such questions is crucial for condensed
matter physics, since polydispersity is found both in artificial (synthetic
colloids, polymers) and natural systems, from supercooled liquids on the
atomic scale up to biological fluids such as blood.
\begin{figure}
\centering
\includegraphics[width=0.7\columnwidth,trim=0 0 0 0]{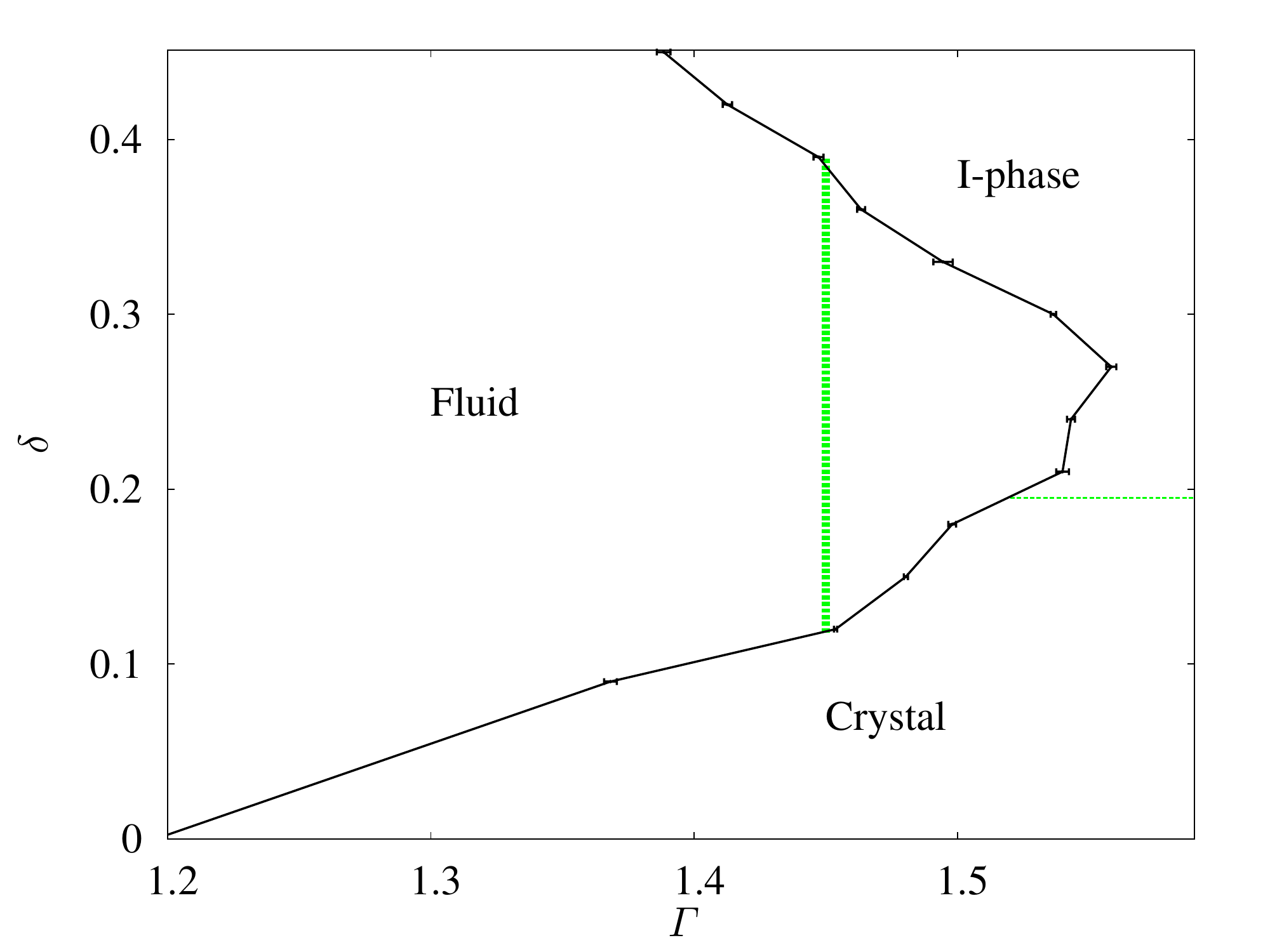} 
\caption{Phase diagram $\delta-\beta$. The green vertical line corresponds to
  the dynamical glass transition, while the black one refers to the freezing
  transition. 
Figure taken from \cite{poly:Fernandez07}.} 
\label{PHASE-DIAG0}
\end{figure}

An attempt to rationalize the experimental findings is the so-called
terminal polydispersity scenario where a characteristic value $\dt
\sim 0.12$ exists above which the homogeneous crystal becomes
thermodynamically unstable (see the phase diagram in Figure \ref{PHASE-DIAG0}). There is not consensus however about what
kind of structure should replace such single phase crystal. Density
functional analysis~\cite{poly:Chaudhuri05} predicts the instability
of any crystal structure (even partial) above $\dt$, thus leaving the
amorphous ones (either liquid or solid) as the only possibility. Yet,
the moment free-energy approach~\cite{poly:Fasolo04} predicts {\em
  fractionation}: phase separation between many crystal phases [though
  of the same ordering, \ac{FCC} for instance], each one with a much
narrower size dispersion than $\delta$. Fractionation is supported by
a recent numerical simulation that found that a first-order
fluid-solid transition actually occurs at any polydispersity
\cite{poly:Fernandez07}. However, the found solid phase is quite complex, at
least in the high polydispersity region.  In fact, for $\delta > 0.19$
the transition regards only a fraction of the particles and the
ordered state is inhomogeneous. Such state has been previously
referred to as I-phase~\cite{poly:Fernandez07} (I stands for inhomogeneous).

In this part of the thesis, the final goal is to study the
phase diagram for this high polydispersity region and the properties of this
new solid phase using computer simulations. Approaching equilibrium in this
kind of systems is very discouraging, the equilibration times become
excessively large for numerical purposes even for rather small systems. In
fact, the typical numerical equilibrium studies cover only $N\lesssim 500$
particles which must be compared with the $N=2^{22}$ (Chapter
\ref{chap:hypercube}) or $32^3$ (Chapter \ref{chap:chaos}) spins studied for
lattice systems. The difficulties we must face can be summarized in three
points:
\begin{itemize}
\item First, at variance with lattice systems, such as the spin models studied
  in the Part \ref{part:sg}, the MC updates imply a random three dimensional
  displacement instead of just a spin flip. In addition, particles can
  diffuse, and then, the nearest neighbors are no longer fixed. This means
  that computing energies for the Metropolis updates would, in principle,
  imply $N^2/2$ operations. However, if the interaction is short-ranged, one
  can reduce the problem to $O(N)$ as we shall discuss below.
\item Second, when $\delta>0$, the system suffers from the dynamic glass
  transition. This fact results in a divergence of the relaxation times, and thus the
  equilibration times as well. Simulated typical Brownian dynamics get
  completely stuck as it happens in experiments, and optimized MC methods
  (implying not physical moves) must be introduced. In fact, the {\em swap} algorithm~\cite{grigera:01,glass:Fernandez06,glass:Fernandez06c} accelerates
  the dynamics and makes possible to thermalize large systems below the
  kinetic glass temperature.

\item Third, the freezing transition is first order, and then, suffers
  from \ac{EDSD}. Indeed, in the vicinity of the transition two or more phases
  are metastable at the simulation conditions. Therefore, all these phases 
  should be found in an equilibrium simulation. The system tunnels between
  these phases by building interfaces of size of order $L^2$ whose free-energy
  cost is $\beta\gamma L^2$ (being $\gamma$ the surface tension).\footnote{The
    number of particles is proportional to the overall volume by means of the
    relation $N=\rho L^3$, where $\rho$ is the density of particles. Hence, if
    $\rho$ is fixed, $L\sim N^{1/3}$.} Then, the probability of creating such
  an interface is $\exp\caja{-\beta\gamma L^{2}}$ and thus the natural time
  scale for the simulation grows exponentially with $N^{2/3}$. This effect
  could be neutralized if one could constrain the simulation to one single
  phase. This can be done by choosing properly the order parameters so that there
  are not multiple metastable states. In other words, if one could avoid jumps
  between phases (they are informally named as {\em flip-flops}). This
  \ac{EDSD} is, by far, the hardest of the three problems presented in here,
  and going beyond it has been one of our main purposes of this thesis.
\end{itemize}

Finding computational strategies to speed up the simulations is specially
important in off-lattice systems, where the amount of available efficient
methods is very reduced as compared to its lattice counterpart. For this
reason, a large part of our efforts have been put in proposing new
algorithms to study this kind of systems. We will devote more
time to algorithm description than in Part \ref{part:sg} of the thesis, where
we used rather standard methods. In particular, it was necessary to move among
many statistical ensembles, some of them standard, some of them not: we go
from the regular ones (canonical, isobaric and microcanonical) to end up with
new ones, original from this thesis: the isocorical and the tethered ensembles
as applied to a first order transition. We include a summary of all them in
Appendix \ref{app:ensembles}.

\section{Research outline: Beating the exponential dynamic slowing down}\label{sec:first-order}
This Part of the thesis is based on three
papers~\cite{fernandez:09e,martin-mayor:11,fernandez:12}, as well as some
unpublished failed trials. Our break through in this problem is a story
of an underestimated problem, unsuccessful approaches, upcoming new questions
and steps back to simpler problems that were finally solved.  For this reason,
I decided to present this part the thesis, not only as a summary of the
physical results we obtained, but also as a description on the research path we
followed. In other words, I will structure this part of the thesis in a
chronological order.

The starting point was the problem discussed above, the study of
the phase diagram in polydisperse systems but paying special attention to the
high-polydispersity region, where the standard simulation methods fail to
thermalize even for very small systems. The origin of this divergence in
equilibration times is precisely the fluid-solid first-order transition and
its associated \ac{EDSD} mentioned above. 

If one seeks to mimic the experiments, the appropriate ensemble should be the
$NpT$ ensemble (constant pressure). Of course, in equilibrium one expects
ensemble equivalence in the large-$N$ limit, but the convergence to it can be
significantly different from one statistical ensemble to other. For this
reason, our first naive proposal was that, since in nature the $NpT$ situation
are preferred, so should they be in numerical simulations. The results clearly
contradict this statement, we find strong metastabilities that leads to very
long thermalization times. One could anticipate this result easily. Indeed,
the fluid and the solid phase have different characteristic volumes, and since
the volume fluctuates in this ensemble, both phases can be accommodated at the
same pressure.

According the last naive explanation, the direct solution would be to fix the
total volume in the system. However, the same problem was studied with $NVT$
simulations in~\cite{poly:Fernandez07} and the same behavior was
observed. Indeed, one can define the pressure using the virial equation, for
instance, and also a different pressure can accommodate the two involved
phases in the same volume. At that moment is clear that not all the magnitudes
that suffer a discontinuity at the transition are proper {\em reaction
  coordinates} (magnitudes that describe univocally the reaction path).

Then, our goal from that point on was to identify an ensemble that forbids
metastabilities at all simulation conditions.  In fact, if phase-tunneling can
be avoided, there are no reasons to expect \ac{EDSD}. Since this objective has been
fulfilled in simulation studies of first order transitions in lattice magnetic
systems, we can try to export their solutions.  Now, our bet was that the
microcanonical ensemble, that was determinant to prove the first-order nature
in the disordered Potts model~\cite{fernandez:08}, should also split up the
fluid and the solid phase here.

Following this intuition, we implemented the microcanonical Monte Carlo me\-thod
strategy~\cite{algorithm:lustig98,martin-mayor:07} to the problem
studied in~\cite{poly:Fernandez07} for $\delta=0.24$ (large polydispersity, in
the region where no crystal is stable). This study is presented in
Chapter~\ref{chap:micro}, which is based on
Ref.~\cite{fernandez:09e}. Unfortunately, as we shall discuss, the energy
turned out not to be a good reaction coordinate in this particular problem. In
plain words, we still suffered from \ac{EDSD}. Even though the main strategy
to thermalize had failed, we still could improve over previous work and to
thermalize in the solid phase thanks to the combination of this microcanonical
\ac{MC} algorithm with a modified version of the \acf{PT}
algorithm~\cite{hukushima:96,marinari:98b}.

At that point, the physics of the problem was clearer to us. However, we still
could not make a clear breakthrough in the size of the systems that could be
thermalized. Indeed, we could equilibrate samples but more with brute force
(very long simulations) than thanks to a clever election of the simulation
methods. Nevertheless, we were not placed at the same point that we were at the
beginning: now we knew that the microcanonical strategy failed because the
first-order transition actually corresponded to a phase separation in our
problem. We needed an order parameter that controlled the size of the
segregated regions of solid phase growing in the fluid.

However, if these metastabilities really arose from a phase separation, this
very same problem should come up in any kind of solidification/melting
problem, not necessarily related to disorder. In particular, it should arise
in the simplest possible problem: the crystallization of monodisperse hard
spheres. The \ac{HS} freezing transition is well established since
1968~\cite{hoover:68}, but, to our surprise when reviewing thoroughly the
literature, not even at \ac{HS} level the \ac{EDSD} problem was controlled.
At this stage, we decided to step backward, and to seek a method that truly
controls the crystallization in this simple model. This goal is achieved in
Chapter~\ref{chap:HS}, which is based on Refs.~\cite{martin-mayor:11,fernandez:12}.

Once the mechanism is fully understood, we would return to the original and more
interesting problem. This last step is beyond this thesis, but we would like
to emphasize that the tools developed here will be extremely useful for
further studies not only in polydisperse soft spheres, but also for any
problem involving a first-order transition.

\section{Crystalline order parameters}\label{sec:Q6-intro}

In this section we introduce the standard crystalline order parameters used in
the modern crystallization studies. The parameter discussed here was
introduced by Steinhardt {\em et al.} in 1983 ~\cite{steinhardt:83}. It will
be studied in the two following chapters, and for this reason, we decided to
place its discussion in a common section. As we shall see, some details in
the definition need to be tuned, so we will take up the discussion again in
each related chapter.  In addition, we want to note that in Chapter
\ref{chap:HS} we will introduce an extra crystalline order parameter, but we
leave its definition to that moment.

The main task of the parameter discussed here is to measure both the local and
the extended orientational symmetries. For this reason, these kind of
parameters are also called bond-orientational order parameters. The idea
underlying its definition is to consider a ``bond'' joining each couple of
``nearest neighbors'' (even though they do not necessary interact, as in hard
spheres).\footnote{In an off-lattice system, the definition of nearest
  neighbor is, of course, arbitrary. In fact, we will use different
  definitions for monodisperse and polydisperse particles, but we postpone the
  discussion to the following chapters.} This ``bond'' has the direction of
the vector that joins the centers of the particles $i$ and $j$, i.e.
$\V{r}_j-\V{r}_i$.  We do not work with perfect lattices in general, then
each particle $i$ will have a different number of neighbors, namely
$N_b(i)$.  

For each particle, we associate a spherical harmonic
$Y_{lm}\paren{\theta(\hat{\V{r}}),\phi(\hat{\V{r}})}$ to each of its outgoing
bonds, where $\hat{\V{r}}$ is the unit vector along the bond direction. We are
only interested in the bonds' orientation, not in their direction. For this
reason, we will only consider even values of the quantum number $l$.  Now,
summing up over all the $N_b(i)$ bonds, we obtain each particle
contribution,\be q_{lm}(i) \equiv \sum_{j=1}^{N_b(i)}
Y_{lm}({\hat{\V{r}}_{ij}})\,.\ee Using this approach, the bonds that belong to
a crystal structure will add up constructively, while the total contribution
coming from random ordered neighbors would cancel out. As in the rest
of magnitudes, we are interested in the overall structure, so we also average
over all the particles in the volume,
\begin{equation}
Q_{lm} \equiv \frac{\sum_{i=1
  }^{N}\, q_{lm}(i)}{\sum_{i=1 }^{N} N_b(i)}\,.
\end{equation}
Finally, we sum up over all the rotationally invariant combinations to get a
rotationally invariant operator, 
\begin{equation} \label{eq:def-Q6}
Q_l \equiv \left( \frac{4 \pi}{2l +1} \sum_{m = -l}^{l} \left|
Q_{lm} \right|^2 \right)^{1/2}\,.
\end{equation}
These $Q_l$ are quasi-order parameters, in the sense that they are ${\cal
  O}(1)$ (independent of $N$) in a crystalline phase and ${\cal
  O}\big(1/\sqrt{N}\big)$ in a disordered phase. In particular we will be
interested in the case where $l=6$. This $\q$ has well defined values in
perfect lattices,\footnote{Defining the nearest neighbors as the particles in
  the first shell of neighbors.} in particular, $Q_6\approx0.574$ in a
\ac{FCC} and $0.510$ in a \ac{BCC}.  For defective crystals we should expect
smaller values ($Q_6\approx 0.4$ is fairly common).

\chapter{Polydisperse soft spheres\label{chap:micro}}

We devote this section to the problem largely described in
Section~\ref{sec:problem} in the previous chapter. We use a microcanonical
strategy that will be detailed in the following sections. This Chapter is
based on~\cite{fernandez:09e}.

\section{The Model}~\label{SECT:MODEL}

We take as a paradigm for polydisperse off-lattice systems the \ac{PSS}
model.  We consider particles of radius $\sigma_i\,,$ with
$i=1,2,\ldots,N\,$. The particle size $\sigma_i$ is drawn from a
\acf{pdf} $P(\sigma)$. Size polydispersity is in general characterized
by a single parameter, $\delta$, defined as the ratio among the
standard deviation and the mean of $P(\sigma)$:
\begin{equation}
\delta = \frac{\sqrt{\langle \sigma^2\rangle-\langle\sigma\rangle^2}}{\langle \sigma\rangle}\,.\label{DEF:DELTA}
\end{equation}
At least for small polydispersity, $\delta$ seems to be the only
feature of $P(\sigma)$ that controls the physical results.

Our particles interact via a continuous pair potential:
\begin{equation}\label{potential}
V(x_{ij})    = \left\{\begin{array}{lc}
\epsilon \caja{f(x_{ij})-f(x_\mathrm{c})} &\mathrm{if}\  x_{ij}<x_\mathrm{c}\,,\\\nonumber
0 \ &\mathrm{if}\  x_{ij}> x_\mathrm{c}\,,\nonumber
\end{array}\right.
\end{equation}
with,
\begin{eqnarray}
f(x)=\frac{1}{x^{12}}+x\,,&  \displaystyle x_{ij}= \frac{|\V{r}_i-\V{r}_j|}{\sigma_i+\sigma_j} \,,&\mathrm{and}\,x_\mathrm{c}= 12^{\frac{1}{13}}.
\end{eqnarray}

We take $\epsilon$ as energy unit. Note that we use the long distance cut-off
of Refs. \cite{glass:Fernandez06c,algorithm:yan04}. The existence of this
cut-off allows us to divide the system in boxes so that the energy computation
is only $O(N)$. Indeed, Eq. \eqref{potential} tells us that two
  particles with radius $\sigma_i$ and $\sigma_j$ interact as long as
  $|\V{r}_i-\V{r}_j|\le x_c(\sigma_i+\sigma_j)$. This has a straight-forward
  consequence: no couple of particles would interact for separations
  $|\V{r}_i-\V{r}_j|\ge 2x_c\sigma_i^\text{max}=r^\text{max}$. We can use this
  fact to divide our total volume in cubic cells of side $a\gtrsim
  r^\text{max}$. Within this division, a given particle would only interact at
  most with the particles in each its $9$ neighboring cells. Then, if one
  keeps a count on the cell in which each particle is contained, the number of
  total interactions to compute the total energy becomes $O(N)$ instead of the
  number of pairs, $N(N-1)/2$.

Although ~(\ref{potential}) generalizes well
known models for simple liquids~\cite{hansen}, its scale-invariant form
suggests that it may describe as well colloids, whose size is in the
micrometer range. In fact, the interaction in~(\ref{potential}) is
short-ranged as it is appropriate for colloidal systems.

Our length unit, $\sigma_0$, is fixed by
\begin{equation}
\sigma_0^3 = \int \mathrm{d}\sigma_i
  \mathrm{d}\sigma_j P(\sigma_i) P(\sigma_j) (\sigma_i +
  \sigma_j)^3\ .
\end{equation}
We simulated $N$ particles in a box with periodic boundary conditions
at density $\rho=\sigma_0^{-3}$.  Due to the scale invariance of the
potential, the thermodynamic parameter that controls the problem is
the combination $\varGamma \equiv \rho\: T^{-1/4}$ ($T$ is the
temperature).

Here we study the case where the size distribution is flat (constant
in the range
$[\sigma_\mathrm{min},\sigma_\mathrm{max}]$). Sample-to-sample
fluctuations, as discussed for spin glasses in Part~\ref{part:sg}, are eliminated by picking the diameters in a deterministic way~\cite{poly:Santen01,
  poly:Fernandez07},
\begin{equation}
\sigma_i =
  \sigma_\mathrm{min} + (i-1)\frac{\sigma_\mathrm{max}-\sigma_\mathrm{min}}{N-1}\,.
\end{equation}
The polydispersity of the system is thus given by
\begin{equation}
\delta=\frac{1}{\sqrt{3}}\frac{(r-1)}{(r+1)},\ \mathrm{with}\ r=\frac{\sigma_\mathrm{max}}{\sigma_\mathrm{min}}\,.
\end{equation}
Hence, $\sigma_\mathrm{max}/\sigma_\mathrm{min} \to \infty$ at
$\delta_{\infty}=1/\sqrt{3}\approx 0.57735$.  

\begin{figure}
\centering
\includegraphics[width=0.8\columnwidth,trim=0 0 0 0]{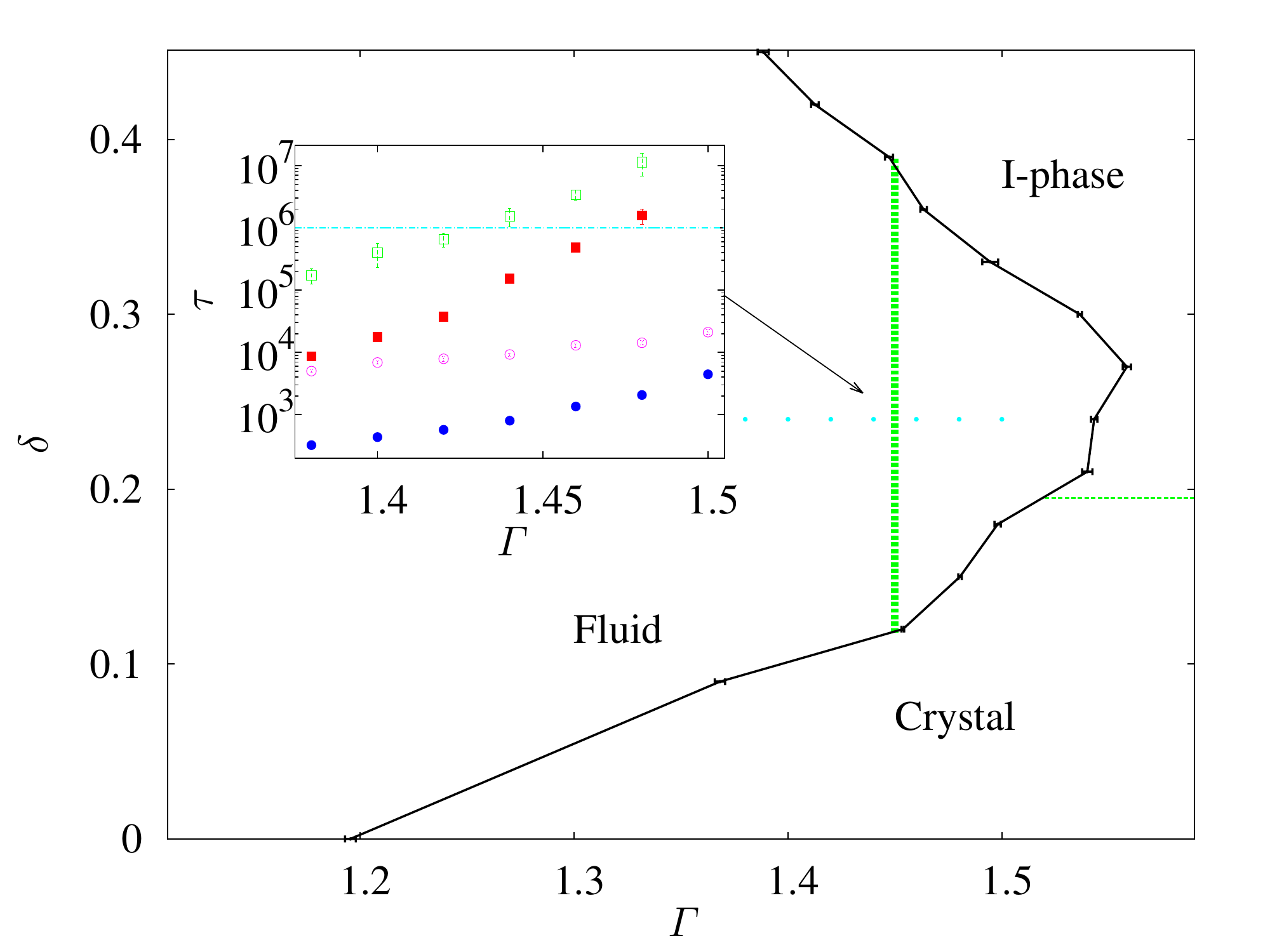} 
\caption{Phase diagram $\delta-\beta$. The green vertical line corresponds to
  the dynamical glass transition, while the black one refers to the freezing
  transition. (Inset) Integrated relaxation time for several observables for
  standard (squares) and local swap (squares) MC updates.
Figure taken from \cite{poly:Fernandez07}.} 
\label{PHASE-DIAG}
\end{figure}

The phase diagram for this model is shown in Figure \ref{PHASE-DIAG}. It was
obtained in \cite{poly:Fernandez07} with simulations in the $(N,V,T)$
ensemble. Let us sketch the main features obtained in this previous
work. First, a fluid-solid transition (in black lines in Figure
\ref{PHASE-DIAG}) is always found for any polydispersity $\delta$. This last
fact rules out the final polydispersity scenario. However, even though there
is a solid phase thermodynamically stable for each $\delta$, it might be
dynamically inaccessible in experimental times due to the presence of the
kinetic glass transition (in green). The exact location of the kinetic glass
transition can be obtained using the same criterion than in a experiment. That
is, simulating Brownian motion dynamics (standard MC steps), and locating it
at the point where the relaxation time $\tau$ reaches the $10^6$ MC steps (see
inset in Figure \ref{PHASE-DIAG}). For colloids a standard \ac{MC} step
corresponds roughly to 0.01 seconds of experimental
time~\cite{poly:Simeonova04}, then, this choice is equivalent to relaxations
of $\sim 3$ hours of physical time.  Both for $N\!=\!500$ and $864$ particles,
we find that $\varGamma_\mathrm{g}= 1.455(5)$. With this definition, there is
a region in polydispersities $\delta\in[0.12,0.38]$, where the the
dynamic glass transition occurs in the stable rather than in the metastable
fluid region. In this work, we focus precisely in this high polydispersity
region, in particular, we will fix $\delta=0.24$.

In previous studies, it was shown that the local swap algorithm (a modified
version of the global swap discussed in Section~\ref{SECT:FIXED-E}) accelerated by
several orders of magnitude (see inset in Figure \ref{PHASE-DIAG}) the
dynamics below $\varGamma_\mathrm{g}$. However, the situation above this value
is rather more difficult. Indeed, the local swap helps to avoid the cage
effect that origins the glass transition.  In the case of solidification, the
effect of the swap is not enough to counteract the effect of the previously
discussed exponential dynamical slowing down associated to the first-order
transition. Actually, the
thermalization deep into the solid phase was not even attempted in
\cite{poly:Fernandez07}. 

\section{The constant energy ensemble}
As we already discussed in Chapter \ref{chap:intro-colloids}, our first
proposal to overcome the \ac{EDSD} was to work in the microcanonical,
$(N,V,E)$, ensemble (see Section~\ref{SECT:MICRO} for a description of the
statistical ensemble). Specifically, we use the Lustig's microcanonical
Monte Carlo~\cite{algorithm:lustig98} in the formulation
of~\cite{martin-mayor:07}. 

In this ensemble, the total energy per particle, $e$, is fixed. Within this
description, the microcanonical average of an arbitrary function of the
particle positions $\{ {\boldsymbol r}\}_i$ and of the energy density $e$,
$O(\{ {\boldsymbol r}\}_i;e)$ is obtained using the following expression
\begin{eqnarray}
\langle O\rangle_e&\equiv&   \frac{\int \prod_{i=1}^N \,\mathrm{d}{\boldsymbol r}_i \, O(\{ {\boldsymbol r}\}_i;e)  \omega_N (\{ {\boldsymbol r}\}_i;e)}
{\int \prod_{i=1}^N \,\mathrm{d}{\boldsymbol r}_i \, \omega_N (\{ {\boldsymbol r}\}_i;e)}\,,
\end{eqnarray}
where the weight is given by
\begin{eqnarray}
\omega_N (\{ {\boldsymbol r}\}_i;e)&=& (e-u)^{\frac{N}{2}
-1} \theta(e-u)\,,\label{eq:microweight}
\end{eqnarray}
with $u=U/N$, with $U$ the total potential energy, defined as
\begin{equation}\label{eq:total-epot}
  U(\{\V{r}_i\})=\sum_{i<j} V\paren{ \frac{|\V{r}_i-\V{r}_j|}{\sigma_i+\sigma_j}}\,,
\end{equation}
where $V(r)$ is the soft-spheres interaction introduced in \eqref{potential}.

\subsection{Observables}\label{SECT:OBSERVABLES}

\subsubsection{The inverse temperature}
As it is discussed and obtained in Appendix~\ref{app:ensembles}, the main
observable in a microcanonical simulation is the inverse temperature, computed
as a microcanonical expectation value at fixed energy $e$:
\begin{equation}
\beta(e)\equiv \langle\hat\beta\rangle_e,\quad \hat\beta
=\frac{N-2}{2N (e-u)}\,.
\label{eq:defbeta}
\end{equation}

The function $\beta(e)$ holds the key to connect the microcanonical
formalism with the canonical one. Indeed, the {\em canonical}
probability density for $e$,
\be P_{\beta}^{(N)}(e)\propto\mathrm{exp}[N(s_N(e)-\beta(e) e)],\ee can be recovered
from $\beta(e)$:
\begin{equation}
\log P_{\beta}^{(N)}(e_2)-\log P_{\beta}^{(N)}(e_1)=
N\int_{e_1}^{e_2}\mathrm{d}e\, \left(
\beta(e) -\beta\right)\,.\label{LINK}
\end{equation}
In the {\em thermodynamically stable region} (i.e.  $\mathrm{d}
\beta(e)/\mathrm{d}e <0$), there is a single root of $\beta(e)=\beta$,
located at the value of $e$ where $P_{\beta}^{(N)}(e)$ is maximum. Instead, at
phase coexistence there are several solutions for
$\beta(e)=\beta$. Their interpretation is explained in
Sect.~\ref{SECT:MAXWELL}. 

\subsubsection{The particle-density field}\label{eq:pd-fieldmicro}
Preceding studies~\cite{poly:Fernandez07} suggested that a very heterogeneous
solid phase would replace the crystal for high polydispersities. For this
reason, we need to define an observable that tracks this property.  With this
aim, we compute explicitly the particle density fluctuations. In particular,
we do it along three perpendicular directions at the smallest, non-vanishing
wavenumber allowed by the periodic boundary conditions,
i.e. $\V{q}=(2\pi/L,0,0) $, $(0,2\pi/L,0)$ and $(0,0,2\pi/L)$. Then \be\label{eq:F}
\mathcal{F}(\V{q})=|\hat \rho (\V{q})|^2\,, \ee $L$ being the linear dimension
of the cubic simulation box and the Fourier field is \be \hat\rho
(\V{q})=\frac{1}{N}\sum_{i=1}^N e^{\I \V{q}\cdot\V{r}_i}, \ee where $\V{r}_i$
is the position of the $i$-th particle. In Chapter \ref{chap:HS} we shall be
interested in the spatial distribution of these fluctuations, but here we just
want to investigate the overall inhomogeneity. For this reason, we will only
consider the averaged value over the three directions \be
\mathcal{F}\equiv=\frac{1}{3}\mathcal{F}(2\pi/L,0,0)+\mathrm{permutations}.
\ee In the homogeneous phases, in a fluid or in a crystal, for instance,
$\mathcal{F}$ must vanish as $1/N$. On the contrary, in an
inhomogeneous phase, one would expect $\mathcal{F}$ to remain $O(1)$.

\subsubsection{Crystalline order parameters}\label{sec:crystalline-micro}
\begin{figure}
\centering
\includegraphics[angle=270,width=0.8\columnwidth,trim=0 0 0 0]{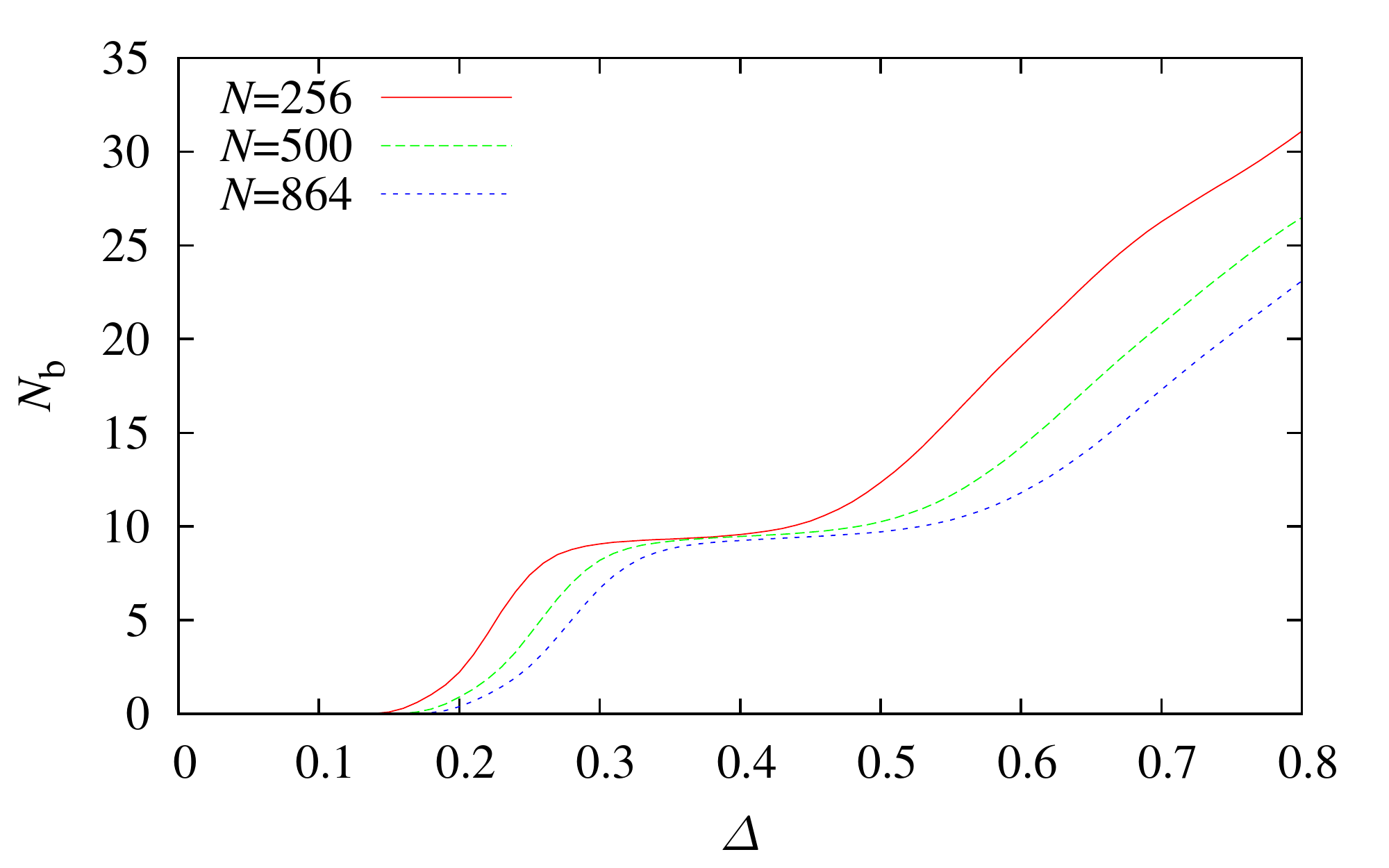} 
\caption{Averaged number of neighbors $N_b(i)$ over the half of biggest
  particles ($i\ge N/2$), as a function of $\varDelta$.} 
\label{fig:Nvec-delta}
\end{figure}

We are not only interested in heterogeneity, but also in crystallinity. We
want to distinguish whether the new solid is disordered or, on the other hand,
there are crystals on it. In addition, we need a tool to distinguish
different kinds of crystals in order to investigate fractionation.  With this
purpose, we generalize the (rotationally invariant) standard crystal order
parameter introduced in Section~\ref{sec:Q6-intro},
by measuring the crystal order
only within a given set of particles ${\cal I}(x)$ (namely, particles whose
index $i$ verifies $|i-xN|<0.05N$, hence only particles of similar size are
considered):
\begin{equation} 
Q_l(x) \equiv \left( \frac{4 \pi}{2l +1} \sum_{m = -l}^{l} \left|
Q_{lm}(x) \right|^2 \right)^{1/2},
\label{Qs}
\end{equation}
where ($Y_{lm}$ are the spherical harmonics):
\begin{equation}
Q_{lm}(x) \equiv 
\frac{\sum_{\sigma_i \in {\cal I}(x)}\, q_{lm}(i)}{\sum_{\sigma_i \in {\cal I}(x)}
N_b(i)},\, q_{lm}(i) \equiv \sum_{j=1}^{N_b(i)} Y_{lm}({\hat r_{ij}}).
\end{equation}
The index $j$ in the latter sum runs over the $N_b(i)$ neighbors of the
particle $i$ and $\hat r_{ij}$ is the unit vector linking the position of
particles $i$ and $j$. Particles $i$ and $j$ are said to be neighbors if
$||\bm r_i - \bm r_j||<\varDelta$. In order to meaningfully fix the scale
$\varDelta$, we considered the average number of neighbors as a function of
$\varDelta$ in Figure \ref{fig:Nvec-delta} for the half of the biggest
particles (which we shall see that are the ordered ones). We find a well
defined plateau along which the number of neighbors remains constant. The
height of this plateau is remarkably $N$-independent, although
its {\em width} increases with $N$ (then, the particular choice of
$\varDelta$ becomes less critical as $N$ grows). Our choice was to fix
$\varDelta=0.35$ (in units of the maximum cut-off for the potential
  $2\sigma_\mathrm{max}\, x_c$). This selection guarantees that all the values
of $N$ used in our simulations lie on the plateau  for all the energies in the solid phase.

\section{Numerical Algorithms and thermalization tests}\label{SECT:NUM}

In order to study the fluid-solid phase transition we implement a
microcanonical \ac{MC}
strategy\cite{martin-mayor:07,algorithm:lustig98}.  Fixing the total
energy density $e$, while the temperature and the potential energy fluctuate
(see \eqref{eq:defbeta} and \eqref{eq:total-epot}), we follow the evolution from
one phase to the other by studying $e$ in the {\em energy gap} between the two
phases.  This strategy turned out to be essential to assess the first-order
nature of the phase transition in disordered Potts
models~\cite{Potts:Fernandez08}. Being the freezing transition a first order
as well, we expected this method to be also promising for this problem.

The peculiarity of the polydisperse models addressed here, as compared with
Potts and similar models, is in that the phase transition actually corresponds
to a phase separation.  In fact, our low energy state is
inhomogeneous~\cite{poly:Fernandez07}. Thus moving $e$ from large values
(fluid) to small ones (partly solid) we gently {\em accompany} the system
during the growth of the spatially segregated regions. Because of that, the
internal energy will not be the only reaction coordinate (see below). However,
the combination of this algorithm with a modified \acf{PT}
algorithm~\cite{hukushima:96,marinari:98b} has allowed us to thermalize in the
solid phase.

For the sake of clarity, we divide the remaining part of this Section
in three paragraphs: particle movements at fixed energy
(Sect. \ref{SECT:FIXED-E}), Parallel Tempering (Sect.~\ref{SECT:PT}),
and thermalization checks (Sect.~\ref{SECT:THERMALIZATION}).

\subsection{Particle movements at fixed energy}\label{SECT:FIXED-E}

The particle moves at fixed energy were, with $50\%$ probability,
either standard Metropolis single-particle moves, or {\em global swap}
attempts (modified for a polydisperse system). Let us recall that in a
swap move, one attempts to exchange the position of two particles of
different sizes~\cite{algorithm:grigera01}.

Both for single-particle and for swap moves we compute the ratio of the
microcanonical weights, defined in (\ref{eq:microweight}), for the new and
the old configuration $\omega_N^\mathrm{old}/\omega_N^\mathrm{new}$. The new
configuration is accepted with Metropolis probability
$\mathrm{min}\{1,\omega_N^\mathrm{old}/\omega_N^\mathrm{new}\}$.

To fully describe the swap algorithm, we need to discuss how we choose
the pair of particles, $A$ and $B$, whose position we are trying to
interchange. Note that one needs to balance two effects in
polydisperse systems.  The acceptance is larger the closer the two
particle sizes are. However, exchanging very different particles
produces a more significant effect when trying to equilibrate the
system. Our compromise has been the following. We pick particle $A$
with uniform probability over the $N$ possibilities.  We pick $B$ with
uniform probability among particles such that $|\sigma_B
-\sigma_A|<0.2 (\sigma_\text{max}-\sigma_\text{min})$ . Particle $B$
is accepted with probability 1 if $|\sigma_B -\sigma_A|> 0.1
(\sigma_\text{max}-\sigma_\text{min})$ or with probability 0.2 in the
opposite case. In case of rejection, a new particle $B$ is selected
until a suitable candidate is picked.

In contrast to~\cite{poly:Fernandez07}, we used here a modified version of the
global swap instead of {\em local swap}. The difference between both
algorithms consists on the way of selecting the two particles whose positions
we try to interchange. In the local swap once chosen one particle, the swap
update is only tried with a particle in its vicinity. On the contrary, for us,
the selection of the two particles does not depend on their separation
distance, but on their relative size. In this work we favored this second kind
of move. The reason for this choice is that, as we shall see, in the
heterogeneous solid phase, the particles tend to crowd only with particles of
similar size, and then, the local swap has little effect. 

We check that on the coexistence-line, the swap moves reduced by three orders of
magnitude the tunneling time between the fluid and the solid
phase.

\subsection{The microcanonical parallel tempering}\label{SECT:PT}

In our Parallel Tempering simulations,\footnote{Parallel tempering is also
  known by Replica exchange \ac{MC}. In the habitual formulation one tries to
  interchange replica configurations at different temperatures, from there
  comes the term ``tempering''~\cite{hukushima:96,marinari:98b}. Here instead of temperatures, we have energy
  interchange attempts.  } several statistically independent copies of the
system at different energies are simulated.

Each Monte Carlo time unit consists of two steps:
\begin{enumerate}
\item For each copy of the system, we perform $10^5\times N$ particle
  move attempts at fixed energy (either single-particle displacements
  or particle-swap attempts). During this stage, each copy of the
  system is completely independent from the others.

\item Copies of the system at neighboring energies try to exchange
  their particle configuration. We first try to sweep the two
  configurations at the lowest energy, afterwards the second lowest
  with third lowest, etc.  In this way, the particle-configuration at
  the lowest energy has a chance of getting to the highest energy in a
  single sweep.

For the sake of clarity let us name $A,B$ the two systems that are
currently attempting to exchange their particle configuration. The
exchange is accepted with probability
\begin{equation}
\mathrm{min}\left[1\ ,\ \frac{\omega_N(\{{\boldsymbol r}_i^{(A)}\};e^{(B)})\,\omega_N(\{{\boldsymbol r}_i^{(B)}\};e^{(A)})}
{\omega_N(\{{\boldsymbol r}_i^{(A)}\};e^{(A)})\, \omega_N(\{{\boldsymbol r}_i^{(B)}\};e^{(B)})}\right]\,.
\end{equation}
The microcanonical weights $\omega_N$ are given in
(\ref{eq:microweight}).
\end{enumerate}
Further details on the simulation are summarized in Table~\ref{tabla}.

Let us finally note that the here used Monte Carlo method is quite similar to
that of Refs.~\cite{algorithm:yan03,algorithm:yan04}. We briefly mention the
main differences. First, particle swap at fixed energy was not used in
Refs.~\cite{algorithm:yan03,algorithm:yan04}. Second, phase coexistence (and
the related Maxwell construction) was not studied. Third, in the formulation
of~\cite{algorithm:yan03}, one has a single copy of the system that performs a
random-walk in energy space: it is a sort of simulated annealing
simulation~\cite{marinari:98b}, rather than our parallel tempering. Besides,
the approximation $\beta(e)\approx (N-2)/[2N \langle (e-u)\rangle]$ is used,
which coincides with Eq.~(\ref{eq:defbeta}) only up to corrections of order
$1/N$. The formulation of~\cite{algorithm:yan04} is somehow intermediate
between simulated annealing and parallel tempering. The energy range of
interest is spliced into non-overlapping subranges. Each copy of the system is
assigned to an energy subrange, where it performs a simulated annealing. From
time to time one uses parallel tempering to exchange the copies of the system
attached to neighboring energy subranges.

\subsection{Thermalization checks}\label{SECT:THERMALIZATION}
\begin{figure}
\centering
\includegraphics[angle=270,width=0.7\columnwidth,trim=0 0 0 0]{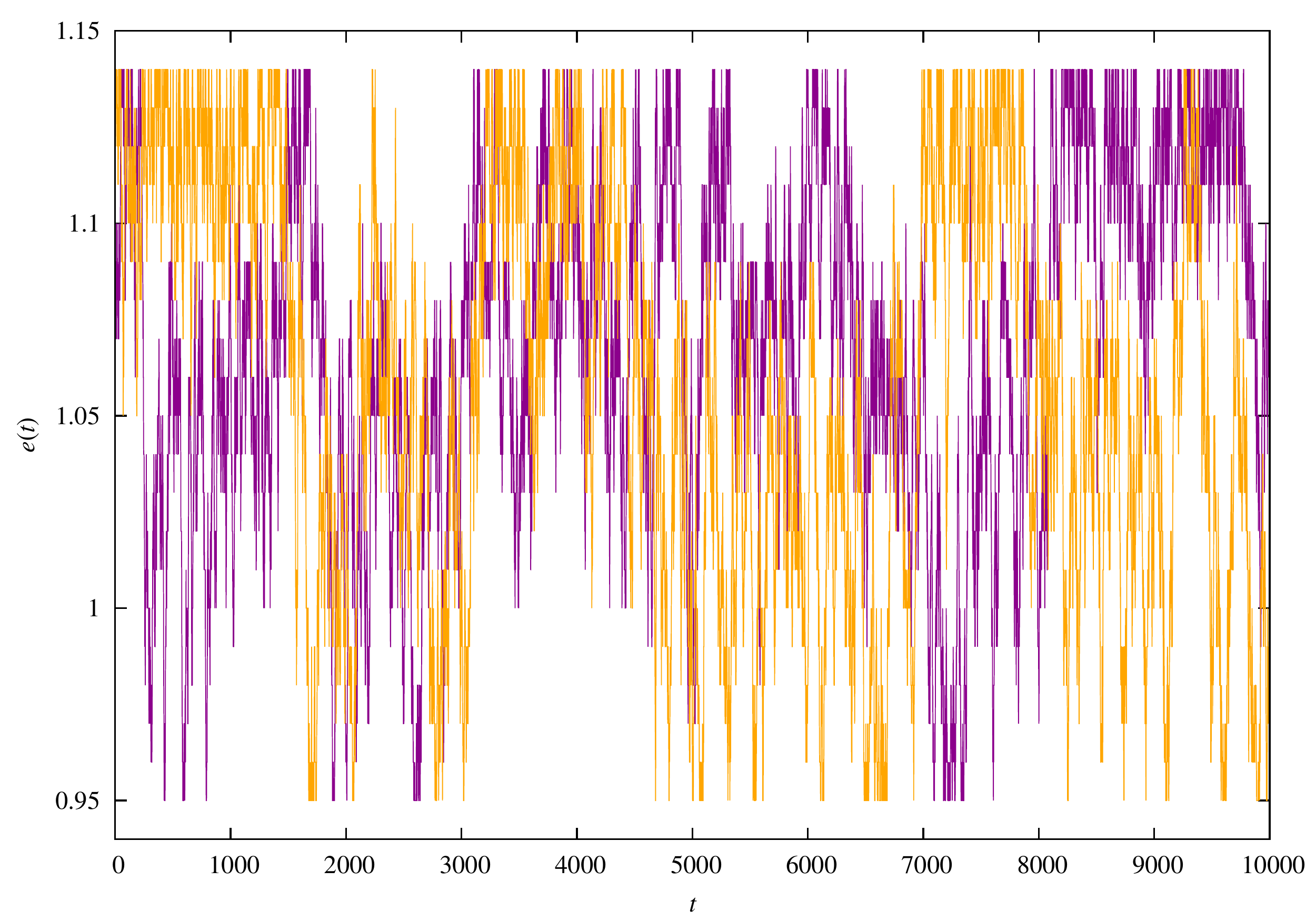} 
\includegraphics[angle=270,width=0.7\columnwidth,trim=0 0 0 0]{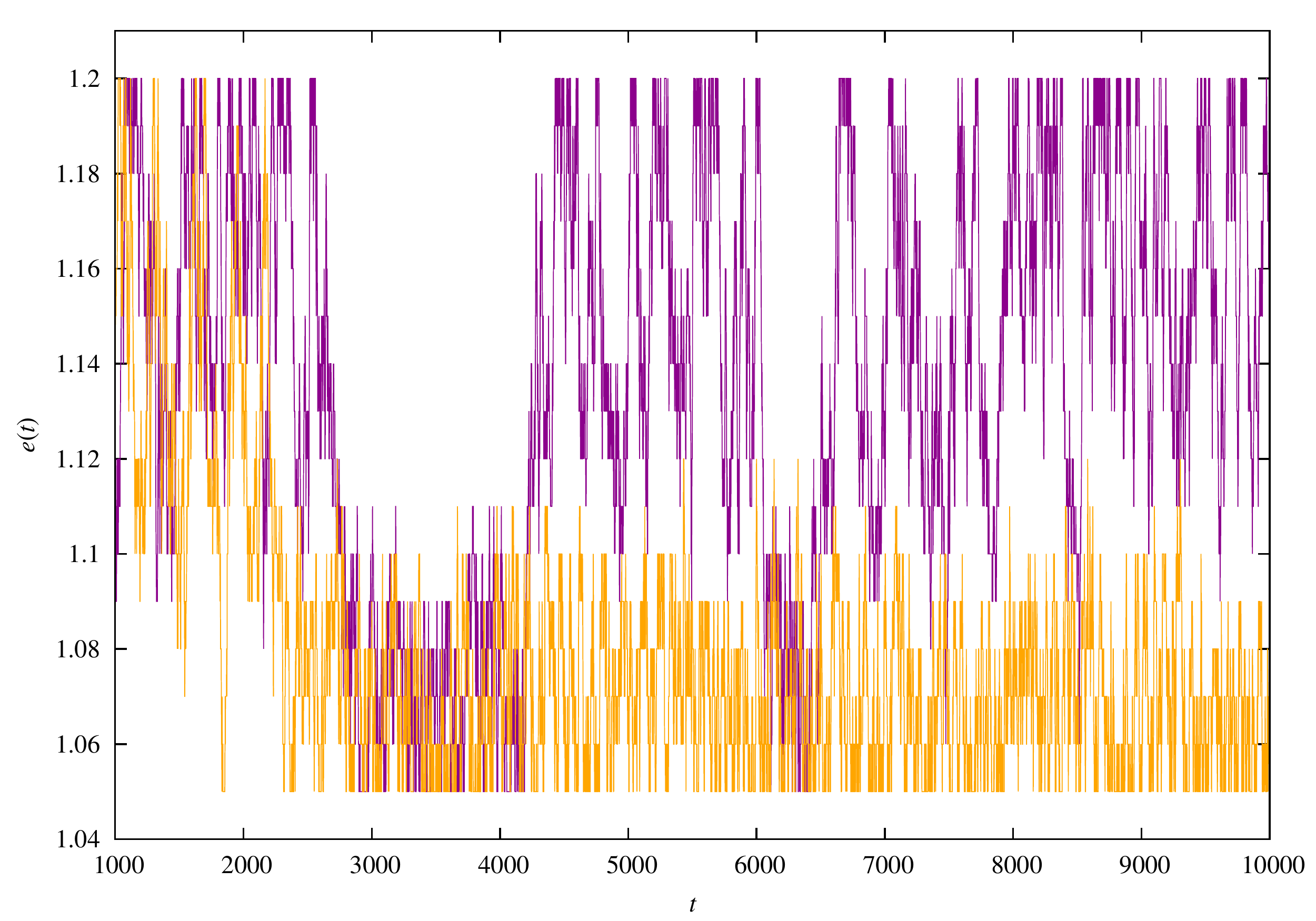} 
\includegraphics[angle=270,width=0.7\columnwidth,trim=0 0 0 0]{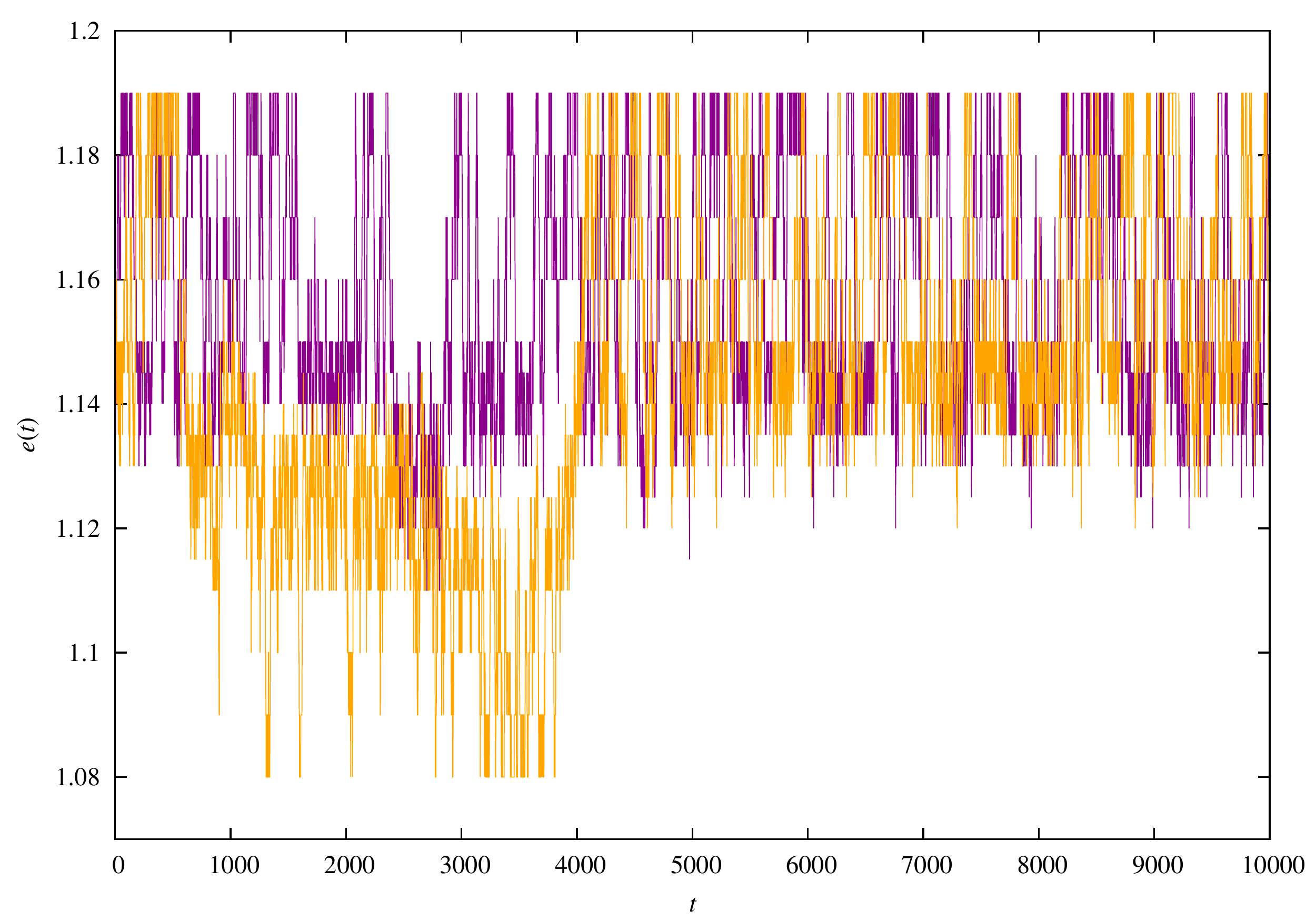} 
\caption{PT random walks for $2$ copies of the system in (top) $N=256$,
  (center) $N=500$ and (bottom) $N=864$.}
\label{fig:PT-RW}
\end{figure}
The PT algorithm has proved to be a very powerful tool for minimization
problems. Indeed, its update consists on proposing jumps from one valley of
the function one wants to minimize to other. However, this flow between
relative minima can be frustrated when the barriers between them are very high
(as also happens with the simulated annealing discussed in
Section~\ref{sec:introQA}). In such case, the PT random walk for each
configuration could get stuck in one region of the state space for a long time
(or indefinitely in the case the barriers were infinitely high) thus making
the algorithm inefficient. Our minimization problem here is to obtain the
equilibrium configuration for each fixed energy $e$ (those with maximum
entropy in our ensemble). One can guarantee that the system is thermalized as
long as each configuration visits uniformly all the energies, this would mean
that we have sampled all the state space. In other words, even when it is not
efficient, the PT algorithm gives an easy way to check the
thermalization~\cite{PTcheck:fernandez09}.

We summarize in Table \ref{tabla:PT} the technical data used for our PT
simulations. For each system size we consider $N_e$ copies of the system each
at a different energy in the intervals $[e_\mathrm{min},e_\mathrm{max}]$. As
an illustration of the problem, we display in Figure \ref{fig:PT-RW} these PT
random walks in energies for two configurations. Clearly, the probability of
tunneling from the fluid phase (high energies) to the solid phase (low
energies) and vice-versa becomes more and more difficult the higher $N$
is. Furthermore, the characteristic tunneling times, even for $N=256$ are
significant long as compared with the total simulation length (see Table
\ref{tabla:PT}). The combination of these two features shows that the PT
strategy is failing and the worsening with $N$ suggests phase coexistence
between the fluid and the solid at intermediate energies. Indeed, the barriers
between both states grow with $N$ shooting up the tunneling times. These
non ergodic random walks point out that the microcanonical strategy is not
fulfilling our final goal, to avoid jumps between phases and its corresponding
exponential dynamic slowing-down.
\begin{table*}
\centering
\begin{tabular*}{0.8\columnwidth}{@{\extracolsep{\fill}}ccccc}
\hline
$N$ & $N_e$ & $e_\mathrm{min}$& $e_\mathrm{max}$&$L_\mathrm{sim}$\\
\hline
$256$ &$20$&$0.95$&$1.14$&$5\!\times 32000$\\
$500$ &$16$&$1.05$&$1.2$&$2\!\times\!30000$\\
$864$ &$16$&$1.08$&$1.19$&$1\!\times\!12000$\\
\hline
\end{tabular*}
\caption{Simulation details. For each number of particles, $N$, we
  perform $10^5 N$ \ac{MC} steps at fixed energy, then try a \ac{PT} sweep. We
  also report the total length of our simulations in units of \ac{PT} sweeps
  ($5\times 32000$ stands for 5 independent runs of $32000$ \ac{PT} sweeps
  each). The energies chosen for the \ac{PT} were evenly spaced
  $e_{i+1}-e_i=0.01$, in the intervals $[e_\mathrm{min},e_\mathrm{max}]$. For
  $N=864$ we added to the \ac{PT} energy list the values $1.115,1.125,1.135$
  and $1.145$ in the fluid-solid energy gap.}
\label{tabla:PT}
\end{table*}

\begin{figure}
\centering
\includegraphics[angle=270,width=0.7\columnwidth,trim=28 40 21
  40]{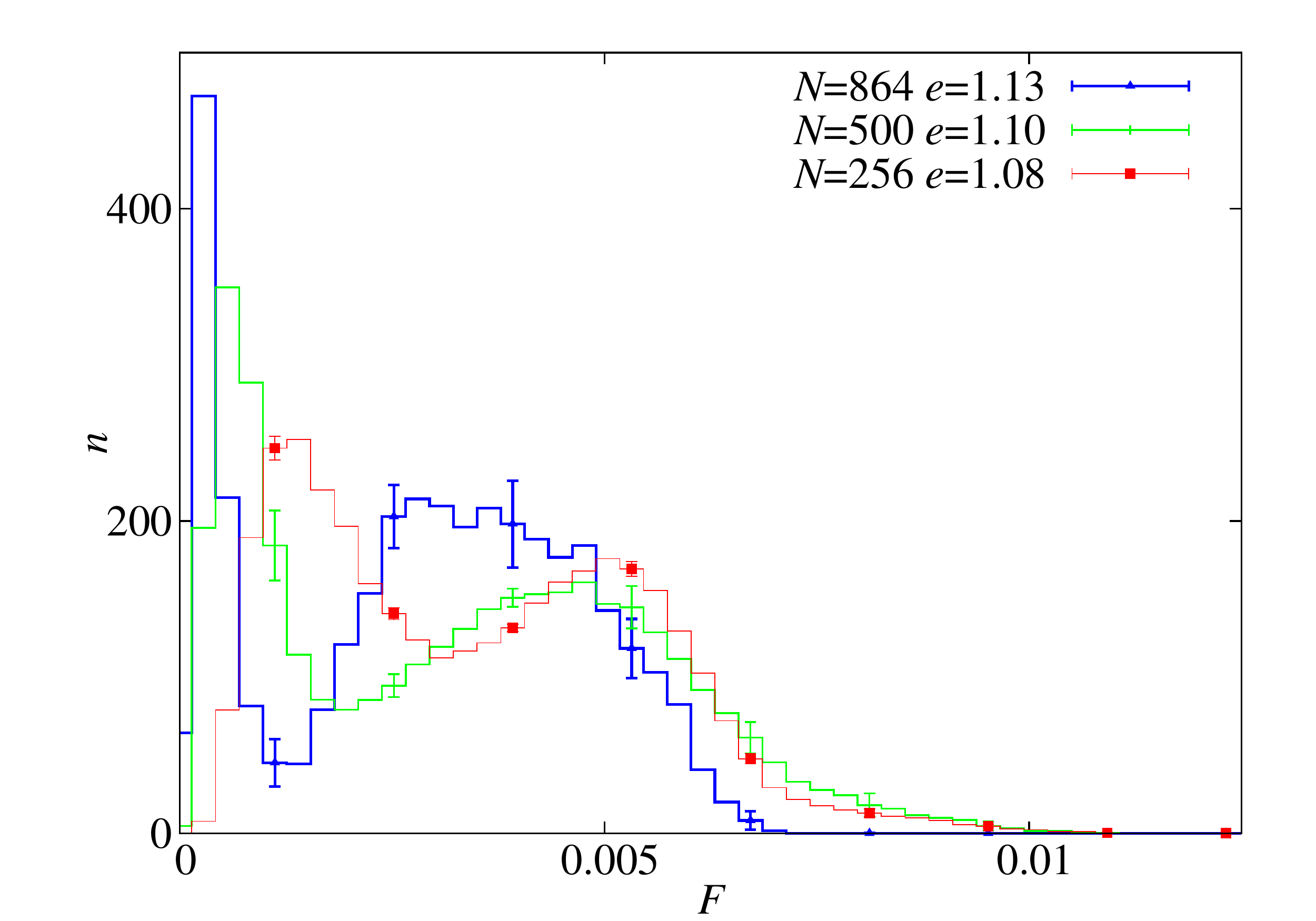}
\includegraphics[angle=270,width=0.7\columnwidth,trim=8 40 21
  40]{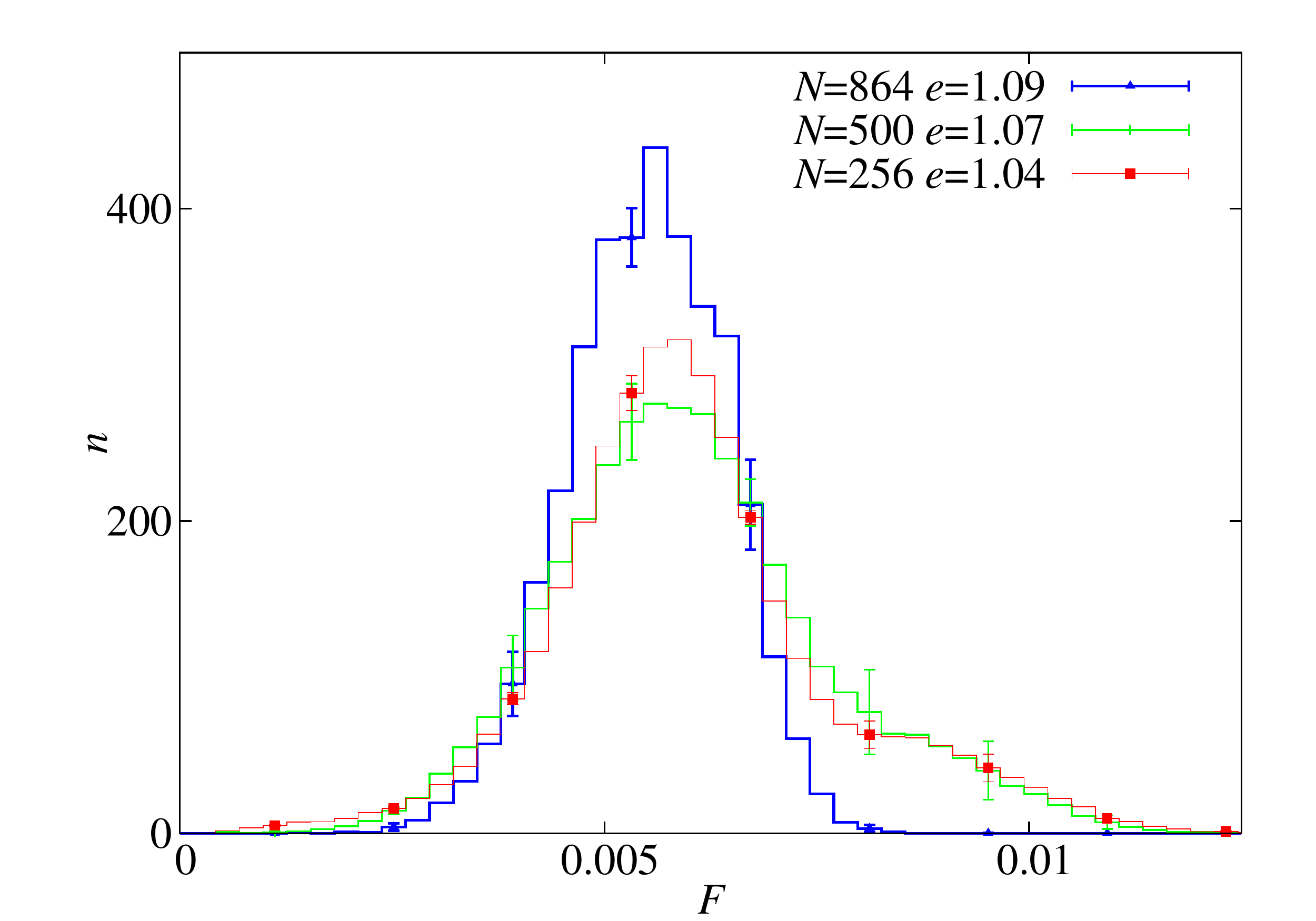}
\caption{\ac{pdf} of ${\cal F}$, \eqref{eq:F} at
  various representative values of $e$. Data in the top panel are
  computed at energy densities in the energy gap between the fluid and
  the solid phases. The double peak structure reveals phase
  coexistence (the position of the leftmost peak scales as
  $1/N$). Data in the bottom panel are computed for $e$ in the solid
  phase (the $e$-dependency there is very mild).}
\label{histogramas}
\end{figure}

We can make this last statement quantitative by looking at the probability
distribution function of ${\cal F}$, defined in ~\eqref{eq:F}. Our results are
shown in Figure~\ref{histogramas}. At values of $e$ close to the transition
(see Figure~\ref{histogramas}--top), we identify two coexisting peaks. One of
them is located at ${\cal F}\sim 1/N$, as expected for an homogeneous fluid
phase. On the other hand, the position of the large ${\cal F}$ maximum becomes
$N$-independent (this is clearer at lower energies, see bottom panel in
Figure~\ref{histogramas}), as it should occur for an inhomogeneous solid.  As
discussed above, such phase coexistence makes us to expect a large growth with
$N$ of the autocorrelation times\cite{LandauBinder}. Actually, the \ac{pdf}
for ${\cal F}$ at low energies (Figure~\ref{histogramas}--bottom) displays a
shoulder at large ${\cal F}$, which corresponds to even more inhomogeneous
solids. Hence, the \ac{PT} dynamics is ruled by two different processes:
tunneling from fluid to solid, and a second tunneling to even more
inhomogeneous configurations.

\begin{figure}
\centering
\includegraphics[angle=270,width=0.8\columnwidth,trim=28 40 21 40]{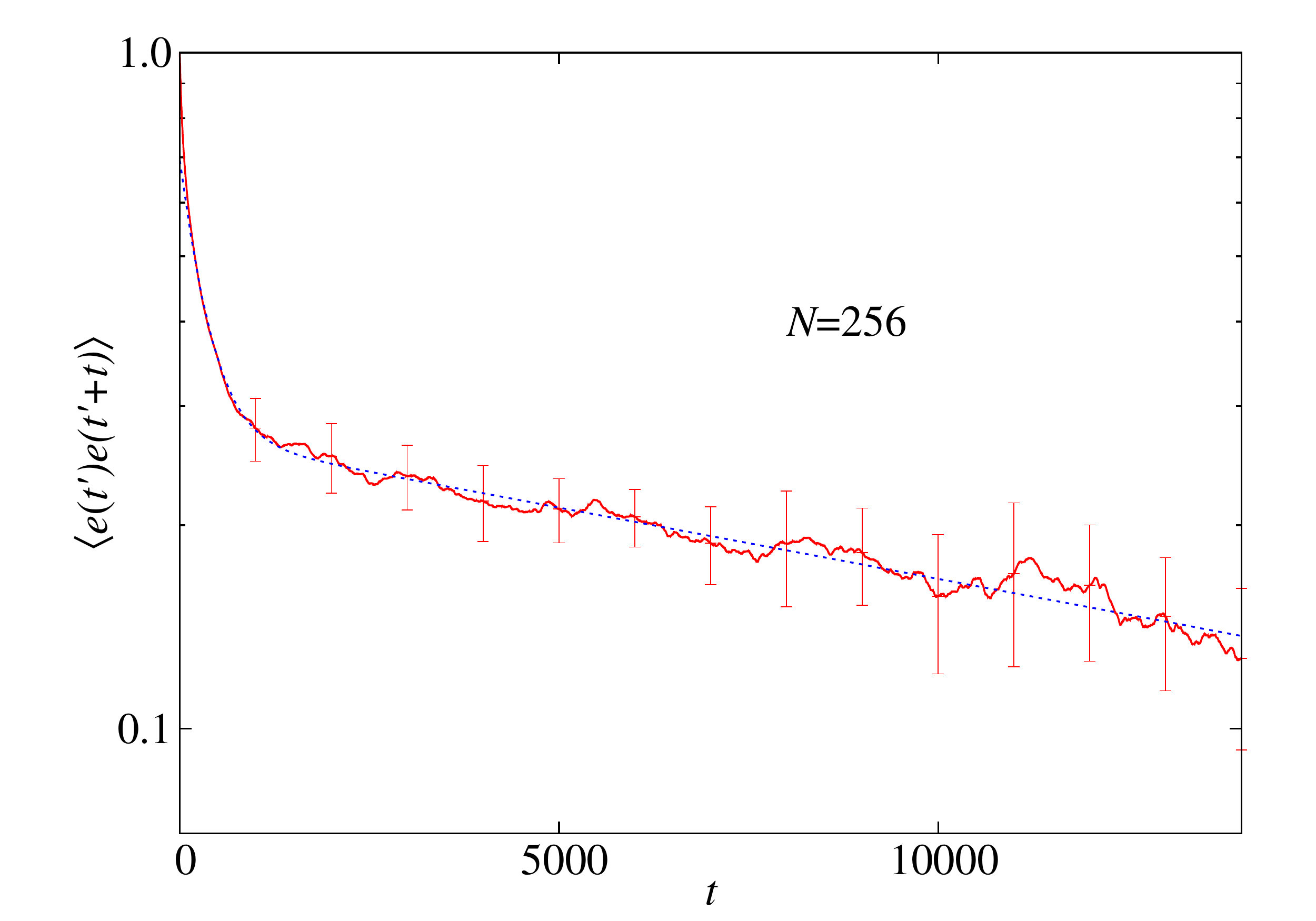} 
\caption{The
  (connected) time autocorrelation function for the energy in the \ac{PT} for $N=256$
  can be fitted (dotted line) as $\langle e(t') \, e (t' +t)\rangle=
  a_{FS}\mathrm{e}^{-t/\tau_{FS}}+ a_{SS}
  \mathrm{e}^{-t/\tau_{SS}}$. Fitted values are displayed in Table~\ref{tabla}.}
\label{correlaciones}
\end{figure}

The random-walk in the energy space shown in Figure \ref{fig:PT-RW} is best
described through a \ac{PT} time autocorrelation function (defined in
Eq.~\eqref{eq:C-O-t} in Appendix~\ref{app:thermalization}). One can fit these functions to a
double exponential for $N=256$ and $N=500$, see Figure~\ref{correlaciones}. Mind
that the {\em time} in this correlation functions correspond to the time-unit
defined in Sect.~\ref{SECT:PT}. It is not related to any physical
time-correlation.

As expected from the above discussion, we identify two different time scales
in Table~\ref{tabla}, one associated to the coexistence of the homogeneous and
inhomogeneous phase, $\tau_{FS}$, and a larger time, $\tau_{SS}$, related to
the more inhomogeneous configurations. For $N=864$, we could only identify the
$\tau_{FS}$ scale. Probably, $\tau_{SS}$ is larger than the total time in our
simulation. We remark that $\tau_{FS}$ for $N\!=\!256$ can be estimated with a
$5\%$ accuracy, while only the order of magnitude of $\tau_{SS}$ is
determined. We have explicitly checked that the effects of these very
inhomogeneous configurations on the Maxwell construction is fortunately
smaller than our statistical errors.\footnote{Indeed, we could compute
  $\beta(e)$ conditioned to a given value of $\mathcal{F}$. Since the more
  heterogeneous phase had higher values of $\mathcal{F}$, we could compare the
  Maxwell construction including all data or only the data corresponding to
  the first peak and the liquid. We could not find any difference beyond the
  statistical errors.}  Furthermore, from the point of view of our measured
crystalline order parameters (see below), the more inhomogeneous
configurations are not distinguishable from the main peak in the \ac{pdf}.
\section{Numerical Results}\label{SECT:RESULTS}

\subsection{The Maxwell construction}\label{SECT:MAXWELL}
\begin{figure}
\centering
\includegraphics[angle=270,width=0.6\columnwidth,trim=27 25 21 33]{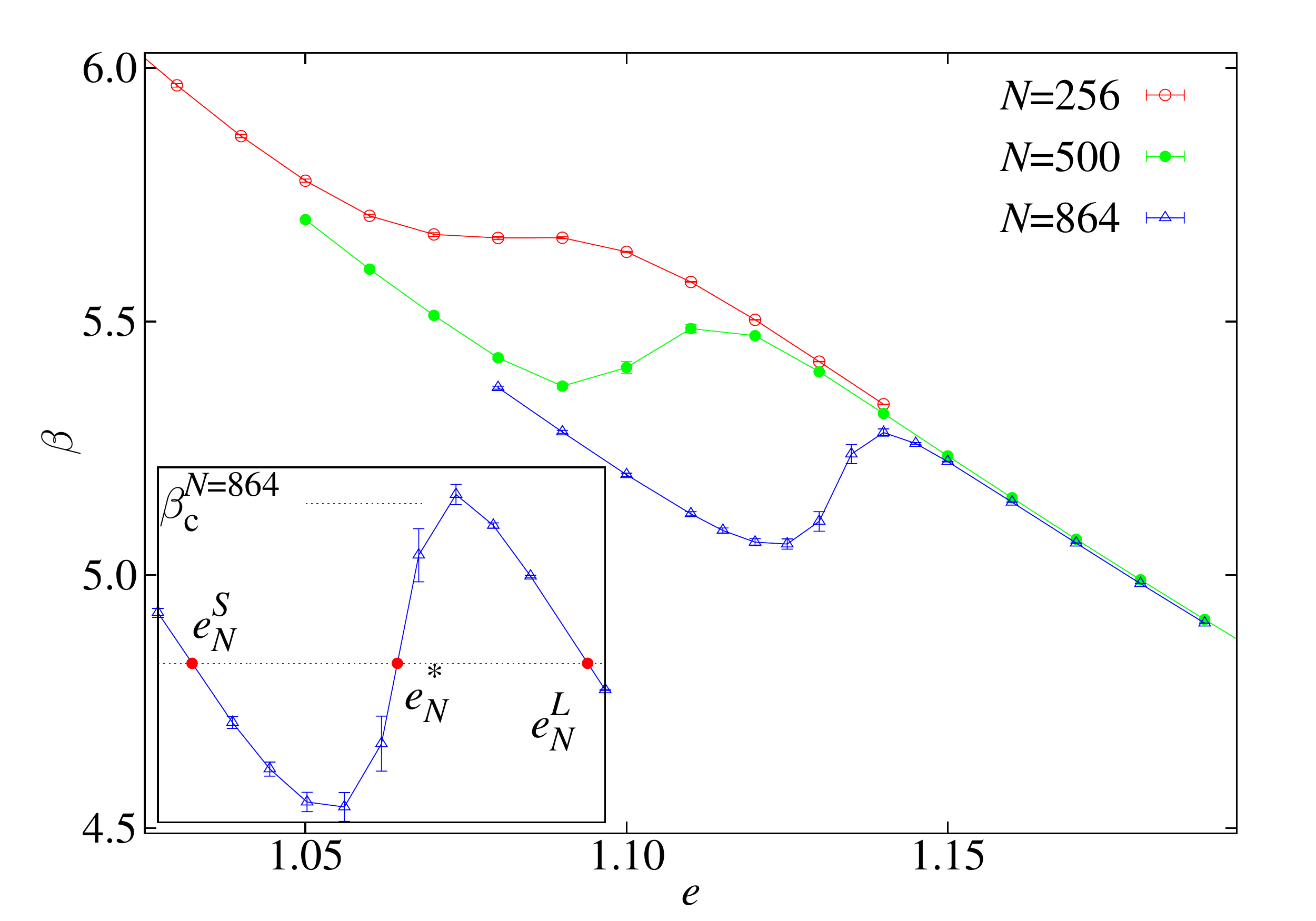} 
\caption{Finite size effects in the Maxwell construction. (Main panel) The inverse temperature $\beta(e)$ as a function of the energy
  density $e$ for various sizes of the sample. (Inset) Enlarged plot of
  $\beta(e)$ for $N=864$, including $\beta_\mathrm{c}^{N=864}$ and the three
  solutions for $\beta(e)=\bcN$: solid $\eNsol(\bcN)$, midpoint $e^*_L(\bcN)$ and fluid $\eNliq(\bcN)$.
} 
\label{SPINODAL}
\end{figure}

As was mentioned in Sec.~\ref{SECT:OBSERVABLES}, in a microcanonical
simulation, a quantity of major interest is the (inverse) temperature,
$\beta(e)$, see ~(\ref{eq:defbeta}). Thermodynamic stability requires that
$\beta(e)$ be a decreasing function (i.e. positivity of the specific
heat). Yet, see main panel in Figure~\ref{SPINODAL}, this is not the case
close to a first-order phase transition. The lack of monotonicity can be used
to obtain the critical temperature, surface tension, etc. through the Maxwell
construction (see below, and Ref.\cite{martin-mayor:07} for
details). Generally speaking, $\beta(e)$ has two distinct branches, one
describing the fluid and the other the solid phase, where the specific heat
$C_v \equiv - \beta^2 d e /d \beta$ is positive. The two branches connected by a
thermodynamically instable line where $C_v <0$. Although at finite $N$ the
system does not undergo a real phase transition, there are various criteria to
define an (inverse) critical temperature, $\bcN$, where the two different
phases coexist with the same thermodynamic weight. Here we utilize the Maxwell
construction, which amounts to obtain $\bcN$ as a solution of:
\begin{equation}
0=\int_{\eNsol(\bcN)}^{\eNliq(\bcN)}
\mathrm{d}e\, \left( \beta(e) -\bcN\right)\,, \label{MAXWELL}
\end{equation}
where the energy $\eNliq(\bcN)$ ($\eNsol(\bcN)$) in turn corresponds to the
rightmost (leftmost) root of the equation $\beta(e)=\bcN$ (see inset in Figure~\ref{SPINODAL}). The relation of the
$\beta$ integrals and the canonical probability (\ref{LINK}) shows that the
Maxwell constructions amounts to the famous equal-height rule for the
canonical probability-distribution function $P_\beta(e)$.

In Fig.~\ref{SPINODAL} we show the function $\beta(e)$ for $N=256,500,864$.
At odds with other models displaying a first order transition, as $N$ grows,
both the supercooled fluid (fluid branch with $\beta > \bcN$) and the
overheated solid (solid branch with $\beta < \bcN$) lines become longer.

As for the values of $\bcN$ reported in Table~\ref{tabla}, they decrease with
$N$. Asymptotically, finite $N$ corrections are of order $1/N$ (see
\cite{martin-mayor:07} and references therein). A fit
$\bcN=\beta_\mathrm{c}^{\infty}+ a_1/N$ fails badly the $\chi^2$ test. In
other words, our estimates for $\bcN$ are accurate enough to resolve
sub-leading scaling corrections in $1/N$. Thus, we have used a different
approach. Let us assume that scaling corrections take the form of a smooth
function in $1/N$, $\bcN=\beta_\mathrm{c}^{\infty}+ a_1/N+a_2/N^2+\ldots$. If
we have at our disposal three values of $N$, we may compute a quadratic
estimator (exact, up to corrections of order $1/N^3$):
\begin{eqnarray}
\beta_\mathrm{c}^{\infty,\mathrm{quad}}&=&\beta_\mathrm{c}^{N_1}\frac{N_1^2}
{(N_1-N_2)(N_1-N_3)}+\nonumber\\ 
&+&\beta_\mathrm{c}^{N_2}\frac{N_2^2}
{(N_2-N_1)(N_2-N_3)}+\nonumber\\ 
&+&\beta_\mathrm{c}^{N_3}\frac{N_3^2}
{(N_3-N_1)(N_3-N_2)}\,.
\end{eqnarray}
Computing the {\em statistical} error in
$\beta_\mathrm{c}^{\infty,\mathrm{quad}}$ is trivial, since
$\beta_\mathrm{c}^{N_1}$, $\beta_\mathrm{c}^{N_2}$ and
$\beta_\mathrm{c}^{N_3}$
are statistically independent random variables. Using the data in
Table~\ref{tabla} we get
\begin{equation}
\beta_\mathrm{c}^{\infty,\mathrm{quad}}= 4.624(20)\,,\quad
\varGamma_\mathrm{c}^{\infty,\mathrm{quad}}=1.4664(15)\,.
\end{equation}
However, the quadratic polynomial in $1/N$ that interpolates our values
$\beta_\mathrm{c}^{N_1}$, $\beta_\mathrm{c}^{N_2}$ and
$\beta_\mathrm{c}^{N_3}$ displays a maximum by $N\approx 256$, and decreases
for smaller $N$. Hence, $\beta_\mathrm{c}^{\infty,\mathrm{quad}}$ probably
overemphasizes curvature effects. On the other hand, a  linear (in $1/N$) extrapolation from $N_1=864$
and $N_2=500$ yields
\begin{equation}
\beta_\mathrm{c}^{\infty,\mathrm{linear}}= 4.791(11)\,,\quad
\varGamma_\mathrm{c}^{\infty,\mathrm{linear}}=1.4795(9)\,.
\end{equation}
The correct thermodynamic limit probably lies in between of the two estimators
$\varGamma_\mathrm{c}^{\infty,\mathrm{quad}}$ and
$\varGamma_\mathrm{c}^{\infty,\mathrm{linear}}$, above the kinetic glass
transition at $\gmc=1.455(5)$.

Furthermore, $\beta(e)$ also allows us to compute the surface tension. Indeed,
the quotient in the canonical probabilities between the fluid root
$\eNliq(\bcN)$ and the central point in the spinodal curve $e^*_L(\bcN)$ (were
we expect a strip configuration at least for a homogeneous system, see Section
\ref{sec:hs-interfacial} for a detailed description) will be given
precisely by the inverse of the probability of creating the two involved
interfaces, \be P_{\bcN}^{(L)}(\eNliq(\bcN))/P_{\bcN}^{(L)}(e^*_L(\bcN))=\E^{2
  \bcN\gamma^{(N)}\sigma_{0}^2 L^{2}}.  \ee Then, using (\ref{LINK}) one gets
\begin{equation}
  \bcN\sigma_0^2\gamma^{(N)}=\frac{N}{2L^{2}}
    \int_{e^*_L(\bcN)}^{\eNliq(\bcN)}
    \mathrm{d}e\, \left( \beta(e) -\bcN\right)\,.\label{SIGMAEQ}
\end{equation}
Data is shown in Table~\ref{tabla}.
\begin{table*}
\centering
\begin{tabular*}{0.8\columnwidth}{@{\extracolsep{\fill}}ccccc c}
\hline
$N$ & $\beta_c$ & $\varGamma_c$ & $\gamma^{(N)}\bcN\sigma_0^2$&$\tau_{FS}$&$\tau_{SS}$\\
\hline
$256$ & 5.665(3) &  1.5428(2)  &   --- &317(15) &$\sim$20000\\
$500$ & 5.432(5)  &  1.5267(2)  & 0.0035(2)&$\sim$1000& $\sim$15000\\
$864$ & 5.162(4)  &  1.5073(2)  & 0.0088(4)&$\sim$7000&---\\
$\infty$ & 4.624(2) & 1.4664(15)   &    & &\\
\hline
\end{tabular*}
\caption{ The (inverse) critical temperature (and the associated
  $\varGamma_\text{c}=\rho \beta_\text{c}^{1/4}$), as well as the
  dimensionless surface tension $\gamma\bcN\sigma_0^2$, as computed
  from Maxwell's construction.}
\label{tabla}
\end{table*}

\subsection{Fractionation and crystalline ordering}
\begin{figure*}
\centering
\includegraphics[angle=270,width=0.7\textwidth]{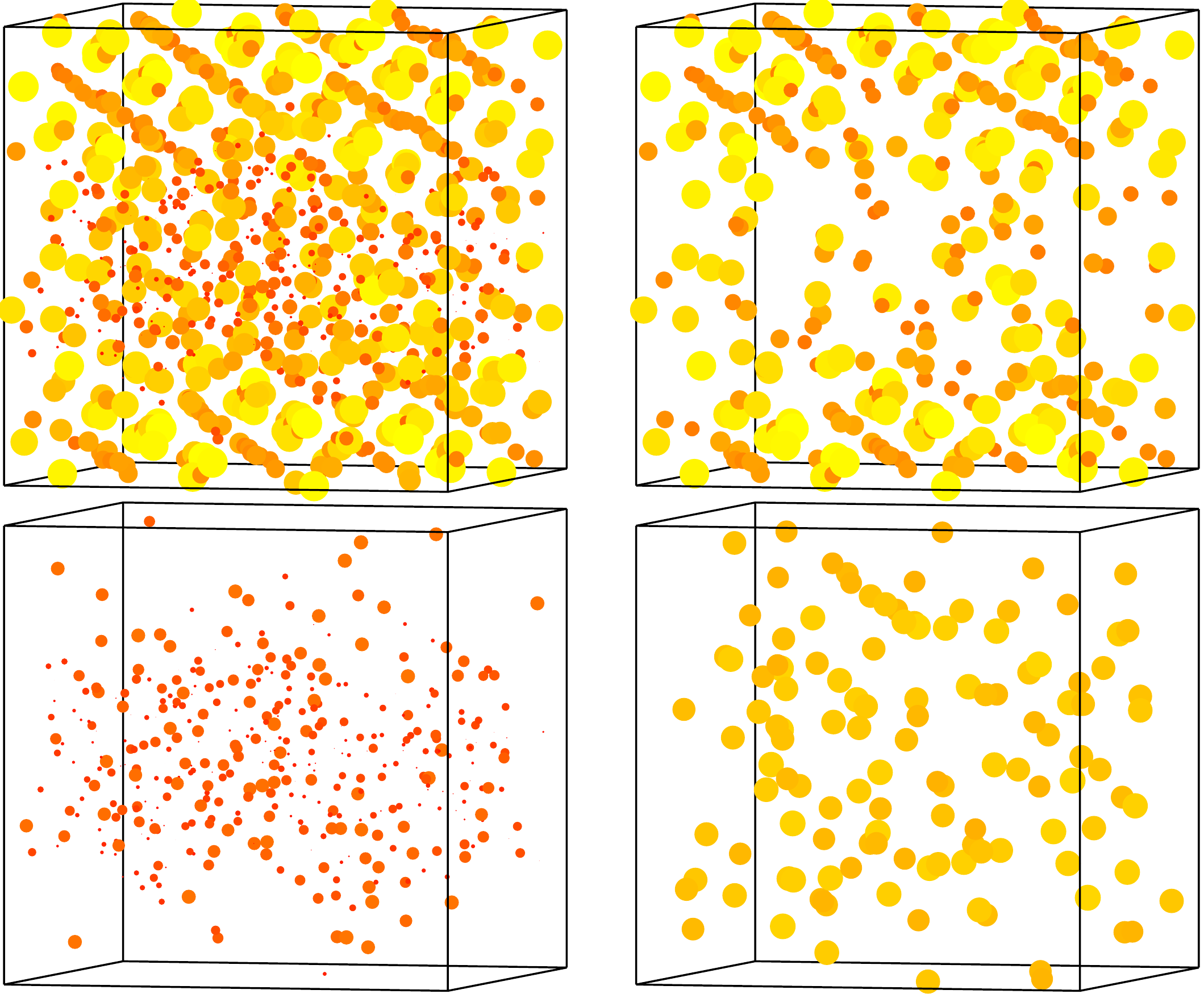} 
\caption{Snapshot of a typical low energy configuration ($N\!=\!864$,
  $e\!=\!1.01$). Top-left: whole system. Top-right: particles with index
  $i\!>\!725$ and $i \in \left[400,600\right]$. Bottom-left: particles
  $i\!<\!400$. Bottom-right: particles $i \in \left[600,725\right]$. The size
  of the circles are proportional to the particle sizes.}
\protect{\label{PICTURES}}
\end{figure*}
Finally, we study the solid phase structure. For the discussion it is
interesting a visual inspection of a typical $N=864$ low-energy configuration,
see Figure~\ref{PICTURES}. In fact, the smallest $400$ particles
(particle index $i <400$) and some of the intermediates ($i \in \left[
  600, 725\right]$) show no sign of spatial order (bottom), while
particles with $i>725$ and $i \in \left[ 400, 600\right]$ form
crystalline planes. Ordered and disordered particles fill different
regions of the sample.

\begin{figure}
\centering
\includegraphics[angle=270,width=0.8\columnwidth,trim=28 25 21 29]{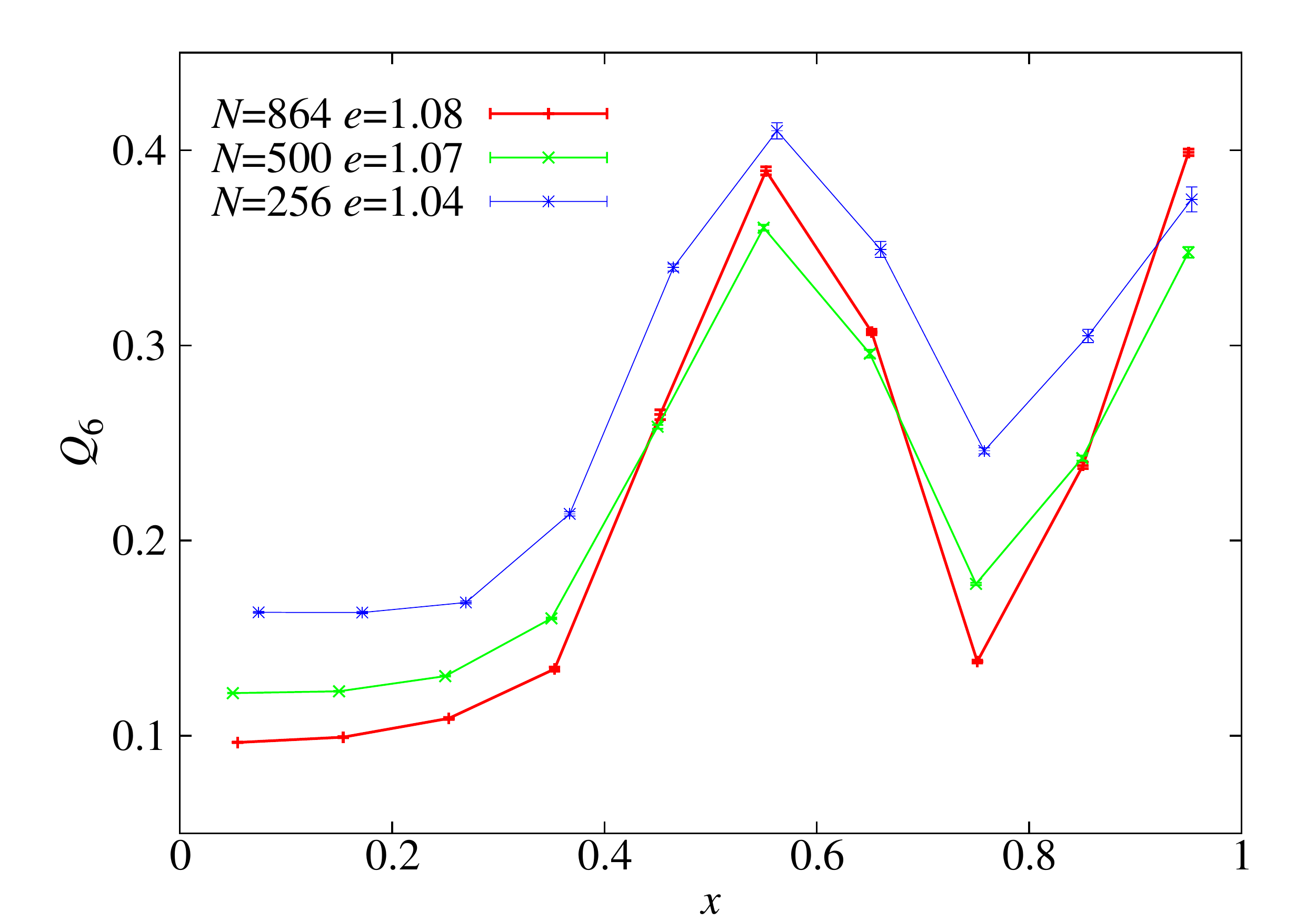} 
\caption{The crystal order parameter $Q_6(x)$, \eqref{Qs} as a
  function of the particles size $x$, for different $N$ values.}
\label{FRACTIONATION}
\end{figure}

\begin{figure}
\centering
\includegraphics[angle=270,width=0.8\columnwidth,trim=28 25 21 29]{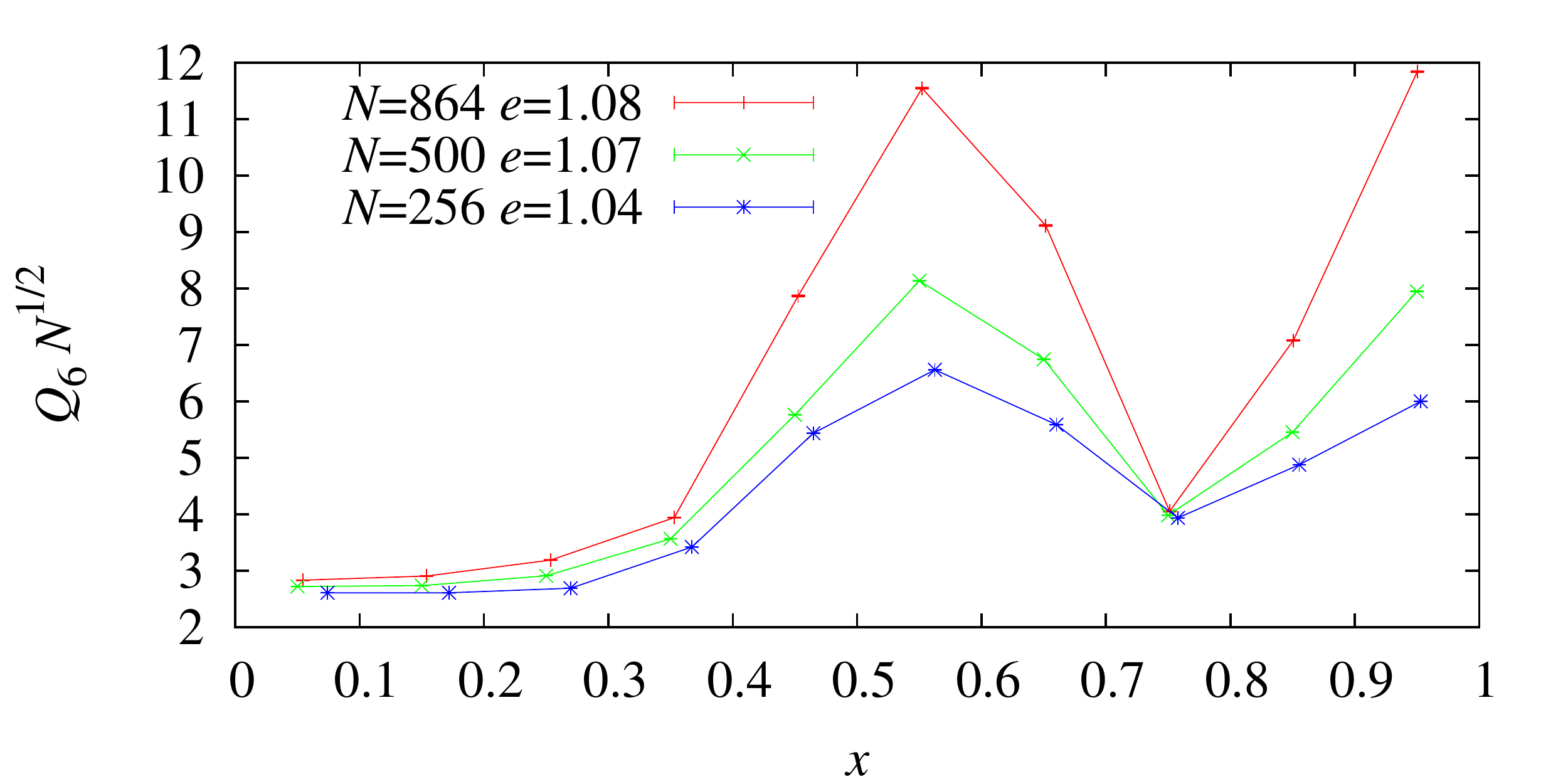} 
\caption{Same data displayed in Figure \ref{FRACTIONATION} but normalized by $\sqrt{N}$.}
\label{FRACTIONATION-N12}
\end{figure}

We can confirm this picture by means of the crystalline parameter $Q_6(x)$
introduced in Section \ref{sec:crystalline-micro}. As discussed above, we
compute the order within sets of particles of similar size, in fact, between
those whose index $i$ satisfies $|i-xN|<0.05N$. We show this $Q_6(x)$ in
Figure~\ref{FRACTIONATION}. For $x<0.45$ the crystalline order parameters
decay as $1/\sqrt{N}$ (see Figure in
Figure~\ref{FRACTIONATION-N12}), while for $x\!=\!0.55$ and $x\!=\!0.95$ we obtain
results roughly $N$ independent. Thus, while the latter group of particles
form a crystal ($Q_6$ is somewhat smaller than expected for \ac{FCC}
ordering), the former one remains amorphous. As for polydispersities, in the
two-components crystal we estimate that $\delta\sim0.15$, while in the fluid
$\delta\sim 0.24$.

In summary,  at low
energies the system divides spatially into an amorphous and a
crystalline part. Particles distribute themselves according to their
size following a complex pattern not described by any fractionation
scenario known to us.

\chapter{Hard spheres crystallization \label{chap:HS}}

As discussed in Chapter~\ref{chap:intro-colloids}, we now step back to the
simplest possible case of solidification, with the aim of beating the
exponential dynamic slowing down (EDSD) associated with a first order
fluid-solid transition. With this aim, we study the hard spheres (HS)
crystallization problem with a novel approach based on the tethered ensemble
that allows us to obtain unprecedentedly high-accuracy estimates of the
fluid-solid coexistence pressure and the interfacial free energy. This chapter
is concerned with the results published in Refs.~\cite{fernandez:12,martin-mayor:11}.

This chapter is organized as follows. In Section \ref{sec:intro-hs} we review
some of the previous approaches available in the literature concerning the
crystallization of hard spheres. The hard spheres model is described in
Section~\ref{set:hs-model}. In Sections~\ref{sec:prelude} we try to apply the
tethered approach to the problem by constraining the mean value of the
bond-order parameter $\q$. In the process, we shall understand why not one but
two bond-order parameters are necessary, and devote
Section~\ref{sec:hs-order-parameters} to introduce the new one. At this point,
we start the discussion of our final approach. The tethered formalism and
simulation details are discussed in Section~\ref{sec:hs-tethered}. In
Section~\ref{sec:p_co-hs} we explain the fluctuation dissipation approach that
allows us to draw a Maxwell construction to obtain the coexistence pressure
from it. Finally, we devote Section~\ref{sec:hs-interfacial} to the geometric
transitions observed for the larger system sizes, as well as to the
computation of the interfacial free-energy. The details concerning to the
thermalization checks are quoted in Appendix~\ref{app:thermalization}.

\section{Background on hard spheres crystallization}\label{sec:intro-hs}

Up to now, numerical simulations of crystallization phase transitions have
been well behind their fluid-fluid counterpart (e.g. vapor-liquid
equilibria~\cite{allen:89}).  Actually, HS are the preferred benchmark for
numerical approaches to crystallization. Yet, the lack of exact solutions
enhances the importance of accurate numerical and/or experimental studies.

However, for preexisting numerical methods, a simulation whose starting
configuration is a fluid  never reaches the equilibrium crystal. Much as in
experiments~\cite{pusey:89}, the simulation gets stuck in a metastable
crystal, or a defective crystal (or even a glass~\cite{zaccarelli:09}). The
proliferation of metastable states defeats optimized Monte Carlo (MC) methods
that overcome free-energy barriers in simpler
systems~\cite{berg:92,wang:01,martin-mayor:07}. Besides, experimental and
numerical determinations of the interfacial free energy are plainly
inconsistent (maybe due to a small electrical charge in the colloidal
particles~\cite{anderson:02}).

Since feasible numerical methods~\cite{vega:08} could not form the correct
crystalline phase spontaneously, choosing the starting particle configuration
became an issue (e.g. crystalline or a carefully crafted mixture of solid and
fluid phases).  Methods can be classified as {\em equilibrium} or {\em
  nonequilibrium.\/} In the phase switch MC~\cite{wilding:00}, one tries to
achieve fluid-crystal equilibrium (only up to $N=500$ HS~\cite{errington:04}).
An alternative to compute the coexistence pressure is the separate computation
of the fluid and solid free energies, supplemented with the conditions of
equal pressure, temperature and chemical potential. For the fluid's free
energy, one resorts to thermodynamic integration, while choices are available
for the crystal (Wigner-Seitz~\cite{hoover:68}, Einstein
crystal~\cite{frenkel:84,polson:00}, Einstein molecule~\cite{vega:07}). On the
other hand, the nonequilibrium \textit{direct coexistence}
method~\cite{ladd:77,noya:08} handles larger systems~\cite{zykova-timan:10}.

As for the accuracy, in equilibrium computations the coexistence
pressure $\p$ was obtained with precisions of $\sim 0.1\%$. Yet, the
$N$ values that can be simulated are rather small. An $N\to\infty$
extrapolation is mandatory, which degrades the final accuracy to $\sim
1\%$~\cite{errington:04,wilding:00,vega:07} (results are summarized in
 Table~\ref{tab:pN}). The situation improves by an order of
magnitude for the direct-coexistence method.  With the exception
of~\cite{errington:04}, the different estimations of $\p$ are
compatible, although with widely differing accuracies. 

The computation of the interfacial free energy, $\gamma$, is more
involved, since the issue of spatially heterogeneous mixtures of fluid
and solid can no longer be skipped (as done in equilibrium
computations of $\p$). Indeed, recent estimations are precise
but mutually incompatible~\cite{davidchack:10,cacciuto:03,hartel:12}, or of
lesser accuracy~\cite{mu:05}.

In this context, it is useful to summarize what has been achieved in this
thesis. We introduce a tethered MC~\cite{fernandez:09,martin-mayor:11}
approach to HS crystallization. The correct crystal appears in our simulation
by constraining the value of two order parameters. At variance with
preexisting methods, the crystal found is independent from the starting
particle configuration. Tethered MC provides a major simplification for the
standard umbrella sampling
method~\cite{torrie:74,torrie:77,bartels:00,tenwolde:95}: chemical-potential
differences among fluid and crystal are very precisely computed from a
thermodynamic integration.  In fact, our method resembles studies of
liquid-vapor equilibria~\cite{schrader:09,binder:11}. We go continuously from
the fluid to the crystal by varying a reaction coordinate that labels the
intermediate states.  Rather than particle density, our reaction coordinate is
a blend of bond-orientational crystal order parameters with different
symmetries~\cite{steinhardt:83,duijneveldt:92,angioletti:10}.  Very accurate
determinations of the coexistence pressure and the interfacial free energy
follow.  The number of HS ranges $108\!\le\!  N\!\!=\!\!4 n^3\! \le \!4000$,
($n$ integer), is large enough to undergo surface-driven geometric
transitions~\cite{biskup:02,binder:03,macdowell:06}, which entitles us to
safely extrapolate to $N=\infty$.

\section{The hard spheres model}\label{set:hs-model}
\begin{figure}
\centering
\includegraphics[width=0.5\columnwidth,trim=0 20 0 20]{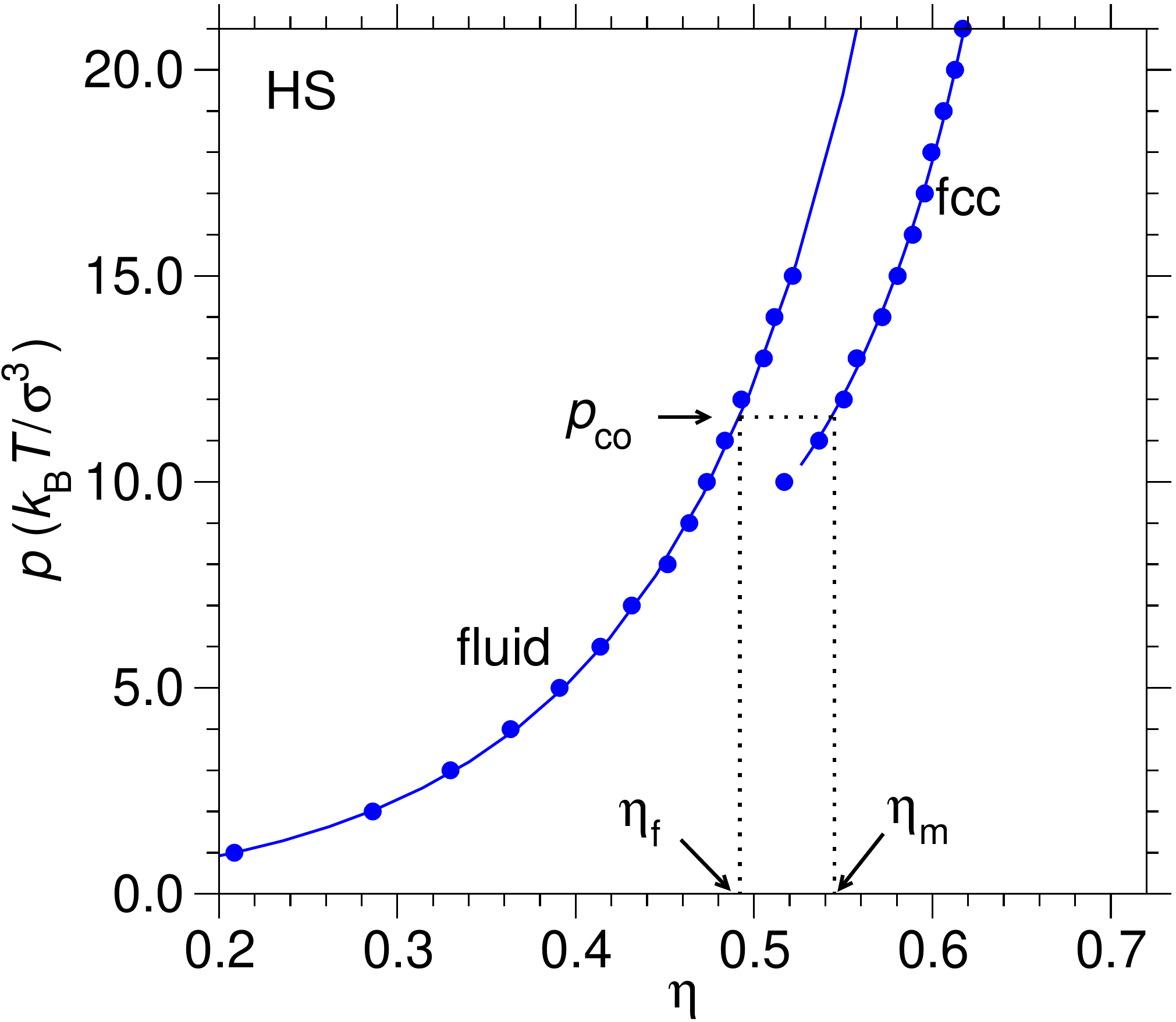}
\caption{The phase diagram for the hard sphere system 
  taken from~\cite{zykova-timan:10} [$\eta$ is the packing density
  $\eta=\pi\sigma_0^3 N/(6 V)$]. Dots correspond to simulation data and
  lines are analytical expression estimates for the fluid and the solid
  branches. The coexistence pressure is designed  by $\p$. Note the presence
  of metastable liquid and solid branches.}
\label{PHASE-DIAG-HS}
\end{figure}

We consider a collection of $N$ hard spheres, of diameter $\sigma_0$.  They are
contained in a cubic simulation box, with periodic boundary conditions. The
system is held at constant pressure $p$ (hence the simulation box may change
its volume, but remaining always cubic). 

Let us introduce the shorthand $\V{R}$ for the set of particle positions,
$\{\V{r}_i\}_{i=1}^N$. The constraint of no overlapping spheres is expressed
with function $H(\V{R})$, which vanishes if any pair of spheres overlaps
($H(\V{R})= 1$ otherwise).

The $NpT$ ensemble is discussed in Appendix~\ref{app:NPT}. For hard-spheres,
the Gibbs free-energy density, $g(p,T)$ (which is the chemical potential), and
the partition function are given by
\begin{equation}\label{eq:YNPT}
Y_{NpT}=\mathrm{e}^{-N \beta g(p,T)}=\frac{p\beta}{ N!
  \Lambda^{3N}}\int_0^\infty\mathrm{d} V \mathrm{e}^{-\beta p V}\int
\mathrm{d} \V{R}\, H(\V{R})\,,
\end{equation}
where $\Lambda$ is the de Broglie thermal wavelength, while $\beta=1/(k_\mathrm{B}T)$. The phase
diagram for this model is shown in Figure \ref{PHASE-DIAG-HS}.

Since the hard spheres cannot overlap, the most stable crystal is given by the
structure with the highest possible packing fraction. This condition is
fulfilled by the \ac{FCC}, which is actually the equilibrium crystal of our
problem. However, the HCP crystal has exactly the same packing
fraction. Because of that, it has been a problem of decades to proof which of
both was the correct structure in this problem. This dilemma was solved in the
90's, and not analytically, in fact, it was necessary to apply sophisticate
numerical methods~\cite{bolhuis:97}.

\section{Prelude}\label{sec:prelude}
In this Section we shall discuss some of the preliminary approaches that were
crucial to end up with the tethered strategy that will be explained in Section~\ref{sec:hs-tethered}.

The simplest possible simulation study of this transition would be, for
instance, to use standard MC moves at constant pressure. However, reaching the
equilibrium within a $NpT$ simulation in the vicinity of the coexistence
pressure, $\p$ (see Figure~\ref{PHASE-DIAG-HS}) is even harder that what was
discussed for soft polydisperse spheres in the previous chapters. Although at
a given pressure both the fluid and the \ac{FCC} phase are metastable, one
does not find {\em flip-flops} between these two phases (one needs to reduce
the system size below $N\sim 30$, to find any), even after a relatively
long simulation time.  Indeed, if the simulation started in the fluid phase,
it would stay forever there. Or even worse, it might form a metastable
\ac{BCC} crystal, but not a \ac{FCC}. The situation is not better when one
starts from a perfect \ac{FCC} lattice, the stochastic dynamics is not able to
melt the crystal structure.

We have tried to use more sophisticated ensembles, for instance, the
microcorical one (see Appendix~\ref{app:microcor}). The situation is exactly
as before, at a given volume, we find both crystal and fluid depending on the
starting configuration: the FCC structures do not melt, and the random initial
configurations crystallize to another metastable defective crystal structure.
Among all the observables computed during the simulation, only the ones
related to the crystalline structure ($\q$ or the number of neighbors) seem to
really distinguish the three phases involved here: fluid, \ac{FCC} and
\ac{BCC}.

Then, we thought of using crystalline parameters as reaction coordinates.  We
started with just one order parameter, in particular, with $\q$ defined in
Section~\ref{sec:Q6-intro}. The goal was then to control the growth of the
crystalline domains by tuning the value of
$\q$.\footnote{\label{foot:neighbors} $\q$ was defined in
  Section~\ref{sec:Q6-intro} but for the technical definition of nearest
  neighbor. This definition is taken here different to the one discussed for
  polydisperse systems in Section \ref{sec:crystalline-micro}. Two particles
  $i$ and $j$ are considered neighbors iff $r_{ij}<1.5\ \sigma$. This choice
  ensures that we enclose only the first-neighbors shell in the FCC structure,
  for all the densities of interest here. Indeed, we need a radius that
  includes all the first nearest neighbors and excludes the second nearest
  ones in the \ac{FCC} structure. The theoretical radius of the perfect
  lattice depends on the total volume it takes up. However, the total volume
  fluctuates in our simulations, but we need a fixed value for the definition
  of the crystalline parameters. Nevertheless, in a perfect \ac{FCC}, the
  first and second nearest neighbors shell in a lattice of volume $V$ are
  placed at a distance (in units of $\sigma$)
$$R_\mathrm{FCC}^{(1)}=\sqrt{2}\paren{\frac{V}{2N}}^{1/3},\ R_\mathrm{FCC}^{(1)}=2\paren{\frac{V}{2N}}^{1/3}.$$
Then, we seek for a value of the radius that defines $N_b(i)$ that is always
in between these two values for all the volumes studied, and $1.5$ does
fulfill this requirement.}  This idea of using $\q$ to
govern the crystallization process in a MC simulation is not new, in
fact~\cite{tenwolde:95,moroni:05,chopra:06} are well examples of works
exploiting that idea. Previous works constrain the value of $\q$ using the
umbrella sampling technique~\cite{torrie:74}. This method, broadly used in the
chemical physics community, consists on ``pressing'' the usual 
$NpT$ probability used for the Metropolis test with a weight associated with
the order parameter. With this idea, one can reconstruct an effective free
energy by means of \be G(\q)=\mathrm{const}-k_\mathrm{B}T\log\caja{P(\q)}, \ee
with $P(\q)$ the probability to find the order parameter around a given value
of $\q$. This probability can be measured directly from the simulation history
by making histograms of the instantaneous $\q(t)$. Nearby the transition one
expects to find two minima in $G(\q)$ and the phase coexistence is then
identified when the two minima are equally deep.

In this work we are using the tethered MC method (see
~\cite{fernandez:09,martin-mayor:11} and Appendix~\ref{app:tetheredO}) rather
than umbrella sampling to constrain the bond-order parameters. This method is
a refinement over the umbrella sampling. It was initially proposed in a
different context, but formally, when applied to crystallization, the tethered
ensemble leads to the same MC weights than umbrella sampling when concerning
the simulation method in this problem. The differences between both methods
appear in the way of analyzing the simulation data. Indeed, in the tethered
formalism, the effective free energy is obtained in a simpler way using a
fluctuation-dissipation formalism~\cite{martin-mayor:07} and time averages of
$\q$. This simplification has strong consequences in the precision for
magnitudes such as the $\p$ or the surface tension achievable with the same
set of simulation data. In fact, the precision in the histograms of $\q$ is
very crude, and because of that, previous
works~\cite{tenwolde:95,moroni:05,chopra:06} are more centered on studying the
structure on the crystalline grains than in determining  precisely the
coexistence point. The situation is even worse when one constrains more than
one order parameter~\cite{moroni:05} using umbrella sampling approach. The
method implies computing bidimensional histograms which damages notably the
accuracy. On the contrary, as we shall see, the tethered approach is not
hampered by the number of constraints one wants to impose.

\subsection{Tethered in $Q_6$}\label{sec:tetheredQ6}

Our first step was then to perform MC simulations in the $\hq NpT$
ensemble (see Appendix~\ref{app:tetheredO}).  We shall see that this
constraint $\hq$ is not sufficient to avoid metastabilities in all the range of
parameters. We will devote this section to justify why it does not. Since
it will not be our final approach, we will not describe here all the formalism
and technical simulation details, but just the necessary tools to give the
reader a clear idea of the problem we encountered.

Thus we employ the tethered ensemble described in Appendix~\ref{app:tetheredO}
for an arbitrary magnitude $O(\V{R})$. In this case we constrain $\q$. In this
ensemble, we let the instantaneous value of the bond-order parameter fluctuate
around a fixed value $\hq$, and the constraint tries to loosely impose
$\mean{Q_6}\approx\hq$.  This can be done in a simulation using the tethered
weight \be\omega_N(\V{R},V;p,\hq)=\sqrt{\frac{\alpha
    N}{2\pi}}H(\V{R})\ \E^{-\beta pV-\frac{\alpha N}{2} \caja{\hq -
    Q_6(\V{R})}^2},\ee for the \ac{MC} updates, which is equivalent to say
that the tethered mean values for a given couple of simulation points
$(\hq,p)$ are given by \be\mean{O}_{\hq,p}=\frac{\int_0^\infty
  db\int\D{\V{R}}\ O(\hq,p;V,\V{R})\:\omega_N(\V{R},V;p,\hq)}{\int_0^\infty
  db\int\D{\V{R}}\ \omega_N(\V{R},V;p,\hq)}.\ee

\begin{figure}
\centering
\includegraphics[angle=270,width=0.7\columnwidth,trim=0 50 0 50]{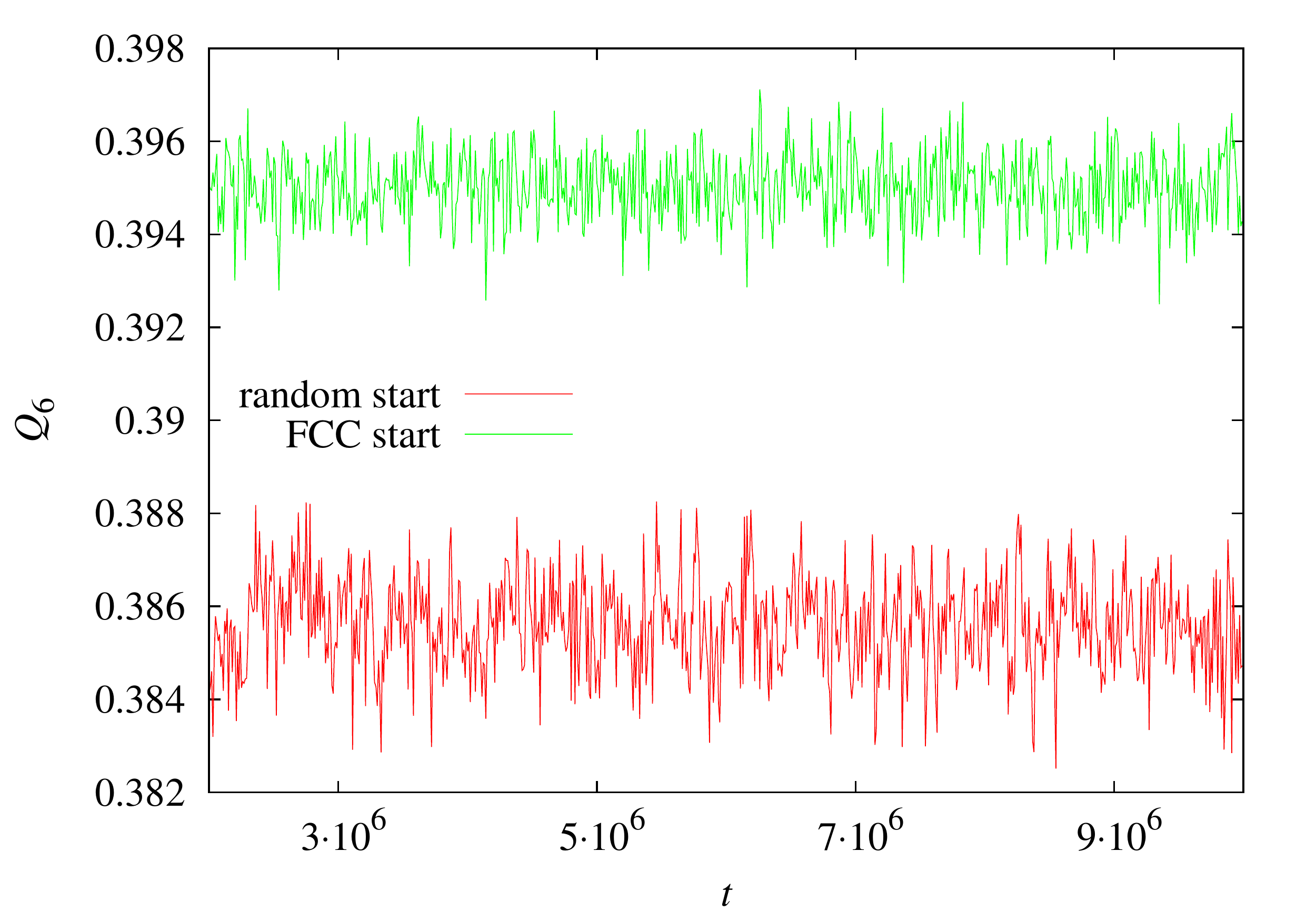}
\caption{Snatch of the evolution of $\q(\V{R},t)$ in a $\hq NpT$ MC simulation
  (for $\hq=0.394$, $N=256$, $p=11.224$). Time
  is measured in units on a elementary MC step: $N$ particles moves, followed
  by a volume update.
 The
  two lines correspond to two identical Markov chains that started from
  (green) FCC or (red) random configurations.}\label{fig:falloQ6}
\end{figure}

\begin{figure}
\centering 
\includegraphics[width=0.45\columnwidth,trim=200 100 300 50]{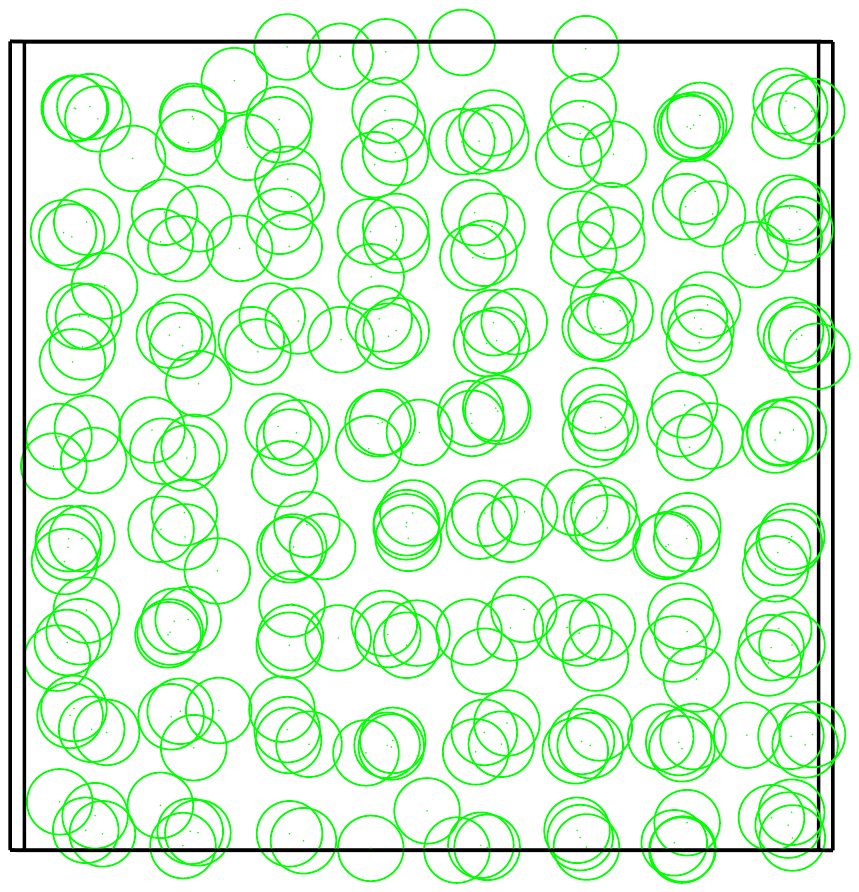}
\includegraphics[width=0.45\columnwidth,trim=200 100 300 50]{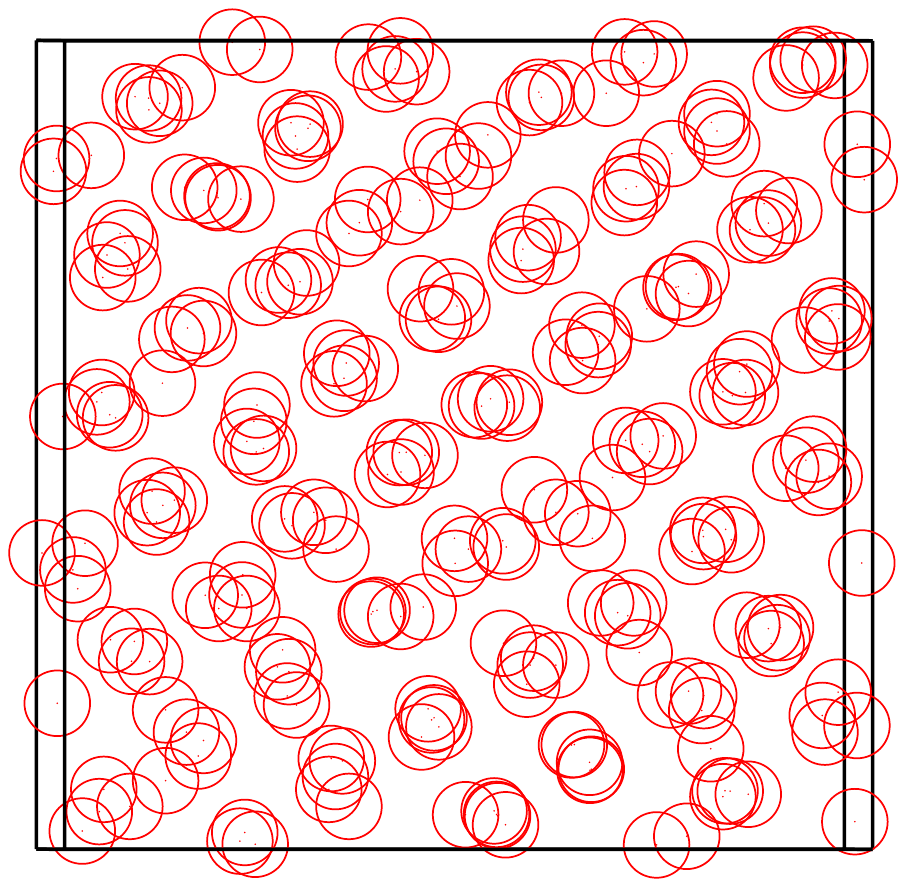}
\caption{Bidimensional projection of the typical configurations found from (green) FCC or
  (red) random start in $\q NpT$ simulations.}\label{fig:configuracionesQ6}
\end{figure}

With this idea, we run simulations at a pressure nearby the freezing
transition in a mesh of values of $\hq$ in between the fluid expectation
value, $\hq^{\mathrm{fluid}}\sim 1/\sqrt{N}$, and the perfect crystal
equilibrium phase (in hard spheres a \ac{FCC} lattice),
$\hq^{\mathrm{FCC}}=0.574$. In order to check the thermalization, as we did
before, we run two simulations at each $\hq$ value, one starting from a random
particle configuration and other from a perfect \ac{FCC} lattice. If the
approach succeeds, both simulations should converge in a very few steps to the
same structure. The method works for most of the $\hq$ points. The $\hq$ value
forces one single phase (in fact, the \ac{pdf}, $p(\hq)$, is unimodal) and no
difference between the two starts in the measured mean values is observed
(within the errors). Let us point out that it is already a great advance from
all our previous approaches, in fact we were not ever able to synthesize a
\ac{FCC} from a fluid nor with $NpT$ nor with $N\hat{V}T$, and now it forms
spontaneously just imposing its mean $\q$ value. However, there are some
points (in the solid phase) at which both simulation runs do not converge to
the same structure (see, for instance, the evolution of $\q(\V{R},t)$ in
Figure \ref{fig:falloQ6}). The situation is less dramatic than it was before,
the $\q(\V{R},t)$ values obtained are very similar thought not equal. The
problem is clarified by the snapshot in Figure~\ref{fig:configuracionesQ6},
the two solid structures are \ac{FCC}-like. However, the \ac{FCC}-start
simulations lead to a defective \ac{FCC} with the planes parallel to the
simulation box walls. On the other hand, the random start freezes in a
helicoidal almost-\ac{FCC} crystal allowed by the periodic boundary conditions
whose planes are misaligned. Since our simulation box is
finite, and cubic, a \ac{FCC} can only be accommodated perfectly with the
planes parallel to the cube faces, which makes this configuration the most
stable one. However, the chances of a \ac{FCC} grain to start to grow in a
fluid with the axis on the right orientation are minimal. Indeed, since the
$\q$ magnitude is rotationally invariant, we have no tool to force a
particular orientation, only the kind of crystal structure.

The straightforward solution would be to consider another bond-orientational
parameter but with cubic symmetry instead of rotationally symmetric one.

\section{The second bond-order parameter: avoiding rotational symmetry}\label{sec:hs-order-parameters}

As justified in the previous section, we need to define a second order
parameter with only cubic
symmetry.  Such a parameter was recently proposed~\cite{angioletti:10}:
\begin{equation}\label{eq:C}
C=\frac{2288}{79}\frac{\sum_{i=1}^{N} \sum_{j=1}^{N_b(i)}c_{\alpha}({\hat{\V{r}}_{ij}})}{\sum_{i=1 }^{N}N_b(i)}-\frac{64}{79} \,,
\end{equation}
where 
\begin{equation}\label{eq:calpha}
c_{\alpha}(\hat {\V{r}})=x^4y^4(1-z^4)+x^4z^4(1-y^4)+y^4z^4(1-x^4)\,.
\end{equation}
Again, $N_b(i)$ represents the number of neighbors of the $i$th particle, here
defined as the number of particles $j$ that fulfill
$|\V{r}_j-\V{r}_i|<1.5$.\footnote{See Footnote~\ref{foot:neighbors} in this
  Chapter.}  Within this definition of nearest neighbors, the expectation
value for $C$ in the different phases is the following: $0.0$ in the fluid,
$1.0$ in the ideal FCC crystal, perfectly aligned with the simulation box, and
$-0.26$ in the perfectly aligned ideal BCC. We include the calculation of $C$
in a perfect lattice in Appendix \ref{app:C}. The difference with the quoted
value in Ref.~\cite{angioletti:10} for the perfect BCC crystal is due to our
smaller threshold for neighboring particles.  For defective structures, we
must expect values for $\q$ and $|C|$, lower than the ones quoted here for
perfect lattices.

\begin{figure}
\centering
\includegraphics[angle=270,width=0.7\columnwidth,trim=0 50 0 50]{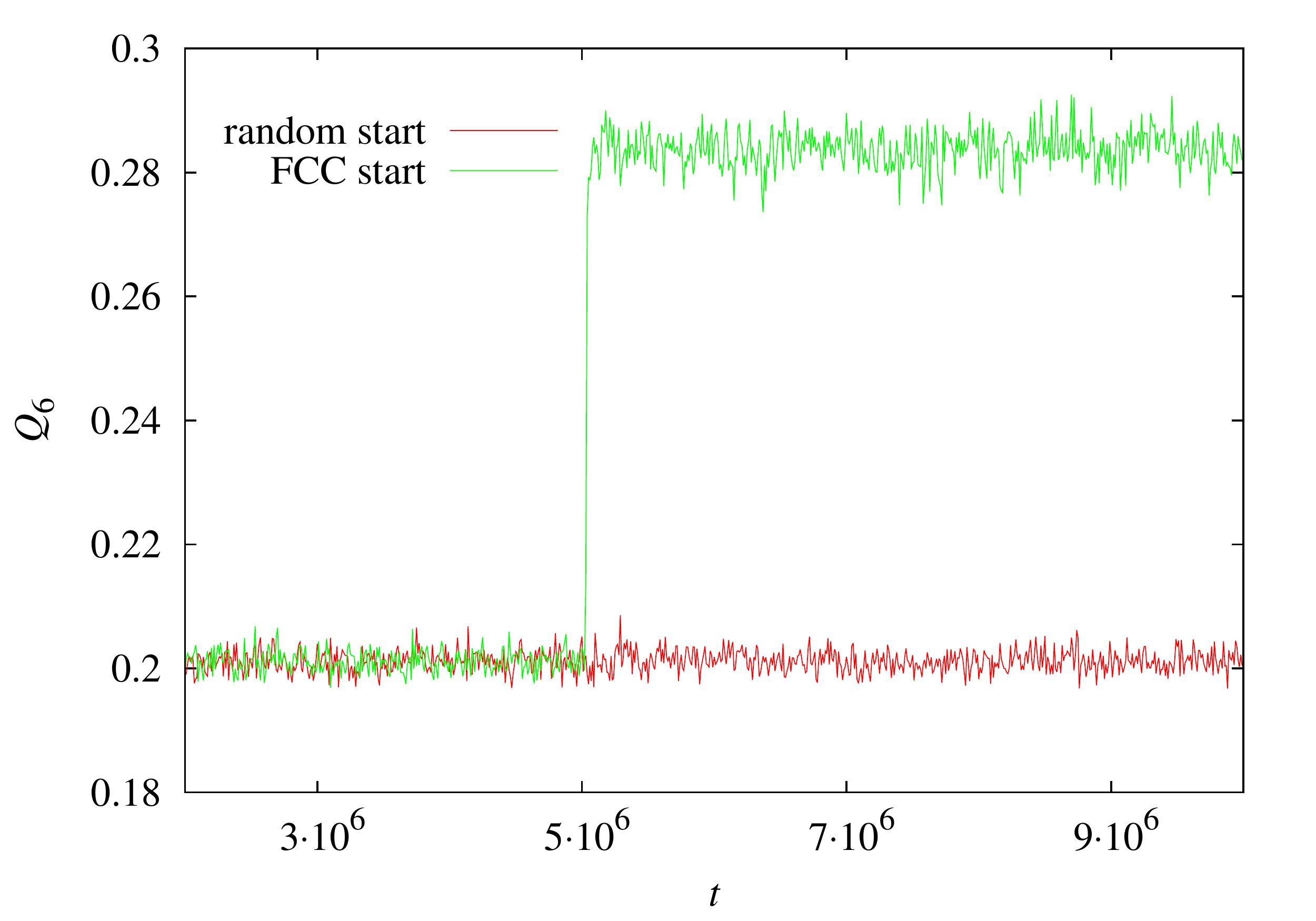}
\caption{History of $\q(\V{R},t)$ in a $\hc NpT$ MC simulation
  (for $\hc=0.3$, $N=256$, $p=11.224$). Time
  is measured in units on a elementary MC step: $N$ particles moves, followed
  by a volume update.}\label{fig:histrun_C}
\end{figure}

Following the previous discussion, we can repeat the previous study but this
time fixing this parameter instead of $\q$. As expected, the problem with
rotated \ac{FCC} lattices does not appear anymore. However, for intermediate
values of $\hc$ we find metastabilities in the simulation history, $C(\V{R})$
is not able to differentiate misaligned crystals and some mixtures of fluid
and crystal. We can distinguish these two phases by looking at $\q$, which is
rotationally invariant. We show this $\q(\V{R},t)$ history in Figure
\ref{fig:histrun_C}. Nevertheless, the region at which these flip-flops appear
in the $\hc NpT$ simulation is not the same one where the $\hq NpT$ fails to
thermalize. This last fact made us wonder what would happen if we fixed $\q$
and $C$ parameters at the same time. We shall see that with this idea we
fulfill our expectations: the runs starting from different configurations
converge quickly and we are finally able to avoid phase coexistence with its
corresponding exponential dynamic slowing down.

\section{Tethered formalism for a hard sphere system}\label{sec:hs-tethered}

As discussed in Section~\ref{sec:prelude}, the cure for the \ac{EDSD} seems to
be constraining simultaneously the values of two bond-order parameters, $\q$
and $C$. We use the tethered ensemble for two parameters, described in
Appendix~\ref{app:tethered2par}. With it, the \textit{tethered averages} of a
generic observable $O(\V{R},V,p)$ are defined as \bea\label{eq:meantethered}
\mean{O}_{\hq,\hc,p}\!=\!\frac{
  \int{d}V\mathrm{d}\V{R}\ O(\V{R},V,p)\ \omega_N(\V{R},V;\hq,\hc,p)}{ \int
  \mathrm{d}V\mathrm{d}\V{R}\ \omega_N(\V{R},V;\hq,\hc,p)}, \eea
with,\footnote{The reader may notice that the $\omega$ presented here is very
  different to the weight deduced in the original work in the Tethered
  method~\cite{fernandez:09}. The explanation regards on our definition for
  the tethered variables. Indeed, in the ensemble definition we add the demons
  linearly, $\hat T=T +\frac{1}{\alpha N}\sum_{i=1}^{\alpha N}\eta_i$, whereas
  in the original work, the demons were added quadratically, i.e.  $\hat T=T
  +\frac{1}{2}\sum_{i=1}^{\alpha N}\eta_i^2$, as an analogy to the momenta in
  the microcanonical ensemble~\cite{martin-mayor:07} (used in
  Chapter~\ref{chap:micro}).  These quadratic sums, introduce a Heaviside step
  function that forbids trial moves with $\hq > Q_6(\V{R})$ and $\hc >
  C(\V{R})$. Note that ascertaining thermalization is an issue in
  crystallization studies. It is very important to compare the outcome of
  simulations with widely differing starting configurations. In this respect,
  the constraints are a major problem, as they prevent us from using the ideal
  FCC crystal as starting configuration. This problem is directly erased if
  one adds the demons linearly as we do here.}  \bea\label{eq:weight}
\omega_N(\V{R},V;\hq,\hc,p)&=\frac{N\alpha}{2\pi }H(\V{R})\E^{-\beta pV}
\E^{-\frac{\alpha N}{2} \caja{\hq - Q_6(\V{R})}^2}\E^{-\frac{\alpha N}{2}
  \caja{\hat{C} - C(\V{R})}^2}.  \eea 

The Metropolis MC simulation of this weight requires
two types of moves: single particle displacements, as well as changes in the
volume of the simulation box. We shall use the short hand Elementary Monte
Carlo Step (EMCS) to the combination of $N$ consecutive single-particle
displacements attempts, followed by a change attempt in the simulation box
volume. For the particle displacements we pick at random a particle-index, say
$i$, and try $\V{r}_i\rightarrow \V{r}_i+\V{\delta}$ with $\V{\delta}$ chosen
with uniform probability within the sphere of radius $\Delta$. We tune
$\Delta$ to keep the acceptance above $30\%$.  We recast $\omega$ in
Eq.~\eqref{eq:weight} as the Boltzmann factor for HS at fixed pressure with a
{\em fictive} potential energy $k_\mathrm{B} T N\alpha\, [(\hq - Q_6(\V{R}))^2
  +(\hat{C} - C(\V{R}))^2]/2$. Since $Q_6(\V{R})$ and $C(\V{R})$ are built out
of sums of local terms, the number of operations needed to compute their
changes after a single-particle displacement does not grow with $N$.

As it is discussed in Appendix~\ref{app:tethered2par}, the Helmholtz effective
potential is here given by
\be\label{eq:probte} 
\E^{-N\Omega_N(\hq,\hc,p)}=\frac{\beta p}{N!\Lambda^{3N}}\frac{\alpha N}{2\pi}\int\RM{d}V
\int \,\mathrm{d}\V{R} \, \E^{-\beta p V}\,\E^{-\beta U(\V{R})}\,\E^{-\frac{\alpha N}{2} \caja{\lazo{\hq -
    \q(\V{R})}^2+\caja{\hc -
    C(\V{R})}^2}}\,.
\ee
Then, it is clear that the ensemble
equivalence with the $NpT$ (in particular Eq. \eqref{eq:YNPT}) is obtained by integrating over all the range of parameters,
\begin{eqnarray}\label{eq:ensemble-independence} Y_{NpT}=e^{-N
  g_N(p,T)}=\int\mathrm{d}\hq\ \mathrm{d}\hc\ e^{-N\varOmega_N(\hq,\hc,p)}\,.\end{eqnarray}

\begin{figure}
\centering
\includegraphics[angle=270,width=0.9\columnwidth,trim=0 0 0 0]{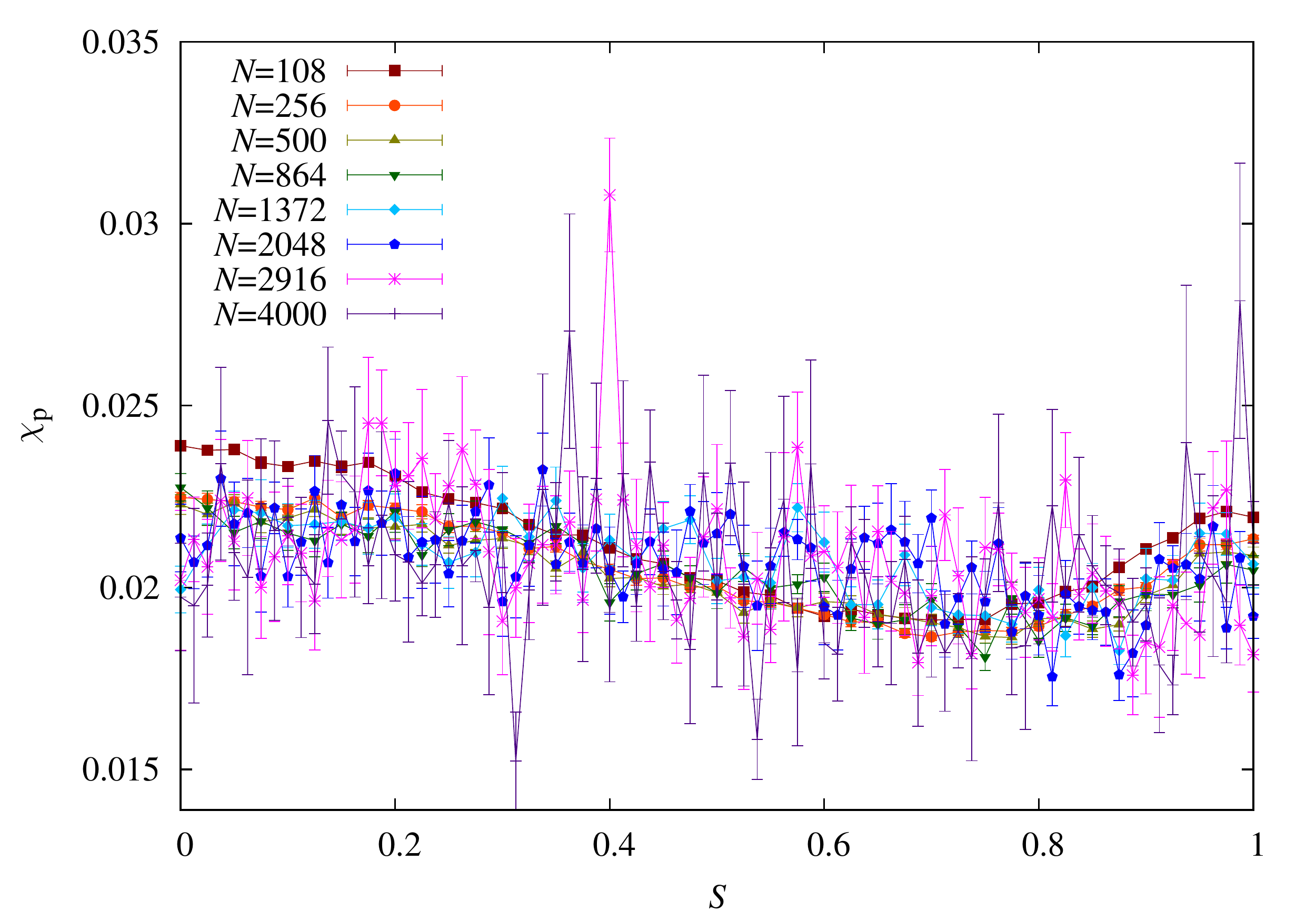} 
\caption{Compressibility of the system at different system sizes as a function
  of $S$. $S$($\in\caja{0,1}$) is the line parameter that covers all the
  $(\hq,\hc)$ points simulated in this work, it will be introduced in
  Section~\ref{sec:p_co-hs}.}
\label{fig:compressibility}
\end{figure}

We shall need to consider the dependency with $p$ in the mean
values~\eqref{eq:meantethered}. We could do it by running many simulations
at different pressures, or alternatively, by taking advantage of our lack of
metastabilities, and using the histogram reweighting
method~\cite{falcioni:82,ferrenberg:88}. Indeed, this method let us to
extrapolate mean values at $p+\delta p$ using simulation data obtained at $p$
using the following equality:
\begin{equation}\label{eq:reweighting-hs}
\langle O \rangle_{\hq,\hc,p+\delta p}=\frac{\langle
 O\ \mathrm{e}^{-\beta\delta p V}\rangle_{\hq,\hc,p}}
{\langle\mathrm{e}^{-\beta\delta p V}\rangle_{\hq,\hc,p}}.
\end{equation}
Although this equation is formally exact, our simulation data is finite and
the stochastic path visits mainly only the volume region relevant for pressure
$p$. The extrapolation will be  safe as long as the probability
distribution functions for the specific-volume, $v=V/N$, at both pressures
overlap (thus having sampled some of the relevant region for $p+\delta
p$). Then, we can compute the maximum safe extrapolation, $\delta p$, making
quantitative this idea.  Indeed, this condition is roughly equivalent to the
following statement. The displacement $\delta v=\mean{v}_{p+\delta p}-\mean{v}$
 should be smaller than the mean deviation of
$v$, i.e.  \be\delta v <
\sqrt{\mean{v^2}_{\hq,\hc,p}-\mean{v}_{\hq,\hc,p}^2}\,.\ee Besides, since the
distribution of volumes is unimodal, we can assume that the response is linear
\be \delta v= \frac{\partial v}{\partial p}\ \delta p= \chi_p\ \delta p,\ee
and the compressibility, $\chi_p$, can be obtained using the \textit{fluctuation-dissipation}
theorem \be \chi_p= \left.\frac{\partial \langle v\rangle}{\partial
  p}\right|_{(\hq,\hc,p)}=N\bigl[\langle v^2\rangle_{\hq,\hc,p} -\langle
  v\rangle^2_{\hq,\hc,p}\bigr]\,. \ee Then, we should restrict ourselves to
\begin{equation}
\delta p\lesssim \frac{1}{\sqrt{N\chi_p}}\,.
\end{equation}
Hence, it is crucial that the pdf for $v$ be unimodal (i.e. single-peaked),
and with an $N$-independent $\chi_p$, for all points considered here. In other
words, it is important that the integration path (see
Section~\ref{sec:p_co-hs}) to be free of metastabilities. This condition
holds very well as shown in Figure~\ref{fig:compressibility}, then, we can be
confident to use extrapolated data.
We include the actual $\delta p$ we used for our computations in
Table~\ref{tab:datos}.

We summarize the simulation technical details in Table \ref{tab:datos}. $N_S$
accounts for the number of $(\hq,\hc)$ points simulated ($S$ is the line
parameter of the integration curve that joins the fluid with the solid phases,
as we shall see in Section~\ref{sec:p_co-hs}).  We run simulations at a
central pressure $p_\text{sim}$, and we extrapolate data to $p\pm\delta p$. We
also include in the table the $(\q,C)$ mean values expected for the \ac{FCC}
and fluid phases. We will refer to these points as the {\em saddle points}
later on, and their technical computation will be explained in
Section~\ref{sec:hs-extrema}.
\begin{table*}
\centering
\begin{tabular*}{\textwidth}{@{\extracolsep{\fill}}ccccccccc}
\hline
$N$ & $N_S$ &$N_\text{sim}\times t_\text{max}$ & $p_\text{sim}$&$\delta p$&$\hq^\text{FCC}$&$\hq^\text{fluid}$ &$\hc^\text{FCC}$&$\hc^\text{fluid}$\\\hline
108  &42&$2\times 10^6$&10.920&0.40&0.3997998&0.0746256&0.6640012&$-0.0076329$\\
256  &42&$2\times 10^6$&11.224&0.40&0.399293&0.0486370&0.662729&$0.0007850$\\
500  &42&$2\times 10^6$&11.363&0.24&0.3993689&0.0349778&0.6627378&$-0.0000134$\\
864  &42&$2\times 10^6$&11.441&0.16&0.3995549&0.0268013&0.6629474&$-0.0009323$\\
1372  &42&$2\times 10^6$&11.487&0.16&0.3996055&0.0213669&0.6630886&$-0.0005104$\\
2048  &82&$2\times 10^6$&11.514&0.08&0.3997456&0.0175258&0.6633223&$-0.0002546$\\
2916  &82&$2\times 10^6$&11.529&0.08&0.3997110&0.0146926&0.6632560&$-0.0001866$\\\hline
4000  &82&$4\times 10^6$&11.540&0.08&0.3997886&0.0125658&0.6633856&$-0.0001238$\\\hline
\end{tabular*}
\caption{Technical details of the simulations. The length of each simulation,
  $t_\text{max}$, is measured in units of \ac{EMCS} ($N$ attempts of particle
  displacements followed by a change attempt of the simulation volume). $N_S$
  represent the number of $(\hq,\hc)$ points studied and $N_\text{sim}$ the
  amount of independent runs we studied at each $(\hq,\hc)$-point. In all
  cases but $N=4000$, $N_\text{sim}=2$, which corresponds a start from a
  random and from a perfect \ac{FCC} configuration. In the case of $N=4000$,
  three independent runs began from a random configuration.}
\label{tab:datos}
\end{table*}

\section{The coexistence pressure: computing differences in the effective
  potential}\label{sec:p_co-hs}

The tethered approach, at variance with the umbrella sampling, presents a
direct way to obtain the effective potential by means of a thermodynamic
integration using only mean values. In this section we explain step
by step how to use this approach to obtain the coexistence pressure $\p$. 

We start with the relationship between the effective potential,
$\Omega_N(\hq,\hc,p)$, and the Gibbs free-energy. The Eq.
\eqref{eq:ensemble-independence} can be simplified using a saddle-point
approximation,
\begin{eqnarray} \label{eq:geqomega} g_{N}(p,T)=\varOmega_N(\hq^*,\hc^*,p)+O(1/N)\,,
\end{eqnarray} 
where $(\hq^*,\hc^*,p)$ is the $p$-dependent absolute minimum of
$\varOmega_N(\hq,\hc,p)$, regarded as a function of $\hq$ and $\hc$.
Coordinates $(\hq^*(p),\hc^*(p))$ are then located in
Section~\ref{sec:hs-extrema} through $\grad \varOmega_N=0$.

Therefore, up to corrections vanishing as $1/N$, the chemical potential $\beta
g(p,T)$ is the absolute minimum of $\varOmega_N(\hq,\hc,p)$, see
\eqref{eq:geqomega}.  Yet, close to phase coexistence, $\varOmega_N$ has two
relevant minima (i.e. the fluid and the FCC crystal). Therefore, the
coexistence pressure $p_\mathrm{co}^{(N)}$ follows from
$\varOmega_N^\mathrm{fluid}=\varOmega_N^\mathrm{FCC}$ (i.e. the standard
condition of equal chemical potential).

Now, this gradient of the Helmholtz effective potential,
$\varOmega_N(\hq,\hc,p)$, is obtained by taking derivatives
in ~\eqref{eq:probte}. Using a Fluctuation-Dissipation formula, it leads to
\begin{eqnarray}\label{eq:hyomega} 
\grad{\varOmega_N(\hq,\hc,p)}=\paren{\frac{\partial \varOmega_N
    (\hq,\hc)}{\partial\hq},\frac{\partial \varOmega_N
    (\hq,\hc)}{\partial\hc}}\\\nonumber=\paren{\bigl\langle\alpha\paren{\hq-\q}\bigr\rangle_{\hq,\hc,p},\bigl\langle\alpha\paren{\hc-C}\bigr\rangle_{\hq,\hc,p}}. \end{eqnarray}
Furthermore, the differences in effective potential between to points,
$\varOmega_N({\hq}^{b},\hc^{b})-\varOmega_N(\hq^{a},\hc^{a})$ at fixed $p$ are
computed as the line integral of this $\grad \varOmega_N$ along any convenient
path joining $({\hq}^{a},\hc^{a},p)$ with $(\hq^{b},\hc^{b},p)$ in the
$(\hq,\hc)$ plane.
\begin{figure}
\centering
\includegraphics[angle=270,width=0.8\columnwidth,trim=0 0 0 0]{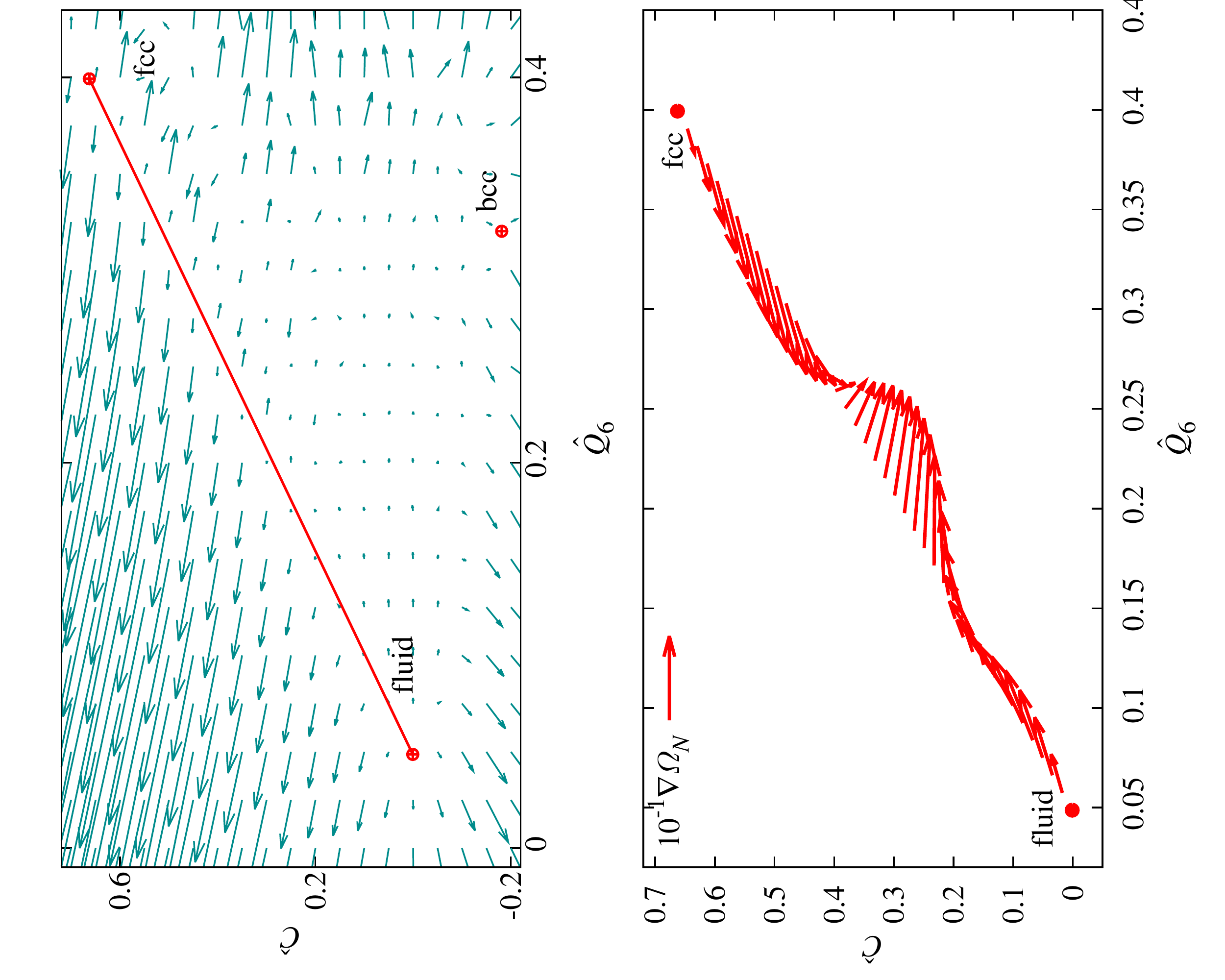}
\caption{ {\bf Top:} map of the gradient field $\grad\varOmega_{N=256}$,
  including the points corresponding to the fluid, FCC and BCC potential
  minima. For the sake of visibility we have divided the gradient by a factor
  $\alpha=200$.  {\bf Bottom:} $\grad\varOmega_{N=256}$, as computed from
  Eq.~\eqref{eq:hyomega} for $N=256$ hard spheres, along the straight path that
  joins the fluid and the FCC minima of the effective potential.  To improve
  visibility, we have divided $\grad\varOmega_{N=256}$ by a factor of 10 (mind
  the different normalization as compared with the top panel). The
  simulation pressure is the phase-coexistence one for both figures.  }
\label{fig:grid}
\end{figure}
Then, the coexistence pressure, $p_\mathrm{co}$, follows from the difference in
effective potential between the pressure-dependent coordinates of
the coexisting pure phases:
\begin{equation}\label{eq:delta-omega-hs}
\Delta\Omega_N(p)=\Omega_N\bigl(\hq^\mathrm{FCC}(p),\hc^\mathrm{FCC}(p),p\bigr)-
\Omega_N\bigl(\hq^\mathrm{fluid}(p),\hc^\mathrm{fluid}(p),p\bigr)\,.
\end{equation}
The scope of the game is finding the coexistence pressure,
$p_\mathrm{co}^{N}$, such that $\Delta\Omega_N(p_\mathrm{co}^{N})=0$.  Indeed,
the saddle-point condition \eqref{eq:geqomega}, tells us that, at
$p_\mathrm{co}^{N}$, the chemical potential for the two phases coincides.

Our framework is illustrated in Fig.~\ref{fig:grid}, where we show $\grad
\varOmega_N(\hq,\hc)$ at $p=\p^{(N)}$.  We identify two local minima where
$\grad \varOmega_N=0$ (the fluid, close to $(\hq,\hc)=(1/\sqrt{N},0)$, and the
FCC minimum where both parameters are positive, and are summarized in Table
\ref{tab:datos}).  Note their distance to other local minima of $\varOmega_N$,
such as the body centered cubic (BCC).

\begin{figure}
\centering
\includegraphics[angle=270,width=0.8\columnwidth,trim=0 0 0 0 ]{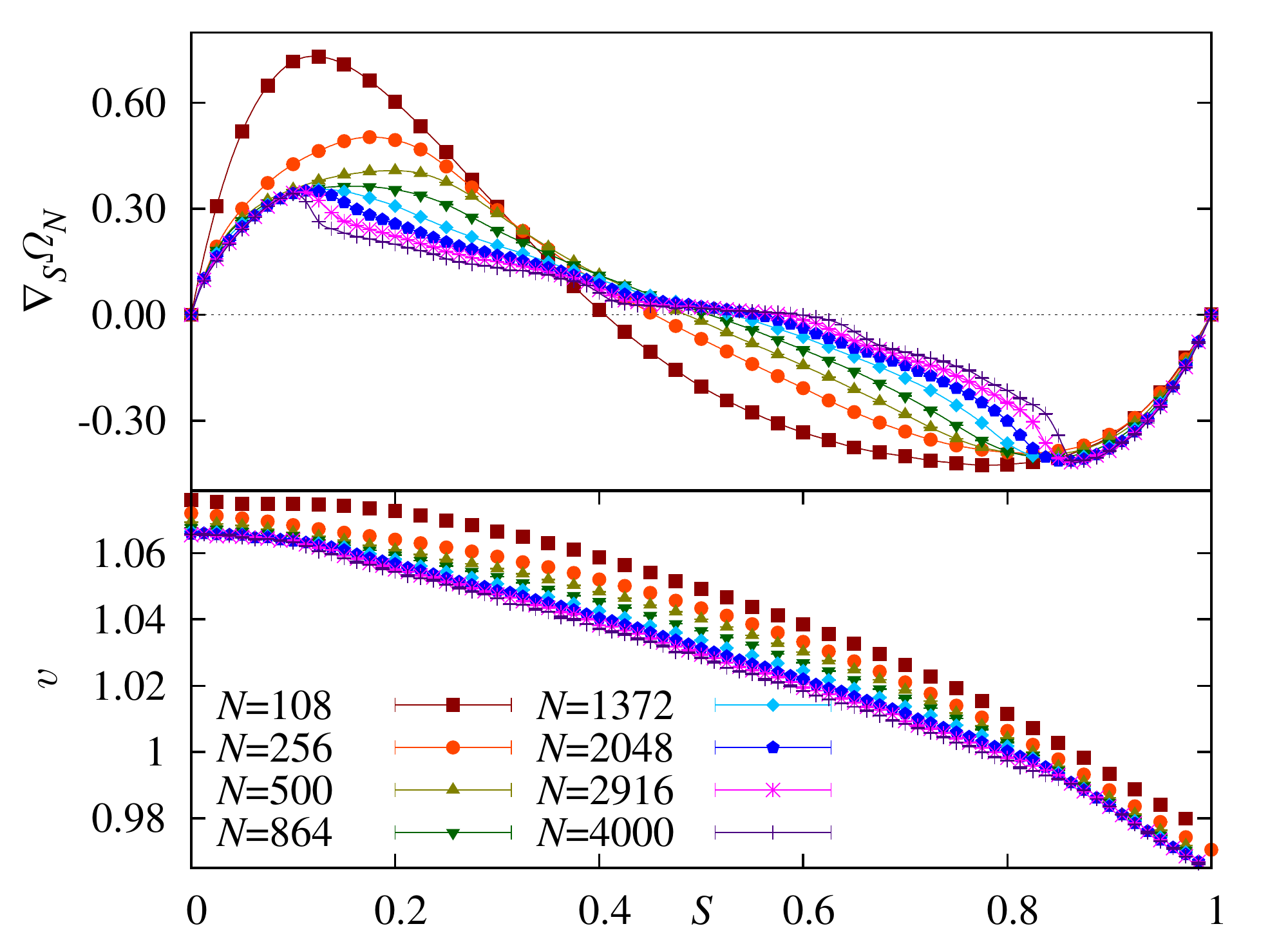} 
\caption{({\bf Top}) $\grad \varOmega_N$ projected over the
  fluid-FCC line, $\grad_S \varOmega_N$, vs. the line parameter $S$ ($S=0$:
  fluid, $S=1$: FCC), for all our system sizes at the simulation
  pressures. ({\bf Bottom}) Specific volume $v=V/N$ as a function of line
  parameter $S$. At large $N$, $v$ becomes a linear function, as expected for
a convex combination of pure phases~\cite{ruelle:69}.}
\label{fig:h}
\end{figure}

Our main goal is to compute
$\Delta\varOmega(p)\!=\!\varOmega^\mathrm{FCC}-\varOmega^\mathrm{fluid}$,
choosing the straight segment in Fig.~\ref{fig:grid} as integration
path. The path is parameterized by our \emph{reaction coordinate}, $S$
($S\!=\!0$: fluid, $S\!=\!1$: FCC). Actually, due to the {\em
  additivity} of $\q$ and $C$,\footnote{A magnitude $A$
  is {\em additive} if $N A$ is extensive: gluing together systems 1,2
  (with $N^{(i)}$ particles and $A=A^{(i)}$, $i=1,2$), results in a
  total system with $N=N^{(1)}+N^{(2)}$ particles and $NA=N^{(1)}
  A^{(1)} +N^{(2)}A^{(2)}$.  $C$ is additive to a great accuracy for
  coexisting fluid and FCC phases, because the average number of
  neighbors $N_b$ is very similar in both phases ($5\%$ difference,
  with negligible effects on additivity in our $N$ range, as compared
  with surface effects $\sim 1/N^{1/3}$). $\q$ is additive only if one
  of the subsystems, say $i=1$, is a liquid so that $\q^{(1)}\sim
  1/\sqrt{N^{(1)}}$ ($\q$ is a pseudo-order parameter, i.e. a strictly
  positive quantity which is of order $1/\sqrt{N}$ in a disordered
  phase). For studies of interfaces on larger systems, it would be
  advisable to choose exactly additive order parameters.} choosing this segment is a must if we
are to compute the interfacial free energy.  Indeed,
physical fluid-solid coexistence is a convex combination of the two
pure phases~\cite{ruelle:69}, which provides a physical interpretation
for $S$ as the fraction of particles in the coexisting solid phase: in
the large $N$ limit, $v$, $C$ and $\q$ vary linearly with $S$ (see
Fig.~\ref{fig:h}---bottom).

Our simulation set up is as follows. We start by locating $(\hq,\hc)$ for the
FCC and liquid minima at $p\approx \p^{(N)}$. The first guess is obtained from
$NpT$ simulations with crystalline/disordered starting configurations. We
later refine by solving for $\grad\varOmega_N=0$ as we will discuss it in
detail in Section \ref{sec:hs-extrema}.

Now, at variance with umbrella sampling, $\Delta\varOmega_N$ follows from the
integral \be\label{eq:deltaomega} \Delta\varOmega_N=\int_{\cal
  C}\grad\varOmega_N\cdot\D{\V{l}}=\int_0^1 \grad_S\varOmega_N\,\D{S}, \ee
with $\grad_S\varOmega_N$, the projection of $\grad \varOmega_N$ along the
straight-line, Fig.~\ref{fig:h}---top. In addition, we
use~\eqref{eq:reweighting-hs} to extrapolate $\grad\varOmega_N$ to different
pressures, which allow us to obtain  $\Delta\varOmega(p)$ as a function of pressure. Then, it is easy to
locate $\p^{(N)}$, Fig.~\ref{fig:omega}. Statistical errors are estimated
using standard Jack-Knife blocks~\cite{amit:05}.

In order to compute $p_\mathrm{co}^N$, we may initially neglect the pressure
dependence of the end points for the integration path in Fig.~\ref{fig:grid}.
One may easily correct for end-points displacements, as explained in
Sect.~\ref{sec:hs-extrema}, which induces a correction in $p_\mathrm{co}^N$
negligible with respect to our statistical errors.

The problem of thermalization is fully tackled in
Appendix~\ref{app:thermalization}, nonetheless we give here a few strokes of the
brush about how we can be confident of it.  We introduce a uniform $S$ grid on
the liquid-FCC line and perform {\em independent} simulations at fixed
$(\hq,\hc,p)$ (see Table \ref{tab:pN} for simulation details). As a test for
equilibration, achieved for all $N$ but $N=4000$, every run was performed
twice (starting from an ideal gas and from an ideal FCC crystal). Furthermore,
our runs for $N\leq 2916$ are, at least, $100 \tau$ long ($\tau$ is the
integrated autocorrelation time~\cite{sokal:97}, computed for $\q$ and
$v$). For $N=2916$, but only at $S=0.4$, we find metastability with a
helicoidal configuration (however, its contribution to final quantities is
smaller than statistical errors).  Metastabilities arise often for $N=4000$,
at intermediate $S$ (yet, a careful selection of starting configurations
yields a $\grad\varOmega_N$ with smooth $S$ dependency).

\subsection{Results}\label{sec:hs-results}

\begin{figure}
\centering
\includegraphics[angle=270,width=0.9\columnwidth,trim= 0 0 0 0]{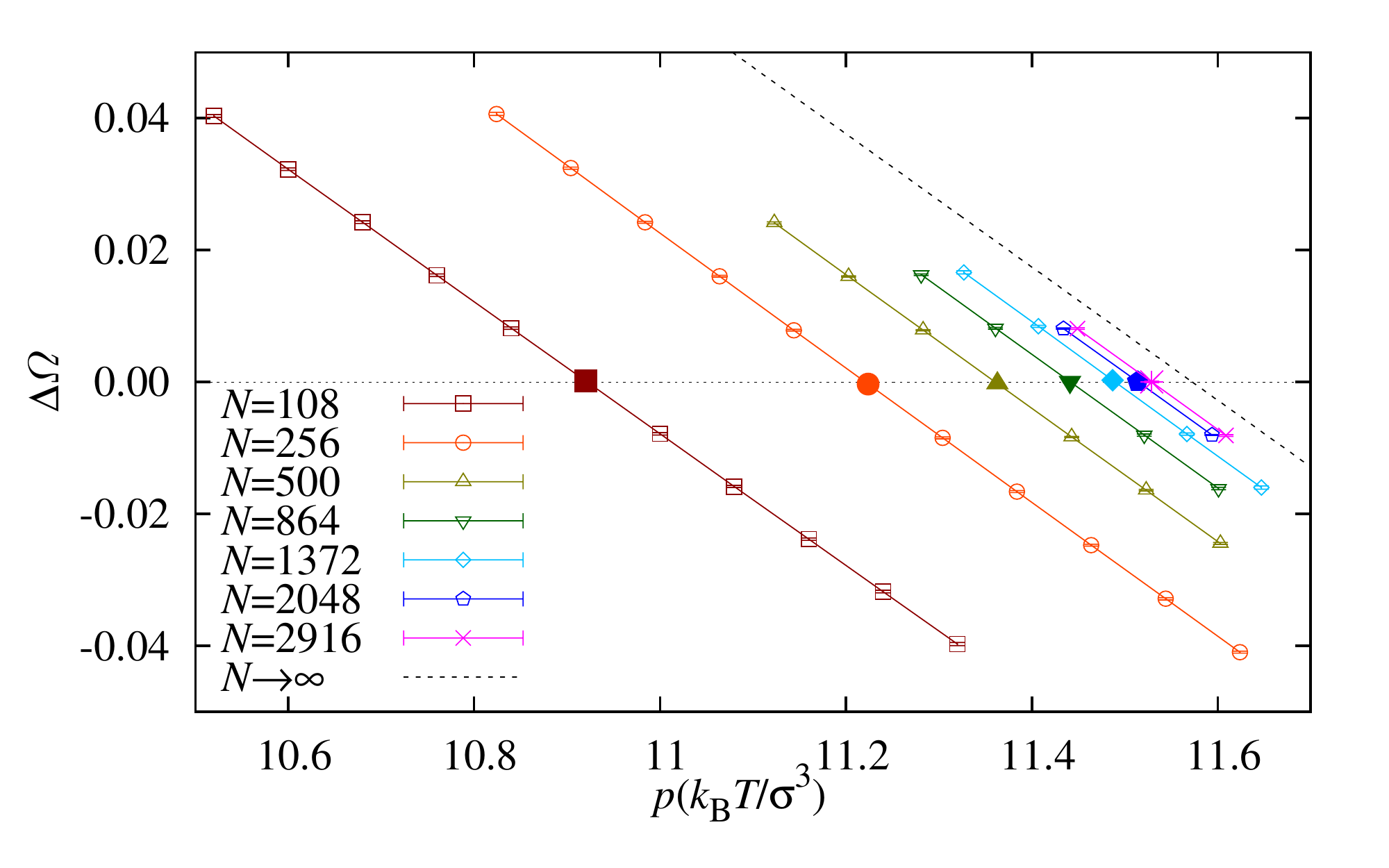} 
\caption{Effective-potential difference
  $\Delta\varOmega(p)\!=\!\varOmega^\mathrm{FCC}-\varOmega^\mathrm{fluid}$,
  as a function of pressure. At $p_\mathrm{co}^N$, $\Delta
  \varOmega_N=0$. The large $N$ limit stems from Eq. \eqref{eq:omegap}. The simulated
  pressures (see Table~\ref{tab:datos}) correspond to the larger, filled
  symbols.}
\label{fig:omega}
\end{figure}

\begin{figure}
\centering
\includegraphics[angle=270,width=0.8\columnwidth,trim=0 0 20 40]{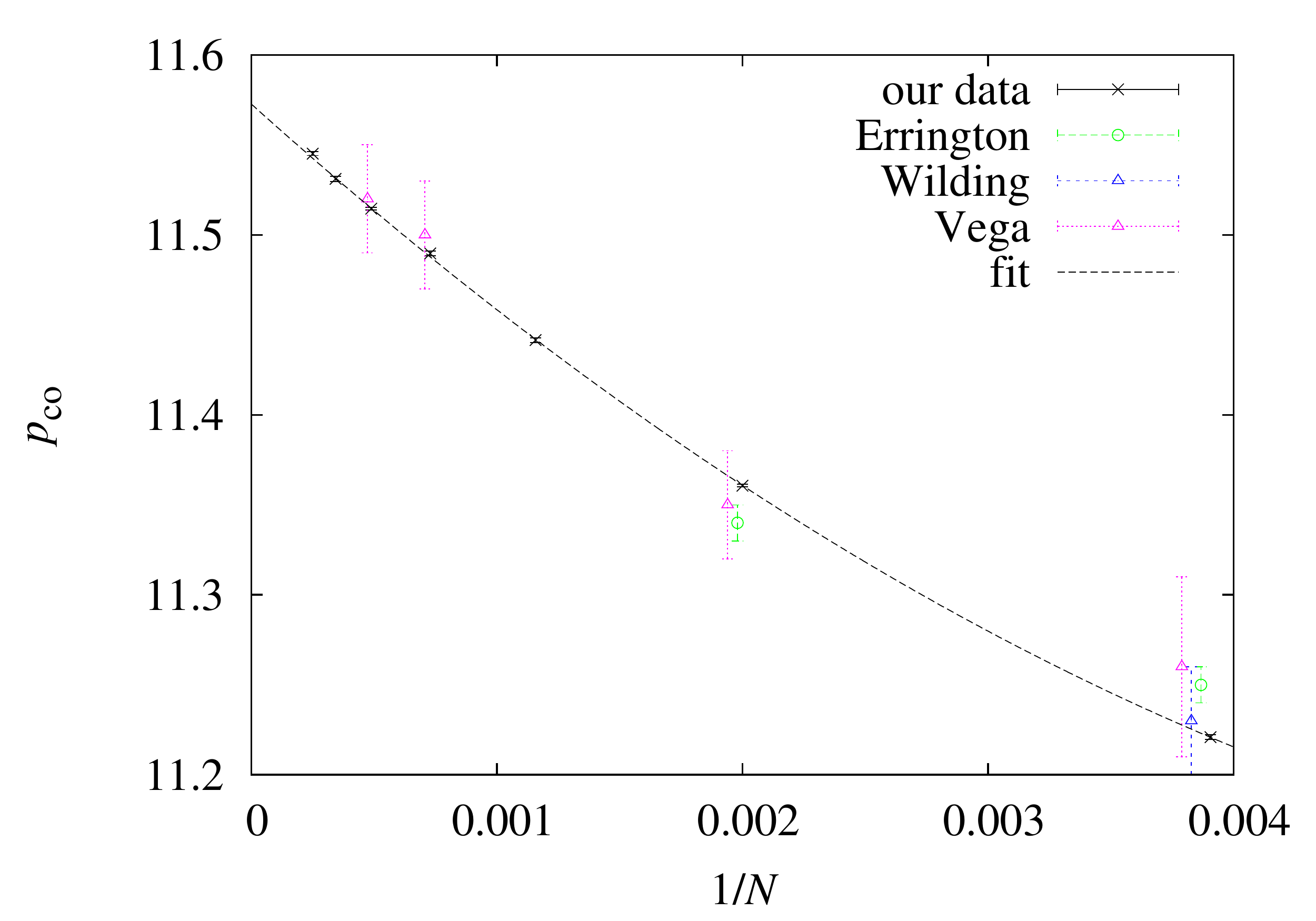} 
\caption{Finite size estimations of $\p$ plotted vs. $1/N$ obtained in this
  work, together with previous determinations using phase-switch \ac{MC},
  Errington~\cite{errington:04} and Wilding~\cite{wilding:00}, and Einstein
  molecule method, Vega~\cite{vega:07}. To improve visibility, estimations by
  other authors are slightly displaced to the left. We plot as well the
  quadratic fit of our data.}
\label{fig:Pc}
\end{figure}

By now, we have all the tools to compute $\Delta\Omega_N(p)$. These potential
differences as a function of $p$ are shown in Figure~\ref{fig:omega}. Once
this effective potential is known, $p_\mathrm{co}^{(N)}$ is obtained as the
pressure at which $\Delta\Omega_N\big(\p^{(N)}\big)=0$. We show in
Table~\ref{tab:pN} and Figure~\ref{fig:Pc} the results for each system size.
As usually, we are interested in the large $N$ limit
$p_\mathrm{co}^{\infty}$. Figure~\ref{fig:Pc} suggests that we need a second
order polynomial to fit the data $p_\mathrm{co}^{(N)}$. We try a fit
$p_\mathrm{co}^{(N)}= p_\mathrm{co}^{\infty} +a_1/N + a_2/N^2$~\cite{borgs:92}
for $256\! \leq\! N\!  \leq\! 2916$ (fitting data and curve are also in
Table~\ref{tab:pN} and Figure~\ref{fig:Pc}), obtaining
\begin{center}
$p_\mathrm{co}^{\infty}=11.5727(10)$\,.
\end{center}
For this extrapolation, we left out the $p_\mathrm{co}^{N=4000}$ value because
of the doubtful thermalization. Nevertheless, we
would like to point out that the $p_\mathrm{co}^{N=4000}$ is compatible with
the fitted curve.

\begin{table*}
\centering
\begin{tabular*}{\textwidth}{@{\extracolsep{\fill}}ccccccccc}
\cline{1-9} 
&\multicolumn{1}{c}{This work} && \multicolumn{1}{c}{\cite{wilding:00}}&\multicolumn{1}{c}{\cite{errington:04}}&&\cite{zykova-timan:10}&&\multicolumn{1}{c}{\cite{vega:07}}\\
 $N$ &$\p$&&\multicolumn{2}{c}{Phase switch}&
&Direct coexistence&&\multicolumn{1}{c}{E. M.}\\
\cline{1-1}\cline{2-2}\cline{4-6}\cline{7-7}\cline{9-9}
108 &10.9216(18)& &10.94(4)&11.00(6)& & & &11.02(5)     \\ 
256 &11.2209(13)& &11.23(4)&11.25(1)&&&&11.26(5) \\ 
500 &11.3607(8) & &        &11.34(1)&&&&11.35(3) \\ 
864 &11.4416(13)& &        &        &         &&&     \\
1372&11.4897(13)& &        &        & &&&11.50(3)\\ 
2048&11.5146(7) & &        &        & &&&11.52(3)\\ 
2916&11.5311(15)& &        &        &         &&&     \\
\cline{1-2}
4000&11.5452(11)\\
\cline{1-1}\cline{2-2}\cline{4-6}\cline{7-7}\cline{9-9}
$\infty$&11.5727(10)&&11.49(9)&11.43(2)&&11.576(6)&&11.54(4)\\ 
\multicolumn{1}{c}{$\chi^2/$dof}&$2.61/3$\\
\cline{1-9}
\end{tabular*}
\caption{For each $N$, we report the phase-coexistence pressure
  $p_\mathrm{co}^N$  in units of
  $k_\mathrm{B}T/\sigma^3$ (which is compared with work by other authors
  using different methods: phase switch Monte Carlo, the
  non-equilibrium direct coexistence method, and the Einstein Molecule
  approach).}
\label{tab:pN}
\end{table*} 

We compare our results in Table~\ref{tab:pN} and Figure~\ref{fig:Pc} with
previous estimates. They are more precise (and compatible with) independent
determinations by other authors, both at finite and in the large $N$ limit.
The best previous equilibrium estimate seems to be the rather crude
$p_\mathrm{co}^\infty=11.50(9)$~\cite{wilding:00}, obtained using phase-switch
Monte Carlo. In fact, the only previous method accurate enough to provide a
meaningful comparison is the non-equilibrium direct-coexistence:
$p_\mathrm{co}^{\infty}=11.576(6)$~\cite{zykova-timan:10}. Note, however, that
in order to achieve such a small error (but still six times larger than the
error in our tethered computation), systems with up to $N=1.6\times 10^5$
particles were simulated~\cite{zykova-timan:10}.

In addition, we can compute the specific volumes for the fluid and the
\ac{FCC} phase averaging the volume data at $S=0$ and $1$ respectively (the
saddle points quoted in Table~\ref{tab:datos}). We
show the results obtained in Table~\ref{tab:datosv} together with an
extrapolation in $1/N$.
\begin{table}
\centering
\begin{tabular*}{\textwidth}{@{\extracolsep{\fill}}ccc}
\hline
$N$&$\mean{v}^\text{FCC}$&$\mean{v}^\text{fluid}$\\\hline
108&0.97580(7)&1.07611(8) \\
256&0.97049(6)&1.07202(7) \\
500&0.96796(10)& 1.06932(7)\\
864&0.96796(10)& 1.06932(7)\\
1372&0.96549(14)&1.06659(13)\\
2048&0.96500(14)&1.06577(15)\\
2916&0.96468(14)&1.06545(19)\\\hline
4000&0.96461(13)&1.06556(15)\\\hline
$\infty$&0.96405(3)&1.06448(10)\\
$\chi^2/\text{dof}$&0.32/3&0.61/2\\
$N_\text{min}$&256&500\\
$N_\text{max}$&2916&2916\\\hline
\end{tabular*}
\caption{Specific-volumes of the FCC crystal and the fluid phase as function
  of the system size. We include the extrapolation to $N=\infty$ together with
  the details of the linear fit to $v^\infty+a_1/N$.}
\label{tab:datosv}
\end{table} 

Finally, the reader might wonder about the linear relation of
$\Delta\varOmega$ vs. $p$ in Figure~\ref{fig:omega}. It follows from
Eq.~\eqref{eq:geqomega}. The potential at each extrema is
$\varOmega^*(p)=g(p)+O(1/N)$, where $g(p)$ is the Gibbs free-energy
density. Then, its derivate is \be\left.\frac{\partial \varOmega}{\partial
  p}\right|_*=\frac{\partial g}{\partial p}+O(1/N)=v_*+O(1/N),\ee where $v_*$
is the intrinsic volume at the extremal point.  Thus, the effective potential
at $p$ close to $p^*$, is:
\be\varOmega(p)=\varOmega^*+v^*(p-p^*)+O\paren{(p-p^*)^2}+O(1/N),\ee and since
the effective potential at the extremal points must be equal in the two phases
at the coexistence pressure, $\p$, the difference in effective potential
between the fluid and the \ac{FCC} phase at $p$, will be determined by
\be\label{eq:omegap}\varOmega_\mathrm{FCC}(p)-\varOmega_\mathrm{f}(p)=(v_\mathrm{FCC}-v_\mathrm{fluid})(p-\p)+O(1/N),\ee
and thus, presents a linear dependency in $p$. We include in
Fig. \ref{fig:omega} the prediction for the thermodynamic limit that follows
from this last relation using the large-$N$ extrapolations for the specific
volume displayed in Table~\ref{tab:datosv}.

\subsection{Calculation of the extremal points and corrections}\label{sec:hs-extrema}
We had postponed the discussion about the computation of the extremal points
shown in Table~\ref{tab:datos}, as well as the issue of considering the same
integration curve for all values of $p$. We devote this section to both
problems.

We need to locate the two extremal points in the straight path in
Fig.~\ref{fig:grid}, which correspond to the fluid or to the FCC crystal. The
two points are local minima of $\Omega_N$, regarded as a function of $\hq$ and
$\hc$ but at fixed pressure.  Our procedure has been as follows.

We first obtain a crude estimate from standard simulations in the $NpT$
ensemble (without any constrain in the crystal parameters). Note that the
autocorrelation time for such simulations is unknown, but larger than any
simulation performed to date. Hence, these standard simulations get stuck at
the local minimum of $\Omega_N$ which is most similar to their starting
configuration. Starting the simulation either from an ideal gas, or from a
perfect FCC crystal, we approach the pure-phases we are interested in. The
Monte Carlo average of $\q(\V{R})$ and $C(\V{R})$ provides our first guess.

To refine the search of either of the two local minima $(\hq^*,\hc^*)$, we
note that, up to terms of third order in $\hq -\hq^*$ or $\hc -\hc^*$,
\begin{equation}\label{eq:gaussian-aprox-hs}
\Omega_N(\hq,\hc)=\Omega_N^*+
\frac{A_{QQ}}{2}(\hq-\hq^*)^2+A_{QC}(\hq-\hq^*)(\hc-\hc^*)+\frac{A_{CC}}{2}(\hc-\hc^*)^2\,.
\end{equation}
The shorthand $\Omega_N^*$ stands for $\Omega_N(\hq^*,\hc^*)$.  Incidentally,
Eq.~\eqref{eq:gaussian-aprox-hs} tells us that the computation in
Sect.~\ref{sec:p_co-hs} is intrinsically stable. An error of order $\epsilon$
in the location of $(\hq^*,\hc^*)$ will result in an error of order
$\epsilon^2$ in the coexistence pressure.

Yet, the tethered computation does not give us access to $\Omega_N$, but to its
gradient:
\begin{equation}\label{eq:gaussian-gradient-hs}
\grad \Omega_N(\hq,\hc)= \bigl(A_{QQ} (\hq-\hq^*)+A_{QC}(\hc-\hc^*), A_{CC}(\hc-\hc^*)+A_{QC}(\hq-\hq^*)\bigr)\,.
\end{equation}
Eq.~\eqref{eq:gaussian-gradient-hs} holds up to corrections quadratic in $\hq
-\hq^*$ or $\hc -\hc^*$. We thus compute the expectation value of the field
$\grad\varOmega_N$, in a grid of nine points $(\hq,\hc)$ that surround our
first guess for $(\hq^*,\hc^*)$, and fit the results to
Eq.~\eqref{eq:gaussian-gradient-hs}. We iterate this procedure until an
accuracy $\sim 10^{-6}$ in both coordinates $(\hq^*,\hc^*)$ is
reached.

Actually, Eq.~\eqref{eq:reweighting-hs}, shows how one extrapolates the
expectation values for the gradient field from the simulated pressure, $p$ to
a nearby $p+\delta p$. The corresponding fit to
Eq.~\eqref{eq:gaussian-gradient-hs} provides the new coordinates
$\bigl(\hq^*(p+\delta p),\hc^*(p+\delta p)\bigr)$.

At this point, one could worry because the integration path in
Fig.~\ref{fig:grid} is no longer appropriate at pressure $p+\delta p$. In
fact, the extremal points in the integration path are pressure-dependent.
However, some reflection shows that this is not a real problem. In fact,
\begin{equation}\label{eq:deltaomegapath}
\Delta\Omega_N(p+\delta p)= \Delta^\mathrm{FCC}(p,p+\delta p)+
\Delta^\mathrm{path}(p,p+\delta p)-\Delta^\mathrm{fluid}(p,p+\delta p)\,.
\end{equation}
The different pieces in Eq.~\eqref{eq:deltaomegapath}
\begin{eqnarray}
\Delta^\mathrm{FCC}(p,p+\delta p)&=&
\Omega_N\bigl(\hq^\mathrm{FCC}(p+\delta p),\hc^\mathrm{FCC}(p+\delta
p);\,p+\delta p\bigr)-\\\nonumber
&-&\Omega_N\bigl(\hq^\mathrm{FCC}(p),\hc^\mathrm{FCC}(p);\,p+\delta
p\bigr)\,,
\end{eqnarray}
the correction due to the shift of order $\delta p$ in the coordinates of the
FCC minimum,
\begin{eqnarray}
\Delta^\mathrm{path}(p,p+\delta p)&=&
\Omega_N\bigl(\hq^\mathrm{FCC}(p),\hc^\mathrm{FCC}(p);\,p+\delta p\bigr)-\\\nonumber
&-&\Omega_N\bigl(\hq^\mathrm{fluid}(p),\hc^\mathrm{fluid}(p);\,p+\delta
p\bigr)\,,
\end{eqnarray}
the line-integral sketched in Fig.~\ref{fig:grid} as computed at pressure
$p+\delta p$, and
\begin{eqnarray}
\Delta^\mathrm{fluid}(p,p+\delta p)&=&
\Omega_N\bigl(\hq^\mathrm{fluid}(p+\delta p),\hc^\mathrm{fluid}(p+\delta
p);\,p+\delta p\bigr)-\\\nonumber
&-&\Omega_N\bigl(\hq^\mathrm{fluid}(p),\hc^\mathrm{fluid}(p);\,p+\delta
p\bigr)\,,
\end{eqnarray}
the correction due to the shift in the coordinates of the fluid minimum.

Now, one expects that the pressure-induced changes in the minima coordinates
as well as on the coefficients $A_{QQ}$, $A_{Q,C}$ and $A_{CC}$ will of order
$\delta p$. Hence, Eq.~\ref{eq:gaussian-aprox-hs} implies that both
$\Delta^\mathrm{FCC}(p,p+\delta p)$ and $\Delta^\mathrm{fluid}(p,p+\delta p)$
are of order $(\delta p)^2$. This is the rationale behind the simplifying
assumption made in Sect.~\ref{sec:p_co-hs}.

At any rate, $\Delta^\mathrm{FCC}(p,p+\delta p)$ and
$\Delta^\mathrm{fluid}(p,p+\delta p)$ can be numerically computed from
Eq.~\ref{eq:gaussian-aprox-hs}. For all values of $N$ simulated, their
combined effect on the determination of the coexistence pressure turns out to
be smaller than 1\% of the statistical error bars as shown in
Table~\ref{tab:correcciones}. Then, at least in our systems, this kind of
refinement seems not to be necessary (partly because we did short simulations
that yielded working estimates of $\p^{(N)}$).

However, we cannot forget that before
running simulations we had an idea of the value of the coexistence pressure
for each system size, and we did not need to displace too much in $p$, but
this is not the normal case in the most interesting systems, and these
corrections might become important as long as one gets further away from the
simulation pressure.
\begin{table*}
\centering
\begin{tabular*}{\columnwidth}{@{\extracolsep{\fill}}ccccccccc}
\hline
$N$ & $\p$&$\p^\text{corrected}$&$(\p-\p^\text{corrected})/error$\\\hline\hline
108&10.9216(18)&10.9216(18)&-0.0046\\
256&11.2209(13)&11.2209(13)&-0.0106\\
500&11.3607(8)&11.3607(8)&-0.0132\\
864&11.4416(13)&11.4416(13)&-0.0018\\
1372&11.4897(13)&11.4897(13)&-0.0062\\
2048&11.5146(7)&11.5146(7)&-0.0020\\
2916&11.5311(10)&11.5311(10)&-0.0082\\\hline
\end{tabular*}
\caption{Study of the effect of the corrections in the determination of $\p^N$.}
\label{tab:correcciones}
\end{table*}

\section{Geometric transitions and the interfacial
  free-energy}\label{sec:hs-interfacial}

The interfacial free energy is the free-energy cost per unit area of a
liquid-to-crystal interface. Its computation has been rather difficult for
hard spheres. In fact, different authors finding mutually incompatible
results~\cite{davidchack:00,mu:05,davidchack:10}.

\begin{figure}
\centering
\includegraphics[angle=270,width=0.8\columnwidth,trim=0 0 0 0]{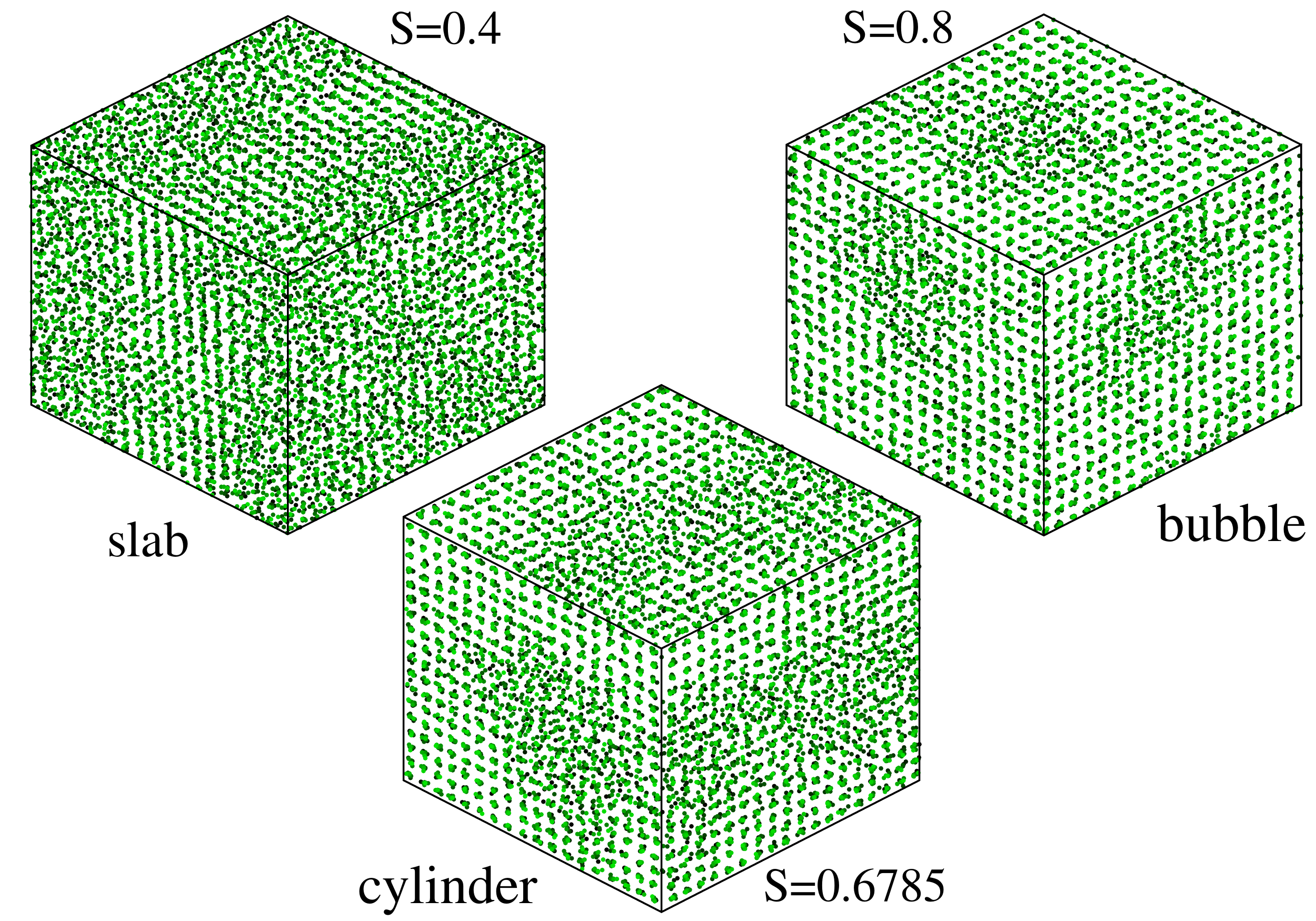}
\caption{Snapshots of mixed configurations for $N=2916$ particles
  found as the line parameter $S$ varies. We present projections in
  the three Cartesian directions. To improve visibility, the radii are
  a fraction of the real ones, and the darkness is an increasing function of
  the distance to the projection plane.}
\label{fig:configuraciones}
\end{figure}

As for the interfacial free energy, $\gamma$, he difficulties are due to the
need of considering inhomogeneous configurations.~\footnote{The tethering
  approach does not induce artificial interfaces. In fact, mathematically, the
  interfacial free-energy is defined though the ratio of two partition
  functions with different boundary conditions. But the tethered potential does
  not change the partition function [with any boundary conditions, see
    Eq.~\eqref{eq:ensemble-independence}].} In a system with periodic boundary
conditions, geometrical transitions arise when the line parameter $S$ varies
from the liquid to the solid. In fact, the system struggles to minimize the
surface energy while respecting the global constraints for $\q$ and
$C$. Depending on the fraction of crystal phase, which is fixed by $S$, the
minimizing geometry can be either a bubble, a cylinder or a slab of liquid in
a crystal matrix (or vice versa). An example of each type of configuration is
displayed in Fig. \ref{fig:configuraciones}. As $S$ varies, the minimizing
geometry changes at definite $S$ values. This phenomenon is named {\em
  geometric transition}, and has been previously studied in simpler models
(for instance, first-order transitions in lattice magnetic
systems~\cite{martin-mayor:07}, or fluid-gas
phase-coexistence~\cite{macdowell:06,binder:11}).  These transitions result in
the cusps and steps that appear for large $N$ in $\grad_S\varOmega_N$,
Fig.~\ref{fig:h}---top.

The physical situation is as follows. When we go from the liquid to the solid,
Fig.~\ref{fig:grid}, the homogeneous fluid becomes unstable at a value of the
linear coordinate $S\propto N^{-1/(D+1)}$, which means that a macroscopic
droplet of crystal forms. This has been established for all types of
first-order phase
transitions~\cite{biskup:02,binder:03,macdowell:04,nussbaumer:06}, and
explicitly verified for crystallization here. As $S$ grows the mass of the
crystal droplet increases, which costs surface energy. At a certain point, the
periodic boundary conditions allow reducing the surface energy by turning the
crystal droplet onto a crystal cylinder. At still larger $S$, the cylinder
becomes a slab. Of course another three analogous geometrical transitions
arise when $S$ keeps increasing as we approach the FCC minimum. All six
geometric transitions appeared in our simulations of large enough hard-spheres
systems. We are interested in identifying systems large enough to form a slab
of crystal surrounded by fluid to be able to compute the interfacial free-energy.

\begin{figure}
\centering
\includegraphics[angle=270,width=0.75\columnwidth,trim=0 0 0 0]{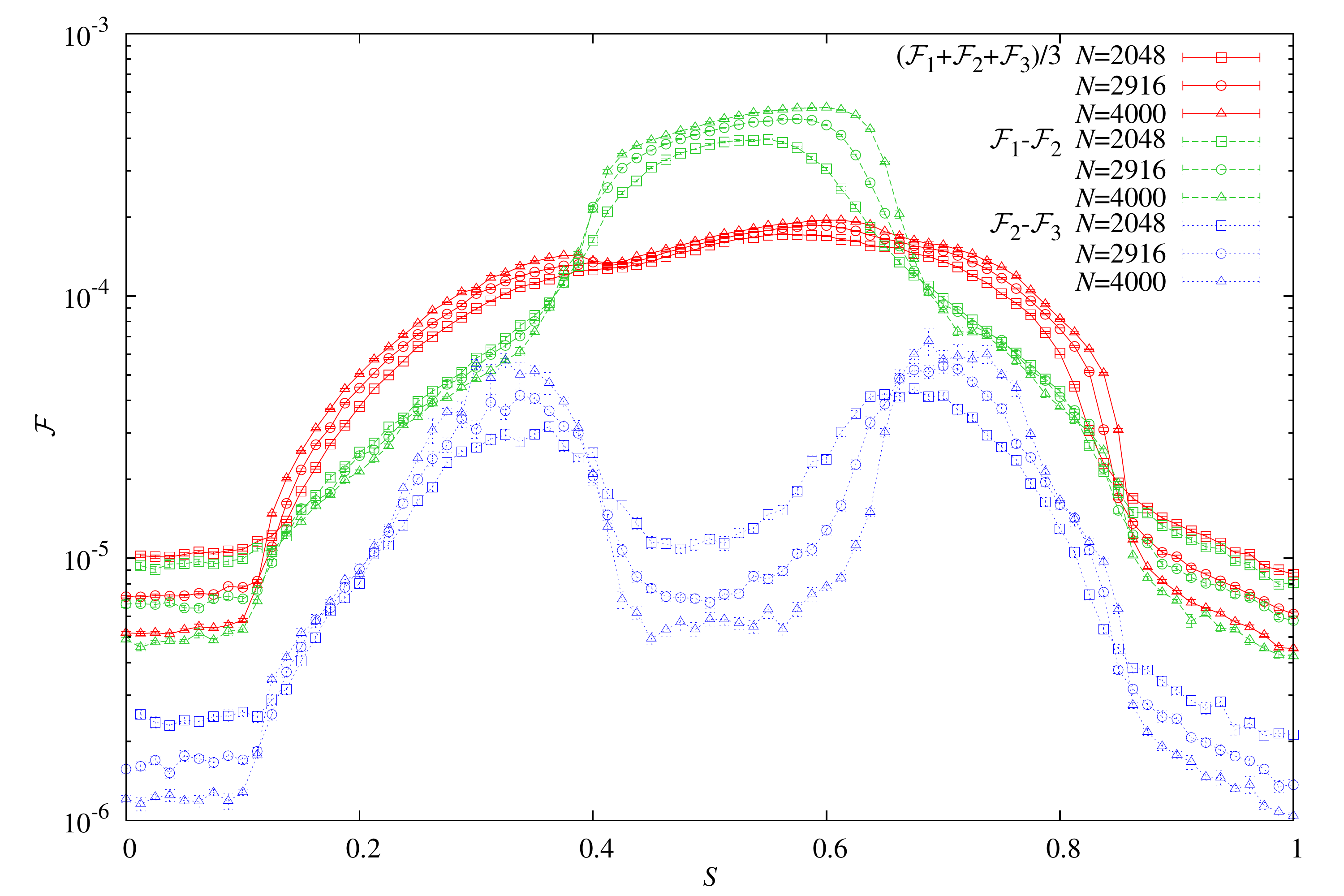}
\caption{For systems of $N$ hard spheres at their phase-coexistence pressure,
  we show, as a function of the line parameter $S$, different linear
  combinations of the particle-density fluctuations, Eq.~\eqref{eq:hs-F-def},
  computed for the minimal wave vectors allowed by periodic boundary conditions
  and ordered in such a way that ${\cal F}_1>{\cal F}_2>{\cal F}_3$. For the
  phase-separated states all three ${\cal F}$ are of order $1$ (order $1/N$
  for homogeneous systems). The slab phase is the only one with ${\cal
    F}_1-{\cal F}_2$ of order one. The cylinder phase is identified by ${\cal
    F}_2-{\cal F}_3$ of order one.}
\label{fig:F}
\end{figure}

In order to follow these geometrical transitions, it is useful to
look at the inhomogeneity of the system. As shown in Figure
\ref{fig:configuraciones}, we deal with phase separation between fluid
and FCC crystal, then it is interesting to 
consider the particle-density fluctuations (recall Section~\ref{eq:pd-fieldmicro}) quantified through
\begin{equation}\label{eq:hs-F-def}
{\cal F}(\V{q})=\frac{1}{N^2} \biggl|\sum_{i=1}^N \E^{\mathrm{i} \V{q}\cdot\V{r}_i}\biggr|^2\,,
\end{equation}
As we are interested in the largest wavelength, we consider the smallest
$\V{q}$ allowed by periodic boundary conditions, $\Vert \V{q}\Vert= 2\pi/L$,
where $L$ is the linear size of the simulation box. There are three such
minimal wave vectors in a cubic box, $(2\pi/L,0,0)$, $(0,2\pi/L,0)$ and $(0,0,2\pi/L)$. 
Given a particle configuration, we define ${\cal
  F}_1$ as the maximum over the three directions, ${\cal F}_3$ as the minimum,
and ${\cal F}_2$ as the intermediate one. As the droplet, cylinder and slab
geometries have different symmetries the natural order parameters are
\begin{itemize}
\item Whenever the system is phase separated,
  $(\mathcal{F}_1+\mathcal{F}_2+\mathcal{F}_3)/3$ is of order 1, (order $1/N$
  otherwise).
\item For a cylinder, two of the ${\cal F}$'s are of order 1, while the ${\cal
  F}$ along the cylinder axis is small. Hence, $\mathcal{F}_2-\mathcal{F}_3$
  is of order 1 in the cylinder phase, but it vanishes (for large $N$) both in
  the droplet and the slab phase.
\item For a slab the only ${\cal F}$ of order 1 is that transverse to
  it. Hence $\mathcal{F}_1-\mathcal{F}_2$ is of order 1 for a slab, but not
  for the cylinder nor the droplet.
\end{itemize}
All these behaviors are identified in Fig.~\ref{fig:F}. We thus conclude
that $N\geq 2048$ is sufficient to attempt a computation of the interfacial free-energy.

\begin{figure}
\centering
\includegraphics[angle=270,width=0.75\columnwidth,trim=0 0 0 0]{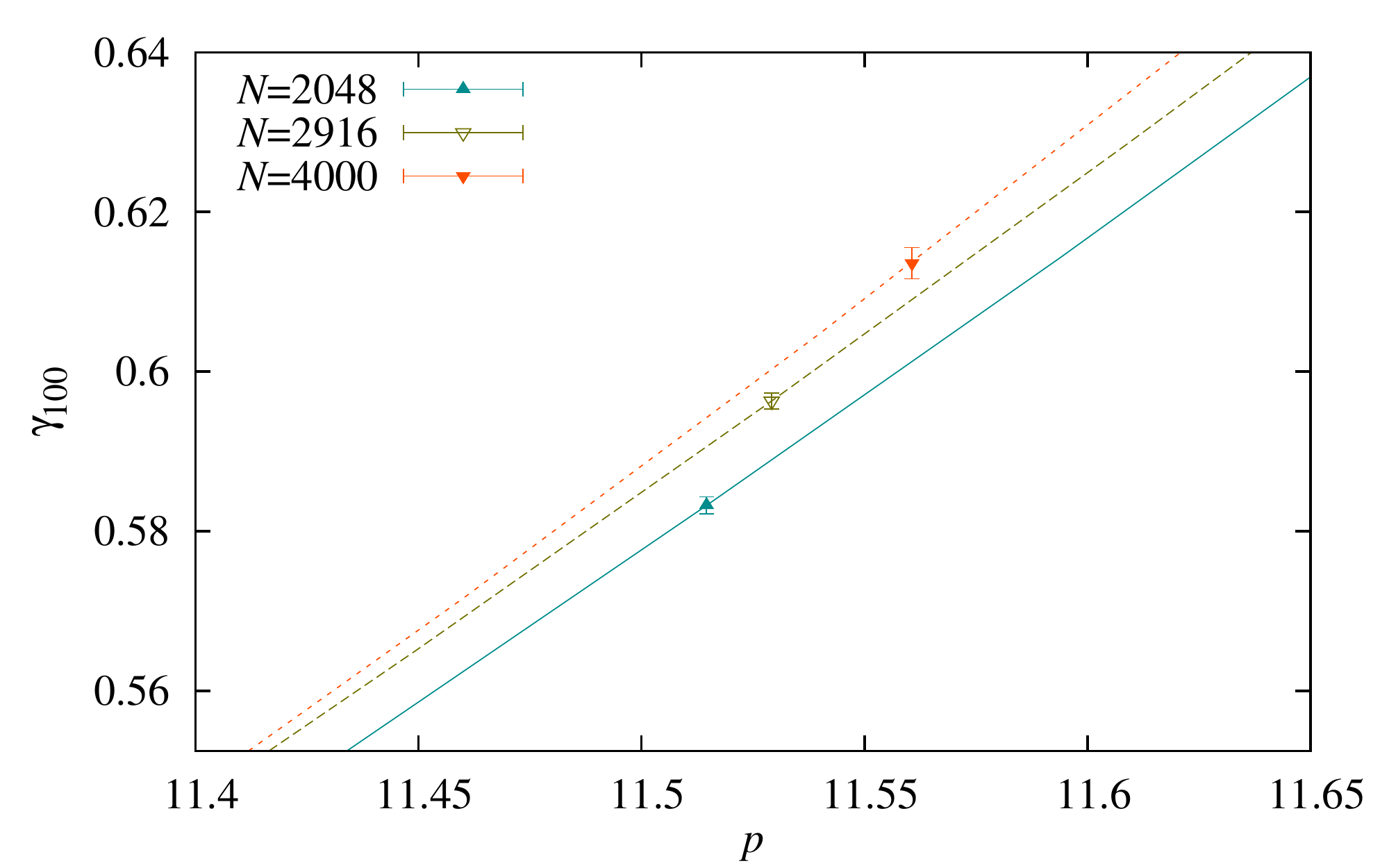}
\caption{Interfacial free-energy for the $(100)$ lattice crystalline direction
  $\gamma_{\{100\}}$ as a function of pressure, for a system of $N$ hard-spheres,
  for $N=2048,2916$ $N=4000$. We estimated $\gamma_{\{100\}}$ from
  Eq.~\ref{eq:hs-gamma}}
\label{fig:sigma_p}
\end{figure}

The effective potential has a local maximum
along the line that joins the FCC and the fluid (the solution of
$\grad_S\varOmega_N=0$ at $S^*\approx 0.5$,
Fig.~\ref{fig:h}---top). The excess free energy is due to the {\em
  two} interfaces that the fluid presents with a crystalline slab
parallel to the simulation box ($\{100\}$ planes). Then the
interfacial free energy at $\p^{(N)}$ is
\begin{equation}\label{eq:hs-gamma}
\gamma_{\{100\}}^{(N)}=
k_\mathrm{B}T\,N\paren{\varOmega_{s^*}-\varOmega_\text{FCC}}/(2
\mean{Nv}_{S^*}^{2/3})\,.
\end{equation}
The $\gamma_{\{100\}}^{(N)}$ (listed in Table~\ref{tab:sigma}) are extrapolated
as~\cite{billoire:94}
\begin{equation}\label{eq:extrapol-gamma}
\frac{\gamma_{\{100\}}^{(N)}\sigma^2}{k_\mathrm{B}T}=\frac{\gamma_{\{100\}}\sigma^2}
      {k_\mathrm{B}T}+ \frac{a_2-\mathrm{log} N}{6N^{2/3}}+
\frac{a_3}{N}+\frac{a_4}{N^{4/3}}+\ldots
\end{equation}
A fit for $256\le N\le 2916$ yields $\gamma_{\{100\}}=0.636(11)$ in units of
$k_\mathrm{B}T/\sigma^2$ ($\chi^2=0.14$ for two degrees of freedom). We remark
that the difference among the fit and $\gamma_{\{100\}}^{(N=4000)}$ (not
included in the fit) is one
fifth of the error bar. Also, the extrapolation for $500\le N\le 2916$ merely
doubles the final error estimate. Our result is compatible with
$\gamma_{\{100\}}=0.64(2)$~\cite{mu:05},
$\gamma_{\{100\}}=0.619(3)$~\cite{cacciuto:03} and
$\gamma_{\{100\}}=0.639(11)$~\cite{hartel:12}, but not with
$\gamma_{\{100\}}=0.5820(19)$~\cite{davidchack:10}.  A peculiarity of the
tethered approach is that one may control the dependence of the estimate of
$\gamma_{100}$ on the actual estimate used for the coexistence pressure. One
simply computes $\gamma_{100}$ as a function of pressure, using
~\eqref{eq:hs-gamma}, as it is shown in Fig.~\ref{fig:sigma_p}. It turns out
that the slope of the curve is of order $0.4$, hence an error of order
$\epsilon$ in the determination of $p^\infty_\mathrm{co}$ results in an error
of order $\sim 0.4 \epsilon$ in $\gamma_{100}$. To our knowledge, such effects
have not been taken into account in previous
computations~\cite{davidchack:00,mu:05,davidchack:10}. In fact, in recent
works~\cite{hartel:12} using the coexistence method, the interfacial
free-energy was computed at the coexistence pressure
$p_\mathrm{co}^{\infty}=11.576(6)$~\cite{zykova-timan:10} very close to our
own computation. Not surprisingly, these authors obtain an almost identical
interfacial free energy.

A final warning is in order. Not much is known about the effect of the
cusps and steps in $\grad_S \varOmega_N$, Fig.~\ref{fig:h}---top, in
the large-$N$ extrapolation $\gamma_{\{100\}}^{(N)}\to\gamma_{\{100\}}$.  This
non-smoothness is a consequence of the geometric transitions that
arise in our larger systems. However, as far as the $\p^{(N)}\to\p$
extrapolation is concerned,  the analogy with simpler
models~\cite{martin-mayor:07} (e.g. the $D\!=\!2$ Potts model, where
comparison with exact solutions is possible), strongly suggests that
these cusps and steps are inconsequential.

\begin{table}
\centering
\begin{tabular}{cl}
\cline{1-2} 
 $N$ &$\gamma_{\{100\}}$\\
\cline{1-2}
108 &0.4063(12)\\ 
256 &0.4243(8)\\ 
500&0.4798(8)\\ 
864 &0.5285(12)\\
1372&0.5611(14)\\ 
2048&0.5832(10)\\ 
2916&0.5971(12)\\
\cline{1-2}
4000&0.607(2)\\
\cline{1-2}
$\infty$&0.636(11)\\ 
\multicolumn{1}{c}{$\chi^2/$dof}&0$.14/2$\\
\cline{1-2}
\end{tabular}
\caption{For each $N$, we report  \{100\} interfacial free energy $\gamma_{\{100\}}$ (in
  $k_\mathrm{B}T/\sigma^2$ units). The large-$N$ limit is obtained using \eqref{eq:extrapol-gamma}.}
\label{tab:sigma}
\end{table}

\part{Quantum Annealing\label{part:QA}}
\chapter[Many-body transverse interactions  in the quantum annealing]{Many-body transverse interactions  in the quantum annealing}\label{chap:QA}

\section{Introduction}\label{sec:introQA}

In all previous chapters, our initial objective was to reach the equilibrium
configuration given certain conditions. This is nothing but an optimization
problem: the task of finding the configuration that optimizes a given
free-energy (or cost in a more general problem) function ${\cal
  H}(\lazo{S_i})$ dependent on a large number $N$ of variables
$S_1,\ldots,S_N$ (often subjected to constraints). This kind of question
appears often in physics when one wonders about equilibrium or ground states,
but it is a whole research field by itself, common to many fields in
science. Finding the minimum energy or cost often becomes a hard task when the
constraints in the system, or the interactions between variables, induce
frustration because there is no way to find a minimum configuration that
minimizes the problem locally (we discussed the concept of frustration in spin
glasses in Figure~\ref{fig:SG-plaquette}). As we have discussed all over this
thesis, the frustration leads to a rugged free-energy landscape of many
relative minima, and an exhaustive search for the absolute minimum is just not
feasible for the interesting sizes (the dimension of the system often grows
exponentially with $N$). As examples of these optimization problems, one can
cite the traveling sales problem~\cite{papadimitriou:98} or the $k$-SAT
problem~\cite{garey:79} in computer science, or 
finding the equilibrium configuration in a glass, as we tried in this thesis.

Complexity in optimization problems is commonly classified as P if an
algorithm is known to solve the problem in a time that grows
polynomially with $N$. On the contrary, if it is not the case, and the
time scales faster with $N$, these problems are normally labeled NP-hard
(Non-deterministic Polynomial-time hard problems) and considered as hard
problems.  Among all the NP problems, there is a subgroup named NP
complete so that any possible NP problem can be reduced to one of them
by means of a polynomial algorithm. Thus, if one algorithm were found
that solved polynomially an NP complete problem, the whole family of
problems would also become easy.  The problems mentioned above belong
all to the NP-complete class.\footnote{With the exception of the 2-SAT
  problem and the 2$D$ Ising spin glass~\cite{barahona:82} that can be
  solved polynomially.}

Statistical mechanics, based on physical intuition, has contributed a lot in
the development of new strategies for optimization problems: parallel
tempering or replica exchange~\cite{hukushima:96}, and simulated
annealing~\cite{kirkpatrick:83} are the two popular and widely used examples
even outside the physics' world. We have also studied and introduced here new
algorithms like the microcanonical or tethered algorithms in
Part~\ref{part:colloids} of this thesis with the same aim.  For the algorithm
we are studying in this Chapter, the quantum annealing, it is interesting to
first discuss the the temperature annealing, its classical counterpart. In
this method, fluctuations are introduced in the problem through a fictitious
temperature. This temperature favors the jump over barriers and thus
encourages the system to visit other possible minima. The system is then
simulated at a temperature $T(t)$ that decreases slowly with time until it is
finally switched off at the end of the simulation. We will refer to this
simulated annealing as classical annealing (CA) in contrast to the quantum
annealing (QA)~\cite{kadowaki:98,finnila:94,das:08,santoro:06}, where
fluctuations are induced also in the system but this time quantum
ones. Quantum perturbations allow tunneling effects, and thus, if narrow
enough, barriers can be crossed instead of surpassed.

In the traditional QA formulation, a time-dependent Hamiltonian is introduced
\be\label{eq:annalgo} \hat H(t)=s(t)\hat H_0+\caja{1-s(t)}\hat V, \ee where $\hat H_0$ is the
target Hamiltonian (or the cost function that one wants to minimize) and
$\hat V$ represents the quantum perturbations. In the field we are working in, the Hamiltonian $\hat H_0$ represents the magnetic interaction between
spins. For the sake of simplicity, we will consider that $\hat H_0$ only
depends on the $z$ components of the Pauli matrix $\hsiz$, where
$i(=1,\ldots,N)$ labels the index of each spin in the system. As normally, we
are interested in finding the lowest energy spin configuration, i.e. the
ground state. Now we introduce the quantum fluctuations through a spin driver
term $\hat V$. In principle, this term is arbitrary, as long
as it does not commute with $\hat H_0$. In
addition, we impose that $\hat V$ has a single, trivial ground state.
A typical example of a driver Hamiltonian is the transverse-field operator
\be\hat V_\mathrm{TF}\equiv -\sump\hsix,\ee where the $\hsix$ $(i=1,\ldots,N)$ are the
$x$ components of the Pauli matrix. This perturbation is very intuitive, since
it represents nothing but the interaction with a magnetic field along the $x$
direction that induces quantum transitions between the eigenstates of $\hat\siz$, whose
modulus is tuned through the control parameter $s(t)$. Initially, at $t=0$, the control parameter
$s(t)$ starts at $s(0)=0$, with $\hat H(0)=\hat V$, and increases
monotonically with time until it reaches unity at time $\tau$ and $\hat
H(\tau)=\hat H_0$. Let us choose the simplest possible scheme where the control
parameter grows linearly with time, i.e. $s(t)=t/\tau$. 

 The evolution of the system, $\ket{\Phi(t)}$, is determined by the
 Schr\"odinger equation, \be\label{eq:Scheq} \I
 \frac{\mathrm{d}}{\mathrm{d}t}\ket{\Phi(t)}=\hat H(t)\ket{\Phi(t)},\,\,0\le
 t\le\tau. \ee The initial state $\ket{\Phi(0)}$ is the ground state of the
 driver Hamiltonian $\hat V$ and is thus known. If the parameter $s(t)$ is
 changed very slowly ($\tau$ is very long), the state will be at every time
 very close to the instantaneous ground state. If it so, by tuning the
 parameters, one will move adiabatically from the initial ground state to the
 ground state of $\hat H_0$.

 The adiabatic theorem states that the system stays close to the instantaneous
 ground state as long as $\tau\gg \Delta^{-2}_\mathrm{min}$ where
 $\Delta_\mathrm{min}$ is the minimum energy gap from the ground state. Of
 course, in order for the above argument to be of general use, this
 $\Delta_\mathrm{min}$ cannot decrease with $N$ too fast. In fact, if the
 energy gap decays exponentially with the system size, as happens generally in
 first-order transitions, the running time will increase exponentially with
 $N$ and the QA would not help to solve the problem efficiently.

This vanishing exponential gap present in many first-order transitions is
sometimes considered to be one of the most important drawbacks of quantum
annealing. Its presence was somehow shadowed for certain time by the
preasymptotic behavior displayed in the small system sizes feasible in
simulations~\cite{farhi:01,hogg:03,young:08}. Indeed, in the last years, an
increasing number of first-order transitions in the annealing parameters are
being found~\cite{young:10,hen:11,jorg:08,jorg:10a,jorg:10b}. It has thus been
suggested that the presence of these quantum first-order transitions when
tuning the transverse field is an intrinsic property of the systems with
complicate free energy landscape, i.e. the hard problems, leading a
pessimistic scenario for the QA algorithm~\cite{young:10,hen:11,jorg:08,jorg:10a,jorg:10b}.

Recently, it was found that the ferromagnetic $p$-spin model, a model without
disorder and with a simple free energy landscape, also suffers from this kind of
first-order transition~\cite{jorg:10a}.  Due to its simplicity, this model
constitutes a perfect benchmark to study the QA performance. Indeed, it was
recently shown~\cite{seki:12} that, at least for finite values of $p$ and
$p\neq 3$, it is possible to avoid this first-order transition by appending an
additional antiferromagnetic driver term and performing the annealing along a
curve in a space of two annealing parameters instead of just one. This study
changes the paradigm about first-order transitions in QA, since the failure of
QA strategies observed up to now could be a failure of the standard
formulation of QA with a transverse field, not a failure of the algorithm
itself.

Here we go deeper into this problem, studying a family of alternative driver
terms, displaying different symmetries. We show analytically the existence of
paths that cross only a second-order transition and thus the speed of QA is
not exponentially damped.  Indeed, in a second order transition the gap 
  vanishes only polynomially with the number of particles, which must be
  compared with the exponential damping observed in the first order transition. The solution to the problem is not unique
and we study the properties of these new driver terms, reaching the conclusion
that the structure of the ground state of the additional Hamiltonians is not
the main important feature that makes the whole algorithm success as argued
in~\cite{bapst:12}.

\section{Problem}
Our starting point is the ferromagnetic $p$-spin model ($p=2,\,3,\,4\ldots$)
\begin{equation}\label{eq:H0}
\hat{H}_0=-N\paren{\frac{1}{N}\sump\hsiz}^p.
\end{equation}
The ground state for this model, $\ket{\Phi_0}$, corresponds to the state of
all the spins aligned along the $z$ direction. In order to avoid the
degeneracy of the up and down configurations present in even powers of $p$, we
consider here only the odd values of $p$ and $p\ge 3$.  In the limiting
$p\to\infty$ case, this model is nothing but the Grover
problem~\cite{jorg:10a,grover:97}.  Although the Grover's quantum
  algorithm,  whose reformulation in quantum annealing is given
  in~\cite{roland:03}, is considered a success of the quantum
  algorithm (provides a square-root gain with respect to the
  classical search~\cite{grover:96}) it remains being a hard problem even with
quantum algorithms.  Now we consider the problem of finding
this already known ground state $\ket{\Phi_0}$ of \eqref{eq:H0} with the QA
algorithm using two driving terms.

As usual, we consider the traditional transverse field
operator, 
\be\label{eq:VTF}\hat V_\mathrm{TF}\equiv -\sump\hsix,\ee 
whose ground state,
$\ket{\Phi^\mathrm{TF}}$, is the one where all the $N$ spins are pointing to the
positive direction along the $x$ axis. 
We next introduce a second Hamiltonian inspired in the antiferromagnetic
interaction suggested in \cite{seki:12},
\begin{equation}\label{eq:Vk}
\hat{V}_k=+N\paren{\frac{1}{N}\sump\hsix}^k,
\end{equation}
that depends on a parameter $k(>1)$. When $k=2$, we recover the
antiferromagnetic interaction studied in \cite{seki:12}.  The ground state for
this Hamiltonian, namely $\ket{\Phi_k}$, depends on the value of the power
$k$. When $k$ is odd, the energy is minimum when all spins are aligned along
the $x$ axis but pointing to the negative direction.  On the contrary, when
$k$ is even, the ground state corresponds to the state with total
$\sump\six=0$ if $N$ is even, or $\sump\six=\pm 1$ for $N$ odd. One of
the goals of the present paper is to clarify whether the value $k=2$ is
essential to avoid the first-order transition.

If we sum up  \eqref{eq:annalgo}, \eqref{eq:VTF} and \eqref{eq:Vk}, the new Hamiltonian of the problem reads as
\begin{equation}
\label{eq-qa:H}
\hat{H}(s,\lambda)=s\caja{\lambda\hat{H}_0+(1-\lambda)\hat{V}_{k}}+(1-s)\hat{V}_{\mathrm{TF}}.
\end{equation}
Here there are two annealing parameters, $s$ and $\lambda$. These parameters
will be tuned slowly during the annealing process so that, at the final time,
$\tau$, $s(\tau)=\lambda(\tau)=1$ and the target Hamiltonian \eqref{eq:H0} is
thus recovered. In that way, one can explore the annealing process following
infinitely different paths. It might resemble the idea of nondeterministic
Turing machines, but one must always keep in mind that, even though many paths
are possible,  only one is chosen in each particular realization.

The traditional QA is one of the infinite possible paths in \eqref{eq-qa:H}. In
fact, one can remove the influence of $\hat V_k$, just by fixing
$\lambda(t)=1$. Then, the annealing is performed by tuning $s$ from 0 to 1.
If one looks at the configurations, at $t=0$ all spins should be aligned with
the $x$ axis, and at the end, with the $z$ axis. In this case, we know that
the system suffers from a quantum first-order phase transition between
these two states. This transition ruins the efficiency of the algorithm as it
becomes exponential~\cite{jorg:10a}. The idea of introducing this
two-parameter space $(\lambda,s)$ is precise to try avoid this transition by
following an alternative route.  Seki and Nishimori succeeded in finding
ingenious paths~\cite{seki:12} with antiferromagnetic interactions, and here,
we generalize that method to check how the value of $k$ affects the
conclusion.

\section{Analysis by a semi-classical approach}\label{sec:CA}

The QA strategy will succeed if we are able to find a path in the space of
parameters $(\lambda,s)$ that avoids crossing any first-order transition. With
this aim, we compute in this section the phase diagram correspondent to the
new Hamiltonian \eqref{eq-qa:H}, as a function of the parameter $k$. The
$N\to\infty$ limit can be computed analytically using a semi-classical
approximation (method to be explained below) or the Trotter-Suzuki
decomposition formula~\cite{suzuki:76} and the static approximation (see
Appendix \ref{sec:SA}), leading to equivalent results.

\subsection{General Properties}

As a starting point, let us rewrite the Hamiltonian \eqref{eq-qa:H} in terms of the total 
spin variables ($S^\alpha=\frac{1}{2}\sump\sigma_i^\alpha$ with $\alpha=x,\,y$ and $z$),
\be
\label{eq:Hnew}
\hat{H}(s,\lambda)=-s\lambda
N\paren{\frac{2}{N}S^z}^p+s\,(1-\lambda)N\paren{\frac{2}{N}S^x}^k-2(1-s)S^x.
\ee This Hamiltonian commutes with the total squared spin, $S^2$. Since
  the total spin is conserved and the initial state in the annealing process
  is the one with all spins aligned with the $x$ axis, we are only interested
in studying the maximum possible $S$ value, i.e. $S=N/2$.

Now, consider the normalized variables 
$m^\alpha=S^\alpha/S$, with $\alpha =x,\,y$ and $z$. The commutation
relations for these variables are
\begin{eqnarray}
[m^x,m^y]=\I\frac{2}{N}m^z,
\end{eqnarray}
and cyclic permutations. The normalized variable $m^\alpha$ can take
$N+1$ values within the interval $[-1,1]$. Thus, in the large $N$ limit, these
variables commute, and we can consider them as the components of a classical
unit vector,
i.e. $\V{m}=(\cos\theta,\sin\theta\sin\varphi,\sin\theta\cos\varphi)$,
being $\theta$ the polar angle measured from the $x$ axis, and
$\varphi$ the azimuthal one measured from the $z$ axis.

Considering the system now as classic, we can write the energy per spin as \be
e=-s\lambda
(\sin\theta\cos\varphi)^p+s\,(1-\lambda)\cos^k\theta-(1-s)\cos\theta.  \ee The
equilibrium state will be determined by the minimum of $e$. Since $p$ is odd,
the minimum lies on the plane with $\varphi=0$, which we call  $XZ^+$ plane.  The energy on this
plane is labeled only by the polar angle $\theta$ \be\label{eq:mine}
e=-s\,\lambda \sin^p\theta+s\,(1-\lambda)\cos^k\theta-(1-s)\cos\theta.  \ee We
search the $\theta_0\in[0,\pi]$  that minimizes \eqref{eq:mine}\footnote{Negative magnetizations along $z$
  axis have always higher free energy due to the change of sign in the
  $\sin^p\theta$ term in \eqref{eq:mine} (remember that we only consider the $p$ odd case in
  this work).}.  The condition for the minimum
is \be\label{eq:condmin} \frac{\partial e}{\partial\theta_0}=-p\,s\,\lambda
\sin^{p-1}\theta_0
\cos\theta_0-k\,s\,(1-\lambda)\cos^{k-1}\theta_0\sin\theta_0+(1-s)\sin\theta_0=0,
\ee whose solutions are the angles $\theta_0$ that satisfy either
$\sin\theta_0=0$ or \be\label{eq:eqtheta0} p\,s\,\lambda \sin^{p-2}\theta_0
\cos\theta_0+k\,s\,(1-\lambda)\cos^{k-1}\theta_0-1+s=0. \ee These two
equations have more than one solution, and each one corresponds to a different
phase. We will consider them as ferromagnetic if $m^z(=\sin\theta_0)>0$, and
quantum paramagnetic if $m^z=0$.  The most stable one at each point
$(\lambda,s)$ will be the absolute minimum of $e$.

We begin with the quantum paramagnetic solutions. The equation
$\sin\theta_0=0$ is satisfied for $\theta_0=0$ or $\pi$. The case $\theta_0=0$
corresponds to positive $x$ magnetization, $m^x=1$. We name this phase QP$^+$.
Its energy is obtained by inserting this angle in \eqref{eq:mine},
\be\label{eq:fQPpca} e_{\mathrm{QP}^+}(s,\lambda)=s(1-\lambda)-1+s.  \ee The
other paramagnetic solution, $\theta_0=\pi$, corresponds to negative
magnetization, $m^x=-1$. We call this phase QP$^-$. This phase is only stable
for odd values of $k$ and its energy is \be \label{eq:fQPmca}
e_{\mathrm{QP}^-}(s,\lambda)=-s(1-\lambda)+1-s. \ee This phase will not appear
in the phase diagrams for $k$ even, since its energy is always positive in
the range of parameters $0\le s,\,\lambda\le 1$.

We consider next the ferromagnetic solutions ($\theta_0> 0$). The purely
ferromagnetic solution $\sin\theta_0=1$ is only a valid solution on the line
$s=1$. Apart from this line, equation \eqref{eq:eqtheta0} cannot be
explicitly solved for any value of $p$, but it can be done in the $p\to\infty$
limit. We study below all the solutions for this limit
and discuss their validity for $p$ finite.

\subsection{Phase diagram for $p\to\infty$}

In this limit, \eqref{eq:eqtheta0} has two possible ferromagnetic solutions.
The parameter $p$ appears in \eqref{eq:eqtheta0} through
$p\sin^{p-2}\theta_0$. We consider the two possible limits for the sine power,
$\sin^{p-2}\theta_0\to$1 (for the F phase) and 0 (for the F' phase), always
keeping $\theta_0> 0$.

We begin the discussion with the F phase. With this aim, we assume
\be\label{eq:limF} \sin^{p}\theta_0\to1,\ee and substitute it in
\eqref{eq:eqtheta0}, \be\label{eq:eqF1}
p\,s\,\lambda\cos\theta_0+k\,s\,(1-\lambda)\cos^{k-1}\theta_0-1+s=0.  \ee In
the $p\to\infty$ limit, this equation can only be satisfied if either the
cosine vanishes, i.e.
$\theta_0=\pi/2$ (but only on the line $s=1$), or $p\cos\theta_0$ tends to a
constant. Let us investigate this second case. We consider $\cos\theta_0=c/p$,
with $c$ a $p$-independent constant, and introduce it in \eqref{eq:eqF1}, and
taking the $p\to\infty$ limit, the equation reads \be\label{eq:eqF2}
s\,\lambda\,c-1+s=0, \ee whose solution is $c=(1-s)/s\lambda$. Thus,
\be\label{eq:solFpinfty} \cos\theta_0=\frac{1-s}{s\,p\,\lambda}\to 0, \ee is a
solution to \eqref{eq:eqtheta0}. Still we need to check that this
$\theta_0$ agrees with the initial assumption \eqref{eq:limF}. Indeed,
\begin{eqnarray}\nonumber
\lim_{p\to\infty} \sin^{p-2}\theta_0=\lim_{p\to\infty}
\caja{1-\paren{\frac{1-s}{2\,s\,p\,\lambda}}^2}^{p}=1.
\end{eqnarray}
We obtain the energy for this phase introducing \eqref{eq:solFpinfty} in \eqref{eq:mine}
\be\label{eq:fFpinftyca}\left. e_\mathrm{F}(s,\lambda)^{k}\right|_{p\to\infty}=-s\,\lambda.\ee

On the other hand, the F' solution is obtained assuming the opposite limit,
\be\label{eq:limF'} p\sin^{p}\theta_0\to0.\ee Under this assumption,
\eqref{eq:eqtheta0} reduces to \be k\,
s\,(1-\lambda)\cos^{k-1}\theta_0-1+s=0, \ee whose solution is
\be\label{eq:solFpCA} \cos\theta_0=\caja{\frac{1-s}{k\,
    s\,(1-\lambda)}}^{\frac{1}{k-1}}. \ee 
Note that if $k$ is odd, the negative solution for the cosine is also a
valid solution. However, it has always  higher energy than its positive
counterpart, so we will not consider it for further discussions.

The energy for the F' phase when $p\to\infty$ is
\be\label{eq:freeFpCAifnty}
\left. e_{\mathrm{F'}}(s,\lambda)\right|_{p\to\infty}=-\frac{k-1}{k}\caja{\frac{1-s}{k\,s\,(1-\lambda)}}^\frac{1}{k-1}(1-s).\ee

\begin{figure}
\begin{center}
 \includegraphics[angle=270,width=0.45\columnwidth,trim=10 20 10
   20]{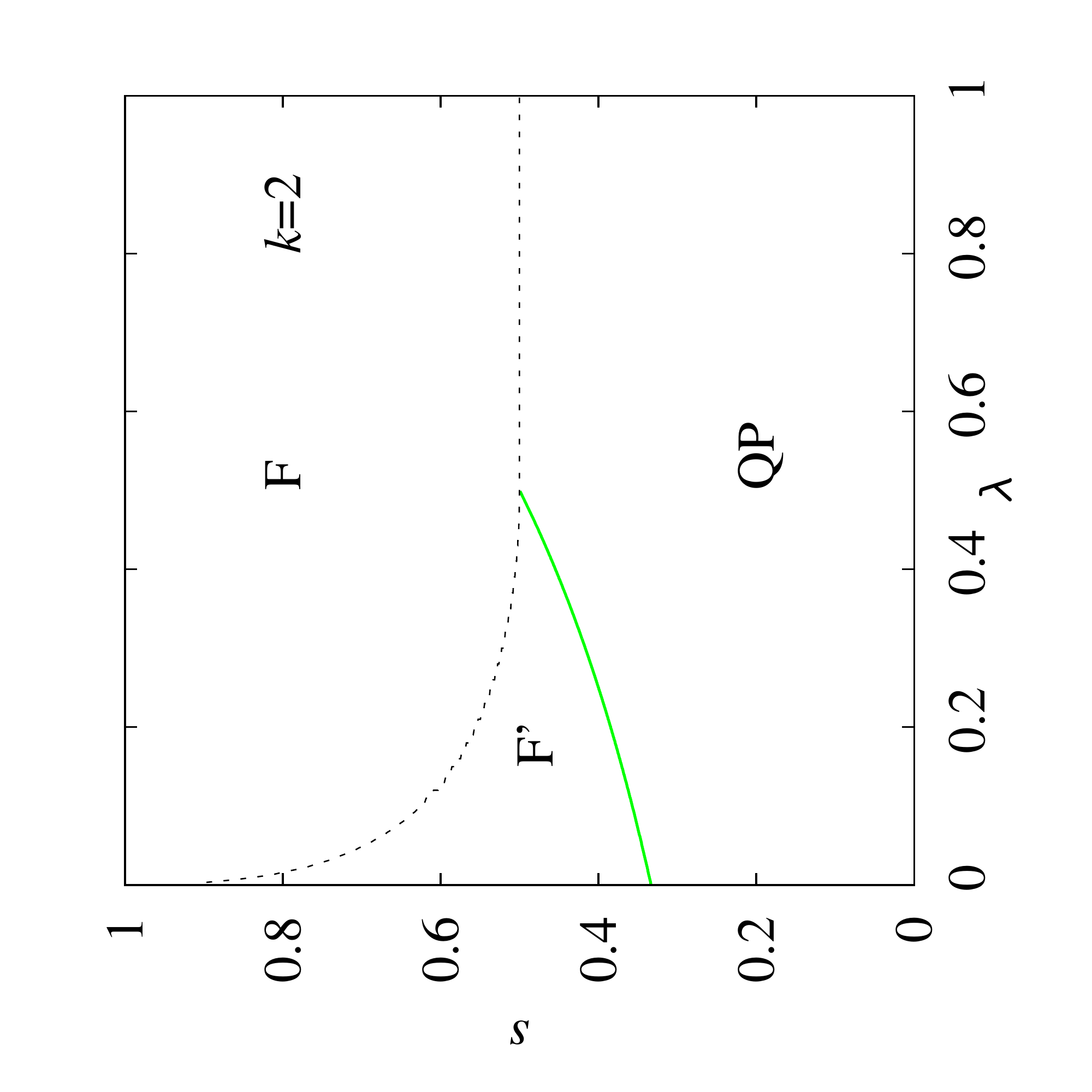}
 \includegraphics[angle=270,width=0.45\columnwidth,trim=10 20 10
   20]{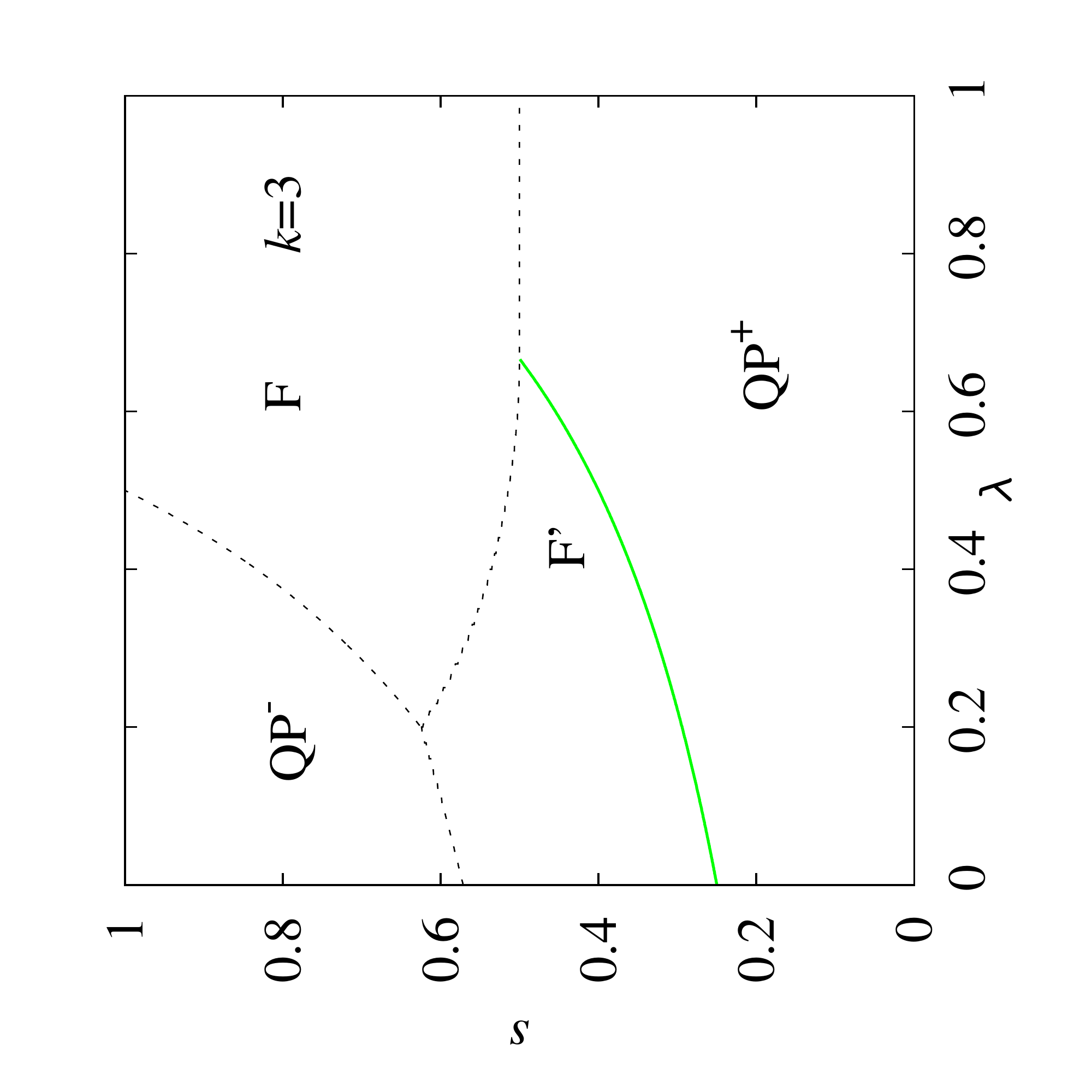}
 \includegraphics[angle=270,width=0.45\columnwidth,trim=10 20 10
   20]{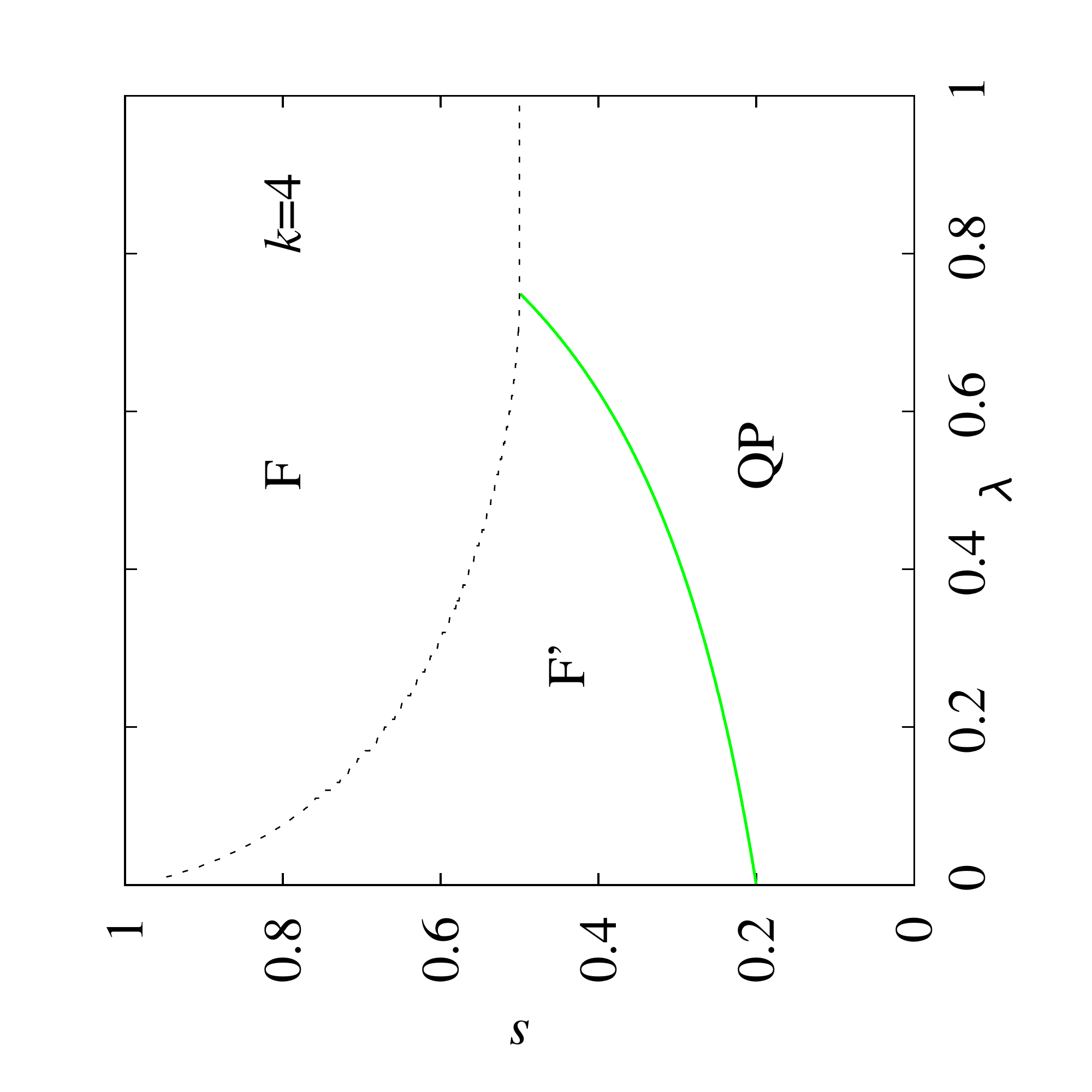}
 \includegraphics[angle=270,width=0.45\columnwidth,trim=10 20 10
   20]{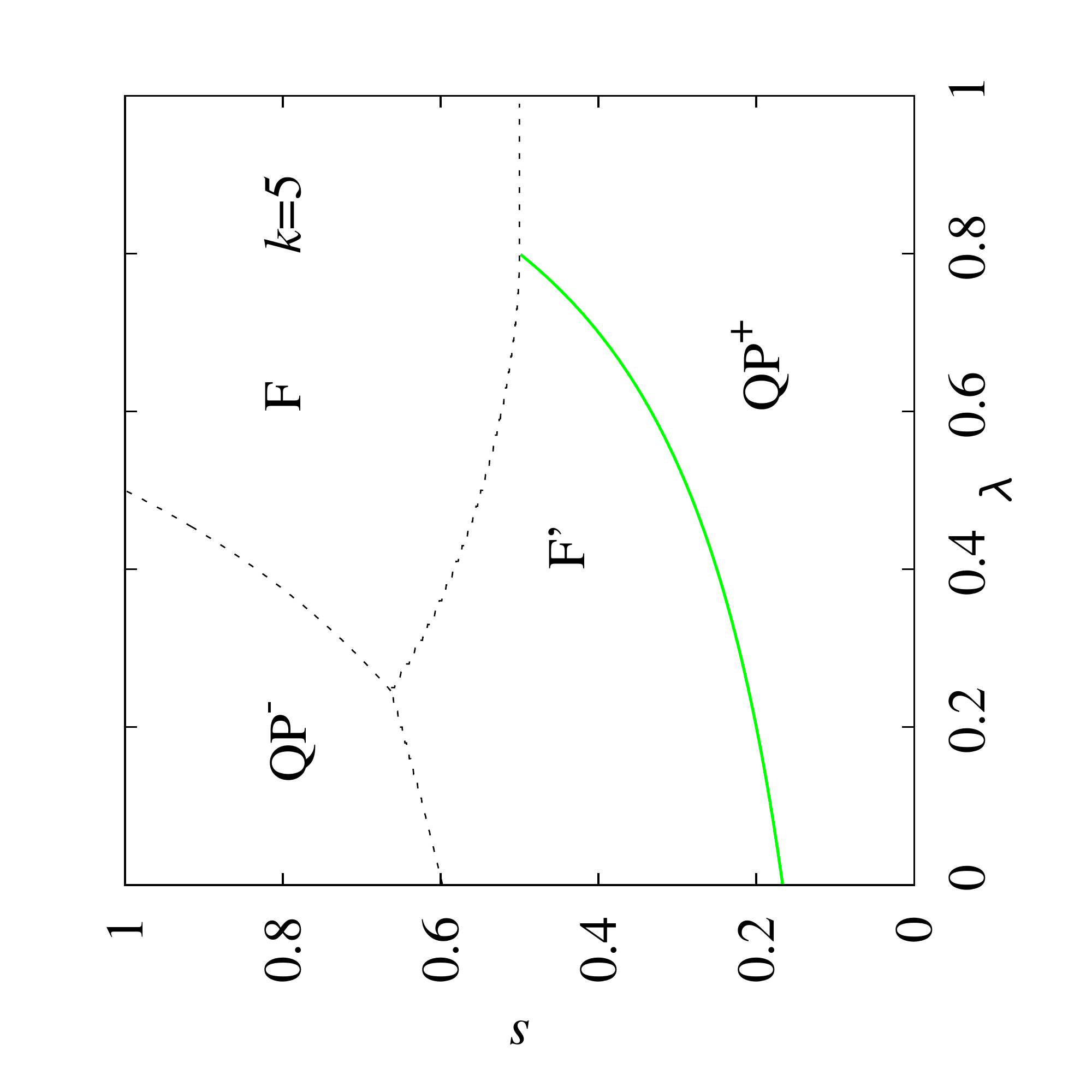}
\caption{Phase diagram for $p\to\infty$. Dashed black lines represent first-order transitions, whereas the solid line in light green accounts for the
  second-order transition.}\label{fig:pd-pifty}
 \end{center}
\end{figure}

Up to this point, we have obtained all the possible solutions to
\eqref{eq:eqtheta0} in the $p\to\infty$ limit: three (for even $k$) and four (for odd $k$) phases. We can use
the energies to determine which phase is the most stable at each point $(\lambda,s)$. We show in figure \ref{fig:pd-pifty} several phase diagrams for $k=2,\,3,\,4$ and $5$.
Let us analyze the nature of each transition. We begin with the transition line
between the F' and  QP$^+$ phases. This line is obtained by solving
$e_\mathrm{F'}-e_\mathrm{QP}=0$ using the expressions \eqref{eq:freeFpCAifnty}
and \eqref{eq:fQPpca}. This equality is fulfilled on the line
$s=1/[1+k(1-\lambda)]$. On this line, $m^x=\cos\theta_0=1$ in both phases,
which corresponds to a second-order transition.  On the other hand, the
transition between the F and the QP$^+$ phases lies on the $s=1/2$ line and,
since magnetization is discontinuous, it is first-order. The
second-order transition extends from $(\lambda,s)=(0,1/(k+1))$ to
$(\lambda,s)=((k-1)/k,1/2)$, the point where these two kinds of transitions
cross. According to that, the higher $k$ is, the broader the second-order line
and the smaller the QP$^+$ region are. Furthermore, in the $k\to\infty$ limit,
the QP$^+$ region completely disappears.

Still there is a first-order transition between the F and  F' phases,
determined by the solution of $e_\mathrm{F'}-e_\mathrm{F}=0$ using
\eqref{eq:freeFpCAifnty} and \eqref{eq:fFpinftyca}. We solve this equation
numerically and obtain the curve displayed in figure  \ref{fig:pd-pifty}.
On this line, the magnetizations are discontinuous but at the point
$(\lambda,s)=(0,1)$ where they two become equal, 
$m^z=\sin\theta_0=1$. The transition is then first-order, but in the mentioned 
point, where it would be second-order.

Up to this point, the discussion is common for even and odd values of
$k$. However, in this latter case the QP$^-$ phase also exists.  Thereby, two
additional transitions between F or F' phases and the QP$^-$ phase appear. In
both cases the $x$ magnetization changes the sign on the transition, and then, they
are first-order. The transition lines are obtained by solving the equations
$e_\mathrm{QP^-}-e_\mathrm{F}=0$, leading to $s=1/(2(1-\lambda))$, and
$e_\mathrm{QP^-}-e_\mathrm{F'}=0$ which must be solved numerically. We display
all the transition lines in figure \ref{fig:pd-pifty}. 

According to these results, when we consider the $p\to\infty$ limit, there is
only one single path that succeeds in avoiding first order transitions. This
is the straight line that joins the initial point $(\lambda,s)=(0,0)$ with the
left upper corner, $(0,1)$, and the final state $(1,1)$. However, even though
this path only crosses second order transitions, along this way there is no
quantum annealing process, as can be seen by an insertion of these parameter
values into the Hamiltonian \eqref{eq-qa:H}, and thus this path is meaningless.

\section{Phase Diagram}
\begin{figure}
\begin{center}
 \includegraphics[angle=270,width=0.6\columnwidth,trim=10 20 10
   20]{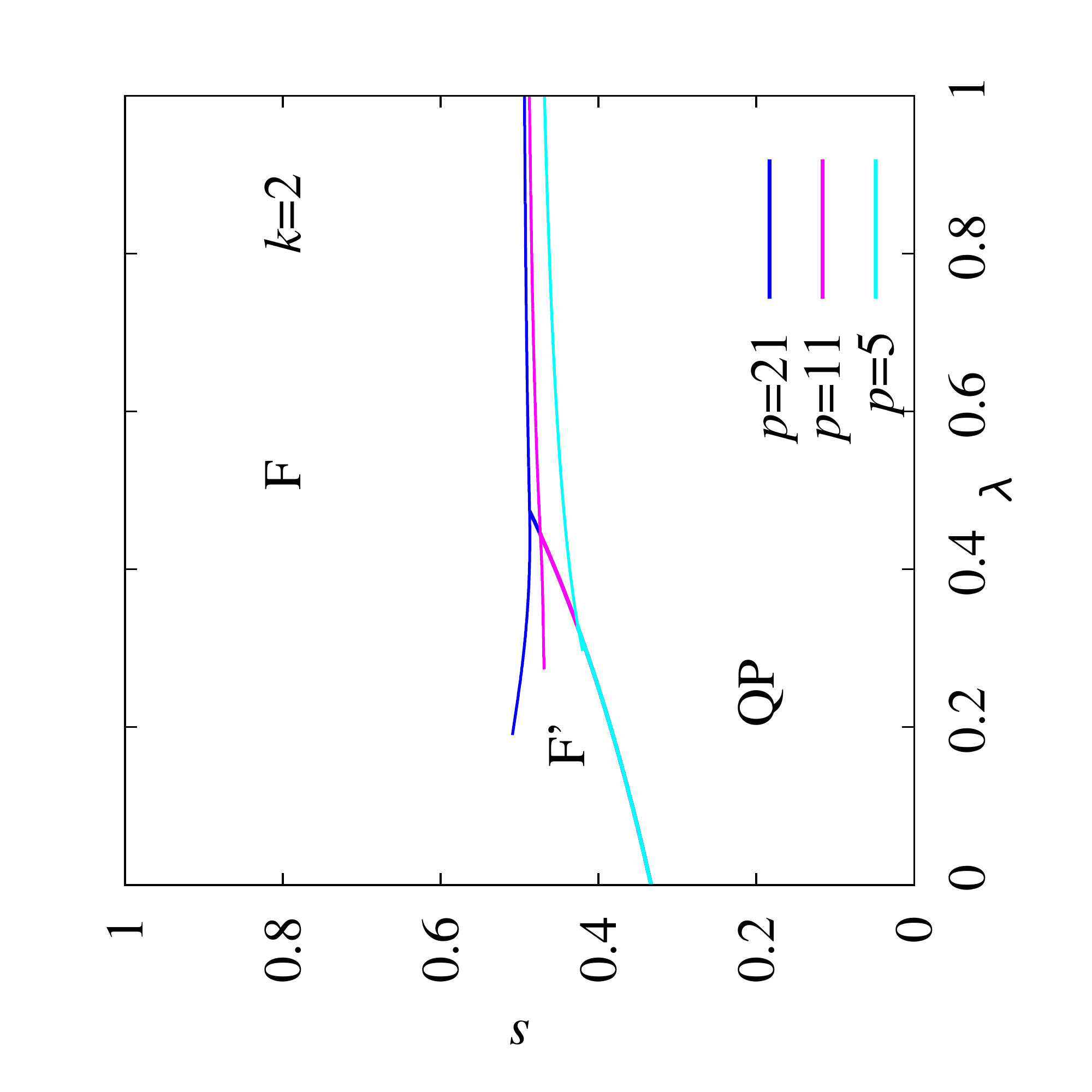}
\caption{Phase diagram for $k=2$. This is the same phase diagram as in reference \cite{seki:12}.  The transition between the F' and QP phases is of second order, and the F-QP and F-F' transitions are of first order. }\label{fig:phase-diag-k2}
 \end{center}
\end{figure}

\begin{figure}
\begin{center}
 \includegraphics[angle=270,width=0.6\columnwidth,trim=10 20 10
   20]{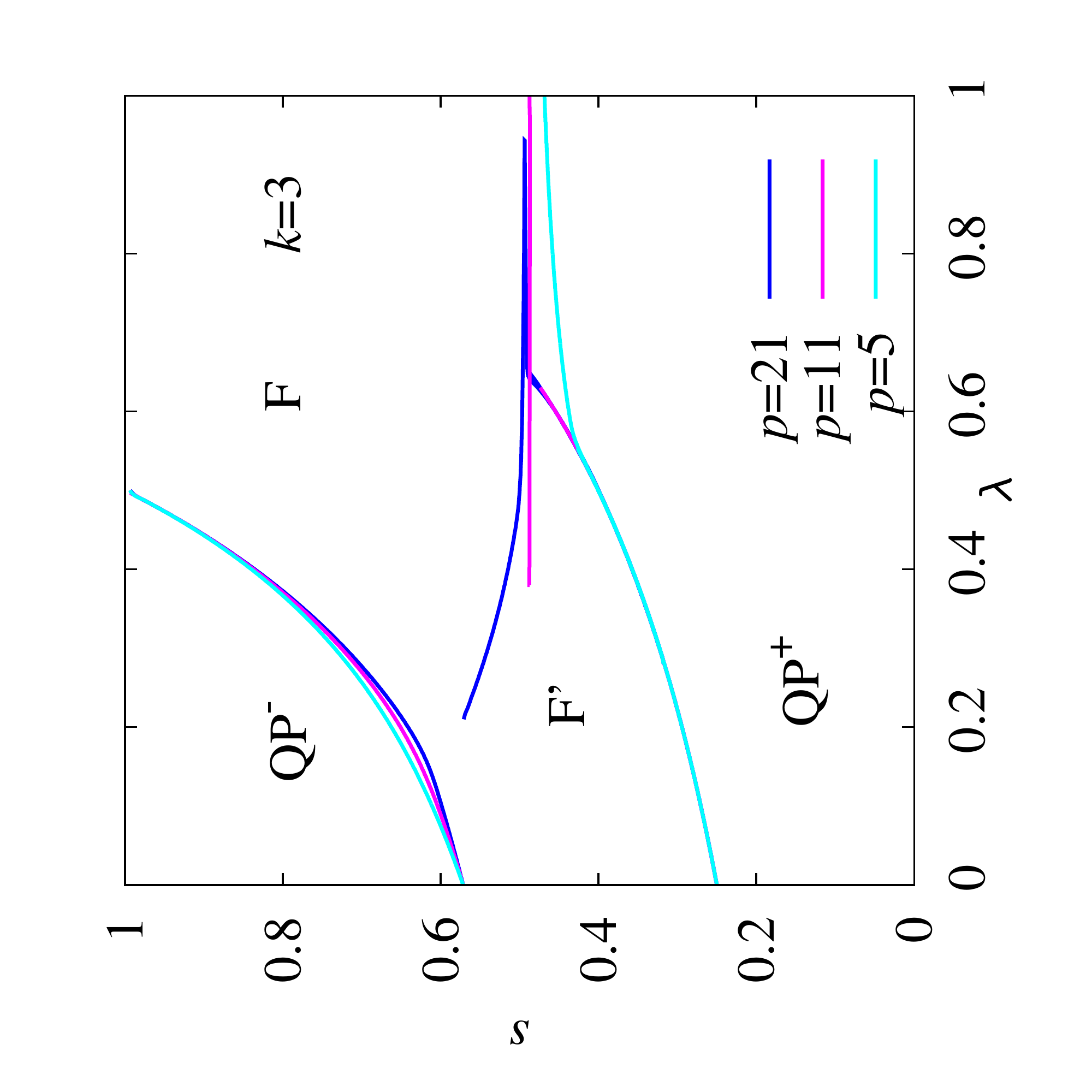}
\caption{Phase diagram for $k=3$. Only the F'-QP$^+$ transition is of second order.}\label{fig:phase-diag-k3}
 \end{center}
\end{figure}

\begin{figure}
\begin{center}
 \includegraphics[angle=270,width=0.6\columnwidth,trim=10 20 10
   20]{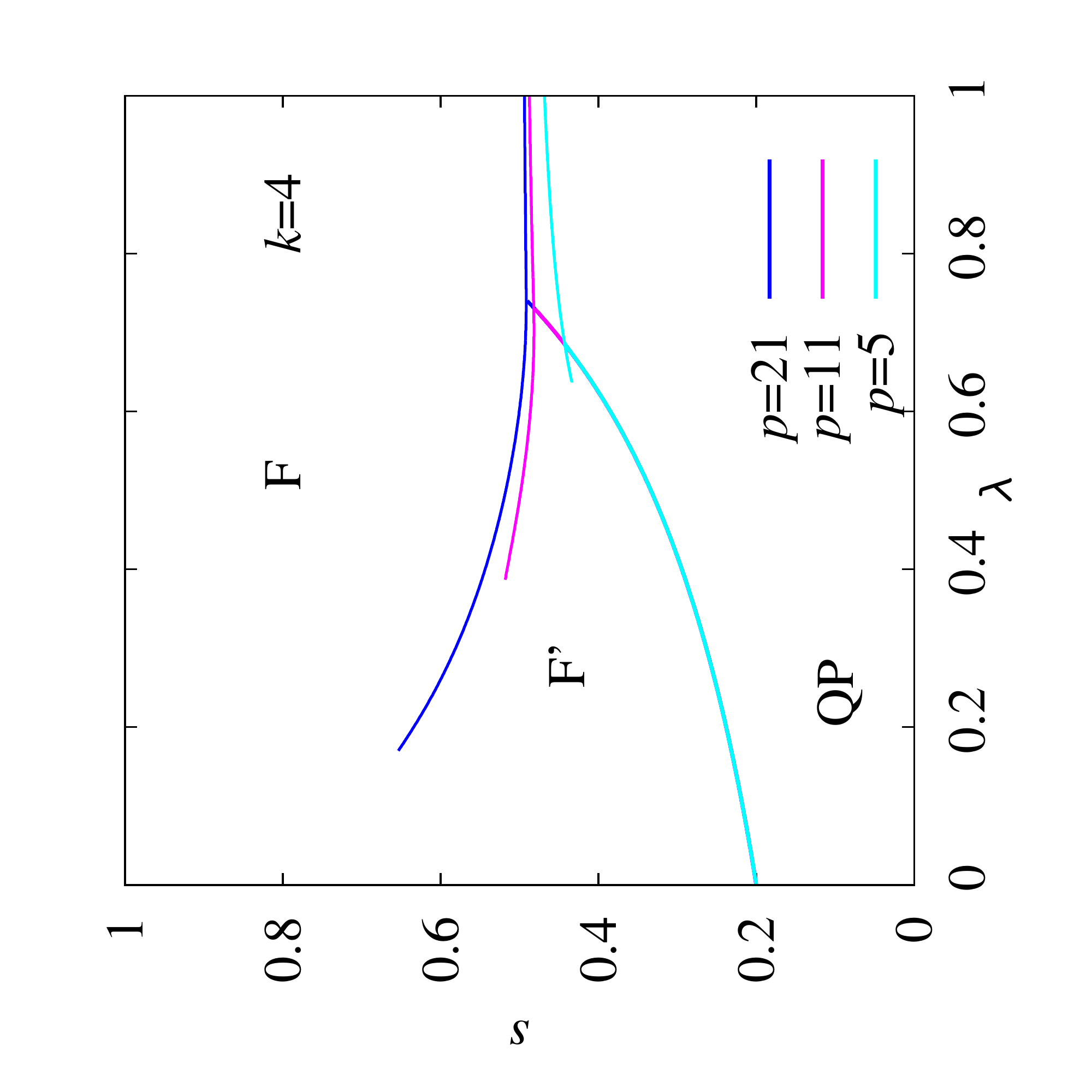}
\caption{Phase diagram for $k=4$. The structure is qualitatively the same as for $k=2$. }\label{fig:phase-diag-k4}
 \end{center}
\end{figure}

\begin{figure}
\begin{center}
 \includegraphics[angle=270,width=0.6\columnwidth,trim=10 20 10
   20]{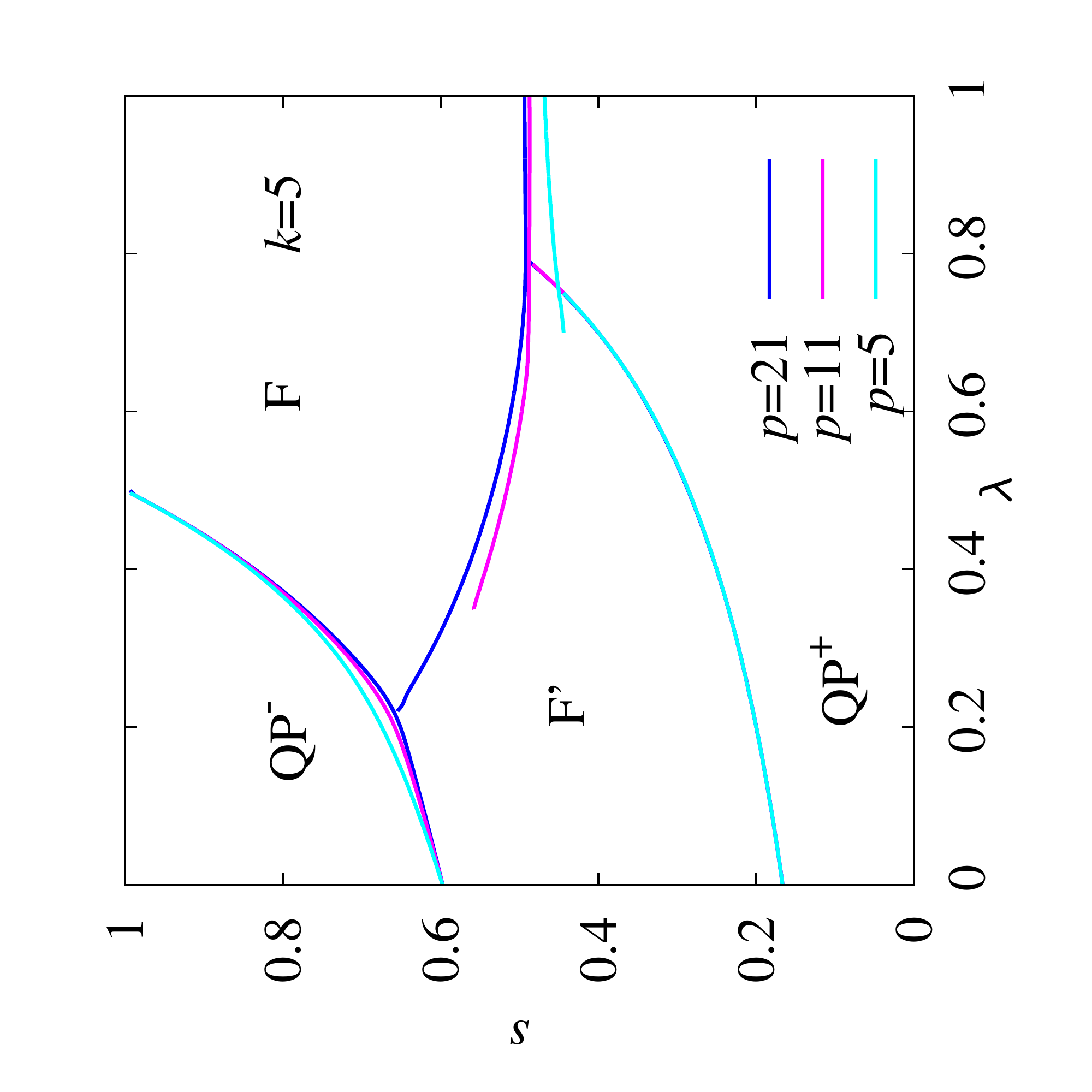}
\caption{Phase diagram  $k=5$. The F'-QP$^+$ transition is of second order, and the other transitions are all of first order.}\label{fig:phase-diag-k5}
 \end{center}
\end{figure}

The phase diagram for finite $p$ is different. Now, there appear regions where
first-order transitions disappear, leaving more space for annealing
trajectories. We display the corresponding diagrams in figures
\ref{fig:phase-diag-k2}, \ref{fig:phase-diag-k3}, \ref{fig:phase-diag-k4} and
\ref{fig:phase-diag-k5} for $k=2$, 3, 4 and 5, respectively. Again, the shape
of the phase diagram strongly depends on whether $k$ is even or odd. In the
former, there are only three phases and in the latter the extra QP$^-$ phase
appears. Besides, the higher $k$ is, the longer is the second-order transition
line.


\begin{figure}
\begin{center}
 \includegraphics[angle=270,width=0.45\columnwidth,trim=0 0 0
   0]{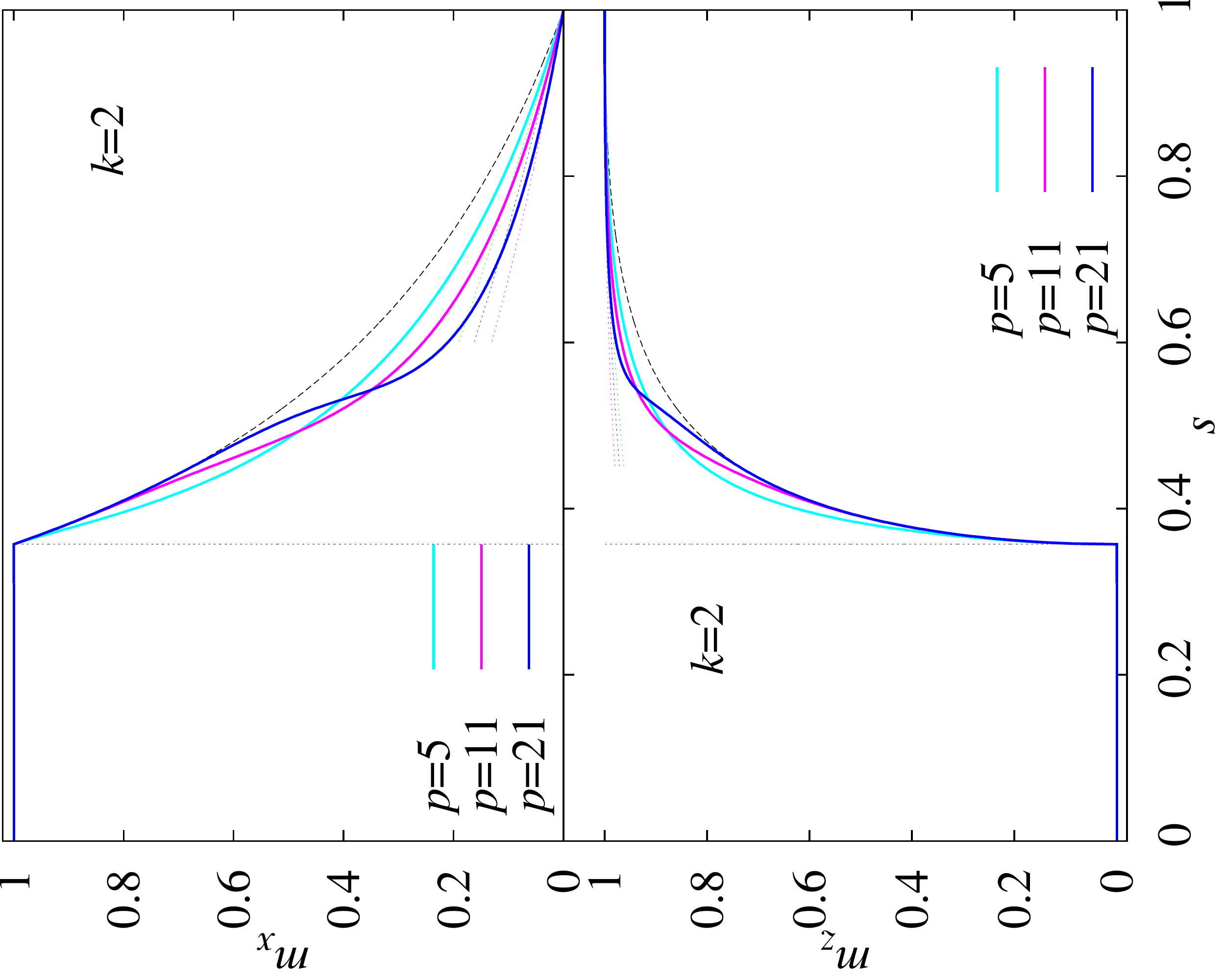}
 \includegraphics[angle=270,width=0.45\columnwidth,trim=0 0 0 0]{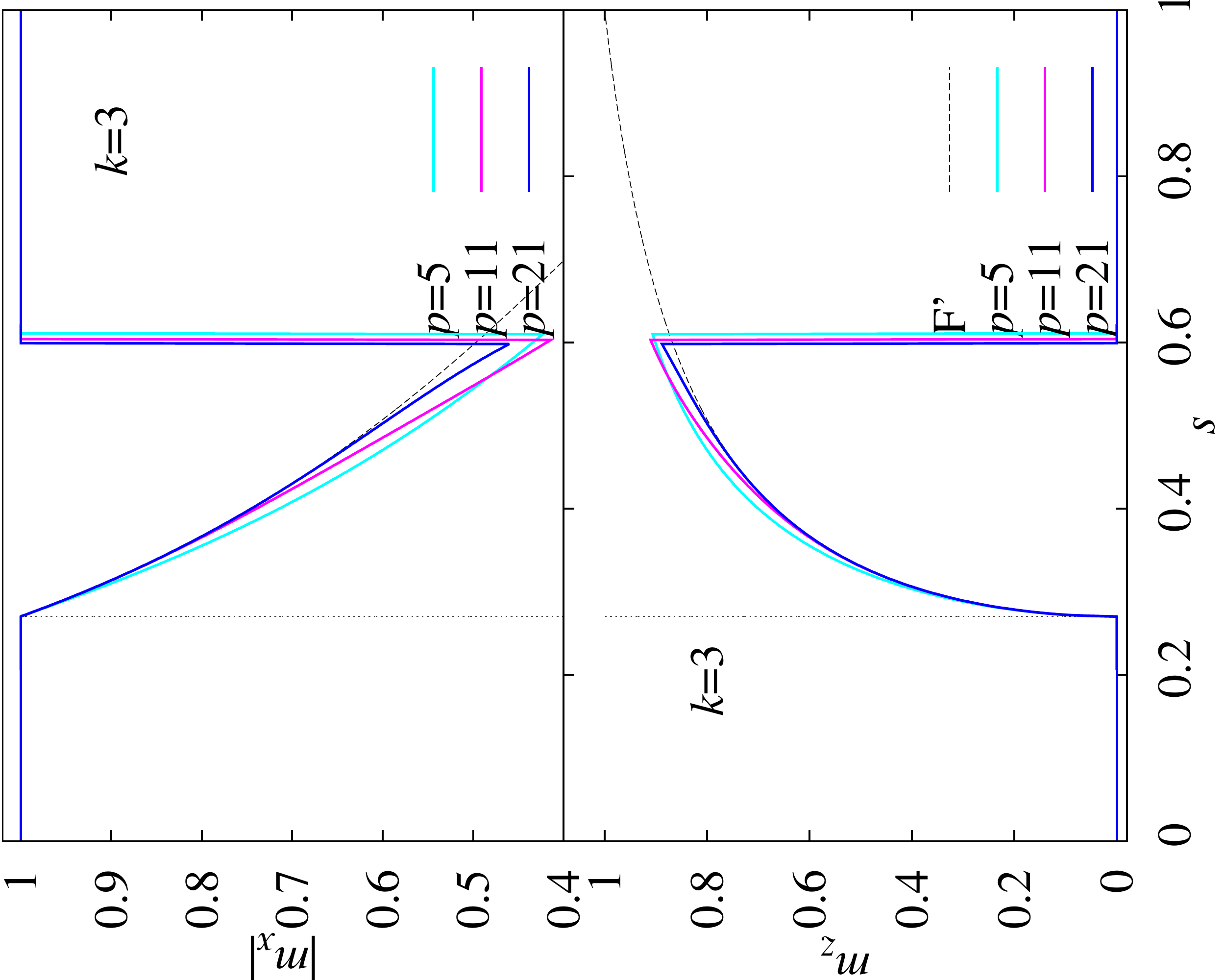}
\caption{Magnetization obtained with the semi-classical approach as a
  function of $s$ for $\lambda=0.1$ and for $k=2$ and 3. The dashed lines
  correspond to the analytical predictions for the QP$^\pm$, F
  \eqref{eq:solFk2} and F' \eqref{eq:solFpCApfin} solutions.}\label{fig:magk2_class_0.1}
 \end{center}
\end{figure}

\begin{figure}
\begin{center}
\includegraphics[angle=270,width=0.45\columnwidth,trim=0 0 0
  0]{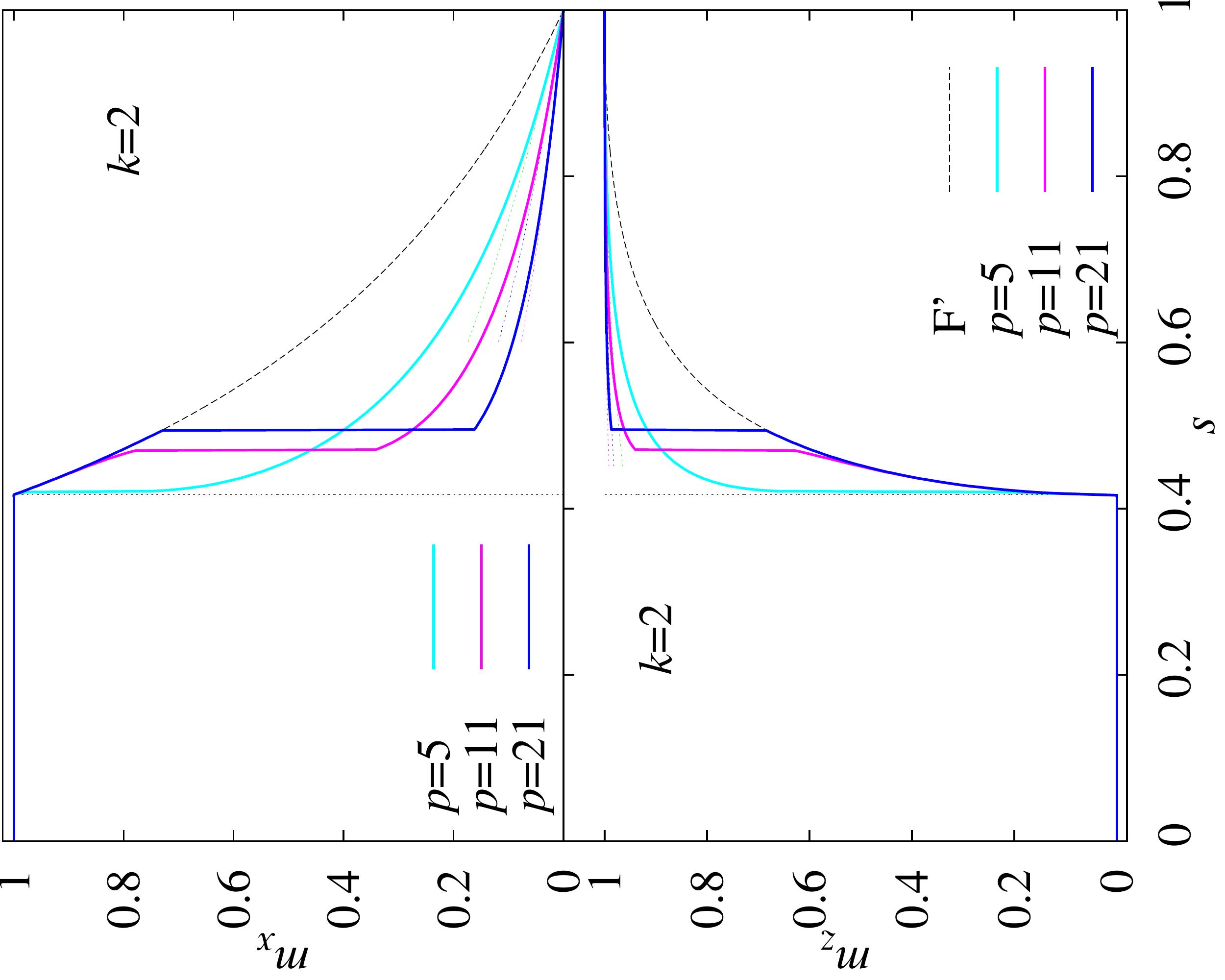}
\includegraphics[angle=270,width=0.45\columnwidth,trim=0 0 0 0]{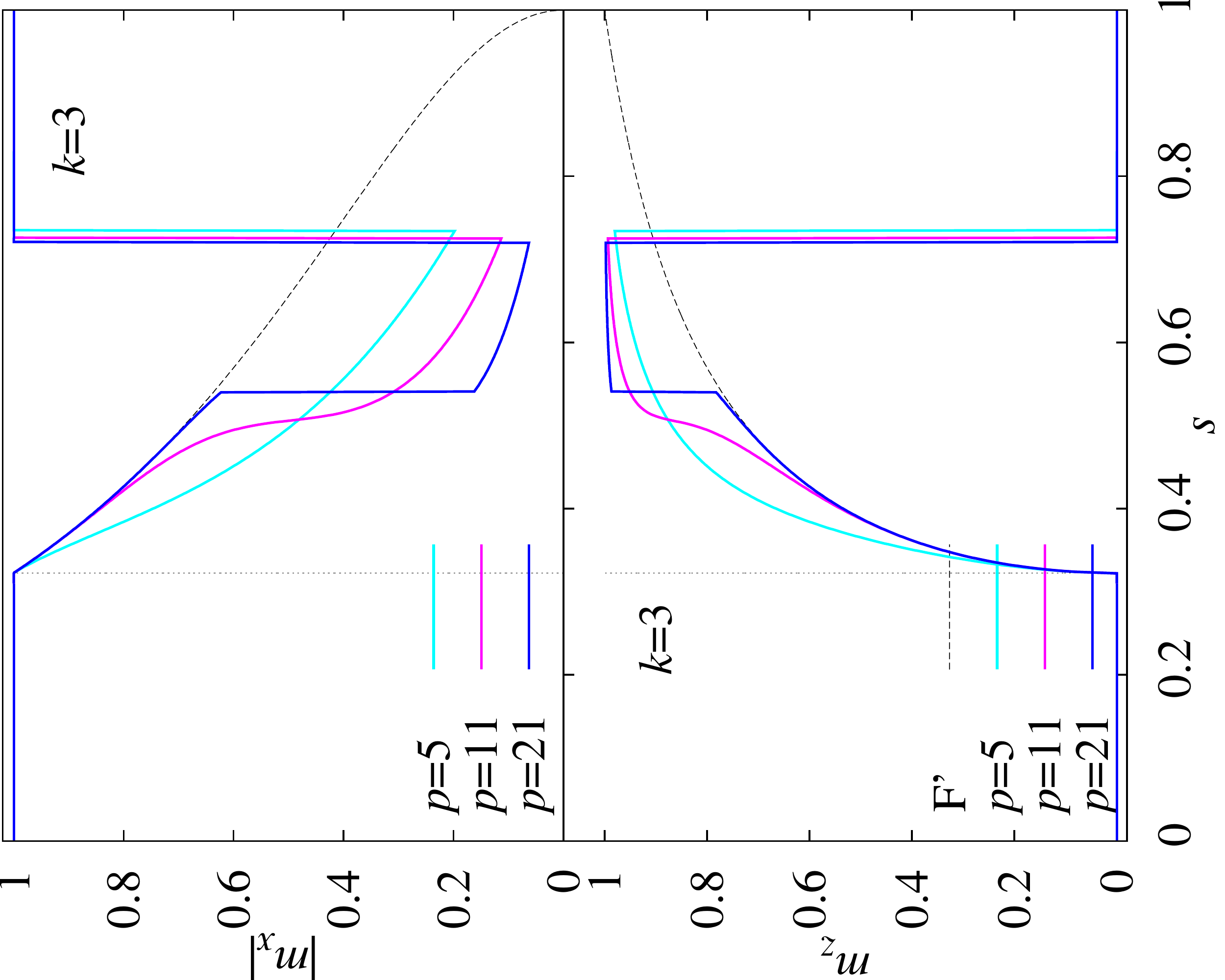}
\caption{Magnetization obtained with the semi-classical  approach as
a  function of $s$ for $\lambda=0.3$ for $k=2$ and 3. The dashed lines
  correspond to the analytical predictions for the QP$^\pm$, F
  \eqref{eq:solFk2} and F' \eqref{eq:solFpCApfin} solutions.}\label{fig:mag_class_0.3}
 \end{center}
\end{figure}

The picture of the ferromagnetic phase for finite $p$ is rather
complicated. When one solves numerically \eqref{eq:eqtheta0} and looks at the
$\theta_0>0$ solutions, the situation is the following: in a wide region, one
finds two possible alternative solutions that look very much alike to the F
and F' phases discussed for the $p\to\infty$ limit. However, near the left and
upper corner in the phase diagram, there is one single ferromagnetic solution
which is neither F nor F' but something intermediate. In fact, for $k$ even,
one can find paths through which the magnetization evolves continuously from
the F' to the F magnetizations without crossing any transition on the way, see
figure \ref{fig:magk2_class_0.1}. However, when $p$ is high and $k$ is odd, transitions between the F and F' phases
cannot be avoided, see figures \ref{fig:phase-diag-k5}, \ref{fig:magk2_class_0.1} and \ref{fig:mag_class_0.3}.

All this effect can be understood quantitatively coming back to
the discussion of the $p\to\infty$ ferromagnetic solutions. Each of the phases
were derived using the assumptions \eqref{eq:limF} for the F phase, and
\eqref{eq:limF'} for the F' phase. Now we discuss the validity of these
approximations for $p$ finite.

We begin with the F phase.  This phase was obtained by introducing
\eqref{eq:limF} in \eqref{eq:eqtheta0}. Since this equality is not strictly
true, we introduce it as an approximation $\sin^{p-2}\theta_0\approx 1$,
thus obtaining a new approximate equation \be
p\,s\,\lambda\cos\theta_0+k\,s\,(1-\lambda)\cos^{k-1}\theta_0-1+s\approx0.
\ee If we assume $\cos\theta_0\ll 1$, the solution is
\be\label{eq:solFk2}
\cos\theta_0\approx\frac{1-s}{s\caja{p\,\lambda+k(1-\lambda)}}, \ee for $k=2$,
and \be \cos\theta_0\approx\frac{1-s}{s\,p\,\lambda}, \ee for $k>2$. That
means, that the F solution found for the $p\to\infty$ limit also appears for
finite $p$ whereas $\cos\theta_0\ll 1$, or \be\label{eq:cociente}
\frac{1-s}{s\,p\,\lambda}\ll 1.  \ee In particular, the smaller this quotient
\eqref{eq:cociente} is, the better approximation the F solution is.  We can
obtain the energies for finite $p$ by introducing this solution in
\eqref{eq:mine}. For $k=2$,
\begin{eqnarray}\nonumber
e^{k=2}_\mathrm{F}(s,\lambda)\approx-s\,\lambda\caja{1-\paren{\frac{1-s}{s\caja{p\,\lambda+2\,(1-\lambda)}}}^2}^{\frac{p}{2}-1}\\+s\,(1-\lambda)\paren{\frac{1-s}{s\caja{p\,\lambda+2\,(1-\lambda)}}}^{k}-(1-s)\paren{\frac{1-s}{s\caja{p\,\lambda+2\,(1-\lambda)}}},
\end{eqnarray}
and for $k>2$
\begin{eqnarray}\label{eq:fFca}\nonumber
e^{k}_\mathrm{F}(s,\lambda)\approx-s\,\lambda\caja{1-\paren{\frac{1-s}{\,s\,p\,\lambda}}^2}^{\frac{p}{2}-1}\\+s(1-\lambda)\paren{\frac{1-s}{s\,p\,\lambda}}^{k}-(1-s)\paren{\frac{1-s}{s\,p\,\lambda}}.
\end{eqnarray}

Next we study the F' solution. We consider the following approximation
\be\label{eq:limitsin} 
p\sin^{p-2}\theta_0\approx0.
\ee As before, if this is a good approximation,
\be\label{eq:solFpCApfin} \cos\theta_0\approx\caja{\frac{1-s}{k\,
    s\,(1-\lambda)}}^{\frac{1}{k-1}} \ee 
is one solution to \eqref{eq:eqtheta0}.  This solution is equal to
the one obtained for $p\to\infty$, 
\eqref{eq:solFpCA}. In other words, at this order of approximation, the
solution is exact at this limit.

We briefly discuss the range of validity of this F' solution
\eqref{eq:solFpCApfin} for $p$ finite. The approximation \eqref{eq:limitsin} is
valid for small values of $\theta_0$. With this idea we expand separately
the two terms in \eqref{eq:eqtheta0} around $\theta_0=0$, the
first term being
\begin{eqnarray}\nonumber
\sin^{p-2}\theta_0\cos\theta_0&=&p\,s\,\lambda\,\theta_0^{p-2}\caja{1-\frac{p+1}{6}\theta_0^2+O(\theta_0^{4})},
\end{eqnarray}
and the second term
\begin{eqnarray}\nonumber
k\,s\,(1-\lambda)\cos^{k-1}\theta_0-1+s
\\=k\,s\,(1-\lambda)\caja{1-\frac{k-1}{2}\theta_0^2+O(\theta_0^4)}-1+s.
\end{eqnarray}
The dependency on $\theta_0$ in the first term becomes irrelevant when
$p>3$, thus recovering the F' solution \eqref{eq:solFpCApfin}. When $p=3$, the
lowest power of $\theta_0$ appears in the first term,
leading to a different ferromagnetic solution, but not the F'. Clearly, the
higher $p$ (and the smaller $\theta_0$) is, the better is approximation
\eqref{eq:limitsin}.

In general, for intermediate values of $s$ and $\lambda$, the higher $p$ is,
the more exact the two ferromagnetic solutions, F and F', are. Then, since
both approximations represent opposite cases in the value of $m^x$ (or $m^z$),
a new first-order transition between both phases will appear on the line when
their two free energies become equal. However, for low values of $p$, or
alternatively for $s\to1$ or $0$, there will only be one ferromagnetic
solution, somewhere in between these two F and F' phases. This idea is well
illustrated in figures \ref{fig:magk2_class_0.1} and \ref{fig:mag_class_0.3},
where both the numerical solution to \eqref{eq:eqtheta0} and the analytical
predictions  \eqref{eq:solFk2} and \eqref{eq:solFpCA} are displayed.

This has straightforward consequences on the performance of the quantum
annealing algorithm: the higher $p$ is, the narrower will be the region where
annealing paths can avoid a first-order transition. In the limit of
$p\to\infty$, as was discussed before, there will be only one possible path,
but not effective as quantum annealing.

Concerning the transitions between the QP and ferromagnetic phases, we can
distinguish two kinds of transitions. First of all, the transitions between
the F and QP$^\pm$ phases will be first order, since the F phase is
characterized by a high value of $m^z$ whereas the paramagnetic solution has
$m^z=0$.  On the other hand, there is another transition between the F' and QP
phases that lies on the line where their two free energies become equal,
i.e. $s=1/[1+k(1-\lambda)]$. On this line, $m^x=1$ ($m^z=0$) for the two
phases. Furthermore, the F' solution is exact for $m^x=1$. Since the
magnetizations are continuous on this line, the transition between F' and QP is
of second order. Besides, it can be checked that there is a wide range of
this line where $e_\mathrm{F'}<e_\mathrm{F}$. Thus, this phase is the stable
one in the ferromagnetic phase. This second-order transition does not hamper
the QA performance and gives us a way to avoid the F-QP phase transition that
appeared when using the traditional QA approach.  It is important to point out
that this second-order transition appears for any value of $k$.

In Appendix \ref{sec:SA}, we describe a different, quantum-mechanical method to derive the same results.

\section{Energy gap}
As discussed in Introduction, the efficiency of the QA algorithm is closely
related to the behavior of the gap between the ground and first excited
states. As usual, this gap can be computed by direct diagonalization of the
problem Hamiltonian \eqref{eq-qa:H}.  Indeed, since the total spin $\V{S}$ is
 conserved during the evolution, the dimension of the problem is $N+1$. That
 means that the diagonalization matrices grow polynomially with the system
 size instead of exponentially as for generic quantum problems. However, still 
 computer resources limit this
computation to moderate sizes although such computations are useful
for some purposes~\cite{seki:12,jorg:10a}.   Here, we adopt an
  alternatively approach, this gap can be
computed in the thermodynamic limit $N\to\infty$ by the method described
in~\cite{filippone:11}. The main idea is to extend the semi-classical scheme
for the ground state by the consideration of quantum fluctuations around the
classical ground state. It is important to point out that this method can
only be applied in the case of finite gaps in the thermodynamic limit, as it
is the case away from the transition points themselves. In case of exponentially small ones, other methods such as
instantonic or WKB methods should be used~\cite{jorg:10a,bapst:12}.

It is most convenient to rotate the system by an angle $\theta_0$ around the
$y$ axis in order to bring the $x$ axis parallel to the semi-classical
magnetization, i.e.
\begin{equation}
\paren{\begin{array}{c} S_x\\S_y\\S_z \end{array}}=\paren{\begin{array}{ccc}
    -\sin{\theta_0}&0&\cos{\theta_0}\\0&1&0\\\cos{\theta_0}&0&\sin{\theta_0}\end{array}}\paren{\begin{array}{c}
    \tilde S_x\\\tilde S_y\\\tilde S_z \end{array}}.
\end{equation}
We rewrite the Hamiltonian \eqref{eq:Hnew} in terms of these new variables
$\tilde S^\alpha$, obtaining
\begin{eqnarray}\label{eq:Hrot}\nonumber
&\hat{H}(s,\lambda)=-s\,\lambda\, N\caja{\frac{2}{N}\paren{\cos\theta_0\,\tilde
    S_x+\sin\theta_0\,\tilde S_z}}^p&\\\nonumber&+s\,(1-\lambda)\,N\,\caja{\frac{2}{N}\paren{-\sin\theta_0\,\tilde S_x+\cos\theta_0\,\tilde S_z}}^k&\\&-2\,(1-s)\paren{-\sin\theta_0\,\tilde S_x+\cos\theta_0\,\tilde S_z}.&
\end{eqnarray}

Now, we add quantum fluctuations to the system by means of the
Holstein-Primakoff transformation
\begin{eqnarray}\label{eq:HP}
\tilde S_z=\frac{N}{2}-a^\dagger a,\,\,\,\,\,\,&\tilde S_+=\displaystyle(N-a^\dagger
a)^{1/2} a=\tilde S_-^\dagger,
\end{eqnarray}
where $a$ is a boson annihilation operator that satisfies $[a,a^\dagger]=1$.
When quantum fluctuations are small relative to the classical state, i.e. for $N\gg\mean{a^\dagger a}$, we can use a simpler expression \be\tilde
S_x\approx\displaystyle\frac{\sqrt{N}}{2}(a+a^\dagger).\ee We introduce these
transformations into the Hamiltonian \eqref{eq:Hrot} and expand the three
different terms in powers of $1/N$. Thanks to the previous rotation, the
coefficient in $1/\sqrt{N}$ vanishes. We keep terms up
to $1/N$ and group together all the coefficients with the same power of $N$.
The result is 
\be\label{eq:Hhp}
H(\gamma,\delta)=N\,e+\gamma+\gamma\caja{(a^\dagger)^2+a^2}+\delta a^\dagger
a.  \ee 
The term for $N^1$ is nothing but the ground energy obtained before in
\eqref{eq:mine}, \be
e\equiv-s\lambda\sin^p\theta_0+s(1-\lambda)\cos^k\theta_0-(1-s)\cos\theta_0.
\ee 
The coefficients $\delta$ and $\gamma$ are given as
\begin{eqnarray}\nonumber
\delta&\equiv&-s\,\lambda\caja{p(p-1)\sin^{p-2}\theta_0\cos^2\theta_0-2\,p\,\sin^p\theta_0}\\&+&s\,(1-\lambda)\caja{k(k-1)\sin^{2}\theta_0\cos^{k-2}\theta_0-2\,k\,\cos^k\theta_0}+2(1-s)\cos\theta_0,
\end{eqnarray}
and 
\begin{eqnarray}
\gamma\equiv-\frac{s\,\lambda\,p(p-1)}{2}\sin^{p-2}\theta_0\cos^2\theta_0+s\,(1-\lambda)
  \frac{k(k-1)}{2}\sin^{2}\theta_0\cos^{k-2}\theta_0.
\end{eqnarray}
We need to diagonalize this Hamiltonian in order to compute the first excited
state by the Bogoliubov transformation
\begin{eqnarray}
a=\cosh\frac{\Theta}{2}\,b+\sinh\frac{\Theta}{2}\,b^\dagger,&\displaystyle
a^\dagger=\cosh\frac{\Theta}{2}\,b^\dagger+\sinh\frac{\Theta}{2}\,b,
\end{eqnarray}
where $b$ is a new bosonic annihilation operator satisfying $[b,b^\dagger]=1$.
Using this transformation, we can eliminate the coefficient of
$\caja{(b^\dagger)^2+b^2}$ by choosing the angle $\Theta$ as
\begin{eqnarray}\nonumber
 \tanh\Theta=-\frac{2\gamma}{\delta}\equiv\epsilon.
\end{eqnarray}
With this choice, the Hamiltonian can be written as
\be\label{eq:HB}
H(\gamma,\delta)=N\,e+\gamma+\frac{\delta}{2}\paren{\sqrt{1-\epsilon^2}-1}+\Delta\ b^\dagger b,
\ee
with \be\label{eq:Delta}
\Delta=\delta\sqrt{1-\epsilon^2}.
\ee
The Hamiltonian is diagonal in
$b^\dagger b$. The energy gap in the $N\to\infty$ limit between the ground 
and first excited states is  $\Delta$.

\begin{figure}
\begin{center}
 \includegraphics[angle=270,width=0.45\columnwidth,trim=10 20 10
   20]{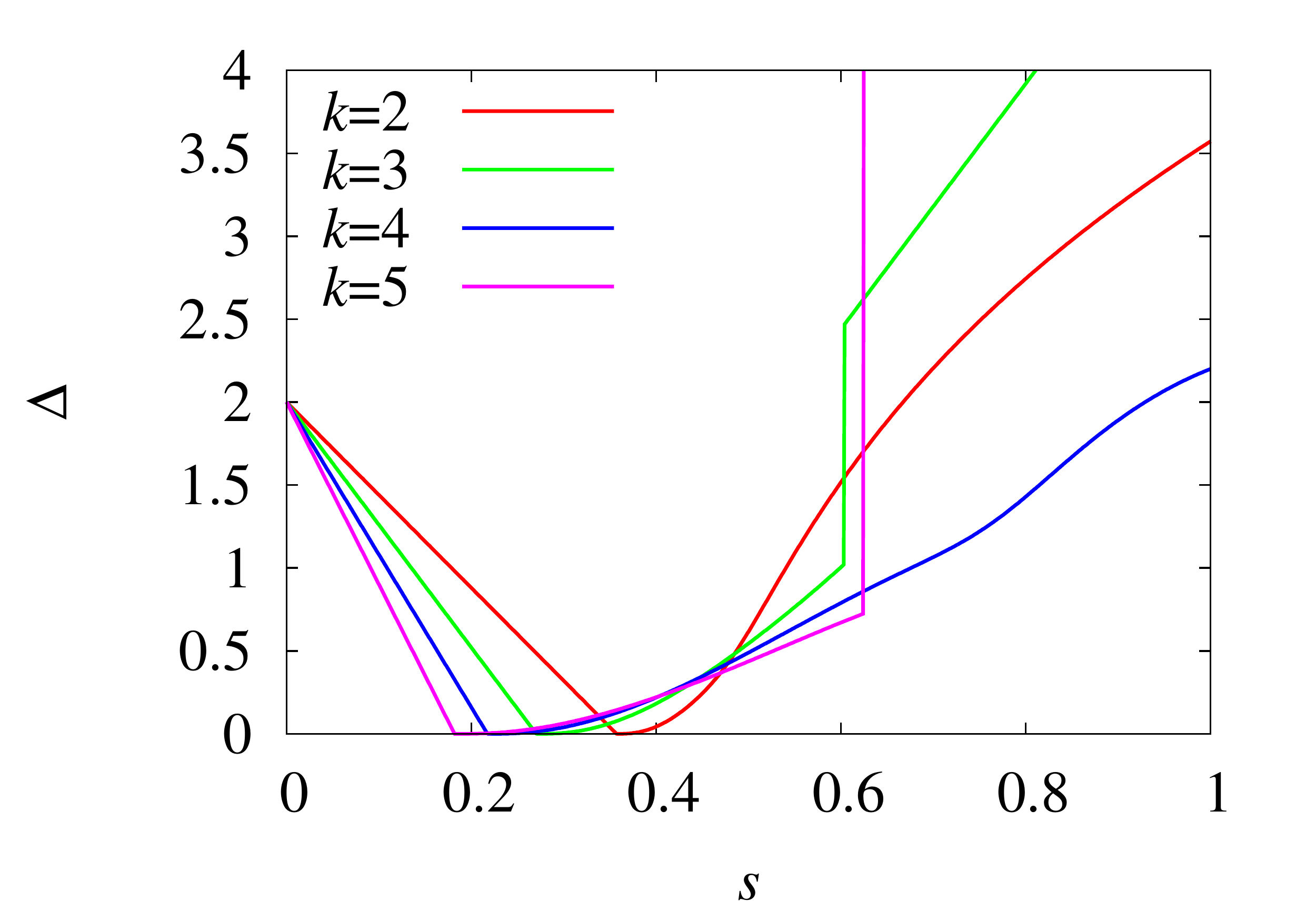}
 \includegraphics[angle=270,width=0.45\columnwidth,trim=10 20 10
   20]{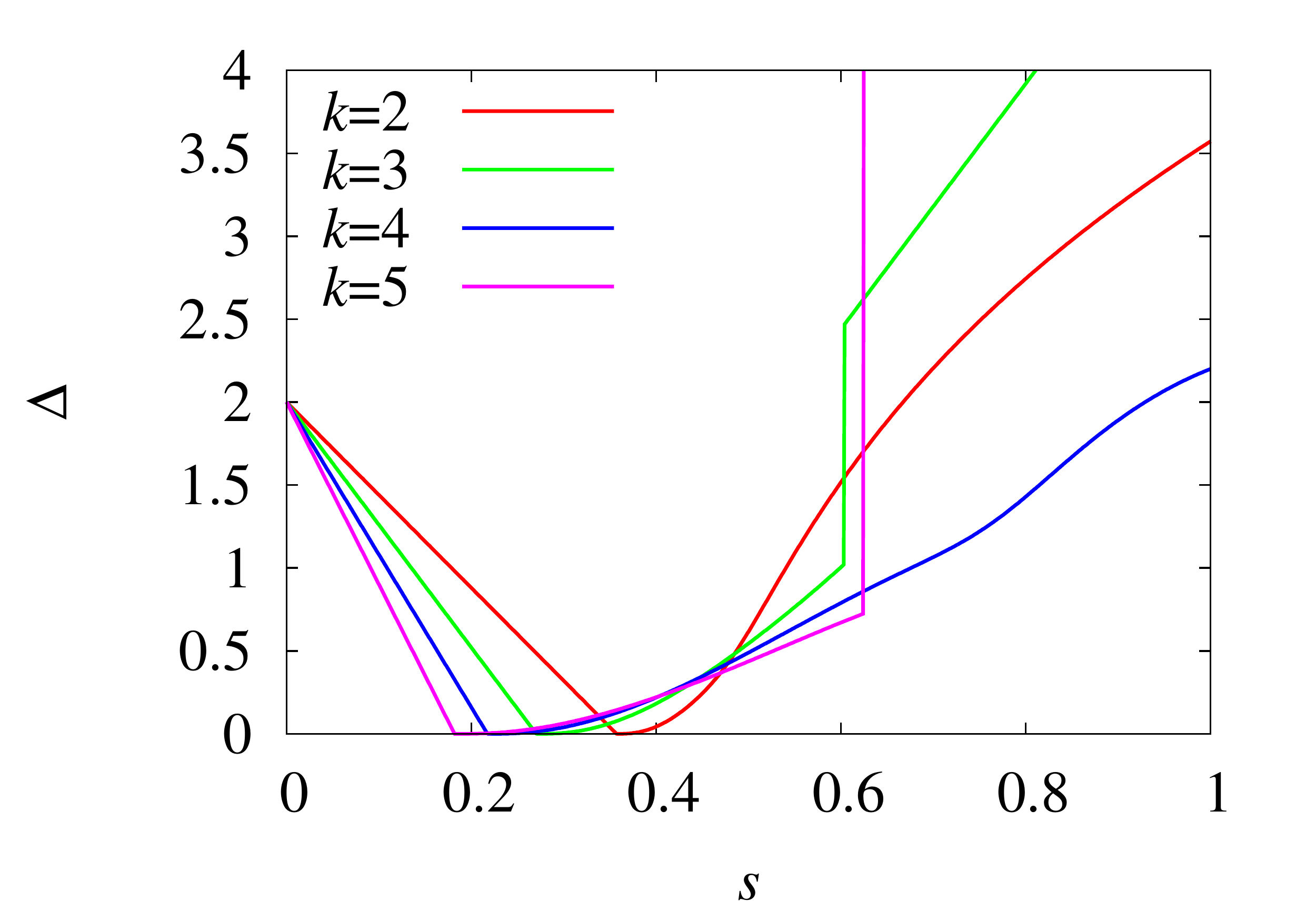}
\caption{Energy gap for $p=11$ as a function of $s$ for $\lambda=0.1$ (left) and
  $\lambda=0.3$ (right) for several values of $k$.}\label{fig:overlap}
 \end{center}
\end{figure}

Using the values $\theta_0$ previously obtained solving
 \eqref{eq:eqtheta0}, we can compute the energy gap for our
system. We show the data for $p=11$ and $\lambda=0.1$ and $0.3$  for
different values of $k$ in figure \ref{fig:overlap}. As was suggested in the
magnetization data in the previous section for $\lambda=0.1$ (figures
\ref{fig:phase-diag-k2} to
\ref{fig:phase-diag-k5}), no first-order
transition F-F' is observed through the energy gap. The gap vanishes continuously on the
second-order transition line but present no further jumps later, but the ones
related to the F-QP$^-$ that always take place in the odd-$k$ cases. On the
contrary, when $\lambda=0.3$, the jumps in the gap appear for all the $k$'s at
the place where we observed the F-F' transition before.

In the thermodynamic limit, the gap vanishes at a single point of first-order transition and remains finite away from this point.  The single point of vanishing gap is hard to see by the present method, which results in an apparent simple jump in the gap at a first-order transition as seen in figure \ref{fig:overlap}.

\section{Overlap of the ground-state wave functions}
It has been suggested in~\cite{bapst:12} that
the reason for the antiferromagnetic interaction, the $k=2$ case in
 \eqref{eq:Vk}
introduced by Seki and Nishimori in \cite{seki:12}, to work better than the
transverse field interaction only is related to the large overlap
between the ground states of the  Hamiltonians $\hat{V}_{k=2}$ and $\hat{H}_0$.
In this section, we will discuss the properties of these
different states, concluding that, even thought the overlap is important,
it is not the decisive factor that makes the strategy to succeed.

The ground state of $\hat V_\mathrm{TF}$ is the one where
all the spins are aligned along the $x$ axis,
$\ket{\phi_\mathrm{TF}}=\otimes_{i=1}^N\ket{\uparrow}_i^x$. If we denote the
ground state of $\hat{H}_0$, as
$\ket{\phi_0}=\otimes_{i=1}^N\ket{\uparrow}_i^z$,  the overlap between $\ket{\phi_\mathrm{TF}}$
and $\ket{\phi_0}$ decreases exponentially with $N$ as $2^{-N}$, as can easily
be seen from the elementary relation $\ket{\uparrow}_i^x=\paren{\ket{\uparrow}_i^z+\ket{\downarrow}_i^z}/\sqrt{2}$.

The overlap computation becomes a little more complicated for the ground state
of $\hat{V}_k$. The ground state for this term depends on the value of
$k$. Indeed, if $k$ is odd, the ground state is the one where
all the spins are aligned along the $x$ axis, but towards the negative
direction, i.e.  $\otimes_{i=1}^N\ket{\downarrow}_i^x$. Then, the overlap with
$\ket{\phi_0}$ for the $k$ odd case will be exponentially suppressed as
$2^{-N}$ as in the case of $\hat{V}_\mathrm{TF}$. Thus, the argument
in~\cite{bapst:12} does not apply directly since we can avoid first-order
transitions even in this case of $k$ odd, in spite of the very small overlap
of the ground state for $\hat{H}_0$ and $\hat{V}_k$.

The ground state for the $k$ even case needs some care to be analyzed. We
compute it in Appendix \ref{sec:Vkground}. We show there that that the overlap
is indeed higher for $k$ even. The antiferromagnetic interactions is a
particular case, $k=2$. In fact, the overlap displays an algebraic decay as
the system size increases, i.e. $\sim 1/\sqrt{N}$.

We conclude that the overlap is not the main ingredient that makes
the present method to succeed.

\part{Conclusions}
\chapter{Conclusions \label{chap:conclusiones}}

In this thesis, we have tackled the general problem of describing complex
systems. The name complex refers to a large amount of degrees of freedom and
the difficulty of finding simple recipes to describe them. The extremely large
amount of possible states draws a complex free energy landscape, which has a
common consequence for many diverse systems: an excessive slow dynamics. This
sluggish evolution has a direct consequence in experiments: these systems must
be regarded to be always out of equilibrium. Finding a rational way to
approach this kind of problems is one of the central problems in the modern
theory of condense matter physics.

As discussed many times in this dissertation, nature lives in a nonequilibrium
world, which crashes with the standard theoretical approach, that needs
equilibrium in order to cancel out the chaotic individual behavior. For this
reason, computer simulations are requested to establish a bridge between these
two worlds.  Besides, in the last years it has been proposed a novel approach
that provides a quantitative relation between both worlds by interchanging some
degrees of freedom that one can control in a computer: finite times in
nonequilibrium simulations with finite sizes in equilibrium
simulations. Indeed, what was regarded as an annoying inconvenience for many
years can be used now for writing a real dictionary between the theoretical
calculations based on an equilibrium eternally unachievable and experiments
that last finite times.

In this thesis we have focused on this final goal, with emphasis on one of the
two parts, that is, on laying the foundations of this dictionary by
characterizing precisely the equilibrium phase at finite system sizes. Due to
the extreme slowness of the dynamics, this mission is extraordinarily complex
and we had to face it by considering several perspectives:
\begin{enumerate}
\item By model building. Indeed, as theoretical physicists, it is of major
  importance to find models simple enough to allow some analytical
  predictions but complex enough to still suffer the phenomenon we are
  interested on.
\item By brute force, that is, by means of large-scale simulations, with the
  help of large computational facilities.
\item By the design of optimized algorithms that allow us to explore the
  topography of the complex landscape. This approach offers as well a major
  practical advantage. It helps us to find the flattest
  simulation path that allows us to speed the simulation.
\end{enumerate}
Regarding the definition of new models, we presented in
Chapter~\ref{chap:hypercube} a new mean field model, which at variance with
the rest of this kind, allows a natural definition of distance. Then, being
mean field, it provides a direct way of investigating the coarsening process
in a replica symmetry breaking scenario, with magnitudes that can be compared
with experiments, like the magnetic domain's size.

Concerning the extensive simulations, let us note that the results presented
all over the thesis demanded large computational facilities (conventional
computer clusters, supercomputing facilities and dedicated computers) as well
as the implementation of modern simulation techniques like multispin
coding. It specially remarkable, that for the temperature chaos work
(Chapter~\ref{chap:chaos}), we reanalyzed data obtained with one year of
non-stop production of Janus, a special-purpose computer many thousands faster
than a conventional computer. These brute-force simulation gave us access to
unprecedentedly large configurations in the $3D$ Edwards-Anderson model
thermalized up to very low temperatures, which was crucial to identify without
any doubt the temperature chaos effect in a simulation, as well as to lay the
foundations for the size dependency characterization.

Apart from large computation facilities, we also followed an alternative
approach to speed up the simulations: to design clever optimized algorithms
that speed up the dynamics. The key lies precisely on the rugged free-energy
landscape. Then, if one were able to identify the topography this landscape, one could find
the best path to go from one minimum to the other. This was precisely our aim in
all Part~\ref{part:colloids} of the thesis and in work on the quantum
annealing algorithm described in Chapter~\ref{chap:QA}.

Indeed, when we started working with colloids, about the beginning of my PhD,
our objective was to describe the phase diagram of highly polydisperse
systems. Previous works had failed in characterizing the disordered solid
equilibrium phase, and our idea was to apply the successful microcanonical
algorithm~\cite{martin-mayor:07} to this problem (this approach was described
in Chapter~\ref{chap:micro}). Initially, the we thought that the difficulty
was the glass transition, but after some months of intense simulation, we
identified another harder problem and unsolved in the literature, the
free-energy barriers in first order transitions in off-lattice systems. With
this idea in mind, we moved back to the simplest possible system of this kind,
to identify the order parameters that could allow us to explore softly the
free-energy landscape, and thus, to find a flat trajectory free from
metastabilities. Our successful solution to the problem was discussed in
Chapter~\ref{chap:HS} in the context of hard spheres crystallization, a
simpler model but carrying still the same problem.

The problem with the quantum annealing algorithm is rather different but still
very related to the rest of the thesis. The main problem to build a quantum
computer based on this kind of computation is precisely the adiabaticity
condition of the algorithm. Indeed, the times needed to keep the system
permanently in equilibrium grow exponentially with the system size if a
quantum first order transition is found on the annealing trajectory. This
problem with the adiabatic condition is the same problem considered all along
the rest of the thesis when talking about thermalization times. The solution,
this time, was to add an additional driver term, and control the trajectory
with two parameters. With this idea, we could map the free energy, and show the
existence of annealing trajectories that avoids the first order transition.

After this general discussion, we extend separately in the following sections
the conclusions for each of the chapters presented in this dissertation.

\section{Spin Glasses}
\subsection{Hypercube model}
We have studied a spin glass model in the $D$-dimensional unit hypercube in
the limit of large $D$, but with finite coordination number. We have shown
that any short range model in such a lattice will behave as a mean field model
in the thermodynamic limit (that coincides with the large $D$ limit). An
important advantage of this model is that it has a natural notion of spatial
distance.

We have argued that any statistical mechanics model on the hypercube with
random connectivity would be afflicted by huge finite size effects, for purely
geometrical reasons. The obvious cure has consisted in restricting the
connectivity graphs to those with a fixed number of neighbors. Unfortunately,
constructing such graphs is far from trivial. We have generated a subset of
them by means of a simple dynamic Monte Carlo. In this way, we obtain sets of
graphs that are isotropic. We have checked that the Edwards-Anderson model
defined over these finite connectivity hypercubes verify some consistency
checks, including comparison with the analytically computable correlation
function in the paramagnetic phase.

We have numerically studied the nonequilibrium dynamics in the spin glass
phase. The three main features found were: (i) aging dynamics consists in the
growth of a coherence length, much as in 3D systems, (ii) the scaling of the
two times correlation function implies infinitely many time-sectors, and (iii)
the $p^4$ propagator has been observed. In addition, we have studied the
finite size effects in our model, finding that a naive finite size scaling
ansatz accounts for our data.

From the static point of view, it is most probable, almost a theorem, that
our model suffers replica symmetry breaking. Hence, it provides an
interesting playground to study nonequilibrium dynamics on RSB systems.
An interesting possible extension of the present study would be the
computation of quantities that are directly measurable in experiments, and/or
of experimental cooling protocols.

\subsection{Temperature chaos}

We have characterized the temperature chaos in the $D\!=\!3$ Ising spin glass
as a rare-event driven phenomenon. When it occurs, its effects are strong, and
can be felt even at the shortest length scales, as confirmed by the
two-temperatures spatial correlation function. We argue that this
characterization was inaccessible to the statistical analysis employed in
previous works. In fact, two ingredients were crucial to obtain this
conclusion. First, the JANUS supercomputer gave us access to unprecedentedly
large configurations, well thermalized up to very low temperatures (remarkable
both for system sizes up to $L=32$ and for the low
temperatures~\cite{janus:10,janus:10b}). And second, we introduce new tools of
statistical analysis, based on a large-deviations functional. 

With this approach, we were able to quantify the size-dependencies using this
large-deviation functional. This step is crucial to find a time-length
dictionary \cite{franz:98,janus:10,janus:10b,barrat:01} for
temperature-varying protocols, which paves the way to design a protocol that
allows to detect the temperature chaos in a real experiment.

A surprising outcome of our finite size analysis is that the chaotic length
scales with system size as $\xi_\mathrm{C} \!\propto\! L^a$, with $a\approx
0.4$: divergent in the thermodynamic limit, yet much smaller than $L$. This
duality will probably be important to interpret the somehow contradictory
memory and rejuvenation effects~\cite{jonason:98}.  In fact, although the
$\xi_\mathrm{C}\!\propto\!L^a$ scaling follows from a $L\to\infty$
extrapolation (which is tricky even in mean-field) we now know that the
extrapolation relevant for experiments is rather to $L\sim 100$ lattice
spacings~\cite{janus:10}.

\section{Colloids}
\subsection{Polydisperse soft spheres}
We have studied in the microcanonical ensemble a
soft-spheres model for liquids and colloids with a $24\%$
polydispersity. Extrapolating by \ac{FSS} to the thermodynamic
limit the results obtained from the Maxwell construction in finite
systems, we show that the critical temperature for the
amorphous-crystal phase-separation is {\em below} the dynamic glass
transition, which makes dynamically difficult (although not impossible
\cite{colloids:Zaccarelli09}) to observe such phase-separation. 

At low temperatures the system divides spatially into an amorphous and
a crystalline part, in agreement with previous
findings~\cite{poly:Fernandez07}. The phase-separated amorphous is a
stable fluid {\em below its dynamic glass temperature,\/} which is an
optimal candidate to suffer a thermodynamic glass transition. On the
other hand, the phase-separated solid displays crystalline
order. Polydispersities on the coexisting amorphous and solid are
smaller than in the fluid. In fact, particles distribute spatially
according to their size following a complex pattern not described by
any fractionation scenario known to us.

We were able to obtain the equilibrium solid phase, but only for smaller
system sizes that what we were seeking. Indeed, we applied the microcanonical
algorithm with the hope of avoiding metastabilities, but the existence of
phase separation ruined our approach. In fact, it was precisely the solution
to this problem which encouraged our research on hard spheres
crystallization.

\subsection{Hard spheres crystallization}

We have introduced a tethered
MC~\cite{fernandez:09,martin-mayor:11} approach to HS crystallization.
We go continuously from the fluid to the crystal by varying a reaction
coordinate (a blend of two {\em global} bond-orientational order
parameters).  Tethered MC provides a major simplification to umbrella
sampling, which makes it possible to study multi-constrained free
energies.  At variance with previous methods, our simulations
equilibrate (i.e.  we find results independent of the starting
particle configuration), not only for the formation of the
space-filling crystal, but even for the more difficult case of mixed
states with fluid-crystal interfaces.  Our estimation of the
coexistence pressure is, by far, the most accurate to date. That of
the interfacial free energy is compatible with most (but not all)
recent determinations. Should one wish to reach larger $N$, the
tethered strategy would easily accommodate additional order
parameters.  The method can also be generalized to other simple
liquids, or to investigate the glass transition.

\section{Quantum Annealing}
We have analyzed the reason for the failure of the traditional annealing with
a transverse-field term in the infinite-range ferromagnetic $p$-spin model.
We have shown that it is possible to find annealing trajectories that avoid
the crossing of first-order transitions thanks to the introduction of a second
driver term in the problem, which may be due to the multiple spin flips in the $z$-basis caused by the second term as was the case in~\cite{suzuki:07}.  This additional term favors the appearance of a
second-order transition that does not hamper the annealing performance.  A
whole family of possible candidates has been studied and we conclude that the
solution to the problem presented by Seki and Nishimori~\cite{seki:12} is a
special case of a more general additional quantum term. The main properties of these additional terms have also been
discussed with the conclusion that the properties of the ground states of the
diverse terms in the Hamiltonian are not a decisive factor to make the quantum
annealing fail or succeed.

\addtocontents{toc}{\protect\setcounter{tocdepth}{1}}
\appendix
\titleformat{\chapter}[display]
{\bfseries\LARGE} {\filleft\MakeUppercase{\chaptertitlename}
\Huge\Alph{chapter}} {2ex} {\titlerule
\vspace{1.5ex}%
\filright}
[\vspace{1.5ex}%
]
\renewcommand{\thesection}{\texorpdfstring{\textsc{\Alph{chapter}}.\oldstylenums{\arabic{section}}}{\Alph{chapter}.\arabic{section}}}
\renewcommand{\thesubsection}{\texorpdfstring{\textsc{\Alph{chapter}}.\oldstylenums{\arabic{section}.\arabic{subsection}}}{\Alph{chapter}.\arabic{section}.\arabic{subsection}}}
\renewcommand{\thesubsubsection}{\texorpdfstring{\textsc{\Alph{chapter}}.\oldstylenums{\arabic{section}.\arabic{subsection}.\arabic{subsubsection}}}{\Alph{chapter}.\arabic{section}.\arabic{subsection}.\arabic{subsubsection}}}
\renewcommand{\theequation}{\textsc{\alph{chapter}}.\oldstylenums{\arabic{equation}}}
\renewcommand{\thefigure}{\textsc{\alph{chapter}}.\oldstylenums{\arabic{figure}}}
\renewcommand{\thetable}{\textsc{\alph{chapter}}.\oldstylenums{\arabic{table}}}

\part{Appendices}
\renewcommand{\sectionmark}[1]{\markright{\textit{\Alph{chapter}.\oldstylenums{\arabic{section}}}\ --- #1}}
\chapter{Analytical calculations on the hypercube}

\section{On the Bethe approximation in a ferromagnet}\label{ap:bethe}
\subsection{The Bethe approximation}
 The Bethe approximation is a refinement over the mean field,
 for this reason, we will begin the discussion applying the \ac{MF} approximation to
 an Ising ferromagnet.

In the standard \ac{MF} approximation, spins are assumed to be uncorrelated. The
probability distribution function is thus factorized ($m_i\equiv\mean{S_i}$)
\be\label{eq:probfac} P(S)=\prod_i P_{m_i}(S_i),\ee where
\be \label{eq:Pm}P_{m_i}(S)=\frac{1+m_i}{2}\ \delta_{S,1}+\frac{1-m_i}{2}\ \delta_{S,-1}.\ee
As usual, the equilibrium solution will be the one that minimizes the
free-energy functional, defined as follows ($\beta=1/T$)\be \Phi
[P]=\mean{H}_P-\frac{S[P]}{\beta},\ee where \be\mean{H}_P=-J\sum_{<i.j>}m_i
m_j,\ee and
\be\label{eq:entropy} S[P]=-\mean{\ln P(S)}_P=-\sum_i \mean{\ln P_i(S)}_P=-\sum_i
s(m_i),\ee with \be
s(m_i)=\frac{1+m_i}{2}\ln\frac{1+m_i}{2}+\frac{1-m_i}{2}\ln\frac{1-m_i}{2}.\ee
Note that the entropy \eqref{eq:entropy} is additive because the probability
\eqref{eq:probfac} is factorized.

The condition for minima leads us to \be\frac{\partial\Phi}{\partial
  m_i}=0\,\Rightarrow\, m_i=\tanh\paren{\beta \sum_{j}J_{ij}\ m_j}.\ee In an
Ising ferromagnet, the $J_{ij}$ are equal to $J$ if spins $i$ and $j$ are
nearest neighbors and zero elsewhere.  The actual minimum of $\Phi$
corresponds to constant magnetization $m_i=m$ for all spins $i$. Then, the
magnetization satisfies the equation \be\label{eq:mf} m=\tanh\paren{\beta J z
  m},\ee where $z$ is the coordination number. This equation predicts a a
transition at the critical point $\beta_c=1/z$. However, this solution is not
very satisfactory since it predicts a phase transition no matter the dimension
of the system, and we know that there is no transition in the one dimensional
Ising model.

This problem can be surpassed by looking at the system locally and applying
the cavity approach. The magnetization of spin $\sigma$ can be computed as a
function of the nearby spins $\tau_i,\ i=1,\ldots,n$, being $n$ the
coordination number of $\sigma$. Let us remove the spin $\sigma$. There is now
a cavity in the system surrounded by the spins $\tau$. We assume that the
spins are not correlated (which is the Bethe approximation) and the
magnetization in the cavity, $m_\mathrm{C}$, is obtained with
\eqref{eq:Pm}. Now, we add back the spin $\sigma$. Thus, the probability for
this spin is given by
\be \label{eq:psigma}P(\sigma)=\frac{F(\sigma)}{F(1)+F(-1)},\ee where \be
F(\sigma)= \sum_{\lazo{\tau}} P_{m_\mathrm{C}}[\tau] e^{\beta
  J\sigma\sum_{i=1}^z \tau_i}=\caja{\cosh (\beta J
  \sigma)+m_\mathrm{C}\sinh(\beta J\sigma)}^n.\ee The probability
\eqref{eq:psigma} is of the form \eqref{eq:Pm}. Hence, it suffices to compute
$\mean{\sigma}$.  Since $\sigma=\pm 1$ in the Ising model, \be\cosh (\beta J
\sigma)+m_\mathrm{C}\sinh(\beta J\sigma)=\cosh(\beta J)\frac{e^{\sigma
    A(\beta,J,m_\mathrm{C})}}{\cosh A(\beta,J,m_\mathrm{C})},\ee where \be
A(\beta,J,m_\mathrm{C})=\tanh^{-1}\caja{\tanh(\beta J)m_\mathrm{C}}.\ee
Thereby, the magnetization of spin $\sigma$ can be obtained as usual
\be\label{eq:m} m=\mean{\sigma}=\tanh
\caja{nA(\beta,J,m_\mathrm{C})}=\tanh\lazo{n \tanh^{-1}\caja{\tanh\paren{\beta
      J}m_\mathrm{C}}}.\ee If we now increase the cavity removing one more
spin, $\tau_i$, keeping fixed the magnetization to $m_\mathrm{C}$ and still
considering the spins uncorrelated, we can obtain a relation for
$m_\mathrm{C}$ using eq. \eqref{eq:m} and imposing a self-consistent condition
\be\label{eq:mc}m_\mathrm{C}=\tanh\lazo{(n-1)
  \tanh^{-1}\paren{\tanh\paren{\beta J}m_\mathrm{C}}}.\ee Once $m_\mathrm{C}$
is known, $m$ can be calculated by means of \eqref{eq:m}.  \eqref{eq:mc} is
more satisfactory than \eqref{eq:mf}, in fact, it predicts no transition in
one dimension (if the connectivity is $n=2$, no $\beta$ would ever satisfy
eq. \eqref{eq:mc}).

We can now come back to the previous discussion on distance between spins, see
eq. \eqref{eq:loopshy}. It now becomes clear that the Bethe approximation is
correct for the hypercubes (and for random Poisson lattices). Then, in the
paramagnetic phase, once a bond has been removed, the two neighboring nodes
are separated by a large distance, i.e. $O(D)$. In fact, it becomes exact when
$D\rightarrow\infty$ since both spins are infinitely far away and are
statistically uncorrelated.
\subsection{Calculation of the critical temperature $K_\mathrm{c}$ }\label{sec:Kc}
We fix the starting point in eq. \eqref{eq:mc}, $(K\equiv \beta J)$
\be\label{eq:mcav}m_\mathrm{C}=\mean{\tanh\lazo{(n-1)
    \tanh^{-1}\paren{\tanh\paren{K_\mathrm{c}}m_\mathrm{C}}}}_1,\ee where the
$\mean{\ }_1$ refers to an average over the coordination number (remember that
$n$ is not necessarily fixed in our model). We seek solutions for
$m_\mathrm{C}$. These solutions must be around $m_\mathrm{C}\rightarrow 0$ we
can expand the right-hand side term in powers of $m_\mathrm{C}$. Introducing
\begin{eqnarray}\tanh x=x+{\cal O}(x^3) &\text{and}&\tanh^{-1}x=x+{\cal O}(x^3),\end{eqnarray}
in \eqref{eq:mcav}, we get
\be m_\mathrm{C}\approx\ m_\mathrm{C}(\mean{n}_1-1)\tanh K_\mathrm{c}.\ee
This equation can be solved, leading to
\be\label{eq:kcinter}
K_\mathrm{c}=\tanh^{-1}\paren{\frac{1}{\mean{n}_1-1}}.
\ee
Although it seems really counterintuitive, $\mean{n}_1$ is different in the two sets of graphs we have been discussing. While in the fixed-$n$ graphs it is 6, in the random-$n$ graphs it is function of $D$, let us explain why.

\subsubsection{Random connectivity graphs}
In order to compute $\mean{n}_1$, let us describe  carefully how eq. \eqref{eq:mc} was obtained. We picked one spin and selected one of its neighbors. Hence, we are asking which is the mean number of neighbors of a spin of which we happen to know for sure that it has a particular neighbor. We write down these ideas in the following way: the coordination number is given by
\be n=1+m, \ee
where $m$ is the number of active links among the $D-1$ remaining
ones. Thereby the
coordination number $n$ is necessarily higher or equal to one. The distribution function
of $m$ is then
\be p(m)=\paren{\begin{array}{c}D-1\\ m\end{array}}
\paren{\frac{z}{D}}^m\paren{1-\frac{z}{D}}^{D-1-m},\ee
where $z=6$.
One can average $n$ using this probability distribution function,
getting:\footnote{Note that, in the large $D$ limit, the number of neighbors of a site picked at random is $\mean{n}=6$, yet for one of its neighbors is $\mean{n}_1=7$.} 
\be\mean{n}_1(D)=1+\overline{m}=1+(D-1)\frac{z}{D}=z+1-\frac{z}{D}.\ee
Now we have all the necessary ingredients to calculate $K_\mathrm{c}$ for a given
dimension $D$, just plugging the value for $\mean{n}_1(D)$ in
eq. \eqref{eq:kcinter} one gets
\be \label{eq:KcDap}K_\mathrm{c}(D)=\tanh^{-1}\paren{\frac{D}{z\paren{D-1}}}.\ee
We present values of $K_\mathrm{c}$ for certain dimensions in Table
\ref{tab:KcD}.\footnote{Note that if we had chosen $\mean{n}_1=6$,
the  transition should be in $K_\mathrm{c}(D)=0.20273$ for all $D$. Then, both
connectivity descriptions, as far as $K_\mathrm{c}^\infty$ is concerned, are not
  equivalent in the thermodynamic limit.}

\begin{table}
\begin{center}
\begin{tabular}{cc}
$D$&$K_\mathrm{c}$\\\hline
6&0.20273\\
8&0.19283\\
10&0.18735\\
12&0.18386\\
14&0.18145\\
16&0.17969\\
18&0.17834\\
$\vdots$&$\vdots$\\\hline
$\infty$&0.16824\\
\end{tabular}
\end{center}
\caption{Values for $K_\mathrm{c}$ for certain dimensions $D$ obtained using the Bethe approximation.}\label{tab:KcD}
\end{table}
One must recall that relation \eqref{eq:mc} is only exact for infinite
dimension. Since correlations between spins vanish with $O\paren{D^{-1}}$, we
must expect corrections of the same order  to eq. \eqref{eq:mc}. Then there will also
be additional corrections to $K_\mathrm{c}^{\infty}=\tanh\paren{1/z}$ with $D$
than those presented in \eqref{eq:KcDap} and Table \ref{tab:KcD}. Then, it
would be certainly more appropriate to work with the following expression for
$K_\mathrm{c}$ instead \be
K_\mathrm{c}(D)=K_\mathrm{c}^{\infty}+\frac{a_1}{D}+\frac{a_2}{D^2}+\cdots.\ee
If we notice that dimension is related to the number of spins by means of \be
D=\frac{\log N}{\log 2},\ee we realize that we must expect logarithmic
corrections to $K_\mathrm{c}^{\infty}$ in the number of spins $N$. This
problem is then really hard. Depending on the actual values of coefficients
$a_i$ these corrections can be huge. Then, this random connectivity model
suffers from such strong finite size effects that make it not suitable for
numerical computations at finite $D$.
\subsubsection{Fixed connectivity graphs}
This strong dependency of $K_\mathrm{c}$ with $D$ in the random connectivity
graphs is the reason that encouraged us to study the systems within the
$n$-fixed ensemble, although its graphs are more difficult to generate. In
these graphs $n=z$ and then eq. \eqref{eq:kcinter} reads as
\be\label{eq:kczfixedap} K_\mathrm{c}=\tanh^{-1}\paren{\frac{1}{z-1}},\ee if
$z=6$, we get $K_\mathrm{c}\approx 0.20273$.  This means that the expectation
number for $K_\mathrm{c}$ provided by the Bethe approximation does not depend
on $D$. For this reason, we should expect less corrections with $D$ in
$K_\mathrm{c}^\infty$ than in the previous case.  In fact, we should only find
corrections associated to the validity of the Bethe approximation for finite
$D$. In other words, the corrections are only due to the short loops.

\section{High temperature expansion}\label{ap:HTE}
For sake of clarity, we will firstly discuss the calculations for the random connectivity hypercube. Results for the fixed connectivity model will be then obtained by minor changes.

Using the identity ($\beta=1/T$)
\be e^{\beta J_{\V{x}\V{y}}\sigma_{\V{x}}\sigma_{\V{y}}}=\cosh{\beta}\paren{1+J_{\V{x}\V{y}}\sigma_{\V{x}}\sigma_{\V{y}}\tanh{\beta}},\ee
we can write the partition function and the spin propagator as ($N_l$ is the
total number of links in the graph): 
\be\begin{array}{c}
 \displaystyle\frac{Z}{2^N(\cosh{\beta})^{N_l}}\!=\! \displaystyle\sum_{\lazo{\sigma}}\prod_{\mean{\V{z}\V{w}}}\paren{1\!+\!J_{\V{z}\V{w}}\sigma_{\V{z}}\sigma_{\V{w}}\tanh{\beta}},\\[0.6cm]
 \displaystyle\mean{\sigma_{\V{x}}\sigma_{\V{y}}}=\displaystyle\frac{\displaystyle\sum_{\lazo{\sigma}}\sigma_{\V{x}}\sigma_{\V{y}}\prod_{\mean{\V{z}\V{w}}}\paren{1+J_{\V{z}\V{w}}\sigma_{\V{z}}\sigma_{\V{w}}\tanh{\beta}}}{\displaystyle\sum_{\lazo{\sigma}}\prod_{\mean{\V{z}\V{w}}}\paren{1+J_{\V{z}\V{w}}\sigma_{\V{z}}\sigma_{\V{w}}\tanh{\beta}}}\,.
\end{array}\ee
The high-temperature expansion (see, for instance~\cite{parisi:88}), expresses the propagator as a sum over lattice paths that join the points $\V{x}$ and $\V{y}$, $\gamma_{\V{x}\rightarrow\V{y}}$:
\be\mean{\sigma_{\V{x}}\sigma_{\V{y}}}=Z^{-1}\sum_{\gamma_{\V{x}\rightarrow\V{y}}}Z_\gamma J(\tanh\beta)^{l_\gamma},\ee
where $l_\gamma$ represents the length of the path
$\gamma_{\V{x}\rightarrow\V{y}}$, $J$ is the product of the couplings,
$J_{\V{z}\V{w}}$, along the path, and $Z_\gamma$ is a restricted partition
function obtained by summing only over all closed paths that do not have any
common link with the path $\gamma_{\V{x}\rightarrow\V{y}}$.

However, when averaging over disorder, due to the randomness in the coupling signs,
$\overline{\mean{\sigma_{\V{x}}\sigma_{\V{y}}}}=0$.
The spin glass propagator is obtained instead by averaging over disorder $\mean{\sigma_{\V{x}}\sigma_{\V{y}}}^2$. Clearly, the sum will be dominated by those diagrams where the go and return path are the same (thus, $J_{\V{z}\V{w}}^2=1$):
\be\overline{\mean{\sigma_{\V{x}}\sigma_{\V{y}}}^2}=Z^{-2}\sum_{\gamma_{\V{x}\rightarrow\V{y}}}
Z_\gamma^2\caja{\tanh^2{\beta}}^{l_\gamma}=Z^{-2}\sum_{\gamma_{\V{x}\rightarrow\V{y}}}\kappa^{l_\gamma}\, Z_\gamma^2,\ee
where $\kappa=\tanh^2{\beta}$. In Bethe lattices, due to their cycle-less nature, $Z_\gamma^2/Z^2=1$ in the thermodynamic limit. Hence, we are left with the problem of counting the average number of paths of length $l_\gamma$ that join $\V{x}$ and $\V{y}$, $p(l_\gamma)$. From it, we obtain 
\be\label{eq:C4Ap} \hat{C}_4(r)=\binom{D}{r}\sum_{l_\gamma\ge r}p(l_\gamma)\kappa^{l_\gamma}.\ee
The sum is restricted to $l_\gamma\ge r$ because the length of the shortest path that joins $\V{x}$ and $\V{y}$ is given by their postman distance $r$.

In order to count the average number of paths, $p(l_\gamma)$, let us distinguish two cases: $l_\gamma=r$ and $l_\gamma>r$. The first will give the leading contribution in the large $D$ limit.

 The number of paths joining $\V{x}$ and $\V{y}$ in precisely $r$ steps is $r!$, because the $r$ steps are all taken along different directions and in a random order. For a given path, the probability of all the $r$ links be active is $(z/D)^r$. Hence
\be p\paren{l_\gamma=r}=\frac{z^r}{D^r}r!\,.\ee
Note that the $D^{-r}$ factor compensates exactly the divergence of the
$\binom{D}{r}$ in  \eqref{eq:C4Ap} (for large $D$).

In the case of $l_\gamma>r$, one has $l_\gamma=r+2k$, with $k>0$. Note that when $l_\gamma=r$ the path contains $r$ different directions (namely, the Euclidean components in which $\V{x}$ and $\V{y}$ differ). Each of these directions appear only once. However, when $l_\gamma>r$, other directions must be included, we call them unnecessary. Note that, if the path is to end at the desired point, any unnecessary step must be undone later on. Hence, $l_\gamma-r$ is always an even number $2k$. Clearly, the number of such paths is bounded by $\Gamma\paren{r,k}D^k$, where $\Gamma\paren{r,k}$ is a $D$-independent amplitude. On the other hand, the probability of finding all the links active is $(z/D)^{r+2k}$. Thus, we conclude that 
\be p\paren{l_\gamma=r+2k}=O\paren{\frac{1}{D^{k+r}}},\ee
that results in a $O\paren{D^{-k}}$ contribution to $\hat{C}_4 (r)$.

Then, in the large $D$ limit we obtain ($A=z\kappa$):
\be \hat{C}_4(r)=A^r=e^{r \log A},\ee
with finite size corrections of $O\paren{D^{-1}}$.
Thus, we encounter an exponential decay with an exponential correlation length given by
\be\xi^\text{exp}=\frac{1}{|\log A|}.\ee

Summing all up, we can compute the spin-glass susceptibility for the large $D$ limit:
\be\chi=\sum_{r=0}^\infty\hat{C}_4(r)=\sum_{r=0}^\infty A^r=\frac{1}{1-A}.\ee
We see that when $A=1$ the correlation no longer decays with distance, and the susceptibility diverges. Of course, one gets $A=1$ precisely at the critical temperature, $T_\mathrm c$, reported in  \eqref{eq:thouless:86}.

The computation for the fixed connectivity model is 
very similar. One only needs to notice that, whereas the probability for the first link in a lattice path to be active is $z/D$, the probability for the next link is roughly $(z-1)/D$ (this is only accurate for large $D$). It follows that, again, the $l_\gamma=r$ paths are the only relevant paths in the high temperature expansion. We find that 
\be p(l_\gamma=r)=\left\{\begin{array}{ll}
 1&\text{ if  }\ r=0,\\
\frac{z}{D}\paren{\frac{z-1}{D}}^{r-1}r!&\text{ if  }\ r>0.
\end{array}\right.\ee
Again, we can use it to compute $ \hat{C}_4(r)$. In the
large $D$ limit, up to corrections of $O\paren{D^{-1}}$, it is given by:
\be\label{eq:hatCap}
\hat{C}_4(r)=\left\{\begin{array}{lc}1&\text{ if  }\  r=0,\\\frac{z}{z-1}\caja{\paren{z-1}\kappa}^r&\text{ if  }\ r>0,\end{array}\right.\ee
which, taking $\tilde A=(z-1)\kappa$, also shows an exponential decay with 
\be\xi^\text{exp}=\frac{1}{|\log \tilde A|}.\ee

Using this spatial correlation function, we can either compute the SG-susceptibility in the fixed connectivity hypercube,
\be\label{eq:chi_inftyAP}\chi=\sum_{r=0}^\infty\hat{C}_4(r)=1+\frac{z}{z-1}\frac{\tilde A}{1-\tilde A}\,,\ee
or the integral correlation length, defined as \eqref{eq:xi},
\be\label{eq:xi_infty}\xi=\frac{\sum_{r=0}^\infty
  r\,\hat{C}_4(r)}{\sum_{r=0}^\infty\hat{C}_4(r)}=\frac{\chi-1}{\chi}\frac{1}{1-\tilde
  A}\, .\ee

Again, when $\tilde A=1$, we find a critical point. The corresponding
$T_\mathrm c$ matches \eqref{eq:thouless:86}. The critical exponents,
$\gamma=1$, $\nu=1$, can be read directly from \eqref{eq:chi_inftyAP} and
\eqref{eq:xi_infty}. The reader might be puzzled by a mean field model with
$\nu\neq 1/2$. The solution to the paradox is in our chosen metrics. Recall
that the postman distance in the hypercube is the square of the Euclidean
one. Hence, the correlation length in \eqref{eq:xi_infty} is the
\textit{square} of the Euclidean correlation length.

\chapter{Multi-spin coding}\label{app:multispin}
In this appendix, we discuss the multispin techniques used in
Chapters~\ref{chap:hypercube} and ~\ref{chap:chaos}.

\section{The computer code for the hypercube model}\label{app:multispin-hy}
Multi-spin coding is a kind of parallel computation that codes independent
systems on each one of the bits in a computer word. The fixed-connectivity
hypercube displays two features that allows for efficient multi-spin coding.
First, spins are located in the nodes of a unit hypercube, that means that, in
a $D$-dimensional hypercube, each spin position can be encoded in a word of
$D$-bits, and the same for the links. And second, both the spins and the
couplings are binary variables, and thus, can be codified in one bit. This
last property allow us to simulate many systems in parallel. Indeed, since
bitwise operations in a computer act at the same time over all the bits in a
word, using multi-spin coding, one can simulate $64$ systems (encoding spins
in an \texttt{unsigned long long} variable)\footnote{Indeed, the traditional
  natural processor word is $64$-bit-long. However, although we did not use
  them, we cite that the SSE lets to work with $128$ or even $256$ (in the
  newest computers) bits words.  } at the same time, thus multiplying the
efficiency by almost this factor.

In this appendix we will discuss, first, some technical details about the
implementation of the hypercube in a computer using C language, discussing the
graph generation, and second, we will explain how to implement the
Metropolis algorithm taking full advantage of multi-spin coding.

\subsection{Bitwise operations}\label{sec:bitwise}
Before anything else, it is interesting to discuss the bitwise operations we
need to use in the program. We begin with the simplest one, the NOT operator
(\texttt{!} in C). Its effect is inverting the bit, i.e. $\texttt{!}0=1$ and
$\texttt{!}1=0$.  We also need the AND operator ($\texttt{\&}$ in C), the OR
operator ($\texttt{|}$ in C) and the exclusive OR or XOR (\texttt{\^{}} in
C). We present their truth tables in Table \ref{tab:true}.  It is interesting
to note that the XOR operator leaves unchanged the second bit in case the
first one is a $0$, and behaves like the NOT operator in the case it is a $1$.
\begin{table}
\begin{center}
\begin{tabular}{|c@{$\texttt{\&}$}c@{$=$}c|}\hline
0&0&0\\
0&1&0\\
1&0&0\\
1&1&1\\\hline
\end{tabular}
\begin{tabular}{|c@{$\texttt{|}$}c@{$=$}c|}\hline
0&0&0\\
0&1&1\\
1&0&1\\
1&1&1\\\hline
\end{tabular}
\begin{tabular}{|c@{\^{}}c@{$=$}c|}\hline
0&0&0\\
0&1&1\\
1&0&1\\
1&1&0\\\hline
\end{tabular}
\end{center}\caption{Truth tables for {\bf (left)} AND ($\texttt{\&}$), {\bf(center)} OR
  ($\texttt{|}$) and {\bf (right)}  XOR (\^{}) operator.}\label{tab:true}
\end{table}

In addition, we will use the shifts operators \texttt{>>} and
\texttt{<<}. They shift bits to the right or to the left.

Now that we have defined all the operators we are going to need, we can use
them to move along the hypercube or the different samples. We begin with the
implementation of the hypercube in a computer.

\subsection{The hypercube in a computer}\label{sec:hypercubecomputer}
Being our problem an unit hypercube, its spatial coordinates are
$\V{x}\!=\!(x_1,\ldots,\ x_D)\!=\!(\{0,1\},\ldots,\ \{0,1\})$ and then, can be
directly coded in a $D$-bit word (let us call this word \texttt{site}) just
writing the $i$-th coordinate ($x_i$) as the $i$-th bit (\texttt{site}$^i$) in
the word. One can easily recover
the $x_i$ coordinate by performing the following operation
$$\texttt{(site>> i)\& 1},$$ that is, the bit located in the
$i$-th bit of \texttt{site} is obtained by displacing the bits in
\texttt{site} by $i$ bits, and finally recovered by means of an AND operator with
$1$. 

In the hypercube, the nearest neighbors of spin in $\V{x}$ are located in
$\V{x}+\V{\hat\mu}$ where $\V{\hat\mu}$ are the Cartesian unit
vectors in a $D$-dimensional space. Again, we can write these unit vectors using $D$-bit words [all bits
will be zero but the one located in the $\texttt{mu}(=0,\ldots,D-1)$ position].   Then, using bitwise
operations, and taking into account the periodic boundary conditions, the
nearest neighbor of \texttt{site} in direction $\V{\hat\mu}$ will be given by
$$\texttt{site\^{}(1<<mu)}.$$

As an example, let us consider a $D=4$ dimensional hypercube and the spin
located in site $14$. The spatial representation in bits is thus
$(1,1,1,0)$. The first neighbor in the $0$ direction is the $(1,1,1,1)=15$-th
spin. Note that the Euclidean coordinates are written from right to left (on
the opposite than usually) to keep the equivalence with the binary
representation. Thereby, the first neighbor in direction $3$ is the
$(0,1,1,0)=6$, while the first neighbor along direction $2$ is $(1,0,1,0)=10$.

\subsection{Connectivity matrix}\label{ap:connmat}

In the next section, we will explain how to parallelize the Metropolis
algorithm to simulate $64$ samples at the same time. However, for practical
reasons, we need to consider the same connectivity matrix,
$n_{\V{x},\V{\hat\mu}}$ in \eqref{eq:Hhy}, for all the samples. The difference
between samples within the same simulation is thus introduced only through the
randomness of the couplings $J_{\V{x},\V{\hat\mu}}$. That means that the
configuration of interacting neighbors will be same at each run, but the
nature of the interaction between spins, will not.

We include here a simplified version for the program used to generate the
fixed connectivity graphs. In Listing \ref{list:hypercube_con} we detail
the dynamic Monte Carlo program used to simulate the fixed connectivity graph, based
on the plaquette transformation, written in Listing \ref{list:plaquette}. This
transformation was explained in detail in Section \ref{sec:fixed-con} and
schematized in Figure \ref{fig:plaquettetrans}.

As one can read in Listing \ref{list:hypercube_con}, we have defined some
vectors. First, we introduce the connectivity matrix,
$\{n_{\V{x},\V{\hat\mu}}\}$ as a $N$-dimensional vector. The index
$\texttt{site}$ labels the starting vertex in the hypercube, and the direction
of the link is encoded in the first $D$ bits of the word. The
link between the spin located at $\V{x}\equiv\texttt{site}$ with its neighbor
in direction $\V{\hat\mu}\equiv\texttt{mu}$ is stored in
$\texttt{n[site\^{}(1<<mu)]}$, and will be $1$ if the two spins are
connected and $0$ otherwise. 
In addition, since the interaction is very diluted, in order to speed the simulation, we store in vector
$\texttt{neighbor\_list[site*6+mu]}$, where \texttt{mu} labels the $6$
neighbors of the spin in $\texttt{site}$.

\begin{lstlisting}[label=list:hypercube_con,caption=hypercube
  connections]
void hypercube_connections(void)
{
  unsigned n_ini;
  int i,k,site,dir1,dir2,mu,indice;
  
  n_ini=63;                             // initial configuration 
                                        // of links
                                        // 0-5  = active   = 1
                                        // 6-.. = inactive = 0
  for (i=0;i<N;i++)                     // only N/2 will be used 
    n[i]=n_ini;                         // (but copies are retained
                                        // to simplify the code)
  
  //  Start Monte Carlo
  
  for (k=0;k<100;k++){                  // run the plaquete 
    for (i=0;i<N;i++){                  // transformations k*N times
      site=MYRANDOM>>(64-D);            // the D most significant bits   
      dir1=MYFRANDOM*D;                 // a random direction
      while ((dir2=MYFRANDOM*D)==dir1); // a different (random) direction
      permut_links(site,dir1,dir2);     // plaquette transformation
    }
  }
  
  for (site=0;site<V;site++){
    indice=0;
    for(mu=0;mu<D;mu++){
      if((n[site]>>mu)&1){              // stores which are the 
	neighbor_list[site*6+indice]=site^(1<<u); 
	indice++;                       // neighbors of 
	                                // the spin in site
      }
    }
  }
}
\end{lstlisting}
\begin{lstlisting}[label=list:plaquette,caption=plaquette transformation]

void permut_links(int i, int mu, int nu)
{
  int J1,J2,J3,J4;
  int mask;
  //                              Boundary conditions
  //  i+nu i+nu,mu  i+mu+nu       'i+mu'=i^(1<<mu)
  //      * ------ *
  //      |   J3   |
  // i,nu |J4    J2| i+mu,nu
  //      |   J1   | 
  //      * ------ *
  //    i    i,mu    i+mu
  
  J1=(n[i]>>mu)&1;
  J4=(n[i]>>nu)&1;
  
  if (J1^J4){ // links are different
    
    J2=(n[i^(1<<mu)]>>nu)&1;
    J3=(n[i^(1<<nu)]>>mu)&1;
    
    if (J2^J3 && J1^J2){ // also are different
      // 8 bits are changed (4 original and 4 copies)
      mask=(1<<mu)|(1<<nu);
      n[i]^=mask;
      n[i^(1<<mu)]^=mask;
      n[i^(1<<nu)]^=mask;
      n[i^mask]^=mask;
    }
  }
}
\end{lstlisting}

\subsection{Multi-spin coding}
As we briefly discussed before, it is possible to take benefit of the
simultaneity of the bitwise operations to simulate at the same time many
systems if they are all coded together in the same word. With this aim, we
define a vector of \texttt{unsigned long long} variables of $N$ spins,
$\texttt{S[N]}$. In this scheme, each
bit in the word $\texttt{S[site]}$ accounts for the spin state placed at \texttt{site} in each of
the $64$ samples simulated. 

As an example, using the operations discussed above, the spin placed in
position $\V{x}=(0,\ldots,0,1,1)$ of the $10$-th sample, for instance, would
be recovered with
$$\texttt{(S[3]>>9)\&1},$$ i.e. the position in the hypercube is recovered by
considering the coordinates of the vector $\V{x}$ as the bits of the word
\texttt{site}, then $\texttt{site=0}\cdots\texttt{11}\equiv 3$. $\texttt{S[3]}$ is an
\texttt{unsigned long long} variable where each of its 64 bits represent the
state of the spin placed at $\V{x}$ in each sample. In order to extract the
corresponding bit of the $10$-th sample, we displace all the bits of $\texttt{S[3]}$,
$10-1=9$ positions, and then perform and \texttt{\&} with $1$ to isolate this value.

Before entering in the algorithm, we need to establish an equivalence between
bits and spins or couplings.  For the spins we consider the following change
of variables:
\begin{eqnarray}\label{eq:transspin}
\sigma=+1\rightarrow s=1,&\mathrm{and},& \sigma=-1\rightarrow s=0.
\end{eqnarray}
In the same way that we did with the spins, we also need to encode the active
coupling constants $J_{\V{x},\V{\hat\mu}}=\pm1$ of the $64$ samples also in a
 \texttt{unsigned long long} vector. Since each spin is connected always
with only 6 spins, the dimension of this vector will be $6N$ (one per site in
the lattice and one per occupied link). We choose
the equivalence between the coupling variables and the bits in the opposite
way than before
\begin{eqnarray}
J=+1\rightarrow j=0,&\mathrm{and},& J=-1\rightarrow j=1.
\end{eqnarray}
The reason for this arbitrary election is to absorb the negative sign in the
definition of the Hamiltonian \eqref{eq:Hhy}. Indeed, using these
transformations, the logic operation (using the $\{0,1\}$ basis)
\be\label{eq:logic} \texttt{s1\^{}j\^{}s2}, \ee and the product of
the original $\{-1,+1\}$ variables, \be\label{eq:mult} -J \sigma_1 \sigma_2, \ee leads
to the same result [using the spin transformation \eqref{eq:transspin}], as is
shown in Table \ref{tab:interaction}.
\begin{table}
\begin{center}
\begin{tabular}{|ccc|c|}\hline
$J$&$\sigma_1$&$\sigma_2$&$-J \sigma_1 \sigma_2$\\
-1&1&1&+1\\
-1&1&-1&-1\\
-1&-1&1&-1\\
-1&-1&-1&+1\\\hline
+1&1&1&-1\\
+1&1&-1&+1\\
+1&-1&1&+1\\
+1&-1&-1&-1\\\hline
\end{tabular}
\begin{tabular}{|ccc|c|c|}\hline
\texttt{s1}&\texttt{j}&\texttt{s2}&\texttt{j\^{} s2}&\texttt{s1\^{} j\^{} s2}\\
1&1&1&0&1\\
1&1&0&1&0\\
0&1&1&0&0\\
0&1&0&1&1\\\hline
1&0&1&1&0\\
1&0&0&0&1\\
0&0&1&1&1\\
0&0&0&0&0\\\hline
\end{tabular}
\end{center}
\caption{Comparison between the product \eqref{eq:mult} and the logic
  operation \eqref{eq:logic}.}\label{tab:interaction}
\end{table}
Summing up, with this election, sample to sample (or bit to bit), the result
will be $0$ if the coupling is satisfied, and $1$ if it is unsatisfied.  Since
the bitwise operations act over all the bits in a word at the same time, this
product is computed for the $64$ samples at once.

At usual, for the Metropolis test, we need to compute the energy gain or lost
of flipping one selected spin, $\sigma_i$. We only consider nearest
neighbor interactions with exactly 6 neighbors, then, the energy difference
will be \be\label{eq:deltaE} \Delta E=-2\sigma_i \sum_{j-\text{neighbor}}^6
J_j\sigma_j, \ee where the sum on $j-\text{neighbor}$ runs only over the 6
connected spins in the graph. In the $\{-1,+1\}$ basis, this $\Delta E$ can
only take 7 different values $-12,$ $-8,$ $ -4,$ $ 0,$ $+4,$ $+8,$ $ +12$. Then, the
flip will be directly accepted if $\Delta E\le 0$ or with probability
$\exp(-\beta\Delta E)$ if $\Delta E> 0$.

Now, we need to compute this $\Delta E$ in the bit basis.
If we consider the equivalence between operations discussed before, for each
sample, the flip will be directly accepted if the number of unsatisfied
couplings, $n_{\text{unsat}}$, is higher or equal to 3. The $\Delta E$,
\eqref{eq:deltaE}, is thus recovered using the relation $\Delta E=12-2
n_{\text{unsat}}$.

The problem now is how to compute this $\Delta E$ and to perform the Metropolis
test without breaking the parallelism between samples. The idea is to store
bit by bit the energy for each sample.  The maximum number of unsatisfied
couplings is $6$ per sample, in binary representation $\texttt{100}$, which means that
we need three bits per sample to store it. Thus, we introduce three
\texttt{unsigned long long} variables: \texttt{bit2}, \texttt{bit1} and
\texttt{bit0}, so that the number of unsatisfied couplings for the $i$-th
sample will be the binary number composed by the
$\texttt{bit2}^i\texttt{bit1}^i\texttt{bit0}^i$, where $\texttt{bit}^i$
represents the $i$-th bit of word $\texttt{bit}$ (representing the $i$-th sample).

 The process to compute the number of unsatisfied links for each samples would be the following:
\begin{enumerate}
\item Select one spin in \texttt{site}, $\texttt{S[site]}$.
\item Select its {\em first} connected neighbor. Its position was stored in the
  vector  \texttt{neigh\-bor\_list} defined in Section \ref{ap:connmat}. Then,
  its position is 

  \texttt{site\_0=neighbor\_list[site*6+0]}, 

  and the coupling,
  
  \texttt{J\_0=J[site*6+0]}.
\item Compute \texttt{link0=S[site]\^{}J\_0\^{}S[site\_0]}. It is the first link we
  count so the total number of unsatisfied couplings can only be $0$ or $1$ so
  far. We only need one bit to keep it, so
  
  \texttt{bit0}=\texttt{link0}.

\item Select the {\em second} neighbor. As before,

  \texttt{site\_1=neighbor\_list[site*6+1]}, 
  
  \texttt{J\_1=J[site*6+1]}.

  And compute
  \texttt{link1=S[site]\^{}J\_1\^{}S[site\_1]}. The total number of
  unsatisfied couplings can be, so far, $0,\ 1$ or $2$ (in binary $\texttt{00,\ 01}$
  and $\texttt{10}$). The possible
  combinations of \texttt{bit0} and \texttt{link1} are
  \begin{equation}\nonumber
\begin{array}{cc|cc}
    \texttt{bit0}&\texttt{link1}&\texttt{bit1\_new}&\texttt{bit0\_new}\\\nonumber
    0&0&0&0\\\nonumber
    0&1&0&1\\\nonumber
    1&0&0&1\\\nonumber
    1&1&1&0\\\nonumber
\end{array}
  \end{equation}
  Then, 

  \texttt{bit1\_new}=\texttt{bit0\&link1},

  and

  \texttt{bit0\_new}=\texttt{bit0}\^{}\texttt{link1}.

\item Select the {\em third} neighbor. Again,

  \texttt{site\_2=neighbor\_list[site*6+2]}, 
  
  \texttt{J\_2=J[site*6+2]},

  and
  \texttt{link2=S[site]\^{}J\_2\^{}S[site\_2]}. The total number of
  unsatisfied couplings can be now $0,\ 1,\ 2$ or $3$ (in binary
  $\texttt{00,\ 01,\ 10}$
  and $\texttt{11}$). The possible
  combinations are now
  \begin{equation}\nonumber
\begin{array}{ccc|c|cc}
    \texttt{bit1}&\texttt{bit0}&\texttt{link2}&\texttt{bit0\&link2}&\texttt{bit1\_new}&\texttt{bit0\_new}\\\nonumber 
    0&0&0&0&0&0\\\nonumber
    0&0&1&0&0&1\\\nonumber
    0&1&0&0&0&1\\\nonumber
    0&1&1&1&1&0\\\nonumber
    1&0&0&0&1&0\\\nonumber
    1&0&1&0&1&1\\\nonumber
\end{array}
  \end{equation}
  Then, 

  \texttt{bit1\_new}=\texttt{bit1}\^{}(\texttt{\texttt{bit0}\&link2}),

  and

  \texttt{bit0\_new}=\texttt{bit0}\^{}\texttt{link2}.

\item Select the {\em forth} neighbor. Again,

  \texttt{site\_3=neighbor\_list[site*6+3]}, 
  
  \texttt{J\_3=J[site*6+3]},

  and \texttt{link3=S[site]\^{}J\_3\^{}S[site\_3]}. The total number of
  unsatisfied couplings can be now $0,\ 1,\ 2,\ 3$ or $4$ (in binary
  \texttt{000,\ 001,\ 010,\ 011} and \texttt{100}). Now we need three bits
  to store all. As before, the possible combinations are now
  \begin{equation}\nonumber
\begin{array}{ccc|c|ccc}
    \texttt{bit1}&\texttt{bit0}&\texttt{link3}&\texttt{bit0\&link3}&\texttt{bit2\_new}&\texttt{bit1\_new}&\texttt{bit0\_new}\\\nonumber 
    0&0&0&0&0&0&0\\\nonumber
    0&0&1&0&0&0&1\\\nonumber
    0&1&0&0&0&0&1\\\nonumber
    0&1&1&1&0&1&0\\\nonumber
    1&0&0&0&0&1&0\\\nonumber
    1&0&1&0&0&1&1\\\nonumber
    1&1&0&0&0&1&1\\\nonumber
    1&1&1&1&1&0&0\\\nonumber
\end{array}
  \end{equation}
  Then, 
  \texttt{bit2\_new}=\texttt{bit1\&(bit0\&link3)},

  \texttt{bit1\_new}=\texttt{bit1}\^{}(\texttt{\texttt{bit0}\&link3}),

  and

  \texttt{bit0\_new}=\texttt{bit0}\^{}\texttt{link3}.

\item Select the {\em fifth} neighbor. Again,

  \texttt{site\_4=neighbor\_list[site*6+4]}, 
  
  \texttt{J\_4=J[site*6+4]},

  and
  \texttt{link4=S[site]\^{}J\_4\^{}S[site\_4]}. The total number of
  unsatisfied couplings can be now $0,\ 1,\ 2,\ 3,\ 4$ or $5$ (in binary
  \texttt{000,\ 001,\ 010,\ 011,\ 100}
  and \texttt{101}). Now both \texttt{bit0} and \texttt{bit1} can saturate,
  for the sake of abbreviation we name \texttt{A=bit0\&link4} and \texttt{B=bit1\&(bit0\&link4)=bit1\&A},
  \begin{equation}\nonumber
\begin{array}{cccc|cc|ccc}
    \texttt{bit2}&\texttt{bit1}&\texttt{bit0}&\texttt{link4}&\texttt{A}&\texttt{B}&\texttt{bit2\_new}&\texttt{bit1\_new}&\texttt{bit0\_new}\\\nonumber 
    0&0&0&0&0&0&0&0&0\\\nonumber
    0&0&0&1&0&0&0&0&1\\\nonumber
    0&0&1&0&0&0&0&0&1\\\nonumber
    0&0&1&1&1&0&0&1&0\\\nonumber
    0&1&0&0&0&0&0&1&0\\\nonumber
    0&1&0&1&0&0&0&1&1\\\nonumber
    0&1&1&0&0&0&0&1&1\\\nonumber
    0&1&1&1&1&1&1&0&0\\\nonumber
    1&0&0&0&0&0&1&0&0\\\nonumber
    1&0&0&1&0&0&1&0&1\\\nonumber
\end{array}
  \end{equation}
  Then, 
  \texttt{bit2\_new}=\texttt{bit2\^{}[bit1\&(bit0\&link4)],}

  \texttt{bit1\_new}=\texttt{bit1}\^{}(\texttt{\texttt{bit0}\&link4}),

  and

  \texttt{bit0\_new}=\texttt{bit0}\^{}\texttt{link4}.

\item Finally we select the {\em sixth} neighbor. Again,

  \texttt{site\_5=neighbor\_list[site*6+5]}, 
  
  \texttt{J\_5=J[site*6+5]},

  and
  \texttt{link5=S[site]\^{}J\_4\^{}S[site\_4]}. The total number of
  unsatisfied couplings can be now $0,\ 1,\ 2,\ 3,\ 4,\ 5$ or $6$ (in binary
  \texttt{000,\ 001,\ 010,\ 011,\ 100,\ 101}
  and \texttt{110}). Again, we abbreviate \texttt{A=bit0\&link5} and \texttt{B=bit1\&A},
  \begin{equation}\nonumber
\begin{array}{cccc|cc|ccc}
    \texttt{bit2}&\texttt{bit1}&\texttt{bit0}&\texttt{link5}&\texttt{A}&\texttt{B}&\texttt{bit2\_new}&\texttt{bit1\_new}&\texttt{bit0\_new}\\\nonumber 
    0&0&0&0&0&0&0&0&0\\\nonumber
    0&0&0&1&0&0&0&0&1\\\nonumber
    0&0&1&0&0&0&0&0&1\\\nonumber
    0&0&1&1&1&0&0&1&0\\\nonumber
    0&1&0&0&0&0&0&1&0\\\nonumber
    0&1&0&1&0&0&0&1&1\\\nonumber
    0&1&1&0&0&0&0&1&1\\\nonumber
    0&1&1&1&1&1&1&0&0\\\nonumber
    1&0&0&0&0&0&1&0&0\\\nonumber
    1&0&0&1&0&0&1&0&1\\\nonumber
    1&0&1&0&0&0&1&0&1\\\nonumber
    1&0&1&1&1&0&1&1&0\\\nonumber
\end{array}
  \end{equation}
  Then, 
  \texttt{bit2\_new}=\texttt{bit2\^{}[bit1\&(bit0\&link5)],}

  \texttt{bit1\_new}=\texttt{bit1}\^{}(\texttt{\texttt{bit0}\&link5}),

  and

  \texttt{bit0\_new}=\texttt{bit0}\^{}\texttt{link5}.
\end{enumerate}
Up to this point, we know the amount of unsatisfied links for each
  sample (and then the energy difference) we need to decide whether the flips
  of the spins in \texttt{S[i]} are accepted or not, but for all the $64$ samples
  at the same time. 

The instruction for inverting one spin is equivalent to make an XOR with \texttt{1}
(indeed, \texttt{1\^{}0=1} and \texttt{1\^{}0=1}). On the contrary, the spin
will be unaltered if the XOR is made with a \texttt{0} in the first place
(i.e. \texttt{0\^{}0=0} and \texttt{0\^{}1=1}). With this idea in mind, we
define a new \texttt{unsigned long long} variable, called \texttt{flip}, that
carries in each of its bits the information about flipping the spin in each of
the samples. That is, for example, if its $10$-th bit is \texttt{1}, the spin in
sample $10$, will be inverted. On the contrary, if it is \texttt{0}, it will
continue as it was. We have to find a way to store the information about
flipping all the samples at once.

We come back to the number of unsatisfied links and its equivalent energy
barrier. The possible results are
  \begin{equation}\nonumber
    \begin{array}{cc|ccc}
      \Delta E & n_{\text {unsat}} &\texttt{bit2}&\texttt{bit1}&\texttt{bit0}\\\nonumber 
      -12 & 6& 1&1&0\\\nonumber 
      -8 & 5& 1&0&1\\\nonumber 
      -4 & 4& 1&0&0\\\nonumber 
      0 & 3& 0&1&1\\\nonumber 
      4 & 2& 0&1&0\\\nonumber 
      8 & 1& 0&0&1\\\nonumber 
      12 & 0& 0&0&0\\\nonumber 
    \end{array}
  \end{equation}
The flip of $\texttt{S[site]}^i$ is directly accepted if $\Delta E\le0$. In terms of
$\texttt{bit2}^i\texttt{bit1}^i\texttt{bit0}^i$, this will occur whereas
$\texttt{bit2}^i=\texttt{1}$ (for the negative values of $\Delta E$) or if
\texttt{bit0}$^i$\texttt{\&}\texttt{bit1}$^i$ \texttt{=1} (for the $\Delta E =0$ case). In the rest
of cases, the flip will be accepted conditioned to the Metropolis test.

Nevertheless, even though many flips will be accepted directly, it is
presumably that the flip will not be accepted simultaneously for all the $64$
samples in the simulation, then, we always need to through a random number,
$0\le R<1$, and to check if $R<\exp(-\beta\Delta E)$ is fulfilled for each
sample to accept the change.  If we use the same random number for all the
samples, we can check if it surpass or not a barrier of $\exp(-4\beta)$,
$\exp(-8\beta)$ or $\exp(-12\beta)$ at once. With this aim, we define another
three \texttt{unsigned long long} variables, \texttt{jump4}, \texttt{jump8}
and \texttt{jump12} that will be a variable with $64$ bits equal to \texttt{1}
(\texttt{!0} in C, the highest possible number)
if the barrier is surpassed, or all them equal to \texttt{0} if it is not. Clearly if
\texttt{jump12=!0}, all samples will be flipped (if jumped the highest
barrier, jumped all). If not, we must decide
which samples are updated and which not. For this aim, it is useful to check this
other combination of variables
 \begin{equation}\nonumber
    \begin{array}{cc|cc|cc|c}
      \Delta E & n_{\text {unsat}} &\texttt{bit1}&\texttt{bit0}&\texttt{jump4}& \texttt{jump8}&\text{Accept}\\\nonumber 
      4 & 2& 1&0&1&1&\text{Y}\\\nonumber 
        &  &  & &1&0&\text{Y}\\\nonumber 
        &  &  & &0&0&\text{N}\\\nonumber 
      8 & 1& 0&1&1&1&\text{Y}\\\nonumber 
        &  &  & &1&0&\text{N}\\\nonumber 
        &  &  & &0&0&\text{N}\\\nonumber 
    \end{array}
  \end{equation}
Then, when \texttt{jump16=0},  if \texttt{jump8=!0} only the samples with
\texttt{bit0}$^i$\texttt{=1} be updated. On the contrary, if \texttt{jump8=0},
but \texttt{jump4=!0}, the flipped ones will
be only the ones with $\texttt{bit1}^i=1$. If none of the \texttt{jump}
variables is \texttt{!0}, no spin will be updated.

Summing all the conditions up, the variable \texttt{flip} will be given by

\texttt{flip=bit2|(bit1\&bit0)|jump12|(jump8\&bit0)|(jump4\&bit1)}

Finally, the whole collection of samples will be updated at once by means of
the instruction

\texttt{S[site]\^{}=flip}.

\section{Multi-spin coding for correlation functions}\label{app:multispin-chaos}

In this appendix, we face up the technical problem of computing an extremely
large number of overlaps in a reasonable computer time. Indeed, all the study
performed in Chapter~\ref{chap:chaos} involves computing overlaps for
$N_\mathrm{s}$ samples, with $4$ independent sets of equilibrium
configurations of $V=L^3$ spins each (obtained with independent Monte Carlo
simulations) at $N_T$ different temperatures. In addition, we consider for the
equilibrium mean values, $N_t=100$ independent times, evenly spaced in the
whole Monte Carlo time. The data for each system size was summarized in Table
\ref{tab-chaos:parameters-eq}.

That means that, in order to compute the averaged $\mean{q_{T_1,T_2}^2}_J$ for
each sample and couple of temperatures $\lazo{T_1,T_2}$, we need to average
over all the
\begin{equation}
q_{T_1,T_2}^{(a,b)}(t_A,t_B;J)=\frac{1}{V}\sum_{\V{x}} s_{\V{x}}^{(a),T_1}(t_A)
s_{\V{x}}^{(b),T_2}(t_B)\,,
\end{equation}
$12 N_t^2 $ overlaps. Indeed, since the temperatures are different, there are
$N_r(N_r-1)=12$ ways of combining two replicas $a$ and $b$, and $N_t^2$ pairs
of times.  In addition, we need to compute overlaps for $N_T^2/2$ couples of
temperature (choosing $T_1\le T_2$) and $N_\mathrm{s}$ samples. Summing all
up, we need to compute $6 N_\mathrm{s} N_t^2 N_T^2$ overlaps, which only for
$L=32$ is $46240000000\ L^3$ operations.  The situation is even worse if we
consider the spatial correlation function $c_4(\V{r})$ since, in addition, we
must consider all the possible displacements. Clearly, a direct computation
would take months if no parallelization is considered.

Our solution to the problem was again to use multispin coding, as we did for
the Hypercube model and detailed in this appendix in Section
\ref{app:multispin-hy}. Indeed, if the $L^3$ spin variables, and overlap
fields are coded in words of $64$ bits, we can reduce the total time by a
factor $64$. We will discuss here only the ideas necessary for the
parallelization, not the whole analysis program.

The approach is analogous to the one discussed in Section
\ref{app:multispin}. That is, to code the $\lazo{-1,+1}$ spin or overlap
values in the $\lazo{\texttt{0,1}}$ bits of a $64$-bit word, and to take
advance of the simultaneously of the bitwise operations to parallelize the
problem.  This time, instead of considering different systems coded in the
same word, we will locate all the $N$ spins sequentially in different size
words, as we will explain below.

The spin configurations from JANUS were written in the following format
$$\texttt{char u[Nr][NT][Nt][V8],}$$ where the indexes \texttt{Nr},
\texttt{NT} and \texttt{Nt} refers to $N_r$, $N_T$ and $N_t$ respectively. Not
as clear is the meaning of this \texttt{V8}. First, \texttt{V8} means $V/8$,
which is always an integer since our system sizes have all even $L$. The
reason for this division by $8$ is that we are using \texttt{char} variables,
which are $8$-bit words.  Then, we can pack the $V$ spins in $V/8$ groups of
$8$ spins each coded together in the bits of the same word. Concerning this
packing, we need to discuss how to move along the lattice points, indeed, for
the spatial correlations functions we will need to now the spatial position
$\V{x}$ of each spin. According to our program the spin at
$\V{x}=(\texttt{x},\texttt{y},\texttt{z})$ is recovered (leaving aside the
indexes for replicas, temperatures or times) as
\be\label{eq:confz}\texttt{(u[z*S8+y*L8+x/8]>>(x\&7))\&1}\ee As before,
\texttt{S8} and \texttt{L8} mean $L^2/8$ and $L/8$ respectively. An
explanation of the bitwise operations can be found in Section
\ref{sec:bitwise}. This relation is valid for all our values of $L$ but
$L=12$.\footnote{Indeed, the $L=12$ case is more difficult since $L$ is not
  divisible by 8. Packing is then a bit less straight-forward. We will not
  fully discuss this case here because the ideas are exactly the same but the
  calculations are more tedious. The underlying idea is that, although a whole
  row in each plane does not hold exactly in an even number of 8-bit words (as
  happens in the other system sizes), two neighboring rows do fit perfectly in
  2 words. Then, if one wants to sum over all the row, must take into account
  whether the index is even or odd. In the case the index is even, one can sum
  all the bits in the word without worries because they belong to the same
  row. In the case it is odd, only 4-bits in the word belong to the desired
  row.}

Now, the process to compute the overlap field \eqref{Q-FIELD-DEF} between the
configurations \texttt{u[ir][iT1][it]} and \texttt{u[ir2][iT2][it2]}
(\texttt{ir}, \texttt{iT} and \texttt{it} accounts for the replica,
temperature and iteration indexes) is summarized in
List~\ref{list:calcula_overlap}.
\begin{lstlisting}[label=list:calcula_overlap,caption=Overlap field
  calculation]

char overlap[V8];

conf1=u[ir][iT1][it];
conf2=u[ir2][iT2][it2];	   

calcula_overlap(conf1, conf2,overlap);   

...


void calcula_overlap(char *u1,char *u2, char *q)
{
  int i;
  long long *ul1,*ul2,*ql;
    
  ul1=(long long *)u1;
  ul2=(long long *)u2;
  ql=(long long *)q;
  

  for (i=0;i<V64;i++){
    ql[i]=ul1[i]^ul2[i];
  }
  
}
\end{lstlisting} Indeed, $V/8$ groups of $8$
bits can be always be packed in $V/64$ ($=$\texttt{V64}) groups of $64$ and
thus parallelize the computation. This is, as displayed, just performed by a
changing the word type. As seen, the exclusive OR bitwise operator (see
Table:xor) is used for computing the spin multiplications. Indeed, as
discussed in Section~\ref{app:multispin} it has the same multiplication table
using bits, than the multiplication of signs.

Up to this point, we only computed the overlap field, in order to compute the
whole overlap we need to sum up all the components. Since the computation is
linear, it can be directly obtained by counting the number of bits equal to
\texttt{1} in \texttt{overlap}, and returning to the original $\lazo{-1,+1}$
basis, see List~\ref{list:total_overlap}
\begin{lstlisting}[label=list:total_overlap,caption=Total overlap sum
]
  long long *miq=(long long *) overlap;
  q12=0;
  for (i=0;i<V64;i+=1)
    q12+=SUM_64BITS(miq[i]);

//return to the original {+1,-1} basis
  mq12=(1.-2.*q12/(double)V);


\end{lstlisting} For the summing bits' function
\texttt{SUM\_64BITS(x)} one can use, either the built-in function
\texttt{\_mm\_popcnt\_u64(x)} or a table initialized at the beginning of the
program counting the amount of bits \texttt{1} for all long long numbers.

The computation for the spatial correlation function is a bit more
complicated. To illustrate it we begin with the simplest case, when
displacements are only considered along the $z$-axis, i.e.
$\V{r}=(0,0,r)$. Note that  the \texttt{z} coordinates are the first indexes in
the overlap field \texttt{overlap} just computed, then, computing sums between
the different planes is straight-forward as explained in List~\ref{list:corrx}.
\begin{lstlisting}[label=list:corrx,caption=Spatial correlation function for
  displacements in $z$]

for(z=0;z<L;z++)
  for(zp=0;zp<=L/2;zp++){
    new_z=(zp+z)%L;	  
    corr=1.-2.*corr_points(&overlap[z*S8],&overlap[new_z*S8])/(double)S;
    my_corr[zp]+=corr;
  }

...

int corr_points(char *q1,char *q2)
{
    int i;
    int sum;
    long long *ql1,*ql2;
    long long l;

    sum=0;
    ql1=(long long *)q1;
    ql2=(long long *)q2;
    for (i=0;i<S64;i++)
      {
	l=ql1[i]^ql2[i];
	sum+=SUM_64BITS(l);
      }
    return sum;
}
\end{lstlisting}

Now, the idea is to compute the other two directions in the same way. In order
to do it so,  we need to rewrite the configurations so that the $x$ or alternatively
the $y$ coordinate are placed on the first index, as $z$ was in
\eqref{eq:confz}. We thus define two alternative rotated
configurations
$$\texttt{char uY[Nr][2][NT][V8]},$$
with where coordinates $\V{r}=(\texttt{x},\texttt{y},\texttt{z})$ are
recovered as
$$\texttt{(uY[y*S8+z*L8+x/8]>>(x\&7))\&1}.$$
And
$$\texttt{char uX[Nr][2][NT][V8]},$$
where the indexes run, this time,  as follows
$$\texttt{(uX[x*S8+y*L8+z/8]>>(z\&7))\&1}.$$ Interchanging $y\leftrightarrow z$ is very
easy, since there is no spin coding and only implies a change of variables.  In
order to get the variable $x$ we need to decode it from the bits and afterwards
to encode $z$ in its place.  We present in List~\ref{list:rotations} the
two different rotations we need.
\begin{lstlisting}[label=list:rotations,caption=Rotations for the spatial
  correlation function]
//first we rotate the spin configurations for each couple of temperatures
for(ir=0; ir< Nr; ir++)
  for(it=0; it < NT ; it++)
    {
      rotate_conf_yz(u[ir][ibeta1][it],uY[ir][0][it]);
      rotate_conf_xz(u[ir][ibeta1][it],uX[ir][0][it]);
	
      rotate_conf_yz(u[ir][ibeta2][it],uY[ir][1][it]);
      rotate_conf_xz(u[ir][ibeta2][it],uX[ir][1][it]);
    }

...



void rotate_conf_yz(char *u1, char *u2) // interchange (y<->z) 
{
  int x8,y,z,i1,i2;   
  bzero(u2,V8);
  
  for (z=0;z<L;z++)
    for (y=0;y<L;y++){
      i1=y*L8+z*S8;  //we just put y in the old place of z
      i2=z*L8+y*S8;		
      for (x8=0;x8<L8;x8++){ 
	u2[i2]=u1[i1];
	i1++;
	i2++;
      }
    }
}


void rotate_conf_xz(char *u1, char *u2) // interchage (x<->y) 
{
    int x,y,z,i1,i2,shift1,shift2,bit;
    bzero(u2,V8);
    
    for (z=0;z<L;z++)
	for (y=0;y<L;y++){
	    shift2=z&7;
	    for (x=0;x<L;x++){
		i1=x/8+y*L8+z*S8;
		i2=z/8+y*L8+x*S8;		
		shift1=x&7; 
		bit=(u1[i1]>>shift1)&1;
		u2[i2]|=bit<<shift2;
	    }
	}
}

\end{lstlisting}
Once rotated the configurations, the computation for the overlap follows from \ref{list:corrx}.

\chapter[Scaling and dynamic ultrametricity]{Scaling and dynamic ultrametricity in the hypercube model}\label{ap:Ultrametricity}
\begin{figure}
\begin{center}
\includegraphics[angle=270,width=0.9\columnwidth,trim=40 0 40 0]{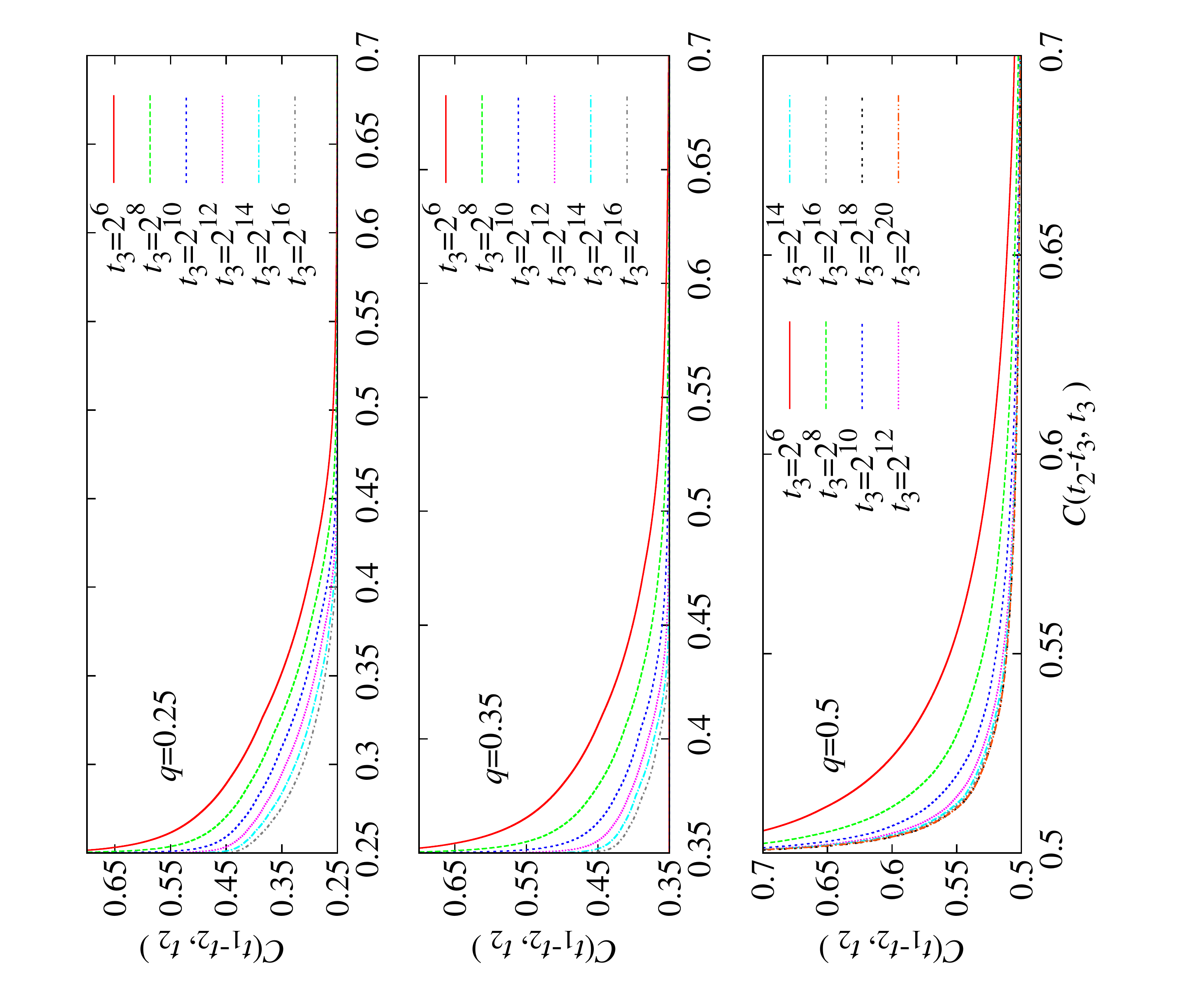}
\caption{ Parametric plot
  $\caja{x(t_2),y(t_2)}=\caja{C(t_1-t_2,t_2),C(t_2-t_3,t_3)}$, $t_1>t_2>t_3$
  with $t_1$ fixed by the condition $C(t_1-t_3,t_3)=q$ and different
  $t_3$. In the presence of dynamic ultrametricity, \eqref{eq:cond}, the parametric plot should
  tend for large $t_3$ to the union of $x=q$ and $y=q$.
The panels correspond to $q=0.25$ (\textbf{top}, nice BB scaling but no
ultrametricity expected), $q=0.35$ (\textbf{middle}, nice BB scaling and
ultrametricity expected) and $q=0.5$
(\textbf{bottom}, supposedly ultrametric but poor BB scaling). Note that there
are not qualitative differences between $q=0.25$ and $q=0.35$.}\label{fig:C12vsC23}
\end{center}
\end{figure}
\begin{figure}
\begin{center}
\includegraphics[angle=270,width=0.8\columnwidth,trim=0 0 0 0]{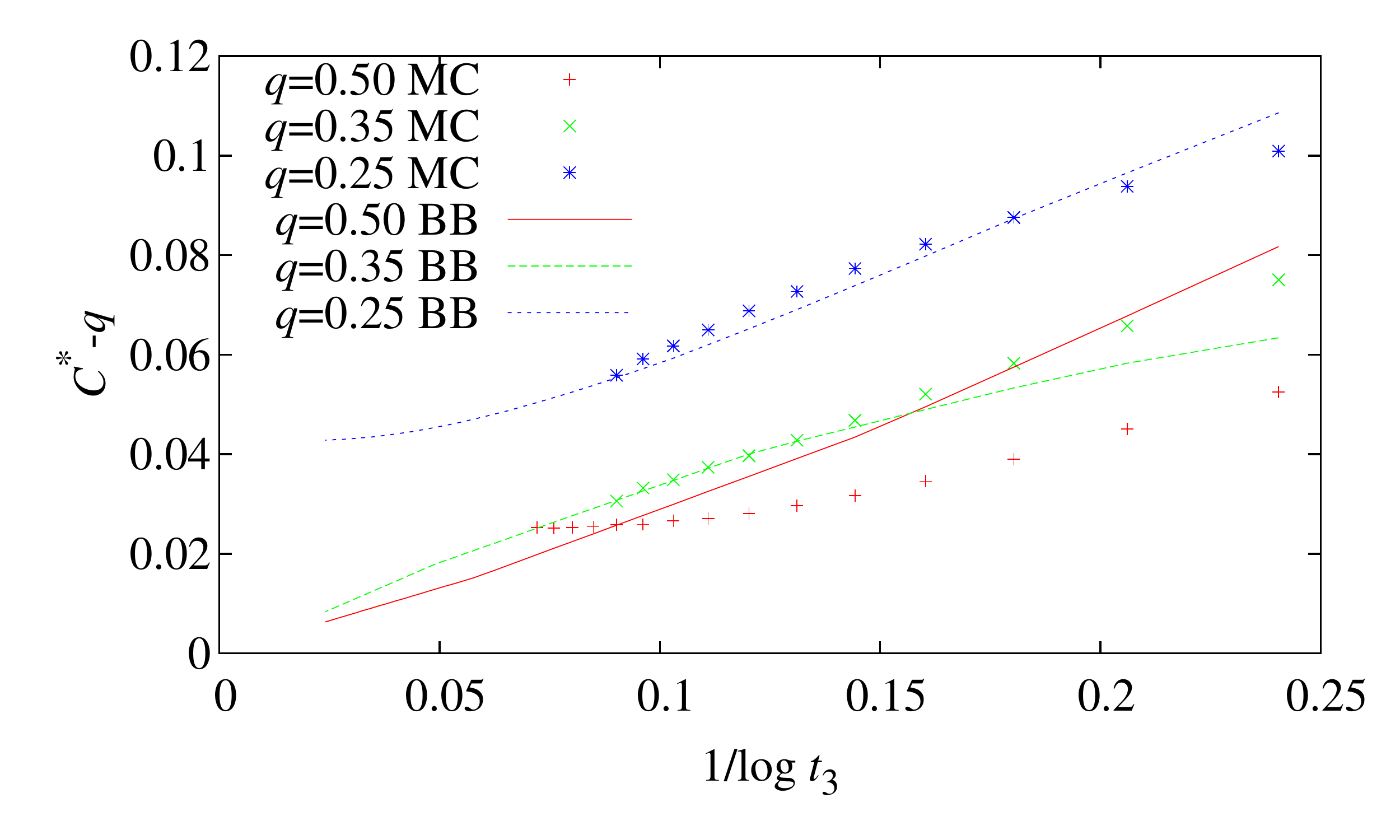}
\caption{ 
\textbf{Dots:} For each $q$ and $t_3$, as in Figure \ref{fig:C12vsC23},
  we take the intercept with $x=y$, i.e. $C^*=C(t_1-t_2,t_2)=C(t_2-t_3,t_3)$,
  and represent $C^*-q$ as a function of $1/\log t_3$. \textbf{Lines:}
   analogous plot for the toy model described in the text,
  where the BB scaling is exact. }\label{fig:C12vst3}
\end{center}
\end{figure}
\begin{figure}
\begin{center}
\includegraphics[angle=270,width=0.8\columnwidth,trim=0 0 0 0]{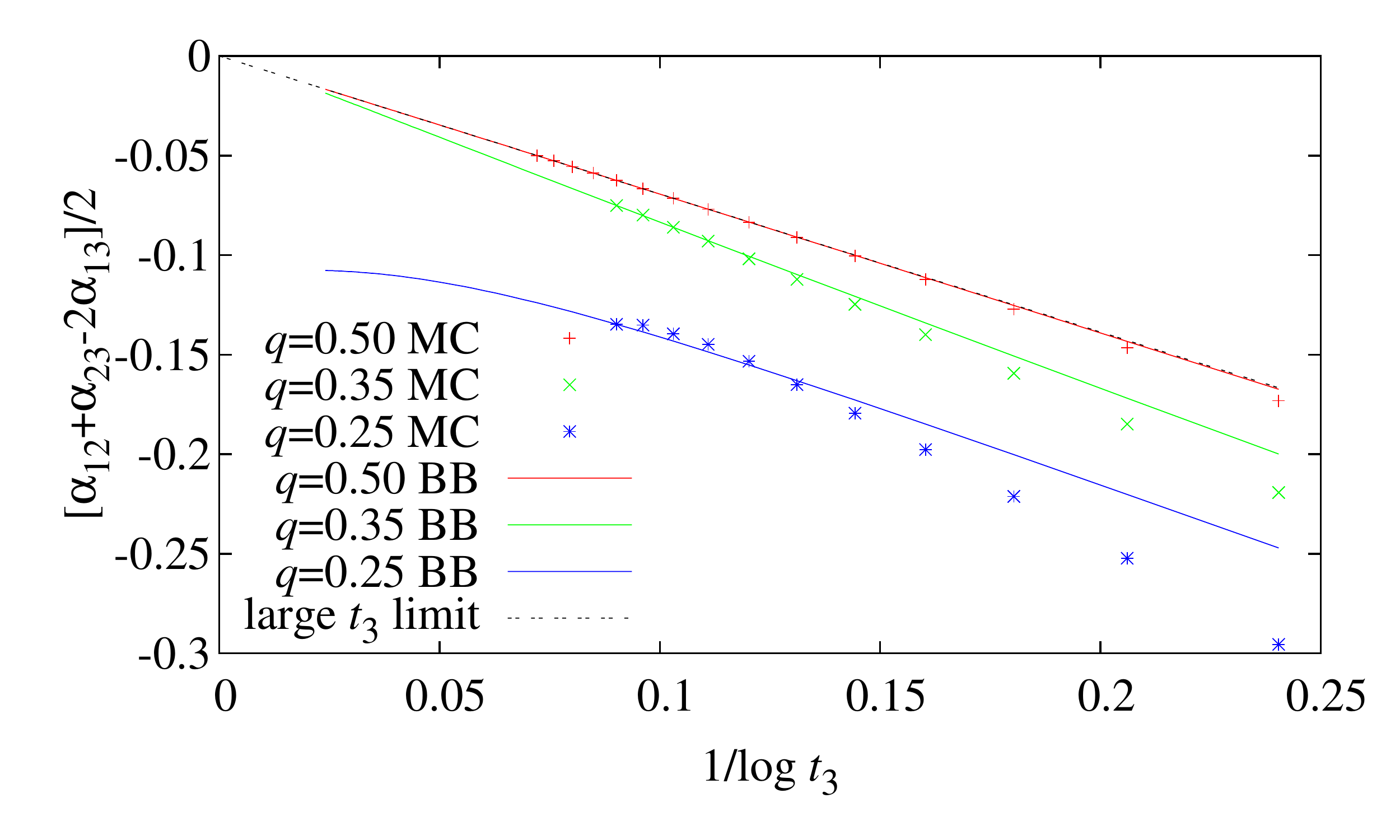}
\caption{ For the data in Figure \ref{fig:C12vst3}, we represent
  $\caja{\alpha(t_1,t_2)+\alpha(t_2,t_3)}/2-\alpha(t_1,t_3)$ vs. $1/\log
  t_3$. The dashed line corresponds to  \eqref{eq:alpha}.   }\label{fig:alpha12}
\end{center}
\end{figure}
As in  \eqref{eq:ultrametric1}, let us assume that the spin time correlation function behaves for large $t_{\mathrm w}$ as 
\be C(t,t_\mathrm{w})=f\paren{\alpha(t,t_\mathrm{w})}\,,\quad \alpha(t,t_\mathrm{w})=\log t/\log t_\mathrm{w}\,,\ee
where the scaling function $f$ is smooth and monotonically decreasing. From
now on, we shall refer to this scaling as BB scaling (after Bertin-Bouchaud).

Let us see under which conditions BB scaling implies the ultrametricity property
\be\label{eq:cond} C(t_1-t_3,t_3)= \min\left\{C(t_1-t_2,t_2),C(t_2-t_3,t_3)\right\},\ee
where $t_1\gg t_2\gg t_3$ and $t_3$ tends to infinity.

The natural time dependency is a power law choice
\begin{eqnarray}
t_1&=&t_3+At_3^{\mu_1},\\
t_2&=&t_3+Bt_3^{\mu_2},
\end{eqnarray}
with $\mu_1>\mu_2$. In that case, the large $t_3$ limit for the argument of
the scaling function are: $\alpha(t_1-t_3,t_3)=\mu_1$,
$\alpha(t_2-t_3,t_3)=\mu_2$ and $\alpha(t_1-t_2,t_2)=\mu_1$ if $\mu_2 < 1$ and
$\alpha(t_1-t_2,t_2)=\mu_1/\mu_2$ if $\mu_2 > 1$. Then, the condition
\eqref{eq:cond} is only satisfied in case $\mu_2 < 1$. If, as it is the case
for the critical trap model~\cite{bertin:02},
$f(\alpha>1)=\text{constant}$,~\footnote{Weak ultrametricity breaking implies
  that $f(\alpha>1)=0$.} the BB scaling would imply dynamic
ultrametricity. This is not the case for a general scaling function $f$ such
as, for instance, the one we get in Figure \ref{fig:ultrametric1}.
Nevertheless, although this analysis implies that the dynamic ultrametricity
is only present in our model in some range of parameters, let us try a more
straight approach. 

We consider a fixed value for the correlation function, $q$. On the view of
the previous considerations and of Figure \ref{fig:ultrametric1}, we should
expect ultrametricity only for $q>f(\alpha=1)\approx0.35$. Now, for each
$t_3$, we find $t_1$ such that $C(t_1-t_3,t_3)=q$. Then, we perform a
parametric plot of $C(t_1\!-\!t_2,t_2)$ vs. $C(t_2\!-\!t_3,t_3)$, for
$t_3<t_2<t_1$. Ultrametricity predicts that, in the large $t_3$ limit, the
curves should tend to a half square (e.g. the intersection of the straight
lines $x=q$ and $y=q$) and, in particular, when $C(t_1\!-\!t_2,t_2)=C(t_2\!-\!t_3,t_3)\!=\!C^*$, $C^*$ should tend to
$q$. 

We present in Figure \ref{fig:C12vsC23} results for three different values
of $q$: 0.5 (ultrametric region, but in our range of $t_\mathrm{w}$ data do
not scale according BB), 0.35 (ultrametric region and good BB scaling) and
0.25 (non ultrametric region but BB scaling works nicely). At the qualitative
level, the parametric curves seem to tend to a corner (but $q\!=\!0.5$), but the convergence is
slow. Furthermore, there are no clear differences between the curves with
$q\!>\!f(\alpha\!=\!1)$ and those with $q\!<\!f(\alpha\!=\!1)$. Hence, due to the failure of
this qualitative approach, we may try a more quantitative analysis.

We obtain numerically $C^*$, the point where
$C^*\!=\!C(t_1-t_2,t_2)\!=\!C(t_2-t_3,t_3)$, and study $C^*\!-\!q$ as function of
$1/\log t_3$. This choice is due to the fact that in
the ultrametric region BB scaling predicts
\be\label{eq:alpha}\alpha(t_1-t_2,t_2)=\alpha(t_1-t_3,t_3)-\frac{\log 2} {\log
  t_3} +\ldots\,.\ee Hence, we expect that $C^*\!-\!q$ will be of order $1/\log
t_3$ if ultrametricity holds. Let us sketch the proof. We define
$y\!=\!f^{-1}(q)\!=\!\alpha(t_1-t_3,t_3)$ (recall that $y<1$ in the
ultrametric region). Hence, the three times are 
\begin{eqnarray}
t_1&=&t_\mathrm{w}+t_\mathrm{w}^y\,,\\
t_2&=&t_\mathrm{w}+A(t_\mathrm{w})t_\mathrm{w}^y\,,\\
t_3&=&t_\mathrm{w}\,.
\end{eqnarray}
The hierarchy of time scales, $t_1\!>\!t_2\!>\!t_3\!\gg\! 1$, implies that, for large
$t_\mathrm{w}$, $A(t_\mathrm{w})$ is bounded. The condition
$\alpha(t_1-t_2,t_2)=\alpha(t_2-t_3,t_3)$ translates to 
\be \label{eq:igualdad}\frac{y\log t_\mathrm{w}+\log\big[1-A(t_\mathrm{w})\big]}{\log
  t_\mathrm{w}+\log\big[1+A(t_\mathrm{w})t_\mathrm{w}^{y-1}\big]}=y+\frac{\log
  A(t_\mathrm{w})}{\log t_\mathrm{w}}.\ee
The above equation can be solved asymptotically for $A(t_\mathrm{w})$ in the
limit of large $t_\mathrm{w}$ as (recall that $y\!<\!1$)
\be A(t_\mathrm{w})=\frac{1}{2}-\frac{y}{8}\;t_\mathrm{w}^{y-1}+\ldots\,.\ee
To obtain \eqref{eq:alpha}, one just notes that $\alpha(t_1\!-\!t_2,t_2)$ is equal
to the right hand side of  \eqref{eq:igualdad}.

The \ac{MC} numerical data in Figure \ref{fig:C12vst3} confirm the expectation of
$C^*\!-\!q\!=\!O(1/\log t_3)$ only partly. For $q\!=\!0.35$ the results are
as expected, yet for $q\!=\!0.25$ the difference is decreasing fast as $t_3$
grows and it is hard to tell whether the extrapolation will be zero or
not. For $q\!=\!0.5$ (where BB scaling is not working for our numerical data)
the behavior is non monotonic.

To rationalize our finding, we consider a
simplified model, where the BB scaling is supposed to hold exactly. The
master curve $f(\alpha)$ is taken from the numerical data for
$C(t,t_3\!=\!2^{16})$ for $D\!=\!22$. This toy model allows us consider ridiculously
large values of $t_3$. As we see in Figure \ref{fig:C12vst3}, the peculiarities
of the master curve cause a non monotonic behavior in $q$ for an ample range
of $t_3$.

The lack of monotonicity in $q$ makes also on interest to focus on $\alpha$,
rather than on the correlation function. With this aim, we consider the time
$t_2$ where $C(t_1\!-\!t_2,t_2)\!=\!C(t_2\!-\!t_3,t_3)\!=\!C^*$, and compute
$\frac12\caja{\alpha(t_1\!-\!t_2,t_2)\!+\!\alpha(t_2-t_3,t_3)}\!-\!\alpha(t_1-t_3,t_3)$. BB
scaling and ultrametricity combined, see  \eqref{eq:alpha}, imply that this
quantity should be of order $1/\log t_3$ (in the non ultrametric region, it
should be of order one).  Our results in Figure \ref{fig:alpha12} basically
agree with these expectations.

\chapter{Statistical ensembles}\label{app:ensembles}
In this Appendix we summarize the statistical ensembles we used along
Part~\ref{part:colloids} of the thesis. We focus on the uncommon ones. The
standard ensembles will be only named and its defining equations will be
defined only as a help to understand the new ensembles.

\section{Common definitions}
We consider $N$ particles, each at the position $\V{r}_i$ with
$i=1,\ldots,N$ in a cubic volume $V=L^3$ with periodic boundary
conditions.  
Let $U$ be the total potential energy of our system,
\begin{equation}
  U(\{\V{r}_i\})=\sum_{i<j} \mathcal{U}(|\V{r}_i-\V{r}_j|)\,,\ (u\equiv U/N)\,,
\end{equation}
with
 $\mathcal{U}(r)$ the pairwise interaction potential. From now on, we will use
the shortcut $\V{R}\equiv\{\V{r}_i\}$.

As it is common in the literature, we label the different ensembles by their
conserved magnitudes. For instance, $NVT$ accounts for the statistical
ensemble with conserved number of particles, $N$, volume, $V$, and
temperature, $T$.

\section{Canonical ensemble ($NVT$)}

The partition function is ($\beta=1/(k_\mathrm{B}T)$)
\be\label{eq:ZNVT}
Z_N(V,T)=\E^{-\beta F_N(V,T)}=\frac{1}{N!\Lambda^{3N}}\int\D\V{R}\,\E^{-\beta U(\V{R})},
\ee 
where $F_N(V,T)$ is the Helmholtz free-energy, $f(v,T)=F_N(V,T)/N$ the
free-energy density and $\Lambda$ the de Broglie thermal wavelength (an
irrelevant constant to make $Z_N$ dimensionless).

The canonical average of a
generic observable $O(\V{R})$ is 
\be
\mean{O}_\beta=\frac{\int\D\V{R}\,O(\V{R})\,\E^{-\beta U(\V{R})}}{\int\D\V{R}\,\E^{-\beta U(\V{R})}}.
\ee
\section{Isobaric ensemble ($NpT$)}\label{app:NPT}
If the pressure $p$ is fixed, the volume fluctuates.  The partition function
is \be\label{eq:YNpT} Y_N(p,T)=\E^{-\beta
  G_N(p,T)}=\frac{p\beta}{N!\Lambda^{3N}}\int \D V\E^{-\beta p
  V}\int\D\V{R}\,\E^{-\beta U(\V{R})}, \ee with $G_N(p,T)$ the Gibbs
free-energy and $Z_N(V,T)$ the $NVT$ partition function defined in
\eqref{eq:ZNVT}. The chemical potential is $g(p,T)=G_N(p,T)/N$.

Again, the isobaric average at fixed $p$ of a function of $V$ and the 
particle positions, $O(V,\V{R})$, is
\be
\mean{O}_p=\frac{\int \D V\,\E^{-\beta p V}\int\D\V{R}\,O(\V{R})\,\E^{-\beta U(\V{R})}}{\int \D V\,\E^{-\beta p V}\int\D\V{R}\,\E^{-\beta U(\V{R})}}.
\ee 
The overlap equivalence is obtained from~\eqref{eq:YNpT}. We rewrite it in
terms of Helmholtz free-energy density, $f(v,T)$, and the intrinsic volume $v=N/N$
\be\label{eq:YNpT2}
\E^{-\beta N g(p,T)}=p\beta N\int \D v\,\E^{-N\beta [p v+f(v,T)]}\,.
\ee 
Then, using a saddle point approximation, we can relate the pressure in the
$NpT$ ensemble with  $NVT$ averages. 
\begin{eqnarray}
p&=&-\displaystyle\left.\frac{\partial f(v,T)}{\partial
  v}\right|_\beta\\&=&\frac{1}{v}\paren{k_\mathrm{B}T+\frac{1}{3N}\mean{\sum_i\V{r}_i\cdot\grad_{\V{r}_i} U(\V{R})}_\beta}
\end{eqnarray}

\section{Microcanonical ensemble ($NVE$)}\label{SECT:MICRO}
In this ensemble, we want to constrain the value of the energy of the
system. Finding standard \ac{MC} moves that satisfy this constraint is rather
difficult. Instead, our proposal is to add a trivial Gaussian bath to the
potential energy, and to conserve the ``total'' joint energy. In order to do
so, we extend the configuration space with $N$ additional momenta $p_i$
(normal variables, they are simply a conceptual device to introduce the
ensemble~\cite{algorithm:lustig98}). Thus, our total energy is
\begin{equation}
E=U+K\,,\ (e\equiv E/N)\,.
\end{equation}
where \be K=\sum_{i=1}^N p_i^2/2\ee is the kinetic energy associated to the
conjugated momenta $\{p_j\}$. In the canonical ensemble, these ${p_i}$ are a Gaussian bath decoupled
from the particles. Here, we are considering just one conjugated
momentum per particle, we will see in Section \ref{app:tetheredO} that
this is not necessarily the best choice.  
 In particular $\mean{e}_\beta=\mean{u}_\beta+1/(2\beta)$.
As the kinetic energy is
non-negative by definition, we should have $E\geq U$.

A quantity of major importance in the microcanonical ensemble is the
entropy density, $s_N(e)$:
\begin{eqnarray}
\mathrm{exp}[N s_N(e)]&=& \int_\infty^\infty\prod_{i=1}^N \D p_i
  \int\,\D \V{R}\ \delta(N e- E).\nonumber
\end{eqnarray}
The conjugated momenta are explicitly integrated out using the Dirac's delta
function,
\begin{eqnarray}
\mathrm{exp}[N s_N(e)]&=& \frac{(2\pi N)^{N/2}}{N \Gamma(N/2)}\int \D \V{R}\ (e-u)^{\frac{N}{2}
-1} \theta(e-u)\,.\nonumber
\end{eqnarray}
The Heaviside step function, $\theta(e-u)$, enforces $e>u$. The
microcanonical average of an arbitrary function of the particle
positions $\V{R}$ and of the energy density $e$,
$O(\V{R};e)$ is
defined as 
\begin{eqnarray}
\langle O\rangle_e&\equiv&   \frac{\int \,\mathrm{d}\V{R} \, O(\V{R};e)  \omega_N (\V{R};e)}
{\int \,\mathrm{d}\V{R} \, \omega_N (\V{R};e)}\,,
\end{eqnarray}
where,
\begin{eqnarray}
\omega_N (\V{R};e)&=& (e-u)^{\frac{N}{2}
-1} \theta(e-u)\,.\label{DEF:MICROWEIGHT}
\end{eqnarray}
The canonical partition function (but for irrelevant constants) can be
recovered from the entropy density $s_N(e)$
\be Z_N(V,T)=\int\D e\,\E^{N[s_N(e)-\beta(e)
  e]}.\ee
Then, using the saddle-point approximation gives us a condition for the
inverse temperature
\begin{equation}
\beta(e) = \frac{\mathrm{d} s_N(e)}{\mathrm{d} e}\,,
\end{equation}
which leads to a microcanonical expectation value at fixed
energy $e$:
\begin{equation}
\beta(e)\equiv \langle\hat\beta\rangle_e,\quad \hat\beta
=\frac{N-2}{2N (e-u)}\,.
\label{BETADEF}
\end{equation}

\section{Microcorical ensemble ($N\hat VT$)}\label{app:microcor}
This ensemble is analogous to the microcanonical ensemble, but less
intuitive. Now we let the volume fluctuate but constrain it as well. It is then
very similar to the $NVT$ ensemble, but gives us more control of the
simulation. 

The fluctuations in the volume are introduced via $N$ Gaussian demons $\eta_i$
analogous to the momenta in the microcanonical approach, that is
\be\label{eq:hatV} \hat{V}=V+\sum_{i=1}^N \eta_i^2/2,\:(\hat{v}=\hat{V}/N).  \ee As we did with the
entropy, we compute the number of states that fulfill the imposed condition
$V+\sum_{i=1}^N \eta_i^2/2=\hat V$, \be \hat{Z}_N(\hat{v},T)=\int \D
V Z_N(V,T)\int \prod_{i=1}^{N} \D\eta_i\ \delta\paren{N\hat{v}-V
  -\sum_i^N\eta_i^2/2}.  \ee Again, these demons are decoupled from the rest
of variables and can be integrated out (now the trick is even clearer than in
the microcanonical case), \be \hat{Z}_N(\hat{V},T)=\E^{-N\beta
  \hat{f}_N(\hat{v},T)}=
\displaystyle\frac{\paren{2\pi}^{N/2}}{\Gamma\paren{N/2}}\int \D V Z_N(V,T)\paren{N\hat{v}-V}^{N/2-1}\varTheta\paren{N\hat{v}-V}.  \ee
This $\hat{f}_N(\hat{v},T)$ is our new Helmholtz free-energy density. 

The microcorical average of an arbitrary function of the particle positions
$\V{R}$ and of the  $\hat{v}$, $O(\{ {\boldsymbol
  r}\}_i;\hat{v})$ is
\begin{eqnarray}
\langle O\rangle_{\hat{v}}&\equiv&   \frac{\int \,\mathrm{d}\V{R} \, O(\V{R};\hat{v})  \omega_N (\V{R};\hat{v})}
{\int \,\mathrm{d}\V{R} \, \omega_N (\V{R};\hat{v})}\,,
\end{eqnarray}
with
\begin{eqnarray}
\omega_N (\V{R};\hat{v})&=& (N\hat{v}-V)^{\frac{N}{2}
-1} \theta(N\hat{v}-V)\,.\label{DEF:MICROVWEIGHT}
\end{eqnarray}
We can relate this
ensemble with the $NpT$ one just integrating over all the $\hat{v}$. Then, the
partition function is recovered
\be
Y_N(p,T)=\int \D \hat{v}\,\E^{-N\beta[p(\hat{v}) \hat{v}-
  \hat{f}_N(\hat{v},T)]}\,.
\ee 
Again, the saddle point approximation lets us to compute the pressure
\begin{equation}
p(\hat{v}) = \frac{\mathrm{d} f_N(\hat{v},T)}{\mathrm{d} \hat{v}}\,,
\end{equation}
which gives us the  microcorical expectation value at fixed
 $\hat{v}$:
\begin{equation}
p(\hat{v})\equiv \langle \hat{p} \rangle_{\hat{v}},\quad \hat{p}
=\frac{N-2}{2(N\hat{v}-V)}\,.
\label{PDEF}
\end{equation}
\section{Tethered ensemble}
\subsection{For one magnitude $O$ ($\hat{O}NpT$)}\label{app:tetheredO}
The tethered ensemble allows us to build an ensemble constraining the mean value
of any desired quantity. Here we develop the formalism for an arbitrary
magnitude $\aO(\V{R})=N \aob(\V{R})$ keeping fixed $N$, $p$ and $T$.

We first note that in the $NpT$ ensemble, the probability of getting certain value $o$ for
the observable $ \aob(\V{R})$ at a given pressure $p$ is \be
p_1( o,p)\propto\Int{0}{\infty}\RM{d}V\ \E^{-\beta p
  V}\int \D\V{R}\,\E^{-\beta U(\V{R})}\,\delta\paren{o-\aob(\V{R})}.  \ee

On the other hand, we consider a Gaussian bath of $\alpha N$
demons.\footnote{In the previous works to this thesis on this
  algorithm~\cite{fernandez:09,martin-mayor:09}, $\alpha$ was taken always
  equal to $1$. However, previous works were performed always in spin systems
  where the normal system sizes simulated are far larger than in colloidal
  systems, which is the case we are interested in applying the method. Indeed, the tethered method is introduced via a convolution of the
  physical ensemble probability with a Gaussian of weight $1/\sqrt{\alpha
    N}$, see Eq.~\eqref{eq:convo}. For the system sizes we studied in Chapter~\ref{chap:HS} ($N\le
  4000$) these Gaussian were too broad to resolve the different peaks if one
  took $\alpha=1$. The problem  could be directly solved by reducing this
  Gaussian weight, or in other words, increasing the amount of demons.  } The probability of $\sum_{i=1}^{\alpha N}\eta_i/\alpha
N$ to be equal to $s$ is \be p_2(s)\propto\Int{-\infty}{+\infty}\,
\prod_{i=1}^{\alpha N}\RM{d}\eta_i\, \E^{-\sum_{i=1}^{\alpha N}\eta_i^2/2}\,
\delta\paren{s-\frac{1}{\alpha N}\sum_{i=1}^{\alpha N}\eta_i}.\ee We introduce
a new variable \be\ho=o+s.\ee The probability distribution function for $\ho$
can be obtained with the convolution of these two last probabilities
\be \label{eq:convo}p(\ho,p)=\Int{0}{+\infty}\D o\Int{0}{+\infty}\D
r\ p_1(o,p) \ p_2(s)\ \delta\paren{\ho - o - s}.\ee As above, in the
microcorical case, the demons can be integrated out.  Then, the tethered mean
value of a generic observable $A(\V{R})$ fixed $\hat{o}$ is then given by
\begin{eqnarray}
\langle A\rangle_{\hat{o}}&\equiv&   \frac{\int\RM{d}V
\int \,\mathrm{d}\V{R} \, A(\V{R})  \omega_N (\V{R},p;\hat{o})}
{\int\RM{d}V
\int \,\mathrm{d}\V{R} \,   \omega_N (\V{R},p;\hat{o})}\,,
\end{eqnarray}
where,
\begin{eqnarray}
  \omega_N (\V{R},p;\hat{o})=\sqrt{\frac{\alpha N}{2\pi}}\,\E^{-\beta p V}\,\E^{-\beta U(\V{R})}\E^{-\frac{\alpha N}{2} \caja{\ho -
    \aob(\V{R})}^2}.\label{DEF:TETHERED}
\end{eqnarray}
In close analogy with the other ensembles, we can define a Helmholtz
effective potential
\be
\E^{-N\Omega_N(\ho,p)}=\frac{\beta p}{N!\Lambda^{3N}}\sqrt{\frac{\alpha N}{2\pi}}\int\RM{d}V
\int \,\mathrm{d}\V{R} \, \E^{-\beta p V}\,\E^{-\beta U(\V{R})}\,\E^{-\frac{\alpha N}{2} \caja{\ho -
    \aob(\V{R})}^2}.
\ee
The most important tethered average is the $\hat{o}$-derivative of this
effective potential, the tethered field,
\be
\frac{\partial \Omega_N}{\partial \hat{o}}=\mean{\hat{h}}_{\hat{o}},
\ee
with 
\be
\hat{h}=\alpha\caja{\hat{o}- \aob(\V{R})}.
\ee
\subsection{For several conserved magnitudes}\label{app:tethered2par}
In the previous section, we considered an ensemble with just one tethered
quantity. However, as we discuss in Chapter~\ref{chap:HS}, sometimes it is
necessary to consider several reaction coordinates at the same time. The
construction of the ensemble is analogous to what described for one
coordinate. We start by coupling the observables $\aO_i(\V{R})=N
\aob_i(\V{R})$, with $i=1,\ldots,n$, with $\alpha N$ demons each,
\begin{eqnarray}\ho_1=\aob_1+s_1,&\ldots,&\ho_n=\aob_n+s_n,\end{eqnarray}
and then follow the same steps of Section~\ref{app:tetheredO}. As a
consequence, we have now a $n$-dimensional effective potential
$\varOmega_N(\V{\ho})$, \be \E^{-N\Omega_N(\V{\ho},p)}=\frac{\beta
  p}{N!\Lambda^{3N}}\paren{\frac{\alpha N}{2\pi}}^{n/2}\int\RM{d}V \int
\,\mathrm{d}\V{R} \, \E^{-\beta p V}\,\E^{-\beta U(\V{R})}\,\E^{-\frac{\alpha
    N}{2} \caja{\sum_i\caja{\ho_i - \aob_i(\V{R})}^2}}\,.  \ee Now, the
gradient field is given by
\begin{align}
\grad\Omega_N(\V{\ho},p)\equiv\paren{\frac{\partial\Omega_N(\V{\ho},p)}{\partial
\ho_1},\ldots,\frac{\partial\Omega_N(\V{\ho},p)}{\partial
\ho_n}}=\\
=\displaystyle\paren{\mean{\alpha\paren{\ho_1 -
    \aob_1}}_{\V{\ho}},\ldots,\mean{\alpha\paren{\ho_n -
    \aob_n}}_{\V{\ho}}}.
\end{align}


\chapter[Thermalization checks]{Thermalization checks in the hard spheres
  crystallization}\label{app:thermalization}

In this appendix, we tackle the problem of thermalization of the systems we
worked with in Chapter~\ref{chap:HS}.
\section{Time-autocorrelation functions}\label{sec:hs-taus}

\begin{figure}
\includegraphics[angle=270,width=0.95\columnwidth,trim=0 0 0 0]{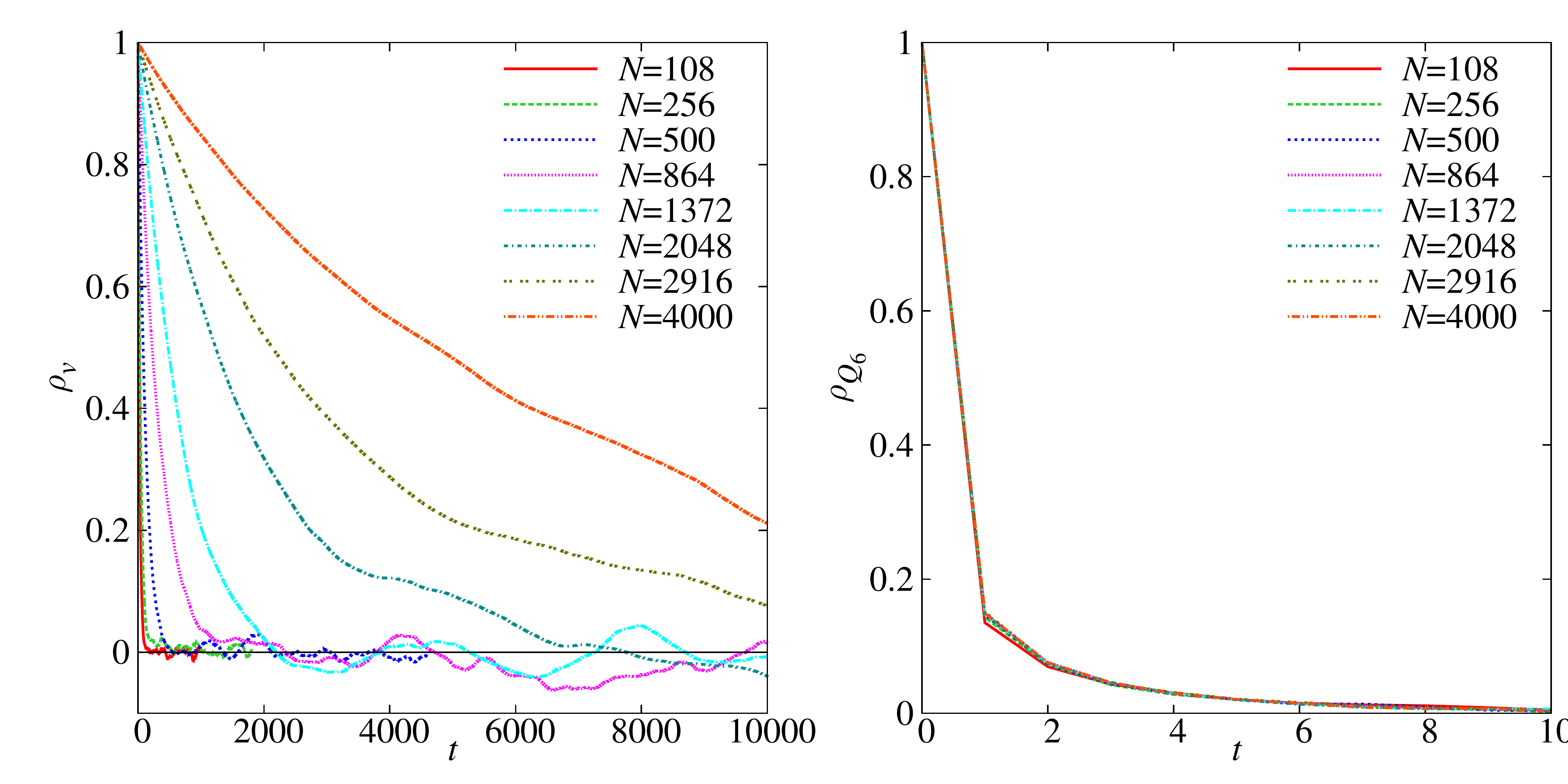}
\caption{Normalized time autocorrelation function, Eq.~\ref{eq:C-O-t}, for the
  specific volume (left) and the crystal order parameter (right), as computed
  for a system of $N$ hard spheres, in the fluid minimum of the effective
  potential (labeled $S=0$). Time is measured in units of EMCS (see
  text). Mind the different time scale for the left and right panels. Each
  value of $N$ was simulated very close to (but not precisely at) its
  phase-coexistence pressure $p_\mathrm{co}^N$ obtained in Section~\ref{sec:hs-results}.}
\label{fig:autocorr}
\end{figure}

We will begin the discussion studying briefly the time-autocorrelation
functions. These functions carry the information about the time it takes the
system to forget a particular configuration. For an observable $O(t)$, it is
defined as \bea \label{eq:C-O-t}\rho_O=\frac{C_{OO}(t)}{C_{OO}(0)}&\text{ with
}& C_{OO}(t)=\mean{O_sO\paren{s+t}}-\mean{O}^2.  \eea One should like to consider
the time autocorrelation functions for the components of the gradient field,
$\grad\varOmega_N$. Yet, its definition~\eqref{eq:hyomega} tells
us that these correlation functions are identical to those of $Q_6(\V{R})$ and
$C(\V{R})$. Eq.~\eqref{eq:reweighting-hs} suggests as well that the time
autocorrelation function for the specific volume $v$ is of interest. An
example of these autocorrelation functions is shown in
Fig.~\ref{fig:autocorr}, for the $S=0$ point (recall Figs.~\ref{fig:grid} and \ref{fig:h}). We note that $v$ plays the role
of the algorithmic slow mode, with a strong $N$ dependence. On the other hand,
the autocorrelation function for $Q_6$ decreases very fast, and it is barely
$N$-dependent. The autocorrelation function for $C$ is qualitatively identical
to that of $Q_6$, and will thus be skipped.

\begin{figure}
\centering
\includegraphics[angle=270,width=0.85\columnwidth,trim=0 50 0 50]{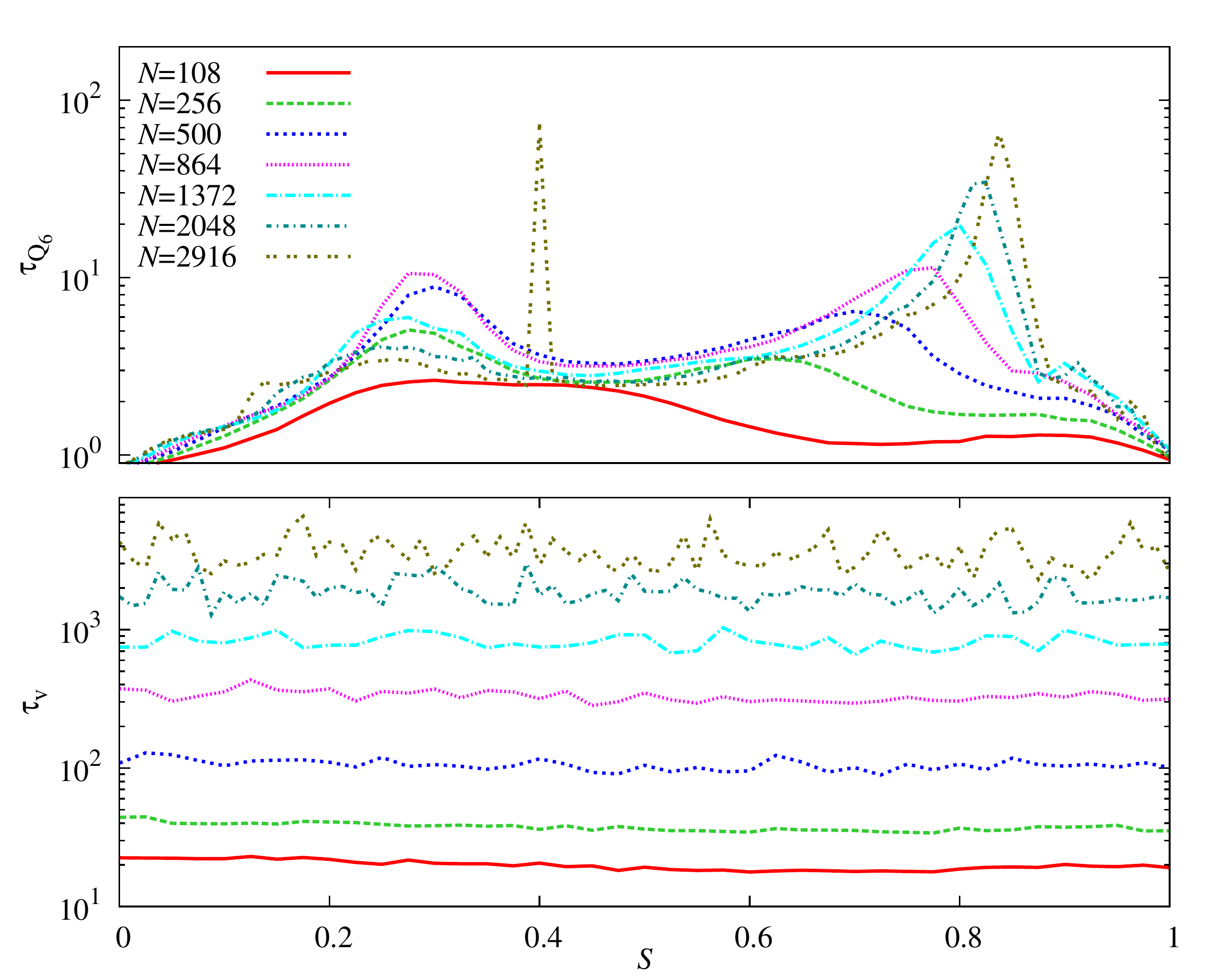}
\caption{Integrated autocorrelation times, defined in~\eqref{eq:tint}, for $\q$ (top) and $v$ (bottom), as a function of
  $S$ (the linear coordinate that labels the integration path in
  Fig.~\ref{fig:grid}, where $S=0$ stands for the fluid minimum and $S=1$
  represents the homogeneous FCC phase). Time is measured in units of
  EMCS.}
\label{fig:tiempos}
\end{figure}

The analysis is made quantitative by considering the integrated
autocorrelation times, \be\label{eq:tint}
\tau_{\text{int},O}=\frac{1}{2}+\sum_{t=1}^{\infty}\rho_{OO}(t),\ee see
Fig.~\ref{fig:tiempos}.\footnote{In practical situations, when times become
  long in comparison with this $\tau_\mathrm{int}$ itself, the signal-to-noise
  in function $\rho_{OO}$ becomes low, which results in large contributions
  to the sum (\ref{eq:tint}) from very noisy data. The solution to this
  problem, is to establish a large-time cutoff and determine
  $\tau_{\text{int},O}$ self-consistently. In our particular calculation, we
  replaced the $\infty$ by $6\tau_{\text{int},O}$.  }  We notice that the
dynamics of $v$ is considerable slower than that of $\q$, and featureless as a
function of $S$. Data for the specific volume scales as $\tau_v\sim N^{5/3}$
(quite worse than standard critical slowing down in three dimensions,
$\tau\sim N^{2/3}$, yet much better than exponential dynamic
slowing-down). There is a clear anomaly in the behavior of $\tau$ for a single
simulation point in $N=2916$. We will discuss this point in
Section~\ref{sec:grandes}, where we focus on the $N=4000$ and $2916$ systems.

Using these tools, we can be confident that all
simulations were, at least, $100\tau$ long. Besides, all simulations were
performed twice, with different starting configurations (either an ideal FCC
crystal, or an ideal gas). We check systematically the compatibility between
the two sets of investigations in the next section.

\section{Independence of results from different runs}\label{sec:independence}
Precisely to control the system equilibration, we run two independent
simulations, each starting from a completely different configuration: one
ordered, which is, in addition, the stable one in the crystal region, and one
disordered, the stable one in the fluid region. We can be confident about the
equilibration of the system if after certain time, we obtain the same mean
values (within error bars) with both startings. Thus, checking that the two
simulations are compatible is the goal of this section.

We can perform a systematic study of this compatibility through the
$\grad_S\varOmega_N$, obtained as the projection of (\ref{eq:hyomega}) on the
simulated straight line in Figure~\ref{fig:grid}, which is central in the
calculation of the main quantities obtained in this work: $\p$ and
$\gamma_{\{100\}}$. The procedure is following: we obtain this
$\grad_S\varOmega_N$ separately in simulations starting from a random
configurations, namely, $\grad_S\varOmega_N^\text{fluid}$ and from FCC
configurations, namely, $\grad_S\varOmega_N^\text{FCC}$, and we compute the
following quotient, \be
y_S=\frac{\grad_S\varOmega_N^\text{FCC}-\grad_S\varOmega_N^\text{fluid}}{\sqrt{\sigma_{\grad_S\varOmega_N^\text{FCC}}^2+\sigma_{\grad_S\varOmega_N^\text{fluid}}^2}}.\ee

Since the two $\grad_S\varOmega_N$ variables are mean values obtained from a Monte Carlo
simulation, they are Gaussian distributed with the same mean (if the
simulation is ergodic). Then, the expected quotient $y_S$ should be normal
distributed. In particular, $\mean{y_S}=0$, $\mean{y_S^2}=1$ and
$\mean{y_S^4}=3$. We can check if this is the case or not.

We start by studying if the mean of these $y_S$ values
is indeed $0$ for all $S$-points. With this aim we perform a $\chi^2$ test to
check this assumption.  We present in the first columns of Table
\ref{tab:chi2}, the $\chi^2$ per \ac{dof}, obtained as,
\be\chi^2=\sum_{i=1}^{N_S}y_S^2,\ee with $N_S$ degrees of freedom (dof in the
Table). As usually, if data $y_S$ are indeed normally distributed,
$\chi^2/\text{dof}$ should be close to 1. In addition, we compute the
probability of obtaining (for a set of $N_S$ perfect normal distributed
variables) a higher value of $\chi^2/\text{dof}$ than $\chi^2_0$ (the value
quoted in the table). We refer to this probability as $Q$, and it is defined
as \be\begin{array}{ll}
Q&=p(\chi^2>\chi^2_0)\!=\!\frac{1}{\Gamma\paren{N_S/2}}\int_{\sum_i
  y_i^2>\chi^2_0}\prod_j \text{d}y_j\ \E^{-\frac{1}{2}\sum_j y_j^2
}=\\&=\displaystyle\frac{1}{\Gamma\paren{N_S/2}}\int_{\chi^2_0/2}^\infty
\E^{-u}u^{N_S-1}\text{d}u,
\end{array}\ee
where $\Gamma$ is the Euler gamma-function.
\begin{table*}
\centering
\begin{tabular*}{\textwidth}{@{\extracolsep{\fill}}|ccc|ccc|cccc|}
\cline{1-10}
$N$&$\chi^2/\text{dof}$&$Q$&$y^\text{min}$&$y^\text{max}$&$p(y^\text{min},y^\text{max})$&$N^{1\sigma}$&$N^{2\sigma}$&$N^{3\sigma}$&$N^{>3\sigma}$\\\cline{1-10}
108&48.09/41&0.21&-2.88&2.49&0.71&25(28.0)&39(39.1)&41(40.9)&0(0.1)\\
256&42.40/41&0.41&-2.42&2.37&0.50&29(28.0)&37(39.1)&41(40.9)&0(0.1)\\
500&47.85/41&0.21&-2.79&1.95&0.31&27(28.0)&39(39.1)&41(40.9)&0(0.1)\\
864&57.84/41&0.04&-2.84&2.94&0.85&27(28.0)&38(39.1)&41(40.9)&0(0.1)\\
1372&37.62/41&0.6&-2.77&1.56&0.07&29(28.0)&40(39.1)&41(40.9)&0(0.1)\\
2048&85.77/81&0.34&-2.66&2.28&0.29&53(55.3)&77(77.3)&81(80.8)&0(0.2)\\\cline{1-10}
2916&97.25/81&0.11&-1.53&6.24&0.01&65(55.3)&78(77.3)&80(80.8)&1(0.2)\\
\cline{1-10}
\end{tabular*}
\caption{Study of the distribution function of $y_S$.}
\label{tab:chi2}
\end{table*}
 
The $N=2916$ data is presented separately because the integrated
correlation times (plotted in Figure~\ref{fig:tiempos})  cast some
doubts about its thermalization. The data for $N=4000$ is not included in this
study, since from direct visualization we can see it is not thermalized (see
Figure~\ref{fig:Maxwellgrandes} below).

One naive approach to the numbers shown in Table \ref{tab:chi2} might lead to
a wrong feeling about the compatibility of mean values for $N\le2048$. But
this idea does not survive a deeper inspection. Indeed, in statistics the rare
events (in the sense that they have low probability to occur) must appear, one
must worry about them if they happen too often.  Let us discuss the worst
$\chi^2$ test case (analogously for the other cases). For
$N=864$ we get a very low value $Q$, in fact, there is only a $4\%$
probability of getting a worst test. However, one must
recall that we computed $7$ values of $\chi^2$ here. Thus, the probability
for the lowest $Q$ being $4\%$ or smaller, becomes as large as 
 $25\%$. This is equivalent to say that if we performed this very
same study 4 times, such a bad result should be expected to occur at least
once. Because of that, we are not concerned by the
thermalization of the system of $N=864$. Besides, this ``bad'' result was
obtained for a middle-sized system, in a region of $N$ where we can be
confident about having a many exponential times in the simulation.

However, if we are not confident enough, the $\chi^2$ test is not the only
check we can perform about the normality of the data. If the $y_S$ are indeed
normal distributed, we can compute the theoretical probability that all the
$y_S$ points lied in the interval in between the minimum and the maximum $y_S$
obtained for each $N$, i.e.  \be\label{eq:probyminymax} p(y_\text{min}<
y_i<y_\text{max})=\caja{\frac{1}{\sqrt{2\pi}}\int_{y_\text{min}}^{y_\text{max}}\E^{-y^2/
    2}\text{d}y}^{N_S}\,.\ee We present these values also in Table
\ref{tab:chi2}, in the second block of columns. 

Finally, we also can compute the number of $y_S$ values we got separated from
the zero mean value by less than one, two and three mean deviations, as well
as the number of data we got beyond 3 sigmas. We can compare these numbers
with the theoretical predictions of a normal distribution computed as
(\ref{eq:probyminymax}) (see third block of columns in Table \ref{tab:chi2},
numbers in between parenthesis represent the theoretical predictions).

As a summary of all the numbers presented in this table, we can conclude that
fluctuations in $y_S$ seem to be completely Gaussian for $N\le2048$, the two
starting points seem to lead to same mean results, which make us feel
confident about the correct thermalization of our samples. In addition,
concerning the case of $N=2916$, all results shown in the table seem
reasonable but the $\grad_S\varOmega_N$ point where the prediction for the random start and
the ordered start differed by $y_S=6.26$ (we know there is only one point with
$|y_S|> 3$). This fact makes us feel confident about that the problem in
thermalization of $N=2916$ is restricted a single $S$ value.

A different check regards the computation of $\p^N$. Indeed, we can check that
its determination does not depend on the initial configuration. We compute the
difference between the two estimations of $\p^N$ and divide this number by its
error (see Table \ref{tab:Pcomp}).
\begin{table}
\centering
\begin{tabular*}{\columnwidth}{@{\extracolsep{\fill}}|ccc|c|}
\hline
$N$&$\p^\text{FCC}$&$\p^\text{fluid}$&$(\p^\text{FCC}-\p^\text{ran})/\text{error}$\\\hline
108&10.9222(22)&10.9206(26)&0.4828\\
256&11.2192(16)&11.2225(18)&-1.4081\\
500&11.3628(13)&11.3589(15)&2.0350\\
864&11.4399(13)&11.4429(18)&-1.3450\\
1372&11.4910(16)&11.4886(16)&1.0530\\
2048&11.5151(12)&11.5143(14)&0.4228\\\hline
2916&11.5267(12)&11.5329(19)&-2.7624\\\hline
\end{tabular*}
\caption{Comparison between the $\p^N$ estimations for simulations starting
  from a random distribution or a FCC perfect lattice.}\label{tab:Pcomp}
\end{table} 
The differences between the two estimations are very reasonable, but for the 
$N=2916$ point, where it is far too large. We will devote next section to the
study of this problem.

Finally, the reader might have noticed that, although both kind of simulations
have exactly the same length, the errors of $\p$ in Table \ref{tab:Pcomp} are
systematically larger when the simulation started from a fluid
configuration. This fact stems from the maximum change in volume, $\delta v$,
allowed for the Metropolis test. Indeed, quite annoyingly, this $\delta v$
depended on the kind of start we were considering since it was associated to
the initial density in the computer program.\footnote{Technically, the random
  configuration was obtained in a larger simulation box (very low density), in
  order to minimize the number of particles whose radius superposed after
  proposing random positions for each particle. With such a density, the
  \ac{FCC} lattice would melt instantaneously, thus running the simulations
  from the fluid phase as well.}

\section{$N=2916$ and $N=4000$ particle systems}\label{sec:grandes}

The anomaly at $S=0.4$ for $N=2916$ in Figure \ref{fig:tiempos} is due to the
emergence of a metastability.  At this point, we expected to find a
spatially segregated state (a slab of FCC crystal in a liquid matrix). This
state appeared indeed, but the simulation tunnels back and forth from it to an
helicoidal crystal (a similar crystal to the one illustrated in
Fig.~\ref{fig:configuracionesQ6}--right, when we 
tethered only  $\q$).

\begin{figure}
\centering
\includegraphics[angle=270,width=0.8\columnwidth,trim=0 0 20 40]{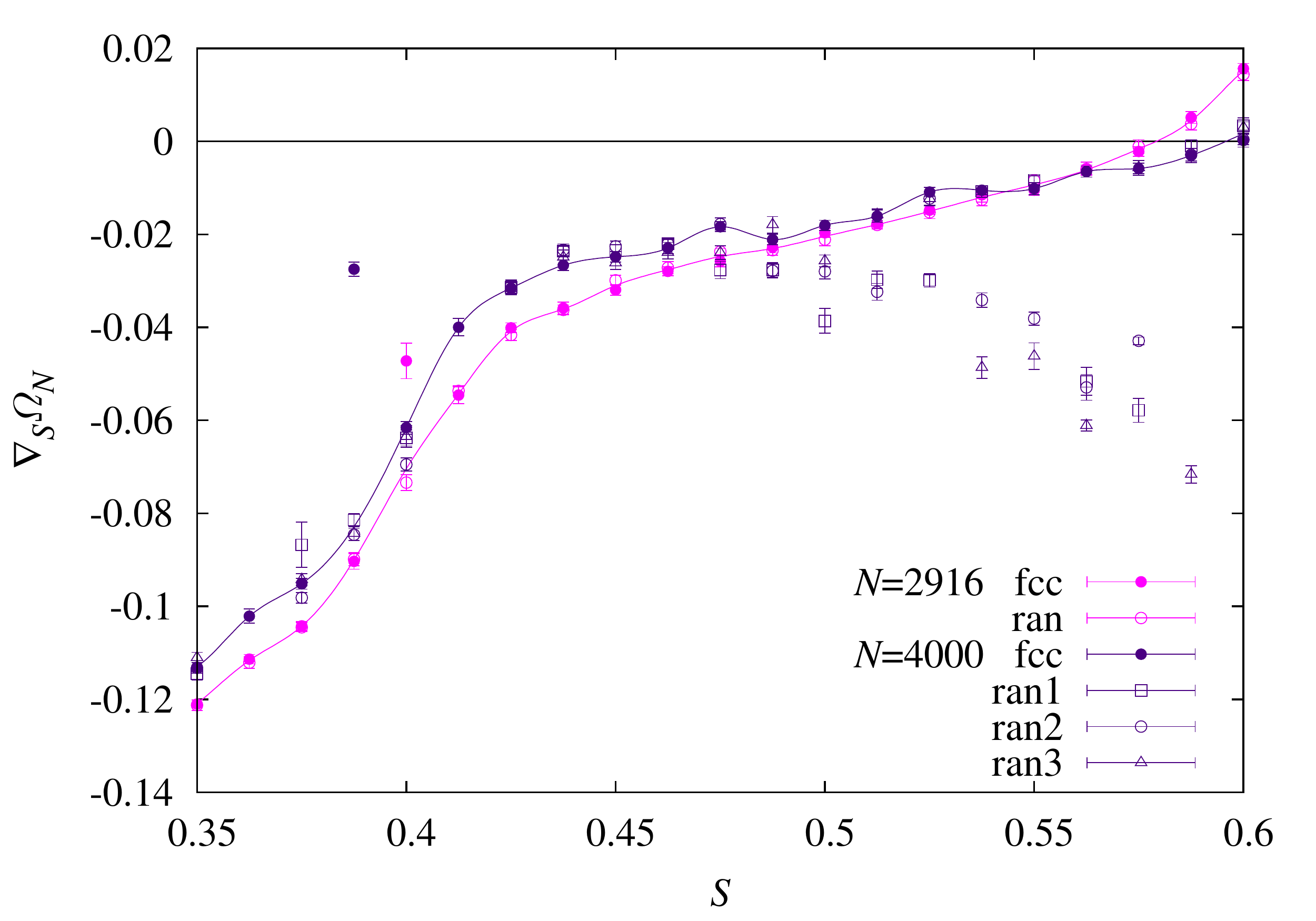} 
\caption{Enlarged central part of the spinodal curve for the two biggest system sizes, $N=2916$ and $4000$.}
\label{fig:Maxwellgrandes}
\end{figure}

We show in Fig. \ref{fig:Maxwellgrandes} both $\grad_S\varOmega_N$ for the two independent runs in $N=2916$ and in $N=4000$ particles. 
From the figure, it is clear that these helicoidal crystals appear much more
often for $N=4000$ and intermediate $S$. Nevertheless, selecting carefully the
starting particle configuration for the simulation at each $S$, one may obtain
a gradient field with a smooth $S$-dependency (represented in a solid line in
Fig. \ref{fig:Maxwellgrandes}). However, it is clear that these $N=4000$
results, although plausible, cannot be regarded as well equilibrated. For this
reason, although we presented results in the Tables all over the Chapter~\ref{chap:HS}, we
did it just as a hint, in the sense that these values are not used for any
large-$N$ extrapolation.

The situation is more subtle for $N=2916$ particles, because the
metastabilities are only observed at $S=0.4$. At this point, both runs find a
solid-fluid mixed state, as happens at nearby points. However, for the
simulation starting from a \ac{FCC} we find a metastability of this mixed
state with an helicoidal crystal, with significantly higher $\q$ and
$C$. Although we extended the length of this random-start run, this crystal
was never found. In order to check how important this new phase was, we ran
some extra new independent simulations (from both kind of startings) at this
point but no one but the original FCC-starting one visited this phase. In
other words, this phase seemed to be very rare. However, even being rare, we
cannot be sure about its statistical weight, it could be the most stable phase
at this point. With the aim of refusing this hypothesis, we also ran some new
simulations starting from one of the configurations corresponding to this
phase (using different random numbers). In all the cases, all the runs
``fell-back'' after some time to the mixed state found in the rest of runs.

Now, after being sure that this helicoidal crystal is not the most stable
state, we can try to delimit its probability of appearance, and mix the data
for $\grad\varOmega_N$ coming from the different runs, according to their
relative probability, to obtain the equilibrium estimate. This we can do it
using the detailed balance condition. Indeed, if
we have two states $i$ and $j$, \be W_{j\to i}\ p_j=W_{i\to j}\ p_i \ee where
$W_{i\to j}$ is the probability of hopping from state $i$ to state $j$, and
$p_i$ the probability of being at state $i$. That means that one can compute
the relative probability between these two phases by computing the flip-flop
probabilities. With this idea, and our numerous simulations with jumps in both
directions, we could estimate that the probability of the helicoidal phase was
upper bounded by $10\%$. Our number of runs is limited, thus, in order to not
underestimate this phase we also include an error in the determination of the
probability of roughly $10\%$, which we know for sure that is an
upper-estimation.

We now mix the values of $\grad\varOmega_N$ obtained in each of the two phases
accordingly to their relative probability. Afterwards, we obtain $\p$
following the same procedure all over the Chapter~\ref{chap:HS}.

In Table~\ref{tab:Pcomp} we obtained incompatible values for $\p$ obtained
with the different runs. Now, mixing the data of both simulations only at
$S=0.4$, we see that the origin of this divergence was nothing but this
described metastability. In order to justify this statement, we mix the data
at $S=0.4$ for $\grad\varOmega_N$ from the two phases using different relative
probabilities. For the rest of the points we mix the data in the same way done
all over the Chapter, that is $50\%-50\%$ from the two starts's
simulations. After obtaining the whole $\grad\varOmega_N(S)$ curve with this
procedure, we can compute $\p$. We display these $\p$ values in Table
\ref{tab:Pcomp2916}. Clearly, the differences found in Table~\ref{tab:Pcomp}
are a direct consequence of the lack of thermalization in $S=0.4$.
\begin{table}
\centering
\begin{tabular*}{\columnwidth}{@{\extracolsep{\fill}}|ccc|}
\hline
Percentage & $\p^N$ & $\gamma_{100}^ N$\\\hline
$p_{0\%}$&11.5314(10)&0.5972(10)\\
$p_{10\%}$&11.5311(9)&0.5971(10)\\\hline
$p_{20\%}$&11.5305(9)&0.5968(10)\\
$p_{50\%}$&11.5292(10)&0.5963(10)\\
$p_{100\%}$&11.5283(11)&0.5959(11)\\\hline
\end{tabular*}
\caption{Comparison between the $\p^N$ and $\gamma_{100}$ between the three
  different ways of mixing the conflicting point.}\label{tab:Pcomp2916}
\end{table} 

Finally, the estimations of $\p^{(2916)}$ quoted in Table~\ref{tab:pN} and
$\gamma_{\{100\}}^{(2916)}$ in Table~\ref{tab:sigma} are obtained with a
relative mixture of $10\%$ helicoidal crystal at $S=0.4$. In
addition, the error is taken as the sum of each inner statistical error plus a
systematic error coming from our uncertainty in the determination of the
probability of the helicoidal crystal. We consider this error as 
the difference between $p_{10\%}$ and $p_{20\%}$.

\chapter{$C$ values in a perfect lattice}\label{app:C}
At variance with  $\q$, it is very easy to obtain the $C$
(defined in \eqref{eq:C}) for a perfect lattice. We compute here the
predictions for the two phases of interest, namely the \ac{FCC}
and the \ac{BCC}.

\section{FCC}
We consider a perfect \ac{FCC} lattice. In it, each particle $\V{r}_0$ has
twelve possible nearest-neighbors at positions $\V{r}_i$. In units of the
characteristic length of the lattice, the $12$ neighbors are placed at \bea
&\displaystyle\paren{\V{r}_i\!-\!\V{r}_0}\!=\!\paren{0,\frac{1}{2},\pm\frac{1}{2}},\paren{0,-\frac{1}{2}\!,\!\pm\frac{1}{2}},\paren{\frac{1}{2},0,\pm\frac{1}{2}},&\\\nonumber&\displaystyle\paren{-\frac{1}{2},0,\pm\frac{1}{2}},\paren{\frac{1}{2},\pm\frac{1}{2},0},\paren{-\frac{1}{2},\pm\frac{1}{2},0}.&
\eea 
Then, plugging these positions in the definition of $c_\alpha(\V{r})$ given in
(\ref{eq:calpha}), we
obtain for all of them,
 \be c_\alpha
(\V{r})=\frac{1}{(1/\sqrt{2})^8}\frac{1}{2^4}\frac{1}{2^4}=\frac{1}{16}\,. \ee
As mentioned, $N_b=12$ for all particles. Then,  \be
C=\frac{2288}{79}\frac{1}{16}-\frac{64}{79}=1\,.
\ee

\section{BCC}
We perform the same calculation for the BCC structure. In this case all
particles have 8 nearest neighbors, located at \be
\paren{\V{r}_i-\V{r}_0}=\paren{\pm\frac{1}{2},\pm\frac{1}{2},\pm\frac{1}{2}}.
\ee Then $r=\sqrt{3}/2$ in all the cases, and \be c_\alpha
(\V{r})=\left.3\frac{1}{2^8}\paren{1-\frac{1}{9}}\right/\frac{3^
  4}{2^8}=\frac{2^3}{5^5}.  \ee We introduce this result in \eqref{eq:C}, \be
C=\frac{2288}{79}\frac{2^3}{5^5}-\frac{64}{79}=0.143\cdots.  \ee This last
result is the one reported in~\cite{angioletti:10}.  However, in
Section~\ref{sec:hs-order-parameters}, we discussed the notion of nearest
neighbor for our systems. We did it terms of the \ac{FCC} radius only. Indeed,
we wanted a definition that could guarantee that we only counted the first
shell of neighbors in this case, no matter the total volume of our simulation
box. It turns out, that the actual definition reaches the second shell of
neighbors in the case of a perfect \ac{BCC}. Then, we need to include 6 extra
neighbors in the calculus, placed at \be \paren{\V{r}_i-\V{r}_0}=\paren{\pm 1,
  0 ,0},\paren{0,\pm 1 ,0},\paren{0,0,\pm 1}.  \ee 
It turns out that $c_\alpha (\V{r})$  is zero for all these vectors, but still
we need to average over all the neighbors \be \frac{\sum_{i=1}^{N} \sum_{j=1}^{N_b(i)}c_{\alpha}({\hat{\V{r}}_{ij}})}{\sum_{i=1 }^{N}N_b(i)}=\frac{6\cdot
  0+8\cdot\frac{8}{3^5}}{8+6}=\frac{2^5}{3^5\cdot 7}, \ee which results in the final
value \be C=-0.2657\cdots.\ee

\chapter{Analysis with the Suzuki-Trotter formula}\label{sec:SA}
We investigate the properties of $\hat H(s,\lambda)$, defined in~\eqref{eq-qa:H},
the phase diagram in particular, using the decomposition formula
\cite{suzuki:76} and the static approximation. This approach, although
quantum, leads to the same results as the semi-classical method described in
section \ref{sec:CA}. The method here is analogous to the one explained in
detail in \cite{seki:12,jorg:10a}, but we leave the power $k$ as a free
parameter in all the calculus. The purpose of this appendix is to confirm
consistency between the method of the main text and that in
\cite{seki:12,jorg:10a}.

The starting point is the partition function, \be\label{eq:Z} Z=\Tr
e^{-\beta\hat{H}(s,\lambda)}.  \ee We use the  decomposition
formula to express it as
\begin{eqnarray}\label{eq:Z0}\nonumber
Z&=&\lim_{M\to\infty}Z_M\equiv\lim_{M\to\infty}\Tr
\lazo{e^{-\frac{\beta}{M}
    s\lambda\hat{H}_0}e^{-\frac{\beta}{M}\caja{s\,(1-\lambda)\hat{V}_{\mathrm{AFF}}+(1-s)\hat{V}_{\mathrm{TF}}}}}^M\\\nonumber
&=&\lim_{M\to\infty}\sum_{\{\sigma^z\}}\bra{\{\sigma^z\}}\left\{\exp\caja{\frac{\beta
      s\lambda N}{M}\paren{\frac{1}{N}\sump\hsiz}^p}\right.\\\nonumber
&&\left.\times\exp\caja{-\frac{\beta s\,(1-\lambda) N}{M}\paren{\frac{1}{N}\sump\hsix}^k+\frac{\beta (1-s) }{M}\sump\hsix}\right\}^M\ket{\{\sigma^z\}},\\
\end{eqnarray}
where $\sum_{\{\sigma^z\}}$ refers to the summation over all the $2^N$
possible spin configurations in the $z$ basis, and
$\ket{\{\sigma^z\}}\equiv\otimes_{i=1}^N\ket{\siz}$.  

We introduce $M$ closure relations, each one labeled by $\alpha(=1,\ldots,M)$,
\be \hat{\mathbb{I}}(\alpha)\equiv
\sum_{\{\sigma^z(\alpha)\}}\ket{\{\sigma^z(\alpha)\}}\bra{\{\sigma^z(\alpha)\}}\times\sum_{\{\sigma^x(\alpha)\}}\ket{\{\sigma^x(\alpha)\}}\bra{\{\sigma^x(\alpha)\}},
\ee just before the $\alpha$th exponential operator involving $\hsix$ in
\eqref{eq:Z0}. The trace over the product of quantum operators is thus reduced
to the product of numbers that commute and can be reordered,
\begin{eqnarray}\label{eq:ZM}\nonumber
Z_M&=&\prod_{\alpha=1}^M
  \sum_{\{\sigma^z(\alpha)\}}\sum_{\{\sigma^x(\alpha)\}} 
\exp\caja{\frac{\beta s \lambda N }{M}\paren{\frac{1}{N}\sump\siz(\alpha)}^p}\\\nonumber
&&\times\exp\caja{-\frac{\beta s (1-\lambda) N
  }{M}\paren{\frac{1}{N}\sump\six(\alpha)}^k+\frac{\beta (1-s)
  }{M}\sump\six(\alpha)}\\
&&\times\prod_{i=1}^N\left\langle\siz(\alpha)\right.\ket{\six(\alpha)}\left\langle\six(\alpha)\right.\ket{\siz(\alpha+1)},
\end{eqnarray}
where $\ket{\siz(M+1)}\equiv\ket{\siz(1)}$.

We write the product in terms of the total $x$ and $z$ magnetizations in each
copy of the system, i.e.  $m^x(\alpha)\equiv\frac{1}{N}\sump\six(\alpha)$ and
$m^z(\alpha)\equiv\frac{1}{N}\sump\siz(\alpha)$, using the integral definition
of the delta distribution \be
f\paren{\frac{1}{N}\sump\sigma_i(\alpha)}=\int\mathrm{d}m\,\delta\paren{m(\alpha)-\frac{1}{N}\sump\sigma_i(\alpha)}f\paren{m(\alpha)}.
\ee After a few simplifications, we introduce the static approximation to
remove the $\alpha$ dependence of the magnetizations. Under this
approximation, we can compute the $M\to\infty$ limit using again the
decomposition formula.  The partition function \eqref{eq:Z} then reduces to
\be
Z=\int\mathrm{d}m^z\,\mathrm{d}m^x\,\exp\caja{-N\beta\,f(\beta,s,\lambda;m^z,m^x)},
\ee where $f(\beta,s,\lambda;m^z,m^x)$ is the pseudo free-energy defined as
follows:
\begin{eqnarray}\label{eq:freebeta}\nonumber
f(\beta,s,\lambda;m^z,m^x)=(p-1)\, s\,
    \lambda(m^z)^p-\, (k-1)\, s\,
      (1-\lambda)(m^x)^k\\-\frac{1}{\beta}\log\lazo{2 \cosh \beta\sqrt{\caja{p\, s\,\lambda\, (m^z)^{p-1}}^2+\caja{1-s-s\,(1-\lambda)\,k\,(m^x)^{k-1}}^2}}.
\end{eqnarray}
Again, one can apply the saddle-point method, obtaining two self-consistent equations
for the two magnetizations,
\begin{eqnarray}\label{eq:mz}
m^z&=&\frac{p\, s\,\lambda\, (m^z)^{p-1}}{\sqrt{\caja{p\, s\,\lambda\,
      (m^z)^{p-1}}^2+\caja{1-s-s\,(1-\lambda)\,k\,(m^x)^{k-1}}^2}}\\\nonumber
&\times& \tanh \beta\sqrt{\caja{p\, s\,\lambda\,
      (m^z)^{p-1}}^2+\caja{1-s-s\,(1-\lambda)\,k\,(m^x)^{k-1}}^2},\\\label{eq:mx}
m^x&=&\frac{1-s-s\,(1-\lambda)\,k\,(m^x)^{k-1}}{\sqrt{\caja{p\, s\,\lambda\,
      (m^z)^{p-1}}^2+\caja{1-s-s\,(1-\lambda)\,k\,(m^x)^{k-1}}^2}}\\\nonumber
&\times& \tanh \beta\sqrt{\caja{p\, s\,\lambda\,
      (m^z)^{p-1}}^2+\caja{1-s-s\,(1-\lambda)\,k\,(m^x)^{k-1}}^2}.
\end{eqnarray}

In this work we are only interested in the purely quantum transitions, not
in the thermodynamical ones. For this reason, and with the sake of
simplification, we remove the dependence of physical quantities on $\beta$ from now on by
considering the low-temperature limit,  $\beta\to\infty$.
In this limit, if $\caja{p\, s\,\lambda\,
  (m^z)^{p-1}}^2+\caja{1-s+s\,(1-\lambda)\,k\,(m^x)^{k-1}}^2\neq 0$, the
hyperbolic tangent in  \eqref{eq:mz} and \eqref{eq:mx} tends to unity,
and thus the self consistent equations simplify
\begin{eqnarray}\label{eq:mzbetainfty}
m^z&=&\frac{p\, s\,\lambda\, (m^z)^{p-1}}{\sqrt{\caja{p\, s\,\lambda\,
      (m^z)^{p-1}}^2+\caja{1-s-s\,(1-\lambda)\,k\,(m^x)^{k-1}}^2}},\\\label{eq:mxbetainfty}
m^x&=&\frac{1-s-s\,(1-\lambda)\,k\,(m^x)^{k-1}}{\sqrt{\caja{p\, s\,\lambda\,
      (m^z)^{p-1}}^2+\caja{1-s-s\,(1-\lambda)\,k\,(m^x)^{k-1}}^2}}.
\end{eqnarray}
The magnetization lies on the unit radius circumference, i.e.
$(m^x)^2+(m^z)^2=1$. This result agrees with the approach in section
\ref{sec:CA}, where the magnetization was a unit vector constrained to the
$XZ$ plane.  The pseudo free energy \eqref{eq:freebeta} becomes
\begin{eqnarray}\label{eq:free}\nonumber
&f(\beta,s,\lambda;m^z,m^x)=(p-1)\, s\,
    \lambda(m^z)^p-(k-1)\, s\,
      (1-\lambda)(m^x)^k&\\&-\sqrt{\caja{p\, s\,\lambda\, (m^z)^{p-1}}^2+\caja{1-s-s\,(1-\lambda)\,k\,(m^x)^{k-1}}^2}.&
\end{eqnarray}
Equations \eqref{eq:mzbetainfty} and \eqref{eq:mxbetainfty} have 
ferromagnetic (F) solutions with $m^z>0$ and quantum paramagnetic (QP) ones
satisfying $m^z=0$ and $m^x\neq0$. Let us begin with the latter case.
\section{Paramagnetic solutions}
Substituting $m^z=0$ in \eqref{eq:mxbetainfty}, we get
\begin{equation}\label{eq:mxQP}
m^x=\frac{1-s-k\,s\,(1-\lambda)(m^x)^{k-1}}{|1-s-k\,s\,(1-\lambda)(m^x)^{k-1}|},
\end{equation} 
which leads to $m^x=\pm 1$. The solution $m^x=-1$ is obtained if the
numerator in \eqref{eq:mxQP} is negative, that is, if $1-s-k\,s\,(1-\lambda)(-1)^{k-1}< 0$, which, in the range of parameters
$0\le s\le 1$ and $0\le \lambda\le 1$ considered, can only be satisfied if $k$ is 
  odd and in the region $1/[1+k(1-\lambda)]< s\le 1$. This phase is
precisely the $\mathrm{QM}^-$ phase discussed in the text. Its free energy is 
\begin{equation}\label{eq:freeQPneg}
  f_{\mathrm{QP}^{-}}(s,\lambda)=1-2s+s\lambda, \end{equation} which coincides
with equation \eqref{eq:fQPmca}.

The other quantum paramagnetic solution with $m^x=+1$ (the $\mathrm{QP}^+$
phase) can be satisfied only if the numerator is positive, i.e. if
$1-s-k\,s\,(1-\lambda)\geq 0$, which can be fulfilled for any value of $k$ as
long as $s$ lies in the region $0\leq s\leq1/[1+k(1-\lambda)]$. The free
energy of this phase is \begin{equation}\label{eq:feQP}
  f_{\mathrm{QP}^{+}}(s,\lambda)=-1+2s-s\lambda, \end{equation}
and is also equal to \eqref{eq:fQPpca}.

There is still one additional paramagnetic solution. In order to obtain it, we
need to come back to the discussion about the $\beta\to\infty$ limit.  The
hyperbolic tangent in  \eqref{eq:mz} and \eqref{eq:mx} could tend to a
finite value in the $\beta\to\infty$ limit, as long as the term in the square
root vanishes. Mathematically,\footnote{\label{fn:mxneg} In the $k$-odd case, the limit 
$$m^z\to0,\,\,m^x\to-\displaystyle\caja{\frac{1-s}{k\,s\,(1-\lambda)}}^\frac{1}{k-1}$$
also makes the square root in \eqref{eq:limtanh} vanish, but it leads to a positive
free energy in \eqref{eq:freeQP2}, and thus it is not relevant.}
\begin{equation}\label{eq:limtanh} \lim_{\beta\to\infty} \tanh \beta\sqrt{\caja{p\,
    s\,\lambda\, (m^z)^{p-1}}^2+\caja{1-s-s\,(1-\lambda)\,k\,(m^x)^{k-1}}^2}=
\tanh c,  \end{equation} when
\begin{eqnarray}\label{eq:limQP2}
m^z\to0,&m^x\to\displaystyle\caja{\frac{1-s}{k\,s\,(1-\lambda)}}^\frac{1}{k-1}.
\end{eqnarray}

In order to find a non-trivial solution, it is also necessary in this limit
that $m^z$ tends to zero faster than the bracketed term of $m^x$ in
\eqref{eq:mx}, i.e.  
\begin{equation}\label{eq:condQP2}
\frac{p\,s\,\lambda (m^z)^{p-1}}{1-s-k\,s\,(1-\lambda)(m^x)^{k-1}}\to 0. \end{equation}
Under these assumptions,  \eqref{eq:mz} and \eqref{eq:mx} imply $m^z=0$ and
$m^x=\tanh c$, where $\tanh c=[(1-s)/k\,s\,(1-\lambda)]^\frac{1}{k-1}$, in order to
be consistent with the limit \eqref{eq:limQP2}. This correspondence
determines the region in the space where this phase can appear. In fact, as any hyperbolic
tangent, $|\tanh c|\le 1$, which is  true only if $1/[1+k\,(1-\lambda)]\le
s\le 1$. Besides, the condition \eqref{eq:condQP2} forces $p>3$.\footnote{
Indeed, using $(m^x)^2+(m^z)^2=\tanh^2
c=[(1-s)/k\,s\,(1-\lambda)]^\frac{2}{k-1}$ and computing the limit
\eqref{eq:limQP2} when $m^x\to\tanh c$, one can check that it vanishes only  as
long as $p>3$.
}

Since the magnetization in the $z$ direction vanishes, we call this phase
QP2. The free energy is obtained with  \eqref{eq:freebeta},
\begin{equation}\label{eq:freeQP2}
f_{\mathrm{QP2}}(s,\lambda)=-\frac{k-1}{k}\caja{\frac{1-s}{k\,s\,(1-\lambda)}}^\frac{1}{k-1}(1-s).
\end{equation}

This last phase was not predicted by the semi classical approach. However, we will
see below that it is irrelevant to the problem, since the F' phase has
always a smaller value of the free energy.

\section{Ferromagnetic solutions}
We next consider the possible solutions with $m^z>0$.\footnote{No negative
  value for $m^z$ can satisfy \eqref{eq:mzbetainfty} for odd values of $p$.}
As before, the ferromagnetic solutions cannot be computed explicitly for a
given value of $p$ but for certain limiting cases.

The solution $m^z=1$ (and $m^x=0$) is  exact only
on the line $s=1$. However, we can see that an approximate solution
$m^z\approx 1$ and $m^x\approx 0$ is valid in a wider space of parameters.
Indeed, the solution
\begin{eqnarray}
m^x=\frac{1-s}{s\,p\,\lambda},&\,\mathrm{and}\,\,&m^z=\sqrt{1-\paren{\frac{1-s}{s\,p\,\lambda}}^2}
\end{eqnarray}
fulfills  \eqref{eq:mzbetainfty} and \eqref{eq:mxbetainfty} when
$(1-s)/p\,s\,\lambda\to 0$. This is the F phase we obtained before in equation
\eqref{eq:fFca}. The free energy is obtained plugging these values into equation
 \eqref{eq:free}. For the $p\to\infty$ limit,
\begin{equation}\label{eq:freeFinfty}
f_\mathrm{F}(s,\lambda)|_{p\to\infty}=-s\lambda.  \end{equation}

We consider an alternative solution for $0<m^z<1$.  With this aim, we
rewrite  \eqref{eq:mzbetainfty} in the following way \begin{equation}\label{eq:mzsq}
\caja{(m^z)^2-1}\caja{p\,s\,\lambda
  (m^z)^{p-1}}^2+\lazo{m^z\caja{1-s-s(1-\lambda)k(m^x)^{k-1}}}^2=0.  \end{equation}
In the $p\to\infty$ limit, $p(m^z)^{p-1}\to
0$, and 
\begin{eqnarray}\label{eq:solFp}
m^x=\caja{\frac{1-s}{k\,s\,(1-\lambda)}}^\frac{1}{k-1}& m^z=\sqrt{1-\caja{\frac{1-s}{k\,s\,(1-\lambda)}}^\frac{2}{k-1}}
\end{eqnarray}
is an exact solution to  \eqref{eq:mzsq}, and similarly of
 \eqref{eq:mxbetainfty}, as long as $(1-s)/k\,s\,(1-\lambda)\!<\!s\!\le 1$, or
$1/[1+k\,(1-\lambda)]< s< 1$.\footnote{Again,  the negative solution for $m^x$ is also a valid solution in
  the odd $k$ case but has a higher free energy than \eqref{eq:solFp} due to the change of sign in
  the $(m^x)^k$ term in  \eqref{eq:free}.} This is precisely the F' phase
discussed in section \ref{sec:CA}.
Again, we compute the free energy by plugging the solution
\eqref{eq:solFp} in \eqref{eq:free} and taking the $p\to\infty$ limit
\begin{equation}\label{eq:freeFpinfty}
\left. f_{\mathrm{F`}}(s,\lambda)\right|_{p\to\infty}=-\frac{k-1}{k}\caja{\frac{1-s}{k\,s\,(1-\lambda)}}^\frac{1}{k-1}(1-s),
\end{equation}
which is exactly equal to the one obtained for the QP2
phase \eqref{eq:freeQP2}. 

The solution \eqref{eq:solFp} is also a good approximate solution for $p$
finite (but $p>3$) when $(m^z)^p\to 0$. 
The free energy for this phase is
\begin{equation}\label{eq:freeFp}
 f_{\mathrm{F`}}(s,\lambda)\approx-s\,\lambda\caja{1-\paren{\frac{1-s}{s\,k\,(1-\lambda)}}^{\frac{2}{k-1}}}^{p}-\frac{k-1}{k}\caja{\frac{1-s}{k\,s\,(1-\lambda)}}^\frac{1}{k-1}(1-s),\end{equation}
which, for finite $p$, is always smaller than $f_\mathrm{QP2}$. According to
this observation, except for the $p\to\infty$ limit, the F' phase is always stabler than the QP2
phase. 

We have therefore reproduced the results of section \ref{sec:CA} by a
completely different method. The present method is nevertheless better suited
for generalizations to more complicate problems where the target Hamiltonian
$\hat{H}_0$ cannot be expressed in terms of simple total spins.

\chapter[Ground state of $\hat V_k$]{Ground state of $\hat V_k$ and its overlap with the ground state of $\hat{H}_0$}\label{sec:Vkground}
In this Appendix, we derive the properties of the ground state of
$\hat{V}_{k}$, defined in~\eqref{eq:Vk}, for
$k$ even.
Let us first consider the case with $N$ even.
The ground state of $\hat{H}_0$, $\ket{\phi_0}=\otimes_{i=1}^N\ket{\uparrow}_i^z$,
can be expressed as
\begin{eqnarray}
\ket{\phi_0}&=&\otimes_{i=1}^N\big(\ket{\uparrow}_i^x+\ket{\downarrow}_i^x\big)/\sqrt{2}\nonumber\\
&=&\frac{1}{2^{N/2}}\Big(\ket{\uparrow}_1^x\ket{\uparrow}_2^x\cdots \ket{\uparrow}_N^x
+\ket{\uparrow}_1^x\ket{\uparrow}_2^x\cdots \ket{\uparrow}_{N-1}^x\ket{\downarrow}_N^x\nonumber\\
&&+\cdots +\ket{\downarrow}_1^x\ket{\downarrow}_2^x\cdots \ket{\downarrow}_N^x \Big).
\end{eqnarray}
This last expression has $2^N$ terms, in which the partial sum of terms with a half of the sites
having $\ket{\uparrow}_i^x$ and the other half $\ket{\downarrow}_i^x$ is nothing but the
ground state of $\hat{V}_k$ in the $S=N/2$ sector $\ket{\phi_k}$, up to a normalization,
\begin{eqnarray}
\ket{\phi_k}&=&a \Big(\ket{\uparrow}_1^x\ket{\uparrow}_2^x\cdots\ket{\uparrow}_{N/2}^x
 \ket{\downarrow}_{N/2+1}^x\cdots\ket{\downarrow}_N^x\nonumber\\
&&+\cdots +\ket{\downarrow}_1^x\ket{\downarrow}_2^x\cdots \ket{\downarrow}_{N/2}^x\ket{\uparrow}_{N/2+1}^x
\cdots\ket{\uparrow}_N^x \Big).
\end{eqnarray}
It is easy to check from the number of terms in the above equation
that the normalization condition is $a^2\displaystyle{N \choose N/2}=1$.
We thus have
\be
\left\langle\phi_0|\phi_k\right\rangle=\frac{a}{2^{N/2}}{N\choose N/2}
=\frac{1}{2^{N/2}}\sqrt{{N\choose N/2}}.
\ee
For large $N$, 
\be
\log
|\left\langle\phi_0|\phi_k\right\rangle |^2=\log\caja{2^{-N}\frac{N!}{\paren{\frac{N}{2}!}^2}}\approx
-\frac{1}{2}\log N +\log\sqrt{\frac{2}{\pi}},
\ee
which means that the overlap decreases only polynomially with $N$ as $\sim
N^{-1/2}$.

The case of odd $N$ can be analyzed similarly but in this case $\displaystyle
{N\choose N/2}$ is replaced
by  $\displaystyle {N\choose (N+1)/2}$ or $\displaystyle {N\choose (N-1)/2}$.



\cleardoublepage
\phantomsection
\nolinenumbers
\addcontentsline{toc}{chapter}{\protect\numberline{}Bibliography}
{\small\bibliography{biblio}}
\bibliographystyle{tesisalphnum}



\end{document}